\documentclass[12pt,dlspace,a4paper]{report}%
\usepackage{a4}
\usepackage{lscape}
\usepackage{amsfonts}
\usepackage{eepic}
\usepackage{ulem}
\usepackage{epic}
\usepackage{graphicx,epsfig}
\usepackage{texdraw}
\usepackage{young}
\usepackage{bm}
\usepackage{dcolumn}
\usepackage{epsfig}
\usepackage{color}
\usepackage{cite}
\usepackage{fancyhdr}
\usepackage{longtable}
\usepackage{amssymb}
\usepackage{amsmath}
\usepackage{setspace}
\newcommand{\be}{\begin{equation}}
\newcommand{\ee}{\end{equation}}
\newcommand{\ba}{\begin{eqnarray}}
\newcommand{\ea}{\end{eqnarray}}
\newcommand{\nn}{\nonumber\\ }

\begin{document}
\baselineskip 20pt
\pagenumbering{roman}
\abovedisplayskip=0mm
\belowdisplayskip=5mm
\abovedisplayshortskip=0mm   
\belowdisplayskip=5mm
\abovedisplayshortskip=0mm
\belowdisplayshortskip=5mm  
\parindent=10mm
\parskip=0.25cm
%
\begin{titlepage}
\begin{center}
\vspace{3cm}
{\bf \Large Sur les Corrections de la G\'eom\'etrie Thermodynamique des Trous Noirs}\\
\vspace{2.5cm}
{\bf Bhupendra Nath Tiwari}\\
\vspace{1.5cm} 
\textbf{Mots-cl\'{e}s:}\\ \textit{Les g\'eom\'etries thermodynamiques,
la th\'eorie de la gravit\'e des d\'eriv\'ees sup\'erieures,
le principe d'incertitude g\'en\'eralis\'ee,
la physique des trous noirs et des branes noirs.} \\
\vspace{0.5cm} 
\textit{\textbf{PACS:} 
04.70.-s, 04.70.Dy, 11.25.-w}
\end{center}
\end{titlepage}

\clearpage







\thispagestyle{plain}
\thispagestyle{empty}
\newpage  {\color{white} . }

\clearpage
\renewcommand{\contentsname}{Table des Mati\`eres}
\tableofcontents
\clearpage
\thispagestyle{empty}
\vspace*{2.0in}
\begin{center}
{\huge{\it \`{A} mes parents et mes professeurs}}
\end{center}
\clearpage
\clearpage







\thispagestyle{plain}
\thispagestyle{empty}
\newpage  {\color{white} . }

\clearpage

\pagestyle{plain}
\begin{center}
\underline{{\Large {\bf Remerciements}}}
\end{center}

Je remercie le professeur Ashoke Sen pour les discussions utiles 
sur la fonction de l'entropie et l'entropie des trous noirs 
au cours de $\langle\langle$ String School- 2006 $\rangle\rangle$ \`a 
l'$\langle\langle$ Institute of Physics, Bhubaneswar, India $\rangle\rangle$
et les issues de l'AdS/CFT lors de sa visite \`a 
l'$\langle\langle$ IIT-Kanpur-2008 $\rangle\rangle$.

Je suis reconnaissant au professeur Jean de Boer pour les discussions
de l'AdS/CFT et pour les calculs de l'entropie microscopique de 
certains branes noirs en $\langle\langle$ String School- 2006 $\rangle\rangle$ 
\`a l'$\langle\langle$ Institute of Physics, Bhubaneswar, India $\rangle\rangle $ 
et \`a l'$\langle\langle$ ICTP Spring String School-2007 $\rangle\rangle $ 
au ICTP, Italie.

Je souhaite remercier Yogesh K. Srivastava pour les int\'eressantes discussions
sur l'entropie microscopique de trous noirs et branes noirs au cours du
$\langle\langle$ Indian String Meeting-2007, Harish-Chandra Research Institute,
Allahabad, India $\rangle\rangle $.

Je voudrais remercier professeur Amandine Almarche, Vincent R\'egnier,
Vinod Chandra, professeur Shubha Karnick et Arjun Basu pour l'aide de la 
langue fran\c{c}aise ce qu'ils m'ont apport\'ee pendant la pr\'eparation 
de ce travail.

Je remercie \'egalement le $\langle\langle$ CSIR- New Delhi, India
$\rangle\rangle $ pour la bourse de recherche au titre des subventions:
$\langle\langle$ CSIR-SRF-9/92(343)/2004-EMR-I $\rangle\rangle $.\\


\clearpage







\thispagestyle{plain}
\thispagestyle{empty}
\newpage  {\color{white} . }

\clearpage

\thispagestyle{plain}
\begin{center}
\large{{\bf AVANT PROPOS}}
\end{center}

Nous \'{e}tudions la g\'{e}om\'{e}trie thermodynamique des certains trous noirs 
et branes noirs avec ou sans les corrections du principe d'incertitude g\'{e}n\'{e}ralis\'{e}e 
et les corrections de $ \alpha^{\prime} $ \`{a} l'entropie. Dans cette perspective, 
nous analysons la g\'{e}om\'{e}trie Ruppenier des trous noirs de Reissner-Nordstr\"om et 
demontrons que cella est bien d\'{e}finie et correspond \`{a} un syst\`{e}me sans interaction statistique globale.  
Dans la domaine de Poincar\'{e} de $AdS_4$, nous \'{e}tudions les propi\'et\'es de la g\'{e}om\'{e}trie  
Weinhold des trous noirs dilatoniques, et montrons que cella est partout r\'{e}guli\`{e}re 
que des patches de la grande masse des trous noirs de Reissner-Nordstr\"om, qui contient des certains nombres restreints, 
et donc la configuration sous-jacente de ces trous noirs est thermodynamiquement instable dans une r\'{e}gion de l'espace d'\'{e}tat. 
Nous obtenons que la g\'{e}om\'{e}trie de Ruppenier corrig\'{e}e par le principe d'incertitude g\'{e}n\'{e}ralis\'{e}e 
des trous noirs de Reissner-Nordstr\"om correspond \`{a} un syst\`{e}me statistique sans interactions, 
et cella est au contraire des propri\'{e}t\'{e}s d'un ensemble des trous noirs magn\'{e}tiques. 
Nous montrons que les corrections de $ \alpha^{\prime} $ de la th\'{e}orie des cordes 
n'introduisent pas de singularit\'{e} dans la g\'{e}om\'{e}trie de l'espace d'\'{e}tat
des trous noirs non-supersym\'{e}triques extr\'{e}maux en dimensions $ D = 4 $. Int\'{e}ressement, 
le degr\'{e} de la courbure scalaire et ceci du d\'eterminant de la g\'{e}om\'{e}trie de 
Ruppenier peut \^{e}tre \'{e}crit comme un entier multiple de l'ordre des corrections 
de $ \alpha^{\prime} $. De plus, nous montrons que la g\'{e}om\'{e}trie 
d'espace d'\'{e}tat des trous noirs supersym\'{e}triques extr\'{e}maux est partout 
r\'{e}guli\`{e}re et corrig\'{e}e par des corrections de Gauss-Bonnet en dimensions $ D = 4$, 
et ainsi que celle des branes noirs non-extr\'{e}maux $ D_1D_5 $ et $ D_2D_6NS_5$ corrig\'ee 
par les corrections de $ \alpha^{\prime} $ en dimensions $ D = 10 $. En outre, aux trous noirs en rotation
en quatre dimensions, la g\'{e}om\'{e}trie thermodynamique des trous noirs extr\'{e}maux de Kerr-Newman 
dans la th\'{e}orie d'Einstein-Maxwell est partout mal d\'{e}finie, et celle des trous noirs de Kaluza-Klein 
dans la th\'{e}orie d'Einstein-Maxwell ou bien des trous noirs obtenus par la compactification 
de la th\'eorie des cordes h\'et\'erotiques est mal d\'{e}finie seulement aux succursales ergos.


\clearpage







\thispagestyle{plain}
\thispagestyle{empty}
\newpage  {\color{white} . }

\clearpage

\thispagestyle{plain}
\begin{center}
\large{{\bf Abstract}}
\end{center}

We study thermodynamic geometry of certain black holes
and black branes with and without generalized uncertainty principle 
or stringy $ \alpha^{\prime} $-corrections to the entropy.
From this perspective, we analyze Ruppenier geometry of Reissner-Nordstr\"om black holes
and show that it is well defined and corresponds to a non-interacting statistical system.
We investigate that the Weinhold geometry of dilatonic black holes
is regular everywhere and that of large mass Reissner-Nordstr\"om black
holes in the Poincar\'e patch of $ AdS_4 $ contains certain narrow range
of thermodynamically unstable regions in the state-space.
We obtain that the generalized uncertainty principle corrected
Ruppenier geometry of Reissner-Nordstr\"om black holes correspond to 
a non-interacting statistical system unlike the magnetically charged black holes.
We show that the stringy $ \alpha^{\prime} $-corrections do not
introduce singularity in the statespace geometry of non-supersymmetric
extremal black holes in $ D= 4 $.
Interestingly, the degree of scalar curvature and that of the determinant
of this Ruppenier geometry can be written as an integer multiple of the
order of $ \alpha^{\prime} $-correction.
We further show that the statespace geometry of Gauss-Bonnet
corrected supersymmetric extremal black holes in $ D=4 $ as well as
that of the non-extremal $D_1D_5$ and $D_2D_6NS_5$ black branes in $ D=10 $ 
under the $\alpha^{\prime} $-correction is regular everywhere.
Furthermore, the thermodynamic geometry of four dimensional rotating
Kerr-Newman extremal black holes in Einstein-Maxwell theory is everywhere
ill-defined and that of the Kaluza-Klein black holes in Einstein-Maxwell
theory or the one arrising from the heterotic string compactification is
ill-defined only at the points of ergo-branch.

\vspace{0.50cm}
{\bf Keywords}:{ Thermodynamic Geometries, Higher Derivative Gravity,
Generalized Uncertainty Principle, Black Hole Physics. }

{{\bf PACS numbers}:
04.70.-s: Physics of black holes;
04.70.Dy: Quantum aspects of black holes, evaporation, thermodynamics;
11.25.-w: Strings and branes. }

\clearpage
\clearpage







\thispagestyle{plain}
\thispagestyle{empty}
\newpage  {\color{white} . }

\clearpage
\renewcommand{\listfigurename}{Liste des Figures}
\listoffigures
\cleardoublepage
\renewcommand{\chaptername}{Chapitre}
\pagestyle{fancy}
\pagestyle{fancyplain}
\lhead[\fancyplain{}{\sf\thepage}]
{\fancyplain{}{\sf\rightmark}}
\rhead[\fancyplain{}{\sf\leftmark}]
{\fancyplain{}{\sf\thepage}}
\cfoot{}
\cfoot[\fancyplain{\rm\thepage}{}]{\fancyplain{\rm\thepage}{}}
\pagenumbering{arabic}
\pagestyle{fancyplain}
\clearpage
\chapter{Introduction}

Motiv\'es par la m\'ethode de la fonction de l'entropie de Sen
et le principe d'incertitude g\'en\'eralis\'ee,
nous examinons la g\'eom\'etrie thermodynamique associ\'ee
\`a l'entropie ou \`a la masse des diff\'erents trous noirs ou branes noirs.
Nous avons consid\'er\'e une s\'erie de syst\`emes de trous noirs et
analysons leur g\'eom\'etrie thermodynamique.
Notre perspective de l'\'etude g\'eom\'etrique est divis\'{e}e en deux parti\'es:
la premi\`ere est les corrections dans l'entropie des trous noirs 
en raison du principe d'incertitude g\'en\'eralis\'ee et
la seconde est les corrections sup\'erieures \`a l'entropie des
trous noirs par la m\'ethode de la fonction de l'entropie de Sen \cite{Sen1}.


Microscopiquement, les propri\'et\'es thermodynamiques, du trou noir de la 
th\'eorie des cordes, sont \'el\'egamment r\'esum\'ees par l'assignment d'une 
entropie du trou noir \cite{Maldacena}.
Il existe une structure riche dans le cadre des trous noirs demi-BPS
dans les th\'eories des supercordes en $\mathcal N = 2 $ et $ D \geq 4 $,
par le biais de certaines compactifications des dimensions sup\'erieures
des th\'eories des supercordes.
En raison de la compactification, certains champs scalaires apparraissent
dans la th\'eorie, avec des valeurs proche de l'horizon sont d\'etermin\'ees
uniquement par les charges port\'ees par le trou noir.
Les valeurs proches de l'horizon sont donc ind\'ependantes de leur
valeur asymptotique.
Ces champs scalaires constituent l'espace de modules scalaires sur lesquels
l'entropie des trous noirs est ind\'ependante \cite{Sen1, Far1, Str1, Str2, Str3}.
Ces trous noirs, des supercordes compactifi\'ees, ont le pouvoir d'\^etre
li\'es au syst\`eme dynamique par ce m\'ecanisme attracteur.
C'est-\`a-dire les \'equations de la structure complexe appel\'ees
les \'equations d'attracteur pour les charges, ont certaines conditions
sur les structures d'Hodge de la vari\'et\'e complexe,
particuli\`erement avec le tore $ T_6 $ ou la vari\'et\'e de Calabi-Yau.

En revanche, l'entropie des diff\'erents trous noirs d\'epend des termes
d\'eriv\'es sup\'erieures se figurant dans le prepotentiel g\'en\'eralis\'e
\cite{Car2, Car3, Car8, Car11}.
De plus, dans certain cas, la partie r\'eelle de l'espace des modules scalaires 
des multiples vecteurs est proportionnelle aux champs magn\'etiques,
alors que la partie imaginaire est proportionnelle aux 
champs \'electriques \`a l'horizon du trou noir \cite{Sen1, Car11}.
Cette propri\'et\'e de d\'ependance de l'entropie des trous noirs
des termes d\'eriv\'es sup\'erieures peut \^etre cod\'ee dans le
nombre de la deuxi\`eme classe de Chern de l'espace topologique
sous-jacent sur lequel la th\'eorie des supercordes est compactifi\'ee.
En outre, \`{a} proximit\'{e} de l'horizon des trous noirs de la th\'eorie
des supercordes de $\mathcal N = 2$, il est bien connu que tous les termes 
de la densit\'e langrangianne s'\'eclipsent, \`a l'exception d'un seul terme 
proportionnel \`a la partie imaginaire du prepotentiel g\'en\'eralis\'e.
L'une des plus int\'eressantes corrections de la densit\'e lagransianne 
dans le cas de la courbure carr\'e de l'espace-temps \`a l'aire de l'horizon
des trous noirs, est le terme: $ 4 \pi Im(\Upsilon F_{\Upsilon}) $ 
\cite{Car2, Car3, Car8, Car11}.


Comme il est d\'esormais bien connu, la m\'ethode de la fonction de Sen
de l'entropie est la meilleure m\'ethode pour calculer les contributions
de $ \alpha^{\prime}$ des d\'eriv\'ees sup\'erieures d'une classe
de trous noirs d\'ecoulant des th\'eories des cordes.
Un exemple de base de notre compr\'ehension du m\'ecanisme attracteur
et de l'entropie d'un trou noir extr\'emal avec les charges \'electriques et
magn\'etiques, est la solution de Reissner-Nordstr\"om.
Cette solution d\'ecrit un trou noir charg\'e et sph\'eriquement sym\'etrique,
dans les quatre dimensions de la th\'eorie d'Einstein-Maxwell.
Il s'agit d'une solution classique exacte pour toute distance finie  $r$
qui d\'ecrit une sph\`ere ordinaire $S^2$ de deux dimensions et un
espace-temps bidimensionel connu $AdS_2$.
Donc, cette solution d\'ecrit \'egalement la gravit\'e d'Einstein en deux
dimensions avec une valeur n\'egative de la constante de cosmologie.

De plus, cette situation a une  isom\'etrie de $SO(3)$ agissant sur la
sph\`ere $S^2$, qui refl\`ete la sym\'etrie sph\'erique du trou noir
original et est pr\'esente \'egalement dans la solution complete de ce
trou noir.
Elle a \'egalement une isom\'etrie de $SO(2,1)$ agissant sur l'espace
de $AdS_2$, alors qu'elle n'\'etait pas pr\'esente initialement dans la
solution complete de ce trou noir.
Dans ce cas, il est facile de montrer que la m\'etrique et les champs de 
jauge peuvent \^etre \'ecrits invariablement par la transformation de 
$SO(2,1) \times SO(3)$.
En fait, cette mani\`{e}re de d\'efinir un trou noir extr\'emal fonctionne
en g\'en\'eral bien avec les d\'eriv\'ees sup\'erieures de la th\'eorie de 
la gravit\'e.
En particulier, $\langle\langle$ dans toute la th\'eorie de la gravit\'e
g\'en\'eralement covariante et conjugu\'ee aux champs de la mati\`ere,
la g\'eom\'etrie proche de l'horizon d'un trou noir extr\'emal pr\'{e}sentent
une sym\'etrie de la sph\`ere en quatre dimensions a l'isom\'etrie de
$SO(2,1) \times SO(3) \rangle\rangle$ \cite{AshokeSen, KunduriLuciettiReall}.


En outre, l'entropie des trous noirs charg\'es extr\'emaux et en rotation est bien connue
dans la th\'eorie des cordes h\'et\'erotiques depuis tr\`es longtemps \cite{CveticYoum}.
Sen et. al. consid\`erent que dans la th\'eorie g\'en\'erale de la gravit\'e des
d\'eriv\'ees sup\'erieures, qui est conjugu\'ee aux champs de jauge et aux champs
scalaires neutres, l'entropie ainisi que l'arri\`ere-plan proche de l'horizon
d'un trou noir extr\'emal en rotation, peuvent \^etre d\'etermin\'es
par extremisation de la fonction d'entropie de Sen.
Actualement, il ne d\'epend que des param\`etres de carat\'erisation de
l'horizon comme par exemple, les charges \'electriques, les charges
magn\'etiques et le moment cin\'etique du trou noir.
Toutes les solutions de trous noirs extr\'emaux, comme dans le cas du
trou noir extr\'emal de Kerr-Newmann ou celui de Kaluza-Klein dans la
th\'eorie d'Einstein-Maxwell, ou bien aussi des trous noirs extr\'emaux
d\'ecoulant de la th\'eorie des cordes h\'et\'erotiques compactifi\'ee
toroidalement, ont \'egalement deux types diff\'erents de limites
extr\'emales que l'on appelle la branche d'ergonomie et la branche
d'ergonomie libre.
Dans la limite extr\'emale correspondante \`a la branche d'ergonomie,
l'expression de l'entropie du trou noir extr\'emal de la th\'eorie des
cordes h\'et\'erotiques compactifi\'ee toroidalement peut \^etre obtenue 
par la m\'ethode de la fonction de l'entropie de Sen qui est donn\'ee par

\ba S(P_1, Q_2, P_3, Q_4, J):= 2 \pi \sqrt{J^2+ P_1 Q_2 P_3 Q_4},\ea

o\`u $P_1, P_3 $ et $Q_2, Q_4 $ sont respectivement des charges 
magn\'etiques et des charges \'electriques dans la th\'eorie des 
cordes h\'et\'erotiques tronqu\'ee \cite{Sen0606244}.
Il est \'egalement bien connu que cette entropie est invariante 
par la transformation de la dualit\'e dans le cadre d'une transformation 
de $SO(2,2)$ pour les vecteurs des charges \'electriques et magn\'etiques 
caract\'{e}risant la solution du trou noir.


Nous pourrions bien s\^ur refaire le m\^eme genre de calcul dans le cas des
trous noirs ou branes noirs non-extr\'emaux.
Pour cela, on a d\'emontr\'e que le formalisme de la fonction de l'entropie
fonctionne de Sen bien pour certains cas sp\'eciaux des trous noirs non-extr\'emaux et
\'egalement pour des branes noirs non-extr\'emaux, malgr\'e le fait que l'horizon
de ces branes n'est pas attracteur.
Selon l'explication du m\'ecanisme d'attracteur consid\'er\'e par 
Kallosh et. al. \cite{Kallosh}, la distance physique \`a partir d'un point
arbitraire de l'horizon attracteur est infinie.
Explicitement, au niveau de la supergravit\'e, cette distance propre d'un
point arbitraire de l'horizon est finie ou infinie, selon le cas des trous
noirs consid\'{e}r\'{e}s comme un trou noir non-extr\'emal ou trou noir extr\'emal.
Par exemple, on a d\'emontr\'e que la fonction de l'entropie de Sen a un extremum
proche de l'horizon d'un trou noir extr\'emal \cite{GarousiGhodsi1, CaiPang}.
C'est-\`a-dire, le formalisme de la fonction de l'entropie de Sen ne doit pas \^etre
quelque chose de sp\'ecifique pour les trous noirs extr\'emaux.
On a aussi sp\'ecul\'e que le formalisme de la fonction de l'entropie de Sen est
utilisable pour les trous noirs/ branes non-extr\'emaux, dont les g\'eom\'etries
proches de l'horizon sont des extensions de l'espace d'AdS, comme le trou noir
de Schwarzschild dans AdS \cite{GarousiGhodsi2}.
Maintenant, compte tenu des corrections sp\'ecifiques des d\'eriv\'ees plus
\'elev\'ees des contributions de la courbure tenseure de Weyl, comme les corrections 
des termes d\'eriv\'es sup\'erieurs de la th\'eorie effective.
Ensuite, pour les cas des termes d\'eriv\'es plus \'elev\'ees qui respectent
la sym\'etrie des solutions au niveau de l'arbre, l'entropie de ces syst\`emes 
de branes est donn\'ee par la valeur de la fonction de l'entropie aux extremums.
En outre, les corrections de $(\alpha^{\prime})^{3} $ n'ont pas d'effets sur la 
temp\'erature du syst\`eme thermodynamique sous-jacente,	
mais elles diminuent la valeur de l'entropie 
\cite{GarousiGhodsi1, FotopoulosTaylor}.
En fait, Pour incorporer les corrections de $\alpha^{\prime}$ \`a l'entropie 
d'un trou noir ou brane noir, la m\'ethode de la fonction de l'entropie de Sen 
est une des techniques les plus efficaces.
Cela n'exige pas que la fonction de l'entropie de Sen doit avoir un minimum 
local \`{a} proximit\'{e} de l'horizon.

D'autre part, afin de d\'ecrire les structures de l'espace-temps \`{a}
petite \'echelle de mani\`{e}re ad\'equate, 
une extension de la m\'ecanique quantique pourrait \^etre n\'ecessaire.
Et afin de tenir compte de la gravit\'e, 
nous avons besoin de modifier la g\'eom\'etrie classique continue,
voir par exemple la g\'eom\'etrie non-commutative de Connes \cite{AlainConnes}.
Et bien, le principe d'incertitude g\'en\'eralis\'ee peut \^etre analys\'e
\`a travers les concepts de base de la limite et de la transformation de Fourier.
Ensuite, la gravit\'e quantique ou la th\'eorie des cordes peut \^etre 
\'etudi\'ee dans la perspective d'une fonction complexe, avec certaines 
modifications des conditions quantificatives dans la th\'eorie quantique.
En particulier, on peut d\'ecrire le principe d'incertitude g\'en\'eralis\'ee
de la th\'eorie des cordes aux conditions d'analyticit\'e d'une certaine 
fonction complexe selon le m\'elange d'UV/IR \cite{BNTiwari}.
Cette consid\'eration est fond\'ee sur le fait que l'\'echelle de Planck est la 
longueur minimale de la nature; ainsi, il existe une longueur maximale de la nature.

En outre, l'existence des sym\'etries de la dualit\'e non-perturbative de la 
th\'eorie des cordes indique que les th\'eories des cordes ne distinguent pas
les petites \'echelles de l'espace-temps \`a partir des grandes \'echelles de 
l'espace-temps.
Cela n\'ecessite une modification du principe d'incertitude 
d'Heisenberg, comme par exemple pour les \'energies au-del\`a de l'\'echelle de 
Planck, la taille de la corde grandit avec le temps au lieu de chuter.
\`A la suite de la th\'eorie des cordes, une introduction sur cette description de 
l'espace-temps T-duale est donn\'ee par Witten \cite{EdwardWitten96, EdwardWitten97}, 
o\`u en dessous de la longueur de Planck, le concept m\^eme de l'espace-temps change son 
sens et le principe d'incertitude d'Heisenberg a besoin d'\^etre modifi\'e.
Le principe d'incertitude g\'en\'eralis\'ee est aussi motiv\'e par
l'\'etude du comportement pour les petites distances de la th\'eorie des cordes
\cite{GrossMende, KonishiPautiProvero, AmatiCiafaloniVeneziano, Veneziano}, 
la physique des trous noirs \cite{Maggiore} et les espaces de de-Sitter \cite{Snyder}.
On peut r\'ev\'eler des indices thermodynamiquement importants 
avec les corrections existentes dans la nature mais aussi l'origine 
g\'eom\'etrique de la M-th\'eorie fondamentale 
\cite{EdwardWitten92, EdwardWitten93}.


De plus, l'analyse des perturbations lin\'earis\'ees des trous noirs de 
Reissner-Nordstr\"om du grand anti-de Sitter en quatre dimensions est 
importante pour avoir la dichotomie de la physique des trous noirs, 
comme l'instabilit\'e thermodynamique. 
Par exemple, au cours des derni\`eres ann\'ees, on a expos\'e l'existence 
de certains modes tachyoniques de ces trous noirs \cite{GubserMitra1, GubserMitra2}.
En outre, dans la limite de grands trous noirs, il existe un \'ecart, et 
lorsque ce trou noir devient thermodynamiquement instable, le tachyon 
appara\^it dans le grand espace d'anti-de Sitter.
Il y a eu des progr\`es remarquables dans la compr\'ehension de la 
m\'ecanique statistique microscopique compte tenu de la thermodynamique 
des trous noirs, en utilisant les constructions de la th\'eorie des 
cordes comme les D-branes \cite{StromingerVafa}.
Il y a une r\`egle g\'en\'erale pour les solutions des trous noirs 
quasi-extr\'emaux obtenus par les compactifications aux dimensions $D= 4, 5$ 
de la th\'eorie des cordes avec plusieurs charges \'electriques, 
magn\'etiques et une masse saturant presque la limite de BPS, 
qui ont sans exception une chaleur sp\'ecifique positive.
C'est parce que la m\'ecanique statistique de l'entropie repose sur 
une th\'eorie des champs des D-branes de basse \'energie \`a partir 
de laquelle les trous noirs sont construits \cite{AharonyGubserMaldacenaOoguriOz, 
Witten, BoerNaqvi, BoerPasquinucciSkenderis, Mathur, MathurSaxenaSrivastava}.

Dans le prolongement de ces trous noirs les plus pertinentes dans 
la th\'eorie des cordes commes les trous noirs d'astrophysiques, 
une \'etape naturelle examiner les variantes des trous noirs 
thermodynamiquement instable pour lesquels la th\'eorie des 
cordes donne une description duale de la th\'eorie des champs conformes.
Le plus simple exemple d'une telle solution est la solution de 
Reissner-Nordstr\"om dans l'espace d'AdS. Cette solution d\'emontre 
son instabilit\'ee thermodynamique et que la solution est instable dans 
une analyse lin\'earis\'ee \cite{Gubser}.
On a aussi conjectur\'e dans le pass\'e qu'il existe une relation g\'en\'erale 
entre l'instabilit\'e thermodynamique et l'instabilit\'e de Gregory-Laflamme 
pour les branes noirs \cite{CsakiOoguriOzTerning, GregoryLaflamme}.
Voir pour plus de d\'etails dans le cas de l'\'evolution des trous noirs 
instables dans l'espace d'anti-de Sitter \cite{GubserMitra1, GubserMitra2}.


Dans l'\'etude g\'eom\'etrique de la thermodynamique, 
il existe deux importantes types de g\'eom\'etries thermodynamiques. 
L'une dans la representaion de l'entropie qui s'appelle la
g\'eom\'etrie de Ruppenier et l'autre dans la repr\'esentation 
de la masse qui est dite la  g\'eom\'etrie de Wienhold.
En fait, il est bien connu que ces deux  g\'eom\'etries thermodynamiques 
sont li\'ees par une transformation conforme avec le facteur conforme 
\`a la temp\'erature du syst\`eme consid\'er\'e \cite{Rupp, Aman}.
De cette mani\`{e}re, nous pouvons calculer la m\'etrique de la g\'eom\'etrie 
thermodynamique dans n'importe quelle repr\'esentation et puis l'obtenir 
dans l'autre repr\'esentation seulement en tenant compte d'un facteur, 
comme en ce cas de la temp\'erature.	
C'est-\`a-dire que les enqu\^etes d'une g\'eom\'etrie sont eqiuvalentes \`a 
celles de l'autre.	
Donc, nous pouvons obtenir la m\'etrique d'une g\'eom\'etrie thermodynamique
par une autre d\'ej\`a connue, et ainsi calculer facilement les quantit\'es 
g\'eom\'etriques dans la repr\'esentation souhait\'ee.
C'est pourquoi dans la plupart des cas de notre \'etude, 
nous avons analys\'e le r\^ole des corrections de la g\'eom\'etrie 
thermodynamique que nous avons examin\'{e}, soit pour la g\'eom\'etrie 
de Ruppenier, soit pour la g\'eom\'etrie de Wienhold.

Dans le Ref. \cite{Sar1}, nous avons d\'ej\`a analys\'e la 
g\'eom\'etrie thermodynamique des trous noirs de BTZ.
Nous avons montr\'e que l'espace d'\'etat n'a pas d'interactions
thermodynamiques et la courbure scalaire de Ruppenier est partout nulle,
et ceci reste \'egalement le cas avec les corrections de Chern-Simons.
De plus, les interactions thermodynamiques sont finies et non-nulles 
lorsque les petites fluctuations thermiques de l'ensemble canonique
sont prises en compte.
Cela reste le cas avec une petite courbure scalaire de Ruppenier,
pour des trous noirs de BTZ ou bien ceux de BTZ-Chern-Simons,
si bien qu'on choisit le param\`etre de rotation $J=0$,
voir \cite{Sar1} pour le d\'etails.
Hormis cela, nous avons \'etudi\'e la g\'eom\'etrie de Ruppenier de
certains trous noirs et branes noirs extr\'emaux.
Nous avons montr\'e que la g\'eom\'etrie thermodynamique des branes 
noirs $D_1D_5$ et $D_2D_6NS_5$ extr\'emaux en $D = 10$ d\'ecoulant 
de la th\'eorie des cordes de type-II, et les petits trous noirs 
en $D = 4$ d\'ecoulant de la th\'eorie des cordes h\'et\'erotiques,
est bien d\'efinie \cite{Sar2}.
Ensuite, il est analys\'e que la courbure scalaire de Ruppenier
est partout r\'eguli\`ere, et la nature reste inchang\'ee,
si on ajoute les corrections de $\alpha^{\prime}$.
Bien que la correction de $\alpha^{\prime}$ de l'ordre premier modifie 
l'entropie des petits trous noirs, la g\'eom\'etrie thermodynamique 
n'est pas bien d\'efinie,
mais les corrections de $\alpha^{\prime}$ des ordres sup\'erieurs la 
rendent bien d\'efinie et partout r\'eguli\`ere.


Dans cet ouvrage, nous \'etudions la g\'eom\'etrie thermodynamique
et les effets des corrections des d\'eriv\'ees sup\'{e}rieures de la 
g\'eom\'etrie thermodynamique.
En particulier, nous \'etudions les corrections de la g\'eom\'etrie
thermodynamique d\^{u}e au principe d'incertitude g\'en\'eralis\'ee, 
et celles en raison des corrections de $ \alpha^{\prime} $ de la th\'eorie 
des cordes.
Le reste de cette recherche est organis\'e en plusieurs chapitres.
Le premi\`er chapitre introduit les probl\`emes et les motivations.
Dans le chapitre $2$, nous avons examin\'e les origines de la g\'eom\'etrie 
thermodynamique dans la m\'ecanique statistique.
Nous avons expliqu\'e que la g\'eom\'etrie thermodynamique se pose 
naturellement dans l'approximation gaussienne de la fonction de partition 
d'un ensemble grand canonique, alors que l'ensemble des canoniques a seulement 
la transformation de l'\'echelle.
Dans le chapitre $3$, nous avons analys\'e la g\'eom\'etrie thermodynamique
de certains trous noirs et branes noirs de la th\'eorie des cordes.
De plus, nous avons donn\'e une reformulation du probl\`eme en termes 
de l'\'energie libre topologique du trou noir et ainsi de la fonction 
de partition d'un ensemble des trous noirs.
Ceci est comptible pour le cas des petits trous noirs que l'ensemble 
doit \^etre un ensemble m\'elang\'e.
En particulier, nous consid\'erons la g\'eom\'etrie de Ruppenier des 
trous noirs de Reissner-Nordstr\"om et \'egalement la g\'eom\'etrie 
de Wienhold des trous noirs dilatoniques et de la solution de 
Reissner-Nordstr\"om dans la nappe de Poincar\'e de $ADS_4$.
Dans le chapitre $4$, nous avons incorpor\'e les corrections d\^{u}es 
au principe d'incertitude g\'en\'eralis\'ee dans la g\'eom\'etrie 
thermodynamique.
Ici, nous analysons la g\'eom\'etrie de Ruppenier corrig\'ee par le
principe d'incertitude g\'en\'eralis\'ee pour le cas de trous noirs 
de Reissner-Nordstr\"om et celui des trous noirs charg\'es 
magn\'etiquement.

Dans le chapitre $5$, en prenant les corrections de $ \alpha ^{\prime} $
de la th\'eorie des cordes, nous consid\'erons la g\'eom\'etrie de
Ruppenier corrig\'ee par les termes de $ \alpha^{\prime} $ des trous
noirs extr\'emaux dans $ D = 4 $.
Nous avons montr\'e que dans le cas des trous noirs non-supersym\'etriques,
les corrections de $ \alpha^{\prime} $ d'ordre diff\'erents n'introduisent pas
la singularit\'e dans la courbure de Ruppenier, et le sous-espace d'\'etat est
partout r\'eguli\`er. En fait, nous trouvons que 
cela reste \'egalement vrai pour les corrections de Gauss-Bonnet aux trous noirs 
supersym\'etriques extr\'emaux dans $ D = 4 $.
En outre, nous avons observ\'e une tendance bien d\'efinie pour la courbure scalaire
de Ruppenier et pour le d\'eterminant de la m\'etrique comme le polyn\^omes.
Aussi, il est int\'eressant de noter que le degr\'e de ces courbures de Ruppenier
et celui des d\'eterminants peuvent \^etre d\'etermin\'es par l'ordre sup\'erieur des
corrections de $ \alpha^{\prime} $, \`a tous les ordres de $ \alpha^{\prime} $
plus grand qu'un.
Dans le chapitre $6$, nous examinons la g\'eom\'etrie de Ruppenier des branes noirs
$ D_1D_5 $ et $ D_2D_6NS_5 $ non-extr\'emaux en $ D = 10 $ et montrons que ces
syst\`emes thermodynamiques sont partout r\'eguliers et bien d\'efinis.
Dans le chapitre $7$, nous concentrons notre attention sur les trous noirs en rotation,
comme les trous noirs de Kerr-Newman, les trous noirs de Kaluza-Klein et les trous noirs
obtenus \`a partir de la th\'eorie des cordes. Dans le cas des trous noirs extr\'emaux
de la th\'eorie des cordes h\'et\'erotiques compactifi\'ee toroidalement, 
nous avons expliqu\'e que la courbure sous-jacente de Ruppenier diverge aux succursales
ergos et en ces points, la g\'eom\'etrie thermodynamique devient mal d\'efinie.
Enfin, le chapitre $8$ contient des questions et des remarques de conclusion pour l'avenir.
%

\clearpage
\chapter{L'origine de la g\'eom\'etrie thermodynamique}
Dans ce chapitre, nous allons faire une petite introduction de la g\'eom\'etrie thermodynamique.
Tout d'abord, le but de ce chapitre est principalement de placer les notations et les conventions 
qui seront suivies dans le reste de cette recherche. 

Commen\c{c}ons en consid\'erant les dispositions 
de bases n\'ecessaires de la m\'ecanique statistique \cite{Huang, Pathria} et donc nous allons expliquer 
les conceptions de la g\'eom\'etrique thermodynamique. \`A la fin de ce chapitre, nous montrons que 
la g\'eom\'etrie thermodynamique d\'ecoule naturellement de la th\'eorie des ensembles.

\section{Ensemble Canonique}

Il est bien connu que la fonction de cloison de l'ensemble 
canonique est donn\'ee par

\ba \mathcal Z&=& \sum_{E} e^{-\frac{E}{kT}}. \ea 

Quand l'\'energie $E$ augmente continuousement, la fonction de cloison 
de l'ensemble canonique $\mathcal Z$ peut \^etre \'ecrite comme

\ba \mathcal Z&=& \int \Gamma(E) e^{-\frac{E}{kT}} dE \nn
&=& \int e^{-\frac{1}{kT} \lbrace E- T S(E)\rbrace } dE, \ea

o\`u $ \Gamma(E) $ est le nombre des micro-\'etats entre l'\'energie $E$ et $E+ dE$. 
Avec l'hypoth\`ese de Boltzmann, on peut voir que l'entropie canonique est d\'efinie par 

\ba S= k \ln \Gamma(E).\ea 

Puisque $ E- T S \sim \heartsuit(N) $ dans la limite $ N \rightarrow \infty $, 
et ainsi cet int\'egral est domin\'e par le minimum de $ E- T S $. Pour une
\'energie $ E= \overline{E} $, nous voyons que la condition d'un minimum de 
$ E- T S $ est obtenue par 

\ba 1= T (\frac{\partial S}{ \partial E})_{\overline{E}}. \ea 

\c{C}a va dire que la temp\'erature est d\'efinie par la relation 

\ba T^{-1}= (\frac{\partial S}{ \partial E})_{\overline{E}}. \ea

Cette relation est  bien connu comme une relation thermodynamique entre l'entropie et 
la temp\'erature du syst\'eme, et donc $ \overline{E} $ est l'\'energie thermodynamique.
L'expansion de Taylor de $ E- T S(E) $ au point $E= \overline{E} $ est simplement
donn\'ee par 

\ba E- TS &=& \overline{E}+ E- \overline{E}- TS(\overline{E})- T 
(\frac{\partial S}{\partial E})_{\overline{E}} (\Delta E) +
\frac{T}{2}(\frac{\partial^2S}{\partial E^2})(\Delta E)^2 +
\heartsuit((\Delta E)^3)+ \ldots \nn
&\simeq& \overline{F}- \frac{T}{2}\frac{\partial}{\partial E}(\frac{1}{T})(\Delta E)^2 \nn
&=& \overline{F}+ \frac{(\Delta E)^2}{2TC}, \ea 

o\`u 

\ba \overline{F}= \overline{E}- TS(\overline{E})\ea 

est l'\'energie libre, c'est-\`a-dire que $ \overline{F} $ 
est une valeure minimume de $ (E- TS) $ quand la correction 
est positive, ce qui se produit, ssi on a une constante $C$ tel que $C> 0$.
En d'autre terme, l'int\'egral est domin\'e ssi la chaleur sp\'ecifique $C$ 
est positive. Donc, on voit que la fonction de cloison de l'ensemble canonique
est donn\'ee par 

\ba \mathcal Z= e^{-\frac{\overline{F}}{kT}} \int e^{-\frac{\Delta E^2}{2kT^2C}}dE.\ea

Il implique que la distribution canonique correspond aux fluctuations gaussiennes 
avec la moyenne \'energie thermodynamique $\overline{E}$. On peut voir facilement que la limite

\ba \mathcal Z \sim \heartsuit(\sqrt C)\sim \heartsuit(\sqrt N) \ea 

entra\^ine

\ba \ln \mathcal Z= -\frac{\overline{F}}{kT}+ \ln (\heartsuit(\sqrt N)).\ea

Donc, en ce cas, on obtient la limite thermodynamique, c'est-\`a-dire que la condition
o\`u le terme $ \ln (\heartsuit(\sqrt N)) $ ne domine pas, parce qu'on a 

\ba -\frac{\overline{F}}{kT} \sim \heartsuit( N ). \ea

Enfin, on a une relation entre la mecanique statistique et la thermodynamique 
qui sont respectivement repr\'esent\'es par les fonctions $Z$ et $F$. 
S\'elon cette relation, l'\'energie libre thermodynamique est donn\'ee par

\ba \overline{F}= kT \ln \mathcal Z. \ea

\section{Ensemble Grand Canonique}

De la m\^eme mani\`ere, la fonction de cloison de l'ensemble grand canonique est donn\'e par 

\ba \mathcal Q= \int \Gamma(E,N) e^{-\frac{1}{kT}(E- \mu N)} dE dN, \ea

o\`u $ \Gamma(E,N) $ est le nombre des \'etats avec l'\'energie $E$ et nombre 
des particules $N$. En ce cas, on peut \'ecrire facilement que

\ba \mathcal Q= \int e^{-\frac{1}{kT}(E- TS- \mu N)} dE dN. \ea

Dans la limite, $ N \rightarrow \infty $, l'int\'egral va dominer par le minimum
de $ (E- TS- \mu N) $, \`a qui les conditions extr\^emaux sont:

\ba T^{-1}&=& (\frac{\partial S}{ \partial E})_{V,N},\ \ et \nn
\mu&=& T(\frac{\partial S}{ \partial N})_{V,E}. \ea

On peut r\'esoudre ces \'equations avec les solutions:
$ E= \overline{E}, N= \overline{N} $ en \'ecrivant

\ba \label{engrand} (E- TS- \mu N)&=& \overline{E}- \mu \overline{N}- TS(\overline{E},\overline{N})-
\frac{T}{2}\lbrace (\frac{\partial^2 S}{\partial E^2})_{N,V}(\Delta E)^2 \nn &&+
2 \frac{\partial^2 S}{\partial E \partial N} (\Delta E \Delta N)+
(\frac{\partial^2 S}{\partial N^2})_{E,V}(\Delta N)^2 + \ldots \rbrace \nn
&\simeq& -PV + \frac{T}{2} \lbrace \alpha (\Delta E)^2+
2 \beta (\Delta E \Delta N)+ \gamma (\Delta N)^2 \rbrace,\ea

o\`u $\{ \overline{E}, \overline{N} \}$ sont respectivement les valeurs thermodynamiques de $\{E, N\}$.
Ici, dans la deuxi\`eme ligne de l'\'equation Eqn. (\ref{engrand}), nous avons d\'efini

\ba \alpha&:=& (\frac{\partial^2 S}{\partial E^2})_{N,V}, \nn
\beta&:=&  (\frac{\partial^2 S}{\partial E \partial N})_{V},\ \ et \nn
\gamma&:=& (\frac{\partial^2 S}{\partial N^2})_{E,V}. \ea

Anfin d'avoir le minimum de la fonction $ (E- TS- \mu N) $, la forme quadratique,
dans la parenth\`ese $ \lbrace .. \rbrace $, doit \^etre positive d\'efinie, ce qui 
est ainsi bien tenue, ssi 

\ba \alpha >0,\ \ \beta>0,\ \ et\ \ \alpha \gamma > \beta^2. \ea 

Notez bien que les conditions $\alpha >0$ et $\beta>0$ r\'epresentent
la stabilit\'e locale d'un syst\`eme statistique. Ici, les param\'etres $\alpha$ 
et $\beta$ s'applent les capacit\'es de chaleurs. En outre, la condition
$\alpha \gamma > \beta^2$ r\'epresente la stabilit\'e globale du syst\`eme
consid\'er\'e. \`A la suite de ce chapitre, nous allons montrer que la 
positivit\'e de la fonction 

\ba g(E, N):= \alpha \gamma- \beta^2 \ea  

donne la stabilit\'e globale du syst\`eme de deux param\'etres $\{E, N\}$. 
Donc, \`a la perspective de la th\'eorie des distributions, nous avons 

\ba \mathcal Q= e^{\frac{PV}{kT}} \int d(E- \overline{E}) d(N- \overline{N}) e^{\frac{-1}{2k}
\lbrace \alpha (\Delta E)^2+ 2 \beta (\Delta E \Delta N)+ \gamma (\Delta N)^2 \rbrace}. \ea

\`A la mesure que $ N \rightarrow \infty $, nous voyons trivialement que

\ba kT \ln \mathcal Q= PV+ \heartsuit(\ln \sqrt{ N }).\ea

Ainsi la distribution grande canonique est une distribution quadratique
des fluctuations dans $ E $ et $ N $. En d'autres termes, la distribution
grande canonique est \'equivalente \`a une distribution gaussienne des
fluctuations de l'\'energie et le nombre des particules.

Et bien, les lois thermodynamiques ne sont pas fondamentales mais elles
viennent des propri\'et\'es microscopiques du syst\`eme. En fait,
il s'av\`ere que nous avons

\ba TdS= dE+ PdV- \mu dN, \ea

o\`u l'entropie $ S(U,V,N) $ joue un r\^ole important. Dans ce cas,
si la d\'ependance de l'entropie $ S(U,V,N) $ aux variables $ U,V,N $ est
connue, alors la connaissance compl\`ete de tous les param\`etres thermodynamiques
peut \^etre obtenue.
En effet, il est facile de voir que l'entropie du syst\`eme caract\'erise les 
fluctuations thermiques de l'\'energie et du nombre de particules du syst\`eme.
D\'efinissons l'\'el\`ement de la distance par

\ba ds^2:= \alpha d E^2+ 2 \beta d E d N+ \gamma d N^2\ea

avec la m\'etrique tenseure thermodynamique

\ba g_{ij}(x)= \left( \begin{array}{rr}
    \alpha & \beta \\
     \beta & \gamma \\
\end{array} \right), \ea

o\`u $ x:= \left (\begin{array}{r}  E \\ N \\ \end{array} \right) $.
Nous allons examiner quand $ g_{ij} $ est sym\'etrique et positive d\'efinie
pour toute la variable thermodynamique $ x \in R^2$. On peut observer que 
l'entropie statistique de ce syst\`eme est 

\ba S= k_B \ln \Gamma (U,V,N),\ea

o\`u $ \Gamma $ repr\'esente le nombre quantique de tous les constituants 
des sous-ensembles. De plus, nous avons $ \Gamma =1 $ si il n'y a aucun d\'esordre.

\section{G\'eom\'etrie Thermodynamique}

Comme dans le cas de l'ensemble grand canonique, 
on peut voir pour un corps macroscopique dans certain \'equilibre 
que les quantit\'es physiques ont g\'en\'eralement de petites d\'eviations 
de la leurs valeurs moyennes. Donc, nous voulons trouver une distribution 
de probabilit\'e de ces fluctuations thermiques en consid\'erant le nombre 
des micro-\'etats

\ba \Gamma= A\ e^S \ea

comme un commencement de la th\'eorie des fluctuations. En g\'en\'erale, 
soit $P$ une distribution de probabilit\'e, c'estr \`a dire que 

\ba P \propto e^S.\ea

En particulier, consid\'erons la distribution gaussienne de plusieurs quantit\'es 
thermodynamiques et leurs fluctuations simultan\'ees de la leurs valeurs moyennes.
Nous pouvons d\'efinir l'entropie $ S(x_1,x_2,\ldots,x_n) $ d'un corps macroscopique 
dans l'\'equilibre en dependant sur $ \lbrace x_i \rbrace $ avec la leurs d\'eviations
par l'expension de Taylor juaqu'\`a le deuxi\`eme ordre: 

\ba S= S_0 - \frac{1}{2}\sum_{i,j=1}^{n} g_{ij}x^i x^j, \ea

o\`u, s\'elon les Refs. \cite{Rupp, Aman, Sar1, Sar2, Wien, Deg, cai1}, 
la m\'etrique tenseure $ g_{ij} $ est d\'efinie par

\ba g_{ij}:= -\frac{\partial^2 S}{\partial x^i \partial x^j}.\ea

Ainsi la distribution de probabilit\'e peut \^etre \'ecrite simplement:

\ba P= A e^{-\frac{1}{2} g_{ij}x^i x^j }. \ea

Avec la normalisation

\ba \int \prod_{i=1}^{n} dx_i P(\lbrace x_j\rbrace_{j=1}^{n}) =1,\ea

il est facile d'obtenir que

\ba A= \frac{\sqrt{g}}{(2 \pi)^{\frac{n}{2}}}, \ea

o\`u $g$ est d\`efinit comme le d\'eterminant de la m\'etrique tenseure

\ba g= \Vert g_{ij}\Vert. \ea

Donc, nous voyons que la limite gaussienne donne la distribution de probabilit\'e

\ba P= \frac{\sqrt{g}}{(2 \pi)^{\frac{n}{2}}} e^{-\frac{1}{2} g_{ij}x^i x^j }.\ea

En fait, nous pouvons d\'efinir la g\'eom\'etrie riemannienne en consid\'erant
l'\'el\'ement de la ligne entre les deux \'etats arbitraires d'un \'equilibre 
comme la somme 

\ba ds^2:= \sum_{i,j} g_{ij} d x^id x^j.\ea

Ainsi, s\'elon la transformation de Legendre

\ba X_i= \frac{\partial S}{\partial x^i}= g_{ij}x_j, \ea

on peut observer que les fonctions de corr\'elations des paires sont donn\'ees par

\ba < x_i x_j>= g^{ij}= < X_i X_j>.\ea

Nous pouvons appeler $ \lbrace X_i \rbrace $ comme variables 
thermodynamiques et assigner $ g_{ij} $ pour \^etre une m\'etrique 
intr\'ensique parce que $X^i$ est conjugu\'e \`a $ x^i $ pour chaque $i$.
Nous pouvons obtenir les propri\'et\'es thermodynamiques en limitant
nous-m\^eme, aux coordonn\'ees qui sont des param\`etres extensives
d'une configuration thermodynamique. Cette consid\'eration est un mod\`ele
de la g\'eom\'etrie riemannienne ordinaire qui est ici bas\'e sur la th\'eorie 
des fluctuations d'un ensemble et les axiomes de la thermodynamique d'\'equilibre. 
Cette g\'eom\'etrie $ ( \mathcal M, g) $ est appel\'ee la g\'eom\'etrie 
thermodynamique de Ruppenier. C'est la g\'eom\'etrie dans laquelle les 
\'etats d'\'equilibre peuvent \^etre repr\'esent\'es par les points d'une 
vari\'et\'e riemannianne $ ( \mathcal M, g) $, et dans cet inclusion, 
la distance entre ces points d'\'equilibre est li\'ee aux fluctuations
thermiques entre eux. De cette fa\c{c}on, nous pouvons avoir une relation 
entre la g\'eom\'etrie et la th\'eorie de probabilit\'e, en particulier, 
nous avons $\langle\langle$ moins probable $ \rangle\rangle $ signifie que 
$ \langle\langle $ dans le lointain $\rangle\rangle $.


Dans chapitre $4$, nous allons consid\'erer la g\'eom\'etrie 
de Ruppenier et celle de Wienhold des syst\`emes de trous noirs. 
Cette g\'eom\'etrie donne une m\'ethode directe pour analyser les points
critiques d'une configuration des trous noirs. Notez aussi que cette 
m\'ethode g\'eom\'etrique est un domaine important de la recherche 
courante pour comprendre la thermodynamique des trous noirs. 
En effet, les observations de cette recherche sont importantes
parce que la courbure scalaire thermodynamique de Ruppenier est 
proportionnelle au volume de la corr\'elation ce qui signifie 
l'interaction du syst\`eme statistique fondamental. C'est-\`a-dire que
nous avons les cas: soit $ \xi $ est la longueur des corr\'elation statistiques, 
puis $\forall (x^1,\ldots, x^n) \in (\mathcal M_n, g)$, on a 

\ba R \sim \xi^d, \ea

o\`u $d$ est la dimension spatiale du syst\`eme statistique \cite{Rupp}.
En fait, c'est la base d'un mod\`ele g\'eom\'etrique qui est bas\'ee sur
l'inclusion de la th\'eorie de fluctuations dans les axiomes de la
thermodynamique d'\'equilibre. C'est pourquoi, dans ce mod\`ele g\'eom\'etrique 
qui est li\'e \`a la th\'eorie de probabilit\'e, o\`u on peut d\'efinir 
un \'el\'ement de la ligne entre les deux \'etats \'equilibres arbitaires
par la m\'etrique tenseure

\ba g_{ij}= -\partial_i \partial_j S(x^1,\ldots, x^n).\ea

Dans le cas particulier des configurations de deux parem\`etres 
$\{x^1, x^2\}$, il y existe des \'etats \'equilibres qui 
peuvent \^etre repr\'esent\'es par les points de la surface 
bidimensionnelle $ \mathcal M_2 $, et ainsi la distance entre 
les points arbitraires sur la vari\'et\'e $ \mathcal M_2 $ est 
li\'ee aux fluctuations entre les \'etats \'equilibres 
correspondants de la configuration statistique.


\clearpage
\chapter{Les g\'eom\'etries thermodynamiques des trous noirs}
Dans ce chapitre, nous avons analys\'e la g\'eom\'etrie thermodynamique 
d'un syst\`eme \'equilibre des \'etats de trous noirs et montrons que 
les configurations de la th\'eorie des cordes et celle de la $M$-th\'eorie 
fondamentale poss\`edent des caract\'eristiques g\'eom\'etriquements int\'eressantes.
En effet, la g\'eom\'etrie thermodynamique peut \^etre appliqu\'ee pour
\'etudier la nature de l'entropie des trous noirs qui sont charg\'es soit
\'electriquement soit magn\'etiquement, ou bien soit par les deux charges.
Et surtout, le produit int\'erieur de l'espace d'\'etat d'une configuration
thermodynamique dans un \'equilibre, dans la repr\'esentation d'\'energie, a \'et\'e 
fournie par Weinhold, ainsi que la matrice Hessianne de l'\'energie int\'erieure ou
celle de la masse d'un syst\`eme des trous noirs, en respectant la vaste charact\'eristique 
thermodynamique de ces variables, a \'et\'e d\'ej\`a expliqu\'ee dans l'introduction.
Apr\`es tout, il s'av\`ere que les corrections thermiques disparaissent dans la limite 
thermodynamique o\`u l'entropie canonique d'un tel syst\`eme et l'ensemble micro-canonique 
sous-jacent d'un trou noir consid\'er\'e deviennent identiques.
Bien que l'origine statistique de l'entropie des trous noirs est encore incertaine,
il va de soi que d'un trou noir en \'equilibre avec le rayonnement thermique d'Hawking 
\`a un terme fixe de la temp\'erature d'Hawking qu'elle est bien d\'ecrite par l'ensemble 
canonique des micro-\'etats. Dans ce chapitre, 
nous allons maintenant discuter de la g\'eom\'etrie thermodynamique des trous 
noirs avec des plusieurs charges et la masse ou l'entropie.
Par-ce que l'entropie d'un trou noir est la valeur extremum de la fonction
de l'entropie de Sen. C'est pourquoi, la g\'eom\'etrie thermodynamique des trous noirs
peut \^etre d\'etermin\'ee par le r\'ef\'erence \`a l'ensemble canonique ce qu'elle vienne
par la fonction de l'entropie de Sen aux points fixes de l'attracteur.
En particulier, l'enqu\^ete de la g\'eom\'etrie thermodynamique covariante
de Ruppeiner et celle de Wienhold d'une configuration des trous noirs peut \^etre 
mise en lumi\`ere aux aspects int\'eressantes comme la transition des phases et 
les g\'eom\'etries de l'espace des modules d'un trou noir consid\'er\'e.

\section{La g\'eom\'etrie thermodynamique de Ruppenier} 

Dans cette section, nous allons expliquer la g\'eom\'etrie thermodynamique 
d'un syst\`eme des trous noirs de deux et trois param\`etres.

\subsection{Les trous noirs de deux param\`etres}

Dans le cas de la g\'eom\'etrie thermodynamique de Ruppenier, 
nous pouvons parametriser l'entropie comme une fonction de la masse $M$ 
et de la charge $Q$ d'un ensemble des trous noirs comme l'application 

\ba S: \mathcal{M}_2 \rightarrow R \ea 

par une valeur de l'entropie $S(M, Q)$. Et alors, nous voyons en g\'en\'erale 
que la m\'etrique de Ruppenier est donn\'ee par l'\'el\'ement de la ligne:

\ba ds^2=
-\frac{\partial^2 S(M,Q)}{\partial M^2} dM^2
-2 \frac{\partial^2 S(M,Q)}{\partial M \partial Q} dM dQ
-\frac{\partial^2 S(M,Q)}{\partial Q^2} dQ^2.\ea

De plus, il est tr\`es facile de voir que le d\'eterminant de la m\'etrique tenseure est

\ba \Vert g \Vert= \frac{\partial^2 S(M,Q)}{\partial M^2} 
\frac{\partial^2 S(M,Q)}{\partial Q^2}-
(\frac{\partial^2 S(M,Q)}{\partial M \partial Q})^2.\ea

Nous voyons que la courbure scalaire de Ruppenier est simplement donn\'ee par:
 
\ba R&=& \frac{1}{2 \Vert g \Vert^2}
\bigg(-\frac{\partial^2 S(M,Q)}{\partial M^2}
(\frac{\partial^3 S(M,Q)}{\partial Q^2 \partial M})^2
+
\frac{\partial^2 S(M,Q)}{\partial M^2} 
\frac{\partial^3 S(M,Q)}{\partial Q \partial M^2}
\frac{\partial^3 S(M,Q)}{\partial Q^3}\nn &&
-
\frac{\partial^2 S(M,Q)}{\partial Q \partial M}
\frac{\partial^3 S(M,Q)}{\partial M^3}
\frac{\partial^3 S(M,Q)}{\partial Q^3}
+
\frac{\partial^3 S(M,Q)}{\partial Q^2 \partial M}
\frac{\partial^3 S(M,Q)}{\partial M^3}
\frac{\partial^2 S(M,Q)}{\partial Q^2}\nn &&
+
\frac{\partial^3 S(M,Q)}{\partial Q \partial M^2}
\frac{\partial^2 S(M,Q)}{\partial M \partial Q}
\frac{\partial^3 S(M,Q)}{\partial Q^2 \partial M}
-
(\frac{\partial^3 S(M,Q)}{\partial Q \partial M^2})^2
\frac{\partial^2 S(M,Q)}{\partial Q^2} \bigg).\ea

\subsection{Les trous noirs de trois param\`etres}
Dans cette sous-section, nous allons examiner le cas g\'en\'erale des trous noirs 
de trois param\`etres. Un exemple typique de cette g\'eom\'etrie de Ruppenier vient
par une configuration des trous noirs en rotations \cite{Sen0606244}. Ici, nous voyons 
que la g\'eom\'etrie thermodynamique de Ruppenier est bien d\'efinie et les quantit\'es 
de la g\'eom\'etrie de Ruppenier peuvent \^etre \'ecrites facilements. Pour la raison 
de la simplicit\'e, par rapport \`a le cas des configurations de trous noirs de deux 
param\`etres, nous pouvons \'ecrire la param\`etrisation de l'entropie 
$S \equiv S(P, Q, J)$ comme l'application 

\ba S: \mathcal{M}_3 \rightarrow R, \ea 

et ainsi par la d\'efinition de la m\'etrique tenseure de Ruppenier, 
il s'a\`ere que l'\'el\'ement de la ligne sous-jacente peut \^etre \'ecrite comme: 

\ba ds^2&:=& - \frac{\partial^2 S}{\partial P^2}dP^2
-2 \frac{\partial^2 S}{\partial P \partial Q} dP dQ
-2 \frac{\partial^2 S}{\partial P \partial J} dP dJ \nn &&
- \frac{\partial^2 S}{\partial Q^2} dQ^2 
-2 \frac{\partial^2 S}{\partial Q \partial J}dQ dJ
- \frac{\partial^2 S}{\partial J^2} dJ^2.\ea

Alors, le d\'eterminant de la m\'etrique tenseure est simplement:

\ba \Vert g \Vert&=& 
- \frac{\partial^2 S}{\partial P^2} 
 \frac{\partial^2 S}{\partial Q^2}
 \frac{\partial^2 S}{\partial J^2}
+ \frac{\partial^2 S}{\partial P^2}
  (\frac{\partial^2 S}{\partial Q \partial J})^2
+(\frac{\partial^2 S}{\partial Q \partial P})^2
  \frac{\partial^2 S}{\partial J^2} \nn &&
-2 \frac{\partial^2 S}{\partial Q \partial P}
  \frac{\partial^2 S}{\partial P \partial J}
  \frac{\partial^2 S}{\partial Q \partial J}
+ (\frac{\partial^2 S}{\partial P \partial J})^2
  \frac{\partial^2 S}{\partial Q^2}.\ea

Nous voyons maintenant que la courbure scalaire de Ruppenier est donn\'ee par:

\ba \label{rup3dg} R= \frac{1}{2} \bigg(\frac{(r_1- \tilde{r}_1)+2(r_2- \tilde{r}_2)
+3(r_3- \tilde{r}_3)+4r_4+ 6r_6}{\Vert g \Vert^2}\bigg), \ea 

o\`u les fonctions $\{ r_i, \tilde{r}_j\ |\ i=1,2,3,4,6, j=1,2,3 \}$ 
sont d\'efinies dans l'annexe $[B]$. 

\subsection{La g\'eom\'etrie thermodynamique de Wienhold} 

Par le biais de la g\'eom\'etrie thermodynamique,
nous pouvons bien s\^ur \'etudier la stabilit\'e des trous noirs 
dilatoniques topologiques. La stabilit\'e d'un syst\`eme thermodynamique 
avec les petites fluctuations thermiques est analys\'e par le comportement 
de l'entropie $S(Q,M) $ autour de l'equlibrie consid\'er\'e.
Pour avoir la stabilit\'e locale dans n'importe quel ensemble,
nous exigeons que $S(Q,M)$ doit \^etre une fonction convexe des variables extensives.
En d'autres termes, c'est la transforme de Legendre o\`{u} $M$
doit \^etre une fonction concave des variables intensives.
La stabilit\'e est \'egalement bien obtenue par le comportement
de l'\'energie $M(S,Q)$ en ce qui concerne les varibles extensives,
ce qui dans ce cas-ci devrait \^etre une fonction convexe.
Ainsi la stabilit\'e locale en principe peut \^etre analys\'ee
en obtenant le determinant de la matrice d'Hessianne de la masse
$M(S, Q)$ avec les variables extensives du trou noir.
En consid\'erant $S$ et $Q$ comme ensemble complet des variables
extensives pour la masse $M(S, Q)$, Ref. \cite{Townsend} montre que
nous pouvons d\'efinir les param\`etres intensifs conjugu\'es aux $S$ 
et $Q$ qui sont respectivement associ\'ees \`a la temp\'erature $T$ 
et au potentiel \'electrique $\phi$, ce qui sont donn\'es par

\ba T&:=& (\frac{\partial M}{\partial S})_Q, et \nn 
\phi&:=& (\frac{\partial M}{\partial Q})_S \ea

se satisfont facilement la premi\`ere loi de la thermodynamique:

\ba dM= TdS+ \phi dQ. \ea

En ce cas, en consid\'erant $\forall x^a= (S, Q)\in \mathcal M_2$, 
comme une application 

\ba S: \mathcal{M}_2 \rightarrow R \ea 

avec la param\'etrisation $M(S, Q)$ de la masse, nous voyons que la m\'etrique 
tenseure de Wienhold peut \^etre \'exprim\'ee comme: 

\ba g_{ab}:= \frac{\partial^2}{\partial x^a \partial x^b}M(S,Q), \ea

et alors que l'\'el\'ement de la ligne de cette vari\'et\'e
thermodynamique $ (\mathcal M_2, g) $ peut \^etre \'ecrite comme:

\ba ds^2= (\frac{\partial^2 M(S,Q)}{\partial S^2}) dS^2 +
2 (\frac{\partial^2 M(S,Q)}{\partial S \partial Q}) dS dQ +
(\frac{\partial^2 M(S,Q)}{\partial Q^2}) dQ^2.\ea

Alors que la masse de ce trou noir charg\'e dilatonique topologique est
donn\'ee par un formule de Smarr, voir le Ref. \cite{Townsend} pous le plus des
d\'etails. De toute fa\c{c}on, il est facile d'obtenir que le d\'eterminant 
de la m\'etrique tenseure de Wienhold est donn\'e par

\ba \Vert g^W \Vert= \frac{\partial^2 M(S,Q)}{\partial S^2} 
\frac{\partial^2 M(S,Q)}{\partial Q^2}-
(\frac{\partial^2 M(S,Q)}{\partial S \partial Q})^2.\ea

Avec la sym\'{e}trie dans les deux premiers indices, nous constatons 
que les symboles de Christoffel du premier type sont donn'es par

\ba
{\Gamma_{S\,S\,S}}&=& \frac{1}{2} \,({\frac{\partial^{3}}{\partial S^{3}}}\,\mathrm{M}(S, \,Q)), \nn
{\Gamma_{S\,S\,Q}}&=& \frac{1}{2} \,({\frac{\partial^{3}}{\partial S^{2}\,\partial Q}}\,\mathrm{M}(S, \,Q)), \nn
{\Gamma_{S\,Q\,S}}&=& \frac{1}{2} \,({\frac{\partial^{3}}{\partial S^{2}\,\partial Q}}\,\mathrm{M}(S, \,Q)), \nn
{\Gamma_{S\,Q\,Q}}&=& \frac{1}{2} \,({\frac{\partial^{3}}{\partial S\,\partial Q^{2}}}\,\mathrm{M}(S, \,Q)), \nn
{\Gamma_{Q\,Q\,S}}&=& \frac{1}{2} \,({\frac{ \partial^{3}}{\partial S\,\partial Q^{2}}}\,\mathrm{M}(S, \,Q)), \nn
{\Gamma_{Q\,Q\,Q}}&=& \frac {1}{2} \,({\frac{\partial^{3}}{\partial Q^{3}}}\,\mathrm{M}(S, \,Q)).
\ea

En ce cas, nous voyons que la courbure scalaire de Wienhold est:
 
\ba R^W&=& -\frac{1}{2 \Vert g^W \Vert^2}
\bigg( - \frac{\partial^2 M(S,Q)}{\partial S^2}
(\frac{\partial^3 M(S,Q)}{\partial Q^2 \partial S})^2
+
\frac{\partial^2 M(S,Q)}{\partial S^2} 
\frac{\partial^3 M(S,Q)}{\partial Q \partial S^2}
\frac{\partial^3 M(S,Q)}{\partial Q^3}\nn &&
-\frac{\partial^2 M(S,Q)}{\partial Q \partial S}
\frac{\partial^3 M(S,Q)}{\partial S^3}
\frac{\partial^3 M(S,Q)}{\partial Q^3}
+
\frac{\partial^3 M(S,Q)}{\partial Q^2 \partial S}
\frac{\partial^3 M(S,Q)}{\partial S^3}
\frac{\partial^2 M(S,Q)}{\partial Q^2}\nn &&
+
\frac{\partial^3 M(S,Q)}{\partial Q \partial S^2}
\frac{\partial^2 M(S,Q)}{\partial S \partial Q}
\frac{\partial^3 M(S,Q)}{\partial Q^2 \partial S}
-
(\frac{\partial^3 M(S,Q)}{\partial Q \partial S^2})^2
\frac{\partial^2 M(S,Q)}{\partial Q^2} \bigg).\ea

Dans la suite de ce chapitre,
nous ferons une br\`eve introduction sur les fondements basics des trous noirs,
puis nous introduirons la g\'eom\'etrie thermodynamique par la fonction de cloison 
d'un ensemble des trous noirs donn\'es, et ainsi l'\'energie libre des cordes topologiques 
trouve sa place dans cet \'etude. En fait, ce sont des deux outils fondements importants que nous 
pouvons \'egalement utiliser avec l'aide de la conjecture d'OSV, donc nous conclurons 
ce chapitre en parlant du lien entre des trous noirs de la th\'eorie des cordes et l'un des 
ces trous noirs de la $M$-th\'eorie et les leurs g\'eom\'etries thermodynamiques de Ruppenier 
et de Wienhold \cite{Rupp, Aman, Sar1, Sar2, Wien, Deg, cai1}. De toute mani\`ere, ces r\'epr\'esentations 
sont aussi bien motiv\'ees par l'\'etude des configurations thermodynamiques \cite{Huang, 
Pathria} comme nous les avons d\'ej\`a pr\'evue dans le chapitre $2$.

\section{L'entropie des trous noirs}

Nous savons qu'un trou noir est une solution classique de la relativit\'e g\'en\'erale,
ce qu'on peuisse pens\'e comme une particule pointuelle de tr\`es grande masse.
Comme, tous les trous noirs ont une grande force d'attraction et notez bien le fait m\^eme 
que la lumi\`ere ne peut pas aussi l'\'echapper une surface associ\'e \`a le trou noir
consid\'ere. Ce fait conduit \`a la notion d'un \'ev\'enement horizon d'un trou noir,
dont pas d'une particule physique peut la franchir. Dans les ann\'ees 1970,
Hawking et Bekenstein ont constat\'e par la voix de la m\'ecanique quantique 
\cite{Bekenstein, Hawking, BardeenCarterHawking} que le trou noir est un objet 
thermique avec certains temperature non-nulle. Comme tout les syst\`emes thermiques, 
un trou noir vient avec certaine entropie li\'ee \`a la domaine de l'\'ev\'enement horizon,
ce qui est donn\'e par la formule c\'el\'ebr\'ee suivante

\ba S_{BH}= \frac{A}{4}. \ea

En outre, du point de vue de la m\'ecanique statistique,
cet entropie est li\'ee \`a un probl\`eme de comptage pour un ensemble des charges 
avec certaine \'energie fixe ou celle d'une masse fixe du trou noir. Pr\'ecis\'ement, 
s\'elon les conceptions de la m\'ecanique statistique \cite{Huang, Pathria},
nous voyons que cette relation est donn\'ee par

\ba S_{micro}= \ln \Omega(\overrightarrow{Q}, \overrightarrow{P}, M),\ea

o\`u $ \Omega(\overrightarrow{Q}, \overrightarrow{P}, M)$
est la d\'eg\'en\'eraence des \'{e}tats avec diverses charges 
\'electriques, magn\'etiques et la masse $M$ du syst\`eme.
En d'autres termes, un trou noir a l'entropie qui r\'eponde
jusqu'\`a la deuxi\`eme loi de la thermodynamique.
De plus, les syst\`emes en ayant une plus grande entropie sont 
thermodynamiquement plus susceptibles de subvenir \`a la nature.

Souvenons-nous, comment un trou noir peut \^etre comprise \`a partir 
des notions de la th\'eorie des cordes.
Il est bien connu que les particules \'el\'ementaires sont caract\'eris\'ees 
par les \'etats vibrationnels d'une corde.
En outre, il n'y a que cinq th\'eories des cordes coh\'erentes 
\cite{GreenSchwarzWitten, Polchinski}, en particulier, ces sont du type-$I$, 
du type-$IIA$, du type-$IIB$, h\'et\'erotique $ E_8 \times E_8 $ et h\'et\'erotique 
$SO(32)$; qui tous vivent dans les dimensions de l'espace-temps $ D= (9+1)$.
Pour venir pr\`es des observations physiques des dimensions $D= (3+1)$,
nous utilisons la proc\'edure de la compactification pour les six dimensions 
int\'erieures et donc les six dimensions compactifi\'ees ne sont pas vues dans 
les acc\'el\'erateurs presents. De plus, chacun de ces spectres de la th\'eorie 
des cordes contient le graviton, qui est le m\'ediateur de la gravit\'e.
Comme nous le savons qu'une corde peut \^etre pens\'ee d'une collection 
des oscillateurs harmoniques de nombre infini et chaqu'une des corde a des 
\'etats quantiques ou excitations d'un nombre infini. C'est-\`a-dire, 
la quantification d'une corde donne une tour de l'infinie des \'etats
qui peuvent \^etre d\'ecrits les diff\'erentes particules \'el\'ementaires.
En particulier, les \'etats d'une grande masse ne sont pas observables
dans les exp\'eriences des laboratoires pr\'esentes.

Maintenant, pour avoir une id\'ee de la d\'eg\'en\'erescence,
on peut la d\'efinir comme le nombre des \'etats ayant certaine 
\'energie fixe ou celle d'une masse fixe. 
On sait que la d\'eg\'en\'erescence augmente rapidement,
dont la masse est augment\'ee.
Pour l'ensemble de diverses charges \'electriques et magn\'etiques
$ (\overrightarrow{Q}, \overrightarrow{P}) $ en caract\'erisant un 
syst\`eme statistique des trous noirs, l'entropie de comptage associ\'ee 
au syst\`eme est d\'efinie comme

\ba S_{micro}= \ln \Omega(\overrightarrow{Q}, \overrightarrow{P}, M).\ea

Une impprtante question pour demander est: Est-ce que on est

\ba S_{micro}= S_{BH} \in R\ ? \ea

Pour avoir une r\'eponse, nous allons examiner la th\'eorie des cordes 
h\'et\'erotiques qu'elle peut \^etre compactifi\'ee soit sur l'\'espace 
$ E_8 \times E_8 $ soit sue l'\'espace  $SO(32)$ avec certains nombre 
d'enroulement $w$ enroul\'ee sur un tore. En g\'en\'eral, cela correspond 
au nombre des modes de gauche et ceux de droite de la th\'eorie des cordes 
qui se d\'eplacent avec les conditions p\'eriodiques de la fronti\`ere des 
champs bosoniques. En outre, ces modes de gauche et ceux de droite peuvent 
s'affronter et s'annihiler au temps ce qui correspond \`a un syst\`eme instable.
C'est la compr\'ehension microscopique des instabilit\'es thermodynamiques  
en termes des \'etats \'el\'ementaires des cordes h\'et\'erotiques.
Afin d'examiner les \'etats stables, nous devons consid\'erer seulement 
les \'etats qui mouvement \'a gauche ou ceux qui mouvement \'a droite, 
ce que l'on appelle les \'etats de BPS.

Ces \'etats sont d\'ecrites par deux nombres quantiques:	
le nombre d'enroulement et l'\'elan total port\'ee par les osscilations 

\ba n/k, n \in Z. \ea

C'est-\`a-dire qu'on a un simple probl\`eme de la
quantification d'une particule dans une bo\^ite.
Ici, le ratio $w:=n/k$ et l'\'elon $n$ sont les deux nombres quantiques 
qui nivelent le niveau des \'etats quantiques d'une corde h\'et\'erotique.
Maintenant, soit $ \Omega(n,w)$ le nombre des \'etats avec des nombres
quantiques $ w $ et $ n $. Puis, dans la limite des grandes charges 

\ba w,\ n \rightarrow \infty, \ea

le Ref. \cite{Sen0505122} montre que la th\'eorie des cordes h\'et\'erotiques 
comporte un trou noir avec la d\'eg\'en\'erescence des \'etats comme

\ba \Omega(n,w)= \exp(4 \pi \sqrt{nw}).\ea

C'est-\`a-dire, l'entropie de comptage de la m\'ecanique statistique 
est donn\'ee par

\ba S_{micro}= 4 \pi \sqrt{nw}.\ea

Afin d'avoir une compr\'ehension microscopique d'un trou noir \`a partir 
de la th\'eorie des cordes, nous avons justement besoin d'avoir 

\ba S_{micro}= S_{BH}.\ea

Il n'est pas surprenant qu'il existe les diff\'erentes corrections, par exemple:

\begin{enumerate}
\item les corrections de $\alpha^{\prime}$ qui arrivent d\^ue au fait que
les cordes ne sont pas des particules pointuelles, et seulement aux grandes
distances, les cordes se comportent comme des particules pointuelles.
\item les corrections quantiques qui arrivent par la consid\'eration que
la gravit\'e elle-m\^eme, est une th\'eorie quantique et pas seulement 
une th\'eorie de la supergravit\'e. C'est-\`a-dire qu'un trou noir est 
consid\'e comme un objet quantique.
\end{enumerate}

Du fait m\^eme que le champ de dilaton peut d\'efinir comme un 
param\`etre de ces effets quantiques, comme le terme $1/\sqrt{nw}$.
Bien que les effets quantiques sont tr\`es faibles dans la limite des grandes charges,
mais les corrections de $\alpha^{\prime}$ sont d'ordre de l'unit\'e.
En fait, dans cette limite, Sen a montr\'e que les sym\'etries de la th\'eorie des cordes 
heterotiques classiques avec les corrections de $\alpha^{\prime}$ donnent lieu \`a: 

\ba S_{micro}= a \pi \sqrt{nw},\ea

o\`u le param\`etre $a$ d\'epend sur les corrections de $\alpha^{\prime}$.
Par cons\'equent, nous voyons que la th\'eorie des cordes fournit une
explication microscopiques de la th\'eorie des trous noirs.
En fait, ces trous noirs consid\'er\'es sont uniquement \'electriquement
charg\'ees et ce sont appel\'es les petits trous noirs.
Dans ce cas, Sen a \'egalement montr\'e qu'au niveau d'arbre de $\alpha^{\prime}$,
l'entropie macroscopique de ces petits trous noirs est:

\ba S_{BH}= A/4= 0, \ea 

pour des d\'etails voir le Ref. \cite{Sen0411255}. En revanche, 
dans la th\'eorie des cordes de type-$II$, il y a des grand trous noirs pour que 
nous pouvons maintenant consid\'erer l'entropie de Bekenstein-Hawking.
En g\'en\'rale, les grand trous noirs viennent avec l'entropie

\ba S_{BH}= A/4 \neq 0, \ea

et ainsi notez bien qu'ils sont charg\'es \'electriquement et magn\'etiquement, 
pour des d\'etails voir les Refs. \cite{StromingerVafa, HorowitzStrominger}. 
Dans ce cas, le comptage microscopique comporte sur certaines D-branes et on peut 
r\'ecup\'erer dans la limite des grandes charges \cite{AharonyGubserMaldacenaOoguriOz,
Witten, BoerNaqvi, BoerPasquinucciSkenderis, Mathur, MathurSaxenaSrivastava} que 

\ba S_{micro}= S_{BH}.\ea

Maintenant, sans la charge magn\'etique, c'est-\`a-dire, quant \`a la valeur $p= 0$, on a 

\ba S_{BH} = 0\ea

ce qui co\"incide avec les petits trous noirs charg\'ees \'electriquement
d\'ecoulant dans la th\'eorie des cordes h\'et\'erotiques.


\section{La th\'eorie des cordes topologiques}

Donc, la th\'eorie effective, dans la limite d'une \'energie faible, 
de la supergravit\'e de $ \mathcal N = 2 $ interagissant avec le multiple des vecteurs,
en suivant de la th\'eorie des cordes, implique qu'on a non-nulles corrections 
des courbures sup\'erieures. En fait, les efects des d\'eriv\'ees sup\'erieures modifient 
la loi d'aire d'Hawking-Bekenstein et introduisent les corrections d'ordres sup\'erieures
\`a l'entropie thermodynamique des trous noirs. \`A l'\'echelle macroscopique, 
l'entropie r\'esulte par le traitement de Walds de la gravit\'e des courbures 
sup\'erieures g\'en\'eralement covariantes, comme une int\'egrante de la surface 
d'horizon du trou noir par une densit\'e de la charge de Noether. 
En outre, les corrections des d\'eriv\'ees sup\'erieures sont encod\'ees 
dans le prepotentiel g\'en\'eralis\'e qui est une fonction homog\`ene,
holomorphes des degr\'es deux des champs scalaires rescalad\'es
et la partie de l'anti-selfduale du champ graviphoton.	
Le m\'ecanisme attracteur continue \`a tenir en pr\'esence 
des d\'eriv\'ee sup\'erieures et une application de l'analyse 
de Walds donne l'entropie macroscopique d'un ces trou noir.	
De plus, les \'equations attracteurs peuvent \^etre r\'esolues, 
et afin de d\'eterminer les champs scalaires en termes des charges,
ce qui garantit \`a l'horizon que l'entropie macroscopique des trous
noirs attracteurs est seulement une fonction des charges port\'ees.
Comme, nous avons expliqu\'e pr\'ec\'edemment, les solutions de ces
trous noirs tombent dans deux distinctes cat\'egories: 

\begin{enumerate}
\item les grands trous noirs qui ont une zone non-nulle au lors de niveaux des deux d\'eriv\'ees,
\item les petits trous noirs qui ont une zone nulle et transportent uniquement des charges \'electriques.
\end{enumerate}

Pour une configuration des grands trous noirs de la th\'eorie des cordes 
de type-$IIA$ compactifi\'ee sur une vari\'et\'e de Calabi-Yau, on a une 
description en termes des certains branes sur les cycles non-triviaux \cite{Str3}. 
L'entropie microscopique est alors d\'etermin\'e en termes de comptage des 
micro-\'etats. Plus pr\'ecis\'ement, \c{c}a vient par la formule de Cardy 
de la th\'eorie bidimensionnelle des champs conformes associ\'ee \`a la 
fronti\`ere sous-jacente d'un ensemble de ces branes. Ceci est perturbativement 
en accord avec la pr\'ecision ordonnance de l'entropie macroscopique. 
De plus, on a une reformulation du probl\`eme en terme de l'\'energie libre 
topologique du trou noir li\'e au logarithme de la fonction de partition d'un ensemble 
des trous noirs qui indique dans le cas des petits trous noirs que l'ensemble sous-jacent
doit \^etre un ensemble m\'elang\'e.

En fait, il est vraiment int\'eressant de savoir la question qu'elles sont les
significations thermodynamiques des grands et des petits trous noirs. S\'elon la
th\'eorie microscopique de ces trous noirs, nous pouvons les caract\'eriser comme les suivants:

\begin{enumerate}
\item la m\'ethode de la fonction d'entropie fournissant certaines
\'equations d'attracteurs ou non-attracteurs en pr\'esence des termes
d\'eriv\'es sup\'erieurs de l'espace-temps.
\item des equations d'attracteurs sous la m\'ethode de la fonction d'entropie
m\`enent pendant que (i) les \'equations de mouvement et l'autre de la th\'eorie
effective viennent comme une configuration non-supersym\'etrique, et (ii) 
les \'etats de la th\'eorie effective suivent les conditions de la supersym\'etrie.
\end{enumerate}

En g\'en\'eral, ceux-ci sont les \'equations qui peuvent nous mener \`a 
comprendre un ensemble possible associ\'e au trou noir correspondant.
Plus rigoureusement, les \'equations d'attracteurs pour les trous noirs 
BPS, dans le cas de $ \mathcal N= 2, D=4 $ avec l'aide de la conjoncture 
d'Ooguri-Strominger-Vafa (OSV) \cite{Str3} d\'efinissent que la fonction de 
cloison de la configuration des trous noirs correspondants est donn\'ee par: 

\ba \mathcal Z_{BH}(p,\phi)&:=& \vert \exp (f_{top}(p+ \frac{i}{2}\phi))\vert^2 \nn
&=& \vert \mathcal Z_{top} \vert^2.\ea

C'est-\`a-dire que pour tout prepotentiel $F$,
l'entropie du trou noir $ S_{BH}(\overrightarrow{p}, \overrightarrow{q})$ 
est \'egal \`a la transformation de Legendre d'une fonction $f:=f_{top}$,
ce qu'elle aux points attracteurs peut \^etre d\'efinie par 

\ba f= Im(F)\vert_{ attracteur }. \ea
 
Donc, nous voyons que 

\ba S_{BH}(\overrightarrow{p}, \overrightarrow{q})= 
f(\overrightarrow{e},\overrightarrow{p})- e^I \frac{\partial}{\partial e^I}
f(\overrightarrow{e},\overrightarrow{p}), \ea

o\`u $ \forall \overrightarrow{Q}:= ( \overrightarrow{q},\overrightarrow{p})$, 
les potentiels \'electriques sont d\'efinis par 

\ba \label{potee} e^I:= -\frac{\partial}{\partial q_I} S_{BH}(\overrightarrow{p}, \overrightarrow{q}), \ea

et les charges \'electriques correspondantes peuvent \^etre exprim\'ees comme

\ba q_I= \frac{\partial}{\partial e^I}f(\overrightarrow{e},\overrightarrow{p}).\ea

En ce cas, le Ref. \cite{Str3} montre que la configuration correspondante de ces 
trous noirs extermaux correspond \`a un ensemble m\'elang\'e. 

Du fait m\^eme, ce syst\`eme statistique correspond \`a un ensemble microcanonique
du point de vue magn\'etique dont les variables $p_i$ sont maintenues fixe, 
tandis qu'on a besoin d'employer l'ensemble canonique aux charges \'electriques 
avec les potentiels \'electriques $ e^I $ ce qui sont d\'efines par Eqn. (\ref{potee}).
Et ainsi, nous pouvons d\'efinir la fonction de cloison de cet ensemble m\'elang\'e 
pour ces trous noirs comme

\ba \mathcal Z_{BH}(\overrightarrow{e},\overrightarrow{p})=
\sum_{\overrightarrow{q}} \Omega(\overrightarrow{p}, \overrightarrow{q})
e^{\overrightarrow{e} \cdot \overrightarrow{p}},\ea

o\`u $ \Omega(\overrightarrow{p}, \overrightarrow{q}) $ est le nombre
entier qui d\'efinit la d\'eg\'en\'erescence des trou noir sous-jacents.
De cette fa\c{c}on nous avons:

\ba \mathcal Z_{BH}(\overrightarrow{e},\overrightarrow{p})&=& 
\sum_{\overrightarrow{q}} e^{\ln \Omega(\overrightarrow{p}, \overrightarrow{q})
+ \overrightarrow{e} \cdot \overrightarrow{p}}\nn
&=& e^{f(\overrightarrow{e}, \overrightarrow{p})}, \ea

o\`u nous avous d\'efini

\ba f(\overrightarrow{e}, \overrightarrow{p})= S_{micro}(\overrightarrow{p},
\overrightarrow{q})+ \overrightarrow{e} \cdot \overrightarrow{p}.\ea

Puisque la fonction de cloison $ \mathcal Z_{BH}(\overrightarrow{e},\overrightarrow{p}) $ est une
fonction de cloison de l'ensemble m\'elang\'ee, tellement a priori il n'est pas assez clair
que la fonction $ f(\overrightarrow{e}, \overrightarrow{p}) $ puisse \^etre interpr\'et\'ee 
comme l'\'energie libre de la configuration statistique sous-jacente de ces trous noirs ou pas.
N'importe comment, l'entropie microscopique d'un ensemble des trous noirs en suivant la th\'eorie
des cordes est donn\'ee par

\ba S_{micro}= \ln \Omega(p^I, q_I).\ea

Donc, la d\'eg\'enérescence peut \^etre obtenue en tant que la transformation 
inverse de Laplace de la fonction $f$ pour \^etre s\'elon l'\'equation

\ba \Omega(\overrightarrow{p}, \overrightarrow{q})= \int e^{f(\overrightarrow{e},
\overrightarrow{p})- \overrightarrow{e} \cdot \overrightarrow{p}} d \overrightarrow{e},\ea

o\`u 

\ba d \overrightarrow{e}:= \prod_I de^I \ea

est une mesure produite.
Nous pouvons noter pour les trous noirs grand qu'on peut utiliser 
l'approximation de point-saddelle ce qui entra\^ine l'equivalance
entr\'e les entropies microscopique et macroscopique comme:

\ba S_{micro}(p^I, q_I)= S_{BH}(p^I, q_I).\ea

Nous savons que l'\'energie libre des cordes topologiques peut \^etre
identifi\'ee avec la fonction de cloison d'un ensemble des trous noirs \cite{Str3}.
Ainsi, nous pouvons voir que l'\'energie libre $ \mathcal F(p,\phi) $ et 
la fonction de cloison $ \mathcal Z_{BH}(p,\phi) $ peuvent \^etre \'ecrites par 

\ba \pi \mathcal F(p,\phi)= \ln \mathcal Z_{BH}(p,\phi). \ea

En fait, par la transformation de Legendre, nous avons

\ba S_{macro}(p,q):= \pi ( \mathcal F(p,\phi)- q_I \phi^I), \ea

o\`u 

\ba q_I:= \frac{\partial \mathcal F}{\partial \phi^I}. \ea

De plus, il existe une fonction holomorphique dans la variabbles $\phi_i$ tels que 
la fonction de cloison d'un ensemble des trous noirs peut \'ecrire comme: 

\ba \mathcal Z_{BH}(p,\phi)&=& \sum_q \Omega(p,q) e^{i q_I \phi^I}\nn
&=& e^{\pi \mathcal F(p,\phi)}.\ea

C'est l'\'energie libre quand d\'efinit thermodynamiquement comme une donction 
de la fonction de cloison de l'ensemble canonique en consid\'erant le volume occup\'e 
dans l'espace de $ \Gamma $ peut \^etre simplement donn\'e par 

\ba \mathcal Q_N(V,T):= e^{-\frac{1}{kT} \mathcal F(V,T)}, \ea

par exemple, il s'av\`ere que l'\'energie libre d'Helmholtz est d\'efinie par

\ba \mathcal F:= M- T S, \ea 

o\`u 

\ba T:= \frac{\partial M}{\partial S}.\ea

Ainsi, nous pouvons \'ecrire simplement la transformation inverse de Legendre comme

\ba \mathcal F= \phi \cdot q+ S \ea

avec 

\ba \phi_I= -\frac{\partial S}{ \partial q^I}.\ea

De cette fa\c{c}on, l'\'el\'ement de la ligne de la g\'eom\'etrie thermodynamique 
peut \^etre param\'etris\'ee en termes d'\'energie libre des cordes topologiques:

\ba ds^2= \frac{1}{k_B} \frac{\partial^2 \mathcal F}{\partial Q^i \partial Q^j} d Q^i d Q^j, \ea

o\`u nous avons pris $ \overrightarrow{Q}= (\overrightarrow{q},\overrightarrow{p}) $.
Donc, la m\'etrique tensuere $ g_{ij}^{ \mathcal F} $ est d\'efinie par

\ba g_{ij}^{ \mathcal F}:= \frac{1}{k_B}
\frac{\partial^2 \mathcal F}{\partial Q^i \partial Q^j}.\ea

Dans le cas des deux variables, nous pouvons voir que cette m\'etrique tenseure 
peut \'ecrire \'ecrite comme 

\ba g^{ \mathcal F }
=\frac{1}{k_B} \left(\begin{array}{rr} 
\frac{\partial^2 \mathcal F}{\partial p^2 } 
& \frac{\partial^2 \mathcal F}{\partial \phi \partial p} \\
\frac{\partial^2 \mathcal F}{\partial \phi \partial p} 
& \frac{\partial^2 \mathcal F}{\partial \phi^2} \\
\end{array}\right).\ea

Nous pouvons \'egalement donner la m\'etrique en termes de la fonction de cloison 
$ \mathcal Z_{BH} $ par-ce que nous savons 

\ba \mathcal F(p,\phi)= \frac{1}{\pi} \ln \mathcal Z_{BH}(p,\phi).\ea

Donc, il est facile de voir en cette terme que la m\'etrique peut \^etre donn\'ee par  

\ba
g^{\mathcal Z}:= \frac{1}{\pi k_B \mathcal Z_{BH}^2}
\left( \begin{array}{rr} 
\mathcal Z_{BH} \partial_p \partial_p \mathcal Z_{BH}- (\partial_p \mathcal Z_{BH})^2
& \mathcal Z_{BH} \partial_{\phi} \partial_p \mathcal Z_{BH}- \partial_p \mathcal Z_{BH} \partial_{\phi} \mathcal Z_{BH} \\
\mathcal Z_{BH} \partial_{\phi} \partial_p \mathcal Z_{BH}- \partial_p \mathcal Z_{BH} \partial_{\phi} \mathcal Z_{BH} 
& \mathcal Z_{BH} \partial_{\phi} \partial_{\phi} \mathcal Z_{BH}- (\partial_{\phi} 
\mathcal Z_{BH})^2 
\end{array} \right). 
\ea

Il est bien connu \cite{AharonyGubserMaldacenaOoguriOz, Witten, BoerNaqvi, BoerPasquinucciSkenderis,
GukovGukovWitten} que la fonction de cloison $\mathcal Z_{BH}(p,\phi) $ d'un sesemble de ces 
trous noirs soit associ\'ee \`a la th\'eorie des cordes topologiques, \`a la g\'eom\'etrie de
Calabi Yau, \`a l'AdS/CFT, \`a l'holographie $ \ldots $ etc. C'est pourquoi nous avons des grands 
int\'er\^ets d'analyser les significations et les interpr\'etations g\'eom\'etriques 
et thermodynamiques en termes des $ \mathcal Z_{BH} $ ou bien $ \mathcal F $.
En particulier, il est vraiment une question importante de la mode 
qu'elles sont les significations thermodynamiques des equations d'attracteur
en pr\'esence des d\'eriv\'ees sup\'erieures de l'espace-temps?

De plus, les r\'esultats de l'entropie macroscopiques et ceux de
microscopiques pour les petites corrections de $\alpha^{\prime}$ sont
perturbativement les m\^emes importance et ainsi nous devons avoir:

\ba S_{BH}> A/4+\ldots,\ea

o\`u $ \ldots $ repr\'esentent les corrections de $\alpha^{\prime}$ 
\`a l'entropie macroscopique du syst\`eme consid\'er\'e des trous noirs.
Pour le cas de z\'ero magn\'etique charge, Dabholker \cite{Dabholkar} 
a montr\'e, dans la limite des grandes charges, que nous avons: 

\ba S_{micro}= 4 \pi \sqrt{nw}.\ea

Pour les trous noirs petits des deux charges, 
nous avons analys\'e la g\'eom\'etrie thermodynamique aux
ordres diff\'erents de corrections de $\alpha^{\prime}$,
voir \cite{Sar2} pour le d\'etails.
Nous avons d\'emontr\'e que la g\'eom\'etrie thermodynamique 
des petits trous noirs corrig\'ee par les d\'eriv\'ees sup\'erieures 
au premier ordre des corrections de $\alpha^{\prime}$ est mal d\'efinie.
Cependant, avec les prochaines ordres de corrections de $\alpha^{\prime}$,
la g\'eom\'etrie thermodynamique des petits trous noirs est bien d\'efinie
et l'espace des \'etats est partout ordinaire.	
Dans l'analyse des branes noirs extr\'emaux $D_1D_5 $ et $ D_2D_6NS_5 $,
nous trouvons que la g\'eom\'etrie thermodynamique de ces branes
est bien d\'efinie et la courbure de Ruppenier reste partout finie
avec et sans les corrections de $\alpha^{\prime}$.
Maintenant, dans la suite de cet ouvrage, 
nous allons analyser la g\'eom\'etrie thermodynamique 
des trous noirs de Reissner-Nordst\"orm, 
des trous noirs dilatoniques topologiques,
des trous noirs de Reissner-Nordstr\"om dans la nappe de Poincar\'e de $ ADS_4 $,
des trous noirs de Reissner-Nordstr\"om corrig\'e par le principe d'incertitude
g\'en\'eralis\'ee et celle des trous noirs magn\'etis\'es.
Ensuite, nous allons voir les corrections de $\alpha^{\prime}$ des trous noirs 
dyoniques extr\'emaux supersym\'etriques et celle des non-supersym\'etriques,
des solutions non-extr\'emales de branes $ D_1D_5 $ et $ D_2D_6NS_5 $, 
des trous noirs extr\'emaux en rotation comme
les trous noirs extr\'emaux de Kerr-Newman dans la th\'eorie d'Einstein Maxwell,
les trous noirs extr\'emaux de Kaluza-Klein dans la th\'eorie d'Einstein-Maxwell 
et les trous noirs extr\'emaux de la th\'eorie des cordes h\'et\'erotiques 
compactifi\'ee toroidalement.
Nous allons montr\'ere que les r\'esultats de la g\'eom\'etrie thermodynamique
sont \'eclairantes et sont en accord avec les notions de la thermodynamique 
et la m\'ecanique statistique des trous noirs et branes noirs.

\section{La g\'eom\'etrie de Ruppenier des trous noirs de Reissner-Nordstr\"om.}

Dans cette section, nous analysons la g\'eom\'etrie de Ruppenier d'un
ensemble des trous noirs de Reissner-Nordstr\"om. C'est le plus simple 
syst\`eme des trous noirs pour lequel il est facilement possible d'analyser 
la g\'eom\'etrie thermodynamique. Du point de vue de la thermodynamique 
des trous noirs, l'enqu\^ete de la g\'eom\'etrie thermodynamique covariante 
de Ruppeiner peut \^etre appliqu\'ee pour \'etudier la nature de l'entropie 
des trous noirs de Reissner-Nordstr\"om. Cela a \'et\'e explor\'ee pour la 
premi\`ere fois dans le contexte des configurations de trous noirs extr\'emaux
charg\'es de BPS de la supergravit\'e de $ \mathcal N = 2 $ \cite{fgk}.
Ensuite, il y avait des plusieurs auteurs qui ont essay\'e de comprendre 
ce sujet \cite{cai1, aman1, aman2, aman3}, \`a la fois pour les trous
noirs supersym\'etriques, et bien aussi quelque fois les trous noirs
non-supersym\'etriques. Les trous noirs charg\'es extr\'emaux de la 
supergravit\'e de $ \mathcal N = 2 $ interagissant avec le multiple 
des vecteurs et celui du multiple des hypers sont d\'ecrits par la 
m\'etrique tenseure de l'espace-temps de Reissner-Nordst\"orm.
Ce sont les BPS-solitons en interpolatant entre l'espace asymptotiquement 
plat de Minkowski et celui de la g\'eom\'etrie proche de l'horizon de
trou noir de Bertotti-Robinson. Du m\'ecanisme attracteur, l'entropie 
macroscopique de ces trous noirs est d\'etermin\'ee uniquement, 
comme une fonction des charges du trou noir et donc est ind\'ependante
des valeurs asymptotiques des modules de champs. Cette solution a un 
point fixe d'attracteur est atteinte \`a l'horizon du trou noir.
Et ainsi, l'aire de l'horizon du trou noir est d\'efinie par l'extremum 
de la charge centrale d'espace des modules.

De toute mani\`ere, l'entropie est une fonction de la masse et des 
charges \'electriques et magn\'etiques du trou noir.
Cela correspond \`a l'entropie d'Hawking-Bekenstein aux deux d\'eriv\'es.
De plus, ce syst\`eme est un exemple typique d'un trou noir extr\'emal 
sph\'eriquement sym\'etrique dans quatre dimensions sans avoir \`a 
proximit\'e de l'horizon singulier, la g\'eom\'etrie de l'horizon est 
$ AdS_2 \times S^2 $. Et donc, le trou noir de Reissner-Nordstr\"om a 
une symm\'etrie de $ SO(2,1) \times SO(3) $. Dans la th\'eorie d'Einstein-Maxwell 
en quatre dimensions, nous savons par la fonction de l'entropie de Sen 
\cite{AshokeSen} que l'entropie du trou noir de Reissner-Nordstr\"om 
est donn\'ee par:

\ba S_{BH}(q,p) = \frac{1}{4}(p^2+ q^2).\ea

Ainsi, la m\'etrique tenseure covariante thermodynamique de Ruppenier
de trou noir de Reissner-Nordstrom est donn\'ee par:

\ba g_{ij}(x)&:=& -\frac{\partial^2 S(x)}{\partial x^i \partial x^j}\nn
&=& \left( \begin{array}{rr}
    -\frac{1}{2} & 0 \\
     0 & -\frac{1}{2} \\
\end{array} \right).\ea

Dans ce cas, le premier r\'esultat est imm\'ediat de voir que le d\'eterminant
de la m\'etrique  tenseure est donn\'e par 

\ba g= 1/4, \ea 

et bien aussi il s'av\`ere que tous les symbols de Christoffel sont trivialement nuls. 
Donc, on voit tr\`es simplement que la courbure scalaire de Ruppenier est aussi nulle.
En conclusion, cette  g\'eom\'etrie thermodynamique est bien d\'efinie 
et d\'ecrit un syst\`eme statistique sans interaction.

\section{La g\'eom\'etrie de Wienhold des trous noirs dilatoniques.}

Dans cette section, nous \'etudions une famille des trous noirs dans la th\'eorie 
de la gravitation. Dans le cadre de la th\'eories de gravitation dilatonique inspir\'ee
par les th\'eories des cordes, nous consid\'erons les effects thermodynamiques de quelques 
nouvelles solutions de trous noirs ou branes noires ce qui sont asymptotiquement non-plates. 
Puis nous calculons les leurs effets thermodynamiques comme l'interaction, la transition des phases,
$\ldots$, etc dans les syst\`emes de ces trous noirs. Ici, les tous ce qu'il concerne \`a nous, 
sont les produits et accessoires de la g\'eom\'etrie thermodynamique, voir la section pr\'ec\'edente
pour la g\'em\'etrie de Ruppenier d'une famille des trous noirs de deux param\`etres.


Ici, nous consid\'erons maintenant la g\'eom\'etrie thermodynamique des
trous noirs charg\'es dilatoniques topologiques \cite{AhmadSheykhi}.
En particulier, nous expliquons une autre g\'eom\'etrie pour le cas de ces 
trous noirs dilatoniques topologiques, qui est associ\'ee conformement \`a 
la g\'eom\'etrie de Ruppenier ce qu'elle est bien importante dans cet \'etude.
Cette g\'eom\'etrie thermodynamique s'appelle la g\'eom\'etrie de Wienhold.
Afin de faire ceci, consid\'erons une $(n+1)$ dimensionnelle configuration 
arbitraire de la gravit\'e dilatonique d'Einstein-Maxwell pour tout $n \geq 3$.
En fait, nous avons besoin d'obtenir la masse $M$ de trou noir comme une fonction 
des quantit\'es extensives $\{S, Q\}$ et alors, nous avons un type du formule de
Smarr \cite{AhmadSheykhi}, ce qui est donn\'e par:

\ba \label{Smarr} M(S,Q)&=& -\frac{k(n-1)(n-2) (\alpha^2+1)b^{-\alpha^2}}{16\pi (\alpha^2-1)
(\alpha^2+ n- 2)} (4S)^{\frac{\alpha^2+ n- 2}{n- 1}}\nn && + \frac{\Lambda}{8\pi}
\frac{(\alpha^2+1)b^{\alpha^2}}
{\alpha^2-n}(4S)^{\frac{n- \alpha^2}{n- 1}}+ 
\frac{2\pi (\alpha^2+1)b^{\alpha^2}}{\alpha^2+ n- 2}
Q^2 (4S)^{\frac{\alpha^2+ n- 2}{1- n}}, \ea

o\`u $k$ d\'etermine la nature de l'horizon du trou noir ou celle de l'horizon cosmologique.
En particulier, les valeurs $k= 0, 1, -1$ respectivement donnent des hypersurfaces de l\'espace-temps 
avec des courbures scalaires constantes plates, elliptiques et hyperboliques. Dans cette consid\'eration,
le param\`etre $ \alpha $ est une constante d'accouplement du dilaton, $b$ est une constante 
arbitraire et le param\`etre libre 

\ba \Lambda:= -\frac{n(n-1)}{2l^2} \ea

joue le r\^ole de la constante de cosmologie.
En fait, par le biais de la g\'eom\'etrie thermodynamique,
nous pouvons bien s\^ur \'etudier la stabilit\'e des trous noirs 
dilatoniques topologiques. En consid\'erant $S$ et $Q$ comme l'ensemble 
complet des variables extensives pour la masse $M(S, Q)$ de trou noir.
Alors que la masse de ce trou noir charg\'e dilatonique topologique est
donn\'ee par un formule de Smarr, voir Eqn. (\ref{Smarr}) ci-dessus.
En ce cas, il est facile voir que les composantes de la m\'etrique 
de Wienhold sont donn\'es par:

\ba g_{SS} &=& \frac{k (n-2) (\alpha^2+1)b^{-\alpha^2}(1-\alpha^2)}{16 \pi (\alpha^2-1)
S^2(n-1)}(4 S)^{\frac{\alpha^2 + n - 2}{n - 1}} \nn &&
+ \frac{\Lambda (\alpha^2+ 1) b^{\alpha^2}(n^2- \alpha^2)(n^2- \alpha^2- n+ 1)}
{8 \pi (\alpha^2- 1) (n - 1)^2 S^2}(4S)^{\frac{n^2- \alpha^2}{n- 1}} \nn &&
+ \frac{2 \pi b^{\alpha^2} (\alpha^2+ 1) Q^2 (\alpha^2+2n- 3)}{(1 - n)^2 S^2}
(4S)^{\frac{\alpha^2+ n- 2}{1 - n}},\nn
g_{SQ}&=& \frac{4 \pi b^{\alpha^2} (\alpha^2+ 1) Q}{(1 - n)S}
(4S)^{\frac{\alpha^2+ n- 2}{1 - n}},\nn
g_{QQ}&=& \frac{4 \pi b^{\alpha^2} (\alpha^2+ 1) }{\alpha^2+ n- 2}
(4S)^{\frac{\alpha^2+ n- 2}{1 - n}}.\ea

\begin{figure}
\hspace*{1.0cm}\vspace*{-6.0cm}
\includegraphics[width=12.0cm,angle=-0]{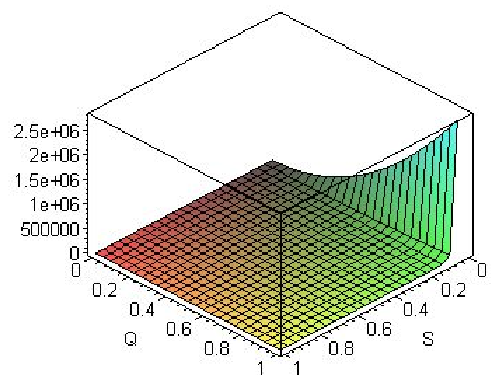}\vspace*{-3.0cm}
\caption{La composante $SS$ de la m\'etrique tenseure trac\'e comme la 
fonction de $\{Q, S\}$, en d\'ecrivant les fluctuations dans la 
configuration des trous noirs dilatoniques.} \label{diltopSS}\vspace*{0.5cm}
\end{figure}

\begin{figure}
\hspace*{1.0cm}\vspace*{-6.0cm}
\includegraphics[width=12.0cm,angle=-0]{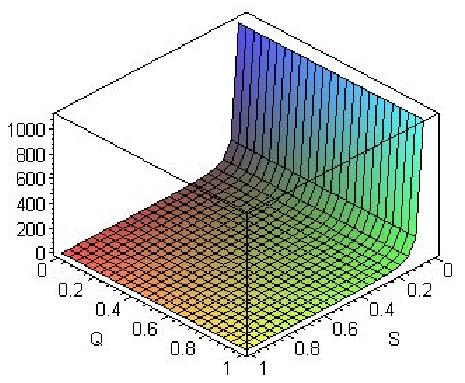}\vspace*{-3.0cm}
\caption{La composante $QQ$ de la m\'etrique tenseure trac\'e comme la 
fonction de $\{Q, S\}$, en d\'ecrivant les fluctuations dans la 
configuration des trous noirs dilatoniques.} \label{diltopQQ}\vspace*{0.5cm}
\end{figure}

\begin{figure}
\hspace*{1.0cm}\vspace*{-6.0cm}
\includegraphics[width=12.0cm,angle=-0]{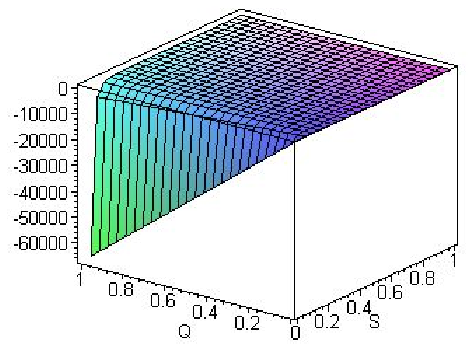}\vspace*{-3.0cm}
\caption{La composante $SQ$ de la m\'etrique tenseure trac\'e comme la 
fonction de $\{Q, S\}$, en d\'ecrivant les fluctuations dans la 
configuration des trous noirs dilatoniques.} \label{diltopSQ}\vspace*{0.5cm}
\end{figure}

Pour la valeur de $n=3$, $k=-1$, $b= 1$, $\alpha= 2$ et $\Lambda= -3$, les Figs.
(\ref{diltopSS}, \ref{diltopQQ}) montrent la nature des composantes 
$\{g_{SS}, g_{QQ}\}$ de la m\'etrique tenseure de Wienhold. Dans le r\'egime 
de $Q \in (0,1)$ et $S \in (0,1)$, on trouve que l'amplitude de $\{g_{SS}\}$ 
prend la valeur \`a l'ordre de $2.5 \times 10^{+06}$. Dans cette gamme de 
$\{S, Q\}$, on observe que la composante $\{g_{QQ}\}$ se situe dans la gamme 
de $(0, 1000)$. Dans ce cas, nous voyons que la gamme de la croissance des amplitudes
de $\{g_{SS}\}$ et $\{g_{QQ}\}$ reste dans la m\^{e}me limite de $S$.
Explicitement, les Figs. (\ref{diltopSS}, \ref{diltopQQ}) indiquent que
la croissance de la premi\`ere composante est \`a lieu dans la limite d'une petite $S$.
Cela signifie que les trous noirs dilatoniques sont thermodynamiquement instables 
dans la limite d'une petite entropie. Du fait m\^{e}me, la Fig. (\ref{diltopSQ})
montre que la nature de la composante $\{g_{SQ}\}$ de la m\'etrique tenseure thermodynamique. 
Nous constatons que la composante mixe $\{g_{SQ}\}$ prend une grande valeur \`a l'ordre 
de $-6000 $ dans la limite d'une grande $Q$ et une petite $S$. Ainsi, les compressibilit\'es 
de chaleurs, telle que repr\'esent\'ees dans les Figs. (\ref{diltopSS}, \ref{diltopQQ}, \ref{diltopSQ}) 
illustrent les propri\'et\'es param\'etrique des fluctuations de la configuration des trous noirs dilatoniques. 
Par la pr\'esente, nous remarquons que les fluctuations d'auto-paires en impliquant $\{S, Q\}$, 
qui sont d\'efinies par la m\'etrique tenseure $\{g_{ii} \ | \ i = S, Q\}$, 
ont des valeurs num\'eriques positives, tandis que cela ne marche pas pour 
la composante mixe $\{g_{SQ}\}$.

On peut voir que le d\'eterminant de la m\'etrique de Wienhold est donn\'e par:

\ba \Vert g_{ab} \Vert&=& g_{SS}g_{QQ}- g_{SQ}^2 \nn 
&=& -\frac{(\alpha^2+1)^2 b^{\alpha^2} (4 S)^{-\frac{\alpha^2 + n - 2}{n - 1}}}
{4 (\alpha^2+ n- 2) (\alpha^2- n) (n - 1)^2 S^2} 
[ 2 \Lambda b^{\alpha^2} f_1(\alpha) (4 S)^{-\frac{\alpha^2- n^2}{n - 1}} \nn &&
+ 32 \pi^2 b^{\alpha^2}Q^2 f_2(\alpha) (4 S)^{-\frac{\alpha^2 + n - 2}{n - 1}}
+ k b^{-\alpha^2} f_3(\alpha)(4 S)^{\frac{\alpha^2 + n - 2}{n - 1}}], \ea

o\`u

\ba f_1(\alpha)&=& -\alpha^4+ \alpha^2- n^4- n^2+ n^3- n \alpha^2+ 2 n^2 \alpha^2, \nn
f_2(\alpha)&=& n+ \alpha^4- n \alpha^2- \alpha^2, \nn 
f_3(\alpha)&=& - 3 n \alpha^2+ 2 \alpha^2+ n^2 \alpha^2- 2 n+ 3 n^2- n^3. \ea

\begin{figure}
\hspace*{1.0cm}\vspace*{-6.0cm}
\includegraphics[width=12.0cm,angle=-0]{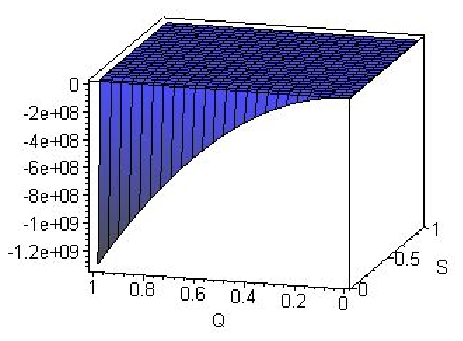}\vspace*{-3.0cm}
\caption{Le d\'eterminant de la m\'etrique tenseure trac\'ee comme la
fonction de $\{Q, S\}$, en d\'ecrivant les fluctuations de la 
configuration des trous noirs dilatoniques.} \label{diltopdetg}\vspace*{0.5cm}
\end{figure}

\begin{figure}
\hspace*{1.0cm}\vspace*{-6.0cm}
\includegraphics[width=12.0cm,angle=-0]{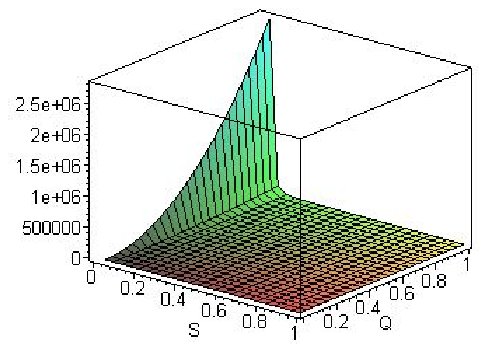}\vspace*{-3.0cm}
\caption{Le premier mineur de la m\'etriquetenseure trac\'ee comme la
fonction de $\{Q, S\}$, en d\'ecrivant les fluctuations de la 
configuration des trous noirs dilatoniques.} \label{diltopminor1}\vspace*{0.5cm}
\end{figure}

\`A ce stade, nous constatons que la stabilit\'e d'un ensemble des 
trous noirs dilatoniques peut \^{e}tre d\'etermin\'e en termes des valeurs
de param\`etres $\{S, Q\}$. Cela d\'ecoule du comportement de d\'eterminant 
de la m\'etrique tenseure thermodynamique. Dans ce cas, nous trouvons que 
le d\'eterminant de la m\'etrique tenseure tend vers une valeure n\'egative.
Pour le cas de $Q \in (0, 1)$ et $S \in (0, 1)$, la Fig.(\ref{diltopdetg}) 
montre que le d\'eterminant de la m\'etrique tenseure r\'eside dans l'intervalle 
$(-1.2 \times 10^{09}, 0)$. Avis que la n\'egativit\'e de $g$ augmente \`a mesure 
que la valeur de $Q$ est augment\'e, en passant de z\'ero \`a $1$. En outre, 
la surface d\'efinie par les fluctuations de $(S, Q)$ est instable en raison d'une 
valeur n\'egative du d\'eterminant de la m\'etrique tenseure correspondante. 
Lorsque seul le param\`etre $S$ est autoris\'e \`a varier, la stabilit\'e des 
configurations trous noirs dilatoniques est d\'etermin\'e par la positivit\'e du 
premier mineur principe $p_1: = g_{SS}$. Une vue rotat\'ee de $p_1$ est montr\'e 
dans la Fig. (\ref{diltopminor1}). Par la pr\'esente, pour une petite valeur de $S$, 
nous voyons que $p_1$ augmente \'a mesure que la fonction de la charge \'electrique $Q$. 
Les propri\'et\'es graphiques, mentionn\'ees ci-dessus, et la positivit\'e des mineurs 
principaux fournent la notion qualitative de la stabilit\'e statistiqu des trous noirs 
dilatoniques de deux param\`etres $\{S, Q\}$.

Comme nous avons fourni les $\Gamma_{abc}$ dans l'annex $[A]$,
maintenant, nous pouvons calculer les $R_{abcd}, R_{ab}$ 
et alors la courbure scalaire de Wienhold est obtenue pour \^etre:

\ba R&=& -\frac{16 \pi (\alpha^2- n)(\alpha^2+ n- 2)}{\alpha^2+ 1} 
\bigg( 2 \Lambda b^{\alpha^2}f_1(\alpha) (4 S)^{-\frac{\alpha^2- n^2}{n - 1}} \nn &&
+ 32 \pi^2 b^{\alpha^2}Q^2 f_2(\alpha)(4 S)^{-\frac{\alpha^2 + n - 2}{n - 1}} 
+ k b^{-\alpha^2}f_3(\alpha)(4 S)^{\frac{\alpha^2 + n - 2}{n - 1}}\bigg)^{-2}
\nn && \times \bigg( k b^{-\alpha^2} 
f_4(\alpha) (4S)^{\frac{\alpha^2+n-2}{n-1}}
+ \Lambda b^{\alpha^2} f_5(\alpha)
(4S)^{-\frac{\alpha^2- n^2}{n-1}} \bigg), \ea

o\`u

\ba
f_4(\alpha)&=& -2 \alpha^2+ 2 \alpha^4+ 2 n- n^3 \alpha^2+ n \alpha^2  \nn &&
- 3 n \alpha^4+ 2 n^2 \alpha^2+ n^2 \alpha^4- 3 n^2+ n^3, \nn
f_5(\alpha)&=& -2n^4+ 2 n^5- n^6+ 2 n^4 \alpha^2- n^2 \alpha^4  \nn &&
+ 2 n^2 \alpha^2- 3 n^3 \alpha^2+ n \alpha^4- n \alpha^2+ n^3. \ea 

Nous voyons pour toutes les param\'etrisations q'elles sont bien d\'efinies par la 
g\'eom\'etrie thermodynamique que cette courbure de Weinhold est partout r\'eguli\`ere.
Donc, il s'agit d'un syst\`eme thermodynamique stable et il n'ya pas
d'instabilit\'e dans l'espace thermodynamique \`a la repr\'esentation
de l'\'energie des trous noirs charg\'es dilatoniques topologiques.

En outre, il y certains cas de $\alpha$ pour que la courbure scalaire
de Wienhold soit \'egale \`a la z\'ero. Ces valeurs de $\alpha$ sont
donn\'ees par les equations suivantes:

\ba (n^2- 3 n+ 2) \alpha^4 -( n^3- 2 n^2- n+ 2) \alpha^2+ n^3- 3 n^2+ 2 n= 0\ea

et

\ba (1- n)\alpha^4+ (2 n^3- 3 n^2+ 2 n- n) \alpha^2+ n^2- 2n^3+ 2 n^4- n^5= 0.\ea

Ce sont les valeurs de la constante d'accouplement du dilaton dont lesquelles
la courbure scalaire de Wienhold est nulle. Ces solutions sont simplement

\ba \vert \alpha \vert= \lbrace 1/2  (n^2- 3 n+ 2)^{-1}( n^3- 2 n^2- n+ 2
\pm \sqrt{\tilde{\alpha}(n)})\rbrace^{1/2}, \ea

o\`u $\tilde{\alpha}(n)= n^6- 8n^5+26n^4+60n^3+41n^2- 8n+4$ et l'autre est

\ba \vert \alpha \vert= \lbrace 1/2 (n-1)^{-1}
( 2n^3- 3n^2+ 2n- 1 \mp \sqrt{n^4- 6n^2- 4n+ 1})\rbrace^{1/2}. \ea

De plus, les cas de 

\ba n^6- 8n^5+26n^4+60n^3+41n^2- 8n+4 \geq 0 \ea

et 

\ba n^4- 6n^2- 4n+ 1 \geq 0 \ea

impliquent des valeurs physiques de la constante d'accouplement du dilaton, 
et ainsi \c{c}a nous permettre de d\'eterminer le r\^ole de la constante 
de cosmologie ou celui de la dimension de l'espace-temps des th\'eories
de la gravit\'e dilatonique dans la cadre d'Einstein-Maxwell.
C'est-\`a-dire que $\forall 2 < n \in Z$, la constante d'accouplement du 
dilaton appartient \`a une configuration statistique sans les interactions, 
dont lesquels $n$ satisfait ces deux \'equations du syst\`eme.

\begin{figure}
\hspace*{1.0cm}\vspace*{-6.0cm}
\includegraphics[width=12.0cm,angle=-0]{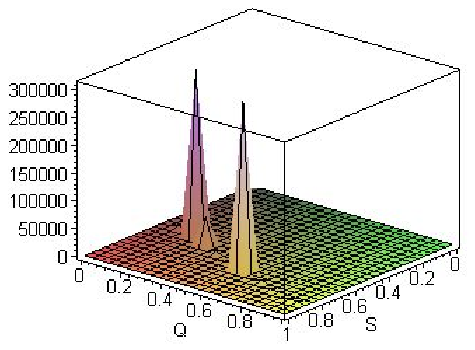}\vspace*{-3.0cm}
\caption{La courbure scalaire trac\'ee en fonction de $\{Q, S\}$,
d\'ecrivant les variations dans la configuration des trous noirs dilatoniques
dans la gamme $S, Q \in (0,1) $.} \label{diltopcur}\vspace*{0.5cm}
\end{figure}

\begin{figure}
\hspace*{1.0cm}\vspace*{-6.0cm}
\includegraphics[width=12.0cm,angle=-0]{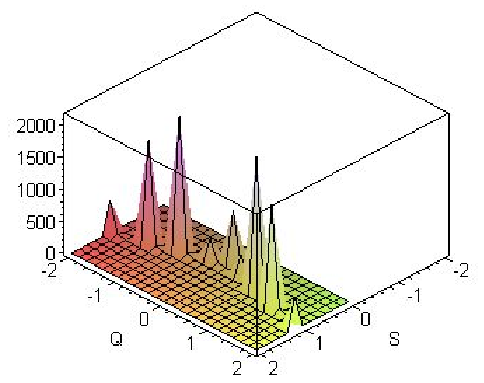}\vspace*{-3.0cm}
\caption{La courbure scalaire trac\'ee en fonction de $\{Q, S\}$,
d\'ecrivant les variations dans la configuration des trous noirs dilatoniques
dans la gamme $S, Q \in (-2, 2)$.} \label{diltopcur2}\vspace*{0.5cm}
\end{figure}

Les propri\'et\'es de la stabilit\'e globale des trous noirs dilatoniques
d\'ecoulent par le comportement de la courbure scalaire thermodynamique.
En particulier, dans la gamme de $S \in (0,1) $ et $Q \in (0,1)$, la
Fig. (\ref{diltopcur}) montre que la courbure scalaire a une grande 
amplitude positive \`a l'ordre de $300 000 $. Cela montre que la configuration 
sous-jacente de ces trous noirs est un syst\`eme statistique fortement r\'epulsif. 
Le signe positif de la courbure scalaire signifie la nature r\'epulsive 
des interactions statistiques. La Fig. (\ref{diltopcur2}) signifie 
la nature de ce qui pr\'ec\`ede de la courbure scalaire dans une range 
\'egale des param\`etres $S, Q \in (-2, 2)$. Dans ce cas, nous remarquons 
de la Fig. (\ref{diltopcur2}) qu'il existe des nombreuses r\'egions globale 
des interactions statistiques. En comparaison des interactions apparaissant 
dans la gamme de $S, Q \in (-2, 2) $, l'amplitude des interactions statistiques 
globales semble \^{e}tre beaucoup plus petites que cele qui figure dans la gamme de 
$S, Q \in (0, 1) $. De plus, nous voyons que la valeur typique du ratio des courbures 
scalaires de l'espace d'\'etat qui appara\^{i}t est \`a l'ordre de $150 $. 
En bref, les chiffres ci-dessus indiquent que les trous noirs dilatoniques correspondent 
\`a une configuration statistique instable, lorsque les deux param\`etres $\{S, Q\}$ 
sont autoris\'es \`a fluctuer.

\section{ La g\'eom\'etrie de Wienhold des solutions de $ M_2$- branes:
Les trous noirs de Reissner-Nordstr\"om dans la nappe de Poincar\'e de $ ADS_4 $.}

Dans cette section, nous analysons la g\'eom\'etrie thermodynamique de Weinhold 
des trous noirs de la $M$-th\'eorie qui d\'ecoulant naturalement comme une
solution des trous noirs de Reissner-Nordstr\"om dans l'espace $AdS_4$.
La relation entre l'instabilit\'e thermodynamique et celle de
Gregory-Laflamme est un probl\`eme important pour comprendre la
condensation de certains bosons ayant en charge globale, par exemple
le cas des trous noirs charg\'es \'electriquement dans l'espace $AdS_5$.
Maintenant, nous allons examiner le plongement de la solution de 
Reissner-Nordstr\"om de $ AdS_4 $ dans la $M$-th\'eorie fondamentale.
Cette incorporation peut \^etre faite comme la suite:
Envisagons un grand nombre des $M_2$-branes qui co\"incidnts de la 
supergravit\'e d'onze dimensions, avec la g\'eom\'etrie de proche 
de l'horizon de $AdS_4  \times S^7 $ du trou noir sous-jacent.
Puis, il y a huit dimensions transversales aux $M_2$-branes, et les quatre 
moments angulaires ind\'ependants ce que les branes peuvent acqu\'erir, 
s'ils sont proches de la limite de l'extremalit\'e. Pour l'ensemble des 
quatre moments angulaires \`a l'\'egale, la solution est un produit 
enroul\'e de la solution de Reissner-Nordstr\"om dans l'espace $AdS_4$  et 
une d\'eformation de $S^7$, voir le Ref. \cite{ChamblinEmparanJohnsonMyers}
pour avoir plus des d\'etails.
	
Pr\'ecis\'ement, la r\'eduction de la filature sur $S^7$, 
pour la solution quasi-extr\'emale des $M_2$-branes noirs 
asymptotiquement plat dans l'espace-temps de la dimension $D= 11$, 
est une limite de la solution de Reissner-Nordstr\"om dans l'espace $AdS_4$. 
Cette solution est la r\'eduction bien connue dans la supergravit\'e
d'onze dimensions qui est asymptotiquement r\'eduite \`a l'espace $AdS_4 \times S^7$.
C'est par-ce que la solution de Reissner-Nordstr\"om dans l'espace $ AdS_4 $ 
est une solution dans la supergravit\'e jaug\'ee de $\mathcal N= 8$,
dont le vide de l'espace $ AdS_4 $ est la r\'eduction supersym\'etrique maximum 
de Kaluza-Klein du vide de l'espace $ AdS_4 \times S^7$ de la $M$-th\'eorie.
De plus, la supergravit\'e jaug\'ee de $\mathcal N= 8$ est une truncation
consistente de la supergravit\'e de la dimension $D= 11$ de l'espace-temps \cite{WitNicolai}.
Donc, toutes les solutions classiques de la th\'eorie de $D= 4$ ont des
ascenseurs \`a une solution exacte classique de la th\'eorie de $D= 11$.
Ainsi, toute les instabilit\'es actuellent en $D= 4$ sont garantiement \`a
persister dans la dimension $D= 11$ de l'espace-temps.
Notez bien que seulement pour une gamme sp\'ecifique des param\`etres,
comme l'entropie, charges, masse et d'autres, s'il y a, nous avons les 
solutions souhait\'ees de ces trous noirs.
Sinon, les solutions ont tendance \`a avoir certaines singularit\'es nues.
Par exemple, nous consid\'erons certaines charges physiques conserv\'ees en
correspondant aux quatre ind\'ependantes moments angulaires des $M_2$-branes
vivant en onze dimensions de l'espace-temps.

En ce qui concerne ici est la correspondence d'AdS/CFT, dont Maldacena a 
conjectur\'e \cite{Maldacena,AharonyGubserMaldacenaOoguriOz} qu'il y a une 
th\'eorie de la supergravit\'e en $D= 11 $ sur l'espace $ AdS_4 \times S^7$
qui est physiquement \'equivalente \'a la limite de grand $N$ de certaine th\'eorie
des champs conformes vivant \`a la fronti\`ere de l'espace $ AdS_4 $ et repr\'esent
la limite de l'\'energie basse de la dynamique de $\langle\langle$ worldvolume
$\rangle\rangle$ des $ M_2 $-branes co\"incidant de nombre $N$.
Donc, les charges \'electriques de la th\'eorie devenues les charges globales
de la $R$-symm\'etrie de la th\'eorie des champs conformes \`a la fronti\`ere.
Pour une masse suffisamment grande, les solutions correspondentes aux \'etats
thermiques de la th\'eorie des champs conformes avec les potentiels chimiques 
lors que les charges globales sont non-nulles.
En fait, pour le cas des trous noirs avec les charges \'electriques dans 
l'espace $AdS_5$, la description duale de l'instabilit\'e thermodynamique est une 
instabilit\'e vers la condensation des bosons transportant les charges 
globaux des certains nombres finis de $U(1)$.
Pour les raisons de la simplicit\'e, nous tournons maintenant
pour le cas de $Q_1= Q_3$ et $Q_2 = Q_4$ et allons de ne pas consid\'erer 
que la limite de grands trous noirs: $ M / L>> 1 $.
Comme $M /L \rightarrow \infty $, on obtient une solution des branes
noirs dans la nappe de Poincar\'e de l'espace $AdS_4$ \cite{GubserMitra1, GubserMitra2}. 
Dans cette limite, la $S^2$ est remplac\'e par $R^2$ dans la m\'etrique de 
l'espace-temps, et ainsi la masse de ces trous noirs de Reissner-Nordstr\"om 
est donn\'ee par une expression simple:

\ba M(S,Q_1,Q_2):= \frac{1}{2 \pi L^2}\sqrt{\frac{S^3}{\pi}+ 
\pi L^2 (Q_1^2+ Q_2^2 ) S+ \frac{\pi^3 L^4}{S} Q_1^2 Q_2^2 }.\ea

Donc, les composantes de la m\'etrique tenseure de Wienhold sont donn\'ees par

\ba g_{SS}&=& -\frac{1}{8} \frac{(\frac{3 S^2}{\pi}+ \pi L^2Q_1^2+ \pi L^2Q_2^2- 
\frac{\pi^3 L^4Q_1^2Q_2^2}{S^2})^2}{\pi L^2 (\frac{S^3}{\pi}+ \pi L^2Q_1^2 S+ 
\pi L^2Q_2^2 S+ \frac{\pi^3 L^4 Q_1^2 Q_2^2}{S})^{3/2}} \nn && + 
\frac{\frac{3S}{2\pi}+\frac{\pi^3L^4Q_1^2Q_2^2}{2S^3}}{\pi L^2 \sqrt{\frac{S^3}{\pi}+ 
\pi L^2 Q_1^2 S+ \pi L^2 Q_2^2 S+ \frac{\pi^3 L^4 Q_1^2 Q_2^2}{S}}},\nn
g_{SQ_1}&=& -\frac{1}{4}\frac{(\pi L^2 Q_1 S +\frac{\pi^3 L^4 Q_1 Q_2^2}{S})(3\frac{S^2}{\pi}+ 
\pi L^2Q_1^2+ \pi L^2Q_2^2- \frac{\pi^3 L^4Q_1^2Q_2^2}{S^2})}{\pi L^2 (\frac{S^3}{\pi}+ 
\pi L^2Q_1^2 S+ \pi L^2Q_2^2 S+ \frac{\pi^3 L^4 Q_1^2 Q_2^2}{S})^{3/2}} \nn &&
+\frac{\frac{\pi L^2 Q_1}{2}- \frac{\pi^3 L^4 Q_1 Q_2^2}{2S^2}}{\pi L^2 \sqrt{\frac{S^3}{\pi}+ 
\pi L^2 Q_1^2 S+ \pi L^2 Q_2^2 S+ \frac{\pi^3 L^4 Q_1^2 Q_2^2}{S}}},\nn
g_{SQ_2}&=& -\frac{1}{4}\frac{(\pi L^2 Q_2 S +\frac{\pi^3 L^4 Q_2 Q_1^2}{S})(3\frac{S^2}{\pi}+ 
\pi L^2Q_1^2+ \pi L^2Q_2^2- \frac{\pi^3 L^4Q_1^2Q_2^2}{S^2})}{\pi L^2 (\frac{S^3}{\pi}+ 
\pi L^2Q_1^2 S+ \pi L^2Q_2^2 S+ \frac{\pi^3 L^4 Q_1^2 Q_2^2}{S})^{3/2}} \nn &&
+\frac{\frac{\pi L^2 Q_2}{2}- \frac{\pi^3 L^4 Q_1^2 Q_2}{2S^2}}{\pi L^2 \sqrt{\frac{S^3}{\pi}+ 
\pi L^2 Q_1^2 S+ \pi L^2 Q_2^2 S+ \frac{\pi^3 L^4 Q_1^2 Q_2^2}{S}}},\ea
\ba g_{Q_1Q_1}&=& -\frac{1}{2}\frac{( \pi L^2 Q_1 S+ \frac{ \pi^3 L^4 Q_1 Q_2^2}{S})^2}
{\pi L^2 (\frac{S^3}{\pi}+ \pi L^2Q_1^2 S+ \pi L^2Q_2^2 S+ \frac{\pi^3 L^4 Q_1^2 Q_2^2}{S})^{3/2}} \nn &&
+\frac{\frac{\pi L^2 S}{2}+ \frac{\pi^3 L^4 Q_2^2}{2S}}{\pi L^2 \sqrt{\frac{S^3}{\pi}+
\pi L^2 Q_1^2 S+ \pi L^2 Q_2^2 S+ \frac{\pi^3 L^4 Q_1^2 Q_2^2}{S}}},\nn
g_{Q_1Q_2}&=&-\frac{1}{2}\frac{(\pi L^2 Q_1 S+\frac{\pi^3 L^4 Q_1 Q_2^2}{S})
(\pi L^2 Q_2 S +\frac{\pi^3 L^4 Q_2 Q_1^2}{S})}
{\pi L^2 (\frac{S^3}{\pi}+ \pi L^2Q_1^2 S+ \pi L^2Q_2^2 S+
\frac{\pi^3 L^4 Q_1^2 Q_2^2}{S})^{3/2} }\nn && +
\frac{\pi^2 L^2 Q_1 Q_2}{S \sqrt{\frac{S^3}{\pi}+ \pi L^2 Q_1^2 S+ \pi L^2 Q_2^2 S+
\frac{\pi^3 L^4 Q_1^2 Q_2^2}{S}}}, \nn
g_{Q_2Q_2}&=&-\frac{1}{2}\frac{( \pi L^2 Q_2 S+ \frac{ \pi^3 L^4 Q_1^2 Q_2}{S})^2}
{\pi L^2 (\frac{S^3}{\pi}+ \pi L^2Q_1^2 S+ \pi L^2Q_2^2 S+ 
\frac{\pi^3 L^4 Q_1^2 Q_2^2}{S})^{3/2}}\nn &&
+\frac{\frac{\pi L^2 S}{2}+ \frac{\pi^3 L^4 Q_1^2}{2S}}{\pi L^2 \sqrt{\frac{S^3}{\pi}
+ \pi L^2 Q_1^2 S+ \pi L^2 Q_2^2 S+ \frac{\pi^3 L^4 Q_1^2 Q_2^2}{S}}}.\ea

\begin{figure}
\hspace*{1.0cm}\vspace*{-6.0cm}
\includegraphics[width=12.0cm,angle=-0]{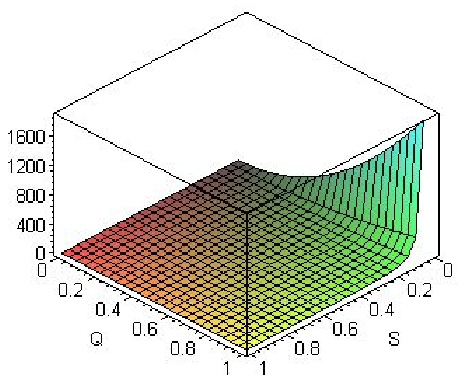}\vspace*{-3.0cm}
\caption{La composante $SS $ de la m\'etrique tenseure trac\'ee en 
fonction de $\{Q, M\}$, en d\'ecrivant les fluctuations dans la configuration 
des trous noirs RN-$AdS_4$.} \label{RNADS4SS}\vspace*{0.5cm}
\end{figure}

\begin{figure}
\hspace*{1.0cm}\vspace*{-6.0cm}
\includegraphics[width=12.0cm,angle=-0]{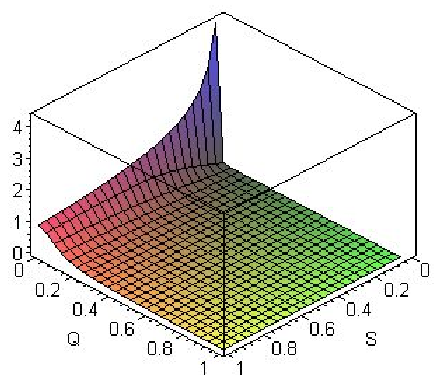}\vspace*{-3.0cm}
\caption{La composante $Q_1Q_1$ de la m\'etrique tenseure trac\'ee en 
fonction de $\{Q, M\}$, en d\'ecrivant les fluctuations dans la configuration 
des trous noirs RN-$AdS_4$.} \label{RNADS4Q1Q1}\vspace*{0.5cm}
\end{figure}

\begin{figure}
\hspace*{1.0cm}\vspace*{-6.0cm}
\includegraphics[width=12.0cm,angle=-0]{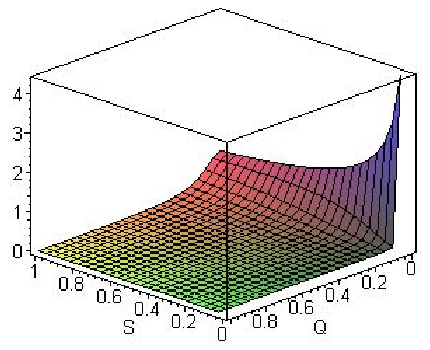}\vspace*{-3.0cm}
\caption{La composante $Q_2Q_2$ de la m\'etrique tenseure trac\'ee en 
fonction de $\{Q, M\}$, en d\'ecrivant les fluctuations dans la configuration 
des trous noirs RN-$AdS_4$.} \label{RNADS4Q2Q2}\vspace*{0.5cm}
\end{figure}

\begin{figure}
\hspace*{1.0cm}\vspace*{-6.0cm}
\includegraphics[width=12.0cm,angle=-0]{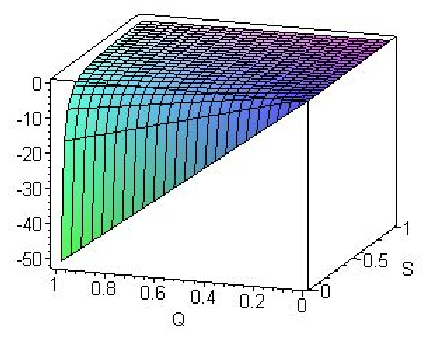}\vspace*{-3.0cm}
\caption{La composante $SQ_1$ de la m\'etrique tenseure trac\'ee en 
fonction de $\{Q, M\}$, en d\'ecrivant les fluctuations dans la configuration 
des trous noirs RN-$AdS_4$.} \label{RNADS4SQ1}\vspace*{0.5cm}
\end{figure}

\begin{figure}
\hspace*{1.0cm}\vspace*{-6.0cm}
\includegraphics[width=12.0cm,angle=-0]{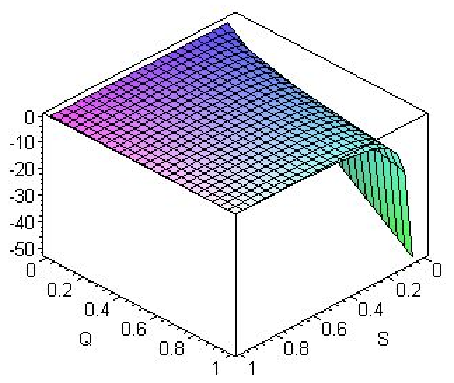}\vspace*{-3.0cm}
\caption{La composante $SQ_2$ de la m\'etrique tenseure trac\'ee en 
fonction de $\{Q, M\}$, en d\'ecrivant les fluctuations dans la configuration 
des trous noirs RN-$AdS_4$.} \label{RNADS4SQ2}\vspace*{0.5cm}
\end{figure}

\begin{figure}
\hspace*{1.0cm}\vspace*{-6.0cm}
\includegraphics[width=12.0cm,angle=-0]{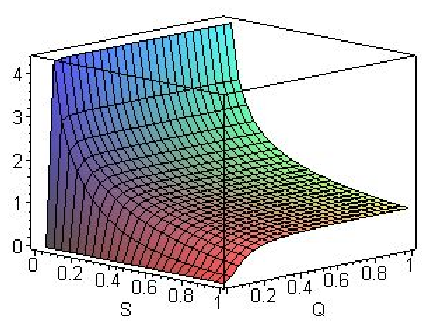}\vspace*{-3.0cm}
\caption{La composante $Q_1Q_2$ de la m\'etrique tenseure trac\'ee en 
fonction de $\{Q, M\}$, en d\'ecrivant les fluctuations dans la configuration 
des trous noirs RN-$AdS_4$.} \label{RNADS4Q1Q2}\vspace*{0.5cm}
\end{figure}

Dans le cas des valeurs de $L = -3$, et des charges $Q_1 = Q$ et $Q_2 = Q$, 
on retrouve dans la gamme de $Q \in (0,1) $ et $S \in (0,1)$ que la composante 
$\{g_{SS}\}$ prend une valeur maximale \`a l'ordre de $1600 $. Alors, nous voyons
que les composantes $\{ g_{Q_1 Q_1}, g_{Q_2 Q_2}\}$ prendent une valeur typique dans 
l'intervalle $(0,4) $. Cela montre que les trous noirs RN-$AdS_4$ correspondent 
localement \`a une configuration statistique stable. En fait, la gamme de la 
croissance de $\{g_{SS}\}$ et $\{g_{Q_1 Q_1}, g_{Q_2 Q_2}\}$ se trouve \^{e}tre 
\`a la limite oppos\'ee des param\`etres $\{S, Q\}$. Explicitement, par les Figs. 
(\ref{RNADS4SS}, \ref{RNADS4Q1Q1}, \ref{RNADS4Q2Q2}), on observe que la croissance 
de $g_{SS}$ est a lieu dans la limite d'une grande $Q$ et d'une petite $S$. 
D'autre part, les Figs. (\ref{RNADS4Q1Q1}, \ref{RNADS4Q2Q2})
montrent que la croissance de $\{g_{Q_1Q_1}, g_{Q_2 Q_2}\}$ est plac\'ee
dans la limite d'une petite $Q$ et une petite $S$. Du fait m\^{e}me, les 
compressibilit\'es de chaleurs en impliquant deux param\`etres distincts 
de la configuration des trous noirs de RN-$AdS_4$ ont \'et\'e repr\'esent\'es 
par les Fig. (\ref{RNADS4SQ1}, \ref{RNADS4SQ2}, \ref{RNADS4Q1Q2}). Dans ce
cas, nous remarquons que la Fig. (\ref{RNADS4Q1Q2}) montre une caract\'eristique 
unique de la composante $Q_1Q_2$ des fluctuations thermodynamiques.
Pour des trous noirs RN-$AdS_4$ donn\'es, les composantes de la m\'etrique
tenseure $\{g_{ij} \ | \ i, j = S, Q_1, Q_2\}$ indiquent que ces fluctuations 
en impliquant des charges $\{Q_1, Q_2\}$ prennent relativement des valeurs 
num\'eriques positives \`a l'\'egard de ceux en impliquant l'entropie $S$ 
des trous noirs de RN-$AdS_4$.

Nous voyons que les fluctuations d'un ensemble des 
trous noirs de Reissner-Nordstr\"om dans la nappe de Poincar\'e de $ ADS_4 $
respectent des conclusions physiquement bien attendues. En particulier, les capacit\'es 
de chaleurs, d\'efinies comme les paires d'auto-corr\'elations, restent positives 
pour des param\`etres $\{S, Q_1, Q_2\}$ venant par la solution des $ M_2$-branes. 
Un calcul simple d\'emontre la nature globale des fluctuations param\'etriques. Il s'`ere
que l'ensemble sous-jacent des trous noirs de Reissner-Nordstr\"om dans la nappe de Poincar\'e de $ ADS_4$ 
est stable dans la limite des fluctuations gaussiennes, si les mineurs principaux $\{p_2, p_3\}$ 
restent cerataines fonctions positives sur la vari\'et\'e $ (M_3, g) $ de \l'espace des \'etats. 
Dans ce cas, il s'ensuit que le mineur de surface est donn\'{e} par

\begin{eqnarray}
p(S, Q_1, Q_2) = {\displaystyle \frac{1}{16\,\pi \,L^{2}\,S^{3}}} \,{\displaystyle \frac{p_n}{p_d}},
\end{eqnarray}

o\`u les fonctions $\{ p_n(S, Q_1, Q_2), p_d(S, Q_1, Q_2)\}$ sont d\'efinies par

\begin{eqnarray}
p_n&=& 3\,S^{6}
+6\,S^{4}\,\pi ^{2}\,L^{2}\,\mathit{Q_2}^{2} 
-S^{4}\,\pi^{2}\,L^{2}\,\mathit{Q_1}^{2}
-\pi^{4}\,L^{4}\, \mathit{Q_2}^{4}\,S^{2} \nn &&
+2\,S^{2}\,\pi^{4}\,L ^{4}\,\mathit{Q_1}^{2}\,\mathit{Q_2}^{2} 
-\pi^{6} \,L^{6}\,\mathit{Q_2}^{4}\,\mathit{Q_1}^{2}, \nn 
p_d&=&S^{4} 
+ \pi ^{2}\,L^{2}\,S^{2}\,\mathit{Q_1}^{2} 
+\pi ^{2}\,L^{2}\,S^{2}\,\mathit{Q_2}^{2} 
+ \pi ^{4}\,L^{4}\,\mathit{Q_1}^{2}\,\mathit{Q_2}^{2}.
\end{eqnarray}

Par la suite, un calcul explicite montre que la contrainte de la stabilit\'e 
sur toute la configuration $(M_3, g)$, comme une fonction de l\'etropie $S$ et 
des charges $\{Q_1, Q_2\}$ est d\'etermin\'ee par le d\'eterminant de la m\'etrique
tenseur $g:=p_3$. Ainsi, le d\'eterminant de la m\'etrique tenseure de Wienhold 
est simplement donn\'e par 

\ba \label{detM2} g&=& -\lbrace \frac{S^4+ \pi^2 L^2 Q_1^2 S^2+ \pi^2 L^2 Q_2^2 S^2+ 
\pi^4 L^4 Q_1^2 Q_2^2}{\pi S} \rbrace^{-5/2} \frac{ \tilde{g} (S, Q_1, Q_2)}{32 \pi^3 L^2 S^6 }, \ea

o\`u la fonction $ \tilde{g} (S, Q_1, Q_2)$ est d\'efinie par

\ba \label{detM2fac} \tilde{g} (S, Q_1, Q_2) &=& \pi^6 L^6 S^6 Q_2^6+ \pi^6 L^6 S^6 Q_2^4 Q_1^2- 
7\pi^4 L^4 S^8 Q_1^2 Q_2^2+ 5 \pi^8 L^8 S^4 Q_1^4 Q_2^4 \nn &&
- \pi^4 L^4 S^8 Q_2^4- \pi^4 L^4 S^8 Q_1^4+ 3 \pi^{10} L^{10} S^2 Q_2^4 Q_1^6+ \pi^6 L^6 S^6 Q_1^4 Q_2^2 \nn &&
+ 3 \pi^8 L^8 S^4 Q_1^6 Q_2^2+ 3 \pi^{10} L^{10} S^2 Q_2^6 Q_1^4+  \pi^6 L^6 S^6 Q_1^6+ \pi^{12} L^{12} Q_2^6 Q_1^6 \nn &&
- 3S^{12}+ 3 \pi^8 L^8 S^4 Q_2^6Q_1^2- 5 \pi^2 L^2 S^{10} Q_1^2- 5 \pi^2 L^2 S^{10} Q_2^2.\ea

Notez bien que le d\'eterminant de la m\'etrique tenseure de Wienhold, 
donn\'e dans l'Enq. (\ref{detM2}), peut avoir une valeur positive, 
si la fonction $\tilde{g} (S, Q_1, Q_2))$, d\'efinie par l'Eqns.
(\ref{detM2fac}), prend une valeur n\'egative sur la vari\'et\'e $(M_3, g)$.

\begin{figure}
\hspace*{1.0cm}\vspace*{-6.0cm}
\includegraphics[width=12.0cm,angle=-0]{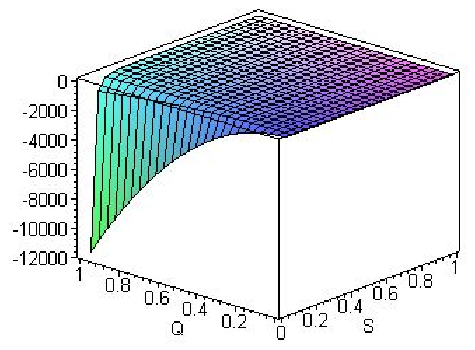}\vspace*{-3.0cm}
\caption{Le d\'eterminant de la m\'etrique tenseure trac\'ee comme la
fonction de $\{Q, M\}$, en d\'ecrivant les fluctuations dans 
la configuration des trous noirs RN-$AdS_4$.} \label{RNADS4det}\vspace*{0.5cm}
\end{figure}

\begin{figure}
\hspace*{1.0cm}\vspace*{-6.0cm}
\includegraphics[width=12.0cm,angle=-0]{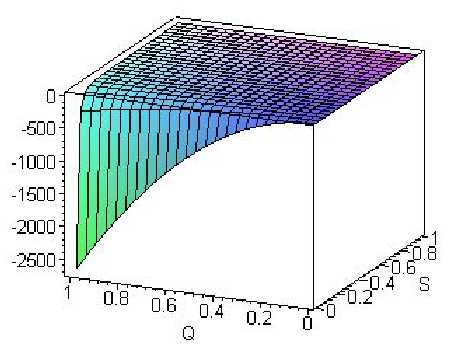}\vspace*{-3.0cm}
\caption{Le mineur de surface de la m\'etrique tenseure trac\'ee comme la
fonction de $\{Q, M\}$, en d\'ecrivant les fluctuations dans 
la configuration des trous noirs RN-$AdS_4$.} \label{RNADS4minor}\vspace*{0.5cm}
\end{figure}

\begin{figure}
\hspace*{1.0cm}\vspace*{-6.0cm}
\includegraphics[width=12.0cm,angle=-0]{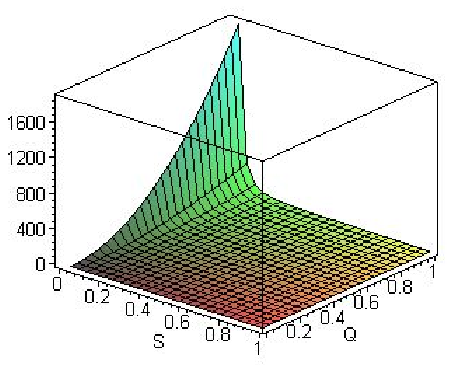}\vspace*{-3.0cm}
\caption{Le premier mineur de la m\'etrique tenseure trac\'ee comme la
fonction de $\{Q, M\}$, en d\'ecrivant les fluctuations dans 
la configuration des trous noirs RN-$AdS_4$.} \label{RNADS4minor1}\vspace*{0.5cm}
\end{figure}

Comme une fonction de $\{S, Q\}$, la condition de la stabilit\'e d'un ensemble 
des trous noirs de RN-$AdS_4$ d\'ecoule par la positivit\'e du d\'eterminant 
de la m\'etrique tenseure. Dans ce cas, nous constatons que le d\'eterminant 
de la m\'etrique tenseure g\'en\'eriquement tend vers une valeur n\'egative. 
Pour une valeur typique de $S \in (0, 1)$ et $Q \in (0, 1) $, la Fig. 
(\ref{RNADS4det}) montre que le d\'eterminant de la m\'etrique tenseure 
r\'eside dans l'intervalle $(0, -12 000)$. Donc, pour une petite $S$,
nous observons que le d\'eterminant de la m\'etrique tenseure a une valeur 
approximative \`a l'ordre de $-12000$. Dans la limite d'une grande $Q$, 
le d\'eterminant de la m\'etrique reste tout d'abord \`a pr\`es d'une constante 
valeur petite, puis, dans la limite de la petite $S$, il augmente brusquement \`a 
une grande valeur n\'egative \`a l'ordre de $-12000 $. La stabilit\'e correspondante 
de la surface d\'efinie par une valeur constante de la charge $Q_2$ est montr\'e 
dans la Fig. (\ref{RNADS4minor}). Dans cette gamme de $\{S, Q\}$, nous voyons que le 
mineur principe $p_2 $ r\'eside dans la gamme de $(-2500, 0)$. Dans la limite d'une petite $S$, 
nous constatons que la n\'egativit\'e de $p_2$ augmente \`a la mesure que la valeur de 
$Q$ est augment\'e de z\'ero \`a $1$. Lorsque $S$ est le seul param\`etre qui est permis
de varier, la stabilit\'e de la configuration sous-jacente est d\'etermin\'ee par la
positivit\'e du premier mineur principe $p_1: = g_{SS} $. Dans ce cas, nous observons 
de la Fig. (\ref{RNADS4minor1}) que $p_1$ a une amplitude positive \`a l'ordre de $2000$. 
En fait, pour une petite valeur de $S$, le premier mineur principe $p_1$ augmente 
continuousement comme la fonction de la charge $Q$. Ainsi, les descriptions graphiques 
ci-dessus des mineurs principaux fournent la notion qualitative de la stabilit\'e 
statistique des trous noirs de RN-$ADS_4$.

De plus, comme nous avons fourni les $\Gamma_{abc}$ dans l'annex $[A]$, 
il n'est pas difficile de calculer une expression exacte de la courbure scalaire
de Wienhold, d\'ecrivant les corr\'elations globales param\'etriques intrins\`eques d'un 
ensemble des trous noirs de Reissner-Nordstr\"om dans la nappe de Poincar\'e de $ ADS_4$. 
Alors, en d\'efinissant un ensemble des fonctions horizons de ces trous noirs, nous pouvons voir
que la courbure scalaire sous-jacente peut \^etre pr\'esent\'ee en g\'en\'eral. Explicitement, 
nous voyons que la courbure scalaire de Wienhold peut \^etre obtenue \`a: 

\ba \label{m2adscur} R&=& -(3\pi^2 L^2 S) \sqrt{\frac{ S^4+ \pi^2 L^2 Q_1^2 S^2+ \pi^2 L^2 Q_2^2 S^2+ 
\pi^4 L^4 Q_1^2 Q_2^2}{\pi S} }  \frac{r_4}{r_1 r_2 r_3}, \ea

o\`u les fonctions $\{r_i(S,Q_1,Q_2)\ | i=1,2,3,4\}$ sont d\'efinies par

\ba 
r_1&=& S^2+ \pi^2 L^2 Q_2^2, \nn
r_2&=& \pi^4 L^4 Q_1^2 Q_2^2+ \pi^2 L^2 Q_1^2 S^2+ \pi^2 L^2 Q_2^2 S^2- 3S^4, \nn
r_3&=& \pi^6 L^6 Q_1^4 Q_2^2+ \pi^4 L^4 Q_1^4 S^2+ 2 \pi^4 L^4 Q_1^2 S^2 Q_2^2- 
2 \pi^2 L^2 Q_1^2 S^4+ \pi^2 L^2 S^4 Q_2^2- 3 S^6, \nn
r_4 &=& \pi^8 L^8 Q_1^4 Q_2^4+ 2 \pi^6 L^6 Q_1^4 S^2 Q_2^2+ 2 \pi^6 L^6 Q_2^4 S^2 Q_1^2+
\pi^4 L^4 Q_1^4 S^4 \nn &&- 12 \pi^4 L^4 S^4 Q_1^2 Q_2^2+ \pi^4 L^4 Q_2^4 S^4-
14 \pi^2 L^2 S^6 Q_1^2- 14 \pi^2 L^2 S^6 Q_2^2+ S^8.\ea

Dans la limite de grande masse, cette solution des branes noirs dans la nappe
de Poincar\'e de $AdS_4$ a l'instabilit\'e thermodynamique locale dont les
comportements peuvent \^etre vus facilement par la courbure scalaire de 
Weinhold, comme ci-dessus nous avons montr\'e par l'Eqn. (\ref{m2adscur}).
Nous voyons que l'instabilit\'e thermodynamique locale peut \^etre \'etudi\'ee
par la g\'eom\'etrie de Weinhold dont la m\'etrique tenseure est d\'efine comme
la matrice d'Hessienne de la masse $ M(S, Q_1, Q_2) $, et donc cette g\'eom\'etrie 
d\'ecrit la convexit\'e de la fonction de la masse de ces trous noirs.
Nous avons observ\'e, si on augmente l'une des charges \'electriques alors l'autres
diminue qui ne peut se produire que lors qu'elle d\'etient localement sur une compte
de la conservation de la charge globale \cite{CveticGubser}.

Ici, nous voyons clairement que la courbure scalaire de Wienhold diverge pour 
tout entropie donn\'ees par les equations suivantes:

\ba \pi^2 L^2 Q_2^2+ S^2&=& 0,\nn
\pi^4 L^4 Q_1^2 Q_2^2+ \pi^2 L^2 (Q_1^2+ Q_2^2) S^2- 3S^4&=& 0, \nn 
\pi^6 L^6 Q_1^4 Q_2^2+ \pi^4 L^4 Q_1^2 (Q_1^2+ 2 Q_2^2) S^2 
+ \pi^2 L^2(Q_2^2- 2 Q_1^2) S^4 - 3 S^6&=& 0. \ea

Il est important de noter qu'il existent certaines gammes qu'elles sont 
thermodynamiquement limit\'ees par les instabilit\'es aux cas des trous noirs 
de Reissner-Nordstr\"om dans la nappe de Poincar\'e de l'espace $AdS_4$. Par exemple, 
les racines r\'eeles d'un entropie ci-dessus des trous noirs sous-jacents 
sont donn\'ees par l'\'equation quadratique:

\ba S= \frac{\pi L}{6} \lbrace Q_1^2+ Q_2^2 +
\sqrt{Q_1^4+ Q_2^4+ 12 Q_1^2 Q_2^2} \rbrace^{1/2}\ea

ou bien celles d'autres sont donn\'ees par les racines r\'eeles de l'\'equation 
cubique ci-dessus, dont lesquelles, on a les fronti\`eres des solutions de l'entropie 
avec lesquelles les singularit\'ees thermodynamiques bien existent.
En outre, la courbure scalaire de Wienhold est nulle dont l'entropie est donn\'ees par,

\ba \pi^8 L^8 Q_1^4 Q_2^4+ 2 \pi^6 L^6 Q_1^4 S^2 Q_2^2+ 2 \pi^6 L^6 Q_2^4 S^2 Q_1^2+ 
\pi^4 L^4 Q_1^4 S^4\nn - 12 \pi^4 L^4 S^4 Q_1^2 Q_2^2+ \pi^4 L^4 Q_2^4 S^4- 
14 \pi^2 L^2 S^6 Q_1^2- 14 \pi^2 L^2 S^6 Q_2^2+ S^8&=& 0.\ea

\begin{figure}
\hspace*{1.0cm}\vspace*{-6.0cm}
\includegraphics[width=12.0cm,angle=-0]{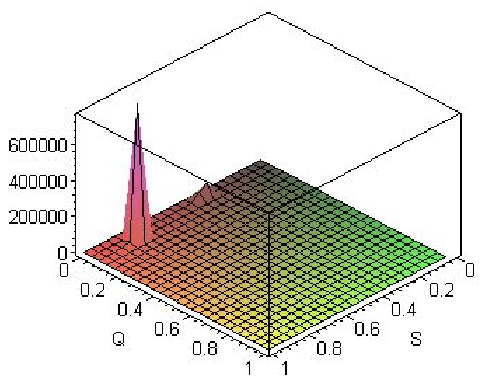}\vspace*{-3.0cm}
\caption{La courbure scalaire trac\'ee en fonction de $\{Q, M\}$,
d\'ecrivant les variations dans la configuration des trous noirs
RN-$AdS_4$ dans la gamme $S, Q \in (0,1) $.} \label{RNADS4cur}\vspace*{0.5cm}
\end{figure}

\begin{figure}
\hspace*{1.0cm}\vspace*{-6.0cm}
\includegraphics[width=12.0cm,angle=-0]{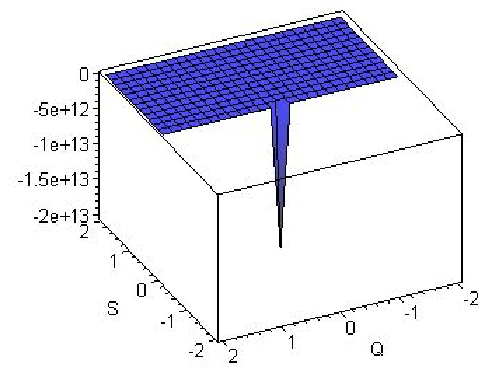}\vspace*{-3.0cm}
\caption{La courbure scalaire trac\'ee en fonction de $\{Q, M\}$,
d\'ecrivant les variations dans la configuration des trous noirs
RN-$AdS_4$ dans la gamme $S, Q \in (-2, 2)$.} \label{RNADS4cur2}\vspace*{0.5cm}
\end{figure}

Sous la fluctuation des charges $\{Q_1, Q_2\}$ et de l'entropie $S$,
les propri\'et\'es de la stabilit\'e globale des trous noirs de RN-$AdS_4$
sont montr\'ees dans les Figs. (\ref{RNADS4cur}, \ref{RNADS4cur2}). 
Dans la gamme de $S \in (0, 1)$ et $Q \in (0, 1) $, il r\'esulte de la
Fig. (\ref{RNADS4cur}) que la courbure scalaire de Winhold acquiert
un grand pic positif. En fait, dans la limite d'une petite $Q$, la
Fig. (\ref{RNADS4cur}) montre que la courbure scalaire a certaines domaines
tr\`es interagissantes. En particulier, dans une petite valeur de $Q$, 
la Fig. (\ref{RNADS4cur}) montre que la courbure scalaire de Winhold a une 
grande amplitude positive \`a l'ordre de $800 000 $. Ainsi, on peut noter que 
la configuration des trous noirs sous-jacente est un syst\`eme statistiques 
interagissant fortement, dans cette limite des charges et de l'entropie. 
Comme mentionn\'e dans le cas de celle des trous noirs dilatoniques, 
les fluctuations des charges et de l'entropie rendent un signe positif 
\`a la courbure scalaire. Cela signifie que les trous noirs de RN-$ADS_4$ 
sont r\'epulsive dans la leur nature. La Fig. (\ref{RNADS4cur2}) d\'ecrit 
la nature de la courbure scalaire de Wienhold dans une range \'egale des 
param\`etres $S, Q \in (-2, 2) $. Notamment, nous voyons de la Fig. (\ref{RNADS4cur2}) 
qu'il y a une moiti\'e coup\'ee pique n\'egative de l'amplitude des interactions globales. 
\`A la comparaison avec des interactions apparaissant dans la gamme de $S \in
(0, 1)$ et $Q \in (0, 1) $, les interactions changent leurs natures et ainsi
ces trous noirs deviennent plut\^{o}t attractifs dans la gamme de $S, Q \in (-2, 2)$.
La valeur typique de l'amplitude des courbures scalaires correspondantes  
semble \^{e}tre \`a l'ordre de $-2\times 10^{13} $. Les propri\'et\'es 
graphiques de cette comparaison suivent des Figs. (\ref{RNADS4cur}, \ref{RNADS4cur2}). 
Qualitativement, les repr\'esentations graphiques ci-dessus des courbures scalaires 
indiquent que les trous noirs de RN-$ADS_4$ sont des configurations instables dans 
la limite des petites param\'etres. Comme pr\'evu, nous constatons que les 
trous noirs de RN-$ADS_4$ sont bien comport\'es, faiblement interactifs, et 
ainsi il s'av\`ere que ces trous noirs correspondent \`a une configuration 
relativement stable dans la limite des grandes charges.

Dans les chapitres suivantes nous allons consid\'er\'e les corrections
\`a la g\'eom\'etrie thermodynamique des trous noirs qu'elles viennent par
(i) le principe d'incertitude g\'en\'eralis\'ee, et (ii) les d\'eriv\'es 
sup\'erieures de la th\'eorie des cordes.

 
\clearpage 
\chapter{Les corrections de $ l_P $ dans la g\'eom\'etrie thermodynamique}
Dans ce chapitre, motiv\'es par la th\'eorie des cordes,
nous \'etudions les effets du principe d'incertitude g\'en\'eralis\'ee
(GUP) sur la g\'eom\'etrie thermodynamique des trous noirs.
En fait, la nature non-commutative de l'espace-temps \`a l'\'echelle 
de Plank \cite{NozariMehdipour, Garay, KempfMangano, CameliaLukierskiNowicki}
implique qu'il existe la distance observable minimale de l'ordre de la 
longueur de Plank o\`u toutes les mesures de la limite d'extr\^eme de la 
gravit\'e quantique sont gouvern\'ees.
Cette distance minimale des observables dite que le principe d'incertitude 
g\'en\'eralis\'ee \cite{BNTiwari} peut \^etre \'ecrit comme:

\ba \label{unsbnt} \Delta x \geq \frac{\hbar}{\Delta p} + \frac{\tilde{\alpha} l_p^2}{\hbar} \Delta p.\ea

Bien que dans toutes les th\'eories g\'en\'erales, les contributions 
de l'ordre sup\'erieur peuvent \^etre non-nulles mais la longueur minimale 
dans la th\'eorie est simplement r\'egie par le param\`etre $\tilde{\alpha}$, 
comme au-dessus voir l'Eqn. (\ref{unsbnt}). 
Le principe d'incertitude g\'en\'eralis\'ee montre qu'il y a une
dispersion minimume $\Delta x$ pour toute valeur $\Delta p$ du moins 
aussi longtemps ce que les deux premiers termes de l'Eqn. (\ref{unsbnt})
expansion soient non-nulles.

Un exemple de cette motivation provient gr\^ace \`a l'\'etude des conditions
d'analyticit\'e d'une fonction complexe conduisant au principe d'incertitude
compl\`etement g\'en\'eralis\'ee, laquelle fait appara\^itre les concepts physiques
bien connues \cite{BNTiwari}.
Nous avons \'etudi\'e ces effets sur la notion physique du principe d'incertitude
et avons expliqu\'e que le principe d'incertitude de la th\'eorie des cordes se pose 
naturellement de l'analyse des fonctions holomorphes et anti-holomorphes.
Ces consid\'erations illustrent le r\'ecit de la forme et de la taille
correspondant du principe d'incertitude d'Heisenberg, lesquelles sont bien 
connues pour toutes fonctions arbitraires de type $L^2$.
De plus, nous pouvons arriver au principe d'incertitude de la th\'eorie des
cordes avec toutes les corrections d'un ordre fini des d\'eriv\'es sup\'erieures, 
la physique de la gravit\'e quantique, la physique des trous noirs, l'existence des 
\'echelles de la longueur minimale et la longuer maximale de la nature, 
la g\'eom\'etrie aux distances courtes versus la th\'eorie des cordes,
la transformation de Fourier par rapport \`a la th\'eorie des distributions.
Voir \cite{BNTiwari} pour les d\'etails de l'\'etat actuel des corrections de $l_P$.
Dans les deux prochaines sections, nous souhaitons analyser les effets de principe 
d'incertitude g\'en\'eralis\'ee sur la g\'eom\'etrie thermodynamique des trous noirs 
de Reissner-Nordstr\"om et ceux des trous noirs charg\'es magn\'etiquements.

Maintenant, nous ferons donc une br\`eve rappelle de la g\'eom\'etrie thermodynamique
de Ruppenier et ensuite l'appliquons \`a l'entropie de ces trous noirs corrig\'e
par les corrections du principe d'incertitude g\'en\'eralis\'ee.
De plus, nous allons voir, les quelles sont les effets des corrections de
$ l_P $ dans la g\'eom\'etrie thermodynamique de Ruppenier?
Pour \c{c}a, d\'efinissons la m\'etrique thermodynamique pour \^etre:

\ba g_{ij}(x)= -\left( \begin{array}{rr}
  S_{MM} & S_{QM} \\
  S_{QM} & S_{QQ} \\
\end{array} \right) \ea

avec $ i, j= M, Q $ \cite{Sar1, Sar2}.
Alors, il s'av\`ere que l'\'el\'ement de la ligne est:

\ba ds^2= - S_{MM} dM^2- 2 S_{MQ} dM dQ- S_{QQ} dQ^2, \ea

dont le d\'eterminant de la  la m\'etrique tenseure est justement donn\'e par

\ba det(g)= S_{MM}S_{QQ}- S_{MQ}^2.\ea

En utilisant cette forme g\'en\'erale de la m\'etrique thermodynamique,
il n'est pas difficile de voir que les symboles de Christoffel, qui sont
d\'efinis comme: 

\ba \Gamma_{ij k}:= g_{ij,k} + g_{ik,j} − g_{j k,i}, \ea

sont donn\'es par les expressions suivantes:

\ba 
\Gamma_{MMM}&=& -\frac{1}{2} S_{MMM},\nn
\Gamma_{QQQ}&=& -\frac{1}{2} S_{QQQ},\nn
\Gamma_{MMQ}&=& -\frac{1}{2} S_{QMM},\nn
\Gamma_{MQM}&=& -\frac{1}{2} S_{QMM},\nn
\Gamma_{MQQ}&=& -\frac{1}{2} S_{MQQ},\nn
\Gamma_{QQM}&=& -\frac{1}{2} S_{QQM};\ea

avec les sym\'etries reliant les autres composantes.
Ensuite, nous pouvons voir facilement qu'il y la seule composante diff\'erente de z\'ero 
de la courbure tenseure de Riemann-Christoffel, laquelle peut \^etre obtenu comme:

\ba R_{MQMQ}&=& \frac{1}{4(S_{MM}S_{QQ}- S_{QM}^2)} (-S_{QM}S_{MMM}S_{QQQ}+
S_{QQM}S_{MMM}S_{QQ}\nn &&- S_{MM}S_{QQM}^2 + S_{MM} S_{QMM} S_{QQQ}+
S_{QMM} S_{QM}S_{QQM}- S_{QMM}^2 S_{QQ}).\ea

Nous pouvons contracter cette courbure de Riemann-Christoffel $R_{MQMQ}$ 
avec la m\'etrique $g^{ij}$ et obtenir $R_{ij}$. En fait, nous voyons facilement 
que la courbure scalaire de Ruppenier, d\'efinie comme $g^{ij}R_{ij}$, est donn\'ee par:

\ba R&=& \frac{1}{2(S_{MM}S_{QQ}- S_{QM}^2)^2}(-S_{QM}S_{MMM}S_{QQQ}+
S_{QQM}S_{MMM}S_{QQ}\nn &&- S_{MM}S_{QQM}^2 + S_{MM} S_{QMM} S_{QQQ}+
S_{QMM}S_{QM}S_{QQM}- S_{QMM}^2 S_{QQ}).\ea

Enfin, dans ce cas, il n'est pas difficile de voir que la courbure scalaire
de Ruppenier est reli\'ee avec la courbure de Riemann-Christoffel par:

\ba R= \frac{2}{ det(g)} R_{MQMQ}.\ea

\section{Les trou noirs de Reissner-Nordstr\"om.}

La m\'etrique de l'espace-temps de trou noir de Reissner-Nordstr\"om
avec la masse $ M $ et la charge \'electrique $ Q $ est donn\'e par:

\ba ds^2= -f(r) dt^2+ \frac{dr^2}{f(r)}+ r^2 d\theta^2+ r^2 sin^2 \theta d \phi^2, \ea 

o\`u 

\ba f(r)= 1- \frac{2M}{r}+ \frac{Q^2}{r^2}.\ea

L'horizon int\'erieur et ext\'erieur sont simplement d\'efinies par,

\ba r_{\mp}= M \mp \sqrt{M^2- Q^2}.\ea

Nous pouvons voir que la temp\'ereture est donn\'ee par

\ba T= \frac{\kappa}{2 \pi}= \frac{r_{+}- r_{-}}{4\pi r_{+}^2},\ea

o\`u $ \kappa $ est la gravit\'e de la surface du trou noir.
On a vu r\'ecemment \cite{YoonHaKim} qu'il n'y a aucune n\'ecessit\'e
de la limite de Brick-Wall dans le calcul de l'entropie par-ce qu'on peut 
avoir un champ scalaire satisfaisant l'\'equation de Klein-Gordon,
avec laquelle dans l'approximation de WKB, t'Hooft \cite{thooft} a montr\'e 
que le nombre des \'etats quantiques est fini, qui reste \'egalement
le m\^eme \`a l'horizon d'un trou noir.

En consid\'erant une couche mince entre $r_{+}$ et $r_{+}+ \epsilon$,
nous allons analyser un trou noir non-extr\'emal de Reissner-Nordstr\"om. 
Puisque la densit\'e des \'etats est dominante pr\`es de l'horizon 
du trou noir de Reissner-Nordstr\"om, quand on fait le comptage des modes.
C'est-\`a-dire que, soit 

\ba x= \frac{\omega}{2k_B T}\rightarrow 0 \ea

et ainsi

\ba x_0= \frac{\mu \sqrt{f}}{2k_B T} \ea

puis pr\`es de l'horizon, il s'av\`ere que la longueur minimume du GUP est 

\ba 2 \sqrt{\lambda}= \sqrt{\frac{2\epsilon}{\kappa}},\ea

ce qui entra\^ine  \`a l'entropie suivante du trou noir de Reissner-Nordstr\"om:

\ba S(M,Q)= \frac{4 \pi r_{+}^2}{3 \lambda}(\frac{\pi}{24}- \frac{25}{32\pi}+
\frac{\zeta(3)}{\pi}), \ea

ce qui est finie dans la limite pr\`es de l'horizon.
Cette entropie peut \^etre \'ecrite comme

\ba S(M,Q)= (\frac{1}{4} A_H)\frac{\delta}{\lambda},\ea

o\`u le param\`etre $\delta$ est d\'efini \cite{YoonHaKim} par

\ba \delta:= \frac{1}{3}[\frac{4}{\pi}\zeta(3)- \frac{25}{8\pi}- \frac{\pi}{6}]. \ea

En mettant la valeur de la $ r_{+} $, nous pouvons \'ecrire l'entropie comme la suivante:

\ba S(M,Q)= \frac{\pi \delta}{\lambda}[2M^2- Q^2+ 2M\sqrt{M^2- Q^2}].\ea

Pour cette entropie de trou noir de Reissner-Nordstr\"om avec des corrections
de $l_P$, les composantes de la m\'etrique de Ruppenier
$ \lbrace g_{ij} \rbrace_{i,j \in \lbrace M,Q \rbrace} $ sont donn\'ees par:

\ba g_{MM}&=& \frac{\pi \delta}{\lambda}(-4- \frac{6M}{\sqrt{M^2- Q^2}}
+ \frac{2M^3}{(M^2- Q^2)^{3/2}}), \nn
g_{MQ}&=& \frac{\pi \delta}{\lambda}(\frac{2Q}{\sqrt{M^2- Q^2}}
- \frac{2M^2Q}{(M^2- Q^2)^{3/2}}), \nn
g_{QQ}&=& \frac{\pi \delta}{\lambda}(2+  \frac{2M}{\sqrt{M^2- Q^2}}
+ \frac{2MQ^2}{(M^2- Q^2)^{3/2}} ). \ea

\begin{figure}
\hspace*{1.0cm}\vspace*{-6.0cm}
\includegraphics[width=12.0cm,angle=-0]{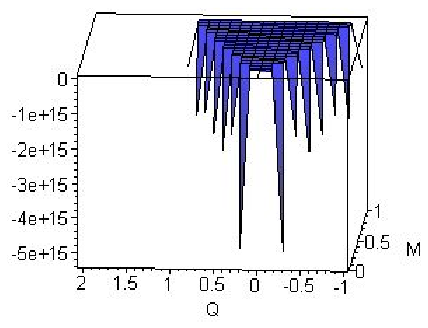}\vspace*{-3.0cm}
\caption{La composante $MM$ de la m\'etrique tenseure trac\'ee comme la
fonction de $\{Q, M\}$, en d\'ecrivant les fluctuations de la 
configuration des trous noirs de RN.} \label{RNgupMM}\vspace*{0.5cm}
\end{figure}

\begin{figure}
\hspace*{1.0cm}\vspace*{-6.0cm}
\includegraphics[width=12.0cm,angle=-0]{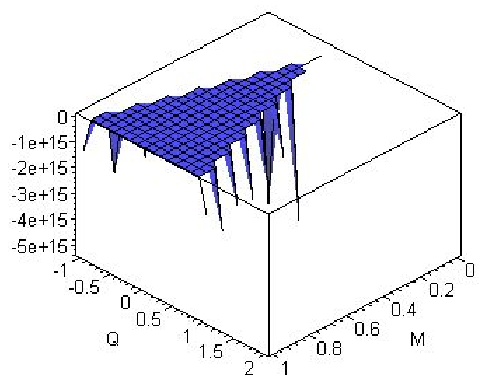}\vspace*{-3.0cm}
\caption{La composante $QQ$ de la m\'etrique tenseure trac\'ee comme la
fonction de $\{Q, M\}$, en d\'ecrivant les fluctuations de la 
configuration des trous noirs de RN.} \label{RNgupQQ}\vspace*{0.5cm}
\end{figure}

\begin{figure}
\hspace*{1.0cm}\vspace*{-6.0cm}
\includegraphics[width=12.0cm,angle=-0]{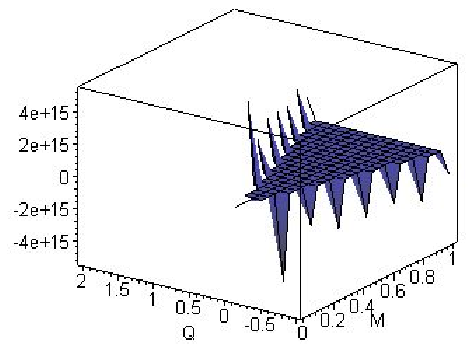}\vspace*{-3.0cm}
\caption{La composante $MQ$ de la m\'etrique tenseure trac\'ee comme la
fonction de $\{Q, M\}$, en d\'ecrivant les fluctuations de la 
configuration des trous noirs de RN.} \label{RNgupMQ}\vspace*{0.5cm}
\end{figure}

Pour $L = 0.01 $, les Figs. (\ref{RNgupMM}, \ref{RNgupQQ}) montrent
la nature des capacit\'es de chaleurs $\{g_{MM}, g_{QQ}\}$ de la m\'etrique
tenseure de l'espace d'\'etat. Dans le r\'egime de $M \in (0, 1)$ et 
$Q \in (-1, 2) $, on observe que l'amplitude de $\{g_{MM}\}$ prend
une valeur \`a l'ordre de $10^{+15} $. Dans cette gamme des param\`etres
$\{M, Q\}$, nous constatons que la composante $\{g_{MM}\}$ r\'eside dans la
gamme de $(-5 \times 10^{+15}, 0)$. Dans ce cas, nous voyons que la
gamme de la croissance de l'amplitude de $\{g_{QQ}\}$ reste presque dans 
la m\^{e}me limite de $\{Q, M\}$. Explicitement, cela signifie que 
les trous noirs de RN sont thermodynamiquement instables dans la
limite de petite charge et de petite masse. Du fait m\^{e}me, la Fig. (\ref{RNgupMQ})
montre que la nature de la composante $\{g_{MQ}\}$ de la m\'etrique tenseure
de l'\'espace d'\'etat. Nous constatons que la composante mixe $\{g_{MQ}\}$ 
prend une grande valeur \`a l'ordre de $\pm 10^{+15}$ dans la limite
d'une petite $Q$ et une petite $M$. Dans cette limite de $\{M, Q\}$, 
les compressibilit\'es de chaleurs, telle que repr\'esent\'ees dans les
Figs. (\ref{RNgupMM}, \ref{RNgupQQ}, \ref{RNgupMQ}) illustrent 
les propri\'et\'es de la fluctuation des trous noirs de RN. En bref, 
les paires pures des fluctuations en impliquant $\{M, Q\}$, tel que 
d\'efinies par la m\'etrique tenseure $\{g_{ii} \ | \ i = M, Q\}$, 
ont des valeurs num\'eriques positives, tandis que la composante mixe 
$\{g_{MQ}\}$ prend les deux signatures.

Dans ce cas, le d\'eterminant de cette m\'etrique de Ruppenier peut \^etre obtenus \`a:

\ba det(g)&=& -\frac{4\pi^2\delta^2}{\lambda^2(Q^2- M^2)^2}
(4M^4- 5M^2Q^2+ Q^4+ 2MQ^2 \sqrt{M^2- Q^2} \nn &&+ 5M(M^2- Q^2)^{3/2}
- M^3 \sqrt{M^2- Q^2} ). \ea

\begin{figure}
\hspace*{1.0cm}\vspace*{-6.0cm}
\includegraphics[width=12.0cm,angle=-0]{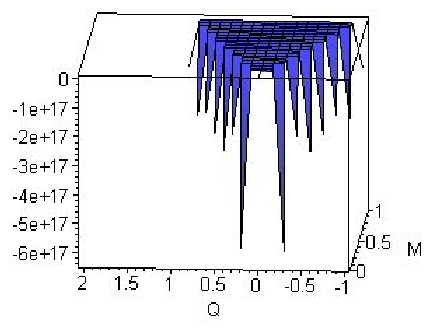}\vspace*{-3.0cm}
\caption{Le d\'eterminant de la m\'etrique tenseure trac\'ee 
comme la fonction de $\{Q, M\}$, en d\'ecrivant les fluctuations de la 
configuration des trous noirs de RN.} \label{RNgupdet} \vspace*{0.5cm}
\end{figure}

\begin{figure}
\hspace*{1.0cm}\vspace*{-6.0cm}
\includegraphics[width=12.0cm,angle=-0]{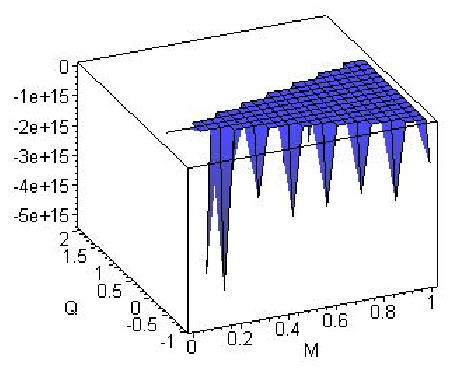}\vspace*{-3.0cm}
\caption{Le premier mineur de la m\'etrique tenseure trac\'ee comme la
fonction de $\{Q, M\}$, en d\'ecrivant les fluctuations de la configuration 
des trous noirs de RN.} \label{RNgupminor1}\vspace*{0.5cm}
\end{figure}

Notez bien que la stabilit\'e d'un ensemble des trous noirs de RN peut \^{e}tre
d\'etermin\'e en termes des valeurs de la masse $M$ et de la charge $Q$.
Cela d\'ecoule du comportement de d\'eterminant de la m\'etrique tenseure
de l'espace d'\'etat. Donc, nous constatons que le d\'eterminant 
de la m\'etrique tenseure tend vers une grande valeur n\'egative. Pour le cas 
de $M \in (0, 1)$ et $Q \in (-1, 2)$, la Fig. (\ref{RNgupdet}) montre que
le d\'eterminant de la m\'etrique tenseure r\'eside dans l'intervalle 
$(-6 \times 10^{+17}, 0)$. En fait, nous constatons que la n\'egativit\'e 
de $g$ augmente \`a mesure que les valeurs de $Q, M$ sont vari\'ees de 
$1$ \`a z\'ero. Ainsi, la surface d\'efinie par les fluctuations de $\{M, Q\}$ 
est instables en raison d'une valeur n\'egative du d\'eterminant de la
m\'etrique tenseure. Lorsque le seul param\`etre $M$ est autoris\'e \`a varier, 
la stabilit\'e de la configuration des trous noirs de RN est d\'etermin\'e par 
la positivit\'e du premier mineur principe $p_1: = g_{MM}$. Une vue rotat\'ee 
de $p_1$ est montr\'ee dans la Fig. (\ref{RNgupminor1}). Ci-dessus, 
les propri\'et\'es graphiques et la positivit\'e des mineurs principaux,
comme les quantit\'es de l'espace d'\'etat fournent la notion qualitative 
de la stabilit\'e statistique des trous noirs de RN de deux param\`etres 
sous les corrections du principe d'incertitude g\'en\'eralis\'e.

Ici, comme nous avons fourni les $\Gamma_{abc}$ dans l'annex $[A]$, 
nous pouvons voir facilement qu'on a la courbure de Ruppenier-Riemann-Christoffel:

\ba R_{MQMQ}= 0 \ea

qui entra\^ine que la courbure scalaire de Ruppenier est 

\ba R= 0.\ea

En conclusion, bien qu'on ajoute les corrections de $l_P$ \`a
l'entropie du trou noir de Reissner-Nordstr\"om, on trouve que
le configuration thermodynamique sous-jacente est bien d\'efini,
et reste \'egalement un syst\`eme statistique sans interaction.
De plus, les propri\'et\'es globales de la stabilit\'e des trous noirs de RN 
d\'ecoulent de la courbure scalaire de l'espace d'\'etat. En g\'en\'eral, 
il s'av\`ere que la courbure scalaire est identiquement nulle pour toute 
valeur de la masse et de la charge. Cela montre que les trous noirs de RN 
sous les corrections du principe d'incertitude g\'en\'eralis\'ee correspondent 
\`a une configuration statistique non-interagissante. En bref, les observations 
ci-dessus de la g\'eom\'etrie de l'espace d'\'etat indiquent que les trous noirs 
de RN sont, bien que non-interagissants, mais correspondent \`a une configuration 
statistiquement instable, lorsque les param\`etres $\{M, Q\}$ sont autoris\'es \`a fluctuer. 
Il est convient de mentionner que l'instabilit\'e survit m\^{e}me \`a l'\'echelle 
locale au niveau des capacit\'es de chaleurs.

\section{Les trous noirs charg\'es magn\'etiquement.}

Maintenant, consid\'erons un syst\`eme non-trivial des trous noirs,
en particulier les trous noirs magn\'etis\'es.
Dans ce cas-ci, soit 

\ba ds^2= -f(r) dt^2+ \frac{dr^2}{f(r)}+ R(r)^2 d\Omega_2^2 \ea

la m\'etrique de l'espace-temps. Puis dans l'approximation entre $ r_{+} $ 
and $ r_{+} + \epsilon $ o\`u $ \epsilon \rightarrow 0 $, nous pouvons \'ecrire:

\ba f(r)= \kappa (r- r_{+}) + \frac{f^{\prime\prime}(R_{+})}{2} (r- r_{+})^2+ \ldots \ea

et 

\ba R(r)^2= R(r_{+})+ \frac{d}{dr}R(r_{+})^2 (r- r_{+}) + \frac{1}{2}
\frac{d^2}{dr^2}R(r_{+})^2 (r- r_{+})^2 + \ldots.\ea

Soit $ \lambda $ le param\`etre de GUP, alors qu'il peut \^etre \'ecrit comme:

\ba 2 \sqrt{\lambda} \simeq \frac{\sqrt{\epsilon}}{\sqrt{\kappa}}
(2- \frac{f^{\prime\prime} \epsilon}{6 \kappa }).\ea

Le trou noir mag\'eétis\'e avec un accouplement arbitraire a:

\ba f(r)&=& 1- \frac{r_{+}}{r}, \nn
R(r)^2&=& r(r-r_{-}), \nn
2M&=& r_{+}, \nn
Q&=& \frac{r_{+} r_{-}}{2}. \ea

Pour le cas non-extr\'emal, nous pouvons voir trivialement que 
la temp\'erature d'Hawking est donn\'e par: 

\ba T_{BH}= \frac{1}{2 \pi r_{+}}.\ea

Et ainsi, avec la superficie de l'horizon d'\'ev\'enement 

\ba A_{H}= 4 \pi (r_{+}- r_{-}), \ea
 
le param\`etre $ \lambda $ de GUP entra\^ine que l'entropie \cite{KimOh} \`a l'ordre de $ \lambda^{0} $ est 

\ba S_{BH}= \frac{a_{H}}{4G_N}+ \frac{4}{9}\zeta(3)+ \frac{2r_{-}}{r_{+}}\zeta(7)+ \heartsuit(\lambda).\ea

Comme dans le cas dernier, en ce cas, nous avons: 

\ba r_{+}&=& 2M, \nn
r_{-}&=& \frac{Q^2}{M}.\ea

Donc, en termes de param\`etres thermodynamiques, c\'est \`a dire la masse $ M $ et 
la charge magn\'etique $Q$, l'entropie du trou noir magn\'etis\'e peut \^etre \'ecrite \`a:

\ba S_{BH}(M,Q)= \frac{4}{9}\zeta(3) + 8 \pi (2M^2- Q^2)+
\frac{Q^2}{M^2} \zeta(7)+ \heartsuit(\lambda).\ea

Pour ce cas de trou noir magn\'etis\'e avec le param\`etre $ \lambda $ du
principe d'incertitude g\'en\'eralis\'ee, les composantes de la m\'etrique
$ g_{ij} $ de Ruppenier sont simplement:

\ba g_{MM}&=& -32 \pi- \frac{6Q^2 \zeta(7)}{M^4},\nn
g_{MQ}&=& \frac{4Q \zeta(7)}{M^3}, \nn
g_{QQ}&=& 16 \pi- \frac{2 \zeta(7)}{M^2}.\ea

\begin{figure}
\hspace*{1.0cm}\vspace*{-6.0cm}
\includegraphics[width=12.0cm,angle=-0]{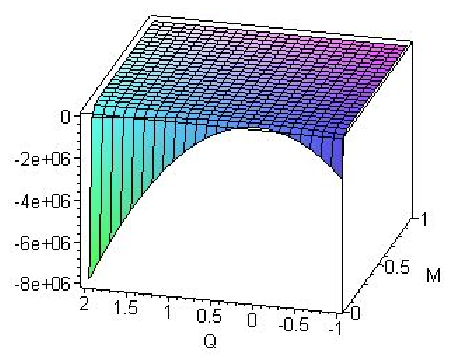}\vspace*{-3.0cm}
\caption{La composante $MM$ de la m\'etrique tenseure trac\'ee comme la
fonction de $\{Q, M\}$, en d\'ecrivant les fluctuations de la configuration 
des trous noirs charg\'es magn\'etiquement.} \label{maggupMM}\vspace*{0.5cm}
\end{figure}

\begin{figure}
\hspace*{1.0cm}\vspace*{-6.0cm}
\includegraphics[width=12.0cm,angle=-0]{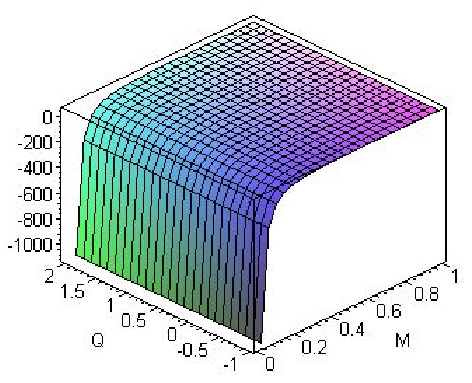}\vspace*{-3.0cm}
\caption{La composante $QQ$ de la m\'etrique tenseure trac\'ee comme la
fonction de $\{Q, M\}$, en d\'ecrivant les fluctuations de la configuration 
des trous noirs charg\'es magn\'etiquement.} \label{maggupQQ}\vspace*{0.5cm}
\end{figure}

\begin{figure}
\hspace*{1.0cm}\vspace*{-6.0cm}
\includegraphics[width=12.0cm,angle=-0]{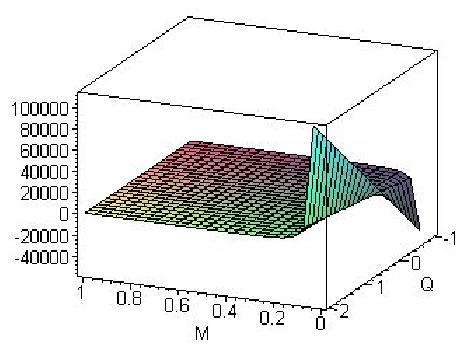}\vspace*{-3.0cm}
\caption{La composante $MQ$ de la m\'etrique tenseure trac\'ee comme la
fonction de $\{Q, M\}$, en d\'ecrivant les fluctuations de la configuration 
des trous noirs charg\'es magn\'etiquement.} \label{maggupMQ}\vspace*{0.5cm}
\end{figure}

La nature des capacit\'es de chaleurs $\{g_{MM}, {g_QQ}\}$ de la
m\'etrique tenseure de l'espace d'\'etat est repr\'esent\'ee 
dans les Figs. (\ref{maggupMM}, \ref{maggupQQ}). Dans le r\'egime de 
$M \in (0, 1)$ et $Q \in (-1, 1) $, nous voyons que l'amplitude de 
$\{g_{MM}\}$ prend une valeur \`a l'ordre de $10^{+06}$. Dans cette gamme 
de param\`etres $\{M, Q\}$, nous constatons que la composante $\{g_{MM}\}$ 
r\'eside dans la gamme de $(-8 \times 10^{+06}, 0)$. Dans ce cas, nous
observons que la gamme de la croissance de l'amplitude de $\{g_{QQ} \}$ 
est assez distinctes, dans la m\^{e}me limite de $\{Q, M\}$. \c{C}a
signifie que les trous noirs charg\'es magn\'etiquement sous les corrections
du principe d'incertitude g\'en\'eralis\'ee sont thermodynamiquement instable 
dans la limite d'une petite charge et une petite masse. De plus, nous remarquons 
de la Fig. (\ref{maggupMQ}) que la composante mixe $\{g_{MQ}\}$ de la m\'etrique 
tenseure de l'espace d'\'etat a \`a la fois les des valeurs positifs et n\'egatifs, 
sous les fluctuations de $\{M, Q\}$. Dans la limite d'une petite $M$, nous trouvons que 
$\{g_{MQ}\}$ prend une grande valeur positive \`a l'ordre de $100000$ et une grande 
valeur n\'egative \`a l'ordre de $-60000$. Dans cette limite des param\`etres, 
nous voyons que les compressibilit\'es thermiques, d\'ecritent dans les Figs. 
(\ref{maggupMM}, \ref{maggupQQ}, \ref{maggupMQ}), illustrent les propri\'et\'es 
graphiques des fluctuations d'un ensemble des trous noirs charg\'es magn\'etiquements. 
En fait, les fluctuations d'auto-paires en impliquant $\{M, Q\}$, tel que d\'efini 
par la m\'etrique tenseure $\{g_{ii} \ | \ i = M, Q\}$, ont seulement certaine valeur 
n\'egative num\'erique, tandis que la composante mixe $\{g_{MQ}\}$ a \`a la fois les 
deux valeurs.

Ainsi, nous voyons imm\'ediatement que le d\'eterminant de cette m\'etrique est,

\ba det(g)&=& -\frac{4}{M^6} \tilde{g}(M,Q),\ea

o\`u la fonction $\tilde{g}(M,Q)$ est d\'efinie par

\ba \tilde{g}(M,Q):= 128 \pi^2 M^6 - 16\pi \zeta(7)M^4+
24 \pi \zeta(7) Q^2 M^2+ \zeta(7)^2 Q^2.\ea

\begin{figure}
\hspace*{1.0cm}\vspace*{-6.0cm}
\includegraphics[width=12.0cm,angle=-0]{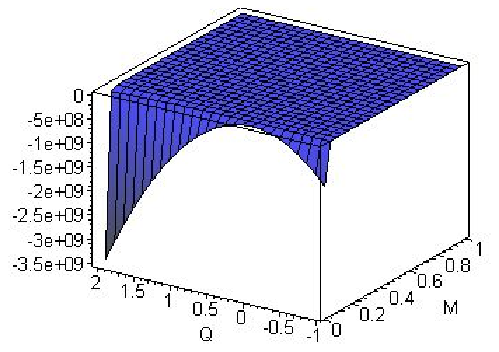}\vspace*{-3.0cm}
\caption{Le d\'eterminant de la m\'etrique tenseure trac\'ee comme la
fonction de $\{Q, M\}$, en d\'ecrivant les fluctuations de la configuration 
de trous noirs charg\'es magn\'etiquement.} \label{maggupdet}\vspace*{0.5cm}
\end{figure}

\begin{figure}
\hspace*{1.0cm}\vspace*{-6.0cm}
\includegraphics[width=12.0cm,angle=-0]{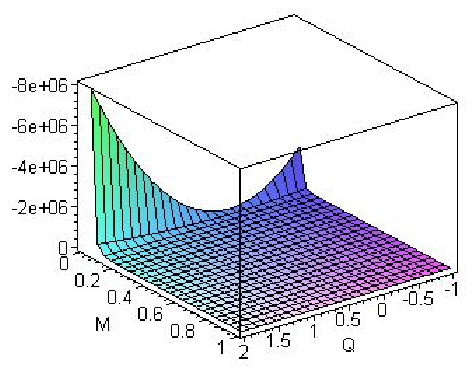}\vspace*{-3.0cm}
\caption{Le premier mineur de la m\'etrique tenseure trac\'ee comme la
fonction de $\{Q, M\}$, en d\'ecrivant les fluctuations de la configuration 
de trous noirs charg\'es magn\'etiquement.} \label{maggupminor1}\vspace*{0.5cm}
\end{figure}

Comme dans le cas des trous noirs de RN, la stabilit\'e d'un ensemble 
des trous noirs charg\'esmagn\'etiquement et corrig\'ees par le principe 
d'incertitude g\'en\'eralis\'ee peuvent \^{e}tre \'egalement d\'etermin\'ee 
en termes des valeurs de la masse $M$ et de la charge $Q$. Par ailleurs, 
cela d\'ecoule du comportement de d\'eterminant de la m\'etrique tenseure
de l'espace d'\'etat. Avis que le d\'eterminant de la m\'etrique tenseure 
tend \`a une grande valeur n\'egative \`a l'ordre de $10^{09}$. Pour le cas de 
$M \in (-1, 1)$ et $Q \in (-1, 2)$, la Fig. (\ref{maggupdet}) montre que le 
d\'eterminant de la m\'etrique tenseure r\'eside dans l'intervalle 
$(-3,5 \times 10^{+09}, 0)$. Dans ce cas, nous constatons que la n\'egativit\'e 
de $g$ augmente, quand la valeur de $Q$ est pass\'e de $-1$ \`a $2$.
Ainsi, la surface d\'efinie par les fluctuations de $\{M, Q\}$ est instables 
en raison d'une valeur n\'egative du d\'eterminant de la m\'etrique tenseure. 
Lorsque le seul param\`etre $M$ est autoris\'e \`a varier, la stabilit\'e des 
trous noirs charg\'es magn\'etiquement et corrig\'ees par le principe 
d'incertitude g\'en\'eralis\'ee est d\'etermin\'ee par la positivit\'e 
du premier mineur principe $p_1:= g_{MM}$. Nous offrons une vue rotat\'ee de 
$p_1$ dans la Fig. (\ref{maggupminor1}). Physiquement, les propri\'et\'es graphiques
ci-dessus des quantit\'es de l'espace d'\'etat, en invoquant les deux param\`etres $\{M, Q\}$,
illustrent l'image qualitative de la stabilit\'e statistique des trous noirs charg\'ees 
magn\'etiquement sous les corrections du principe d'incertitude g\'en\'eralis\'ee.

Avec les $\Gamma_{abc}$ comme nous les avons fournis dans l'annex $[A]$, 
Il n'est pas difficile d'obtenir que la courbure scalaire de Ruppenier est:

\ba R&=& 8 \pi \zeta(7)^2 M^4 (\tilde{g}(M,Q))^{-2} (3Q^2- 2M^2).\ea

Les divergences de cette equation sont donn\'ees par une equation cubique
dans une variable $N:= M^2$ qui sont seulement les points z\'eros du
d\'eterminant, ou bien les solutions de $\tilde{g}(M,Q)=0$, 
lesquels sont donn\'ees par l'\'equation:

\ba 128 \pi^2 N^3 -16 \pi \zeta(7) N^2+ 24 \pi \zeta(7) Q^2 N+
\zeta(7)^2 Q^2= 0,\ea

dont nous n'avons pas beaucoup d'int\'er\^et ici, par-ce que la vari\'et\'e 
$(\mathcal{M}_2, g)$ est d\'eg\'en\'er\'ee avec la condition $det(g)= 0$.
Cependant, il y a des autres int\'er\^ets lesquels nous allons parler dans
une autre recherche. Maintenant, nous pouvons voir facilement qu'aux racines 
de cette equation cubique, la g\'eom\'etrie de Ruppenier n'est pas bien d\'efinie. 
Sinon aux tous les autre points de la vari\'et\'e de l'espace d'\'etat, 
cette curbure scalaire est partout r\'eguli\`ere. En outre, il y a les deux 
cas particuli\`eres de la charge 

\ba Q = \pm \sqrt{\frac{2}{3}} M \ea

dont lesquelles, nous voyons que (i) ce syst\`eme n'a pas d'interaction statistique, 
(ii) la curbure scalaire correspondante de Ruppenier est nulle.

\begin{figure}
\hspace*{1.0cm}\vspace*{-6.0cm}
\includegraphics[width=12.0cm,angle=-0]{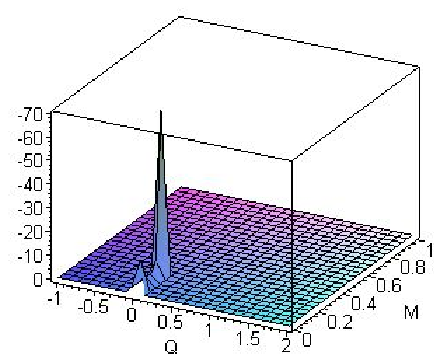}\vspace*{-3.0cm}
\caption{Le courbure scalaire trac\'ee en fonction de $\{Q, M\}$, d\'ecrivant 
les variations dans la configuration des trous noirs charg\'es magn\'etiquement 
dans la gamme $M\in (0, 1)$ et $Q \in (-1, 1) $.} \label{maggupcur}\vspace*{0.5cm}
\end{figure}

\begin{figure}
\hspace*{1.0cm}\vspace*{-6.0cm}
\includegraphics[width=12.0cm,angle=-0]{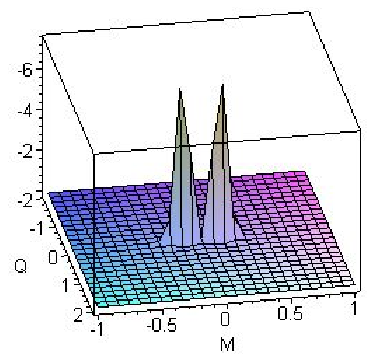}\vspace*{-3.0cm}
\caption{Le courbure scalaire trac\'ee en fonction de $\{Q, M\}$, d\'ecrivant 
les variations dans la configuration des trous noirs charg\'es magn\'etiquement 
dans la gamme $M \in (-1,1)$ et $Q \in (-2, 2)$.} \label{maggupcur1}\vspace*{0.5cm}
\end{figure}

Les propri\'et\'es globales de la stabilit\'e des trous noirs charg\'es
magn\'etiquements et corrig\'ees par le principe d'incertitude g\'en\'eralis\'ee 
d\'ecoulent de la courbure scalaire correspondante de l'espace d'\'etat. 
En particulier, dans la gamme de $M \in (0, 1)$ et $Q \in (-1, 1)$, 
la Fig. (\ref{maggupcur}) montre que la courbure scalaire a une amplitude 
n\'egative \`a l'ordre de $70$. Cela montre que la configuration des trous noirs
sous-tendents correspond \`a un syst\`eme statistique interactif. Le signe n\'egatif 
de la courbure scalaire signifie la nature attractive des interactions statistiques. 
La Fig. (\ref{maggupcur1}) signifie la nature de la courbure scalaire ci-dessus 
dans une range \'egale des param\`etres $\{M, Q\}$. Pour le cas de $M \in (-1,1)$ et $Q \in (-2, 2)$, 
on constate de la Fig. (\ref{maggupcur1}) qu'il y existe deux r\'egions identiques 
des interactions statistiques globales. De plus, en comparaison des interactions apparaissant 
dans la gamme de $M \in (0, 1)$ et $Q \in (-1, 1)$, nous voyons que l'amplitude de 
l'\'echelle des interactions statistiques semble \^{e}tre beaucoup plus petite dans 
la gamme de $M \in (-1, 1)$ et $Q \in (-2,2) $. En outre, nous constatons que le ratio 
d'une valeur typique de ces courbures scalaires de l'espace d'\'etat qui apparaissent 
ci-dessus, est \`a l'ordre de $10$. En bref, lorsque les param\`etres $\{M, Q\}$ sont 
autoris\'es \`a fluctuer, l'analyse ci-dessus de la g\'eom\'etrie thermodynamique d'une 
configuration des trous noirs charg\'es magn\'etiquement et corrig\'ees par le principe 
d'incertitude g\'en\'eralis\'ee correspondent \`a un syst\`eme statistique interactive 
et globalement instable.


\clearpage
\chapter{Les corrections de $\alpha^{\prime}$ dans la g\'eom\'etrie thermodynamique}
Dans ce chapitre, nous consid\'erons la g\'eom\'etrie thermodynamique 
des trous noirs dont les entropies, corrig\'ees par les corrections 
de $\alpha^{\prime}$, sont obtenues par la fonction de l'entropy de Sen \cite{Sen1}. 
En fait, nous avons \'egalement enqu\^et\'e sur l'utilisation de la
g\'eom\'etrie de Ruppenier dans certains bien connu syst\`emes de trous noirs,
et indiquons que la courbure scalaire de cette g\'eom\'etrie fournit quelques 
r\'esultats int\'eressants en cas de la physique des trous noirs.
Ensuite, nous avons obtenu une g\'en\'eralisation de la courbure de cette
g\'eom\'etrie \`a l'ordre sup\'erieur arbitraires des corrections
de $\alpha^{\prime}$ en termes des quantit\'es sous-jacentes de l'espace d'\'etat.
Comme, tous les trous noirs extr\'emaux ont une g\'eom\'etrie proche de la leurs horizons 
d'espace $AdS_2 \times S^{D-2}$, ce qui s'est appell\'ee le vide de Robinson-Berttoti,
et dans le cas habituelle des trous noirs de $D=4$, cette g\'eom\'etrie de l'horizon est
r\'eduite \`a $AdS_2 \times S^{2}$. De plus, il s'av\`ere que tous les champs de la th\'eorie 
de $D=4$ doivent respecter la sym\'etrie de $ SO(2,1) \times SO(3) $. 
Dans ce cadre, l'entropie d'un trou noir extr\'emal est d\'efinie par Sen \cite{Sen1}, 
comme la limite suivante: 

\ba S_{BH}^{(ext)} = \lim_{h \rightarrow 0 } S_{BH},\ea

o\`u

\ba h = r_{+}- r_{-} \ea

est la diff\'erence entre les distances de l'horizon ext\'erieur
et l'int\'erieur du trou noir. Cette limitation dans la proc\'edure de Sen 
est n\'ecessaire car les trous noirs extr\'emaux n'ont pas de l'horizon 
bifurquant de Killing. En g\'en\'eral, quand nous consid\'erons la th\'eorie 
covariante de la gravit\'e des d\'eriv\'ees sup\'erieures de deux, les Refs. 
\cite{Wald1, Wald2, Wald3, Wald4} montrent que la lagrangianne correspondante 
est donn\'ee par:

\ba \mathcal{L}:= \mathcal{L}[g_{\mu\nu},Dg_{\mu\nu},...;\Phi_s,D\Phi_s,...; F^i_{\mu\nu},DF^i_{\mu\nu},...;\gamma] \ea

qui peut \^etre aussi \'ecrite dans une forme manifestement covariante.
Notez aussi qu'il est bien connue par le th\'eorme de remplacement de Thomos
que la lagrangianne $ \mathcal{L} $ est ind\'ependante de la base $\gamma$.
Dans la consideration de Sen, cette th\'eorie de Wald \`a l'int\'er\^et par-ce que 
les d\'eriv\'ees covariantes de tous les champs tenseurs dispara\^issent,
de sorte que la formule de Walds de l'entropie peut \^etre donn\'ee par:

\ba S_{BH} := 8\pi \frac{\partial  \mathcal{L}}{R_{\alpha\beta\gamma\delta}}
g_{\alpha\gamma}g_{\beta\delta}A_H,\ea

o\`u $ A_H $  est la zone d'horizon de l'\'ev\'enement.
De plus, l'entropie d'un n'importe quel trou noir extr\'emal peut \^etre obtenu
par la m\'ethode de la fonction de l'entropie de Sen. \`A la fa\c{c}on en suivante
les Refs. \cite{Sen1, AshokeSen}, d\'efinissions une fonction:

\ba f(\vec{u},\vec{v},\vec{e},\vec{p}):= \int\int_{S^2} \sqrt{-detg}
\mathcal{L}\,d\theta\,d\phi, \ea 

o\`u les param\`etres $ (\theta,\phi) $ d\'efinissent l'\'el\'ement sur $S^2$,
ce qui est l'horizon du trou noir dans les dimensions $D= 4$ de
l'espace-temps. Ensuite, la fonction de l'entropie de Sen reste 
consistante avec les \'equations du mouvement, est donn\'ee par:

\ba F_{BH} := 2\pi(e_i \frac{\partial f}{\partial e_i} - f), \ea
 
o\`u les fonctions $e_i$ sont les $(rt)$-composantes d'une force 
tenseure des champs de jauges, peuvent \^etre definies comme 

\ba F^i_{\mu\nu}:= \partial_{ \mu} A^i_{\nu} -\partial_{ \nu} A^i_{\mu}.\ea

Ici, les charges \'electriques $\{q_i\}$ sont mesur\'ees par la transformation de Legendre

\ba q_i := \frac{\partial f}{\partial e_i}.\ea

Ainsi, la m\'ethode de la fonction de l'entropie de Sen peut \^etre consid\'er\'ee
comme une mani\`ere la plus suggestive du m\'ecanisme attracteur.
Ensuite, l'entropie de ces trous noirs est d\'efinie par l'extremum de
$ F(\vec{u},\vec{v},\vec{e},\vec{p})$ \`a l'\'egard des valeurs horizons 
de param\`etres $ \vec{u},\vec{v}$. C'est-\`a-dire que nous avons

\ba S_{BH}:= F( \overrightarrow{u}, \overrightarrow{v}, \overrightarrow{e}, 
\overrightarrow{p})\vert_{ (\overrightarrow{u_0},\overrightarrow{v_0}) }.\ea


Donc, la g\'eom\'etrie de Ruppenier est d\'efinie comme les fluctuations 
gaussiennes de la fonction de distribution des probabilit\'es  ou la n\'egative 
d'Hessianne de l'entropie \`a l'\'egard des charges invariantes 
$N^a$, o\`u $ a = 1,2 ,..., N $. Ici, nous consid\'erons la g\'eom\'etrie 
thermodynamique de l'entropie par la matrice d'Hessienne de l'entropie d'un 
trou noir extr\'emal obtenu par la fonction de l'entropie de Sen tel que 
la m\'etrique de Ruppenier est d\'efinie comme ci-dessous: 

\ba g^R_{ij} := -\partial_i\partial_j S(M,N^a),\ea

o\`u $a= 1,2,...,N$ et les derivatives partielles $\partial_i$ sont d\'efinies 
sur l'ensemble de la masse $M$ et des charges $N^a$. Nous voyons qu'il s'agit 
une forme bilin\'eaire sym\'etrique et positive d\'efinie. Pour un ensemble donn\'e 
de $\{M, N^1, N^2, \ldots, N^N\}$, l'\'el\'ement de la ligne de la g\'eom\'etrie
de l'espace d'\'etat \cite{Rupp, Aman, Sar1} est simplement donn\'ee par: 

\ba dS^2 := g^R_{ij}(M, N ^ a) dx^idx^j.\ea 

En outre, notez que ce cadre g\'eom\'etrique est attrayant, comme nous allons 
consid\'erer seulement l'entropie, et nous allons travailler habituellement 
aux points fixes d'attracteur de l'espace sous-soujacent des modules.
Cela tient du fait que l'entropie d'un trou noir extr\'emal est l'extremum 
de la fonction de l'entropie de Sen.

\section{La g\'eom\'etrie de Ruppenier des trous noirs dyoniques extr\'emaux 
supersym\'etriques en quatre dimensions.}

Dans cette section, nous examinons les corrections de $\alpha^{\prime}$
dans la g\'eom\'etrie thermodynamique d\^ues aux termes de Gauss-Bonnet de la
th\'eorie effective d'une boucle de l'action de la forme:

\ba \label{gbl} \triangle S= \int d^4 x \sqrt{- det g} \phi(a,s)
({R_{\mu\nu\rho\sigma} R^{\mu\nu\rho\sigma} 
-4 R_{\mu\nu} R^{\mu\nu}+ R^2}),\ea

o\`u $ R_{\mu\nu\rho\sigma} $ est la courbure tenseure de Riemann-Christoffel construite par 
la m\'etrique canonique: 

\ba g_{\mu\nu}= s G_{\mu\nu}, \ea 

et la fonction $ \phi(a,s)$ apparaissant ci-dessus, dans l'Eqn (\ref{gbl}), 
est d\'efinie dans le Ref. \cite{DavidJatkarSen}. Pour calculer cette fonction 
$ \phi(a,s) $, on peut voir dans la revue d'Ashoke Sen \cite{AshokeSen}, 
o\`u le r\'esultat est donn\'e par:

\ba \phi(a,s)= -\frac{1}{64 \pi^2}((k+2)\ln s+ \ln g(a+ is)+ \ln g(−a+ is))+ constant).\ea

Dans cette formule, le param\`etre $k$ est \'egale \`a la moiti\'e du nombre des certaines
formes invariantes harmoniques de type $(1, 1)$ sur $ \mathcal M $ qui d\'epend sur les
d\'etails de la compactification et la fonction $g(\tau)$ est donn\'ee par:

\ba g(\tau)= e^{2 \pi \widehat{\alpha} \tau} \prod_{n=1}^{\infty} \prod_{r=0}^{N-1}
(1- e^{2\pi ir/N} e^{2 \pi in \tau})^{s_r}.\ea

Ici, il est bien connu que les param\`etres $s_r$ comptent les nombres des formes 
harmoniques de type $p$ sur $\mathcal M$ avec les valeurs propres $e^{2\pi ir/N}$ 
pond\'er\'ee par $(-1)^p$ et $\widehat{\alpha}$ est la caractéristique d'Euler de la 
vari\'et\'e $ \mathcal M $ divis\'e par $24$ qui est respectivement \'egal \`a 
$(1, 0)$ pour $\mathcal M$ d'\^etre $(K_3, T^4) $ ce qui est la m\^eme inqui\'etude 
dans la description de la dualit\'e entre la th\'eorie des cordes h\'et\'erotiques
et la th\'eorie des cordes de type-II.

D'autre part, les corrections \`a l'entropie d'un trou noir d\^ue aux termes de
Gauss-Bonnet peuvent \^etre obtenus en consid\'erant des corrections au niveau
d'arbres de la th\'eorie des supercordes \`a certaine action effective.
Puis, par la d\'efinition de la fonction de l'entropie de Sen \cite{AshokeSen},
on peut avoir les corrections en raison des termes de Gauss-Bonnet \`a l'entropie 
de ces trous noirs supersym\'etriques. Par souci de la simplicit\'e, consid\'erons
maintenant une classe sp\'eciale des trous noirs pour lesquels, les vecteurs
de charges \'electriques et magn\'etiques sont donn\'ees par:

\ba
Q= \left( \begin{array}{rrrr} 
n \\
0 \\
w \\
0\\
\end{array} \right),\ \
P= \left( \begin{array}{rrrr} 
0 \\
W \\
0 \\
N\\
\end{array} \right). \ea

Puis dans le cas de $N, W \gg n,w$ avec $N, W > 0$ et $n, w < 0$ 
dont au pr\`es de l'horizon du trou noir, la fonction $\phi(a, s)$ 
devient de sorte qu'on a: 

\ba \phi \simeq \frac{1}{16 \pi} \widehat{\alpha} 
\sqrt{\frac{nw}{NW+ 4 \widehat{\alpha}}}.\ea

Donc, enfin de ce calcul, on obtient l'entropie des trous noirs avec les
inclusions des termes de Gauss-Bonnet, qui per le Ref. \cite{AshokeSen} 
est donn\'ee par

\ba S_{BH}= 2 \pi \sqrt{ nw (NW+ 4 \widehat{\alpha})}.\ea

Maintenant pour voir les \^id\'ees comme les interaction thermodynamiques, 
la transition des phases $ \cdots $ etc des configurations thermodynamiques 
des trous noirs, nous avons besoin d'examiner les fluctuations thermodynamiques 
qui sont bien cod\'ees dans la g\'eom\'etrie thermodynamique pour que l'\'el\'ement 
de la ligne soit d\'efinie par: 

\ba ds^2= g_{ab}dN^a dN^b, \ea

o\`u $ N^a=(n,w,N,W) $ sont les grandes charges comme nous les avons d\'efinies
dans le chapitre $3$. Puis, il n'est pas difficile de d'\'ecrire que les composantes 
de la m\'etrique tenseure de Ruppenier associ\'ees avec cet entropie, peuvent \^etre
donn\'ees simplement par:

\ba g_{nn}&=& \frac{\pi}{2n} \sqrt{\frac{w}{n}(NW+ 4 \widehat{\alpha})},\nn
g_{nw}&=& -\frac{\pi}{2}\sqrt{\frac{NW+ 4 \widehat{\alpha}}{nw}},\nn
g_{nN}&=& -\frac{\pi W}{2} \sqrt{\frac{w}{n(NW+ 4 \widehat{\alpha})}},\nn
g_{nW}&=& -\frac{\pi N}{2} \sqrt{\frac{w}{n(NW+ 4 \widehat{\alpha})}},\ea

\ba g_{ww}&=& \frac{\pi}{2w} \sqrt{\frac{w}{n}(NW+ 4 \widehat{\alpha})},\nn
g_{wN}&=& -\frac{\pi W}{2} \sqrt{\frac{n}{w(NW+ 4 \widehat{\alpha})}},\nn
g_{wW}&=& -\frac{\pi N}{2} \sqrt{\frac{n}{w(NW+ 4 \widehat{\alpha})}},\nn
g_{NN}&=& \frac{\pi W^2 \sqrt{nw}}{2(NW+ 4 \widehat{\alpha})^{3/2}},\nn
g_{NW}&=& -\frac{\pi \sqrt{nw}(NW+ 8 \widehat{\alpha})}{2(NW+ 4 \widehat{\alpha})^{3/2}},\nn
g_{WW}&=& \frac{\pi N^2 \sqrt{nw}}{2(NW+ 4 \widehat{\alpha})^{3/2}}.\ea

\begin{figure}
\hspace*{1.0cm}\vspace*{-6.0cm}
\includegraphics[width=12.0cm,angle=-0]{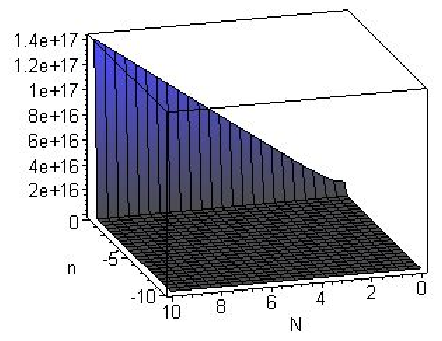}\vspace*{-3.0cm}
\caption{La composante $nn$ de la m\'etrique tenseure trac\'e comme la 
fonction de $\{n, N\}$, en d\'ecrivant les fluctuations dans la 
configuration des trous noirs supersym\'etriques corrig\'es par 
les termes de Gauss-Bonnet.} \label{alphaGBnn1}\vspace*{0.5cm}
\end{figure}

\begin{figure}
\hspace*{1.0cm}\vspace*{-6.0cm}
\includegraphics[width=12.0cm,angle=-0]{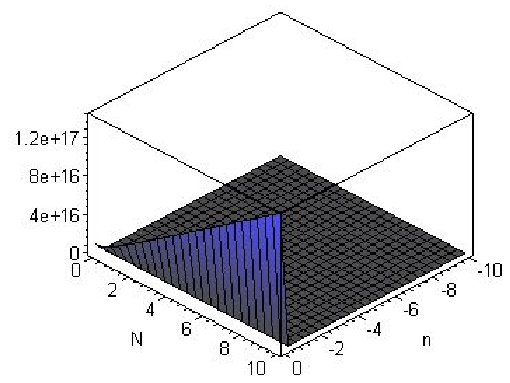}\vspace*{-3.0cm}
\caption{La composante $ww$ de la m\'etrique tenseure trac\'e comme la 
fonction de $\{n, N\}$, en d\'ecrivant les fluctuations dans la 
configuration des trous noirs supersym\'etriques corrig\'es par 
les termes de Gauss-Bonnet.} \label{alphaGBww5}\vspace*{0.5cm}
\end{figure}

\begin{figure}
\hspace*{1.0cm}\vspace*{-6.0cm}
\includegraphics[width=12.0cm,angle=-0]{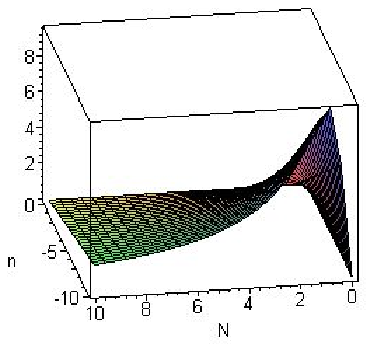}\vspace*{-3.0cm}
\caption{La composante $NN$ de la m\'etrique tenseure trac\'e comme la 
fonction de $\{n, N\}$, en d\'ecrivant les fluctuations dans la 
configuration des trous noirs supersym\'etriques corrig\'es par 
les termes de Gauss-Bonnet.} \label{alphaGBNN8}\vspace*{0.5cm}
\end{figure}

\begin{figure}
\hspace*{1.0cm}\vspace*{-6.0cm}
\includegraphics[width=12.0cm,angle=-0]{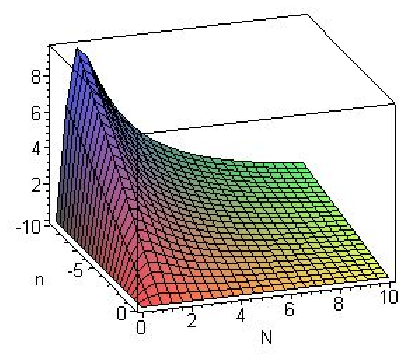}\vspace*{-3.0cm}
\caption{La composante $WW$ de la m\'etrique tenseure trac\'e comme la 
fonction de $\{n, N\}$, en d\'ecrivant les fluctuations dans la 
configuration des trous noirs supersym\'etriques corrig\'es par 
les termes de Gauss-Bonnet.} \label{alphaGBWW10}\vspace*{0.5cm}
\end{figure}

\begin{figure}
\hspace*{1.0cm}\vspace*{-6.0cm}
\includegraphics[width=12.0cm,angle=-0]{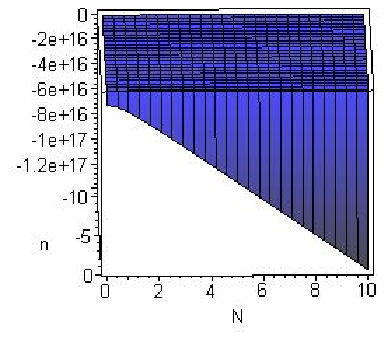}\vspace*{-3.0cm}
\caption{La composante $nw$ de la m\'etrique tenseure trac\'e comme la 
fonction de $\{n, N\}$, en d\'ecrivant les fluctuations dans la 
configuration des trous noirs supersym\'etriques corrig\'es par 
les termes de Gauss-Bonnet.} \label{alphaGBnw2}\vspace*{0.5cm}
\end{figure}

\begin{figure}
\hspace*{1.0cm}\vspace*{-6.0cm}
\includegraphics[width=12.0cm,angle=-0]{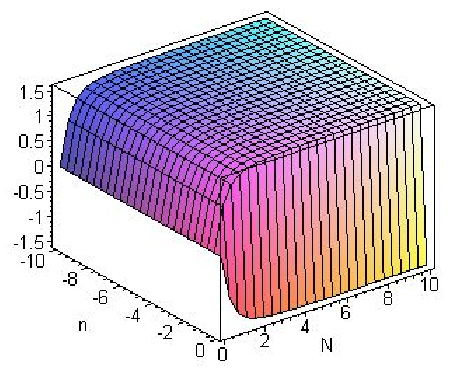}\vspace*{-3.0cm}
\caption{La composante $nN$ de la m\'etrique tenseure trac\'e comme la 
fonction de $\{n, N\}$, en d\'ecrivant les fluctuations dans la 
configuration des trous noirs supersym\'etriques corrig\'es par 
les termes de Gauss-Bonnet.} \label{alphaGBnN3}\vspace*{0.5cm}
\end{figure}

\begin{figure}
\hspace*{1.0cm}\vspace*{-6.0cm}
\includegraphics[width=12.0cm,angle=-0]{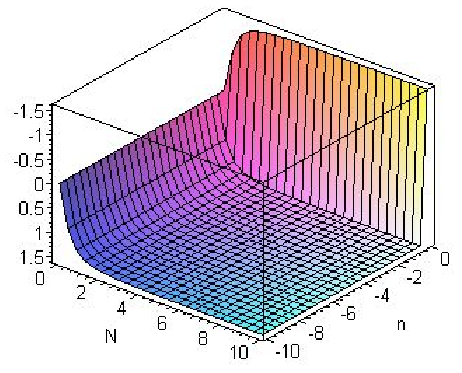}\vspace*{-3.0cm}
\caption{La composante $nW$ de la m\'etrique tenseure trac\'e comme la 
fonction de $\{n, N\}$, en d\'ecrivant les fluctuations dans la 
configuration des trous noirs supersym\'etriques corrig\'es par 
les termes de Gauss-Bonnet.} \label{alphaGBnW4}\vspace*{0.5cm}
\end{figure}

\begin{figure}
\hspace*{1.0cm}\vspace*{-6.0cm}
\includegraphics[width=12.0cm,angle=-0]{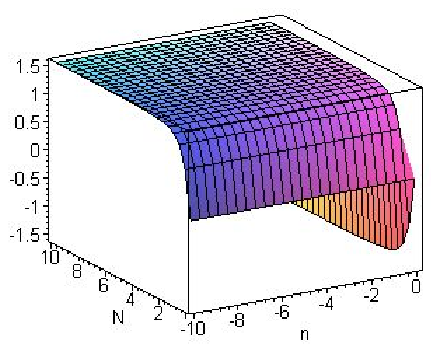}\vspace*{-3.0cm}
\caption{La composante $wN$ de la m\'etrique tenseure trac\'e comme la 
fonction de $\{n, N\}$, en d\'ecrivant les fluctuations dans la 
configuration des trous noirs supersym\'etriques corrig\'es par 
les termes de Gauss-Bonnet.} \label{alphaGBwN6}\vspace*{0.5cm}
\end{figure}

\begin{figure}
\hspace*{1.0cm}\vspace*{-6.0cm}
\includegraphics[width=12.0cm,angle=-0]{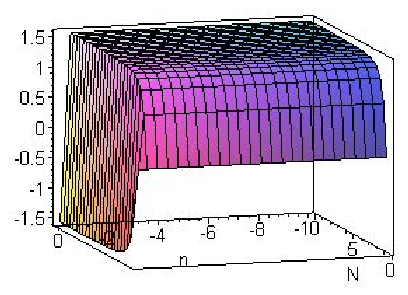}\vspace*{-3.0cm}
\caption{La composante $wW$ de la m\'etrique tenseure trac\'e comme la 
fonction de $\{n, N\}$, en d\'ecrivant les fluctuations dans la 
configuration des trous noirs supersym\'etriques corrig\'es par 
les termes de Gauss-Bonnet.} \label{alphaGBwW7}\vspace*{0.5cm}
\end{figure}

\begin{figure}
\hspace*{1.0cm}\vspace*{-6.0cm}
\includegraphics[width=12.0cm,angle=-0]{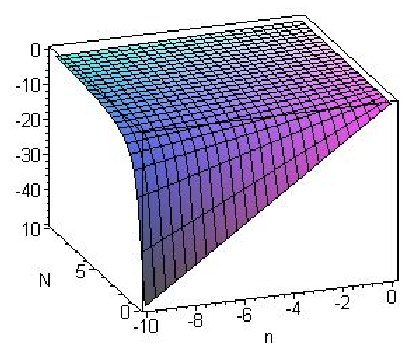}\vspace*{-3.0cm}
\caption{La composante $NW$ de la m\'etrique tenseure trac\'e comme la 
fonction de $\{n, N\}$, en d\'ecrivant les fluctuations dans la 
configuration des trous noirs supersym\'etriques corrig\'es par 
les termes de Gauss-Bonnet.} \label{alphaGBNW9}\vspace*{0.5cm}
\end{figure}

Puisqu'il y a des quatre variables ind\'ependantes, en particulier les nombres de branes
$\{n, w, N, W \}$, donc en afin d'offrir leur vue graphique en trois dimensions, 
nous allons consid\'ere la limite $n = w $ et $N = W $. Pour une correction  donn\'ee
de Gauss-Bonnet aux trous noirs extr\^{e}maux supersym\'etriques, nous allons choisir 
la valeur du param\`etre des corrections d\'eriv\'ees sup\'erieures de la th\'eorie 
des cordes d'\^{e}tre $\widehat{\alpha}= 0.1$ tout au long du pr\'esent chapitre. 
Ainsi, dans le r\'egime de $n \in (-10, 0)$ et $N \in (0, 10)$, nous observons
que l'amplitude des capacit\'es de chaleurs $\{g_{nn}, g_{ww}\}$ corrig\'ees par
les $\alpha^{\prime}$ prend une valeur \`a l'ordre de $10^{17}$. Dans cette gamme de 
$n, N$, les capacit\'es de chaleurs $\{g_{NN}, g_{WW}\}$ relient dans la gamme de 
$(0, 10)$. En conform\'ement \`a la pr\'ediction de la g\'eom\'etrie de l'espace d'\'etat,
nous constatons que la gamme de croissance de la premi\`ere ensemble et de la deuxi\`eme 
ensemble des capacit\'es de chaleur reste dans la limite oppos\'ee des param\`etres
$\{n, N\}$. \`A savoir, la premi\`ere s\'erie de capacit\'es de chaleurs augmente 
avec la valeur de $N$, tandis que l'autre croissante diminue. Du fait m\^{e}me, les Figs. 
(\ref{alphaGBnn1}, \ref{alphaGBww5}) montrent que la croissance de premier ensemble 
des capacit\'es de chaleurs $\{g_{nn}, g_{ww}\}$ prend place dans la limite d'un grand 
$N$ et un petit $n$. En outre, les Figs. (\ref{alphaGBNN8}, \ref{alphaGBWW10}) 
montrent que la croissance de $\{g_{NN}, g_{WW}\}$ a lieu dans la limite d'un
petit $N$ et un grand $n$. De plus, les compressibilit\'es de chaleurs, 
en impliquant deux param\`etres distincts de la configuration de ces trous noirs,
sont repr\'esent\'es dans les Figs. (\ref{alphaGBnw2},\ref{alphaGBnN3}, \ref{alphaGBnW4},
\ref{alphaGBwN6}, \ref{alphaGBwW7}, \ref{alphaGBNW9}). En fait, les fluctuations en
impliquant les param\`etres $\{n, w\}$, tel que d\'efini par la m\'etrique tenseure de 
l'espace d'\'etat $\{g_{ij} \ | \ i, j = n, w \}$, ont relativement des plus grande 
valeurs num\'eriques en comparaison de celles qui impliquant les param\`etres $\{N, W\}$.

Par ailleurs, pour la m\'{e}trique tenseure ci-dessus, il n'est pas difficile 
de montrer que les mineurs principaux $\{\mathit{p_1}, \mathit{p_2}, \mathit{p_3}\}$
sous-jacents peuvent \^{e}tre exprim\'{e} comme

\begin{eqnarray}
\mathit{p_1}&=& {\displaystyle \frac{\pi}{2\, n}} \,{\displaystyle
\sqrt{\frac{w}{n}\,(N\,\mathrm{W} + 4\,\widehat{\alpha})}},  \nonumber \\
\mathit{p_2} &=& 0,  \nonumber \\
\mathit{p_3} &=&  - {\displaystyle \frac{1}{2}} \,{\displaystyle
\frac{\pi ^{3}\,\mathrm{W}^{2}}{\sqrt{n\,w\,(N\,\mathrm{W} + 4\,
\widehat{\alpha})}}}. 
\end{eqnarray}

Nous voyons sans aucune difficult\'e que le d\'eterminant de cette m\'etrique est:

\ba g= -\pi^4 \frac{NW}{NW+ 4 \widehat{\alpha}}.\ea

\begin{figure}
\hspace*{1.0cm}\vspace*{-6.0cm}
\includegraphics[width=12.0cm,angle=-0]{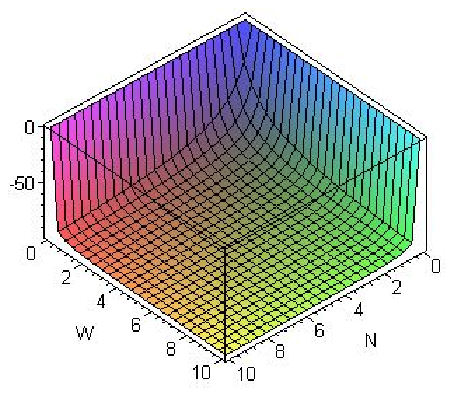}\vspace*{-3.0cm}
\caption{Le d\'eterminant de la m\'etrique tenseure trac\'e comme la 
fonction de $\{n, N\}$, en d\'ecrivant les fluctuations dans la 
configuration des trous noirs supersym\'etriques corrig\'es par 
les termes de Gauss-Bonnet.} \label{alphaGBdetg}\vspace*{0.5cm}
\end{figure}

\begin{figure}
\hspace*{1.0cm}\vspace*{-6.0cm}
\includegraphics[width=12.0cm,angle=-0]{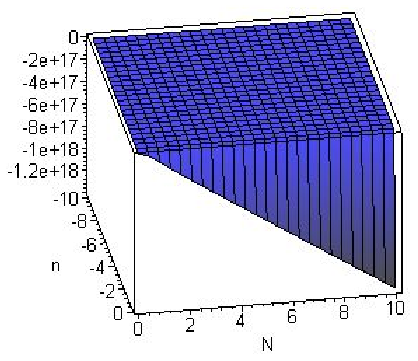}\vspace*{-3.0cm}
\caption{Le mineur d'hypersurface de la m\'etrique tenseure trac\'e comme la 
fonction de $\{n, N\}$, en d\'ecrivant les fluctuations dans la 
configuration des trous noirs supersym\'etriques corrig\'es par 
les termes de Gauss-Bonnet.} \label{alphaGBminor3}\vspace*{0.5cm}
\end{figure}

\begin{figure}
\hspace*{1.0cm}\vspace*{-6.0cm}
\includegraphics[width=12.0cm,angle=-0]{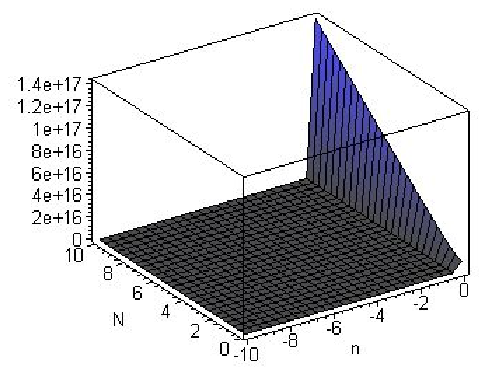}\vspace*{-3.0cm}
\caption{Le premier mineur de la m\'etrique tenseure trac\'e comme la 
fonction de $\{n, N\}$, en d\'ecrivant les fluctuations dans la 
configuration des trous noirs supersym\'etriques corrig\'es par 
les termes de Gauss-Bonnet.} \label{alphaGBminor1}\vspace*{0.5cm}
\end{figure}

Comme une fonction des param\`etres $\{N, W\}$, la stabilit\'e d'un ensemble 
des trous noirs supersym\'etriques extr\'emaux, corrig\'ees par les termes de
Gauss-Bonnet, suivit de la positivit\'e du d\'eterminant de la m\'etrique tenseure.
Dans ce cas, nous remarquons que le d\'eterminant de la m\'etrique tenseure 
$g$ est ind\'epend des param\`etres $\{n, w\}$. En effet, nous constatons que 
$g \in (-100, 0)$ pour une valeur typique de $N \in (0, 10)$ et $W \in (0, 10)$. 
En ce cas, la Fig. (\ref{alphaGBdetg}) offre la nature graphique du d\'eterminant 
de la m\'etrique tenseure $g$. Pour le choix de $n= w $, $N= W$, la stabilit\'e de 
l'hypersurface d\'efinie par une constante valeur de $W$ est montr\'e dans la 
Fig. (\ref{alphaGBminor3}). Par la pr\'esente, nous pouvons voir que le mineur $p_3$ 
r\'eside dans la gamme de $(-1.2 \times 10^{+18}, 0)$. Notez bien en cas de $n \in (-10, 0)$ 
et $N \in (0, 10)$ que la n\'egativit\'e de $p_3$ augmente \`a mesure que la valeur 
de $N$ est pass\'ee de z\'ero \`a $10$. En outre, la surface d\'efinie par les 
fluctuations de $(n, w)$ est instable \`a cause du fait m\^eme que le mineur 
principe correspondant est identiquement nulle, \`a savoir que nous avons $p_2= 0$. 
Enfin, lorsque le seul param\`etre $n$ est autoris\'e \`a varier, 
la stabilit\'e de la configuration des trous noirs sous-jacents est donn\'ee 
par la positivit\'e du premier mineur principe $p_1:= g_{nn}$. Une vue rotat\'ee
de $p_1$ est montr\'ee dans la Fig. (\ref{alphaGBminor1}). Les propri\'et\'es 
ci-dessus de l'espace de l'\'etat et la positivit\'e des mineurs principes 
concern\'es fournent la notion qualitative de la stabilit\'e statistique
des trous noirs supersym\'etriques extr\'emaux corrig\'ees par les termes 
de Gauss-Bonnet.

De plus, comme nous avons fourni les $\Gamma_{abc}$ dans l'annex $[A]$, 
il est aussi facile d'obtenir pour cette $ g_{ab}(n,w,N,W) $
que la courbure scalaire de Ruppenier est simplement:

\ba R= \frac{3}{2 \pi NW} \frac{2 \widehat{\alpha}+ NW}{\sqrt{ nw (NW+ 4 \widehat{\alpha})}},\ea

qu'elle est partout r\'eguli\`ere. 
On voit que cette courbure scalaire de Ruppenier est nulle 
pour tous $(N, W)$ tel que 

\ba NW= -2\widehat{\alpha}, \ea

qui est possible si et seulement si $ \widehat{\alpha} < 0 $, pour le cas de $N, W > 0$. 
Il est aussit\^ot de voir que sans les corrections de $ \alpha^{\prime} $,
le d\'eterminant de la m\'etrique tenseure $ g_{ab}(n,w,N,W) $
et la courbure scalaire sont respectivement donn\'ees par

\ba g= -\pi^4 \ea et 

\ba R= \frac{3}{2 \pi \sqrt{ nwNW }}.\ea

\begin{figure}
\hspace*{1.0cm}\vspace*{-6.0cm}
\includegraphics[width=12.0cm,angle=-0]{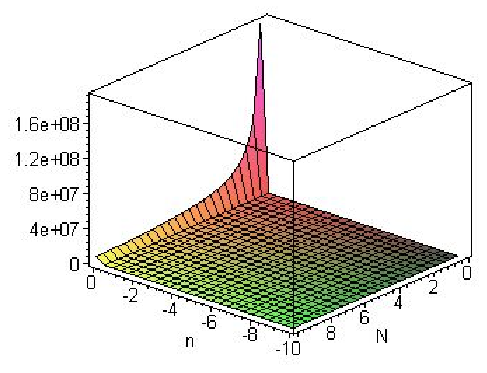}\vspace*{-3.0cm}
\caption{La courbure scalaire trac\'e comme la fonction de $\{n, N\}$, 
en d\'ecrivant les fluctuations dans la configuration des trous noirs 
supersym\'etriques corrig\'es par les termes de Gauss-Bonnet dans la gamme
$n \in (-10, 0)$ et $N \in (0,10) $.} \label{alphaGBR}\vspace*{0.5cm}
\end{figure}

\begin{figure}
\hspace*{1.0cm}\vspace*{-6.0cm}
\includegraphics[width=12.0cm,angle=-0]{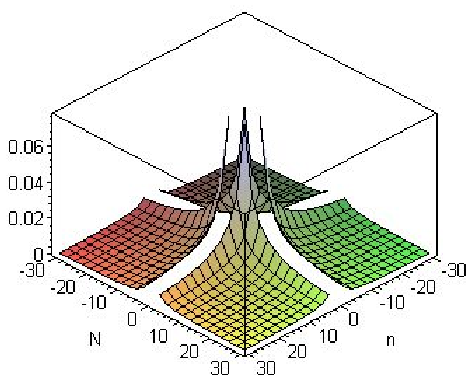}\vspace*{-3.0cm}
\caption{La courbure scalaire trac\'e comme la fonction de $\{n, N\}$, 
en d\'ecrivant les fluctuations dans la configuration des trous noirs 
supersym\'etriques corrig\'es par les termes de Gauss-Bonnet dans la gamme
$n, N \in (-30,30) $.} \label{alphaGBR30}\vspace*{0.5cm}
\end{figure}

En g\'en\'erale, les propri\'et\'es statistiques globales d\'ecoulent du comportement 
de la courbure scalaire de l'espace d'\'etat. Dans la gamme de $n \in (-10, 0)$
et $N \in (0, 10)$, la Fig. (\ref{alphaGBR}) montre que la courbure scalaire 
a une grande amplitude positive \`a l'ordre de $10^{+08} $. Nous observons que 
la configuration sous-jacente des trous noirs est un syst\`eme statistique 
fortement interagissant dans cette gamme des param\`etres. Physiquement, 
le signe positif de la courbure scalaire signifie que les interactions 
statistiques sont r\'epulsives dans la leur nature. La Fig. (\ref{alphaGBR30}) 
illustre le comportement de ce qui pr\'ec\`ede de la courbure scalaire dans
la gamme des param\`etres $n, N \in (-30, 30)$. En fait, lorsque la gamme de 
$n$ et $N$ est prise dans le multiple de $3$, alors nous remarquons de la 
Fig. (\ref{alphaGBR30}) qu'il existe quatre r\'egions disjointes similaires 
des interactions statistiques globales. En comparaison des interactions 
apparaissant dans la gamme de $n \in (-10, 0)$ et $N \in (0, 10)$, l'amplitude 
des interactions statistiques globales se r\'ev\`ele \^{e}tre beaucoup plus petite
dans la gamme de $n, N \in (-30,30) $. Sous les corrections de d\'eriv\'ees
sup\'erieures de $\alpha^{\prime}$, nous constatons que le ratio des valeurs 
typiques de l'amplitude de ces courbures scalaires de l'espace d'\'etat est
\`a l'ordre de $10^{-10}$. La vue graphique de la comparaison mentionn\'ee ci-dessus
de la courbure scalaire est apparente dans les Figs. (\ref{alphaGBR}, \ref{alphaGBR30}). 
Qualitativement, en cas de petite valeur des param\`etres $\{n, N\}$, nous voyons par
les repr\'esentations de l'espace d'\'etat indiquent que la configuration 
des trous noirs supersym\'etriques extr\'emaux corrig\'ees par les termes de 
Gauss-Bonnet correspond \`a un syst\`eme statistique globalement instable
en vertu des variations de param\`etres $\{n, w, N, W\}$.

En fait, il bien connu dans la g\'eom\'etrie de Ruppenier avec quatre charges
que sans les corrections de $ \alpha^{\prime} $, $g$ et $R$ prennent ces formes,
voir \cite{Sar2} pour les d\'etails. En outre, cela reste toute vraie aussi pour 
le cas des trous noirs extr\'emaux non-supersym\'etriques en quatre dimensions de
l'espace-temps ce que nous allons analyser dans la section suivante. Et bien aussi, 
voyez la section, sans les corrections de $ \alpha^{\prime} $, pour les branes noirs 
$D_2D_6NS_5$ non-extr\'emaux en dimensions $D = 10$ de l'espace-temps.


\section{La g\'eom\'etrie de Ruppenier des trous noirs dyoniques extr\'emaux 
non-supersym\'etriques en quatre dimensions.}

Nous somme maintenant en mesure de calculer les contributions perturbatives
de $ \alpha^{\prime} $ \`a la g\'eom\'etrie thermodynamique de Ruppenier pour 
les trous noirs dyoniques extr\'emaux non-supersym\'etriques en quatre dimensions.
Il est bien connu que l'entropie de ces trous noirs extr\'emaux supersym\'etriques
et bien aussi les trous noirs non-supersym\'etriques dans la th\'eorie des supergravit\'es 
en $ \mathcal N = 2 $ en quatre dimensions avec les corrections des d\'eriv\'ees 
sup\'erieures de $ \alpha^{\prime}$ peut \^etre calcul\'ee facilement par la 
m\'ethode de la fonction d'entropie de Sen \cite{SahooSen}.
En particulier, les trous noirs extr\'emaux, dans le mod\`ele de STU avec les trois
multiplets de vecteurs, peuvent \^etre \'ecrits d'un sous-secteur d'\'energie
faible de l'action effective pour la th\'eorie des cordes h\'et\'erotiques au
niveau d'arbre sur le tore $T^6$ ou $K_3 \times T^2$.
Nous savons que ce mod\`ele contient \'egalement les trous noirs extr\'emaux
supersym\'etriques et aussi les trous noirs extr\'emaux non-supersym\'etriques.
Surtout, il s'av\`ere qu'on peut obtenir ces trous noirs des valeurs diff\'erentes 
de charges vecteurs. Il est int\'eressant de noter que l'entropie de ces trous noirs 
peut \^etre obtenue par l'extr\'emisation de la fonction d'entropie de Sen.
Alors, au niveau d'arbre de la th\'eorie des cordes h\'et\'erotiques, nous avons
une forme simple de l'entropie d'un trou noir extr\'emal non-supersym\'etrique
qui est peut \^etre donn\'ee g\'en\'eralement par la m\'ethode de la fonction
d'entropie de Sen \cite{AshokeSen}. Par souci de simplicit\'e, nous allons
consid\'erer les vecteurs charges de la th\'eorie avec les conditions 
$ N^{\prime}, W^{\prime}, \widehat{n}>0 $ et $ \widehat{w}<0 $ aux formes:

\ba 
Q= \left( \begin{array}{rrrr} 
\widehat{n} \\
0 \\
\widehat{w} \\
0\\
\end{array} \right), \ \
P= \left( \begin{array}{rrrr} 
0 \\
W^{\prime} \\
0 \\
N^{\prime}\\
\end{array} \right).\ea

\subsection{\`A l'ordre de $(\alpha^{\prime})^0 $}

Ensuite, avec les normes de la supersym\'etrie de $ \mathcal N = 2 $ 
et celles de $ \mathcal N = 4 $, les relations entres des champs scalaires,  
il s'av\`ere par le Ref.\cite{AshokeSen} que dans ce cas des param\`etres 
$ (\widehat{n},\widehat{w}, N^{\prime}, W^{\prime} )$ le r\'esultat 
de l'entropie d'un trou noir extr\'emal non-supersym\'etrique est: 

\ba S_{BH}^{ns}= 2 \pi \sqrt{\vert \widehat{n}
\widehat{w}\vert N^{\prime} W^{\prime}}.\ea

Maintenant nous d\'efinissions la g\'eom\'etrie thermodynamique avec un vecteur, 
$ \overrightarrow{N}:= (n,w,N,W) $, o\`u $ w= \vert \widehat{w} \vert $.
Ici, le $ \overrightarrow{N} $ parametrise l'entropie ci-dessus comme: 

\ba S_{BH}^{ns}= 2 \pi \sqrt{nw N W}. \ea

Cette repr\'esentation de l'entropie \'ecarte les calculs de la g\'eom\'etrie 
thermodynamique de la configuration sous-jacente. Donc, avec ces quatre charges 
\'electriques et magn\'etiques, on peut \'ecrire simplement que les composantes 
de la m\'etrique de Ruppenier associ\'ees avec cet entropie sont donn\'ees par:

\ba g_{nn}&=& \frac{\pi}{2n} \sqrt{\frac{wNW}{n}},\nn
g_{nw}&=& -\frac{\pi}{2}\sqrt{\frac{NW}{nw}},\nn
g_{nN}&=& -\frac{\pi}{2} \sqrt{\frac{wW}{nN}},\nn
g_{nW}&=& -\frac{\pi}{2} \sqrt{\frac{wN}{nW}},\ea

\ba g_{ww}&=& \frac{\pi}{2w} \sqrt{\frac{nNW}{w}},\nn
g_{wN}&=& -\frac{\pi}{2} \sqrt{\frac{nW}{wN}},\nn
g_{wW}&=& -\frac{\pi}{2} \sqrt{\frac{nN}{wW}},\nn
g_{NN}&=& \frac{\pi}{2N} \sqrt{\frac{nwW}{N}},\nn
g_{NW}&=& -\frac{\pi}{2} \sqrt{\frac{nw}{NW}},\nn
g_{WW}&=& \frac{\pi}{2W} \sqrt{\frac{nwN}{W}}.\ea

\begin{figure}
\hspace*{1.0cm}\vspace*{-6.0cm}
\includegraphics[width=12.0cm,angle=-0]{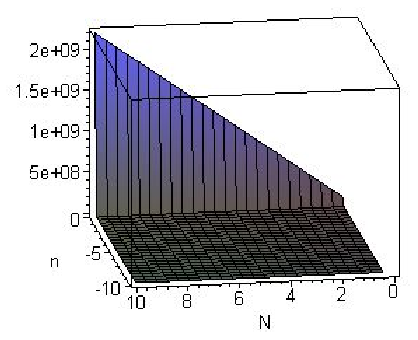}\vspace*{-3.0cm}
\caption{La composante $nn$ de la m\'etrique tenseure trac\'ee 
comme la fonction de $\{n, N\}$, d\'ecrivant les fluctuations dans 
la configuration des trous noirs charg\'es non-supersym\'etriques 
\`a l'ordre dominant.} \label{nonsusy0nn1}\vspace*{0.5cm}
\end{figure}

\begin{figure}
\hspace*{1.0cm}\vspace*{-6.0cm}
\includegraphics[width=12.0cm,angle=-0]{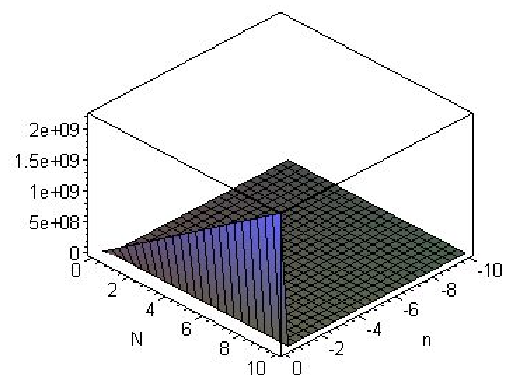}\vspace*{-3.0cm}
\caption{La composante $ww$ de la m\'etrique tenseure trac\'ee 
comme la fonction de $\{n, N\}$, d\'ecrivant les fluctuations dans 
la configuration des trous noirs charg\'es non-supersym\'etriques 
\`a l'ordre dominant.} \label{nonsusy0ww5}\vspace*{0.5cm}
\end{figure}

\begin{figure}
\hspace*{1.0cm}\vspace*{-6.0cm}
\includegraphics[width=12.0cm,angle=-0]{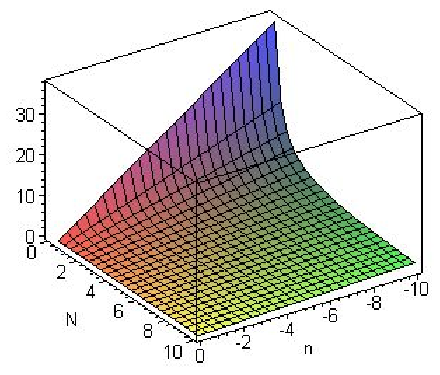}\vspace*{-3.0cm}
\caption{La composante $NN$ de la m\'etrique tenseure trac\'ee 
comme la fonction de $\{n, N\}$, d\'ecrivant les fluctuations dans 
la configuration des trous noirs charg\'es non-supersym\'etriques 
\`a l'ordre dominant.} \label{nonsusy0NN8}\vspace*{0.5cm}
\end{figure}

\begin{figure}
\hspace*{1.0cm}\vspace*{-6.0cm}
\includegraphics[width=12.0cm,angle=-0]{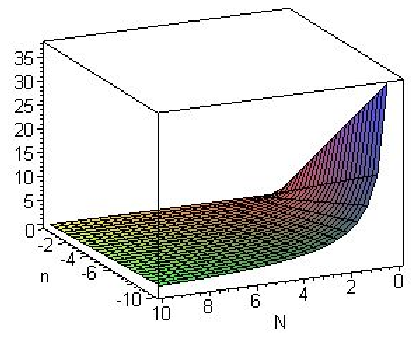}\vspace*{-3.0cm}
\caption{La composante $WW$ de la m\'etrique tenseure trac\'ee 
comme la fonction de $\{n, N\}$, d\'ecrivant les fluctuations dans 
la configuration des trous noirs charg\'es non-supersym\'etriques 
\`a l'ordre dominant.} \label{nonsusy0WW10}\vspace*{0.5cm}
\end{figure}

\begin{figure}
\hspace*{1.0cm}\vspace*{-6.0cm}
\includegraphics[width=12.0cm,angle=-0]{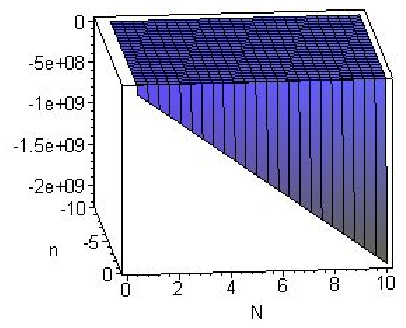}\vspace*{-3.0cm}
\caption{La composante $nw$ de la m\'etrique tenseure trac\'ee 
comme la fonction de $\{n, N\}$, d\'ecrivant les fluctuations dans 
la configuration des trous noirs charg\'es non-supersym\'etriques 
\`a l'ordre dominant.} \label{nonsusy0nw2}\vspace*{0.5cm}
\end{figure}

\begin{figure}
\hspace*{1.0cm}\vspace*{-6.0cm}
\includegraphics[width=12.0cm,angle=-0]{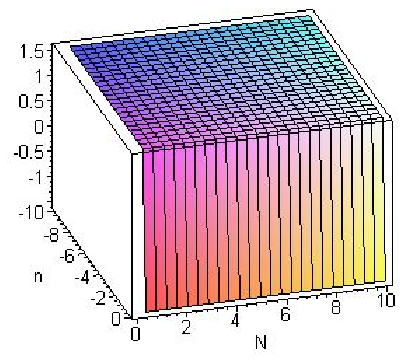}\vspace*{-3.0cm}
\caption{La composante $nN$ de la m\'etrique tenseure trac\'ee 
comme la fonction de $\{n, N\}$, d\'ecrivant les fluctuations dans 
la configuration des trous noirs charg\'es non-supersym\'etriques 
\`a l'ordre dominant.} \label{nonsusy0nN3}\vspace*{0.5cm}
\end{figure}

\begin{figure}
\hspace*{1.0cm}\vspace*{-6.0cm}
\includegraphics[width=12.0cm,angle=-0]{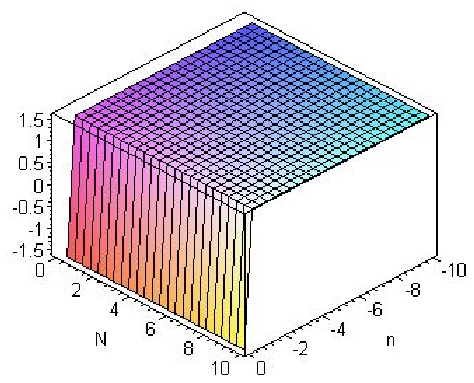}\vspace*{-3.0cm}
\caption{La composante $nW$ de la m\'etrique tenseure trac\'ee 
comme la fonction de $\{n, N\}$, d\'ecrivant les fluctuations dans 
la configuration des trous noirs charg\'es non-supersym\'etriques 
\`a l'ordre dominant.} \label{nonsusy0nW4}\vspace*{0.5cm}
\end{figure}

\begin{figure}
\hspace*{1.0cm}\vspace*{-6.0cm}
\includegraphics[width=12.0cm,angle=-0]{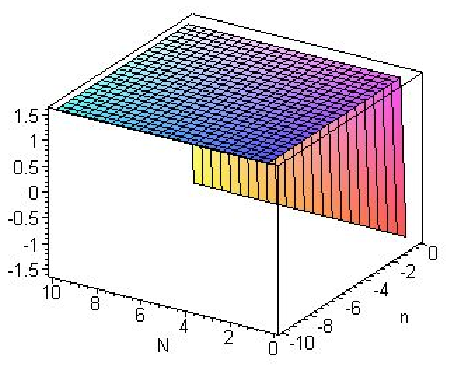}\vspace*{-3.0cm}
\caption{La composante $wN$ de la m\'etrique tenseure trac\'ee 
comme la fonction de $\{n, N\}$, d\'ecrivant les fluctuations dans 
la configuration des trous noirs charg\'es non-supersym\'etriques 
\`a l'ordre dominant.} \label{nonsusy0wN6}\vspace*{0.5cm}
\end{figure}

\begin{figure}
\hspace*{1.0cm}\vspace*{-6.0cm}
\includegraphics[width=12.0cm,angle=-0]{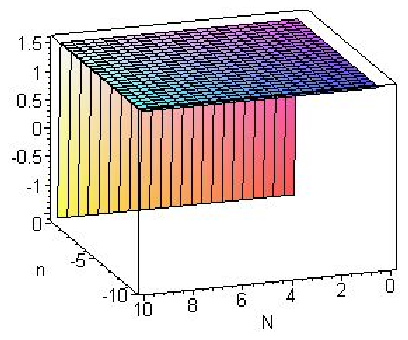}\vspace*{-3.0cm}
\caption{La composante $wW$ de la m\'etrique tenseure trac\'ee 
comme la fonction de $\{n, N\}$, d\'ecrivant les fluctuations dans 
la configuration des trous noirs charg\'es non-supersym\'etriques 
\`a l'ordre dominant.} \label{nonsusy0wW7}\vspace*{0.5cm}
\end{figure}

\begin{figure}
\hspace*{1.0cm}\vspace*{-6.0cm}
\includegraphics[width=12.0cm,angle=-0]{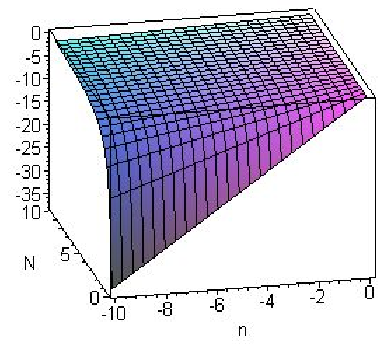}\vspace*{-3.0cm}
\caption{La composante $NW$ de la m\'etrique tenseure trac\'ee 
comme la fonction de $\{n, N\}$, d\'ecrivant les fluctuations dans 
la configuration des trous noirs charg\'es non-supersym\'etriques 
\`a l'ordre dominant.} \label{nonsusy0NW9}\vspace*{0.5cm}
\end{figure}

\`A la limite de $n= w$ et $N= W$, on observe dans l'intervalle de
$n \in (-10, 0)$ et $N \in (0, 10)$ que l'amplitudes des capacit\'es 
de chaleures $\{g_{nn}, g_{ww}\}$ prendent la valeur maximale \`a l'ordre
de $2 \times 10^{09}$. Du fait m\^{e}me, les capacit\'es de chaleurs $\{g_{NN}, g_{WW}\}$ 
prendent une valeur typique entre $(0, 40)$. Nous observons que la gamme
de croissance de la premi\`ere ensemble et celle de la deuxi\`eme ensemble 
des capacit\'es de chaleurs se trouve \^{e}tre \`a la limite oppos\'ee 
des param\`etres $\{n, N\}$. Explicitement, par les Figs. (\ref{nonsusy0nn1},
\ref{nonsusy0ww5}), nous constatons que la croissance des capacit\'es 
de chaleurs $\{g_{nn}, g_{ww}\}$ a lieu dans la limite d'un grand $N$
et d'un petit $n$. D'autre part, par les Figs. (\ref{nonsusy0NN8}, \ref{nonsusy0WW10}), 
nous voyons que la croissance de $\{g_{NN}, g_{WW}\}$ a lieu dans la limite d'un
petit $N$ et un grand $n$. De plus, les compressibilit\'es de chaleures en
impliquant deux param\`etres distincts de la configuration de ces trous noirs 
non-supersym\'etriques ont repr\'esent\'ees dans les Figs. (\ref{nonsusy0nw2},
\ref{nonsusy0nN3}, \ref{nonsusy0nW4}, \ref{nonsusy0wN6}, \ref{nonsusy0wW7}, 
\ref{nonsusy0NW9}). Par la d\'efinition de la m\'etrique tenseure de l'espace
d'\'etat $\{g_{ij} \ | \ i, j = n, w, N, W\}$, nous observer que les fluctuations 
en impliquant $\{n, w\}$ ont relativement une plus grande valeur num\'erique 
de celles qui impliquant $\{N, W\}$.

Par l'expression de la m\'{e}trique tenseure, on constate que les mineurs principaux sont donn\'{e}s par

\begin{eqnarray}
\mathit{p_1}&=& {\displaystyle \frac{\pi}{2\, n}} \,{\displaystyle
\sqrt{\frac{w\,N\,\mathrm{W}}{n}}},  \nonumber \\
\mathit{p_2}&=& 0,  \nonumber \\
\mathit{p_3}&=& - {\displaystyle \frac{\pi^{3}\,\mathrm{W}}{2}} \,{\displaystyle
\sqrt{\frac{\mathrm{W}}{n\,w\,N}}}. 
\end{eqnarray}

Il est tr\`es claire de voir que le d\'eterminant de la m\'etrique tenseure est:

\ba g= -\pi^4.\ea

\begin{figure}
\hspace*{1.0cm}\vspace*{-6.0cm}
\includegraphics[width=12.0cm,angle=-0]{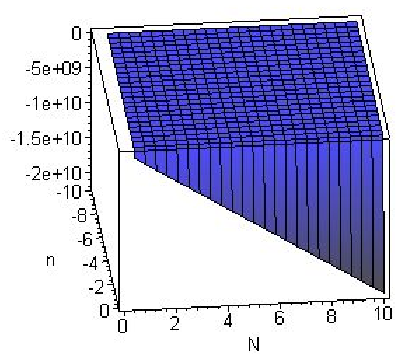}\vspace*{-3.0cm}
\caption{Le mineur d'hypersurface de la m\'etrique tenseure trac\'ee 
comme la fonction de $\{n, N\}$, d\'ecrivant les fluctuations dans 
la configuration des trous noirs charg\'es non-supersym\'etriques 
\`a l'ordre dominant.} \label{nonsusy0minor3}\vspace*{0.5cm}
\end{figure}

\begin{figure}
\hspace*{1.0cm}\vspace*{-6.0cm}
\includegraphics[width=12.0cm,angle=-0]{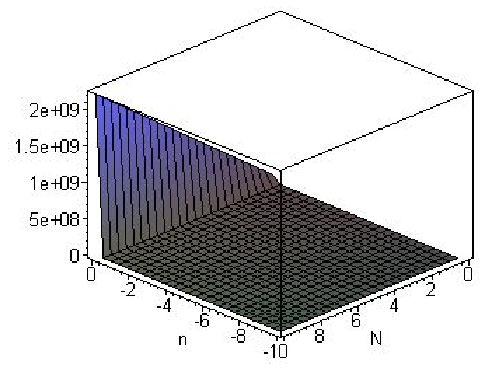}\vspace*{-3.0cm}
\caption{Le premier mineur de la m\'etrique tenseure trac\'ee 
comme la fonction de $\{n, N\}$, d\'ecrivant les fluctuations dans 
la configuration des trous noirs charg\'es non-supersym\'etriques 
\`a l'ordre dominant.} \label{nonsusy0minor1}\vspace*{0.5cm}
\end{figure}

Comme une fonction de $\{n, N\}$, la condition de la stabilit\'e d'un 
ensemble de ces trous noirs non-supersymmetriques d\'ecoule de la positivit\'e 
du d\'eterminant de la m\'etrique tenseure. Dans ce cas, nous voyons que 
le d\'eterminant de la m\'etrique tenseure prend la valeur n\'egative de $-\pi^4$. 
Pour une valeur typique de $n \in (-10, 0)$ et $N \in (0, 10) $, la stabilit\'e 
de l'hypersurface d\'efinie par une valeur constante de $W$ est pr\'esent\'ee 
dans la Fig. (\ref{nonsusy0minor3}). De la Fig. (\ref{nonsusy0minor3}), nous voyons 
que le mineur $p_3$ r\'eside dans l'intervalle $(0, -2 \times 10^{10})$. 
Notez bien que la n\'egativit\'e du mineur $p_3$ augmente quand la valeur de 
$N$ est pass\'ee de z\'ero \`a $10$. Par ailleurs, le surface d\'efinie par 
des fluctuations de $\{n, w\}$ est instable \`a cause du fait que le mineur 
principe correspondant est identiquement nulle, \`a savoir que nous avons $p_2= 0$. 
Lorsque le seul param\`etre $n$ est autoris\'e \`a varier, nous constatons que 
la stabilit\'e de la configuration de ces trous noirs concern\'es est donn\'ee par 
la positivit\'e du premier mineur principe $p_1:= g_{nn}$. Comme montr\'e dens
la Fig. (\ref{nonsusy0minor1}), nous observons que le premier mineur principe $p_1$ 
r\'eside dans la gamme de $(0, 2 \times 10^{19})$. Les propri\'et\'es graphiques 
ci-dessus des composantes et la positivit\'e des mineurs principaux de l'espace 
d'\'etat fournent la notion qualitative de la stabilit\'e statistique des 
trous noirs non-supersym\'etriques \`a l'ordre dominant.

En fait, comme nous avons fourni les $\Gamma_{abc}$ dans l'annex $[A]$, 
nous pouvons \'egalement facilement obtenir pour cette m\'etrique tenseure 
$g_{ab}(n,w,N,W)$ que la courbure scalaire de Ruppenier est donn\'ee par: 

\ba R= \frac{3}{2 \pi \sqrt{nwNW}},\ea

qu'elle est partout r\'eguli\`ere $\forall \overrightarrow{N}$ tel que 
les param\`eres $\{n, w, N, W\}$ sont non-nulles.

\begin{figure}
\hspace*{1.0cm}\vspace*{-6.0cm}
\includegraphics[width=12.0cm,angle=-0]{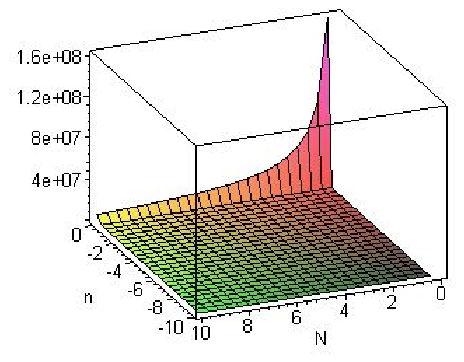}\vspace*{-3.0cm}
\caption{La courbure scalaire trac\'ee en fonction de $\{n, N\}$,
d\'ecrivant les fluctuations de la configuration des trous noirs 
charg\'es non-supersym\'etriques \`a l'ordre dominant dans la gamme
$n \in (-10, 0)$ et $N \in (0, 10)$.} \label{nonsusy0R}\vspace*{0.5cm}
\end{figure}

\begin{figure}
\hspace*{1.0cm}\vspace*{-6.0cm}
\includegraphics[width=12.0cm,angle=-0]{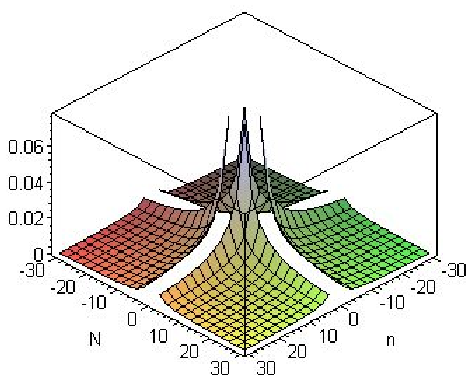}\vspace*{-3.0cm}
\caption{La courbure scalaire trac\'ee en fonction de $\{n, N\}$,
d\'ecrivant les fluctuations de la configuration des trous noirs 
charg\'es non-supersym\'etriques \`a l'ordre dominant dans la gamme
$n, N \in (-30,30)$.} \label{nonsusy0R30}\vspace*{0.5cm}
\end{figure}

Les propri\'et\'es globales de la stabilit\'e d\'ecoulent par le comportement 
de la courbure scalaire de l'espace d'\'etat. En particulier, dans la gamme des
param\`etres $n \in (-10, 0)$ et $N \in (0, 10) $, la Fig. (\ref{nonsusy0R}) montre 
que l'amplitude de la courbure scalaire de l'espace d'\'etat tend vers une tr\`es 
grande valeur positive. On peut remarquer que la configuration sous-jacente des 
trous noirs est un syst\`eme statistique fortement interagissant. Physiquement, 
le signe positif de la courbure scalaire signifie une r\'epulsion de ces interactions.
La Fig. (\ref{nonsusy0R30}) illustre la nature de ce qui pr\'ec\`ede de la courbure 
scalaire dans une range \'egale de $n, N \in (-30, 30) $. En tant que, en effet, 
lorsque la gamme de $n$ et $N$ est prise dans le multiple de $3$, par exemple 
$(-30, 30)$, alors que nous voyons de la Fig. (\ref{nonsusy0R30}) qu'il y a 
quatre r\'egions disjointes similaires de ces interactions statistiques globales. 
En comparaison des interactions apparaissant dans la gamme de $n \in (-10, 0)$
et $N \in (0, 10) $, l'amplitude de ces interactions se r\'ev\`ele \^{e}tre beaucoup 
plus petite dans la gamme des param\`etres $n, N \in (-30,30) $. En ce cas, nous observons 
que le ratio d'une amplitude typique de ces courbures scalaires semble \^{e}tre \`a l'ordre 
de $10^{-10}$. La vue graphique ci-dessus mentionn\'ee de cette comparaison des interactions 
statistiques globales d\'ecoule par les Figs. (\ref{nonsusy0R}, \ref{nonsusy0R30}). 
Qualitativement, dans une petite valeur des $n, N$, les Figs. (\ref{nonsusy0R}, 
\ref{nonsusy0R30}) indiquent que les trous noirs non-supersym\'etriques correspondent 
\`a une configuration statistique interactive qu'elle est globalement instable.

Par cons\'equent, sans les corrections de $ \alpha^{\prime}$,
les trous noirs dyoniques extr\'emaux non-supersym\'etriques ont aussi
le m\^eme d\'eterminant de la m\'etrique et la m\^eme courbure scalaire
de Ruppenier comme les trous noirs dyoniques extr\'emaux supersym\'etriques,
comme nous les avons vus dans la section pr\'ec\'edente.

\subsection{\`A l'ordre de $(\alpha^{\prime})^1 $}

Maintenant, nous allons examiner les corrections de $ \alpha^{\prime} $ dans cette 
g\'eom\'etrie, en consid\'erant la th\'eorie des supercordes au niveau d'arbre de l'action 
effective. Soit $ \widehat{\alpha} $ le coefficient des d\'eriv\'ees sup\'erieures, alors 
les contributions d'ordre premier de la th\'eorie des supercordes \`a l'action effective donnent 
que l'entropie d'un trou noir dyonique non-supersym\'etrique \cite{AshokeSen} est donn\'ee par:

\ba S_{BH}^{ns}= 2 \pi \sqrt{nw N W}+ \frac{5 \pi \widehat{\alpha}}{4} \sqrt{\frac{nw}{N W}}.\ea

On peut voir que les composantes de la m\'etrique de Ruppenier sont: 

\ba g_{nn}&=& \frac{\pi}{2n}\sqrt{\frac{wNW}{n}}
+ \frac{5 \pi \widehat{\alpha}}{16n}\sqrt{\frac{w}{nNW}},\nn
g_{nw}&=& -\frac{\pi}{2} \sqrt{\frac{NW}{nw}}
- \frac{5 \pi \widehat{\alpha}}{16 \sqrt{nw N W}},\nn
g_{nN}&=& -\frac{\pi}{2} \sqrt{\frac{wW}{nN}}
+ \frac{5 \pi w \widehat{\alpha}}{16N \sqrt{nw N W}},\nn
g_{nW}&=& -\frac{\pi}{2} \sqrt{\frac{wN}{nW}}
+ \frac{5 \pi w \widehat{\alpha}}{16W \sqrt{nw N W}},\ea
\ba g_{ww}&=& \frac{\pi}{2w}\sqrt{\frac{nNW}{w}}
+ \frac{5 \pi n \widehat{\alpha}}{16w \sqrt{nw N W}},\nn
g_{wN}&=& -\frac{\pi}{2} \sqrt{\frac{nW}{wN}}
+ \frac{5 \pi n \widehat{\alpha}}{16N \sqrt{nw N W}},\nn
g_{wW}&=& -\frac{\pi}{2} \sqrt{\frac{nN}{wW}}
+ \frac{5 \pi n \widehat{\alpha}}{16W \sqrt{nw N W}},\nn
g_{NN}&=& \frac{\pi}{2N}\sqrt{\frac{nwW}{N}}
- \frac{15\pi \widehat{\alpha}}{16N^2}\sqrt{\frac{nw}{NW}},\nn
g_{NW}&=& -\pi \sqrt{\frac{nw}{NW}}
- \frac{5 \pi \widehat{\alpha}}{16NW} \sqrt{\frac{nw}{NW}},\nn
g_{WW}&=& \frac{\pi}{2W}\sqrt{\frac{nwN}{W}}
- \frac{15\pi \widehat{\alpha}}{16W^2}\sqrt{\frac{nw}{NW}}.\ea

\begin{figure}
\hspace*{1.0cm}\vspace*{-6.0cm}
\includegraphics[width=12.0cm,angle=-0]{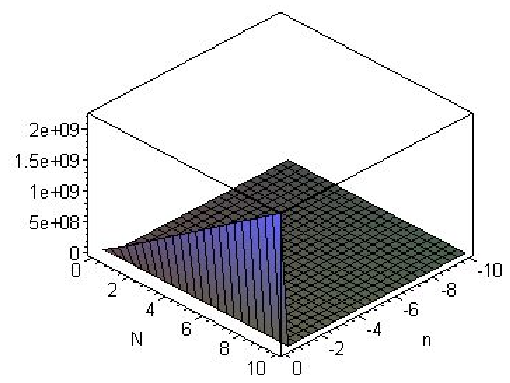}\vspace*{-3.0cm}
\caption{La composante $nn$ de la m\'etrique tenseure trac\'ee comme la
fonction de $\{n, N\}$, d\'ecrivant les fluctuations de la configuration
des trous noirs charg\'es non-supersym\'etriques aux corrections de 
premier ordre de $\alpha^{\prime}$.} \label{nonsusy1nn1}\vspace*{0.5cm}
\end{figure}

\begin{figure}
\hspace*{1.0cm}\vspace*{-6.0cm}
\includegraphics[width=12.0cm,angle=-0]{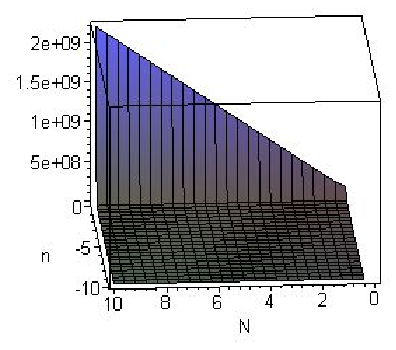}\vspace*{-3.0cm}
\caption{La composante $ww$ de la m\'etrique tenseure trac\'ee comme la
fonction de $\{n, N\}$, d\'ecrivant les fluctuations de la configuration
des trous noirs charg\'es non-supersym\'etriques aux corrections de 
premier ordre de $\alpha^{\prime}$.} \label{nonsusy1ww5}\vspace*{0.5cm}
\end{figure}

\begin{figure}
\hspace*{1.0cm}\vspace*{-6.0cm}
\includegraphics[width=12.0cm,angle=-0]{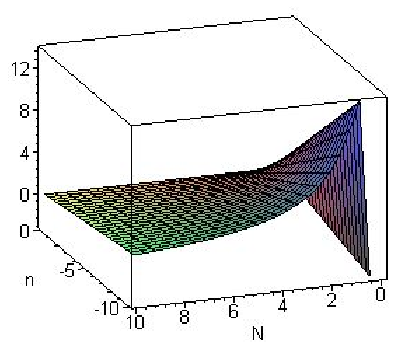}\vspace*{-3.0cm}
\caption{La composante $NN$ de la m\'etrique tenseure trac\'ee comme la
fonction de $\{n, N\}$, d\'ecrivant les fluctuations de la configuration
des trous noirs charg\'es non-supersym\'etriques aux corrections de 
premier ordre de $\alpha^{\prime}$.} \label{nonsusy1NN8}\vspace*{0.5cm}
\end{figure}

\begin{figure}
\hspace*{1.0cm}\vspace*{-6.0cm}
\includegraphics[width=12.0cm,angle=-0]{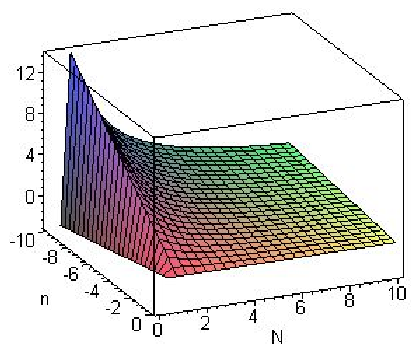}\vspace*{-3.0cm}
\caption{La composante $WW$ de la m\'etrique tenseure trac\'ee comme la
fonction de $\{n, N\}$, d\'ecrivant les fluctuations de la configuration
des trous noirs charg\'es non-supersym\'etriques aux corrections de 
premier ordre de $\alpha^{\prime}$.} \label{nonsusy1WW10}\vspace*{0.5cm}
\end{figure}

\begin{figure}
\hspace*{1.0cm}\vspace*{-6.0cm}
\includegraphics[width=12.0cm,angle=-0]{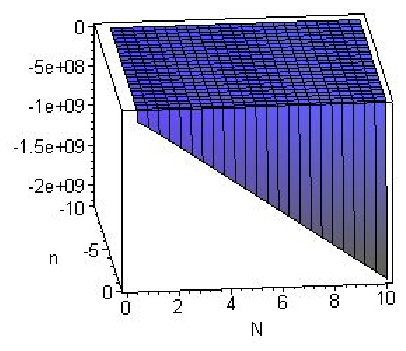}\vspace*{-3.0cm}
\caption{La composante $nw$ de la m\'etrique tenseure trac\'ee comme la
fonction de $\{n, N\}$, d\'ecrivant les fluctuations de la configuration
des trous noirs charg\'es non-supersym\'etriques aux corrections de 
premier ordre de $\alpha^{\prime}$.} \label{nonsusy1nw2}\vspace*{0.5cm}
\end{figure}

\begin{figure}
\hspace*{1.0cm}\vspace*{-6.0cm}
\includegraphics[width=12.0cm,angle=-0]{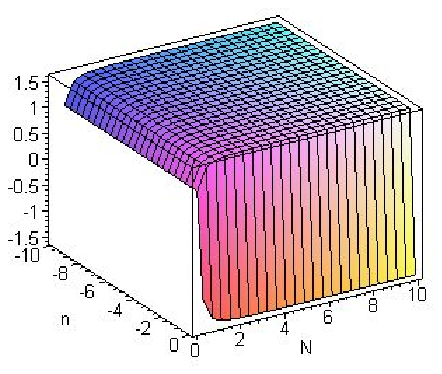}\vspace*{-3.0cm}
\caption{La composante $nN$ de la m\'etrique tenseure trac\'ee comme la
fonction de $\{n, N\}$, d\'ecrivant les fluctuations de la configuration
des trous noirs charg\'es non-supersym\'etriques aux corrections de 
premier ordre de $\alpha^{\prime}$.} \label{nonsusy1nN3}\vspace*{0.5cm}
\end{figure}

\begin{figure}
\hspace*{1.0cm}\vspace*{-6.0cm}
\includegraphics[width=12.0cm,angle=-0]{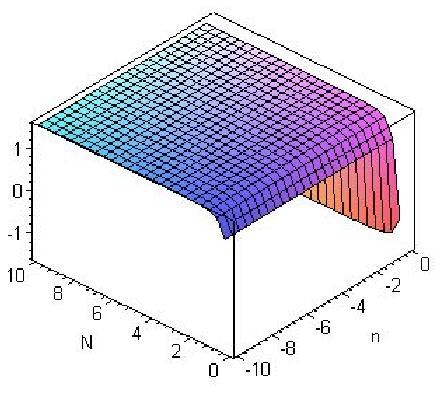}\vspace*{-3.0cm}
\caption{La composante $nW$ de la m\'etrique tenseure trac\'ee comme la
fonction de $\{n, N\}$, d\'ecrivant les fluctuations de la configuration
des trous noirs charg\'es non-supersym\'etriques aux corrections de 
premier ordre de $\alpha^{\prime}$.} \label{nonsusy1nW4}\vspace*{0.5cm}
\end{figure}

\begin{figure}
\hspace*{1.0cm}\vspace*{-6.0cm}
\includegraphics[width=12.0cm,angle=-0]{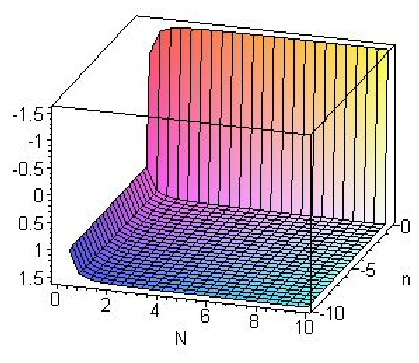}\vspace*{-3.0cm}
\caption{La composante $wN$ de la m\'etrique tenseure trac\'ee comme la
fonction de $\{n, N\}$, d\'ecrivant les fluctuations de la configuration
des trous noirs charg\'es non-supersym\'etriques aux corrections de 
premier ordre de $\alpha^{\prime}$.} \label{nonsusy1wN6}\vspace*{0.5cm}
\end{figure}

\begin{figure}
\hspace*{1.0cm}\vspace*{-6.0cm}
\includegraphics[width=12.0cm,angle=-0]{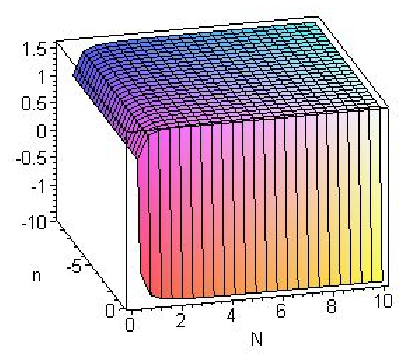}\vspace*{-3.0cm}
\caption{La composante $wW$ de la m\'etrique tenseure trac\'ee comme la
fonction de $\{n, N\}$, d\'ecrivant les fluctuations de la configuration
des trous noirs charg\'es non-supersym\'etriques aux corrections de 
premier ordre de $\alpha^{\prime}$.} \label{nonsusy1wW7}\vspace*{0.5cm}
\end{figure}

\begin{figure}
\hspace*{1.0cm}\vspace*{-6.0cm}
\includegraphics[width=12.0cm,angle=-0]{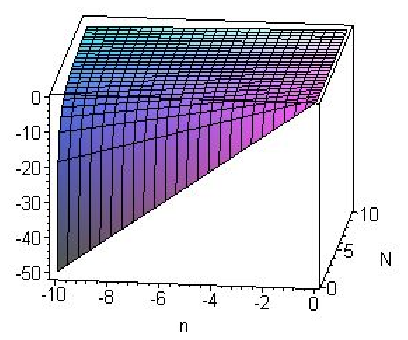}\vspace*{-3.0cm}
\caption{La composante $NW$ de la m\'etrique tenseure trac\'ee comme la
fonction de $\{n, N\}$, d\'ecrivant les fluctuations de la configuration
des trous noirs charg\'es non-supersym\'etriques aux corrections de 
premier ordre de $\alpha^{\prime}$.} \label{nonsusy1NW9}\vspace*{0.5cm}
\end{figure}

Dans la limite de $n = w$, $N= W$ et $\widehat{\alpha}=0.1$, nous
observons pour $n \in (-10, 0)$ et $N \in (0, 10)$ que l'amplitude
des capacit\'es de chaleurs $\{g_{nn}, g_{ww}\}$ prend la valeur maximale
\`a l'ordre de $2 \times 10^{+09}$. Alors, les capacit\'es de chaleurs 
$\{g_{NN},g_{WW}\}$ prendent une valeur typique entre $(-4, 14​​)$. En fait, 
la gamme de la croissance de la premi\`ere ensemble et la deuxi\`eme ensemble
des capacit\'es de la chaleure se trouvent \^{e}tre \`a la limite oppos\'ee 
des param\`etres $\{n, N\}$. Explicitement, par les Figs. (\ref{nonsusy1nn1},
\ref{nonsusy1ww5}), nous constatons que la croissance des capacit\'es 
de chaleurs $\{g_{nn}, g_{ww}\}$ a lieu dans la limite d'un grand $N$
et un petit $n$. D'autre part, par les Figs. (\ref{nonsusy1NN8}, \ref{nonsusy1WW10}), 
nous pouvons voir facilement que la croissance de $\{g_{NN}, g_{WW}\}$ 
a lieu dans la limite d'un petit $N$ et un grand $n$. Du fait m\^{e}me, 
les compressibilit\'es de chaleurs en impliquant deux param\`etres distincts 
de la configuration de ces trous noirs sont repr\'esent\'ees dans les Figs. 
(\ref{nonsusy1nw2}, \ref{nonsusy1nN3}, \ref{nonsusy1nW4}, \ref{nonsusy1wN6}, 
\ref{nonsusy1wW7}, \ref{nonsusy1NW9}). Dans ce cas, on observe que les 
fluctuations en impliquant $\{n, w\}$, tel qu'elle sont d\'efinies 
par les composantes de la m\'etrique tenseure de l'espace d'\'etat
$\{g_{ij} \ | \ i, j = n, w\}$, ont relativement des plus grande valeurs 
num\'eriques par rapport \`a celles qui impliquant $\{N, W\}$.

Par rapport \`a la m\'etrique tenseure ci-dessus, nous trouvons 
que les corrections d\'{e}riv\'{e}s sup\'{e}rieures \`a l'ordre dominant 
conduissent les mineurs principaux suivants

\begin{eqnarray}
\mathit{p_1}&=& {\displaystyle \frac{\pi}{16\,n}}\,
\,{\displaystyle \sqrt{\frac{w}{n\,N\,\mathrm{W}}}} \mathit{\tilde{p}_1}(\widehat{\alpha}),  \nonumber \\
\mathit{p_2}&=& 0,  \nonumber \\
\mathit{p_3}&=&  - {\displaystyle \frac{1}{1024}} 
\frac{\pi^{3}}{(N^{3}\,\mathrm{W}\,\sqrt{n\,w\,N\,\mathrm{W}})}\mathit{\tilde{p}_3}(\widehat{\alpha}). 
\end{eqnarray}

o\`u les fonctions $\{ \mathit{\tilde{p}_1}(\widehat{\alpha}), \mathit{\tilde{p}_3}(\widehat{\alpha}) \}$ sont d\'efinies par

\begin{eqnarray}
\mathit{\tilde{p}_1}(\widehat{\alpha})&:=&  5\,\widehat{\alpha}\,+ 8\,\mathrm{W}\,N, \nonumber \\
\mathit{\tilde{p}_3}(\widehat{\alpha})&:=&  125\,\widehat{\alpha}^{3} - 200 \,N\,\mathrm{W}\,\widehat{\alpha}^{2}
- 320\,N^{2}\,\mathrm{W}^{2}\,\widehat{\alpha}+ 512\,N^{3}\,\mathrm{W}^{3}. 
\end{eqnarray}

Nous pouvons voir que le d\'eterminant de la m\'etrique tenseure est donn\'e par: 

\ba g= \frac{\pi^4}{4096(NW)^4} \tilde{g}(N,W), \ea

o\`u la fonction $\tilde{g}(N,W)$, comme une fonction de $\widehat{\alpha}$ est d\'efinie par

\ba \tilde{g}(N,W):= 625 \widehat{\alpha}^4- 2000 \widehat{\alpha}^3 (NW)
+ 5120 \widehat{\alpha} (NW)^3- 4096 (NW)^4.\ea

Ainsi, nous voyons que la configuration sous-jacente de ces trous noirs 
est assez stable dans les r\'{e}gions o\`{u} le mineur d'hypersurface 
et le d\'eterminant de la m\'{e}trique tenseure sont positifs.
Notamment, \c{c}a pose comme une valeur sp\'{e}cifique du param\`{e}tre
$\widehat{\alpha}$ des corrections tel que les conditions suivantes sont remplies:
(i) le polyn\^{o}me lin\'{e}aire $\mathit{\tilde{p}_1}(\widehat{\alpha})$
a un signe positif, (ii) le polyn\^{o}me cubique $\mathit{\tilde{p}_3}(\widehat{\alpha})$ 
a un signe n\'{e}gatif et (iii) le polyn\^{o}me quartique $\mathit{\tilde{g}}(\widehat{\alpha})$ 
a un signe positif. Pour tout $\widehat{\alpha}$ tels que les conditions $\mathit{\tilde{p}_1}(\widehat{\alpha}) >0$, 
$\mathit{\tilde{p}_3}(\widehat{\alpha}) <0$ et $\mathit{\tilde{g}}(\widehat{\alpha})>0$ sont remplies, 
la solution sous-tendente des trous noirs vienne \^{e}tre relativement plus stables.
En fait, la stabilit\'{e} relative de ces trous noirs au-dessus, c'est-\`a-dire que 
les valeurs sp\'{e}cifiques du param\`{e}tre $\widehat{\alpha}$ peuvent \^{e}tre d\'{e}termin\'{e}es
comme les racines communes des \'{e}quations cubiques et quartiques ci-dessus. 
Notez cependant que la disparition de la surface de mineurs, c'est-\`a-dire que
nous avons $\mathit{p_2}= 0$ \'{e}vite la stabilit\'{e} compl\`{e}te de la 
configuration sous-jacente de ces trous noirs.

\begin{figure}
\hspace*{1.0cm}\vspace*{-6.0cm}
\includegraphics[width=12.0cm,angle=-0]{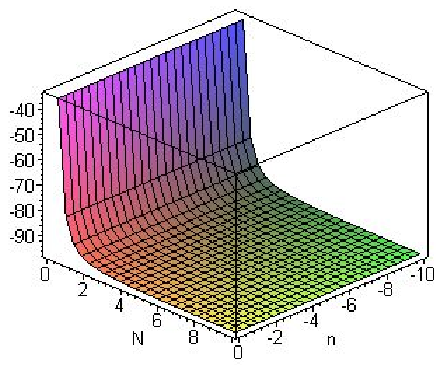}\vspace*{-3.0cm}
\caption{Le d\'eterminant de la m\'etrique tenseure trac\'ee comme la
fonction de $\{n, N\}$, d\'ecrivant les fluctuations de la configuration
des trous noirs charg\'es non-supersym\'etriques aux corrections de 
premier ordre de $\alpha^{\prime}$.} \label{nonsusy1detg}\vspace*{0.5cm}
\end{figure}

\begin{figure}
\hspace*{1.0cm}\vspace*{-6.0cm}
\includegraphics[width=12.0cm,angle=-0]{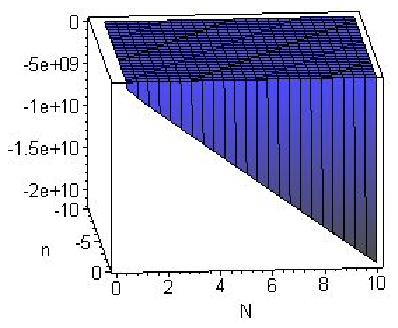}\vspace*{-3.0cm}
\caption{Le mineur d'hypersurface de la m\'etrique tenseure trac\'ee comme la
fonction de $\{n, N\}$, d\'ecrivant les fluctuations de la configuration
des trous noirs charg\'es non-supersym\'etriques aux corrections de 
premier ordre de $\alpha^{\prime}$.} \label{nonsusy1minor3}\vspace*{0.5cm}
\end{figure}

\begin{figure}
\hspace*{1.0cm}\vspace*{-6.0cm}
\includegraphics[width=12.0cm,angle=-0]{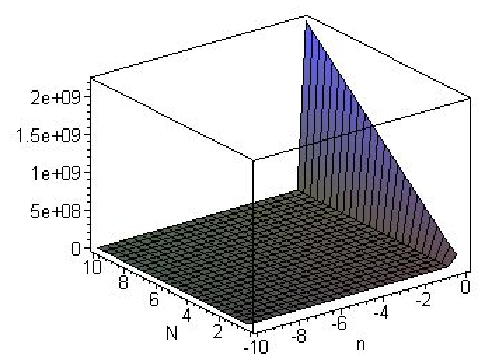}\vspace*{-3.0cm}
\caption{Le premier mineur de la m\'etrique tenseure trac\'ee comme la
fonction de $\{n, N\}$, d\'ecrivant les fluctuations de la configuration
des trous noirs charg\'es non-supersym\'etriques aux corrections de 
premier ordre de $\alpha^{\prime}$.} \label{nonsusy1minor1}\vspace*{0.5cm}
\end{figure}

Comme une fonction de $\{n, N\}$, la condition de la stabilit\'e d'un ensemble
de ces trous noirs d\'ecoule de la positivit\'e du d\'eterminant de la m\'etrique 
tenseure. Dans ce cas, nous voyons que le d\'eterminant de la m\'etrique tenseure 
tend vers une valeur n\'egative. Pour une valeur typique de $n \in (-10, 0)$
et $N \in (0, 10)$, la Fig. (\ref{nonsusy1detg}) montre que le d\'eterminant 
de la m\'etrique tenseure r\'eside dans l'intervalle $(-30, -100)$. En outre,
la stabilit\'e de l'hypersurface d\'efinie par une valeur constante de $W$ est 
montr\'ee dans la Fig. (\ref{nonsusy1minor3}). Par la pr\'esente, nous voyons que 
le mineur $p_3$ r\'eside dans la gamme de $(-2 \times 10^{+10}, 0)$. Notez bien
que la n\'egativit\'e de $p_3$ augmente \`a mesure que la valeur de $N$ est pass\'ee 
de z\'ero \`a $10$. Par ailleurs, il s'av\`ere que la surface d\'efinies par les fluctuations 
de $\{n, w \}$ est instable en raison du fait m\^eme que le mineur principe correspondant 
dispara\^{i}t, \`a savoir que nous avons $p_2= 0$. Quand le seulement param\`etre $n$ 
est autoris\'e \`a varier, la stabilit\'e de la configuration de ces trous noirs est donn\'ee 
par la positivit\'e du premier mineur principe $p_1:= g_{nn}$. Une vue rotat\'ee de $p_1$
est pr\'esent\'ee dans la Fig. (\ref{nonsusy1minor1}). Les propri\'et\'esgraphiques
ci-dessus des mineurs principaux et la positivit\'e des composants de la m\'etrique 
tenseure de l'espace d'\'etat fournent la notion qualitative de la stabilit\'e 
statistique des non-supersym\'etriques trous noirs.

En fait, avec les valeurs des $\Gamma_{abc}$ comme nous les avons fournis dans l'annex $[A]$, 
it n'est pas difficile de voir que la courbure scalaire de Ruppenier est

\ba R= - \frac{96}{\pi}(\frac{NW}{nw})^{1/2} \frac{r_2}{\tilde{g}(N,W) r_1^2}, \ea

o\`u les fonctions $\{ r_1(N,W), r_2(N, W) \}$ sont d\'efinies par:

\ba
r_1&:=& 25 \widehat{\alpha}^2- 80 \widehat{\alpha} NW+ 64(NW)^2,\nn
r_2&:=& 15625 \widehat{\alpha}^6- 150000\widehat{\alpha}^5 NW
+600000 \widehat{\alpha}^4 (NW)^2- 1280000\widehat{\alpha}^3 (NW)^3\nn &&
+1536000 \widehat{\alpha}^2 (NW)^4- 983040 \widehat{\alpha} (NW)^5
+262144 (NW)^6.\ea

Il peut sembler qu'il existe quelques divergences dans cette courbure scalaire
de Ruppenier qui sont donn\'ees par l'\'equation quadratique $r_1=0$. C'est-\`a-dire que

\ba 25 \widehat{\alpha}^2- 80 \widehat{\alpha} NW+ 60(NW)^2= 0.\ea

Mais cet \'equation implique simplement qu'on a $\{N,W\}$ tel que

\ba NW= (-8 \pm \sqrt{59}) \frac{\widehat{\alpha}}{12}.\ea

En fait, pour la configuration des charges que nous avons consid\'er\'e,
\c{c}a n'est pas possible parce que $ N, W > 0 $. Et bien, nous voyons 
que ces divergences de la courbure scalaire devient \^etre vraiment possible 
si et seulement si $ \widehat{\alpha}< 0 $ parce que nous avons consid\'er\'e 
les vecteurs charges $(Q, P)$ tel que $ N^{\prime}, W^{\prime}, \widehat{n}>0 $ 
et $ \widehat{w}<0 $.

\begin{figure}
\hspace*{1.0cm}\vspace*{-6.0cm}
\includegraphics[width=12.0cm,angle=-0]{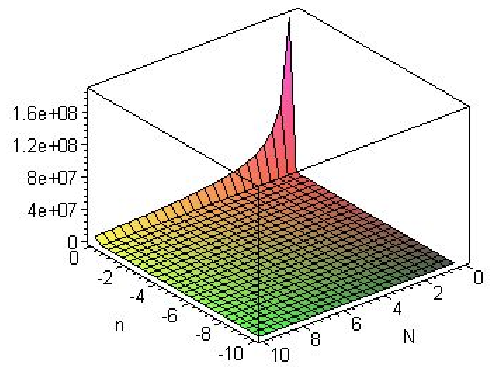}\vspace*{-3.0cm}
\caption{La courbure scalaire trac\'ee comme la
fonction de $\{n, N\}$, d\'ecrivant les fluctuations de la configuration
des trous noirs charg\'es non-supersym\'etriques aux corrections de 
premier ordre de $\alpha^{\prime}$ dans la gamme
$n \in (-10,0)$ et $N \in (0,10) $.} \label{nonsusy1R}\vspace*{0.5cm}
\end{figure}

\begin{figure}
\hspace*{1.0cm}\vspace*{-6.0cm}
\includegraphics[width=12.0cm,angle=-0]{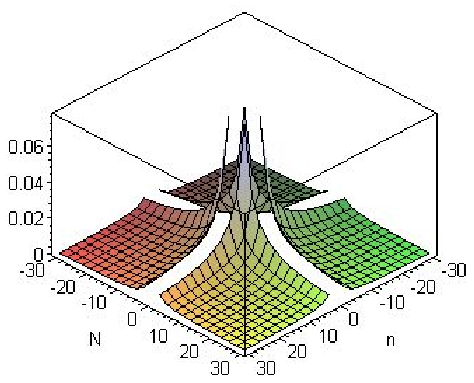}\vspace*{-3.0cm}
\caption{La courbure scalaire trac\'ee comme la
fonction de $\{n, N\}$, d\'ecrivant les fluctuations de la configuration
des trous noirs charg\'es non-supersym\'etriques aux corrections de 
premier ordre de $\alpha^{\prime}$ dans la gamme
$n, N \in (-30,30)$.} \label{nonsusy1R30}\vspace*{0.5cm}
\end{figure}

De plus, les propri\'et\'es de la stabilit\'e globale r\'esultent du
comportement de la courbure scalaire de l'espace d'\'etat. En particulier, 
dans la gamme de $n \in (-10, 0)$ et $N \in (0, 10)$, la Fig. (\ref{nonsusy1R}) 
montre que la courbure scalaire a une grande positive amplitude. On peut noter 
que la configuration des trous noirs sous-tendents est un syst\`eme statistique
fortement interagissant. Le signe positif de la courbure scalaire signifie une 
r\'epulsion de ces interactions. La Fig. (\ref{nonsusy1R30}) illustre la nature
de la courbure scalaire ci-dessus dans une range \'egale de $n, N \in
(-30,30)$. En fait, lorsque la gamme de $n$ et $N$ est prise dans le
multiple de $3$, par exemple $(-30, 30)$, alors que nous voyons de la
Fig. (\ref{nonsusy1R30}) qu'il y a quatre r\'egions disjointes similaires 
des interactions statistiques globales. En comparaison des interactions 
apparaissant dans la gamme de $n \in (-10, 0)$ et $N \in (0, 10)$, 
l'amplitude de l'interaction se r\'ev\`ele \^{e}tre beaucoup plus petit 
dans la gamme de $n, N \in (-30, 30) $. Le ratio de la valeur typique de l'amplitude 
de ces courbures scalaires semble \^{e}tre $10^{-10}$. La vue graphique de la 
comparaison ci-dessus d\'ecoule des Figs. (\ref{nonsusy1R}, \ref{nonsusy1R30}).
Qualitativement, dans une petite valeur des $n, N$, les repr\'esentations 
graphiques ci-dessus indiquent que la configuration des trous noirs non-supersym\'etriques
correspond \`a un syst\`eme statistique instable aux corrections de premier 
ordre de $\alpha^{\prime}$.

\subsection{\`A l'ordre de $(\alpha^{\prime})^2 $}

Dans la suite, nous nous int\'eresserons uniquement \`a la cosideration
d'un ordre sup\'erieur prochain des corrections de $\alpha^{\prime} $.
Maintenant, le Ref. \cite{AshokeSen} montre que nous pouvons \'etudier la 
g\'eom\'etrie thermodynamique avec les corrections de $(\alpha^{\prime})^2 $.
En ce cas, l'entropie d'un trou noir dyonique non-supersym\'etrique est modifi\'ee 
comme la suivante

\ba S_{BH}^{ns}=
2 \pi \sqrt{nw N W}+ \frac{5 \pi \widehat{\alpha}}{4} \sqrt{\frac{nw}{N W}}
- \frac{29 \pi \widehat{\alpha}^2}{64} \frac{\sqrt{nw}}{(NW)^{3/2}}.\ea

Maintenant, les composantes de la m\'etrique de Ruppenier sont: 

\ba g_{nn}&=& \frac{\pi}{2n}\sqrt{\frac{wNW}{n}}
+ \frac{5 \pi \widehat{\alpha}}{16n}\sqrt{\frac{w}{nNW}} 
- \frac{29\pi \widehat{\alpha}^2}{256 nNW} \sqrt{\frac{w}{nNW}},\nn
g_{nw}&=& -\frac{\pi}{2} \sqrt{\frac{NW}{nw}}
- \frac{5 \pi \widehat{\alpha}}{16 \sqrt{nw N W}} 
+ \frac{29\pi \widehat{\alpha}^2}{256 NW \sqrt{nwNW}},\nn
g_{nN}&=& -\frac{\pi}{2} \sqrt{\frac{wW}{nN}}
+ \frac{5 \pi \widehat{\alpha}}{16NW} \sqrt{\frac{wW}{nN}} 
- \frac{87\pi \widehat{\alpha}^2}{256 (NW)^2} \sqrt{\frac{wW}{nN}},\nn
g_{nW}&=& -\frac{\pi}{2} \sqrt{\frac{wN}{nW}}
+ \frac{5 \pi \widehat{\alpha}}{16NW} \sqrt{\frac{w N}{n W}} 
- \frac{87\pi \widehat{\alpha}^2}{256 (NW)^2} \sqrt{\frac{wN}{nW}},\ea
\ba g_{ww}&=& \frac{\pi}{2w}\sqrt{\frac{nNW}{w}}
+ \frac{5 \pi \widehat{\alpha}}{16w} \sqrt{\frac{n}{w N W}} 
- \frac{29\pi \widehat{\alpha}^2}{256 wNW} \sqrt{\frac{n}{w N W}},\nn
g_{wN}&=& -\frac{\pi}{2} \sqrt{\frac{nW}{wN}}
+ \frac{5 \pi \widehat{\alpha}}{16N} \sqrt{\frac{n}{w N W}}
- \frac{87\pi \widehat{\alpha}^2}{256 (N^2W)} \sqrt{\frac{n}{wNW}},\nn
g_{wW}&=& -\frac{\pi}{2} \sqrt{\frac{nN}{wW}}
+ \frac{5 \pi \widehat{\alpha}}{16NW }\sqrt{\frac{nN}{wW}}
- \frac{87\pi \widehat{\alpha}^2}{256 (NW)^2} \sqrt{\frac{nN}{wW}},\nn
g_{NN}&=& \frac{\pi}{2N}\sqrt{\frac{nwW}{N}}
- \frac{15\pi \widehat{\alpha}}{16N^2}\sqrt{\frac{nw}{NW}}
+\frac{435\pi \widehat{\alpha}^2}{256 (N^3W)} \sqrt{\frac{nN}{wW}},\nn
g_{NW}&=& -\pi \sqrt{\frac{nw}{NW}}
- \frac{5 \pi \widehat{\alpha}}{16NW} \sqrt{\frac{nw}{NW}} 
+ \frac{261 \pi \widehat{\alpha}^2}{256 (NW)^2} \sqrt{\frac{nw}{NW}},\nn
g_{WW}&=& \frac{\pi}{2W}\sqrt{\frac{nwN}{W}}
- \frac{15\pi \widehat{\alpha}}{16W^2}\sqrt{\frac{nw}{NW}} 
+\frac{435\pi \widehat{\alpha}^2}{256 (NW^3)} \sqrt{\frac{nw}{NW}}.\ea

\begin{figure}
\hspace*{1.0cm}\vspace*{-6.0cm}
\includegraphics[width=12.0cm,angle=-0]{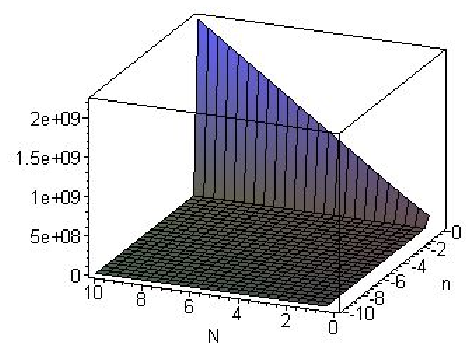}\vspace*{-3.0cm}
\caption{La composante $nn$ de la m\'etrique tenseure trac\'ee comme la
fonction de $\{n, N\}$, d\'ecrivant les fluctuations de la configuration
des trous noirs charg\'es non-supersym\'etriques aux corrections de 
second ordre de $\alpha^{\prime}$.} \label{nonsusy2nn1}\vspace*{0.5cm}
\end{figure}

\begin{figure}
\hspace*{1.0cm}\vspace*{-6.0cm}
\includegraphics[width=12.0cm,angle=-0]{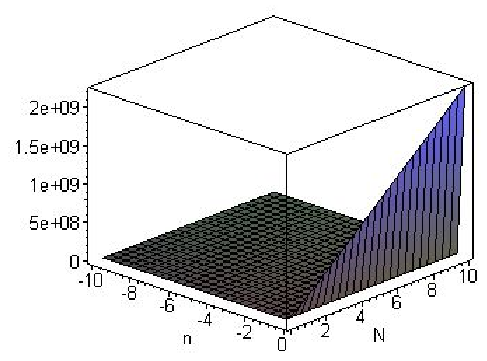}\vspace*{-3.0cm}
\caption{La composante $ww$ de la m\'etrique tenseure trac\'ee comme la
fonction de $\{n, N\}$, d\'ecrivant les fluctuations de la configuration
des trous noirs charg\'es non-supersym\'etriques aux corrections de 
second ordre de $\alpha^{\prime}$.} \label{nonsusy2ww5}\vspace*{0.5cm}
\end{figure}

\begin{figure}
\hspace*{1.0cm}\vspace*{-6.0cm}
\includegraphics[width=12.0cm,angle=-0]{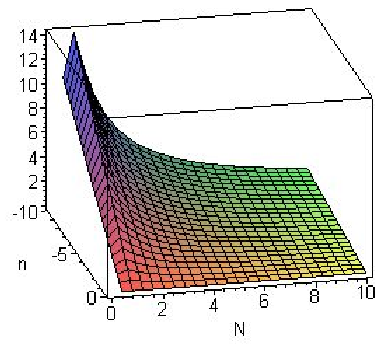}\vspace*{-3.0cm}
\caption{La composante $NN$ de la m\'etrique tenseure trac\'ee comme la
fonction de $\{n, N\}$, d\'ecrivant les fluctuations de la configuration
des trous noirs charg\'es non-supersym\'etriques aux corrections de 
second ordre de $\alpha^{\prime}$.} \label{nonsusy2NN8}\vspace*{0.5cm}
\end{figure}

\begin{figure}
\hspace*{1.0cm}\vspace*{-6.0cm}
\includegraphics[width=12.0cm,angle=-0]{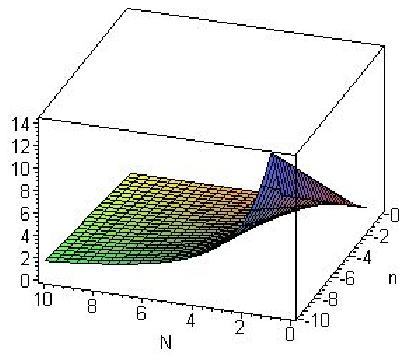}\vspace*{-3.0cm}
\caption{La composante $WW$ de la m\'etrique tenseure trac\'ee comme la
fonction de $\{n, N\}$, d\'ecrivant les fluctuations de la configuration
des trous noirs charg\'es non-supersym\'etriques aux corrections de 
second ordre de $\alpha^{\prime}$.} \label{nonsusy2WW10}\vspace*{0.5cm}
\end{figure}

\begin{figure}
\hspace*{1.0cm}\vspace*{-6.0cm}
\includegraphics[width=12.0cm,angle=-0]{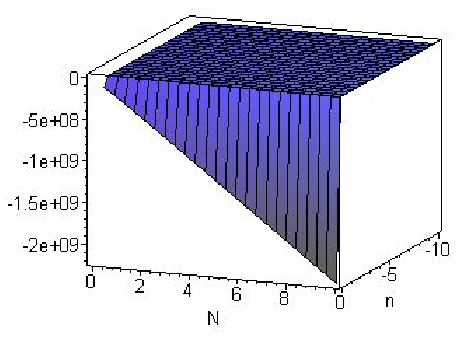}\vspace*{-3.0cm}
\caption{La composante $nw$ de la m\'etrique tenseure trac\'ee comme la
fonction de $\{n, N\}$, d\'ecrivant les fluctuations de la configuration
des trous noirs charg\'es non-supersym\'etriques aux corrections de 
second ordre de $\alpha^{\prime}$.} \label{nonsusy2nw2}\vspace*{0.5cm}
\end{figure}

\begin{figure}
\hspace*{1.0cm}\vspace*{-6.0cm}
\includegraphics[width=12.0cm,angle=-0]{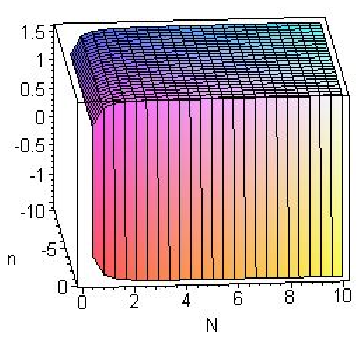}\vspace*{-3.0cm}
\caption{La composante $nN$ de la m\'etrique tenseure trac\'ee comme la
fonction de $\{n, N\}$, d\'ecrivant les fluctuations de la configuration
des trous noirs charg\'es non-supersym\'etriques aux corrections de 
second ordre de $\alpha^{\prime}$.} \label{nonsusy2nN3}\vspace*{0.5cm}
\end{figure}

\begin{figure}
\hspace*{1.0cm}\vspace*{-6.0cm}
\includegraphics[width=12.0cm,angle=-0]{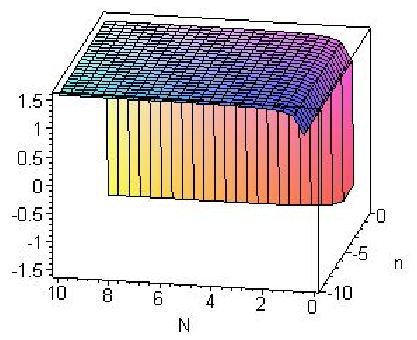}\vspace*{-3.0cm}
\caption{La composante $nW$ de la m\'etrique tenseure trac\'ee comme la
fonction de $\{n, N\}$, d\'ecrivant les fluctuations de la configuration
des trous noirs charg\'es non-supersym\'etriques aux corrections de 
second ordre de $\alpha^{\prime}$.} \label{nonsusy2nW4}\vspace*{0.5cm}
\end{figure}

\begin{figure}
\hspace*{1.0cm}\vspace*{-6.0cm}
\includegraphics[width=12.0cm,angle=-0]{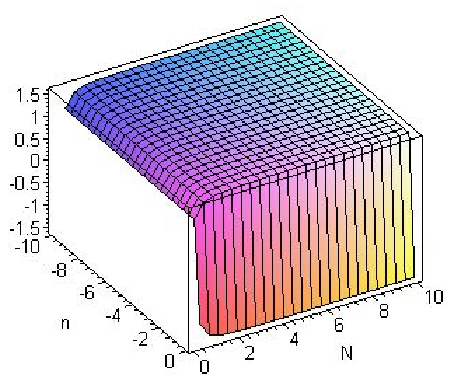}\vspace*{-3.0cm}
\caption{La composante $wN$ de la m\'etrique tenseure trac\'ee comme la
fonction de $\{n, N\}$, d\'ecrivant les fluctuations de la configuration
des trous noirs charg\'es non-supersym\'etriques aux corrections de 
second ordre de $\alpha^{\prime}$.} \label{nonsusy2wN6}\vspace*{0.5cm}
\end{figure}

\begin{figure}
\hspace*{1.0cm}\vspace*{-6.0cm}
\includegraphics[width=12.0cm,angle=-0]{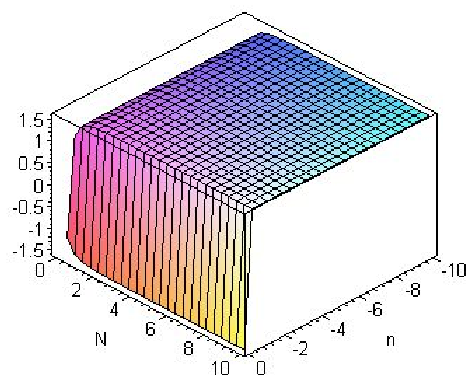}\vspace*{-3.0cm}
\caption{La composante $wW$ de la m\'etrique tenseure trac\'ee comme la
fonction de $\{n, N\}$, d\'ecrivant les fluctuations de la configuration
des trous noirs charg\'es non-supersym\'etriques aux corrections de 
second ordre de $\alpha^{\prime}$.} \label{nonsusy2wW7}\vspace*{0.5cm}
\end{figure}

\begin{figure}
\hspace*{1.0cm}\vspace*{-6.0cm}
\includegraphics[width=12.0cm,angle=-0]{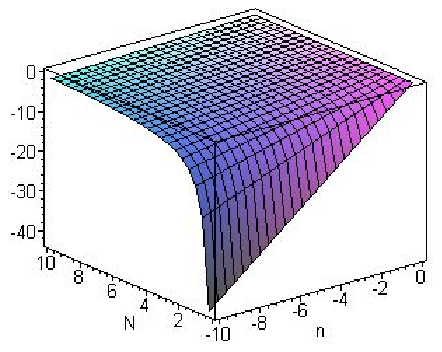}\vspace*{-3.0cm}
\caption{La composante $NW$ de la m\'etrique tenseure trac\'ee comme la
fonction de $\{n, N\}$, d\'ecrivant les fluctuations de la configuration
des trous noirs charg\'es non-supersym\'etriques aux corrections de 
second ordre de $\alpha^{\prime}$.} \label{nonsusy2NW9}\vspace*{0.5cm}
\end{figure}

Dans la limite de $n= w$, $N= W$ et $\widehat{\alpha}= 0.1$, sous le second ordre des
corrections de d\'eriv\'ees sup\'erieures de $\alpha^{\prime}$, nous observons dans 
le r\'egime de $n \in (-10, 0)$ et $N \in (0, 10)$ que l'amplitudes des capacit\'es de
chaleurs $\{g_{nn}, g_{ww}\}$ prend la valeur maximale \`a l'ordre de $2 \times 10^{+09}$. 
Dans cette gamme de $\{n, N\}$, nous constatons que les capacit\'es de chaleurs $\{g_{NN}, g_{WW}\}$ 
restent dans la gamme de $(0, 14) $. Comme pr\'evu au premier ordre des corrections de $\alpha^{\prime} $, 
nous voyons que la gamme de la croissance de la premi\`ere ensemble et celle de la la deuxi\`eme ensemble 
des capacit\'es de chaleurs reste dans la limite oppos\'ee des param\`etres $\{n, N\}$. Explicitement, 
les Figs. (\ref{nonsusy2nn1}, \ref{nonsusy2ww5}), montrent que la croissance de la premi\`ere ensemble des 
capacit\'es de chaleurs $\{g_{nn}, g_{ww}\}$ a lieu dans la limite d'un grand $N$ et un petit $n$. 
De m\^{e}me, les Fig. (\ref{nonsusy2NN8}, \ref{nonsusy2WW10}) montrent que la croissance de 
$\{g_{NN}, g_{WW}\}$ a lieu dans la limite d'un petit $N$ et d'un grand $n$. D'autre part, 
les compressibilit\'es de chaleurs en impliquant deux param\`etres distincts de la configuration
des trous noirs non-supersym\'etriques sont d\'ecrits dans les Figs.(\ref{nonsusy2nw2},\ref{nonsusy2nN3},
\ref{nonsusy2nW4}, \ref{nonsusy2wN6},\ref{nonsusy2wW7}, \ref{nonsusy2NW9}). Dans ce cas, 
nous observons que les fluctuations en impliquant $\{n, w\}$, qui sont d\'efinies par les composantes 
de la m\'etrique tenseure de l'espace d'\'etat $\{g_{ij} \ | \ i, j = n, w, N, W\}$, ont relativement
une grandes valeurs num\'eriques par rapport \`a celles qui impliquent des param\`etres $\{N, W\}$.

Puisque la m\'{e}trique tenseure ci-dessus est modifi\'{e}s par 
les corrections d\'{e}riv\'{e}es sup\'{e}rieures d'ordre prochain, 
nous constatons que les mineurs principaux sont modifi\'{e}s 
comme les suivants

\begin{eqnarray}
\mathit{p_1} &=& - {\displaystyle \frac {1}{256}} \,
{\displaystyle \frac {\pi \, \sqrt{w}}{(n\,N\,\mathrm{W})^{(3/2)}}}\,
\mathit{\tilde{p}_1}(\widehat{\alpha}), \nonumber \\
\mathit{p_2} &=& 0, \nonumber \\
\mathit{p_3}&=& 
{\displaystyle \frac {1}{4194304}} 
\frac{\pi^{3}}{N^{(13/2)} \mathrm{W}^{(9/2)}\,\sqrt{n}\,\sqrt{w}}\,
\mathit{\tilde{p}_3}(\widehat{\alpha}). 
%
\end{eqnarray}

o\`u les fonctions $\{ \mathit{\tilde{p}_1}(\widehat{\alpha}),
 \mathit{\tilde{p}_3}(\widehat{\alpha}) \}$, comme les polyn\^{o}mes du param\`{e}tre 
$\widehat{\alpha}$, peuvent \^etre exprim\'es par

\begin{eqnarray}
\mathit{\tilde{p}_1}(\widehat{\alpha})&:=&
29\,\widehat{\alpha}^{2}- 80\,\widehat{\alpha}\,N\,\mathrm{W} 
- 128\,N^{2}\,\mathrm{W}^{2}, \nonumber \\
\mathit{\tilde{p}_3}(\widehat{\alpha})&:=&
 219501\,\widehat{\alpha}^{6} - 1009200\,N\,\mathrm{W}\,\widehat{\alpha}^{5} 
+ 976256\,N^{2}\,\mathrm{W}^{2}\,\widehat{\alpha}^{4} \nonumber \\ &&
-1105920\,N^{3}\,\mathrm{W}^{3}\,\widehat{\alpha}^{3} 
- 1556480\,N^{4}\,\mathrm{W}^{4}\,\widehat{\alpha}^{2} \nonumber \\ &&
+ 1310720\,N^{5}\,\mathrm{W}^{5}\,\widehat{\alpha} 
- 2097152\,N^{6}\,\mathrm{W}^{6}. 
%
\end{eqnarray}

On voit sans probl\`eme que le d\'eterminant de la m\'etrique tenseure est:

\ba g= \frac{\pi^4}{268435456(NW)^8} \tilde{g}(N,W), \ea

o\`u la fonction $\tilde{g}(N,W)$, comme une fonction de $\widehat{\alpha}$, est d\'efinie par:

\ba \tilde{g}(N,W)&:=& -a^{(2)}_0 (NW)^8 
+a^{(2)}_1 \widehat{\alpha}  (NW)^7
-a^{(2)}_2 \widehat{\alpha}^2 (NW)^6
+a^{(2)}_3 \widehat{\alpha}^3  (NW)^5
+a^{(2)}_4 \widehat{\alpha}^4 (NW)^4 \nn &&
-a^{(2)}_5 \widehat{\alpha}^5 (NW)^3
+a^{(2)}_6 \widehat{\alpha}^6 (NW)^2
-a^{(2)}_7 \widehat{\alpha}^7 (NW)
+a^{(2)}_8 \widehat{\alpha}^8;\ea

o\`u les constantes r\'eelles positives $ \lbrace a^{(2)}_i \rbrace_{i=0}^8 $ 
sont donn\'ees dans l'annex $[C]$.

Sous les corrections de $ \alpha^{\prime}$ \`a l'ordre de deux, 
nous voyons que la configuration sous-jacente de ces trous noirs 
est assez stable dans les r\'egions o\`u les valeurs des mineurs de 
hypersurface et de d\'eterminant de la m\'etrique tenseure sont positifs.
Notamment, lorsque le param\`etre $\widehat{\alpha}$ des corrections est tel que 
(i) l polyn\^{o}me quadratic $\mathit{\tilde{p}_1}(\widehat{\alpha}) $ a un signe n\'egatif, 
(ii) le polyn\^{o}me de degr\'e six $ \mathit{\tilde{p}_3}(\widehat{\alpha}) $ a un signe positif 
et (iii) le polyn\^{o}me de degr\'e huit $\mathit{\tilde{g}}(\widehat{\alpha}) $ a un signe positif. 
Pour tout $\widehat{\alpha}$ tels que les mineurs principaux $\mathit{\tilde{p}_1}(\widehat{\alpha}) <0$, 
$\mathit{\tilde{p}_3}(\widehat{\alpha}) >0$ et le d\'eterminant de la m\'etrique tenseure 
$\mathit{\tilde{g}}(\widehat{\alpha})>0$ sont satisfaites, la solution sous-jacente de ces 
trous noirs est relativement stable. Ainsi, la stabilit\'e relative de ces 
trous noirs au-dessus, c'est-\`a-dire que la valeur sp\'ecifique du param\`etre 
$\widehat{\alpha}$, peut \^{e}tre d\'etermin\'e comme les racines communes des \'equations 
au-dessus des degr\'es six et huit. En outre, l'avis de la disparition du mineur
de la surface, voil\`a, $\mathit{p_2}= 0 $, indique que la configuration de ces 
trous noirs au-dessus reste instable \`a ce niveau des corrections d\'eriv\'ees 
sup\'erieures de $\alpha^{\prime}$ \`a l'entropie de trou noir.

\begin{figure}
\hspace*{1.0cm}\vspace*{-6.0cm}
\includegraphics[width=12.0cm,angle=-0]{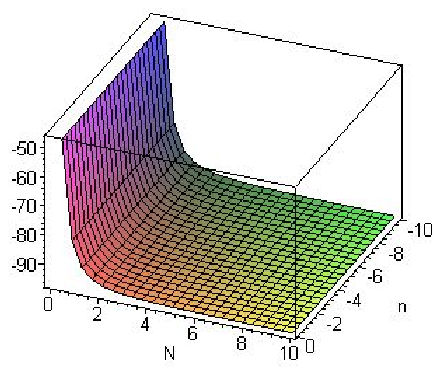}\vspace*{-3.0cm}
\caption{Le d\'eterminant de la m\'etrique tenseure trac\'ee comme la
fonction de $\{n, N\}$, d\'ecrivant les fluctuations de la configuration
des trous noirs charg\'es non-supersym\'etriques aux corrections de 
second ordre de $\alpha^{\prime}$.} \label{nonsusy2detg}\vspace*{0.5cm}
\end{figure}

\begin{figure}
\hspace*{1.0cm}\vspace*{-6.0cm}
\includegraphics[width=12.0cm,angle=-0]{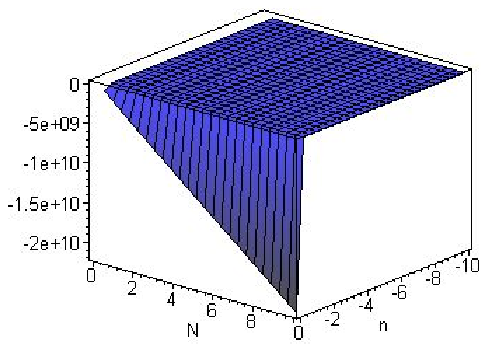}\vspace*{-3.0cm}
\caption{Le mineur d'hypersurface de la m\'etrique tenseure trac\'ee comme la
fonction de $\{n, N\}$, d\'ecrivant les fluctuations de la configuration
des trous noirs charg\'es non-supersym\'etriques aux corrections de 
second ordre de $\alpha^{\prime}$.} \label{nonsusy2minor3}\vspace*{0.5cm}
\end{figure}

\begin{figure}
\hspace*{1.0cm}\vspace*{-6.0cm}
\includegraphics[width=12.0cm,angle=-0]{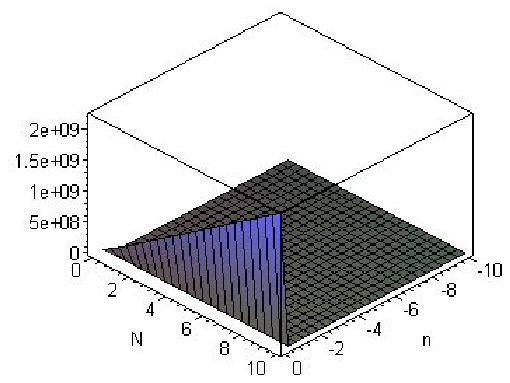}\vspace*{-3.0cm}
\caption{Le premier mineur de la m\'etrique tenseure trac\'ee comme la
fonction de $\{n, N\}$, d\'ecrivant les fluctuations de la configuration
des trous noirs charg\'es non-supersym\'etriques aux corrections de 
second ordre de $\alpha^{\prime}$.} \label{nonsusy2minor1}\vspace*{0.5cm}
\end{figure}

Comme une fonction de $\{n, N\}$, la stabilit\'e de cet ensemble d\'ecoule de la 
positivit\'e du d\'eterminant de la m\'etrique tenseure. Dans ce cas, nous remarquons 
que le d\'eterminant de la m\'etrique tenseure tend vers une valeur n\'egative. 
Pour une valeur typique de $n \in (-10, 0) $ et $N \in (0, 10)$, la Fig. (\ref{nonsusy2detg}) 
montre que le d\'eterminant de la m\'etrique tenseure r\'eside dans l'intervalle $(-30, -100)$.
De plus, la stabilit\'e de l'hypersurface d\'efinie par une valeur constante de $W$
est montr\'e dans la Fig. (\ref{nonsusy2minor3}). Par la pr\'esente, nous voyons que le
mineur $p_3$ r\'eside dans la gamme de $(-2.5 \times 10^{10}, 0)$. Avis que la 
n\'egativit\'e de $p_3$ augmente \`a mesure que la valeur de $N$ est augment\'e
de z\'ero \`a $10$. Par ailleurs, la surface d\'efinie par les fluctuations de
$\{n, w\}$ est instable en raison du fait que le mineur principe correspondant est 
identiquement nulle, \`a savoir que nous avons $p_2= 0$. Lorsque seul le param\`etre
$n$ est autoris\'e \`a varier, la stabilit\'e de la configuration de ces trous noirs
est donn\'ee par la positivit\'e du premier mineur principe $p_1:= g_{nn}$.
Une vue rotat\'ee de $p_1$ est montr\'e dans la Fig. (\ref{nonsusy2minor1}).
Les vues graphiques ci-dessus de l'espace d'\'etat et les propri\'et\'es de 
la positivit\'e des mineurs principaux concern\'es fournent la notion qualitative 
de la stabilit\'e statistique des trous noirs non-supersym\'etriques.

Finalement, comme nous avons fourni les $\Gamma_{abc}$ dans l'annex $[A]$, 
nous pourrions bien s\^ur refaire le m\^eme genre de calcul avec cette 
$ g_{ab}(n,w,N,W) $ corrig\'ee par les corrections de $(\alpha^{\prime})^2 $. 
En fait, nous obtenons que la courbure scalaire de Ruppenier est donn\'ee par:

\ba R= -\frac{192}{\pi} (NW) (\frac{NW}{nw})^{1/2} \frac{r(N,W)}{\tilde{g}(N,W)^3},\ea

o\`u la fonction $r(N,W)$ est d\'efinie par

\ba r(N,W)&:=& b^{(2)}_{0} (NW)^{22}
-b^{(2)}_{1} (NW)^{21} \widehat{\alpha}
+b^{(2)}_{2} (NW)^{20} \widehat{\alpha}^2
-b^{(2)}_{3} (NW)^{19} \widehat{\alpha}^3 \nn &&
+b^{(2)}_{4} (NW)^{18} \widehat{\alpha}^4
-b^{(2)}_{5} (NW)^{17} \widehat{\alpha}^5
+b^{(2)}_{6} (NW)^{16} \widehat{\alpha}^6 
+b^{(2)}_{7} (NW)^{15} \widehat{\alpha}^7 \nn &&
-b^{(2)}_{8} (NW)^{14} \widehat{\alpha}^8
+b^{(2)}_{9} (NW)^{13} \widehat{\alpha}^9
-b^{(2)}_{10} (NW)^{12} \widehat{\alpha}^{10}
+b^{(2)}_{11} (NW)^{11} \widehat{\alpha}^{11} \nn &&
-b^{(2)}_{12} (NW)^{10} \widehat{\alpha}^{12}
-b^{(2)}_{13} (NW)^9 \widehat{\alpha}^{13}
+b^{(2)}_{14} (NW)^8 \widehat{\alpha}^{14}
-b^{(2)}_{15} (NW)^7 \widehat{\alpha}^{15} \nn &&
+b^{(2)}_{16} (NW)^6 \widehat{\alpha}^{16}
-b^{(2)}_{17} (NW)^5 \widehat{\alpha}^{17}
+b^{(2)}_{18} (NW)^4 \widehat{\alpha}^{18}
-b^{(2)}_{19} (NW)^3 \widehat{\alpha}^{19} \nn &&
+b^{(2)}_{20} (NW)^2 \widehat{\alpha}^{20}
-b^{(2)}_{21}  (NW) \widehat{\alpha}^{21}
+b^{(2)}_{22} \widehat{\alpha}^{22}, \ea  

o\`u les constantes r\'eelles positives $ \lbrace a^{(2)}_i \rbrace_{i=0}^8 $ ce qui
apprisent \`a la fonction $\tilde{g}(N,W)$ et $ \lbrace b^{(2)}_i \rbrace_{i=0}^{22} $
et ce qui apprisent \`a la fonction $r(N,W)$ sont donn\'ees dans l'annex $[C]$. 
En ce cas, $\forall \tilde{g}(N,W) \ne 0$ on voit que cette curbure scalaire 
de Ruppenier est partout r\'eguli\`ere.

\begin{figure}
\hspace*{1.0cm}\vspace*{-6.0cm}
\includegraphics[width=12.0cm,angle=-0]{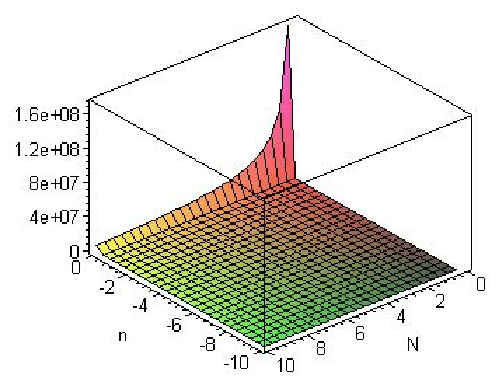}\vspace*{-3.0cm}
\caption{La courbure scalaire trac\'ee comme la
fonction de $\{n, N\}$, d\'ecrivant les fluctuations de la configuration
des trous noirs charg\'es non-supersym\'etriques aux corrections de 
second ordre de $\alpha^{\prime}$ dans la gamme
$n \in (-10,0)$ et $N \in (0,10) $.} \label{nonsusy2R}\vspace*{0.5cm}
\end{figure}

\begin{figure}
\hspace*{1.0cm}\vspace*{-6.0cm}
\includegraphics[width=12.0cm,angle=-0]{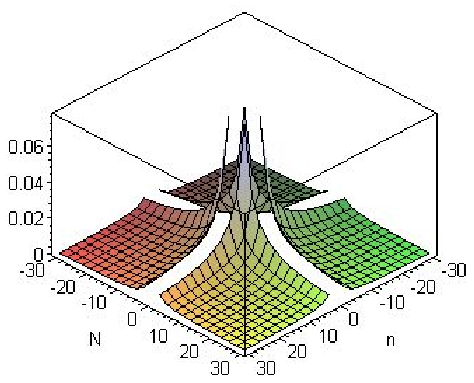}\vspace*{-3.0cm}
\caption{La courbure scalaire trac\'ee comme la
fonction de $\{n, N\}$, d\'ecrivant les fluctuations de la configuration
des trous noirs charg\'es non-supersym\'etriques aux corrections de 
second ordre de $\alpha^{\prime}$ dans la gamme
$n, N \in (-30,30) $.} \label{nonsusy2R30}\vspace*{0.5cm}
\end{figure}

De plus, les propri\'et\'es de la stabilit\'e globale r\'esulte du
comportement de la courbure scalaire de l'espace d'\'etat. En particulier,
dans la gamme de $n \in (-10, 0)$ et $N \in (0, 10)$, la Fig. (\ref{nonsusy2R}) 
montre que la courbure scalaire a une grande amplitude positive. On peut noter 
bien que la configuration des trous noirs sous-tendents est un syst\`eme statistique
fortement interagissant. Le signe positif de la courbure scalaire signifie la nature 
r\'epulsive des interactions statistiques. La Fig. (\ref{nonsusy2R30}) illustre
la nature de la courbure scalaire ci-dessus dans une range \'egale des param\`etres
$n, N \in (-30, 30)$. En fait, lorsque la gamme de $n$ et $N$ est prise en
le multiple de $3$, par exemple $(-30, 30)$, alors que nous voyons de la
Fig. (\ref{nonsusy2R30}) qu'il existe quatre r\'egions disjoints similaires
de l'interaction statistique globale. En comparaison des interactions apparaissant 
dans la gamme de $n \in (-10, 0)$ et $N \in (0, 10)$, l'amplitude des interactions 
statistiques globales se r\'ev\`elent \^{e}tre beaucoup plus petit ce qui figurant 
dans la gamme de $n, N \in (-30, 30)$. Comme mentionn\'e avant pour le cas de premier 
ordre des corrections de $\alpha^{\prime}$, le ration de la valeur typique de l'amplitude 
de ces courbures scalaires de l'espace d'\'etat semble \^{e}tre $10^{-10}$. 
La vue graphique de la comparaison ci-dessus d\'ecoule par les Figs. 
(\ref{nonsusy2R}, \ref{nonsusy2R30}). Qualitativement, dans les petites valeurs 
de $n, N $, les repr\'esentations graphiques ci-dessus indiquent que ces trous noirs 
non-supersym\'etriques de la th\'eorie des cordes correspondent aux configurations 
statistiques instables aux corrections de second ordre de $\alpha^{\prime}$.


\subsection{\`A l'ordre de $(\alpha^{\prime})^3 $}

Ensuite, nous enqu\^etons sur la g\'eom\'etrie thermodynamique avec les
corrections de $(\alpha^{\prime})^3$ d'un autre ordre sup\'erieur prochain,
\`a l'entropie d'un trou noir dyonique non-supersym\'etriques \cite{AshokeSen}
qu'elle est modifi\'e \`a:

\ba 
S_{BH}^{ns}= 2 \pi \sqrt{nw N W}+ \frac{5 \pi \widehat{\alpha}}{4} \sqrt{\frac{nw}{N W}} 
- \frac{29 \pi \widehat{\alpha}^2}{64} \frac{\sqrt{nw}}{(NW)^{3/2}}
- \frac{119 \pi \widehat{\alpha}^3}{512} \frac{\sqrt{nw}}{(NW)^{5/2}}.
\ea

Cette fois, on voit que les composantes de la m\'etrique tenseure de Ruppenier sont:

\ba g_{nn}&=& \frac{\pi}{2n}\sqrt{\frac{wNW}{n}}
         + \frac{5 \pi \widehat{\alpha}}{16n}\sqrt{\frac{w}{nNW}} 
         - \frac{29\pi \widehat{\alpha}^2}{256 nNW} \sqrt{\frac{w}{nNW}}
         - \frac{119\pi \widehat{\alpha}^3}{2048 n(NW)^2} \sqrt{\frac{w}{nNW}},\nn
g_{nw}&=& -\frac{\pi}{2} \sqrt{\frac{NW}{nw}}
           - \frac{5 \pi \widehat{\alpha}}{16 \sqrt{nw N W}} 
           + \frac{29\pi \widehat{\alpha}^2}{256 NW \sqrt{nwNW}}
           + \frac{119\pi \widehat{\alpha}^3}{2048 (NW)^2 \sqrt{nwNW}},\nn
g_{nN}&=& -\frac{\pi}{2} \sqrt{\frac{wW}{nN}}
          + \frac{5 \pi \widehat{\alpha}}{16NW} \sqrt{\frac{wW}{nN}} 
          - \frac{87\pi \widehat{\alpha}^2}{256 (NW)^2} \sqrt{\frac{wW}{nN}} 
          -  \frac{595 \pi \widehat{\alpha}^3}{2048 (NW)^3} \sqrt{\frac{wW}{nN}},\nn
g_{nW}&=& -\frac{\pi}{2} \sqrt{\frac{wN}{nW}}
          + \frac{5 \pi \widehat{\alpha}}{16NW} \sqrt{\frac{w N}{n W}} 
          - \frac{87\pi \widehat{\alpha}^2}{256 (NW)^2} \sqrt{\frac{wN}{nW}} 
          - \frac{595 \pi \widehat{\alpha}^3}{2048 (NW)^3} \sqrt{\frac{wN}{nW}},\ea
\ba g_{ww}&=&  \frac{\pi}{2w}\sqrt{\frac{nNW}{w}}
          + \frac{5 \pi \widehat{\alpha}}{16w} \sqrt{\frac{n}{w N W}} 
          - \frac{29\pi \widehat{\alpha}^2}{256 wNW} \sqrt{\frac{n}{w N W}} 
          - \frac{119 \pi \widehat{\alpha}^3}{2048 w(NW)^2} \sqrt{\frac{n}{w N W}},\nn
g_{wN}&=& -\frac{\pi}{2} \sqrt{\frac{nW}{wN}}
          + \frac{5 \pi \widehat{\alpha}}{16NW} \sqrt{\frac{nW}{w N}}
          - \frac{87\pi \widehat{\alpha}^2}{256 (NW)^2} \sqrt{\frac{nW}{wN}}
          - \frac{595 \pi \widehat{\alpha}^3}{2048 (NW)^3} \sqrt{\frac{nW}{wN}},\nn
g_{wW}&=& -\frac{\pi}{2} \sqrt{\frac{nN}{wW}}
          + \frac{5 \pi \widehat{\alpha}}{16NW }\sqrt{\frac{nN}{wW}}
          - \frac{87\pi \widehat{\alpha}^2}{256 (NW)^2} \sqrt{\frac{nN}{wW}} 
          - \frac{595 \pi \widehat{\alpha}^3}{2048 (NW)^3} \sqrt{\frac{nN}{wW}},\nn
g_{NN}&=& \frac{\pi}{2N}\sqrt{\frac{nwW}{N}}
          - \frac{15\pi \widehat{\alpha}}{16N^2}\sqrt{\frac{nw}{NW}}
          +\frac{435\pi \widehat{\alpha}^2}{256 (N^3W)} \sqrt{\frac{nw}{NW}}
          + \frac{4165 \pi \widehat{\alpha}^3}{2048 (N^4W^2)} \sqrt{\frac{nw}{NW}},\nn
g_{NW}&=& -\frac{\pi}{2} \sqrt{\frac{nw}{NW}}
          - \frac{5 \pi \widehat{\alpha}}{16NW} \sqrt{\frac{nw}{NW}} 
          + \frac{261 \pi \widehat{\alpha}^2}{256 (NW)^2} \sqrt{\frac{nw}{NW}} 
          + \frac{2975 \pi \widehat{\alpha}^3}{2048 (NW)^3} \sqrt{\frac{nw}{NW}},\nn
g_{WW}&=& \frac{\pi}{2W}\sqrt{\frac{nwN}{W}}
          - \frac{15\pi \widehat{\alpha}}{16W^2}\sqrt{\frac{nw}{NW}} 
           +\frac{435\pi \widehat{\alpha}^2}{256 (NW^3)} \sqrt{\frac{nw}{NW}} 
         + \frac{4165 \pi \widehat{\alpha}^3}{2048 (N^2W^4)} \sqrt{\frac{nw}{NW}}.\ea

\begin{figure}
\hspace*{1.0cm}\vspace*{-6.0cm}
\includegraphics[width=12.0cm,angle=-0]{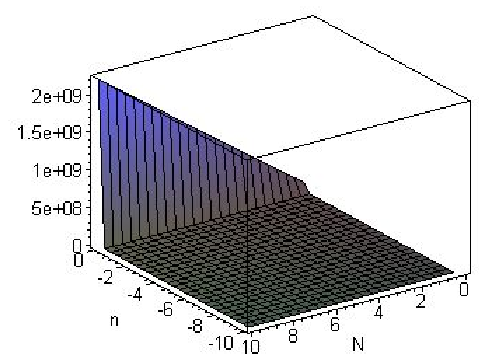}\vspace*{-3.0cm}
\caption{La composante $nn$ de la m\'etrique tenseure trac\'ee comme la
fonction de $\{n, N\}$, d\'ecrivant les fluctuations de la configuration
des trous noirs charg\'es non-supersym\'etriques aux corrections de 
troisi\`eme ordre de $\alpha^{\prime}$.} \label{nonsusy3nn1}\vspace*{0.5cm}
\end{figure}

\begin{figure}
\hspace*{1.0cm}\vspace*{-6.0cm}
\includegraphics[width=12.0cm,angle=-0]{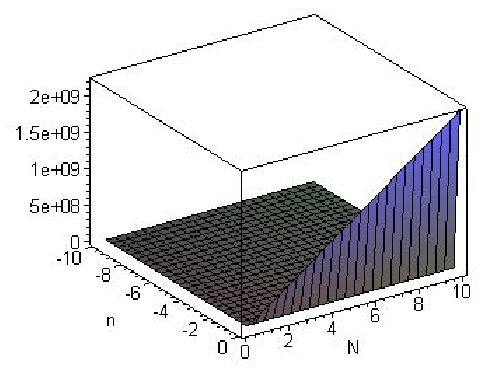}\vspace*{-3.0cm}
\caption{La composante $ww$ de la m\'etrique tenseure trac\'ee comme la
fonction de $\{n, N\}$, d\'ecrivant les fluctuations de la configuration
des trous noirs charg\'es non-supersym\'etriques aux corrections de 
troisi\`eme ordre de $\alpha^{\prime}$.} \label{nonsusy3ww5}\vspace*{0.5cm}
\end{figure}

\begin{figure}
\hspace*{1.0cm}\vspace*{-6.0cm}
\includegraphics[width=12.0cm,angle=-0]{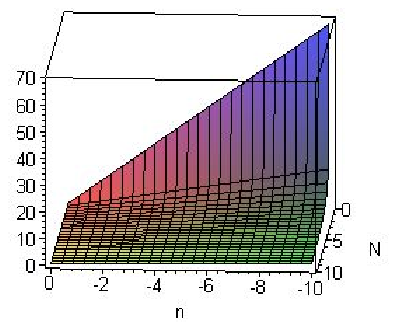}\vspace*{-3.0cm}
\caption{La composante $NN$ de la m\'etrique tenseure trac\'ee comme la
fonction de $\{n, N\}$, d\'ecrivant les fluctuations de la configuration
des trous noirs charg\'es non-supersym\'etriques aux corrections de 
troisi\`eme ordre de $\alpha^{\prime}$.} \label{nonsusy3NN8}\vspace*{0.5cm}
\end{figure}

\begin{figure}
\hspace*{1.0cm}\vspace*{-6.0cm}
\includegraphics[width=12.0cm,angle=-0]{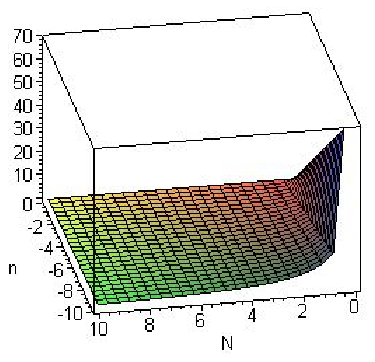}\vspace*{-3.0cm}
\caption{La composante $WW$ de la m\'etrique tenseure trac\'ee comme la
fonction de $\{n, N\}$, d\'ecrivant les fluctuations de la configuration
des trous noirs charg\'es non-supersym\'etriques aux corrections de 
troisi\`eme ordre de $\alpha^{\prime}$.} \label{nonsusy3WW10}\vspace*{0.5cm}
\end{figure}

\begin{figure}
\hspace*{1.0cm}\vspace*{-6.0cm}
\includegraphics[width=12.0cm,angle=-0]{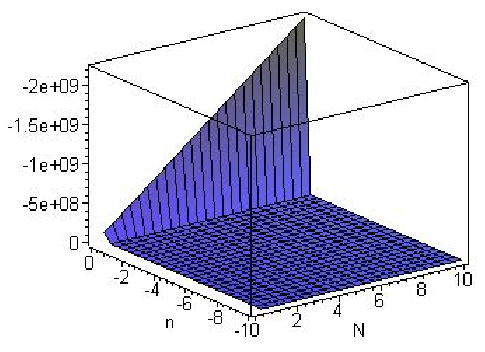}\vspace*{-3.0cm}
\caption{La composante $nw$ de la m\'etrique tenseure trac\'ee comme la
fonction de $\{n, N\}$, d\'ecrivant les fluctuations de la configuration
des trous noirs charg\'es non-supersym\'etriques aux corrections de 
troisi\`eme ordre de $\alpha^{\prime}$.} \label{nonsusy3nw2}\vspace*{0.5cm}
\end{figure}

\begin{figure}
\hspace*{1.0cm}\vspace*{-6.0cm}
\includegraphics[width=12.0cm,angle=-0]{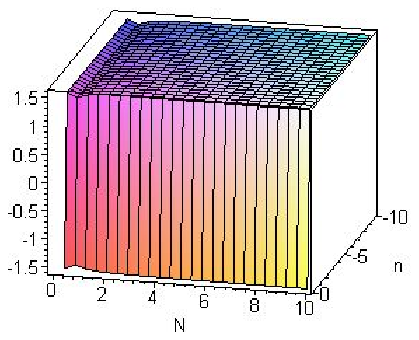}\vspace*{-3.0cm}
\caption{La composante $nN$ de la m\'etrique tenseure trac\'ee comme la
fonction de $\{n, N\}$, d\'ecrivant les fluctuations de la configuration
des trous noirs charg\'es non-supersym\'etriques aux corrections de 
troisi\`eme ordre de $\alpha^{\prime}$.} \label{nonsusy3nN3}\vspace*{0.5cm}
\end{figure}

\begin{figure}
\hspace*{1.0cm}\vspace*{-6.0cm}
\includegraphics[width=12.0cm,angle=-0]{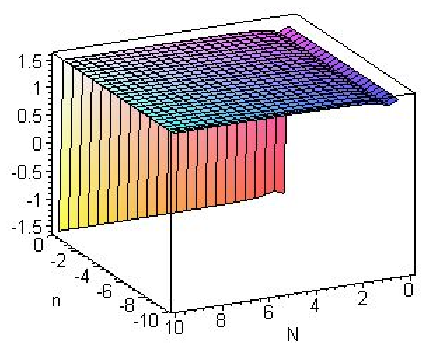}\vspace*{-3.0cm}
\caption{La composante $nW$ de la m\'etrique tenseure trac\'ee comme la
fonction de $\{n, N\}$, d\'ecrivant les fluctuations de la configuration
des trous noirs charg\'es non-supersym\'etriques aux corrections de 
troisi\`eme ordre de $\alpha^{\prime}$.} \label{nonsusy3nW4}\vspace*{0.5cm}
\end{figure}

\begin{figure}
\hspace*{1.0cm}\vspace*{-6.0cm}
\includegraphics[width=12.0cm,angle=-0]{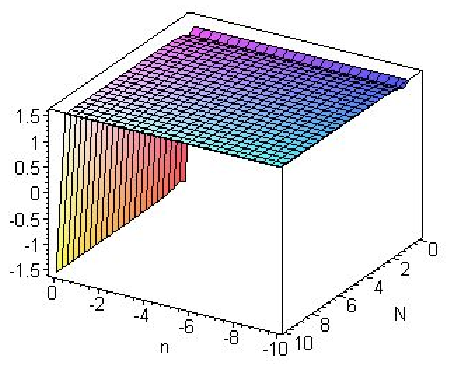}\vspace*{-3.0cm}
\caption{La composante $wN$ de la m\'etrique tenseure trac\'ee comme la
fonction de $\{n, N\}$, d\'ecrivant les fluctuations de la configuration
des trous noirs charg\'es non-supersym\'etriques aux corrections de 
troisi\`eme ordre de $\alpha^{\prime}$.} \label{nonsusy3wN6}\vspace*{0.5cm}
\end{figure}

\begin{figure}
\hspace*{1.0cm}\vspace*{-6.0cm}
\includegraphics[width=12.0cm,angle=-0]{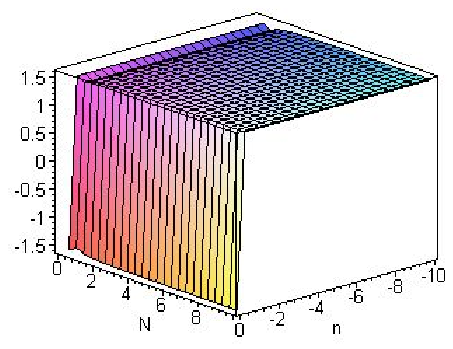}\vspace*{-3.0cm}
\caption{La composante $wW$ de la m\'etrique tenseure trac\'ee comme la
fonction de $\{n, N\}$, d\'ecrivant les fluctuations de la configuration
des trous noirs charg\'es non-supersym\'etriques aux corrections de 
troisi\`eme ordre de $\alpha^{\prime}$.} \label{nonsusy3wW7}\vspace*{0.5cm}
\end{figure}

\begin{figure}
\hspace*{1.0cm}\vspace*{-6.0cm}
\includegraphics[width=12.0cm,angle=-0]{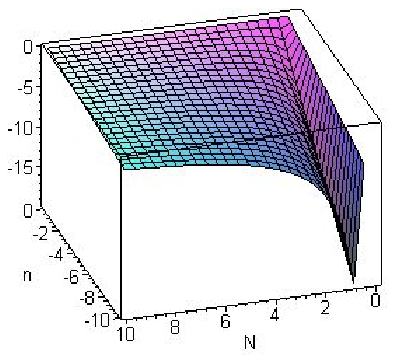}\vspace*{-3.0cm}
\caption{La composante $NW$ de la m\'etrique tenseure trac\'ee comme la
fonction de $\{n, N\}$, d\'ecrivant les fluctuations de la configuration
des trous noirs charg\'es non-supersym\'etriques aux corrections de 
troisi\`eme ordre de $\alpha^{\prime}$.} \label{nonsusy3NW9}\vspace*{0.5cm}
\end{figure}

Comme mentionn\'e dans la section pr\'ec\`edente, nous allons bien continuer 
\`a offrir les repr\'esentations graphiques dans la limite de $n=w$, $N= W$ et
$\widehat{\alpha}=0.1$. Sous les corrections du troisi\`eme ordre de $\alpha^{\prime}$, 
nous trouvons dans le r\'egime de $n \in (-10, 0)$ et $N \in (0,10)$ que l'amplitude 
des capacit\'es de chaleurs $\{g_{nn}, g_{ww}\}$ prend la valeur maximale \`a l'ordre 
de $2 \times 10^{+09} $. Dans cette gamme de $\{n, N\}$, nous observons que les capacit\'es 
de chaleurs $\{g_{NN}, g_{WW}\}$ se situent dans la gamme de $(0, 70)$. Comme pr\'evu au 
second ordre des corrections de $\alpha^{\prime}$, nous voyons en ce cas aussi que la gamme 
de la premi\`ere ensemble et celle de la deuxi\`eme ensemble des capacit\'es de chaleurs 
reste dans la limite oppos\'ee des param\`etres $\{n, N\}$ . Explicitement, les Figs. 
(\ref{nonsusy3nn1}, \ref{nonsusy3ww5}) indiquent que la croissance de premier ensemble 
des capacit\'es de chaleurs $\{g_{nn}, g_{ww}\}$ a lieu dans la limite d'un grand $N$ 
et un petit $n$. Cela signifie que les corrections de d\'eriv\'ees sup\'erieures ne modifient 
pas significativement le comportement thermodynamique de la configuration des trous noirs 
non-supersym\'etriques. De plus, les Figs. (\ref{nonsusy3NN8}, \ref{nonsusy3WW10}) montrent 
que la croissance de $\{g_{NN}, g_{WW}\}$ a lieu dans la limite d'un petit $N$ et d'un grand $n$. 
D'autre part, les compressibilit\'es de chaleurs en impliquant deux param\`etres distincts d'une 
configuration donn\'ee des trous noirs non-supersym\'etriques de la th\'eorie des cordes sont 
repr\'esent\'es dans les Figs. (\ref{nonsusy3nw2}, \ref{nonsusy3nN3}, \ref{nonsusy3nW4}, 
\ref{nonsusy3wN6}, \ref{nonsusy3wW7}, \ref{nonsusy3NW9}). A cet ordre des corrections de 
d\'eriv\'ees sup\'erieures de $\alpha^{\prime}$ aux trous noirs non-supersym\'etriques, 
nous remarquons que les fluctuations, en impliquant $\{n, w\}$, tel qu'elle sont 
d\'efinies par les composantes de la m\'etrique tenseure de l'espace d'\'etat
$\{g_{ij} \ | \ i, j = n, w, N, W\}$, ont relativement des plus grande valeurs 
num\'eriques en comparision de celles qui impliquant $\{N, W\}$.

Par la m\'etrique tenseure donn'ee ci-dessus, nous constatons que 
l'ordre imm\'ediatement des corrections d\'eriv\'ees sup\'erieures
conduisent les mineurs principaux suivants

\begin{eqnarray}
\mathit{p_1} &=& - {\displaystyle \frac{1}{2048}} \,
{\displaystyle \frac{\pi \,\sqrt{w}}{n^{(3/2)}\,(N\,\mathrm{W})^{(5/2)}}}
\,\mathit{\tilde{p}_1}(\widehat{\alpha}), \nonumber \\
\mathit{p_2} &=& 0, \nonumber \\
\mathit{p_3} &=& 
{\displaystyle \frac{1}{2147483648}} \frac{\pi ^{3}}{N^{(19/2)}\,
\mathrm{W}^{(15/2)}\,\sqrt{n\,w}}\,\mathit{\tilde{p}_3}(\widehat{\alpha}).
\end{eqnarray}

o\`u les fonctions $\{ \mathit{\tilde{p}_1}(\widehat{\alpha}), 
\mathit{\tilde{p}_3}(\widehat{\alpha}) \}$ sont d\'efinies comme

\begin{eqnarray}
\mathit{\tilde{p}_1}(\widehat{\alpha})&:=& 
+ 119\,\widehat{\alpha}^{3} + 232\,\widehat{\alpha}^{2}\,N\,\mathrm{W}
- 640\,N^{2}\,\mathrm{W}^{2}\,\widehat{\alpha}
- 1024\,N^{3}\,\mathrm{W}^{3}, \nonumber \\
\mathit{\tilde{p}_3}(\widehat{\alpha})&:=&
+ 42128975\,\widehat{\alpha}^{9}
+ 180694360\,N\,\mathrm{W}\,\widehat{\alpha}^{8}
- 67409216\,N^{2}\,\mathrm{W}^{2}\,\widehat{\alpha}^{7} \nonumber \\ && \mbox{} 
- 917907968\,N^{3}\, \mathrm{W}^{3}\,\widehat{\alpha}^{6} 
- 376332288\,N^{4}\,\mathrm{W}^{4}\,\widehat{\alpha}^{5}  \nonumber \\ && \mbox{} 
+ 343867392\,N^{5}\,\mathrm{W}^{5}\,\widehat{\alpha}^{4} 
- 1689255936\,N^{6}\,\mathrm{W}^{6}\,\widehat{\alpha}^{3} \nonumber \\ && \mbox{}  
- 796917760\,N^{7}\,\mathrm{W}^{7}\,\widehat{\alpha}^{2} 
+ 671088640\,N^{8}\,\mathrm{W}^{8}\,\widehat{\alpha} \nonumber \\ && \mbox{} 
- 1073741824\,N^{9}\,\mathrm{W}^{9}.
\end{eqnarray} 

Il n'est pas aussi difficile de voir que le d\'eterminant de la m\'etrique tenseure est:

\ba g= \frac{\pi^4}{1099 5116 27776(NW)^{12}} \tilde{g}(N,W), \ea

o\`u la fonction $ \tilde{g}(N,W)$, comme une fonction de $\widehat{\alpha}$, est d\'efinie par:

\ba \tilde{g}(N,W)&=& 
-a^{(3)}_0 (NW)^{12} 
+a^{(3)}_1 (NW)^{11} \widehat{\alpha}
-a^{(3)}_2 (NW)^{10} \widehat{\alpha}^2 
-a^{(3)}_3 (NW)^9 \widehat{\alpha}^3 \nn &&
+a^{(3)}_4 (NW)^8 \widehat{\alpha}^4 
-a^{(3)}_5 (NW)^7 \widehat{\alpha}^5
-a^{(3)}_6 (NW)^6 \widehat{\alpha}^6
+ a^{(3)}_7 (NW)^5 \widehat{\alpha}^7 \nn &&
-a^{(3)}_8 (NW)^4 \widehat{\alpha}^8 
-a^{(3)}_9 (NW)^3 \widehat{\alpha}^9
+a^{(3)}_{10} (NW)^2 \widehat{\alpha}^{10} 
+a^{(3)}_{11} (NW) \widehat{\alpha}^{11} \nn &&
+a^{(3)}_{12}  \widehat{\alpha}^{12},\ea

o\`u les constantes r\'eelles positives $ \lbrace a^{(3)}_i \rbrace $ 
sont donn\'ees dans l'annex $[C]$. 

Sous le troisi\`eme ordre des corrections d\'eriv\'ees sup\'erieures 
de $\alpha^{\prime}$, nous constatons que la configuration sous-jacente 
de ces trous noirs est relativement stable dans les r\'egions o\`u les mineurs 
de l'hypersurface correspondante et le d\'eterminant de la m\'etrique tenseure prennent 
des valeurs positives. Notamment, lorsque le param\`etre $\widehat{\alpha}$ des corrections
est tel que (i) le polyn\^{o}me cubique $ \mathit{\tilde{p}_1}(\widehat{\alpha})$ a un signe n\'egatif, 
(ii) le polyn\^{o}me $ \mathit{\tilde{p}_3}(\widehat{\alpha}) $ de degr\'e neuf a un signe positif 
et (iii) le polyn\^{o}me $\mathit{\tilde{g}}(\widehat{\alpha}) $ de degr\'e douze a un signe positif.
Pour tout $\widehat{\alpha}$ tels que les mineurs principaux $\mathit{\tilde{p}_1}(\widehat{\alpha}) <0$,
$\mathit{\tilde{p}_3}(\widehat{\alpha}) >0$ et le d\'eterminant de la m\'etrique tenseure 
$\mathit{\tilde{g}}(\widehat{\alpha})>0$ sont satisfaites, la solution sous-jacente de ces
trous noirs est relativement stable.
Ainsi, la stabilit\'e relative de ces trous noirs au-dessus, c'est-\`a-dire que
la valeur sp\'ecifique du param\`etre $\widehat{\alpha}$ peut \^{e}tre d\'etermin\'ee comme 
les racines communes des polyn\^{o}mes ci-dessus. Par ailleurs, la disparition de mineur 
de surface $\mathit{p_2}= 0$ montre que la configuration sous-jacente de ces trous noirs
reste instable au troisi\`eme ordre des corrections d\'eriv\'es sup\'erieures
\`a l'entropie. Par la suite, nous supposons que les corrections perturbatives 
de la th\'eorie des cordes ne sont pas suffisantes pour produire la stabilit\'e 
thermodynamique des trous noirs non-supersym\'etriques de quatre dimensions.

\begin{figure}
\hspace*{1.0cm}\vspace*{-6.0cm}
\includegraphics[width=12.0cm,angle=-0]{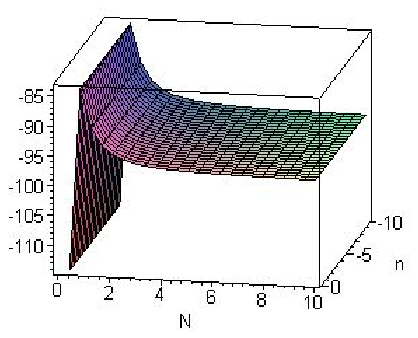}\vspace*{-3.0cm}
\caption{Le d\'eterminant de la m\'etrique tenseure trac\'ee comme la
fonction de $\{n, N\}$, d\'ecrivant les fluctuations de la configuration
des trous noirs charg\'es non-supersym\'etriques aux corrections de 
troisi\`eme ordre de $\alpha^{\prime}$.} \label{nonsusy3detg}\vspace*{0.5cm}
\end{figure}

\begin{figure}
\hspace*{1.0cm}\vspace*{-6.0cm}
\includegraphics[width=12.0cm,angle=-0]{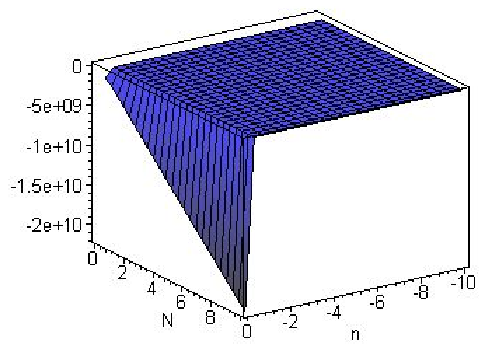}\vspace*{-3.0cm}
\caption{Le mineur d'hypersurface de la m\'etrique tenseure trac\'ee comme la
fonction de $\{n, N\}$, d\'ecrivant les fluctuations de la configuration
des trous noirs charg\'es non-supersym\'etriques aux corrections de 
troisi\`eme ordre de $\alpha^{\prime}$.} \label{nonsusy3minor3}\vspace*{0.5cm}
\end{figure}

\begin{figure}
\hspace*{1.0cm}\vspace*{-6.0cm}
\includegraphics[width=12.0cm,angle=-0]{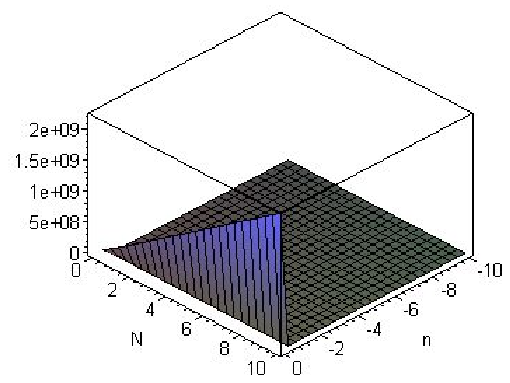}\vspace*{-3.0cm}
\caption{Le premier mineur de la m\'etrique tenseure trac\'ee comme la
fonction de $\{n, N\}$, d\'ecrivant les fluctuations de la configuration
des trous noirs charg\'es non-supersym\'etriques aux corrections de 
troisi\`eme ordre de $\alpha^{\prime}$.} \label{nonsusy3minor1}\vspace*{0.5cm}
\end{figure}

Comme mentionn\'e pr\'ec\'edemment, la stabilit\'e d'un ensemble de syst\`eme de ces 
trous noirs charg\'es non-supersym\'etriques peut \^etre d\'etermin\'ee comme la fonction 
de $\{n, N\}$. Cela d\'ecoule de la positivit\'e du d\'eterminant de la m\'etrique tenseure. 
Dans ce cas, nous constatons que le d\'eterminant de la m\'etrique tenseure tend vers une
valeur n\'egative. Pour une valeur typique de $n \in (-10, 0)$ et $N \in (0, 10) $, 
la Fig. (\ref{nonsusy3detg}) montre que le d\'eterminant de la m\'etrique tenseure r\'eside 
dans l'intervalle $(-80, -120)$. En outre, la stabilit\'e de l'hypersurface d\'efinie par une valeur 
constante de $W$ est montr\'e dans la Fig. (\ref{nonsusy3minor3}). Par la pr\'esente, 
nous observons que les mineurs $p_3$ r\'eside dans la gamme de $(-2.5 \times 10^{10}, 0)$. 
Notez que la n\'egativit\'e de $p_3$ augmente \`a mesure que la valeur de $N$ est pass\'ee 
de z\'ero \`a $10$. En outre, la surface d\'efinie par les fluctuations du $\{n, w\}$ est 
instable en raison du fait que le mineur principe correspondant est identiquement nulle, 
\`a savoir que nous avons $p_2= 0$. Lorsque le seul param\`etre $n$ est autoris\'e \`a varier, 
la stabilit\'e de la configuration des trous noirs charg\'es non-supersym\'etriques est donn\'e 
par la positivit\'e du premier mineur principe $p_1:= g_{nn}$. Une vue rotat\'ee de $p_1$ est 
montr\'ee dans la Fig. (\ref{nonsusy3minor1}). Les propri\'et\'es graphiques ci-dessus 
de l'espace d'\'etat et la positivit\'e des mineurs principaux concern\'es fournent 
la notion qualitative de la stabilit\'e statistique des trous noirs non-supersym\'etriques 
corrig\'ees \`a l'ordre de troisi\`eme ordre de $\alpha^{\prime}$.

Et bien encore \`a cette fois, par les $\Gamma_{abc}$ donn\'es dans l'annex $[A]$, 
nous voyons avec cette $ g_{ab}(n,w,N,W) $ corrig\'ee par les corrections 
de $(\alpha^{\prime})^3 $ que la courbure scalaire de Ruppenier est partout r\'eguli\`ere, 
$\forall \tilde{g}(N,W) \ne 0$, et elle est donn\'ee par:

\ba R= - \frac{3072 (NW)^2}{\pi} \sqrt{\frac{NW}{nw}} \frac{r(N,W)}{\tilde{g}(N,W)^3},\ea

o\`u la fonction $ r(N,W)$ est d\'efinie par:


\ba
r(N,W)&:=&
b^{(3)}_{0} (NW)^{33}
-b^{(3)}_{1} (NW)^{32} \widehat{\alpha} 
+b^{(3)}_{2} (NW)^{31} \widehat{\alpha}^2 
-b^{(3)}_{3} (NW)^{30} \widehat{\alpha}^3  \nn &&
+b^{(3)}_{4} (NW)^{29} \widehat{\alpha}^4 
+b^{(3)}_{5} (NW)^{28} \widehat{\alpha}^5 
-b^{(3)}_{6} (NW)^{27} \widehat{\alpha}^6
+b^{(3)}_{7} (NW)^{26} \widehat{\alpha}^7  \nn &&
+b^{(3)}_{8} (NW)^{25} \widehat{\alpha}^8
-b^{(3)}_{9} (NW)^{24} \widehat{\alpha}^9 
+b^{(3)}_{10} (NW)^{23} \widehat{\alpha}^{10} 
+b^{(3)}_{11} (NW)^{22} \widehat{\alpha}^{11}  \nn &&
-b^{(3)}_{12} (NW)^{21} \widehat{\alpha}^{12} 
+b^{(3)}_{13} (NW)^{20} \widehat{\alpha}^{13} 
+b^{(3)}_{14} (NW)^{19} \widehat{\alpha}^{14} 
-b^{(3)}_{15}  (NW)^{18} \widehat{\alpha}^{15} \nn &&
+b^{(3)}_{16} (NW)^{17} \widehat{\alpha}^{16} 
+b^{(3)}_{17} (NW)^{16} \widehat{\alpha}^{17} 
-b^{(3)}_{18} (NW)^{15} \widehat{\alpha}^{18} 
+b^{(3)}_{19} (NW)^{14} \widehat{\alpha}^{19}  \nn &&
+b^{(3)}_{20} (NW)^{13} \widehat{\alpha}^{20} 
-b^{(3)}_{21} (NW)^{12} \widehat{\alpha}^{21} 
+b^{(3)}_{22} (NW)^{11} \widehat{\alpha}^{22} 
+b^{(3)}_{23} (NW)^{10} \widehat{\alpha}^{23}  \nn &&
-b^{(3)}_{24}  (NW)^{9} \widehat{\alpha}^{24} 
-b^{(3)}_{25} (NW)^{8} \widehat{\alpha}^{25} 
+b^{(3)}_{26} (NW)^7 \widehat{\alpha}^{26} 
-b^{(3)}_{27} (NW)^6 \widehat{\alpha}^{27}  \nn &&
-b^{(3)}_{28} (NW)^5 \widehat{\alpha}^{28} 
-b^{(3)}_{29} (NW)^4 \widehat{\alpha}^{29} 
+b^{(3)}_{30} (NW)^3 \widehat{\alpha}^{30} 
+b^{(3)}_{31}  (NW)^2 \widehat{\alpha}^{31} \nn &&
+b^{(3)}_{32}  (NW) \widehat{\alpha}^{32} 
+b^{(3)}_{33}  \widehat{\alpha}^{33},\ea

o\`u les constantes r\'eelles positives $ \lbrace a^{(3)}_i \rbrace $ et 
$ \lbrace b^{(3)}_i \rbrace $ sont donn\'ees dans l'annex $[C]$.

\begin{figure}
\hspace*{1.0cm}\vspace*{-6.0cm}
\includegraphics[width=12.0cm,angle=-0]{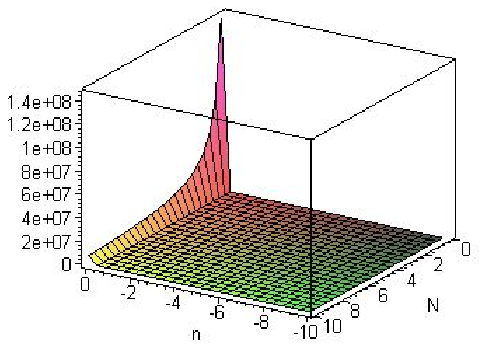}\vspace*{-3.0cm}
\caption{La courbure scalaire trac\'ee comme la fonction de $\{n, N\}$, d\'ecrivant 
les fluctuations de la configuration des trous noirs charg\'es non-supersym\'etriques 
aux corrections de troisi\`eme ordre de $\alpha^{\prime}$ dans la gamme de 
$n \in (-10, 0)$ et $N \in (0,10) $.} \label{nonsusy3R}\vspace*{0.5cm}
\end{figure}

\begin{figure}
\hspace*{1.0cm}\vspace*{-6.0cm}
\includegraphics[width=12.0cm,angle=-0]{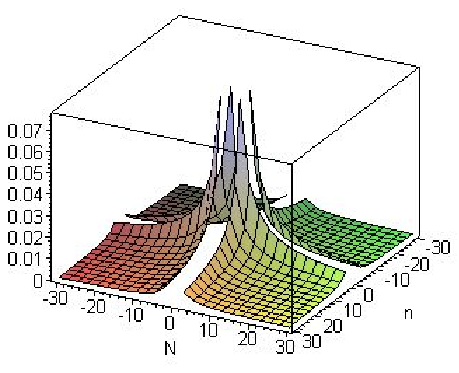}\vspace*{-3.0cm}
\caption{La courbure scalaire trac\'ee comme la fonction de $\{n, N\}$, d\'ecrivant 
les fluctuations de la configuration des trous noirs charg\'es non-supersym\'etriques 
aux corrections de troisi\`eme ordre de $\alpha^{\prime}$ dans la gamme de 
$n \in (-10,0)$ et $N \in (0,10) $.} \label{nonsusy3R30}\vspace*{0.5cm}
\end{figure}

Les propri\'et\'es globales de la stabilit\'e statistique d\'ecoule par le comportement 
de la courbure scalaire de l'espace d'\'etat. En particulier, dans la gamme de 
$n \in (-10, 0)$ et $N \in (0, 10)$, la Fig. (\ref{nonsusy3R}) offre que la courbure 
scalaire a une grande amplitude positive. Nous trouvons que la configuration sous-jacente 
des trous noirs est un syst\`eme statistique fortement interagissant. Le signe positif de
la courbure scalaire signifie une nature r\'epulsive des interactions statistiques. 
La Fig. (\ref{nonsusy3R30}) signifie la nature de la courbure scalaire ci-dessus dans une 
range \'egale des param\`etres $n, N \in (-30,30)$. En fait, lorsque la gamme de $n$ et $N$
est prise dans les multiples de $3$, par exemple $(-30, 30)$, nous remarquons par la Fig. 
(\ref{nonsusy3R30}) qu'il existe quatre r\'egions similaires disjointes des interactions 
statistiques globales. En comparaison des interactions apparaissant dans la gamme de $n \in
(-10, 0)$ et $N \in (0,10) $, les interactions statistiques globales semble \^{e}tre beaucoup 
plus petits de celles qui apparaissant dans la gamme de $n, N \in (-30, 30)$. Comme mentionn\'e 
pr\'ec\'edemment au second ordre des corrections de $\alpha^{\prime}$, nous voyons que le ration de
la valeur typique de l'amplitude de la courbure scalaire de l'espace d'\'etat semble \^{e}tre $10^{-10}$. 
En bref, la conclusion graphique de la comparaison ci-dessus d\'ecoule des Figs. (\ref{nonsusy3R}, 
\ref{nonsusy3R30}). Dans le cas des petite valeur des param\`etres $n, N$, nous voyons que 
les trous noirs non-supersym\'etriques sont int\'eractifs et correspondent \`a une 
configuration statistique instable sous le troisi\`eme ordre de $\alpha^{\prime}$.

Bien que'on ait ajout\'e les correction de $\alpha^{ \prime }$ de l'ordre trois,
nous voyons encore le m\^eme r\'esultat que la curbure scalaire de Ruppenier reste 
partout r\'eguli\`ere, sauf le cas \`ou $\tilde{g}(N,W)= 0$.  Enfin de confirmer
cet observation, nous allons faire la question: comment la g\'eom\'etrie thermodynamique 
d'un ensemble des trous noirs dyoniques non-supersym\'etriques se comporte avec 
un prochain ordre des corrections de $\alpha^{ \prime }$?

\subsection{\`A l'ordre de $(\alpha^{\prime})^4 $}

Pour voir \c{c}a, consid\'erons les corrections suivantes de $\alpha^{ \prime }$ 
dans l'entropie du trou noir dyonique non-supersym\'etrique, s\'elon le Ref. \cite{AshokeSen},
on l'a modifi\'e \`{a}:

\ba S_{BH}^{ns}&=& 2 \pi \sqrt{nw N W}+ \frac{5 \pi \widehat{\alpha}}{4} \sqrt{\frac{nw}{N W}} 
           - \frac{29 \pi \widehat{\alpha}^2}{64} \frac{\sqrt{nw}}{(NW)^{3/2}} \nn &&
           - \frac{119 \pi \widehat{\alpha}^3}{512} \frac{\sqrt{nw}}{(NW)^{5/2}} 
           - \frac{2237 \pi \widehat{\alpha}^4}{16384} \frac{\sqrt{nw}}{(NW)^{7/2}}.\ea

Dans ce cas de l'entropie de ce trou noir, il est egalement facile d'obtenir 
que les composantes de la m\'etrique tenseure de Ruppenier sont:

\ba g_{nn}&=& \frac{\pi}{2n} \sqrt{\frac{wNW}{n}}
         + \frac{5 \pi \widehat{\alpha}}{16n}\sqrt{\frac{w}{nNW}} 
         - \frac{29\pi \widehat{\alpha}^2}{256 nNW} \sqrt{\frac{w}{nNW}} \nn &&
         - \frac{119\pi \widehat{\alpha}^3}{2048 n(NW)^2} \sqrt{\frac{w}{nNW}}
         -\frac{2237 \pi \widehat{\alpha}^4}{65536 n (NW)^3} \sqrt{\frac{w}{nNW}},\nn
g_{nw}&=& -\frac{\pi}{2} \sqrt{\frac{NW}{nw}}
           - \frac{5 \pi \widehat{\alpha}}{16 \sqrt{nw N W}} 
           + \frac{29\pi \widehat{\alpha}^2}{256 NW \sqrt{nwNW}} \nn &&
           + \frac{119\pi \widehat{\alpha}^3}{2048 (NW)^2 \sqrt{nwNW}}
           + \frac{2237 \pi \widehat{\alpha}^4}{65536 (NW)^3 \sqrt{nwNW}},\nn
g_{nN}&=& -\frac{\pi}{2} \sqrt{\frac{wW}{nN}}
          + \frac{5 \pi \widehat{\alpha}}{16NW} \sqrt{\frac{wW}{nN}} 
          - \frac{87\pi \widehat{\alpha}^2}{256 (NW)^2} \sqrt{\frac{wW}{nN}} \nn &&
          -  \frac{595 \pi \widehat{\alpha}^3}{2048 (NW)^3} \sqrt{\frac{wW}{nN}}
         -  \frac{15695 \pi \widehat{\alpha}^4}{65536 (NW)^4} \sqrt{\frac{wW}{nN}},\nn
g_{nW}&=& -\frac{\pi}{2} \sqrt{\frac{wN}{nW}}
          + \frac{5 \pi \widehat{\alpha}}{16NW} \sqrt{\frac{w N}{n W}} 
          - \frac{87\pi \widehat{\alpha}^2}{256 (NW)^2} \sqrt{\frac{wN}{nW}}  \nn &&
          - \frac{595 \pi \widehat{\alpha}^3}{2048 (NW)^3} \sqrt{\frac{wN}{nW}} 
          - \frac{15659 \pi \widehat{\alpha}^4}{65536 (NW)^4} \sqrt{\frac{wN}{nW}},\ea
\ba g_{ww}&=& \frac{\pi}{2w}\sqrt{\frac{nNW}{w}}
          + \frac{5 \pi \widehat{\alpha}}{16w} \sqrt{\frac{n}{w N W}} 
          - \frac{29\pi \widehat{\alpha}^2}{256 wNW} \sqrt{\frac{n}{w N W}}  \nn &&
          - \frac{119 \pi \widehat{\alpha}^3}{2048 w(NW)^2} \sqrt{\frac{n}{w N W}}
          - \frac{2237 \pi \widehat{\alpha}^4}{65536 w(NW)^3} \sqrt{\frac{n}{w N W}},\nn
g_{wN}&=& -\frac{\pi}{2} \sqrt{\frac{nW}{wN}}
          + \frac{5 \pi \widehat{\alpha}}{16NW} \sqrt{\frac{nW}{w N}}
          - \frac{87\pi \widehat{\alpha}^2}{256 (NW)^2} \sqrt{\frac{nW}{wN}} \nn &&
          - \frac{595 \pi \widehat{\alpha}^3}{2048 (NW)^3} \sqrt{\frac{nW}{wN}} 
          - \frac{15659 \pi \widehat{\alpha}^4}{65536 (NW)^4} \sqrt{\frac{nW}{wN}},\nn
g_{wW}&=& -\frac{\pi}{2} \sqrt{\frac{nN}{wW}}
          + \frac{5 \pi \widehat{\alpha}}{16NW }\sqrt{\frac{nN}{wW}}
          - \frac{87\pi \widehat{\alpha}^2}{256 (NW)^2} \sqrt{\frac{nN}{wW}}  \nn &&
          - \frac{595 \pi \widehat{\alpha}^3}{2048 (NW)^3} \sqrt{\frac{nN}{wW}}
          - \frac{15659 \pi \widehat{\alpha}^4}{65536 (NW)^4} \sqrt{\frac{nN}{wW}},\nn
g_{NN}&=& \frac{\pi}{2N}\sqrt{\frac{nwW}{N}}
          - \frac{15\pi \widehat{\alpha}}{16N^2}\sqrt{\frac{nw}{NW}}
          +\frac{435\pi \widehat{\alpha}^2}{256 (N^3W)} \sqrt{\frac{nw}{NW}} \nn &&
          + \frac{4165 \pi \widehat{\alpha}^3}{2048 (N^4W^2)} \sqrt{\frac{nw}{NW}} 
          + \frac{140931 \pi \widehat{\alpha}^4}{65536 (N^5W^3)} \sqrt{\frac{nw}{NW}},\nn
g_{NW}&=& -\frac{\pi}{2} \sqrt{\frac{nw}{NW}}
          - \frac{5 \pi \widehat{\alpha}}{16NW} \sqrt{\frac{nw}{NW}} 
          + \frac{261 \pi \widehat{\alpha}^2}{256 (NW)^2} \sqrt{\frac{nw}{NW}} \nn && 
          + \frac{2975 \pi \widehat{\alpha}^3}{2048 (NW)^3} \sqrt{\frac{nw}{NW}} 
          + \frac{1099613 \pi \widehat{\alpha}^4}{65536 (NW)^4} \sqrt{\frac{nw}{NW}},\nn
g_{WW}&=& \frac{\pi}{2W}\sqrt{\frac{nwN}{W}}
          - \frac{15\pi \widehat{\alpha}}{16W^2}\sqrt{\frac{nw}{NW}} 
           +\frac{435\pi \widehat{\alpha}^2}{256 (NW^3)} \sqrt{\frac{nw}{NW}} \nn && 
         + \frac{4165 \pi \widehat{\alpha}^3}{2048 (N^2W^4)} \sqrt{\frac{nw}{NW}} 
         + \frac{140931 \pi \widehat{\alpha}^4}{65536 (N^3W^5)} \sqrt{\frac{nw}{NW}}.\ea

\begin{figure}
\hspace*{1.0cm}\vspace*{-6.0cm}
\includegraphics[width=12.0cm,angle=-0]{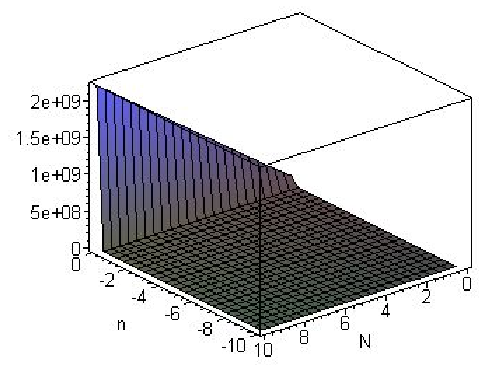}\vspace*{-3.0cm}
\caption{La composante $nn$ de la m\'etrique tenseure trac\'ee comme la
fonction de $\{n, N\}$, d\'ecrivant les fluctuations de la configuration
des trous noirs charg\'es non-supersym\'etriques aux corrections de 
quatri\`eme ordre de $\alpha^{\prime}$.} \label{nonsusy4nn1}\vspace*{0.5cm}
\end{figure}

\begin{figure}
\hspace*{1.0cm}\vspace*{-6.0cm}
\includegraphics[width=12.0cm,angle=-0]{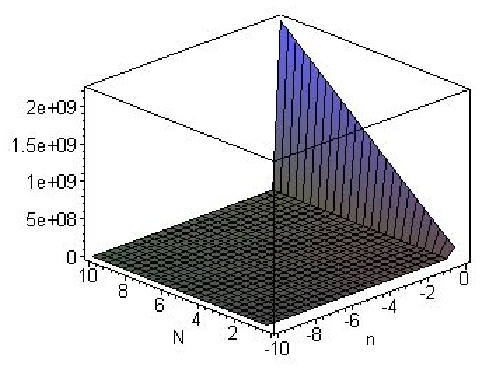}\vspace*{-3.0cm}
\caption{La composante $ww$ de la m\'etrique tenseure trac\'ee comme la
fonction de $\{n, N\}$, d\'ecrivant les fluctuations de la configuration
des trous noirs charg\'es non-supersym\'etriques aux corrections de 
quatri\`eme ordre de $\alpha^{\prime}$.} \label{nonsusy4ww5}\vspace*{0.5cm}
\end{figure}

\begin{figure}
\hspace*{1.0cm}\vspace*{-6.0cm}
\includegraphics[width=12.0cm,angle=-0]{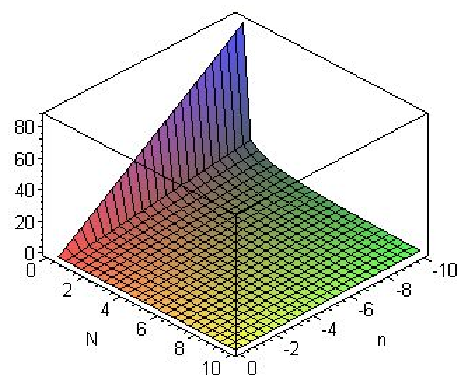}\vspace*{-3.0cm}
\caption{La composante $NN$ de la m\'etrique tenseure trac\'ee comme la
fonction de $\{n, N\}$, d\'ecrivant les fluctuations de la configuration
des trous noirs charg\'es non-supersym\'etriques aux corrections de 
quatri\`eme ordre de $\alpha^{\prime}$.} \label{nonsusy4NN8}\vspace*{0.5cm}
\end{figure}

\begin{figure}
\hspace*{1.0cm}\vspace*{-6.0cm}
\includegraphics[width=12.0cm,angle=-0]{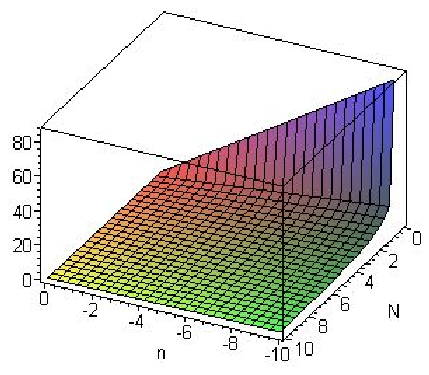}\vspace*{-3.0cm}
\caption{La composante $WW$ de la m\'etrique tenseure trac\'ee comme la
fonction de $\{n, N\}$, d\'ecrivant les fluctuations de la configuration
des trous noirs charg\'es non-supersym\'etriques aux corrections de 
quatri\`eme ordre de $\alpha^{\prime}$.} \label{nonsusy4WW10}\vspace*{0.5cm}
\end{figure}

\begin{figure}
\hspace*{1.0cm}\vspace*{-6.0cm}
\includegraphics[width=12.0cm,angle=-0]{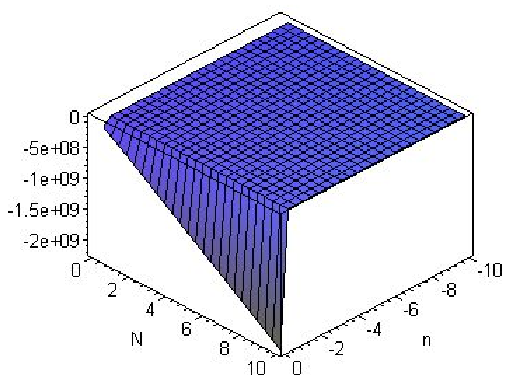}\vspace*{-3.0cm}
\caption{La composante $nw$ de la m\'etrique tenseure trac\'ee comme la
fonction de $\{n, N\}$, d\'ecrivant les fluctuations de la configuration
des trous noirs charg\'es non-supersym\'etriques aux corrections de 
quatri\`eme ordre de $\alpha^{\prime}$.} \label{nonsusy4nw2}\vspace*{0.5cm}
\end{figure}

\begin{figure}
\hspace*{1.0cm}\vspace*{-6.0cm}
\includegraphics[width=12.0cm,angle=-0]{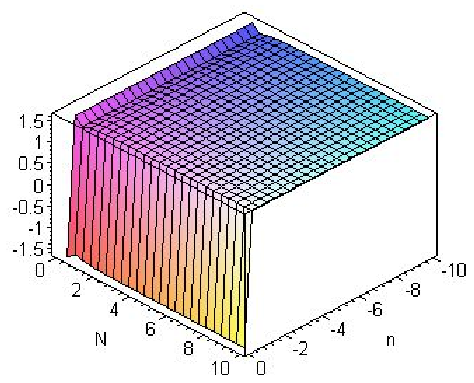}\vspace*{-3.0cm}
\caption{La composante $nN$ de la m\'etrique tenseure trac\'ee comme la
fonction de $\{n, N\}$, d\'ecrivant les fluctuations de la configuration
des trous noirs charg\'es non-supersym\'etriques aux corrections de 
quatri\`eme ordre de $\alpha^{\prime}$.} \label{nonsusy4nN3}\vspace*{0.5cm}
\end{figure}

\begin{figure}
\hspace*{1.0cm}\vspace*{-6.0cm}
\includegraphics[width=12.0cm,angle=-0]{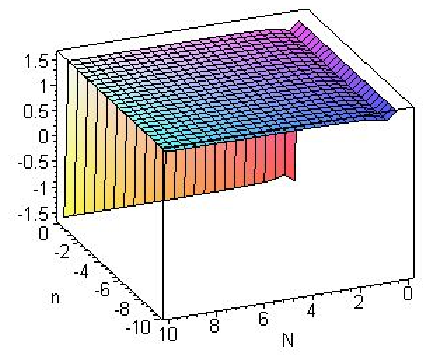}\vspace*{-3.0cm}
\caption{La composante $nW$ de la m\'etrique tenseure trac\'ee comme la
fonction de $\{n, N\}$, d\'ecrivant les fluctuations de la configuration
des trous noirs charg\'es non-supersym\'etriques aux corrections de 
quatri\`eme ordre de $\alpha^{\prime}$.} \label{nonsusy4nW4}\vspace*{0.5cm}
\end{figure}

\begin{figure}
\hspace*{1.0cm}\vspace*{-6.0cm}
\includegraphics[width=12.0cm,angle=-0]{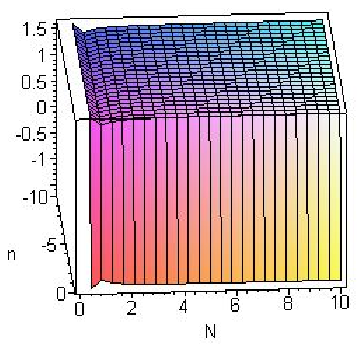}\vspace*{-3.0cm}
\caption{La composante $wN$ de la m\'etrique tenseure trac\'ee comme la
fonction de $\{n, N\}$, d\'ecrivant les fluctuations de la configuration
des trous noirs charg\'es non-supersym\'etriques aux corrections de 
quatri\`eme ordre de $\alpha^{\prime}$.} \label{nonsusy4wN6}\vspace*{0.5cm}
\end{figure}

\begin{figure}
\hspace*{1.0cm}\vspace*{-6.0cm}
\includegraphics[width=12.0cm,angle=-0]{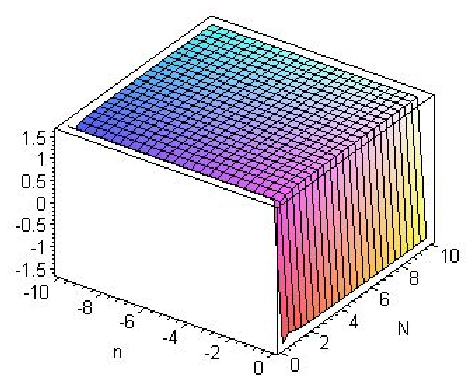}\vspace*{-3.0cm}
\caption{La composante $wW$ de la m\'etrique tenseure trac\'ee comme la
fonction de $\{n, N\}$, d\'ecrivant les fluctuations de la configuration
des trous noirs charg\'es non-supersym\'etriques aux corrections de 
quatri\`eme ordre de $\alpha^{\prime}$.} \label{nonsusy4wW7}\vspace*{0.5cm}
\end{figure}

\begin{figure}
\hspace*{1.0cm}\vspace*{-6.0cm}
\includegraphics[width=12.0cm,angle=-0]{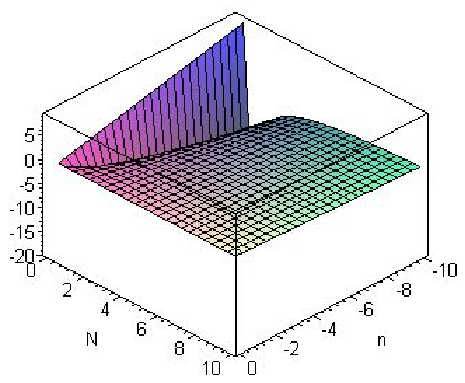}\vspace*{-3.0cm}
\caption{La composante $NW$ de la m\'etrique tenseure trac\'ee comme la
fonction de $\{n, N\}$, d\'ecrivant les fluctuations de la configuration
des trous noirs charg\'es non-supersym\'etriques aux corrections de 
quatri\`eme ordre de $\alpha^{\prime}$.} \label{nonsusy4NW9}\vspace*{0.5cm}
\end{figure}

Comme \`a la correction pr\'ec\'edente des d\'eriv\'ees sup\'erieure de troisi\`eme ordre de 
$\alpha^{\prime}$, pour la cas de $n= w$, $N= W$ et $\widehat{\alpha}= 0.1$, nous observons pour 
$n \in (-10, 0)$ et $N \in (0, 10)$ que l'amplitude des capacit\'es de chaleurs $\{g_{nn}, g_{ww}\}$ 
se prend la valeur maximale \`a l'ordre de $2 \times 10^+{09}$. Alors, les capacit\'es de chaleurs 
$\{g_{NN}, {g_ WW}\}$ prennent une valeur typique dans l'intervalle $(0, 90)$. Cela montre que 
les trous noirs non-supersym\'etriques corrig\'es par des corrections quatri\`eme ordre de 
$\alpha^{\prime}$ correspondent \`a une configuration statistique localement stable. En fait, 
les gammes de la croissance de la premi\`ere ensemble et celle la deuxi\`eme ensemble des capacit\'es 
de chaleurs se trouvent \^{e}tre dans la limite oppos\'ee des param\`etres $\{n, N\}$. 
Explicitement, par les Figs. (\ref{nonsusy4nn1}, \ref{nonsusy4ww5}), nous constatons que la croissance 
des capacit\'es de chaleurs $\{g_{nn}, g_{ww}\}$ a lieu dans la limite d'un grand $N$ et un petit $n$. 
D'autre part, les Figs. (\ref{nonsusy4NN8}, \ref{nonsusy4WW10}) montrent que la croissance des 
$\{g_{NN}, g_{WW}\}$ a lieu dans la limite d'un petit $N$ et un grand $n$. De m\^eme, les compressibilit\'es 
de chaleurs en comprenant deux param\`etres distincts de ces trous noirs sont repr\'esent\'ees dans les 
Figs. (\ref{nonsusy4nw2}, \ref{nonsusy4nN3}, \ref{nonsusy4nW4}, \ref{nonsusy4wN6}, \ref{nonsusy4wW7},
\ref{nonsusy4NW9}). Dans ce cas, on peut bien noter que la Fig. (\ref{nonsusy4NW9}) montre une 
caract\'eristique unique de la composante $NW$ des fluctuations statistiques des trous noirs
non-supersym\'etriques. Les composantes tel qui sont d\'efinies par les composantes de la m\'etrique 
tenseure de l'espace d'\'etat $\{g_{ij} \ | \ i, j = n, w, N, W\}$, le quatri\`eme ordre des corrections 
de $\alpha^{\prime}$ indiquent que les fluctuations, en impliquant les param\`etres $\{n, w\}$, 
prennent relativement des plus grande valeurs num\'eriques par rapport \`a celles qui sont d\'efinies 
ci-dessus en impliquant les param\`etres $\{N, W\}$.

De la m\'etrique tenseure donn\'ee ci-dessus, nous trouvons que 
les corrections de $\alpha^{\prime}$ \'a l'ordre quatre 
conduisent les mineurs principaux suivants

\begin{eqnarray}
\mathit{p_1} &=&  - {\displaystyle \frac{1}{65536}} \,
{\displaystyle \frac{\pi \,\sqrt{w}}{n^{(3/2)}\,N^{(7/2)}\,
\mathrm{W}^{(7/2)}}}\,\mathit{\tilde{p}_1}(\widehat{\alpha}), \nonumber \\ 
\mathit{p_2} &=& 0, \nonumber \\ 
\mathit{p_3} &=& {\displaystyle \frac{1}{70368744177664}}
\, \frac{\pi ^{3}}{N^{(25/2)}\,\mathrm{W}^{(21/2)}\,\sqrt{n}\,\sqrt{w}}
\mathit{\tilde{p}_3}(\widehat{\alpha}).
\end{eqnarray}

o\`u les fonctions $\{ \mathit{\tilde{p}_1}(\widehat{\alpha}), \mathit{\tilde{p}_3}(\widehat{\alpha}),
\mathit{\tilde{g}}(\widehat{\alpha}) \}$, comme les polyn\^{o}mes du param\`{e}tre $\widehat{\alpha}$, 
peuvent \^etre exprim\'es par

\begin{eqnarray}
\mathit{\tilde{p}_1}(\widehat{\alpha})&:=&
2237\,\widehat{\alpha}^{4}
+ 3808\,N\,\mathrm{W} \,\widehat{\alpha}^{3}
+ 7424\,N^{2}\,\mathrm{W}^{2}\,\widehat{\alpha}^{2} \nonumber \\ &&
- 20480\,N^{3}\,\mathrm{W}^{3}\,\widehat{\alpha}  
- 32768\,N^{4}\,\mathrm{W}^{4}, \nonumber \\
\mathit{\tilde{p}_3}(\widehat{\alpha})&:=&
548521976597\,\widehat{\alpha}^{12} 
+ 2267649190688\,N\,\mathrm{W}\,\widehat{\alpha}^{11}  \nonumber \\ && 
+ 6462387622656\,N^{2}\,\mathrm{W}^{2}\,\widehat{\alpha}^{10}
+ 3904169873408\,N^{3}\,\mathrm{W}^{3}\,\widehat{\alpha}^{9}  \nonumber \\ && 
- 9929322037248\,N^{4}\,\mathrm{W}^{4}\,\widehat{\alpha}^{8} 
- 36136556691456\,N^{5}\,\mathrm{W}^{5}\,\widehat{\alpha}^{7}  \nonumber \\ && 
- 27976292892672\,N^{6}\,\mathrm{W}^{6}\,\widehat{\alpha}^{6} 
- 15334106988544\,N^{7}\,\mathrm{W}^{7}\,\widehat{\alpha}^{5}  \nonumber \\ && 
- 19957639282688\,N^{8}\,\mathrm{W}^{8}\,\widehat{\alpha}^{4} 
- 55353538510848\,N^{9}\,\mathrm{W}^{9}\,\widehat{\alpha}^{3}  \nonumber \\ && 
- 26113401159680\,N^{10}\,\mathrm{W}^{10}\,\widehat{\alpha}^{2} 
+ 21990232555520\,N^{11}\,\mathrm{W}^{11}\,\widehat{\alpha}  \nonumber \\ && 
- 35184372088832\,N^{12}\,\mathrm{W}^{12}.  
\end{eqnarray}

On peut avoir facilement que le d\'eterminant de cette m\'etrique tenseure est:

\ba g= \frac {\pi^4}{115 2921 5046 0684 6976 (NW)^{16}} \tilde{g}(N,W),\ea

o\`u la fonction $ \tilde{g}(N,W)$, comme une fonction de $\widehat{\alpha}$ est d\'efinie par:

\ba
\tilde{g}(N,W)&:=& 
- a^{(4)}_0 (NW)^{16} 
+ a^{(4)}_1 (NW)^{15} \widehat{\alpha}
- a^{(4)}_2 (NW)^{14} \widehat{\alpha}^{2}
+ a^{(4)}_3 (NW)^{13} \widehat{\alpha}^{3} \nn && 
- a^{(4)}_4  (NW)^{12} \widehat{\alpha}^{4}
- a^{(4)}_5 (NW)^{11} \widehat{\alpha}^{5}
- a^{(4)}_6 (NW)^{10} \widehat{\alpha}^{6}
- a^{(4)}_7 (NW)^{8} \widehat{\alpha}^{7} \nn && 
- a^{(4)}_8 (NW)^{8} \widehat{\alpha}a^{8}
- a^{(4)}_9 (NW)^{7} \widehat{\alpha}^{9}
- a^{(4)}_{10} (NW)^{6} \widehat{\alpha}^{10}
- a^{(4)}_{11} (NW)^{5} \widehat{\alpha}^{11} \nn && 
+ a^{(4)}_{12} (NW)^{4} \widehat{\alpha}^{12}
- a^{(4)}_{13} (NW)^{3} \widehat{\alpha}a^{13}
+ a^{(4)}_{14} (NW)^{2} \widehat{\alpha}^{14} 
+ a^{(4)}_{15} (NW) \widehat{\alpha}^{15}  \nn && 
+ a^{(4)}_{16} \widehat{\alpha}^{16},\ea

o\`u les constantes r\'eels positifs $ \lbrace a^{(4)}_i \rbrace $
sont donn\'ees dans l'annex $[C]$. 

\`A la perspective des corrections de $\alpha^{\prime}$ de l'ordre quatre \`a
l'entropie des trous noirs non-supersym\'etriques, nous voyons que la configuration 
sous-jacente des trous noirs est assez stable dans les r\'egions o\`u les mineurs 
d'hypersurface et le d\'eterminant de la m\'etrique tenseure sont positifs.
Notamment, lorsque le param\`etre $\widehat{\alpha}$ des corrections est tel que les conditions 
suivantes sont saisfaites: (i) le polyn\^ome quartique $ \mathit{\tilde{p}_1}(\widehat{\alpha}) $
a un signe n\'egatif, (ii) les polyn\^ome $ \mathit{\tilde{p}_3}(\widehat{\alpha})$ de degr\'e douze
a un signe positif et (iii) le polyn\^ome $\mathit{\tilde{g}}(\widehat{\alpha})$ de degr\'e seize 
a signe positif. Pour tout $\widehat{\alpha}$ tels que les mineurs principaux $\mathit{\tilde{p}_1}(\widehat{\alpha}) <0$, 
$\mathit{\tilde{p}_3}(\widehat{\alpha}) >0$ et le d\'eterminant de la m\'etrique tenseure $\mathit{\tilde{g}}(\widehat{\alpha})>0$
sont satisfaites, la solution sous-jacente de ces trous noirs est relativement stable.
Ainsi, la stabilit\'e relative des trous noirs au-dessus, c'est-\`a-dire que la valeur sp\'ecifique 
du param\`etre $\widehat{\alpha}$, peut \^etre d\'etermin\'ee comme les racines communes des degr\'es 
duex, douze et seize des \'equations polyn\^omes. Notez cependant que la disparition de mineur de 
surface, c'est à dire $\mathit{p_2}= 0$ \'{e}vite la stabilit\'{e} compl\`{e}te de la configuration
sous-jacente de ces trous noirs. Donc, l'analyse de la g\'eom\'erire de l'espace d'\'etat 
montre que les corrections perturbative des d\'eriv\'ees sup\'erieures de $\alpha^{\prime}$ 
\`a l'entropie ne rendent pas la stabilit\'e statistiques compl\`ete \`a la configuration 
des trous noirs non-supersym\'etriques.

\begin{figure}
\hspace*{1.0cm}\vspace*{-6.0cm}
\includegraphics[width=12.0cm,angle=-0]{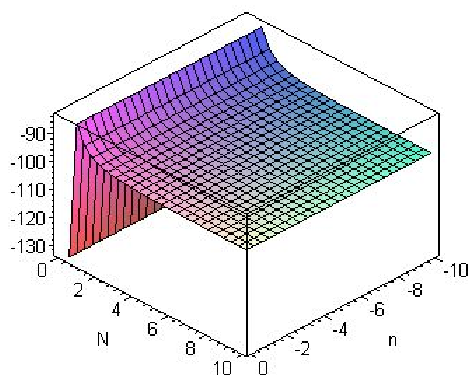}\vspace*{-3.0cm}
\caption{Le d\'eterminant de la m\'etrique tenseure trac\'ee comme la
fonction de $\{n, N\}$, d\'ecrivant les fluctuations de la configuration
des trous noirs charg\'es non-supersym\'etriques aux corrections de 
quatri\`eme ordre de $\alpha^{\prime}$.} \label{nonsusy4g}\vspace*{0.5cm}
\end{figure}

\begin{figure}
\hspace*{1.0cm}\vspace*{-6.0cm}
\includegraphics[width=12.0cm,angle=-0]{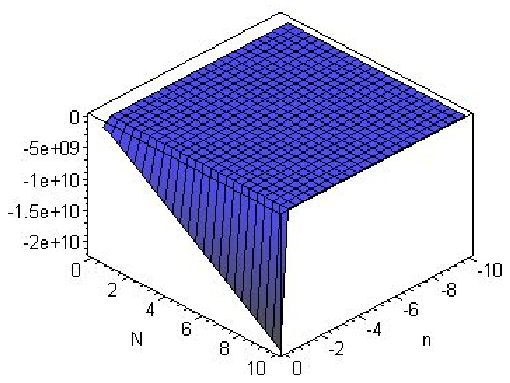}\vspace*{-3.0cm}
\caption{Le mineur d'hypersurface de la m\'etrique tenseure trac\'ee comme la
fonction de $\{n, N\}$, d\'ecrivant les fluctuations de la configuration
des trous noirs charg\'es non-supersym\'etriques aux corrections de 
quatri\`eme ordre de $\alpha^{\prime}$.} \label{nonsusy4minor3}\vspace*{0.5cm}
\end{figure}

\begin{figure}
\hspace*{1.0cm}\vspace*{-6.0cm}
\includegraphics[width=12.0cm,angle=-0]{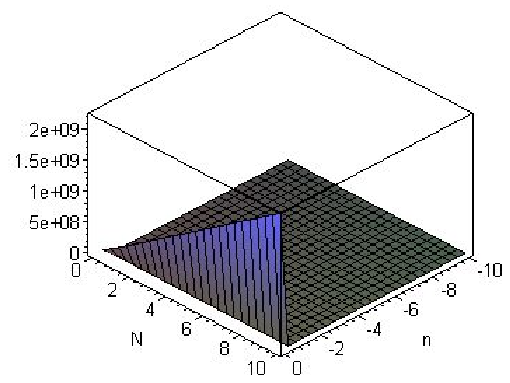}\vspace*{-3.0cm}
\caption{Le premier mineur de la m\'etrique tenseure trac\'ee comme la
fonction de $\{n, N\}$, d\'ecrivant les fluctuations de la configuration
des trous noirs charg\'es non-supersym\'etriques aux corrections de 
quatri\`eme ordre de $\alpha^{\prime}$.} \label{nonsusy4minor1}\vspace*{0.5cm}
\end{figure}

Comme une fonction de $\{n, N\}$, la condition de stabilit\'e d'un ensemble des trous noirs
non-supersym\'etriques corrig\'es au quatri\`eme ordre de $\alpha^{\prime}$ d\'ecoule 
de la positivit\'e du d\'eterminant de la m\'etrique tenseure. Dans ce cas, nous voyons 
que le d\'eterminant de la m\'etrique tenseure tend d'une fa\c{c}on g\'en\'erique \`a une valeur 
n\'egative. Pour une valeur typique de $n \in (-10, 0)$ et $N \in (0, 10) $, la Fig. (\ref{nonsusy4g})
montre que le d\'eterminant de la m\'etrique tenseure r\'eside dans l'intervalle $(-80, -140)$. 
Par la pr\'esente, pour $N \in (2, 10) $, nous voyons que le d\'eterminant de la m\'etrique tenseure 
a une valeur approximative de $-97$. Dans l'intervalle $N \in (0, 2) $, le d\'eterminant de la 
m\'etrique tenseure \`a tout d'abord augmente \`a une valeur de $-80$ et puis il r\'eduit fortement 
\`a une valeur de $-140$. La stabilit\'e correspondante de l'hypersurface d\'efinie par une valeur 
constante de $W$ est montr\'e dans la Fig. (\ref{nonsusy4minor3}). Dans cette gamme de $\{n, N\}$, 
nous voyons que le mineur $p_3$ r\'eside dans la gamme de $(-2 \times 10^{ +10}, 0)$. 
Dans la limite d'un petit $n$, nous constatons que la n\'egativit\'e de $p_3$ augmente 
\`a mesure que la valeur de $N$ est pass\'ee de z\'ero \`a $10$. Par ailleurs, la surface 
d\'efinie par les fluctuations de $\{n, w\}$ est instable par-ce que les mineurs principe 
correspondant dispara\^{i}t identiquement, \`a savoir que nous avons $p_2= 0$. Lorsque le seul 
param\`etre $n$ est autoris\'e \`a varier, la stabilit\'e de la configuration est d\'etermin\'ee
par la positivit\'e du premier mineur principe $p_1:= g_{nn}$. Dans ce cas, pour un petit $n$ et
un grand $N$, on peut observer par la Fig. (\ref{nonsusy4minor1}) que $p_1$ a une amplitude 
positive \`a l'ordre de $10^{+10}$. En fait, pour une petite valeur de $n$, le premier mineur
principe $p_1$ augmente \`a mesure que la fonction de $N$. Ainsi, les descriptions graphiques 
ci-dessus des mineurs principaux de l'espace d'\'etat fournent la notion qualitative de la 
stabilit\'e statistique des trous noirs non-supersym\'etriques sous les quatri\`eme ordre 
des corrections de $\alpha^{\prime}$.

En fait, comme nous avons fourni les $\Gamma_{abc}$ dans l'annex $[A]$, 
il n'est pas tr\`es difficile de voir que la courbure scalaire de Ruppenier est:

\ba  R= -\frac{49152}{\pi}\sqrt{\frac{(NW)^7}{nw}} \frac{r(N,W)}{\tilde{g}(N,W)^3},\ea

o\`u la fonction $r(N,W)$ est d\'efinie par:


\ba r(N,W)&:=&
b^{(4)}_0 (NW)^{44}  
- b^{(4)}_1 (NW)^{43} \widehat{\alpha}  
+ b^{(4)}_2 (NW)^{42} \widehat{\alpha}^{2} 
- b^{(4)}_3 (NW)^{41} \widehat{\alpha}^{3}  \nn && 
+ b^{(4)}_4 (NW)^{40} \widehat{\alpha}^{4}
+ b^{(4)}_5 (NW)^{39} \widehat{\alpha}^{5} 
+ b^{(4)}_6 (NW)^{38} \widehat{\alpha}^{6} 
+ b^{(4)}_7 (NW)^{37} \widehat{\alpha}^{7}  \nn && 
+ b^{(4)}_8 (NW)^{36} \widehat{\alpha}^{8}
+ b^{(4)}_9 (NW)^{35} \widehat{\alpha}^{9} 
+ b^{(4)}_{10} (NW)^{34} \widehat{\alpha}^{10}
+ b^{(4)}_{11} (NW)^{33} \widehat{\alpha}^{11} \nn && 
+ b^{(4)}_{12} (NW)^{32} \widehat{\alpha}^{12}
+ b^{(4)}_{13} (NW)^{31} \widehat{\alpha}^{13}
+ b^{(4)}_{14} (NW)^{30} \widehat{\alpha}^{14}
+ b^{(4)}_{15} (NW)^{29} \widehat{\alpha}^{15} \nn && 
+ b^{(4)}_{16} (NW)^{28} \widehat{\alpha}^{16}
+ b^{(4)}_{17} (NW)^{27}\widehat{\alpha}^{17}
+ b^{(4)}_{18} (NW)^{26} \widehat{\alpha}^{18}
+ b^{(4)}_{19} (NW)^{25} \widehat{\alpha}^{19} \nn && 
+ b^{(4)}_{20} (NW)^{24} \widehat{\alpha}^{20}
+ b^{(4)}_{21} (NW)^{23} \widehat{\alpha}^{21}
+ b^{(4)}_{22} (NW)^{22} \widehat{\alpha}^{22}
+ b^{(4)}_{23} (NW)^{21} \widehat{\alpha}^{23} \nn && 
+ b^{(4)}_{24} (NW)^{20} \widehat{\alpha}^{24}
+ b^{(4)}_{25} (NW)^{19} \widehat{\alpha}^{25}
+ b^{(4)}_{26} (NW)^{18} \widehat{\alpha}^{26}
+ b^{(4)}_{27} (NW)^{17} \widehat{\alpha}^{27} \nn && 
+ b^{(4)}_{28} (NW)^{16} \widehat{\alpha}^{28}
+ b^{(4)}_{29} (NW)^{15} \widehat{\alpha}^{29}
- b^{(4)}_{30} (NW)^{14} \widehat{\alpha}^{30}
- b^{(4)}_{31} (NW)^{13} \widehat{\alpha}^{31} \nn && 
- b^{(4)}_{32} (NW)^{12} \widehat{\alpha}^{32}
- b^{(4)}_{33} (NW)^{11} \widehat{\alpha}^{33}
- b^{(4)}_{34} (NW)^{10} \widehat{\alpha}^{34}
- b^{(4)}_{35} (NW)^{9} \widehat{\alpha}^{35} \nn && 
- b^{(4)}_{36} (NW)^{8} \widehat{\alpha}^{36}
+ b^{(4)}_{37} (NW)^{7} \widehat{\alpha}^{37}
+ b^{(4)}_{38} (NW)^{6} \widehat{\alpha}^{38}
+ b^{(4)}_{39} (NW)^{5} \widehat{\alpha}^{39} \nn && 
+ b^{(4)}_{40} (NW)^{4} \widehat{\alpha}^{40}
+ b^{(4)}_{41} (NW)^{3} \widehat{\alpha}^{41}
+ b^{(4)}_{42} (NW)^{2} \widehat{\alpha}^{42}
+ b^{(4)}_{43} (NW) \widehat{\alpha}^{43} \nn && 
+ b^{(4)}_{44} \widehat{\alpha}^{44},\ea

o\`u les constantes r\'eelles positives $ \lbrace a^{(4)}_i \rbrace $ et
$ \lbrace b^{(4)}_i \rbrace $ sont donn\'ees dans l'annex $[C]$.

\begin{figure}
\hspace*{1.0cm}\vspace*{-6.0cm}
\includegraphics[width=12.0cm,angle=-0]{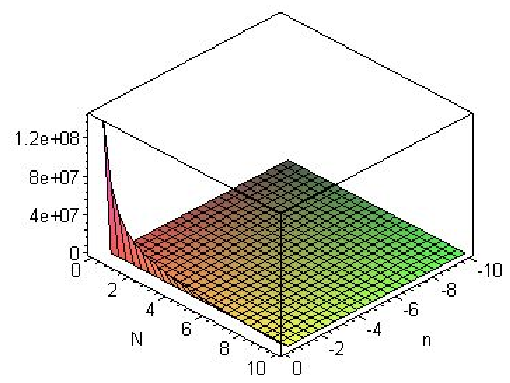}\vspace*{-3.0cm}
\caption{La courbure scalaire trac\'ee comme la fonction de $\{n, N\}$, d\'ecrivant 
les fluctuations de la configuration des trous noirs charg\'es non-supersym\'etriques 
aux corrections de quatri\`eme ordre de $\alpha^{\prime}$ dans la gamme 
$n \in (-10, 0)$ et $N \in (0,10) $.} \label{nonsusy4R}\vspace*{0.5cm}
\end{figure}

\begin{figure}
\hspace*{1.0cm}\vspace*{-6.0cm}
\includegraphics[width=12.0cm,angle=-0]{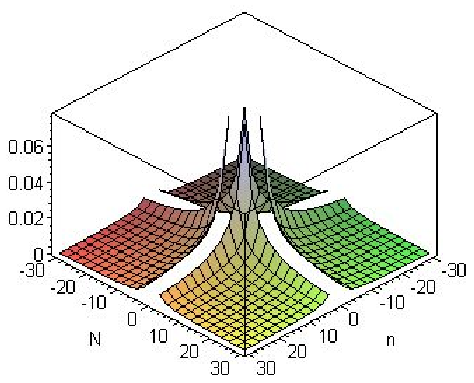}\vspace*{-3.0cm}
\caption{La courbure scalaire trac\'ee comme la fonction de $\{n, N\}$, d\'ecrivant 
les fluctuations de la configuration des trous noirs charg\'es non-supersym\'etriques 
aux corrections de quatri\`eme ordre de $\alpha^{\prime}$ dans la gamme 
$n, N \in (-30,30)$.} \label{nonsusy4R30}\vspace*{0.5cm}
\end{figure}

\`A quatri\`eme ordre des corrections de d\'eriv\'ee sup\'erieures de $\alpha^{\prime}$,
les propri\'et\'es de la stabilit\'e globale des trous noirs non-supersym\'etriques sont 
pr\'esent\'es dans les . Fig. (\ref{nonsusy4R}, \ref{nonsusy4R30}). Dans la gamme de 
$n \in (-10, 0)$ et $N \in (0, 10)$, il r\'esulte de la Fig. (\ref{nonsusy4R}) que la 
courbure scalaire de l'espace d'\'etat acquiert un pic \`a l'ordre de $10^{+08}$ dans la 
limite d'une petite valeur de $n, N$. En fait, dans la limite d'un petit $n$, la Fig. (\ref{nonsusy4R}) 
montre que la courbure scalaire diminue \`a mesure que la fonction de $N$. En particulier,
dans la cas de petite valeur de $\{n, N\}$, la Fig. (\ref{nonsusy4R}) montre que la courbure 
scalaire a une grande amplitude positive \`a l'ordre de $10^{+08}$. Ainsi, on peut noter que 
la configuration des trous noirs sous-tendents est un syst\`eme statistique fortement interagissant
dans cette limite des charges. Comme mentionn\'e dans les cas pr\'ec\'edents, le quatri\`eme ordre 
des corrections de d\'eriv\'es sup\'erieures de $\alpha^{\prime}$ rend le signe positif \`a la 
courbure scalaire. Cela signifie que la configuration statistique de ces trous noirs a des 
interactions r\'epulsives. La Fig. (\ref{nonsusy4R30}) illustre la nature de la courbure 
scalaire ci-dessus dans une range \'egale de $n, N \in (-30,30)$. \`A savoir, quand $n$ et 
$N$ prennent des valeurs en multiple de $3$, par exemple $(-30, 30)$, alors que nous voyons 
de la Fig. (\ref{nonsusy4R30}) qu'il y a quatre r\'egions disjointes similaire des interactions
statistiques globales. En comparaison des interactions apparaissant dans la gamme de $n \in (-10,0)$ 
et $N \in (0, 10) $, l'amplitude des interactions demeure \`a une beaucoup plus petite valeur dans la 
gamme de $n, N \in (-30,30)$. Le ration de la valeur typique de l'amplitude de ces courbures scalaires
semble \^{e}tre \`a l'ordre de $10^{-10}$. La vue graphique de cette comparaison est montr\'e dans
les Figs. (\ref{nonsusy4R}, \ref{nonsusy4R30}). Qualitativement, dans le cas des petites valeurs 
de $\{n, N\}$, les repr\'esentations graphiques ci-dessus de la courbure scalaire indiquent que les 
trous noirs non-supersym\'etriques sont globalement une configuration statistique instable 
bien qu'on aie ajout\'e le quatri\`eme ordre des corrections de $\alpha^{\prime}$.

On voit clairement que la conclusion de la courbure scalaire de Ruppenier 
reste la m\^eme ce que nous avons annonc\'e avant dans le cas de $(\alpha^{\prime})^3$.
On peut aussi facilement continuer ce type de calcul et peut g\'en\'eralement montrer 
que cette conclusion reste la m\^eme, si bien qu'on ajoute les corrections suivantes 
de l'ordre fini de $ \alpha^{\prime} $ \`a l'entropie des trous noirs dyoniques 
non-supersym\'etriques extr\'emaux en quatre dimensions de l'espace-temps. 
\`A la fin de ce chapitre, nous allons montrer ce r\'esultat exactement.

De toute mani\`ere,
sur la base d'un sch\'ema g\'en\'eral ce que nous avons vu au-dessus
aux ordres diff\'erents des corrections de $ \alpha^{\prime} $ \`a la 
g\'eom\'etrie thermodynamique de Ruppenier, comme le d\'eterminant de la 
m\'etrique tenseure et la courbure scalaire, des trous noirs dyoniques 
non-supersym\'etriques extr\'emaux en quatre dimensions, nous pouvons 
voir maintenant qu'elle satisfait l'observation suivante.

\section{La nature de la courbure scalaire de Ruppenier 
d'apr\`es les corrections de $\alpha^{\prime} $ des trous noirs dyoniques
extr\'emaux non-supersym\'etriques en quatre dimensions.}

Dans cette section, nous allons obtenir la nature des corrections 
de $\alpha^{\prime} $ \`{a} la g\`{e}om\`{e}trie d'espace de l'\'{e}tas.
En g\'en\'erale, soit $ l $ le plus grand exposant de $ \alpha^{\prime} $ 
dans l'entropie des trous noirs dyoniques non-supersym\'etriques extr\'emaux 
en quatre dimensions. Alors $ \forall l >1 $, nous pouvons \'ecrire que 
la courbure de Ruppenier d'un trou noir dyonique non-supersym\'etrique 
extr\'emal en quatre dimensions peut \^etre donn\'ee par une formule g\'en\'erale:

\ba R^{(l)}(n,w,N,W)=-\frac{k}{\pi} \frac{(NW)^l}{\sqrt{nwNW}} \frac{f_2(NW)}{f_1(NW)^{3}},\ea

o\`u $ k $ est une constante r\'eelle et le degr\'e des fonctions
polynomiales $ f_1(NW) $ et $ f_2(NW) $ sont d\'etermin\'ees par
le plus grand exposant de $ \alpha^{\prime} $ se figurant \`a
l'entropie de ce trou noir. Pour une r\'epr\'esentation donn\'ee 
des fonctions $\{ f_i\ | i= 1,2 \}$, soient les degr\'es de 
$ f_1(NW) $ et $ f_2(NW) $ sont respectivement $ l_1 $ et 
$ l_2 $, c'est-\`a-dire que nous avons

\ba deg(f_1)&=& l_1,\nn
deg(f_2)&=& l_2 \ea

tels que la courbure scalaire de Ruppenier sans les corrections
des d\'eriv\'ees sup\'erieures de $ \alpha^{\prime} $ peut \^etre
\'ecrite proportionnelle \`a 

\ba R\sim (nwNW)^{-1/2}.\ea

C'est-\`a-dire que, quand on n'ajoute pas les corrections
de $ \alpha^{\prime} $, la courbure scalaire de Ruppenier
doit \^etre comme: 

\ba R= 3/(2 \pi \sqrt{nwNW}).\ea

Ensuite, dans le cas pr\'esent, il s'av\`ere que nous avons: 

\ba l+ l_2- 3l_1= 0.\ea

Nous pouvons observer \`a partir des courbures de Ruppenier
ce qui pr\'ec\`edent que le d\'egre de la fonction $f_1$ est donn\'e
par la r\'elation 

\ba l_1= 4l.\ea

Afin que nous ayons une formule simple pour le degr\'e
de la fonction $ f_1 $, nous voyons que le degr\'e de la
fonction $f_2$ est justement donn\'ee par comme la suivante: 

\ba l_2= 11 l.\ea

Donc, comme une fonction de $NW$, le degr\'e des fonctions $f_i$ sont:

\ba deg(f_1)= 4 l, \ \ deg(f_2)= 11 l.\ea


De cette fa\c{c}on, \`a tout ordre des corrections de $ \alpha^{\prime} $
\`a l'entropie des trous noirs non-supersym\'etriques extr\'emaux,
nous pouvons  d\'eterminer facilement toutes les propri\'et\'es 
de l'espace d'\'etat de la g\'eom\'etrie thermodynamique.
De plus, $\forall l \ge 2$, le degr\'e du d\'eterminant de la m\'etrique 
de Ruppenier et celui de la num\'erateur de la courbure scalaire correspondante 
sont respectivement $4l$ et $11l$. Notez bien aussi que notre observation n'est pas 
valable pour le cas de $l= 1$. En fait, $ \forall \widehat{\alpha} > 0$, nous pouvons 
voir dans le cas de $ l = 1 $ que la configuration sous-jacente de ces trous noirs 
n'est pas stable par-ce que les nombres $ N,W $ sont positives, comme nous les avons
consid\'er\'es thermodynamique g\'eom\'etriquement que l'entropie d'un trou noir 
doit \^etre corrig\'ee, par-ce qu'un trou noir extr\'emal peut \^etre vu comme une 
collection finie des \'etats quantiques de BPS \cite{AshokeSen}, qui est un objet 
assez stable par rapport \`a la th\'eorie des cordes.

En fait, pour tout ordre arbitraire $ l >1$, les corrections de $\alpha^{\prime}$ 
\`a l'entropie des trous noirs non-supersym\'etriques extr\'emaux en quatre dimensions, 
la g\'eom\'etrie de Ruppenier est bien d\'efinie et partout r\'eguli\`ere, $\forall f_1(N,W) \ne 0$. 
Donc, \`a la conclusion, notre observation est comme une conjoncture
ce qu'elle peuisse \^etre \'enonc\'ee comme la suivante.

\subsection*{L'observation}

\`A tout ordre $l$ grand qu'un des corrections de $ \alpha^{\prime}$,
la courbure scalaire de Ruppenier des trous noirs dyoniques
non-supersym\'etriques extr\'emaux en quatre dimensions 
est donn\'ee par une formule g\'en\'erale, comme la suivante: 

\ba R^{(l)}(n,w,N,W)= 
-\frac{k}{\pi} \frac{(NW)^l}{\sqrt{nwNW}} \frac{f_2(NW)}{f_1(NW)^{3}},\ \ k \in R, \ea

o\`u les degr\'es de $ f_1(NW) $ et $ f_2(NW) $ sont 
respectivement $4l$ et $11 l$.

\section{\`A l'ordre arbitraire de $\alpha^{\prime} $}

\`A la perspective de cet examen de la g\'eom\'etrie thermodynamique, 
ce que nous l'avons expliqu\'ee ci-dessus au vers des corrections de $\alpha^{\prime}$, 
il s'av\`ere en g\'en\'eral que l'entropie des trous noirs nonsupersymmetriques 
\cite{AshokeSen} peut \^etre exprim\'ee comme 

\begin{eqnarray}
\mathrm{S}(n, \,w, \,N, \,\mathrm{W})&=& \pi \,\sqrt{n\,w\,N\,
\mathrm{W}}\, \left(  \! {\displaystyle \sum _{k=0}^{r}} \,{c_{k}
}\,({\displaystyle \frac {\widehat{\alpha}}{N\,\mathrm{W}}} )^{k} \!  \right) 
\end{eqnarray}

En g\'en\'eral, les composantes de la m\'etrique tenseure sont

\begin{eqnarray}
{g_{\mathit{nn}}}&=&{\displaystyle \frac {1}{4}} \,{\displaystyle 
\frac {\pi \,w\,N\,\mathrm{W}\, \left(  \! {\displaystyle \sum _{
k=0}^{r}} \,{c_{k}}\,({\displaystyle \frac {\widehat{\alpha}}{N\,\mathrm{W}}} )
^{k} \!  \right) }{n\,\sqrt{n\,w\,N\,\mathrm{W}}}} , \nonumber \\
{g_{\mathit{nw}}}&=& - {\displaystyle \frac {1}{4}} \,
{\displaystyle \frac {\pi \, \left(  \! {\displaystyle \sum _{k=0
}^{r}} \,{c_{k}}\,({\displaystyle \frac {\widehat{\alpha}}{N\,\mathrm{W}}} )^{k}
 \!  \right) \,N\,\mathrm{W}}{\sqrt{n\,w\,N\,\mathrm{W}}}}, \nonumber  \\
{g_{\mathit{nN}}}&=& - {\displaystyle \frac {1}{4}} \,
{\displaystyle \frac {\pi \,w\,\mathrm{W}\, \left(  \!  \left( 
 \! {\displaystyle \sum _{k=0}^{r}} \,{c_{k}}\,({\displaystyle 
\frac {\widehat{\alpha}}{N\,\mathrm{W}}} )^{k} \!  \right)  - 2\, \left(  \! 
{\displaystyle \sum _{k=0}^{r}} \,{c_{k}}\,({\displaystyle 
\frac {\widehat{\alpha}}{N\,\mathrm{W}}} )^{k}\,k \!  \right)  \!  \right) }{
\sqrt{n\,w\,N\,\mathrm{W}}}}, \nonumber  \\
{g_{\mathit{nW}}}&=& - {\displaystyle \frac {1}{4}} \,
{\displaystyle \frac {\pi \,w\,N\, \left(  \!  \left(  \! 
{\displaystyle \sum _{k=0}^{r}} \,{c_{k}}\,({\displaystyle 
\frac {\widehat{\alpha}}{N\,\mathrm{W}}} )^{k} \!  \right)  - 2\, \left(  \! 
{\displaystyle \sum _{k=0}^{r}} \,{c_{k}}\,({\displaystyle 
\frac {\widehat{\alpha}}{N\,\mathrm{W}}} )^{k}\,k \!  \right)  \!  \right) }{
\sqrt{n\,w\,N\,\mathrm{W}}}}, \nonumber  \\
{g_{\mathit{ww}}}&=& {\displaystyle \frac {1}{4}} \,{\displaystyle 
\frac {\pi \,n\,N\,\mathrm{W}\, \left(  \! {\displaystyle \sum _{
k=0}^{r}} \,{c_{k}}\,({\displaystyle \frac {\widehat{\alpha}}{N\,\mathrm{W}}} )
^{k} \!  \right) }{w\,\sqrt{n\,w\,N\,\mathrm{W}}}}, 
\end{eqnarray}
\begin{eqnarray}
{g_{\mathit{wN}}}&=& - {\displaystyle \frac {1}{4}} \,
{\displaystyle \frac {\pi \,n\,\mathrm{W}\, \left(  \!  \left( 
 \! {\displaystyle \sum _{k=0}^{r}} \,{c_{k}}\,({\displaystyle 
\frac {\widehat{\alpha}}{N\,\mathrm{W}}} )^{k} \!  \right)  - 2\, \left(  \! 
{\displaystyle \sum _{k=0}^{r}} \,{c_{k}}\,({\displaystyle 
\frac {\widehat{\alpha}}{N\,\mathrm{W}}} )^{k}\,k \!  \right)  \!  \right) }{
\sqrt{n\,w\,N\,\mathrm{W}}}}, \nonumber  \\
{g_{\mathit{wW}}}&=& - {\displaystyle \frac {1}{4}} \,
{\displaystyle \frac {\pi \,n\,N\, \left(  \!  \left(  \! 
{\displaystyle \sum _{k=0}^{r}} \,{c_{k}}\,({\displaystyle 
\frac {\widehat{\alpha}}{N\,\mathrm{W}}} )^{k} \!  \right)  - 2\, \left(  \! 
{\displaystyle \sum _{k=0}^{r}} \,{c_{k}}\,({\displaystyle 
\frac {\widehat{\alpha}}{N\,\mathrm{W}}} )^{k}\,k \!  \right)  \!  \right) }{
\sqrt{n\,w\,N\,\mathrm{W}}}}, \nonumber  \\
{g_{\mathit{NN}}}&=& - {\displaystyle \frac {1}{4}} \,
{\displaystyle \frac {\pi \,n\,w\,\mathrm{W}\, \left(  \!  - 
 \left(  \! {\displaystyle \sum _{k=0}^{r}} \,{c_{k}}\,(
{\displaystyle \frac {\widehat{\alpha}}{N\,\mathrm{W}}} )^{k} \!  \right)  - 4\,
 \left(  \! {\displaystyle \sum _{k=0}^{r}} \,{c_{k}}\,(
{\displaystyle \frac {\widehat{\alpha}}{N\,\mathrm{W}}} )^{k}\,k \!  \right)  + 
4\, \left(  \! {\displaystyle \sum _{k=0}^{r}} \,{c_{k}}\,(
{\displaystyle \frac {\widehat{\alpha}}{N\,\mathrm{W}}} )^{k}\,k\,(k + 1) \! 
 \right)  \!  \right) }{\sqrt{n\,w\,N\,\mathrm{W}}\,N}},\nonumber \\
{g_{\mathit{NW}}}&=& - {\displaystyle \frac {1}{4}} \,
{\displaystyle \frac {\pi \,n\,w\, \left(  \!  \left(  \! 
{\displaystyle \sum _{k=0}^{r}} \,{c_{k}}\,({\displaystyle 
\frac {\widehat{\alpha}}{N\,\mathrm{W}}} )^{k} \!  \right)  - 4\, \left(  \! 
{\displaystyle \sum _{k=0}^{r}} \,{c_{k}}\,({\displaystyle 
\frac {\widehat{\alpha}}{N\,\mathrm{W}}} )^{k}\,k \!  \right)  + 4\, \left(  \! 
{\displaystyle \sum _{k=0}^{r}} \,{c_{k}}\,({\displaystyle 
\frac {\widehat{\alpha}}{N\,\mathrm{W}}} )^{k}\,k^{2} \!  \right)  \!  \right) 
}{\sqrt{n\,w\,N\,\mathrm{W}}}}, \nonumber  \\
{g_{\mathit{WW}}}&=& - {\displaystyle \frac {1}{4}} \,
{\displaystyle \frac {\pi \,n\,w\,N\, \left(  \!  -  \left(  \! 
{\displaystyle \sum _{k=0}^{r}} \,{c_{k}}\,({\displaystyle 
\frac {\widehat{\alpha}}{N\,\mathrm{W}}} )^{k} \!  \right)  - 4\, \left(  \! 
{\displaystyle \sum _{k=0}^{r}} \,{c_{k}}\,({\displaystyle 
\frac {\widehat{\alpha}}{N\,\mathrm{W}}} )^{k}\,k \!  \right)  + 4\, \left(  \! 
{\displaystyle \sum _{k=0}^{r}} \,{c_{k}}\,({\displaystyle 
\frac {\widehat{\alpha}}{N\,\mathrm{W}}} )^{k}\,k\,(k + 1) \!  \right)  \! 
 \right) }{\sqrt{n\,w\,N\,\mathrm{W}}\,\mathrm{W}}}.
\end{eqnarray}

Nous voyons que les mineurs principaux
$\{ \mathit{p_1}, \mathit{p_2}, \mathit{p_3} \}$ 
sont donn\'es par les expressions 

\begin{eqnarray}
\mathit{p_1} &=& {\displaystyle \frac {1}{4}} \,{\displaystyle 
\frac {\pi \,w\,N\,\mathrm{W}\, \left(  \! {\displaystyle \sum _{
k=0}^{r}} \,{c_{k}}\,({\displaystyle \frac {\widehat{\alpha}}{N\,\mathrm{W}}} )
^{k} \!  \right) }{n\,\sqrt{n\,w\,N\,\mathrm{W}}}},  \nonumber \\ 
\mathit{p2} &=& 0, \nonumber \\ 
\mathit{p_3} &=&  - {\displaystyle \frac {1}{16}} \pi ^{3}\,
\mathrm{W}^{2}\, \left(  \! {\displaystyle \sum _{k=0}^{r}} \,{c
_{k}}\,({\displaystyle \frac {\widehat{\alpha}}{N\,\mathrm{W}}} )^{k} \! 
 \right)  \left( {\vrule height1.44em width0em depth1.44em}
 \right. \!  \!  \left(  \! {\displaystyle \sum _{k=0}^{r}} \,{c
_{k}}\,({\displaystyle \frac {\widehat{\alpha}}{N\,\mathrm{W}}} )^{k} \! 
 \right) ^{2}  \nonumber \\ &&
\mbox{} - 4\, \left(  \! {\displaystyle \sum _{k=0}^{r}} \,{c_{k}
}\,({\displaystyle \frac {\widehat{\alpha}}{N\,\mathrm{W}}} )^{k} \!  \right) \,
 \left(  \! {\displaystyle \sum _{k=0}^{r}} \,{c_{k}}\,(
{\displaystyle \frac {\widehat{\alpha}}{N\,\mathrm{W}}} )^{k}\,k \!  \right) 
\nonumber \\ && + 4\, \left(  \! {\displaystyle \sum _{k=0}^{r}} \,{c_{k}}\,(
{\displaystyle \frac {\widehat{\alpha}}{N\,\mathrm{W}}} )^{k}\,k \!  \right) ^{2
} \! \! \left. {\vrule height1.44em width0em depth1.44em}
\right)  \left/ {\vrule height0.41em width0em depth0.41em}
\right. \!  \! \sqrt{n\,w\,N\,\mathrm{W}}.
\end{eqnarray}

De plus, il n'est pas difficile de voir que le d\'eterminant 
de la m\'etrique tenseure est donn\'e par

\begin{eqnarray}
g&=&{\displaystyle \frac {1}{16}} \pi ^{4}\, g_1 g_3
\end{eqnarray}

o\`u la fonction $g_1$ est la somme 

\begin{eqnarray}
g_1&=& 
{\displaystyle \sum _{k=0}^{r}} \,{c_{k}}\,(
{\displaystyle \frac {\widehat{\alpha}}{N\,\mathrm{W}}} )^{k} 
\end{eqnarray}

et la fonction $g_3$ est la produit des trois sommes 

\begin{eqnarray}
g_3&=& 
-  \left(  \! {\displaystyle \sum _{k=0}^{r}} \,{c_{k}}\,({\displaystyle 
\frac {\widehat{\alpha}}{N\,\mathrm{W}}} )^{k} \!  \right) ^{3} + 4\, \left( 
 \! {\displaystyle \sum _{k=0}^{r}} \,{c_{k}}\,({\displaystyle 
\frac {\widehat{\alpha}}{N\,\mathrm{W}}} )^{k} \!  \right) ^{2}\, \left(  \! 
{\displaystyle \sum _{k=0}^{r}} \,{c_{k}}\,({\displaystyle 
\frac {\widehat{\alpha}}{N\,\mathrm{W}}} )^{k}\,k \!  \right)   \nonumber \\ &&
\mbox{} + 2\, \left(  \! {\displaystyle \sum _{k=0}^{r}} \,{c_{k}
}\,({\displaystyle \frac {\widehat{\alpha}}{N\,\mathrm{W}}} )^{k} \!  \right) ^{
2}\, \left(  \! {\displaystyle \sum _{k=0}^{r}} \,({c_{k}}\,(
{\displaystyle \frac {\widehat{\alpha}}{N\,\mathrm{W}}} )^{k}\,k^{2} + {c_{k}}\,
({\displaystyle \frac {\widehat{\alpha}}{N\,\mathrm{W}}} )^{k}\,k) \!  \right) 
 \nonumber \\ && \mbox{} 
- 8\, \left(  \! {\displaystyle \sum _{k=0}^{r}} \,{c_{k}
}\,({\displaystyle \frac {\widehat{\alpha}}{N\,\mathrm{W}}} )^{k} \!  \right) \,
 \left(  \! {\displaystyle \sum _{k=0}^{r}} \,{c_{k}}\,(
{\displaystyle \frac {\widehat{\alpha}}{N\,\mathrm{W}}} )^{k}\,k \!  \right) \,
 \left(  \! {\displaystyle \sum _{k=0}^{r}} \,({c_{k}}\,(
{\displaystyle \frac {\widehat{\alpha}}{N\,\mathrm{W}}} )^{k}\,k^{2} + {c_{k}}\,
({\displaystyle \frac {\widehat{\alpha}}{N\,\mathrm{W}}} )^{k}\,k) \!  \right) 
  \nonumber \\ &&
\mbox{} + 8\, \left(  \! {\displaystyle \sum _{k=0}^{r}} \,{c_{k}
}\,({\displaystyle \frac {\widehat{\alpha}}{N\,\mathrm{W}}} )^{k} \!  \right) \,
 \left(  \! {\displaystyle \sum _{k=0}^{r}} \,{c_{k}}\,(
{\displaystyle \frac {\widehat{\alpha}}{N\,\mathrm{W}}} )^{k}\,k \!  \right) \,
 \left(  \! {\displaystyle \sum _{k=0}^{r}} \,{c_{k}}\,(
{\displaystyle \frac {\widehat{\alpha}}{N\,\mathrm{W}}} )^{k}\,k^{2} \! 
 \right)   \nonumber \\ &&
\mbox{} + 8\, \left(  \! {\displaystyle \sum _{k=0}^{r}} \,{c_{k}
}\,({\displaystyle \frac {\widehat{\alpha}}{N\,\mathrm{W}}} )^{k}\,k \! 
 \right) ^{2}\, \left(  \! {\displaystyle \sum _{k=0}^{r}} \,({c
_{k}}\,({\displaystyle \frac {\widehat{\alpha}}{N\,\mathrm{W}}} )^{k}\,k^{2} + {
c_{k}}\,({\displaystyle \frac {\widehat{\alpha}}{N\,\mathrm{W}}} )^{k}\,k) \! 
 \right)   \nonumber \\ &&  
- 4\, \left(  \! {\displaystyle \sum _{k=0}^{r}} \,{c_{k}
}\,({\displaystyle \frac {\widehat{\alpha}}{N\,\mathrm{W}}} )^{k} \!  \right) \,
 \left(  \! {\displaystyle \sum _{k=0}^{r}} \,{c_{k}}\,(
{\displaystyle \frac {\widehat{\alpha}}{N\,\mathrm{W}}} )^{k}\,k \!  \right) ^{2
} \mbox{} - 2\, \left(  \! {\displaystyle \sum _{k=0}^{r}} \,{c_{k}
}\,({\displaystyle \frac {\widehat{\alpha}}{N\,\mathrm{W}}} )^{k} \!  \right) ^{
2}\, \nonumber \\ && \times \left(  \! {\displaystyle \sum _{k=0}^{r}} \,{c_{k}}\,(
{\displaystyle \frac {\widehat{\alpha}}{N\,\mathrm{W}}} )^{k}\,k^{2} \! 
 \right) \mbox{} 
\mbox{} - 8\, \left(  \! {\displaystyle \sum _{k=0}^{r}} \,{c_{k}
}\,({\displaystyle \frac {\widehat{\alpha}}{N\,\mathrm{W}}} )^{k}\,k \! 
 \right) ^{2}\, \left(  \! {\displaystyle \sum _{k=0}^{r}} \,{c_{
k}}\,({\displaystyle \frac {\widehat{\alpha}}{N\,\mathrm{W}}} )^{k}\,k^{2} \! 
\right) 
\end{eqnarray}

\`A la fin, comme nous avons fourni les $\Gamma_{abc}$ dans l'annex $[A]$, 
il s'av\`ere que la courbure scalaire prend la forme suivante

\begin{eqnarray}
R&=& - \frac{3}{\pi\,\sqrt{n\,w\,N\,\mathrm{W}}} \frac{r_1- r_2}{g_1 g_3}
\end{eqnarray}

o\`u les fonctions $\{ r_1, r_2 \}$ sont donn\'ees par

\begin{eqnarray}
r_1&=& 
 \left(  \! {\displaystyle \sum _{k=0}^{r}} \,{c_{k}}\,
({\displaystyle \frac {\widehat{\alpha}}{N\,\mathrm{W}}} )^{k} \!  \right) ^{3}
+  \left(  \! {\displaystyle \sum _{k=0}^{r}} \,{c_{k}}\,
({\displaystyle \frac {\widehat{\alpha}}{N\,\mathrm{W}}} )^{k} \!  \right) ^{2}\, 
\left(  \! {\displaystyle \sum _{k=0}^{r}} \,{c_{k}}\,(
{\displaystyle \frac {\widehat{\alpha}}{N\,\mathrm{W}}} )^{k}\,k^{2} \! 
 \right)   \nonumber \\ &&
+ \left(  \! {\displaystyle \sum _{k=0}^{r}} \,{c_{k}}\,
({\displaystyle \frac {\widehat{\alpha}}{N\,\mathrm{W}}} )^{k} \!  \right) ^{2}
\, \left(  \! {\displaystyle \sum _{k=0}^{r}} \,({c_{k}}\,(
{\displaystyle \frac {\widehat{\alpha}}{N\,\mathrm{W}}} )^{k}\,k^{3} + {c_{k}}\,
({\displaystyle \frac {\widehat{\alpha}}{N\,\mathrm{W}}} )^{k}\,k^{2}) 
\!  \right)  \nonumber \\ &&
+  \left(  \! {\displaystyle \sum _{k=0}^{r}} \,{c_{k}}\,
({\displaystyle \frac {\widehat{\alpha}}{N\,\mathrm{W}}} )^{k} \!  \right) ^{2}
\, \left(  \! {\displaystyle \sum _{k=0}^{r}} \,({c_{k}}\,(
{\displaystyle \frac {\widehat{\alpha}}{N\,\mathrm{W}}} )^{k}\,k^{2} + {c_{k}}\,
({\displaystyle \frac {\widehat{\alpha}}{N\,\mathrm{W}}} )^{k}\,k) \!  \right)  \nonumber \\ &&
+ 2\, \left(  \! {\displaystyle \sum _{k=0}^{r}} \,{c_{k}}\,(
{\displaystyle \frac {\widehat{\alpha}}{N\,\mathrm{W}}} )^{k} \!  \right) \,
 \left(  \! {\displaystyle \sum _{k=0}^{r}} \,{c_{k}}\,(
{\displaystyle \frac {\widehat{\alpha}}{N\,\mathrm{W}}} )^{k}\,k^{2} \! 
\right) ^{2} 
\mbox{} 
+ 2\, \left(  \! {\displaystyle \sum _{k=0}^{r}} \,{c_{k}}\,(
{\displaystyle \frac {\widehat{\alpha}}{N\,\mathrm{W}}} )^{k}\,k \!  \right)  \,
 \nonumber \\ && \times
\left(  \! {\displaystyle \sum _{k=0}^{r}} \,{c_{k}}\,(
{\displaystyle \frac {\widehat{\alpha}}{N\,\mathrm{W}}} )^{k} \!  \right) \,
 \left(  \! {\displaystyle \sum _{k=0}^{r}} \,({c_{k}}\,(
{\displaystyle \frac {\widehat{\alpha}}{N\,\mathrm{W}}} )^{k}\,k^{3} + 3\,{c_{k}
}\,({\displaystyle \frac {\widehat{\alpha}}{N\,\mathrm{W}}} )^{k}\,k^{2} + 2\,{c
_{k}}\,({\displaystyle \frac {\widehat{\alpha}}{N\,\mathrm{W}}} )^{k}\,k) \! 
\right)   \nonumber \\ && 
+ 4\, \left(  \! {\displaystyle \sum _{k=0}^{r}} \,{c_{k}
}\,({\displaystyle \frac {\widehat{\alpha}}{N\,\mathrm{W}}} )^{k}\,k \! 
 \right) ^{2}\, \left(  \! {\displaystyle \sum _{k=0}^{r}} \,{c_{
k}}\,({\displaystyle \frac {\widehat{\alpha}}{N\,\mathrm{W}}} )^{k}\,k^{2} \! 
 \right) 
+ 6\, \left(  \! {\displaystyle \sum _{k=0}^{r}} \,{c_{k}
}\,({\displaystyle \frac {\widehat{\alpha}}{N\,\mathrm{W}}} )^{k} \!  \right) \,
\nonumber \\ && \times 
 \left(  \! {\displaystyle \sum _{k=0}^{r}} \,{c_{k}}\,(
{\displaystyle \frac {\widehat{\alpha}}{N\,\mathrm{W}}} )^{k}\,k \!  \right) \,
 \left(  \! {\displaystyle \sum _{k=0}^{r}} \,({c_{k}}\,(
{\displaystyle \frac {\widehat{\alpha}}{N\,\mathrm{W}}} )^{k}\,k^{2} + {c_{k}}\,
({\displaystyle \frac {\widehat{\alpha}}{N\,\mathrm{W}}} )^{k}\,k) \!  \right).
\end{eqnarray}

\begin{eqnarray}
r_2&=& 
\left(  \! {\displaystyle \sum _{k=0}^{r}} \,{c_{k}}\,
({\displaystyle \frac {\widehat{\alpha}}{N\,\mathrm{W}}} )^{k} \!  \right) ^{2}
\, \left(  \! {\displaystyle \sum _{k=0}^{r}} \,({c_{k}}\,(
{\displaystyle \frac {\widehat{\alpha}}{N\,\mathrm{W}}} )^{k}\,k^{3} + 3\,{c_{k}
}\,({\displaystyle \frac {\widehat{\alpha}}{N\,\mathrm{W}}} )^{k}\,k^{2} + 2\,{c
_{k}}\,({\displaystyle \frac {\widehat{\alpha}}{N\,\mathrm{W}}} )^{k}\,k) \! 
\right)   \nonumber \\ && \mbox{} 
+ 2\, \left(  \! {\displaystyle \sum _{k=0}^{r}} \,{c_{k}}\,
({\displaystyle \frac {\widehat{\alpha}}{N\,\mathrm{W}}} )^{k} \! \right) \, 
\left(  \! {\displaystyle \sum _{k=0}^{r}} \,({c_{k}}\,
({\displaystyle \frac {\widehat{\alpha}}{N\,\mathrm{W}}} )^{k}\,k^{2} + {c_{k}
}\,({\displaystyle \frac {\widehat{\alpha}}{N\,\mathrm{W}}} )^{k}\,k) \! 
 \right) ^{2}  \nonumber \\ && \mbox{} 
+2\, \left(  \! {\displaystyle \sum _{k=0}^{r}} \,{c_{k}}\,(
{\displaystyle \frac {\widehat{\alpha}}{N\,\mathrm{W}}} )^{k}\,k \!  \right) \,
 \left(  \! {\displaystyle \sum _{k=0}^{r}} \,{c_{k}}\,(
{\displaystyle \frac {\widehat{\alpha}}{N\,\mathrm{W}}} )^{k} \!  \right) \,
 \left(  \! {\displaystyle \sum _{k=0}^{r}} \,({c_{k}}\,(
{\displaystyle \frac {\widehat{\alpha}}{N\,\mathrm{W}}} )^{k}\,k^{3} + {c_{k}}\,
({\displaystyle \frac {\widehat{\alpha}}{N\,\mathrm{W}}} )^{k}\,k^{2}) \! 
\right)   \nonumber \\ && 
+ 4\, \left(  \! {\displaystyle \sum _{k=0}^{r}} \,{c_{k}}\,
({\displaystyle \frac {\widehat{\alpha}}{N\,\mathrm{W}}} )^{k}\,k \! 
\right) ^{2}\, \left(  \! {\displaystyle \sum _{k=0}^{r}} \,({c
_{k}}\,({\displaystyle \frac {\widehat{\alpha}}{N\,\mathrm{W}}} )^{k}\,k^{2} + {
c_{k}}\,({\displaystyle \frac {\widehat{\alpha}}{N\,\mathrm{W}}} )^{k}\,k) \! 
\right)   \nonumber \\ && 
+ 4\, \left(  \! {\displaystyle \sum _{k=0}^{r}} \,{c_{k}
}\,({\displaystyle \frac {\widehat{\alpha}}{N\,\mathrm{W}}} )^{k} \!  \right) ^{
2}\, \left(  \! {\displaystyle \sum _{k=0}^{r}} \,{c_{k}}\,(
{\displaystyle \frac {\widehat{\alpha}}{N\,\mathrm{W}}} )^{k}\,k \!  \right)   \nonumber \\ &&
+ 6\, \left(  \! {\displaystyle \sum _{k=0}^{r}} \,{c_{k}
}\,({\displaystyle \frac {\widehat{\alpha}}{N\,\mathrm{W}}} )^{k} \!  \right) \,
 \left(  \! {\displaystyle \sum _{k=0}^{r}} \,{c_{k}}\,(
{\displaystyle \frac {\widehat{\alpha}}{N\,\mathrm{W}}} )^{k}\,k \!  \right) \,
 \left(  \! {\displaystyle \sum _{k=0}^{r}} \,{c_{k}}\,(
{\displaystyle \frac {\widehat{\alpha}}{N\,\mathrm{W}}} )^{k}\,k^{2} \! 
\right).
\end{eqnarray}


\clearpage
\chapter{La g\'eom\'etrie de Ruppenier des solutions non-extr\'emales de 
branes $ D_1D_5 $ et $ D_2D_6NS_5 $ en dimensions $ D = 10 $}

Dans ce chapitre, nous consid\'erons les solutions non-extr\'emales de 
branes $D_1D_5$ et $D_2D_6NS_5$ en $D= 10$ de l'action effective de 
la th\'eorie des cordes de type-II. 
Ces solutions non-extr\'emales de branes sont obtenues par l'application de 
la fonction de l'entropie de Sen par Ahmad Ghodsi et. al. \cite{GarousiGhodsi1}.
Ces syst\`emes de branes noirs sont le plus simple syst\`eme pour lesquels,
il est imm\'ediat d'analyser la g\'eom\'etrie thermodynamique. 
Ceux sont respectivement les branes noirs non-extr\'emaux avec trois et quatre charges.
En d'autres termes, nous consid\'erons les solutions non-extr\'emales
de branes $D_1D_5$ et $D_2D_6NS_5$ d\'ecoulant de l'action effective 
de la th\'eorie des cordes de type-II \cite{HorowitzWelch}.

Le contenu des champs de l'action effectif dans le cadre des cordes
sont le champ de dilaton, le champ de NS-NS, le champ d'autoduale de RR.
Ult\'erieurement, le Ref. \cite{GarousiGhodsi1} montre que l'application 
de formalisme de la fonction de l'entropie de Sen peut \^etre faite pour les 
deux solutions non-extr\'emales de branes $ D_1D_5 $ et $ D_2D_6NS_5 $  
dans l'approximation de la gorge, o\`u ces solutions de branes respectivement 
representent les trous noirs de Schwarzschild dans $AdS_3 \times S^3 \times T_4$ 
et $AdS_3 \times S^2 \times S^1 \times T_4$. Donc, \`a la suite, nous voudrons 
voir comment les corrections de $\alpha^{\prime} $ modifient la g\'eom\'etrie 
thermodynamique de ces solutions non-extr\'emales de branes $ D_1D_5 $ et $ D_2D_6NS_5 $.

\section{La g\'eom\'etrie de Ruppenier des solutions non-extr\'emales de branes $D_1D_5$:}

Dans cette section, nous analysons la g\'eom\'etrie thermodynamique de Ruppenier 
des branes $ D_1D_5 $ non-extr\'emaux. Dans la th\'eorie effictive de type-IIB,
la solution non-extr\'emales des branes $ D_1D_5 $ peut \^etre d\'ecrite 
en prenant le champ de dilaton et les deux champs de RR $\{ C_2, C_6 \}$.
En fait, cette configuration  peut \^etre d\'ecrite pour les moments du 
mouvement de droit et ceux de gauche a l'\'egalit\'e et en consid\'erant 
les branes $D_1$ dans la direction compacte de $S^1$ et les branes $D_5$ 
dans la direction compacte de $ S^1 \times T^4 $. En ce cas, le Ref. \cite{GarousiGhodsi1}
montre que l'horizon de cette solution se d\'eveloppe \`a $r=r_0$ avec 
une g\'eom\'etrie de la gorge de $S^3 \times T^4$ dans la limite de 
grande distance: $r \ll Q_1$ et $r \ll Q_2$ qui est un trou noir de 
Schwarzschil dans l'$ AdS_3 $. En outre, notez bien que cette solution 
se permet de r\'eduire \`a la solution  extr\'emale ordinaire des branes 
$D_1D_5$ pour certaine valeur du param\`etre $ r_0 $ pour laquelle l'horizon 
est situ\'e \`a $ r_0=0 $. L'analyse de la fonction de l'entropie de Sen, 
au niveau des deux d\'eriv\'es et bien aussi avec les corrections de $\alpha^{\prime}$ 
aux quelles la g\'eom\'etrie (de l'espace-temps) pr\`es de l'horizon est donn\'ee
par $S^1\times S^3\times T^4$, donne l'entropie de ce syst\`eme des branes noirs.

\subsection{\`A l'ordre de $(\alpha^{\prime})^0 $}

Il est bien connu \cite{GarousiGhodsi1} que l'entropie est proportionnelle
\`a la fonction de l'entropie de Sen et au niveau des deux d\'eriv\'es, 
l'entropie de cette solution des branes $ D_1 D_5 $ est:

\ba S_{BH}(N_1, N_5, N_R):= 4 \pi \sqrt{N_1 N_5 N_R}.\ea

Donc, en consid\`erant $\{N_1, N_5, N_R\}$ comme les variables de
la fluctuation statistique, on peut voir que les composantes de la 
m\'etrique tenseure de Ruppenier sont donn\'ees par:

\ba 
g_{N_1N_1}&=& \frac{ \pi}{N_1} \sqrt{\frac{N_5 N_R}{N_1}},\nn
g_{N_1N_5}&=& - \pi \sqrt{\frac{N_R}{N_1 N_5}},\nn
g_{N_1N_R}&=& - \pi \sqrt{\frac{N_5}{N_1 N_R}},\nn
g_{N_5N_5}&=& \frac{ \pi}{N_5} \sqrt{\frac{N_1 N_R}{N_5}},\nn
g_{N_5N_R}&=& - \pi \sqrt{\frac{N_1}{N_5 N_R}},\nn
g_{N_RN_R}&=& \frac{ \pi}{N_R} \sqrt{\frac{N_1 N_5}{N_R}}.\ea

\begin{figure}
\hspace*{1.0cm}\vspace*{-6.0cm}
\includegraphics[width=12.0cm,angle=-0]{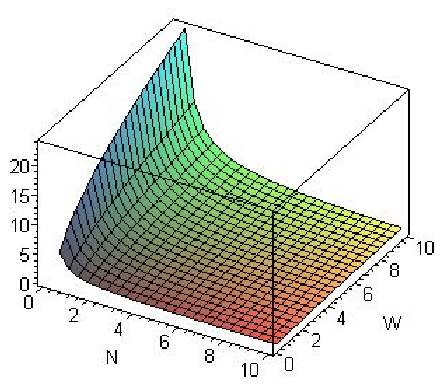}\vspace*{-3.0cm}
\caption{La composante $N_1N_1$ de la m\'etrique tenseure trac\'ee 
comme la fonction de $\{N, W\}$, en d\'ecrivant les fluctuations dans 
la configuration des trous noirs charg\'es nonextremaux $D_1 D_5 $ 
\`a l'ordre dominant.} \label{nonextN1N1}\vspace*{0.5cm}
\end{figure}

\begin{figure}
\hspace*{1.0cm}\vspace*{-6.0cm}
\includegraphics[width=12.0cm,angle=-0]{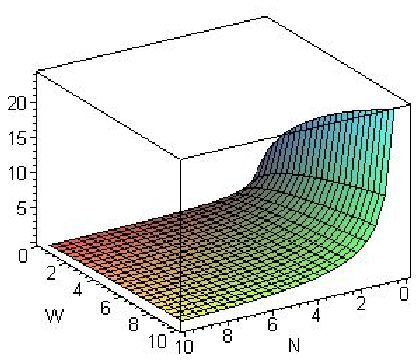}\vspace*{-3.0cm}
\caption{La composante $N_5N_5$ de la m\'etrique tenseure trac\'ee 
comme la fonction de $\{N, W\}$, en d\'ecrivant les fluctuations dans 
la configuration des trous noirs charg\'es nonextremaux $D_1 D_5 $ 
\`a l'ordre dominant.} \label{nonextN5N5}\vspace*{0.5cm}
\end{figure}

\begin{figure}
\hspace*{1.0cm}\vspace*{-6.0cm}
\includegraphics[width=12.0cm,angle=-0]{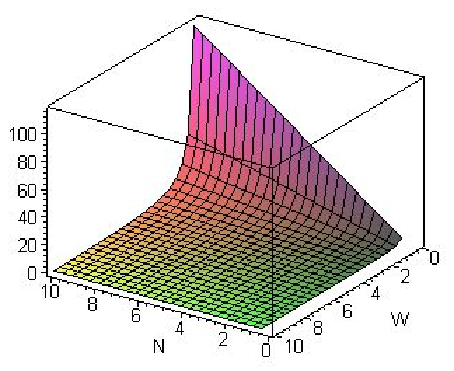}\vspace*{-3.0cm}
\caption{La composante $N_RN_R$ de la m\'etrique tenseure trac\'ee 
comme la fonction de $\{N, W\}$, en d\'ecrivant les fluctuations dans 
la configuration des trous noirs charg\'es nonextremaux $D_1 D_5 $ 
\`a l'ordre dominant.} \label{nonextNRNR}\vspace*{0.5cm}
\end{figure}

\begin{figure}
\hspace*{1.0cm}\vspace*{-6.0cm}
\includegraphics[width=12.0cm,angle=-0]{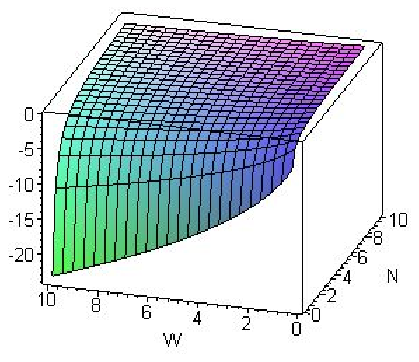}\vspace*{-3.0cm}
\caption{La composante $N_1N_5$ de la m\'etrique tenseure trac\'ee 
comme la fonction de $\{N, W\}$, en d\'ecrivant les fluctuations dans 
la configuration des trous noirs charg\'es nonextremaux $D_1 D_5 $ 
\`a l'ordre dominant.} \label{nonextN1N5}\vspace*{0.5cm}
\end{figure}

\begin{figure}
\hspace*{1.0cm}\vspace*{-6.0cm}
\includegraphics[width=12.0cm,angle=-0]{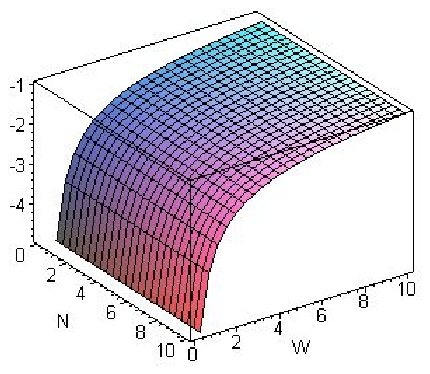}\vspace*{-3.0cm}
\caption{La composante $N_1N_R$ de la m\'etrique tenseure trac\'ee 
comme la fonction de $\{N, W\}$, en d\'ecrivant les fluctuations dans 
la configuration des trous noirs charg\'es nonextremaux $D_1 D_5 $ 
\`a l'ordre dominant.} \label{nonextN1NR}\vspace*{0.5cm}
\end{figure}

\begin{figure}
\hspace*{1.0cm}\vspace*{-6.0cm}
\includegraphics[width=12.0cm,angle=-0]{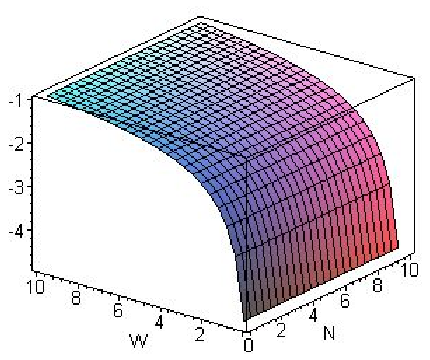}\vspace*{-3.0cm}
\caption{La composante $N_5N_R$ de la m\'etrique tenseure trac\'ee 
comme la fonction de $\{N, W\}$, en d\'ecrivant les fluctuations dans 
la configuration des trous noirs charg\'es nonextremaux $D_1 D_5 $ 
\`a l'ordre dominant.} \label{nonextN5NR}\vspace*{0.5cm}
\end{figure}

Puisqu'il y a trois variables ind\'ependantes, c'est-\`a-dire que nous avons
les nombres $\{N_1, N_5, N_R \}$, et donc en afin d'offrir les vues graphiques 
de trois dimensions, nous allons consid\'erer dans la limite de $N_1= N$, $N_5 = N$ 
et $N_R = W$. Pour une configuration donn\'ee des trous noirs $D_1 D_5$ nonextremaux, 
les propri\'et\'es des fluctuations peuvent \^{e}tre exprim\'ees par les charact\'erisation 
suivantes. Notamment, dans le r\'egime de $N \in (0, 10)$ et $W \in (0, 10)$, on observe que 
l'amplitudes des les capacit\'es de chaleurs $\{g_{N_1 N_1}, g_{N_5 N_5}\}$ prennent une valeur 
\`a l'ordre de $25$. Dans cette gamme de $N, W$, la capacit\'e de chaleur $\{g_{N_R N_R}\}$ 
se situe dans la gamme de $(0, 120)$. Conform\'ement \`a la pr\'ediction de la g\'eom\'etrie 
de l'espace d'\'etat, nous constatons que les gammes de croissance de la premi\`ere ensemble 
et celle de la deuxi\`eme ensemble des capacit\'es de chaleurs restent dans la limite oppos\'ee 
des param\`etres $\{N, W\}$. Plus pr\'ecis\'ement, pour une petite valeur donn\'ee de $N$, 
la premi\`ere s\'erie des capacit\'es de chaleurs augmente quand on augmente la valeur de $W$. 
Cependant, pour une petite valeur donn\'ee de $W$, la composante $g_{N_R N_R}$ augmente 
quand on augmente la valeur de $N$. Dans cette pr\'eoccupation, les Figs. (\ref{nonextN1N1}, 
\ref{nonextN5N5}) montrent que la croissance de premier ensemble des capacit\'es de chaleurs 
$\{g_{N_1 N_1}, g_{N_5 N_5}\}$ est li\'ee dans la limite d'un grand $W$ et un petit $N$. 
Du fait m\^{e}me, la Fig. (\ref{nonextNRNR}) montre que la croissance de $\{g_{N_R N_R}\}$
se situe dans la limite d'un petit $W$ et un grand $N$. De plus, les compressibilit\'es
de chaleurs en impliquant deux param\`etres distincts de la configuration des trous noirs
sous-jacents sont repr\'esent\'es dans les Figs. (\ref{nonextN1N5},\ref{nonextN1NR}, \ref{nonextN5NR}). 
En fait, nous constatons que l'amplitude de (i) $g_{N_1 N_5}$ est \`a l'ordre de $-25$, 
(ii) $g_{N_1 N_R}$ est \`a l'ordre de $-5$ et (iii) $g_{N_5 N_R}$ est aussi \`a l'ordre de $-5$. 
Par la pr\'esente, la m\'etrique tenseure de l'espace d'\'etat $\{g_{ij} \ | \ i, j = N_1, N_5, N_R\}$ 
tel qu'elle est d\'ecrite dans les Figs. (\ref{nonextN1N1}, \ref{nonextN5N5}, \ref{nonextNRNR}) ce qui
est montr\'ee \`a l'ordre dominant de la configuration des trous noirs $D_1 D_5 $ nonextremaux, 
correspond \`a un ensemble statistique localement stable avec une positive valeur des 
capacit\'es de chaleurs.

Dans ce cas, il s'av\`ere que le mineur de surface est 

\begin{equation}
p := {g_{\mathit{N_1N_1}}}\, {g_{\mathit{N_5N_5}}} -
({g_{\mathit{N_1N_5} }})^{2}
\end{equation} 

est identiquement nulle pour toute valeurs des nombres $\{N_1, N_5, N_R\}$.

Nous voyons que le d\'eterminant de la m\'etrique tenseure est:

\ba g= -4 \pi^3 (N_1 N_5 N_R)^{-1/2}.\ea

\begin{figure}
\hspace*{1.0cm}\vspace*{-6.0cm}
\includegraphics[width=12.0cm,angle=-0]{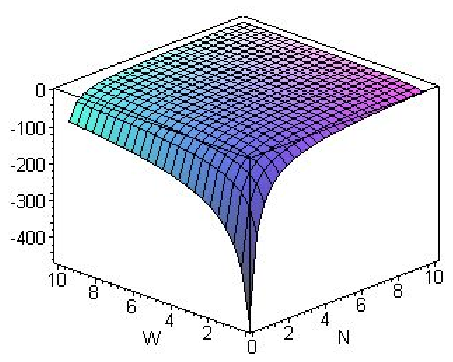}\vspace*{-3.0cm}
\caption{Le d\'eterminant de la m\'etrique tenseure trac\'ee 
comme la fonction de $\{N, W\}$, en d\'ecrivant les fluctuations dans 
la configuration des trous noirs charg\'es nonextremaux $D_1 D_5 $ 
\`a l'ordre dominant.} \label{nonextdetg}\vspace*{0.5cm}
\end{figure}

\begin{figure}
\hspace*{1.0cm}\vspace*{-6.0cm}
\includegraphics[width=12.0cm,angle=-0]{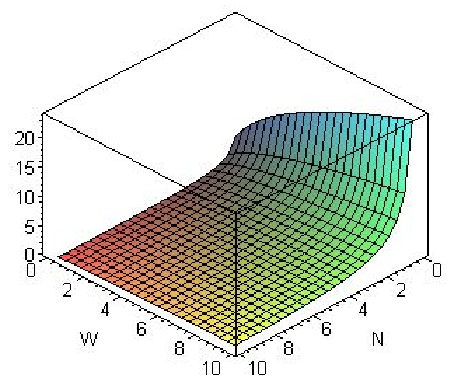}\vspace*{-3.0cm}
\caption{Le premier mineur de la m\'etrique tenseure trac\'ee 
comme la fonction de $\{N, W\}$, en d\'ecrivant les fluctuations dans 
la configuration des trous noirs charg\'es nonextremaux $D_1 D_5 $ 
\`a l'ordre dominant.} \label{nonextminor1}\vspace*{0.5cm}
\end{figure}

Sous les fluctuations des param\`etres $\{N_1, N_5, N_R\}$, la stabilit\'e 
d'un ensemble de la configuration des trous noirs $D_1 D_5$ nonextremaux
d\'ecoule de la positivit\'e du d\'eterminant de la m\'etrique tenseure. 
Pour le choix de $N_1= N$, $N_5= N$ et $W= N_R$, nous trouvons que le 
d\'eterminant de la m\'etrique tenseure $g$ prend une grande valeur n\'egative 
quand nous diminuons les param\`etres $\{N, W\}$. Dans ce cas, nous constatons que 
$g \in (-500, 0)$. Pour une valeur typique de $N \in (0, 10)$ et $W \in (0,10)$, 
la Fig. (\ref{nonextdetg}) offre la nature graphique du d\'eterminant de la m\'etrique 
tenseure $g$. En outre, la surface d\'efinie par les fluctuations de $\{N_1, N_5\}$ 
est instable par-ce que le mineur principe correspondant est identiquement nulle, \`a savoir que
nous avons $p_2= 0$. Enfin d'alors que le seulement le param\`etre $N_1$ est autoris\'e \`a varier, 
la stabilit\'e de la configuration des trous noirs $D_1 D_5$ nonextremaux est donn\'ee par la 
positivit\'e du premier principe mineur $p_1:= g_{N_1 N_1}$. Une vue rotat\'ee de $p_1$ est 
pr\'esent\'ee dans la Fig. (\ref{nonextminor1}). Les propri\'et\'es ci-dessus de l'espace 
d'\'etat et la positivit\'e des mineurs principaux concern\'es d\'ecritent le comportement 
qualitative de la stabilit\'e statistique d'un ensemble des trous noirs $D_1D_5 $ nonextremaux
ce qui sont d\'efinis par la th\'eorie des cordes.

Et bien aussi, comme nous avons fourni les $\Gamma_{abc}$ dans l'annex $[A]$, 
il s'av\`ere que on voit que la courbure scalaire de Ruppenier est simplement:

\ba R= \frac{3}{8 \pi} (N_1 N_5 N_R)^{-1/2}\ea

qu'elle est partout r\'eguli\`ere, 
pour chaque non-nulle valeur des param\`etres $\{N_1, N_5, N_R\}$.

\begin{figure}
\hspace*{1.0cm}\vspace*{-6.0cm}
\includegraphics[width=12.0cm,angle=-0]{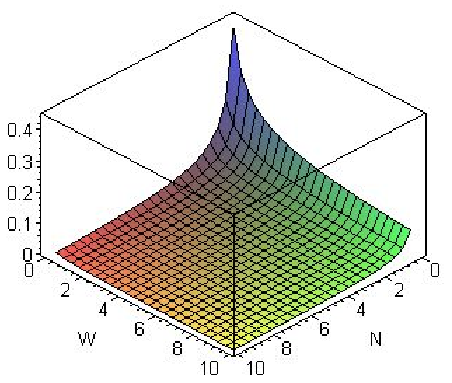}\vspace*{-3.0cm}
\caption{La courbure scalaire trac\'ee comme la fonction de $\{N, W\}$, 
en d\'ecrivant les fluctuations dans la configuration des trous noirs charg\'es 
nonextremaux $D_1 D_5 $ \`a l'ordre dominant dans la gamme des charges
$N, W \in (0, 10)$.} \label{nonextR}\vspace*{0.5cm}
\end{figure}

\begin{figure}
\hspace*{1.0cm}\vspace*{-6.0cm}
\includegraphics[width=12.0cm,angle=-0]{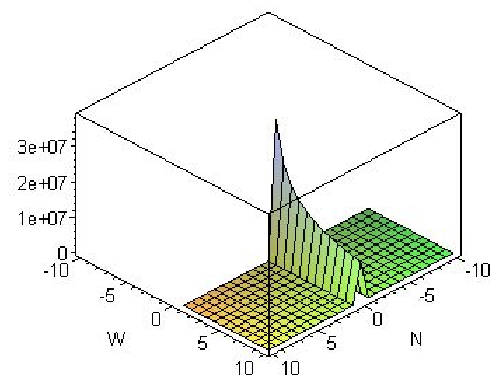}\vspace*{-3.0cm}
\caption{La courbure scalaire trac\'ee comme la fonction de $\{N, W\}$, 
en d\'ecrivant les fluctuations dans la configuration des trous noirs charg\'es 
nonextremaux $D_1 D_5 $ \`a l'ordre dominant dans la gamme des charges
$N, W \in (-10, 10)$.} \label{nonextR10}\vspace*{0.5cm}
\end{figure}

Comme mentionn\'e dans le chapitre pr\'ec\'edent, la courbure scalaire 
de l'espace d'\'etat offre les propri\'et\'es statistiques globales 
sous la fluctuation d'une configuration des trous noirs charg\'es 
$D_1 D_5 $ nonextremaux en terme de $\{N_1, N_5, N_R\}$. Dans la gamme de 
$N \in (0, 10)$ et $W \in (0, 10)$, la Fig. (\ref{nonextR}) montre que 
la courbure scalaire a une petite amplitude positive de l'ordre de $0.5$. 
Dans cette gamme de param\`etres, nous observons que la configuration 
sous-jacente de ces trous noirs $D_1 D_5$ nonextremaux est un syst\`eme 
statistique faiblement interactive. Physiquement, le signe positif de la 
courbure scalaire signifie que les interactions statistiques sont r\'epulsives 
dans la leur nature. La Fig. (\ref{nonextR10}) illustre le comportement de la 
courbure scalaire ci-dessus pour la range des param\`etres $N, W \in (-10, 10)$. 
En fait, nous voyons que les interactions sont largement pr\'esents pr\`es de la 
ligne $N= 0$ et la courbure scalaire de l'espace d'\'etat acquiert un grand 
pic \`a l'ordre de $4 \times 10^{+07}$ pr\`es de l'origine $(N, W) = (0, 0)$. 
Dans la gamme d'un petit $W$ et un petit $N$, nous remarquons de la Fig. (\ref{nonextR10}) 
que les interactions statistiques globales sont sym\'etriques autour de la ligne $N= 0$. 
En comparaison des interactions apparaissant dans la gamme de $N, W \in (0, 10)$, 
l'amplitude des interactions statistiques globales se r\'ev\`elent d'\^{e}tre beaucoup 
plus grande dans la gamme de $N, W \in (-10, 10) $. Pr\'ecis\'ement, nous constatons que 
le ratio des valeurs typiques de l'amplitude de ces courbures scalaires de l'espace d'\'etat
est \`a l'ordre de $10^{-08}$. La vue graphique de cette comparaison ci-dessus mentionn\'ee 
est apparente par les Figs. (\ref{nonextR}, \ref{nonextR10}). Qualitativement, par les 
repr\'esentations de l'espace d'\'etat, nous observons que la configuration des trous 
noirs $D_1D_5$ nonextremaux sont interactives au niveau globale, et elles correspond \`a 
un syst\`eme statistique instable dans des deux d\'eriv\'es de la gravit\'e.


\subsection{\`A l'ordre de $\alpha^{\prime} $}

Maintenant, examinons les effets de la th\'eorie des cordes, en
particulier, les corrections de $(\alpha^{\prime})^3$ venant de la 
courbure tenseure de Wey \cite{GarousiGhodsi1}. L'analyse de la 
fonction de l'entropie de Sen peut \^etre faite pour la contribution 
non-nulle de la courbure tenseure de Weyl \`a l'entropie et ainsi la 
g\'eom\'etrie de l'espace-temps pr\`es de l'horizon peut \^etre obtenues 
\`a l'$AdS_3 \times S^3 \times T^4$, comme dans le cas d'habitude des 
deux d\'eriv\'ees. En particulier, en tenant compte de ces effets 
de la th\'eorie des cordes, nous consid\'erons les corrections des 
d\'eriv\'ees sup\'erieures au niveau de $(\alpha^{\prime })^3$.
Soit 

\ba \gamma = \frac{1}{8}\zeta(3)(\alpha^{ \prime })^3 \ea

le coefficient de la contribution des d\'eriv\'ees sup\'erieures \`a la 
densit\'e lagrangienne effective. Alors, \`a l'aide de la m\'ethode de la 
fonction d'entropie de Sen, le Ref. \cite{GarousiGhodsi1} montre que 
l'entropie de la configuration des branes $ D_1 D_5 $ est:

\ba S_{BH}(N_1, N_5, N_R):= 4 \pi \sqrt{N_1 N_5 N_R}- \frac{4 \pi b \sqrt{N_R}}{N_1 N_5},\ea

o\`u nous avons d\'efini le param\`etre $b$ comme:

\ba b= \gamma (\frac{(2\pi)^3 V_4}{16 \pi G_{10}})^{3/2}.\ea

Donc, les composantes de la m\'etrique de Ruppenier sont donn\'ees par:

\ba g_{N_1N_1}&=& \frac{ \pi}{N_1} \sqrt{\frac{N_5 N_R}{N_1}}+\frac{8 \pi b \sqrt{N_R}}{N_1^3 N_5},\nn
g_{N_1N_5}&=& -\pi \sqrt{\frac{N_R}{N_1 N_5}}+ \frac{4 \pi b \sqrt{N_R}}{N_1^2 N_5^2},\nn
g_{N_1N_R}&=& -\pi \sqrt{\frac{N_5}{N_1 N_R}}- \frac{2 \pi b}{N_1^2N_5 \sqrt{N_R}},\nn
g_{N_5N_5}&=& \frac{ \pi}{N_5} \sqrt{\frac{N_1 N_R}{N_5}}+\frac{8 \pi b \sqrt{N_R}}{N_1 N_5^3},\nn
g_{N_5N_R}&=& -\pi \sqrt{\frac{N_1}{N_5 N_R}}- \frac{2 \pi b}{N_1 N_5^2 \sqrt{N_R}},\nn
g_{N_RN_R}&=& \frac{ \pi}{N_R} \sqrt{\frac{N_1 N_5}{N_R}}- \frac{\pi b}{N_1 N_5 N_R^{3/2}}.\ea

\begin{figure}
\hspace*{1.0cm}\vspace*{-6.0cm}
\includegraphics[width=12.0cm,angle=-0]{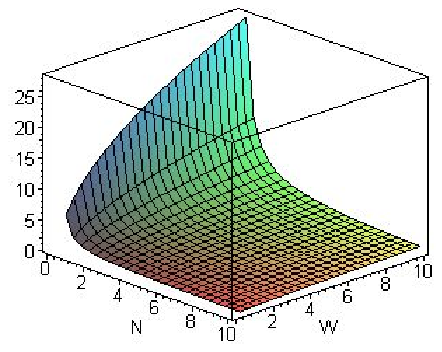}\vspace*{-3.0cm}
\caption{La composante $N_1N_1$ de la m\'etrique tenseure trac\'ee 
comme la fonction de $\{N, W\}$, en d\'ecrivant les fluctuations dans 
la configuration des trous noirs charg\'es nonextremaux $D_1 D_5 $ 
\`a l'ordre de $\alpha^{\prime}$.} \label{nonextalphaN1N1}\vspace*{0.5cm}
\end{figure}

\begin{figure}
\hspace*{1.0cm}\vspace*{-6.0cm}
\includegraphics[width=12.0cm,angle=-0]{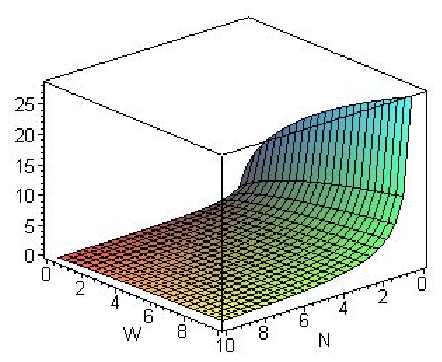}\vspace*{-3.0cm}
\caption{La composante $N_5N_5$ de la m\'etrique tenseure trac\'ee 
comme la fonction de $\{N, W\}$, en d\'ecrivant les fluctuations dans 
la configuration des trous noirs charg\'es nonextremaux $D_1 D_5 $ 
\`a l'ordre de $\alpha^{\prime}$.} \label{nonextalphaN5N5}\vspace*{0.5cm}
\end{figure}

\begin{figure}
\hspace*{1.0cm}\vspace*{-6.0cm}
\includegraphics[width=12.0cm,angle=-0]{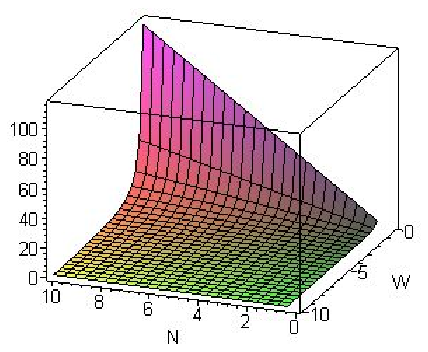}\vspace*{-3.0cm}
\caption{La composante $N_RN_R$ de la m\'etrique tenseure trac\'ee 
comme la fonction de $\{N, W\}$, en d\'ecrivant les fluctuations dans 
la configuration des trous noirs charg\'es nonextremaux $D_1 D_5 $ 
\`a l'ordre de $\alpha^{\prime}$.} \label{nonextalphaNRNR}\vspace*{0.5cm}
\end{figure}

\begin{figure}
\hspace*{1.0cm}\vspace*{-6.0cm}
\includegraphics[width=12.0cm,angle=-0]{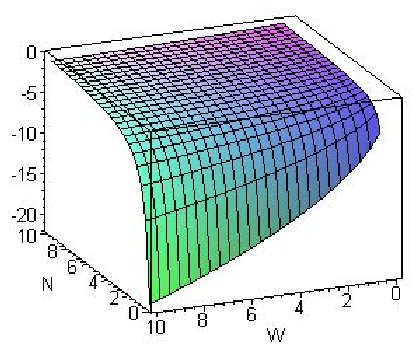}\vspace*{-3.0cm}
\caption{La composante $N_1N_5$ de la m\'etrique tenseure trac\'ee 
comme la fonction de $\{N, W\}$, en d\'ecrivant les fluctuations dans 
la configuration des trous noirs charg\'es nonextremaux $D_1 D_5 $ 
\`a l'ordre de $\alpha^{\prime}$.} \label{nonextalphaN1N5}\vspace*{0.5cm}
\end{figure}

\begin{figure}
\hspace*{1.0cm}\vspace*{-6.0cm}
\includegraphics[width=12.0cm,angle=-0]{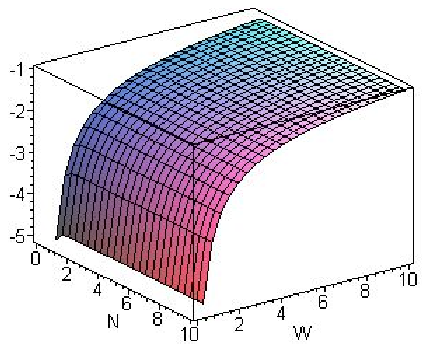}\vspace*{-3.0cm}
\caption{La composante $N_1N_R$ de la m\'etrique tenseure trac\'ee 
comme la fonction de $\{N, W\}$, en d\'ecrivant les fluctuations dans 
la configuration des trous noirs charg\'es nonextremaux $D_1 D_5 $ 
\`a l'ordre de $\alpha^{\prime}$.} \label{nonextalphaN1NR}\vspace*{0.5cm}
\end{figure}

\begin{figure}
\hspace*{1.0cm}\vspace*{-6.0cm}
\includegraphics[width=12.0cm,angle=-0]{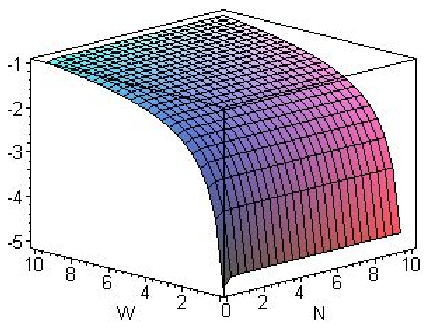}\vspace*{-3.0cm}
\caption{La composante $N_5N_R$ de la m\'etrique tenseure trac\'ee 
comme la fonction de $\{N, W\}$, en d\'ecrivant les fluctuations dans 
la configuration des trous noirs charg\'es nonextremaux $D_1 D_5 $ 
\`a l'ordre de $\alpha^{\prime}$.} \label{nonextalphaN5NR}\vspace*{0.5cm}
\end{figure}

En suivant la convention des graphiques de la section pr\'ec\'edente, 
nous allons continuer \`a la consid\'eration de $N_1= N$, $N_5= N$ et $N_R= W$. 
\c{C}a offre une repr\'esentation des propri\'et\'es de la fluctuation en trois 
dimensions de la configuration des trous noirs $D_1 D_5$ nonextremaux sous les
corrections de $\alpha^{\prime}$. Dans le r\'egime de $N \in (0, 10)$ et $W \in (0, 10)$,
nous trouvons que l'amplitudes des capacit\'es de chaleurs $\{g_{N_1 N_1},
g_{N_5 N_5}\}$ prend une valeur \`a l'ordre de $30$. Dans cette gamme de
$N, W$, les capacit\'es de chaleurs $\{g_{N_R N_R}\}$ est dans l'intervalle $(0, 120)$. 
En conformant \`a la pr\'ediction de lag\'eom\'etrie de l'espace d'\'etat 
sous lec corrections de $\alpha^{\prime}$, on observe que l'augmentation de l'amplitude
de la premi\`ere ensemble et celle de la deuxi\`eme ensemble des capacit\'es de chaleurs
reste dans la limite oppos\'ee des param\`etres $\{N, W\}$. Plus pr\'ecis\'ement, 
nous constatons que les corrections de $\alpha^{\prime}$ ne modifie pas la nature graphique 
des capacit\'es de chaleurs. En fait, pour une petite valeur donn\'ee de $N$, nous pouvons noter 
que la premi\`ere s\'erie des capacit\'es de chaleurs $\{g_{N_1 N_1}, g_{N_5 N_5}\}$ augmente,
quand nous augmentons la valeur de $W$. Du fait m\^{e}me, pour une valeur donn\'ee petite de $W$, 
on observe que la capacit\'e thermique $g_{N_R N_R}$ augmente lorsque nous augmentons la valeur de $N$. 
Dans ce cas, les Figs. (\ref{nonextalphaN1N1}, \ref{nonextalphaN5N5}) montrent que
la croissance de premier ensemble des capacit\'es de chaleurs $\{g_{N_1N_1}, g_{N_5 N_5}\}$ 
a lieu dans la limite d'un grand $W$ et un petite $N$. De plus, la Fig. (\ref{nonextalphaNRNR}) 
montre que l'amplitude de $\{g_{N_R N_R}\}$ a lieu dans la limite d'un petit $W$ et un grand $N$. 
Il est \`a noter que les capacit\'es de chaleurs de cet ensemble trouvent des positives num\'eriques
valeurs. D'autre part, les Figs. (\ref{nonextalphaN1N5}, \ref{nonextalphaN1NR},
\ref{nonextalphaN5NR}) r\'epresentent les compressibilit\'es de chaleurs en impliquant
deux param\`etres distincts de la configuration des trous noirs $D_1 D_5$ nonextremaux sous
les corrections de $\alpha^{\prime}$. Par la pr\'esente, nous remarquons que l'amplitude de (i) 
$g_{N_1 N_5} $ est \`a l'ordre de $-22$, (ii) $g_{N_1N_R}$ est \`a l'ordre de $-5$ et (iii) 
$g_{N_5 N_R}$ est aussi \`a l'ordre de $-5$. Comme nous avons montr\'e dans les Figs. 
(\ref{nonextalphaN1N1}, \ref{nonextalphaN5N5}, \ref{nonextalphaNRNR}), la m\'etrique 
tenseure de l'espace d'\'etat $\{g_{ij} \ | \ i, j = N_1, N_5, N_R\}$ illustre le fait 
que les corrcetions de $\alpha^{\prime}$ \`a la configuration des trous noirs $D_1 D_5$ 
nonextremaux correspondent \`a un syst\`eme statistique localement stable.

Sous les corrections de d\'eriv\'ees sup\'erieures de la th\'eorie des cordes, 
nous constatons que le mineure de la surface 

\begin{equation}
p_2 := {g_{\mathit{N_1N_1}}}\,{g_{\mathit{N_5N_5}}}
- {g_{\mathit{N_1N_5} }}^{2} 
\end{equation}

se r\'eduit \`a une valeur non-nulle. En fait, il n'est pas difficile de voir 
que le mineur de surface comme d\'efini au-dessus est donn\'ee par

\begin{eqnarray}
p_2&=& 24\,{\displaystyle \pi^{2}\,b\, \mathit{N_R}\,
\frac{(\mathit{N_5}\, \mathit{N_1})^{3/2} + 2\,b}{(\mathit{N_1}\,
\mathit{N_5})^{4}}}\
\end{eqnarray}

Il est facile de voir que le d\'eterminant de la m\'etrique tenseure est donn\'e par:

\ba g&=& -4 \pi^3 (N_1 N_5)^{-5}(N_1 N_5 N_R)^{-1/2} \tilde{g}(N_1, N_5), \ea

o\`u la fonction $\tilde{g}(N_1, N_5)$ est d\'efinie par

\ba \label{gtild1d5} \tilde{g}(N_1, N_5):= (N_1 N_5)^5+ 6b^2 (N_1 N_5)^2+ 20 b^3 (N_1 N_5)^{1/2}. \ea

\begin{figure}
\hspace*{1.0cm}\vspace*{-6.0cm}
\includegraphics[width=12.0cm,angle=-0]{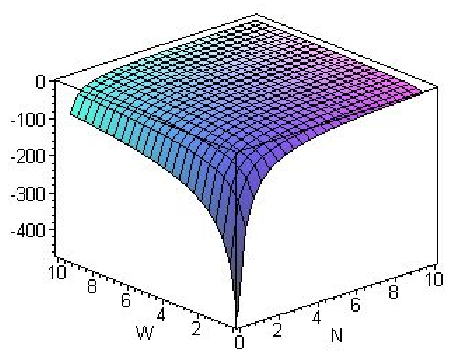}\vspace*{-3.0cm}
\caption{Le d\'eterminant de la m\'etrique tenseure trac\'ee 
comme la fonction de $\{N, W\}$, en d\'ecrivant les fluctuations dans 
la configuration des trous noirs charg\'es nonextremaux $D_1 D_5 $ 
\`a l'ordre de $\alpha^{\prime}$.} \label{nonextalphadetg}\vspace*{0.5cm}
\end{figure}

\begin{figure}
\hspace*{1.0cm}\vspace*{-6.0cm}
\includegraphics[width=12.0cm,angle=-0]{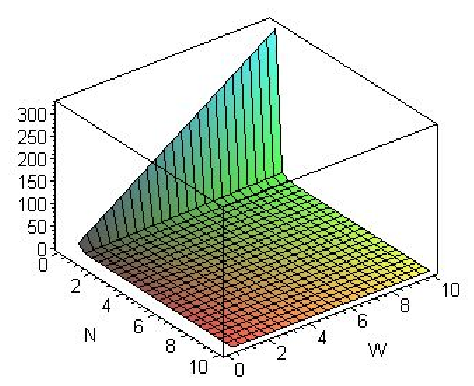}\vspace*{-3.0cm}
\caption{Le mineur de surface de la m\'etrique tenseure trac\'ee 
comme la fonction de $\{N, W\}$, en d\'ecrivant les fluctuations dans 
la configuration des trous noirs charg\'es nonextremaux $D_1 D_5 $ 
\`a l'ordre de $\alpha^{\prime}$.} \label{nonextalphaminor2}\vspace*{0.5cm}
\end{figure}

\begin{figure}
\hspace*{1.0cm}\vspace*{-6.0cm}
\includegraphics[width=12.0cm,angle=-0]{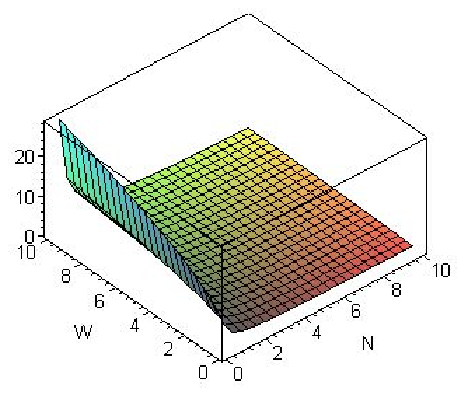}\vspace*{-3.0cm}
\caption{Le premier mineur de la m\'etrique tenseure trac\'ee 
comme la fonction de $\{N, W\}$, en d\'ecrivant les fluctuations dans 
la configuration des trous noirs charg\'es nonextremaux $D_1 D_5 $ 
\`a l'ordre de $\alpha^{\prime}$.} \label{nonextalphaminor1}\vspace*{0.5cm}
\end{figure}

Sous les fluctuations des param\`etres $\{N_1, N_5, N_R\}$, la stabilit\'e d'un
ensemble de la configuration des trous noirs $D_1 D_5$ nonextremaux peut \^{e}tre
d\'ecrite de la positivit\'e du d\'eterminant de la m\'etrique tenseure $g$. 
Sous les corrections de $\alpha^{\prime}$ \`a la configuration des trous noirs
$D_1D_5$ nonextremaux avec $N_1= N$, $N_5= N$ et $W= N_R$, nous constatons que
le d\'eterminant de la m\'etrique tenseure $g$ prend une grande valeur n\'egative, 
quand nous diminuons les param\`etres $\{N, W\}$. \c{C}a montre que les corrections 
des d\'eriv\'es sup\'erieures de la th\'eorie des cordes ne donnent pas la stabilit\'e 
statistique aux trous noirs $D_1D_5$ nonextremaux. Dans ce cas, on peut aussi constater
que $g \in (-500, 0)$. Pour une valeur typique de $N \in (0, 10)$ et $W \in (0, 10)$, 
la Fig. (\ref{nonextalphadetg}) d\'ecrit la vue graphique du d\'eterminant de la m\'etrique 
tenseure $g$. Fait int\'eressant, la surface d\'efinie par les fluctuations de $\{N_1, N_5\}$ 
devient stable sous les corrections de $\alpha^{\prime}$. Cela d\'ecoule de la positivit\'e 
de mineur principe correspondant $p_2 $. La vue graphique de $p_2$ est montr\'e dans la
Fig. (\ref{nonextalphaminor2}). Dans ce cas, nous constatons que $p_2 $ r\'eside dans la 
gamme de $(0, 330)$. Pour une petite valeur donn\'ee de $N$, notez bien que la positivit\'e 
de $p_2$ augmente \`a mesure que nous augmentons la valeur de $W$ de z\'ero \`a $10$. 
Enfin, lorsque le seul param\`etre $N_1$ est autoris\'e \`a varier, la stabilit\'e de la
configuration des trous noirs est donn\'ee par la positivit\'e du premier mineur principe 
$p_1:= g_{N_1 N_1}$. Avec une orientation diff\'erente, la vue graphique du mineur $p_1$
est montr\'e dans la Fig. (\ref{nonextalphaminor1}). Les propri\'et\'es ci-dessus de l'espace 
d'\'etat et la positivit\'e des mineurs principaux d\'ecritent le comportement qualitatif 
des fluctuations statistiques de la configuration des trous noirs $D_1 D_5$ nonextremaux
sous les corrections de $\alpha^{\prime}$. 

Avec les $\Gamma_{abc}$ donn\'es dans l'annex $[A]$,
on peut aussi obtenir que la courbure scalaire de Ruppenier est:

\ba R&=& \frac{3}{8 \pi } \frac{(N_1 N_5)^2}{\sqrt{N_R}} \frac{r(N_1, N_5)}{\tilde{g}(N_1, N_5)^3}, \ea

o\`u la fonction $r(N_1, N_5)$ est d\'efinie par

\ba \label{rtild1d5} r(N_1, N_5)&:=& (N_1 N_5)^{25/2}- 48 b^4 (N_1 N_5)^{13/2}+ 8 b^2 (N_1 N_5)^{19/2}\nn &&
+6520 b^6 (N_1 N_5)^{7/2}+ 1700 b^5 (N_1 N_5)^{5}- 1600 b^3 (N_1 N_5)^{1/2}\nn &&
+1920 b^7 (N_1 N_5)^2+ 256 b^3 (N_1 N_5)^8- 9 b (N_1 N_5)^{11}.\ea

On voit simplement qu'il n'y a pas des divergences dans l'espace d'\'etat et cette 
curbure scalaire de Ruppenier est partout r\'eguli\`ere pour chaque non-nulle $\tilde{g}$.
Et bien aussi, on peut voir que la curbure scalaire de Ruppenier est devenue une fraction 
des deux polyn\^omiales. En fait, les equations Eqns. (\ref{gtild1d5}, \ref{rtild1d5}), 
montrent que ces deux fonctions de deux variable peuvent \^etre exprim\'ees comme une 
fonction dans le produit des nombres de $D_1$ et $D_5$. C\'est-\`a-dire que nous 
avons un param\`etre effectif

\ba N:= N_1 N_5.\ea

qui governe la configuration thermodynamique sous-jacente. De plus, dans le cas des trous noirs 
extr\'emaux des deux charges avec $ N_R= 0 $ avec ou sans les corrections de $\alpha^{\prime} $, 
nous voyons que la g\'eom\'etrie thermodynamique sous-jacente de Ruppenier est partout mal-d\'efinie.
En particulier, le d\'eterminant et la courbure scalaire de la g\'eom\'etrie thermodynamique
de Ruppenier agrandissent dans l'espace d'\'etat des branes noirs.

\begin{figure}
\hspace*{1.0cm}\vspace*{-6.0cm}
\includegraphics[width=12.0cm,angle=-0]{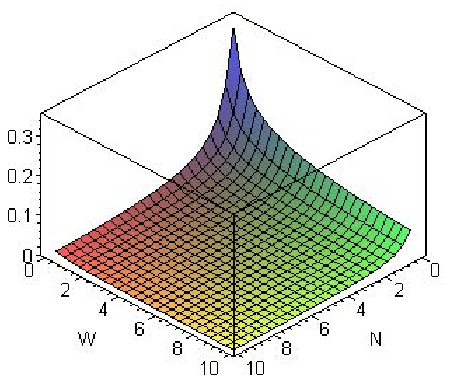}\vspace*{-3.0cm}
\caption{La courbure scalaire trac\'ee comme la fonction de $\{N, W\}$, 
en d\'ecrivant les fluctuations dans la configuration des trous noirs 
charg\'es nonextremaux $D_1 D_5 $ \`a l'ordre de $\alpha^{\prime}$ dans la gamme des charges
$N, W \in (0, 10)$.} \label{nonextalphaR}\vspace*{0.5cm}
\end{figure}

\begin{figure}
\hspace*{1.0cm}\vspace*{-6.0cm}
\includegraphics[width=12.0cm,angle=-0]{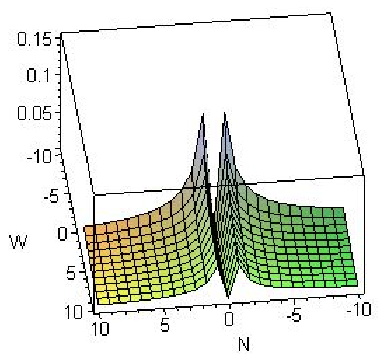}\vspace*{-3.0cm}
\caption{La courbure scalaire trac\'ee comme la fonction de $\{N, W\}$, 
en d\'ecrivant les fluctuations dans la configuration des trous noirs 
charg\'es nonextremaux $D_1 D_5 $ \`a l'ordre de $\alpha^{\prime}$ dans la gamme des charges
$N, W \in (-10, 10)$.} \label{nonextalphaR10}\vspace*{0.5cm}
\end{figure}

Comme mentionn\'e dans le chapitre pr\'ec\'edent, la description de la
g\'eom\'etrie de l'espace d'\'etat reste valable en vertu des corrections 
d\'eriv\'eesup\'erieures de $\alpha^{\prime}$. A savoir, la courbure scalaire 
offre les propri\'et\'es globales des fluctuations statistiques. Dans la gamme de 
$N \in (0, 10)$ et $W \in (0, 10)$, on observe de la Fig. (\ref{nonextalphaR}) que 
la courbure scalaire a une petite positive amplitude \`a l'ordre de $0.4$. 
Dans cette gamme des param\`etres $\{N_1, N_5, N_R\}$, nous constatons que les
trous noirs $D_1 D_5$ nonextremaux corrig\'es par les termes de $\alpha^{\prime}$ 
correspondent \`a une configuration statistique faiblement interactive. 
Comme mentionn\'e pr\'ec\'edemment, le signe positif de la courbure scalaire 
de l'espace d'\'etat signifie que les interactions statistiques sont r\'epulsives 
dans la leur caract\`ere. En outre, la Fig. (\ref{nonextalphaR10}) illustre le 
comportement de ce qui pr\'ec\`ede, voil\`a, la courbure scalaire dans la gamme 
des param\`etres $N, W \in (-10, 10) $. Justement, nous voyons que les interactions 
sont largement pr\'esentes pr\`es de la ligne $N= 0$, et la courbure scalaire sous-jacente 
de l'espace d'\'etat acquiert un petit pic \`a l'ordre de $0.15$ pr\`es de 
l'origine $(N, W)= (0,0)$. Dans une petite gamme de $N$ et $W$, on observe de 
la Fig. (\ref{nonextalphaR10}) que les interactions statistiques globales sont 
sym\'etriques autour de la ligne $N= 0$. En comparaison des interactions 
apparaissant dans la gamme de $N, W \in (0, 10)$, l'amplitude de la courbure scalaire 
globale se r\'ev\'ele plus petits dans la gamme de $n, N \in (-10, 10) $. 
Aussi, \`a partir des Figs. (\ref{nonextalphaR}, \ref{nonextalphaR10}), nous constatons que 
la courbure scalaire de l'espace d'\'etat dans ces deux intervalles ci-dessus reste
pr\`es du m\^{e}me ordre. Qualitativement, les repr\'esentations de l'espace d'\'etat
indiquent que la configuration des trous noirs $D_1D_5$ nonextremaux corrig\'ee par
les termes de $\alpha^{\prime}$  correspond \`a un syst\`eme statistique faiblement 
interagissant et globalement instable. En bref, les corrections de d\'eriv\'ees sup\'erieures
de $\alpha^{\prime}$ am\'eliorent la stabilit\'e de la configuration des trous noirs
$D_1 D_5$ nonextremaux. C'est-\`a-dire que, sous les fluctuation de $\{N_1, N_5, N_R\}$, 
les corrections de $\alpha^{\prime}$ viennet (i) en offrant une surface de l'espace 
d'\'etat avec une positive valeur de mineur $p_2$ et (ii) en r\'eduissant le sommet 
sous-jacent de la courbure scalaire.

\section{La g\'eom\'etrie de Ruppenier des solutions non-extr\'emales de branes $ D_2D_6NS_5 $:}

Dans cette section, il est \'egalement int\'eressant d'analyser la 
g\'eom\'etrie thermodynamique de Ruppenier associ\'ee aux branes 
$D_2D_6NS_5$ non-extr\'emaux. Nous envisagerons maintenant qui les 
solutions non-extr\'emales de branes $D_2D_6NS_5$ qui appara\^issent 
dans l'action effictive de la th\'eorie des cordes de type $IIA$,
pouvent \^etre consid\`ere comme la suite. Pour cette solution,
nous pouvons consid\'erer les branes de $D_2$ dans le sens des directions 
compactes de $ S^1 \times S^{\prime 1}$, les branes de $D_6$ dans des directions
compactes de $S^1\times S^{\prime 1} \times T_4$ et les branes de $NS_5$ dans 
des directions compactes de $ S^1 \times T_4 $,  pour le d\'etails voir
les Refs. \cite{SfetsosSkenderis, Maldacena}. S\'elon cette consid\`eration,
le r\'esultat est int\'eressant que la g\'eom\'etrie de \'espace-temps est le 
produit de $ S^3 \times T_4 $ avec le trou noir de Schwarzschild dans l'$AdS_3$.
Afin d'appliquer le formalisme de la fonction de l'entropie de Sen aux branes
$D_2D_6NS_5$ non-extr\'emaux, nous avons besoin de d\'eformer la g\'eom\'etrie
proche de l'horizon du syst\`eme de ces branes \`a une forme g\'en\'erale du
produit de l'espace de $S^1 \times S^2 \times T_4$ et le trou noir de Schwarzchild
dans l'AdS. En particulier, nous prenons en compte sur les propri\'et\'es des 
solutions non-extr\'emales de branes $D_2D_6NS_5$, et ainsi nous examinons 
la g\'eom\'etrie thermodynamique sous-jacente de ces trous noirs dans le reste
de ce chapitre.

\subsection{\`A l'ordre de $(\alpha^{\prime})^0 $}

Ensuite, la m\'ethode de la fonction d'entropie de Sen donne l'entropie des
branes $D_2D_6NS_5 $ non-extr\'emaux au niveau de l'ordre des deux d\'eriv\'es 
\cite{GarousiGhodsi1}. Donc, s\'elon cette consid\'eration, en termes des nombres, 
ou les charges et les moments de branes, $\{N_2, N_6, N_5, N_R\}$, nous pouvons \'ecrire 
au niveau des deux d\'eriv\'es que l'entropie est donn\'ee par:

\ba S_{BH}(N_2, N_6, N_5, N_R):= 4 \pi \sqrt{N_2 N_6 N_5 N_R}.\ea

Ainsi, il est imm\'ediat que les composantes de la m\'etrique tenseure
sont donn\'ees par:

\ba g_{N_2N_2}&=& \frac{\pi}{N_2} \sqrt{\frac{N_6 N_5 N_R}{N_2}},\nn
g_{N_2N_6}&=& -\pi \sqrt{\frac{N_5 N_R}{N_2 N_6}},\nn
g_{N_2N_5}&=& -\pi \sqrt{\frac{N_6 N_R}{N_2 N_5}},\nn
g_{N_2N_R}&=& -\pi \sqrt{\frac{N_6 N_5}{N_2 N_R}},\ea
\ba g_{N_6N_6}&=& \frac{\pi}{N_6} \sqrt{\frac{N_2 N_5 N_R}{N_6}},\nn
g_{N_6N_5}&=& -\pi \sqrt{\frac{N_2 N_R}{N_6 N_5}},\nn
g_{N_6N_R}&=& -\pi \sqrt{\frac{N_2 N_5}{N_6 N_R}},\nn
g_{N_5N_5}&=& \frac{\pi}{N_5} \sqrt{\frac{N_2 N_6 N_R}{N_5}},\nn
g_{N_5N_R}&=& -\pi \sqrt{\frac{N_2 N_6}{N_5 N_R}},\nn
g_{N_RN_R}&=& \frac{ \pi}{N_R} \sqrt{\frac{N_2 N_6 N_5}{N_R}}.\ea

\begin{figure}
\hspace*{1.0cm}\vspace*{-6.0cm}
\includegraphics[width=12.0cm,angle=-0]{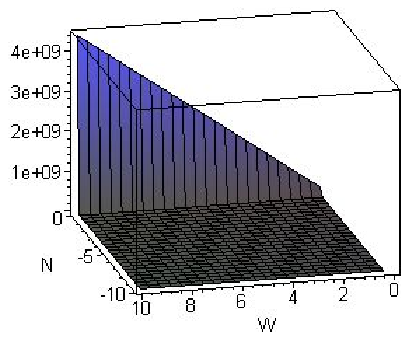}\vspace*{-3.0cm}
\caption{La composante $N_2N_2$ de la m\'etrique tenseure trac\'ee 
comme la fonction de $\{N, W\}$, en d\'ecrivant les fluctuations dans 
la configuration des trous noirs charg\'es nonextremaux $D_2 D_6 NS_5$
\`a l'ordre dominant.} \label{d2d6ns5N2N2}\vspace*{0.5cm}
\end{figure}

\begin{figure}
\hspace*{1.0cm}\vspace*{-6.0cm}
\includegraphics[width=12.0cm,angle=-0]{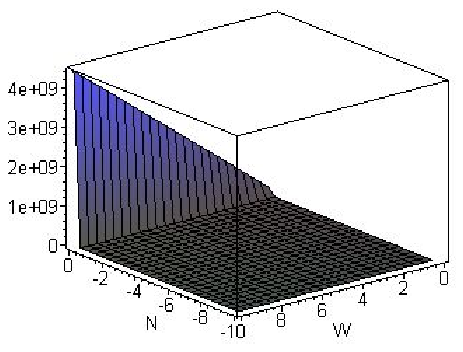}\vspace*{-3.0cm}
\caption{La composante $N_6N_6$ de la m\'etrique tenseure trac\'ee 
comme la fonction de $\{N, W\}$, en d\'ecrivant les fluctuations dans 
la configuration des trous noirs charg\'es nonextremaux $D_2 D_6 NS_5$
\`a l'ordre dominant.} \label{d2d6ns5N6N6}\vspace*{0.5cm}
\end{figure}

\begin{figure}
\hspace*{1.0cm}\vspace*{-6.0cm}
\includegraphics[width=12.0cm,angle=-0]{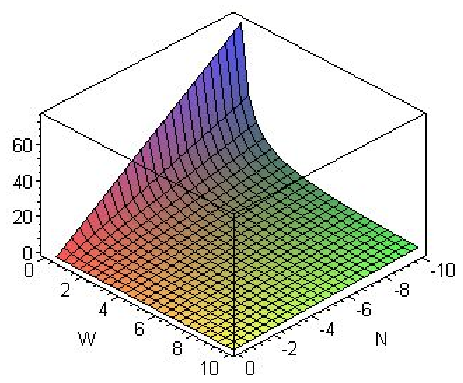}\vspace*{-3.0cm}
\caption{La composante $N_5N_5$ de la m\'etrique tenseure trac\'ee 
comme la fonction de $\{N, W\}$, en d\'ecrivant les fluctuations dans 
la configuration des trous noirs charg\'es nonextremaux $D_2 D_6 NS_5$
\`a l'ordre dominant.} \label{d2d6ns5N5N5}\vspace*{0.5cm}
\end{figure}

\begin{figure}
\hspace*{1.0cm}\vspace*{-6.0cm}
\includegraphics[width=12.0cm,angle=-0]{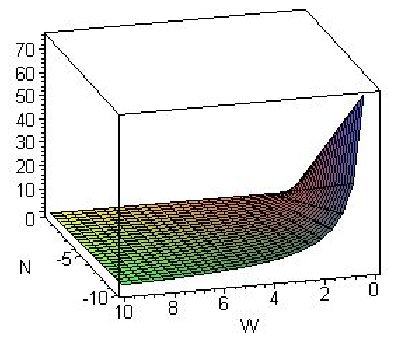}\vspace*{-3.0cm}
\caption{La composante $N_RN_R$ de la m\'etrique tenseure trac\'ee 
comme la fonction de $\{N, W\}$, en d\'ecrivant les fluctuations dans 
la configuration des trous noirs charg\'es nonextremaux $D_2 D_6 NS_5$
\`a l'ordre dominant.} \label{d2d6ns5NRNR}\vspace*{0.5cm}
\end{figure}

\begin{figure}
\hspace*{1.0cm}\vspace*{-6.0cm}
\includegraphics[width=12.0cm,angle=-0]{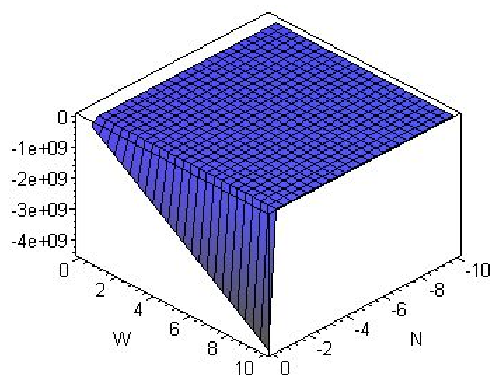}\vspace*{-3.0cm}
\caption{La composante $N_2N_6$ de la m\'etrique tenseure trac\'ee 
comme la fonction de $\{N, W\}$, en d\'ecrivant les fluctuations dans 
la configuration des trous noirs charg\'es nonextremaux $D_2 D_6 NS_5$
\`a l'ordre dominant.} \label{d2d6ns5N2N6}\vspace*{0.5cm}
\end{figure}

\begin{figure}
\hspace*{1.0cm}\vspace*{-6.0cm}
\includegraphics[width=12.0cm,angle=-0]{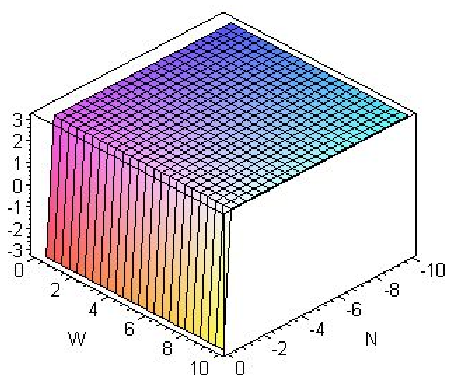}\vspace*{-3.0cm}
\caption{La composante $N_2N_5$ de la m\'etrique tenseure trac\'ee 
comme la fonction de $\{N, W\}$, en d\'ecrivant les fluctuations dans 
la configuration des trous noirs charg\'es nonextremaux $D_2 D_6 NS_5$
\`a l'ordre dominant.} \label{d2d6ns5N2N5}\vspace*{0.5cm}
\end{figure}

\begin{figure}
\hspace*{1.0cm}\vspace*{-6.0cm}
\includegraphics[width=12.0cm,angle=-0]{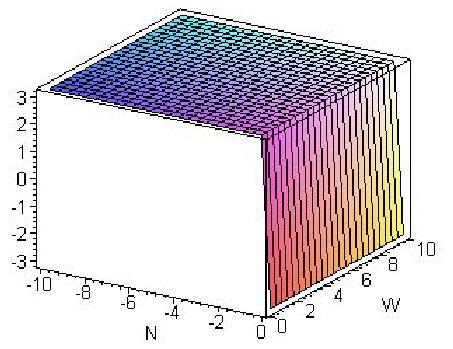}\vspace*{-3.0cm}
\caption{La composante $N_2N_R$ de la m\'etrique tenseure trac\'ee 
comme la fonction de $\{N, W\}$, en d\'ecrivant les fluctuations dans 
la configuration des trous noirs charg\'es nonextremaux $D_2 D_6 NS_5$
\`a l'ordre dominant.} \label{d2d6ns5N2NR}\vspace*{0.5cm}
\end{figure}

\begin{figure}
\hspace*{1.0cm}\vspace*{-6.0cm}
\includegraphics[width=12.0cm,angle=-0]{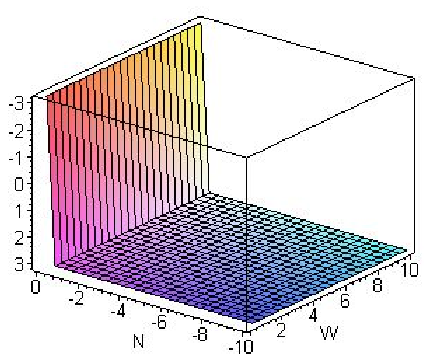}\vspace*{-3.0cm}
\caption{La composante $N_6N_5$ de la m\'etrique tenseure trac\'ee 
comme la fonction de $\{N, W\}$, en d\'ecrivant les fluctuations dans 
la configuration des trous noirs charg\'es nonextremaux $D_2 D_6 NS_5$
\`a l'ordre dominant.} \label{d2d6ns5N6N5}\vspace*{0.5cm}
\end{figure}

\begin{figure}
\hspace*{1.0cm}\vspace*{-6.0cm}
\includegraphics[width=12.0cm,angle=-0]{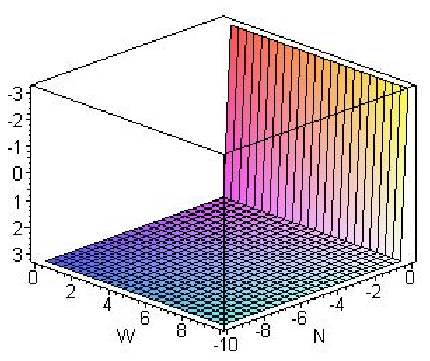}\vspace*{-3.0cm}
\caption{La composante $N_6N_R$ de la m\'etrique tenseure trac\'ee 
comme la fonction de $\{N, W\}$, en d\'ecrivant les fluctuations dans 
la configuration des trous noirs charg\'es nonextremaux $D_2 D_6 NS_5$
\`a l'ordre dominant.} \label{d2d6ns5N6NR}\vspace*{0.5cm}
\end{figure}

\begin{figure}
\hspace*{1.0cm}\vspace*{-6.0cm}
\includegraphics[width=12.0cm,angle=-0]{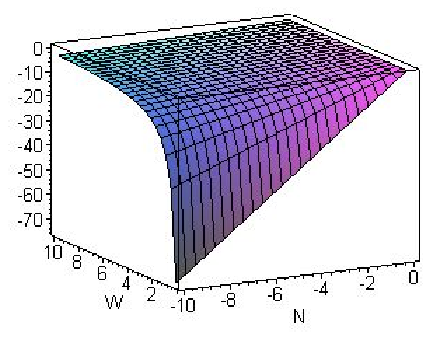}\vspace*{-3.0cm}
\caption{La composante $N_5N_R$ de la m\'etrique tenseure trac\'ee 
comme la fonction de $\{N, W\}$, en d\'ecrivant les fluctuations dans 
la configuration des trous noirs charg\'es nonextremaux $D_2 D_6 NS_5$
\`a l'ordre dominant.} \label{d2d6ns5N5NR}\vspace*{0.5cm}
\end{figure}

Dans ce cas, puisqu'il y a quatre variables ind\'ependantes, c'est-\`a-dire
que nous avous les nombre $\{N_2, N_6, N_5, N_R\}$, et ainsi, dans le but d'offrir 
une vue tridimensionnelle des propri\'et\'es statistiques, nous consid\'erons la 
mise de $N_2= N$, $N_6= N$, $N_5= W$ et $N_R= W$. Pour une configuration donn\'ee 
des trous noirs $D_2 D_6 NS_5$ nonextremaux, la propri\'et\'e des fluctuations peut 
\^{e}tre exprim\'ee par l'ensemble suivant. \`A savoir, dans le r\'egime de 
$N \in (0, -10)$ et $W \in (0, 10)$, nous voyons que l'amplitude des capacit\'es 
de chaleurs $\{g_{N_2 N_2}, g_{N_6 N_6}\}$ prend les valeurs \`a l'ordre de 
$4 \times 10^{+09}$. Dans cette gamme de $N, W$, les capacit\'es de chaleurs 
$\{g_{N_5 N_5}, g_{N_R N_R}\}$ restent dans l'intervalle $(0, 80) $. Conform\'ement, 
\`a la pr\'ediction de la g\'eom\'etrie de l'espace d'\'etat, nous observons que 
les gammes de la premi\`ere s\'erie et celle de la seconde s\'erie des capacit\'es de 
chaleurs restent dans la limite opposite des param\`etres $\{N, W\}$. Plus pr\'ecis\'ement, 
pour une petite valeur donn\'ee de $N$, la premi\`ere s\'erie des capacit\'es de chaleurs 
augmente lorsqu'on augmente le param\'etre $W$. Cependant, pour une petite valeur donn\'ee
de $W$, l'autre s\'erie des composantes augmentent lorsqu'on augmente la valeur $N$. 
Dans ce cas, les Figs. (\ref{d2d6ns5N2N2}, \ref{d2d6ns5N6N6}) montrent que la croissance 
de premier ensemble des capacit\'es de chaleurs $\{g_{} N_2 N_2, g_{N_6 N_6}\}$ a lieu dans 
la limite d'un grand $W$ et un petit $|N|$. Du fait m\^{e}me, les Figs. (\ref{d2d6ns5N5N5}, 
\ref{d2d6ns5NRNR}) montrent que la croissance de $\{g_{N_5 N_5}, g_{N_R N_R}\}$ a lieu dans 
la limite d'un petit $W$ et un grand $|N|$. De plus, les compressibilit\'es de chaleurs en
impliquant deux param\`etres distincts de la configuration des trous noirs sous-jacents 
sont repr\'esent\'es dans les Figs. (\ref{d2d6ns5N2N6},\ref{d2d6ns5N2N5}, \ref{d2d6ns5N2NR}, 
\ref{d2d6ns5N6N5}, \ref{d2d6ns5N6NR}, \ref{d2d6ns5N5NR}). Dans ce cas, nous constatons que 
l'amplitude de (i) $g_{N_2 N_6} $ reste environt de $-4 \times 10^{09}$, 
(ii) $g_{N_2 N_5} $, $g_{N_6 N_5} $, $g_{N_6 N_R}$ et $g_{N_2N_R}$ reste 
dans l'intervalle $(-3, 3)$ et (iii) $g_{N_5 N_R} $ reste dans l'intervalle $(0, -80)$.
Ainsi, la repr\'esentation graphique de la m\'etrique tenseure de l'espace d'\'etat
$\{g_{ij} \ | \ i, j = N_2, N_6, N_5, N_R\}$ montre que la configuration des trous noirs 
$D_2 D_6 NS_5$ nonextremaux, \`a l'ordre dominant de $\alpha^{\prime}$ de la th\'eorie de cordes, 
poss\`ede un ensemble des capacit\'es de chaleurs positives, et donc elle correspond \`a un 
syst\`eme statistique stables au niveau local.

En fait, il s'av\`ere que les mineurs principaux de la configuration
sous-jacente peuvent \^{e}tre simplifi\'es comme les expressions suivantes 

\begin{eqnarray}
\mathit{p_1} &=& {\displaystyle \frac{\pi \,\mathit{N_6}^{2}\,
\mathit{N_5}^{2}\,\mathit{N_R}^{2}}{(\mathit{N_2}\,\mathit{N_6}\,
\mathit{N_5}\,\mathit{N_R})^{(3/2)}}},  \nonumber \\
\mathit{p_2} &=& 0,  \nonumber \\
\mathit{p_3} &=& - 4\,{\displaystyle \frac{\pi^{3}\,
\mathit{N_R}^{3}\,\mathit{N_2}\,\mathit{N_6}\,\mathit{N_5}}{(\mathit{N_2}\,
\mathit{N_6}\,\mathit{N_5}\,\mathit{N_R})^{(3/2)}}}. 
\end{eqnarray}

Ci-joint, nous voyons que le d\'eterminant de la m\'etrique tenseure est:

\ba g&=& -16 \pi^4.\ea

\begin{figure}
\hspace*{1.0cm}\vspace*{-6.0cm}
\includegraphics[width=12.0cm,angle=-0]{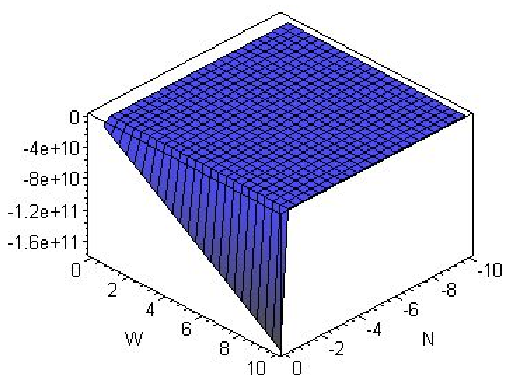}\vspace*{-3.0cm}
\caption{Le mineur d'hypersurface de la m\'etrique tenseure trac\'ee 
comme la fonction de $\{N, W\}$, en d\'ecrivant les fluctuations dans 
la configuration des trous noirs charg\'es nonextremaux $D_2 D_6 NS_5$
\`a l'ordre dominant.} \label{d2d6ns5minor3}\vspace*{0.5cm}
\end{figure}

\begin{figure}
\hspace*{1.0cm}\vspace*{-6.0cm}
\includegraphics[width=12.0cm,angle=-0]{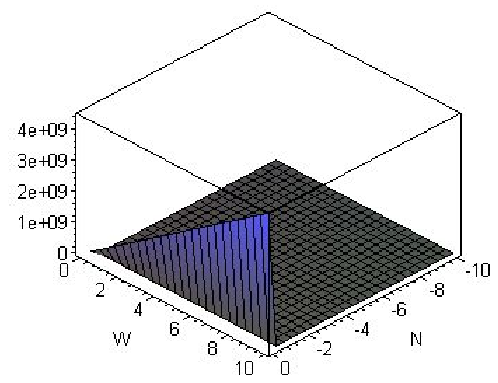}\vspace*{-3.0cm}
\caption{Le premier mineur de la m\'etrique tenseure trac\'ee 
comme la fonction de $\{N, W\}$, en d\'ecrivant les fluctuations dans 
la configuration des trous noirs charg\'es nonextremaux $D_2 D_6 NS_5$
\`a l'ordre dominant.} \label{d2d6ns5minor1}\vspace*{0.5cm}
\end{figure}

Sous les fluctuations des param\`etres $\{N_2, N_6, N_5, N_R\}$,
la stabilit\'e d'un ensemble des trous noirs $D_2 D_6 NS_5$ nonextremaux 
d\'ecoule de la positivit\'e du d\'eterminant de la m\'etrique tenseure.
Pour le choix des param\`etres $N_2= N$, $N_5= N$ et $W= N_R$, nous constatons 
que le d\'eterminant de la m\'etrique tenseure $g$ reste \`a une constante valeur 
n\'egative. Dans ce cas, il s'av\`ere que la valeur num\'erique de $g$ est $-16 \pi^4$.
Par ailleurs, la stabilit\'e de l'hypersurface d\'efinie par une valeur constante de 
$N_R$ est montr\'e dans la Fig. (\ref{d2d6ns5minor3}). Par la pr\'esente, nous voyons 
que le mineur $p_3$ r\'eside dans la gamme de $(-1.6 \times 10^{+11}, 0)$. 
Pour une petite valeur donn\'ee de $N$, on remarque que la n\'egativit\'e de $p_3$
diminu\'e quand la valeur du param\`etre $N$ est pass\'ee de z\'ero \`a $10$. 
Du plus, la surface d\'efinies par les fluctuations de $\{N_2, N_6\}$ est instable 
en raison du fait que le mineur principe correspondant est identiquement nulle, \`a savoir que
nous avons $p_2= 0$. Enfin, lorsque le seul param\`etre $N_1$ est autoris\'e \`a varier, 
nous constatons que la stabilit\'e de la configuration des trous noirs sous-jacents est donn\'ee 
par la positivit\'e de premier mineur principe $p_1:= g_{N_2 N_2}$. Une vue rotat\'ee de $p_1$
est donn'ee dans la Fig. (\ref{d2d6ns5minor1}). Ci-dessus, les propri\'et\'es de l'espace 
d'\'etat et la positivit\'e des mineurs principaux offrent le comportement qualitatif de la
stabilit\'e statistique de la configuration des trous noirs $D_2D_6 NS_5$ nonextremaux.

Donc, comme nous avons offrir les $\Gamma_{abc}$ dans l'annex $[A]$, 
il est aussi imm\'ediat de voir que la courbure scalaire de Ruppenier est: 

\ba R&=& \frac{3}{4 \pi} (N_2 N_6 N_5 N_R)^{-1/2}\ea 

qu'elle est partout r\'eguli\`ere. Dans ce cas, on peut \'egalement voir 
dans les sections $5.1$ et $5.2$ qu'il y a une diff\'erence \'evidente sans les 
corrections des d\'eriv\'ees sup\'erieures de $\alpha^{\prime}$, par example, comme
nous avons montr\'e ici, cela vient aussi dans la g\'eom\'etrie thermodynamique des 
trous noirs dyoniques extr\'emaux supersym\'etriques vivant en quatre dimensions et 
celle des trous noirs dyoniques extr\'emaux non-supersym\'etriques.

\begin{figure}
\hspace*{1.0cm}\vspace*{-6.0cm}
\includegraphics[width=12.0cm,angle=-0]{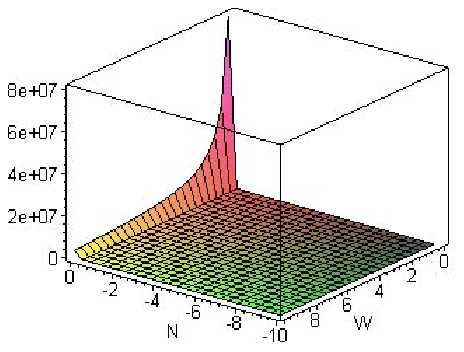}\vspace*{-3.0cm}
\caption{La courbure scalaire trac\'ee comme la fonction de $\{N, W\}$, 
en d\'ecrivant les fluctuations dans la configuration des trous noirs 
charg\'es nonextremaux $D_2 D_6 NS_5$ \`a l'ordre dominant dans la gamme
$n \in (-10,0)$ et $N \in (0,10)$.} \label{d2d6ns5R}\vspace*{0.5cm}
\end{figure}

\begin{figure}
\hspace*{1.0cm}\vspace*{-6.0cm}
\includegraphics[width=12.0cm,angle=-0]{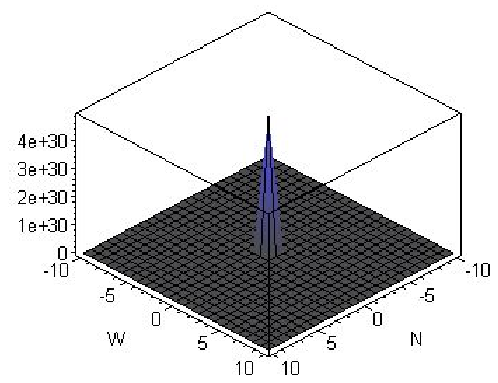}\vspace*{-3.0cm}
\caption{La courbure scalaire trac\'ee comme la fonction de $\{N, W\}$, 
en d\'ecrivant les fluctuations dans la configuration des trous noirs 
charg\'es nonextremaux $D_2 D_6 NS_5$ \`a l'ordre dominant dans la
gamme $n, N \in (-10, 10)$.} \label{d2d6ns5R10}\vspace*{0.5cm}
\end{figure}

Comme mentionn\'e dans le cas pr\'ec\'edent, la courbure scalaire d'espace 
d'\'etat offre les propri\'et\'es statistiques globales sous la fluctuation 
des param\`etres $\{N_2, N_6, N_5, N_R\}$. Dans la gamme de $N \in (0, -10)$
et $W \in (0, 10) $, la Fig. (\ref{d2d6ns5R}) montre que la scalaire courbure 
sous-jacente a une grande amplitude positive \`a l'ordre de $8 \times 10^{+07}$. 
Dans cette gamme des param\`etres, nous observons que la configuration sous-jacente
des trous noirs est un syst\`eme statistique fortement interagissant. Physiquement,
le signe positif de la courbure scalaire signifie que les interactions statistiques
sont r\'epulsives dans la leur nature. La Fig. (\ref{d2d6ns5R10}) illustre le comportement 
de la courbure scalaire ci-dessus pour la range des param\`etres $N, W \in (-10, 10)$. 
Dans la gamme de $N \in (0, -10) $ et $W \in (0, 10) $, nous voyons que les interactions 
sont largement pr\'esents pr\`es de la ligne $N = 0 $. Dans la gamme de $N, W \in (-10, 10)$,
la courbure scalaire de l'espace d'\'etat acquiert un grand pic \`a l'ordre de $4 \times 10^{30}$
pr\`es de l'origine $(N, W) = (0, 0)$. Dans la gamme de petit $N$ et un petit $W$, on constate 
de la Fig. (\ref{d2d6ns5R10}) que les interactions statistiques globales ne sont pr\'esents pr\`es de 
l'origine. En comparaison des interactions apparaissant dans la gamme des param\`etres $N, W \in (0,10)$, 
nous voyons que l'amplitude des interactions statistiques globales se r\'ev\`ele \^{e}tre beaucoup 
plus grande dans la gamme de $N, W \in (-10, 10) $. Pr\'ecis\'ement, il s'ensuit que le ratio des 
valeurs typiques de l'amplitude de ces courbures scalaires de l'espace d'\'etat est \`a l'ordre 
de $10^{-23}$. La vue graphique de la comparaison mentionn\'e ci-dessus est montr\'ee par les Figs. 
(\ref{d2d6ns5R}, \ref{d2d6ns5R10}). Qualitativement, les repr\'esentations de l'espace d'\'etat
indiquent que la configuration des trous noirs $D_2D_6NS_5$ nonextremaux de la th\'eorire des 
cordes correspond \`a un syst\`eme statistique globalement instable et fortement interactive. 


\subsection{\`A l'ordre de $\alpha^{\prime} $}

D'autre part, l'examen des contributions non-nulle de la courbure tenseure 
de Weyl de la th\'eorie effective de type-IIA, en respectant les sym\'etries
de la solution au niveau de l'arbre nous permet d'analyser la configuration
thermodynamique par la fonction d'entropie de Sen avec l'horizon de trou noir
comme le produit $(S^1 \times S^{\prime 1} \times S^2 \times T_4 )$, et ainsi 
s\'elon le Ref. \cite{GarousiGhodsi1}, nous pouvons \'ecrire l'entropie de la 
solution non-extr\'emale des branes $ D_2D_6NS_5 $. D\`es que les termes des 
d\'eriv\'ees sup\'erieures respectent la sym\'etrie de la solution au niveau de l'arbre. 
En cette fois encore, le Ref. \cite{GarousiGhodsi1} montre que l'entropie est corrig\'e comme:

\ba S_{BH}(N_2, N_6, N_5, N_R)&:=& 4 \pi \sqrt{N_2 N_6 N_5 N_R}-
\frac{4 \pi b_1 \sqrt{N_R}}{N_2 N_6 N_5^{5/2}},\ea

o\`u $ b_1 $ est un coefficient des contributions de d\'eriv\'ees sup\'erieures
de $ \alpha^{\prime} $ \`a la densit\'e lagrangienne effective \cite{GarousiGhodsi1}.
Maintenant, soit $ \mathcal M $ une vari\'et\'e riemannienne avec la m\'etrique 
tenseure de Ruppenier $ g_{ij} $ ce qu'elle peuisse \^etre aussi \'ecrite comme:

\ba g_{ij}:= g^{(arbre)}_{ij}+ g^{(b_1)}_{ij}.\ea

On peut naturellement rapporter la correction $ g^{(b_1)}_{ij} $ 
dans la m\'etrique tenseure $ g_{ij} $. C'est-\`a-dire qu'on a besoin d'apporter 
ces corrections \`a la m\'etrique de cette vari\'et\'e de Ruppenier ce que 
la m\'etrique tenseure de la $ \mathcal M $ soit proprement rep\`er\'ee au point 
des corrections d\'eriv\'ees sup\'erieures de $\alpha^{ \prime }$. Ensuite, 
il est ais\'e de constater que les composantes de la m\'etrique avec les 
contributions de $\alpha^{\prime}$ sont donn\'ees par: 

\ba g_{N_2N_2}&=& \frac{\pi}{N_2} \sqrt{\frac{N_6 N_5 N_R}{N_2}}
              +\frac{8 \pi b_1 \sqrt{N_R}}{N_2^3 N_6 N_5^{5/2}},\nn
g_{N_2N_6}&=& -\pi \sqrt{\frac{N_5 N_R}{N_2 N_6}} 
              +\frac{4 \pi b_1 \sqrt{N_R}}{N_2^2 N_6^2 N_5^{5/2}} ,\nn
g_{N_2N_5}&=& -\pi \sqrt{\frac{N_6 N_R}{N_2 N_5}} 
              +\frac{10 \pi b_1 \sqrt{N_R}}{N_2^2 N_6 N_5^{7/2}} ,\nn
g_{N_2N_R}&=& -\pi \sqrt{\frac{N_6 N_5}{N_2 N_R}}
              -\frac{2 \pi b_1}{N_2^2 N_6 N_5^{5/2}\sqrt{N_R}} ,\ea
\ba g_{N_6N_6}&=& \frac{\pi}{N_6} \sqrt{\frac{N_2 N_5 N_R}{N_6}}
              +\frac{8 \pi b_1 \sqrt{N_R}}{N_2 N_6^3 N_5^{5/2}} ,\nn
g_{N_6N_5}&=& -\pi \sqrt{\frac{N_2 N_R}{N_6 N_5}}
              +\frac{10 \pi b_1 \sqrt{N_R}}{N_2 N_6^2 N_5^{7/2}} ,\nn
g_{N_6N_R}&=& -\pi \sqrt{\frac{N_2 N_5}{N_6 N_R}} 
              -\frac{2 \pi b_1}{N_2 N_6^2 N_5^{5/2}\sqrt{N_R}} ,\nn
g_{N_5N_5}&=& \frac{\pi}{N_5} \sqrt{\frac{N_2 N_6 N_R}{N_5}}
              +\frac{35 \pi b_1 \sqrt{N_R}}{N_2 N_6 N_5^{9/2}} ,\nn
g_{N_5N_R}&=& -\pi \sqrt{\frac{N_2 N_6}{N_5 N_R}}
              -\frac{5 \pi b_1}{N_2 N_6 N_5^{7/2}\sqrt{N_R}}  ,\nn
g_{N_RN_R}&=& \frac{ \pi}{N_R} \sqrt{\frac{N_2 N_6 N_5}{N_R}}
             -\frac{\pi b_1}{N_2 N_6 N_5^{5/2}N_R^{3/2}}.\ea

\begin{figure}
\hspace*{1.0cm}\vspace*{-6.0cm}
\includegraphics[width=12.0cm,angle=-0]{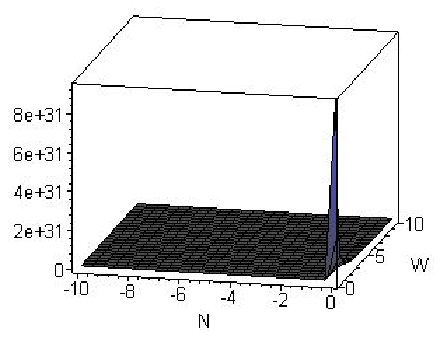}\vspace*{-3.0cm}
\caption{La composante $N_2N_2$ de la m\'etrique tenseure trac\'ee 
comme la fonction de $\{N, W\}$, en d\'ecrivant les fluctuations dans 
la configuration des trous noirs charg\'es nonextremaux $D_2 D_6 NS_5$
\`a l'ordre de $\alpha^{\prime}$.} \label{d2d6ns5alphaN2N2}\vspace*{0.5cm}
\end{figure}

\begin{figure}
\hspace*{1.0cm}\vspace*{-6.0cm}
\includegraphics[width=12.0cm,angle=-0]{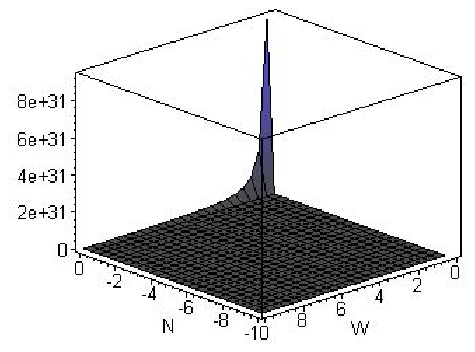}\vspace*{-3.0cm}
\caption{La composante $N_6N_6$ de la m\'etrique tenseure trac\'ee 
comme la fonction de $\{N, W\}$, en d\'ecrivant les fluctuations dans 
la configuration des trous noirs charg\'es nonextremaux $D_2 D_6 NS_5$
\`a l'ordre de $\alpha^{\prime}$.} \label{d2d6ns5alphaN6N6}\vspace*{0.5cm}
\end{figure}

\begin{figure}
\hspace*{1.0cm}\vspace*{-6.0cm}
\includegraphics[width=12.0cm,angle=-0]{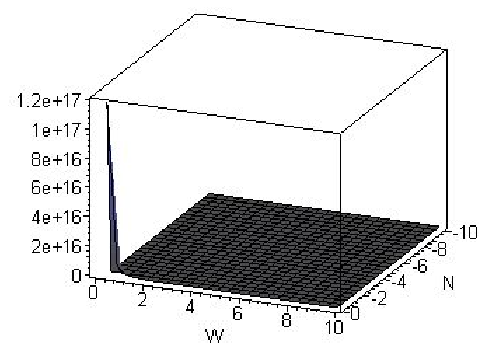}\vspace*{-3.0cm}
\caption{La composante $N_5N_5$ de la m\'etrique tenseure trac\'ee 
comme la fonction de $\{N, W\}$, en d\'ecrivant les fluctuations dans 
la configuration des trous noirs charg\'es nonextremaux $D_2 D_6 NS_5$
\`a l'ordre de $\alpha^{\prime}$.} \label{d2d6ns5alphaN5N5}\vspace*{0.5cm}
\end{figure}

\begin{figure}
\hspace*{1.0cm}\vspace*{-6.0cm}
\includegraphics[width=12.0cm,angle=-0]{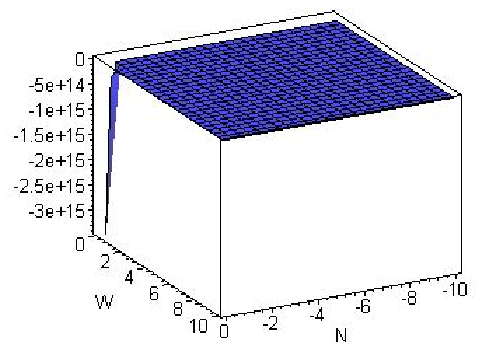}\vspace*{-3.0cm}
\caption{La composante $N_RN_R$ de la m\'etrique tenseure trac\'ee 
comme la fonction de $\{N, W\}$, en d\'ecrivant les fluctuations dans 
la configuration des trous noirs charg\'es nonextremaux $D_2 D_6 NS_5$
\`a l'ordre de $\alpha^{\prime}$.} \label{d2d6ns5alphaNRNR}\vspace*{0.5cm}
\end{figure}

\begin{figure}
\hspace*{1.0cm}\vspace*{-6.0cm}
\includegraphics[width=12.0cm,angle=-0]{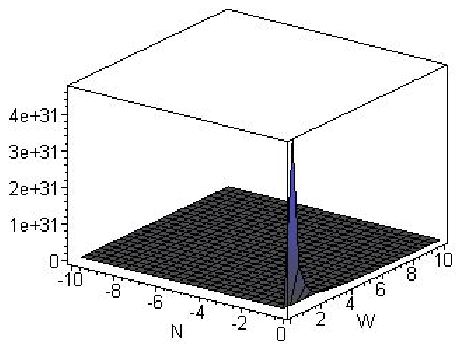}\vspace*{-3.0cm}
\caption{La composante $N_2N_6$ de la m\'etrique tenseure trac\'ee 
comme la fonction de $\{N, W\}$, en d\'ecrivant les fluctuations dans 
la configuration des trous noirs charg\'es nonextremaux $D_2 D_6 NS_5$
\`a l'ordre de $\alpha^{\prime}$.} \label{d2d6ns5alphaN2N6}\vspace*{0.5cm}
\end{figure}

\begin{figure}
\hspace*{1.0cm}\vspace*{-6.0cm}
\includegraphics[width=12.0cm,angle=-0]{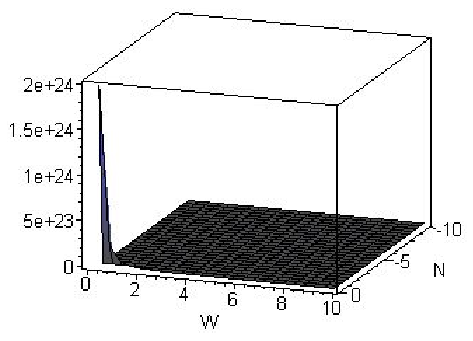}\vspace*{-3.0cm}
\caption{La composante $N_2N_5$ de la m\'etrique tenseure trac\'ee 
comme la fonction de $\{N, W\}$, en d\'ecrivant les fluctuations dans 
la configuration des trous noirs charg\'es nonextremaux $D_2 D_6 NS_5$
\`a l'ordre de $\alpha^{\prime}$.} \label{d2d6ns5alphaN2N5}\vspace*{0.5cm}
\end{figure}

\begin{figure}
\hspace*{1.0cm}\vspace*{-6.0cm}
\includegraphics[width=12.0cm,angle=-0]{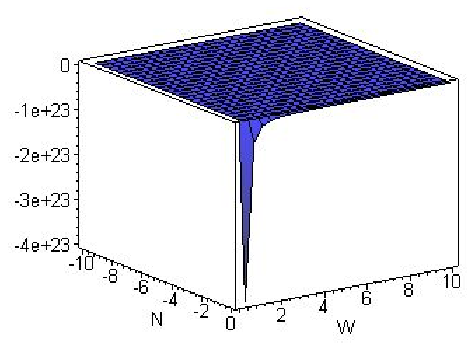}\vspace*{-3.0cm}
\caption{La composante $N_2N_R$ de la m\'etrique tenseure trac\'ee 
comme la fonction de $\{N, W\}$, en d\'ecrivant les fluctuations dans 
la configuration des trous noirs charg\'es nonextremaux $D_2 D_6 NS_5$
\`a l'ordre de $\alpha^{\prime}$.} \label{d2d6ns5alphaN2NR}\vspace*{0.5cm}
\end{figure}

\begin{figure}
\hspace*{1.0cm}\vspace*{-6.0cm}
\includegraphics[width=12.0cm,angle=-0]{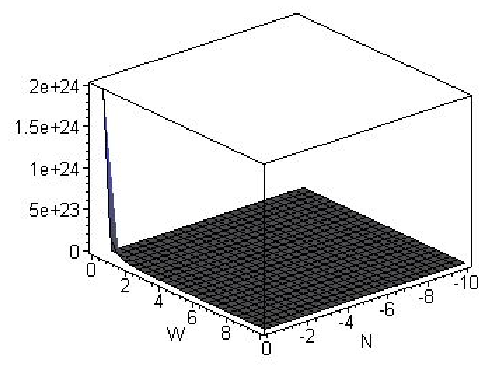}\vspace*{-3.0cm}
\caption{La composante $N_6N_5$ de la m\'etrique tenseure trac\'ee 
comme la fonction de $\{N, W\}$, en d\'ecrivant les fluctuations dans 
la configuration des trous noirs charg\'es nonextremaux $D_2 D_6 NS_5$
\`a l'ordre de $\alpha^{\prime}$.} \label{d2d6ns5alphaN6N5}\vspace*{0.5cm}
\end{figure}

\begin{figure}
\hspace*{1.0cm}\vspace*{-6.0cm}
\includegraphics[width=12.0cm,angle=-0]{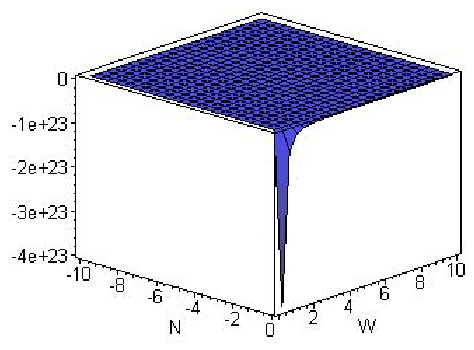}\vspace*{-3.0cm}
\caption{La composante $N_6N_R$ de la m\'etrique tenseure trac\'ee 
comme la fonction de $\{N, W\}$, en d\'ecrivant les fluctuations dans 
la configuration des trous noirs charg\'es nonextremaux $D_2 D_6 NS_5$
\`a l'ordre de $\alpha^{\prime}$.} \label{d2d6ns5alphaN6NR}\vspace*{0.5cm}
\end{figure}

\begin{figure}
\hspace*{1.0cm}\vspace*{-6.0cm}
\includegraphics[width=12.0cm,angle=-0]{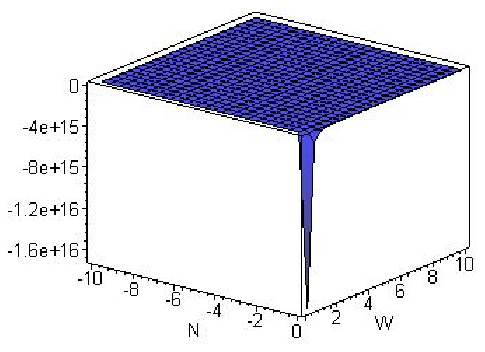}\vspace*{-3.0cm}
\caption{La composante $N_5N_R$ de la m\'etrique tenseure trac\'ee 
comme la fonction de $\{N, W\}$, en d\'ecrivant les fluctuations dans 
la configuration des trous noirs charg\'es nonextremaux $D_2 D_6 NS_5$
\`a l'ordre de $\alpha^{\prime}$.} \label{d2d6ns5alphaN5NR}\vspace*{0.5cm}
\end{figure}

Comme mentionn\'e pr\'ec\'edemment, nous allons continuer avec la mise de 
$N_2= N$, $N_6= N$, $N_5= W$ et $N_R= W$. Afin d'offrir les repr\'esentations 
graphiques de trois dimensions des fluctuations de la configuration des trous noirs 
$D_2D_6 NS_5$ nonextremaux corrig\'es par le dominant terme de $\alpha^{\prime}$, 
nous pouvons choisir la valeur du param\`etre des corrections d\'eriv\'ees sup\'erieures 
de la th\'eorie des cordes comme $b_1= 0.001644670833$. Dans le r\'egime de 
$N \in (-10, 0)$ et $W \in (0, 10)$, nous constatons que l'amplitude des capacit\'es de chaleurs 
$\{g_{N_2 N_2}, g_{N_6 N_6}\}$ prend une valeur dans l'intervalle $(0, 10^{+32}$. 
Dans cette gamme de $N, W$, la capacit\'e thermique $\{g_{N_5 N_5}\}$ est compris\'ee 
entre $(0, 1.2 \times 10^{+17})$. Cependant, la capacit\'e thermique $\{g_{N_R N_R}\}$ 
change sa nature et elle se situe dans la gamme de $(-3,5 \times 10^{15}, 0) $. 
Le signe n\'egatif de $\{g_{N_R N_R}\}$ montre une instabilit\'e locale du syst\`eme 
statistique de ces trous noirs. De ce qui pr\'ec\`ede de la pr\'evision des corrections 
de $\alpha^{\prime}$ \`a la g\'eom\'etrie de l'espace d'\'etat, nous observons que 
la range de toutes les capacit\'es de chaleurs $\{g_{N_2 N_2}, g_{N_6 N_6}, g_{N_5 N_5},
g_{N_R N_R}\}$ appara\^{i}t dans la limite d'une petite valeur des param\`etres $\{N, W\}$.
Notez bien que les deux premi\`eres composantes ont des pics \`a l'ordre de $10^{+32} $, la
composante $g_{N_5 N_5} $ a le pic \`a l'ordre de $10^{+17}$ et la composante $\{g_{N_R N_R}\}$ 
a le pic \`a l'ordre $10^{+15}$. En fait, sous les corrections de $\alpha^{\prime}$, 
nous constatons que la nature graphique des capacit\'es de chaleurs est alt\'er\'e. 
Plus pr\'ecis\'ement, nous observons que les capacit\'es de chaleurs ont un grand pic 
pour certaines petites valeurs de $\{N, W\}$. Dans ce cas, les Figs. (\ref{d2d6ns5alphaN2N2},
\ref{d2d6ns5alphaN6N6}) montrent que la croissance des capacit\'es de chaleurs $\{g_{N_2N_2}, 
g_{N_6 N_6}\}$ a lieu dans la limite d'un petit $W$ et un petit $|N|$. De plus, les Figs. 
(\ref{d2d6ns5alphaN5N5}, \ref{d2d6ns5alphaNRNR}) montrent que la croissance de 
$\{g_{N_5N_5}, g_{N_R N_R}\}$ a lieu dans la limite des petits $W$ et $|N|$. 
Il est \`a noter que le premier trois des capacit\'es de chaleurs trouvent un ensemble 
des valeurs num\'eriques positives, tandis que la quatri\`eme n'a pas. D'autre part, 
les Figs. (\ref{d2d6ns5alphaN2N6}, \ref{d2d6ns5alphaN2N5}, \ref{d2d6ns5alphaN2NR}, 
\ref{d2d6ns5alphaN6N5}, \ref{d2d6ns5alphaN6NR}, \ref{d2d6ns5alphaN5NR}) d\'epeignent 
les compressibilit\'es de chaleurs m\'elang\'es, en impliquant deux param\`etres distincts 
des trous noirs $D_2D_6NS_5$ nonextremaux sous les corrections de $\alpha^{\prime}$.
Par la pr\'esente, nous remarquons que l'amplitude des 
(i) $g_{N_2 N_6}$ est \`a l'ordre de $10{+31}$, 
(ii) $g_{N_2 N_5}$ est \`a l'ordre de $10{+24}$, 
(iii) $g_{N_2 N_R}$ est \`a l'ordre de $-10{+23}$, 
(iv) $g_{N_6 N_5}$ est \`a l'ordre de $10{+24}$, 
(v) $g_{N_6N_R}$ est \`a l'ordre de $-10{+23}$, 
et (vi) $g_{N_5 N_R}$ est \`a l'ordre de $-10{+16}$. 
Comme montr\'e ci-dessus par la repr\'esentation graphique, 
les composantes de la m\'etrique tenseure de l'espace d'\'etat
$\{g_{ij} \ | \ i, j = N_2, N_6, N_5, N_R\}$ illustrent le fait que
les trous noirs $D_2D_6NS_5$ nonextremaux corrig\'ees par les
$\alpha^{\prime}$ correspondent \`a une configuration statistique
localement stable.

En outre, sous les corrections de d\'eriv\'ees sup\'erieures  de la th\'eorie des cordes, 
les mineurs de la ligne, surface et d'hypersurface sont simplifi\'ees comme les suivants

\begin{eqnarray}
\mathit{p_1} &=&  {\displaystyle \frac
{\pi \,(\mathit{N_2}^{3}\,
\mathit{N_6}^{3}\,\mathit{N_R}^{2}\,\mathit{N_5}^{(9/2)} + 8\,
\mathit{b_1}\,\sqrt{\mathit{N_R}}\,(\mathit{N_2}\,\mathit{N_6}\,
\mathit{N_5}\,\mathit{N_R})^{(3/2)})}{(\mathit{N_2}\,\mathit{N_6}\,
\mathit{N_5}\,\mathit{N_R})^{(3/2)}\,\mathit{N_2}^{3}\,\mathit{N_6}\,
\mathit{N_5}^{(5/2)}}},  \nonumber \\
\mathit{p_2} &=&  24\,{\displaystyle \frac{\pi^{2}\,\mathit{b_1}\,
(\mathit{N_2}\,\mathit{N_5}^{(5/2)}\,\sqrt{\mathit{N_R}}\,\mathit{N_6}
\,\sqrt{\mathit{N_2}\,\mathit{N_6}\,\mathit{N_5}\,\mathit{N_R}} +
2\,\mathit{b_1}\,\mathit{N_R})}{\mathit{N_5}^{5}\,\mathit{N_6}^{4}\,
\mathit{N_2}^{4}}},  \nonumber \\
\mathit{p_3} &=&  4\pi ^{3}\,\mathit{N_R}(24\,\mathit{N_2}^{3}\,
\mathit{N_6}^{3}\,\sqrt{\mathit{N_R}}\,\mathit{N_5}^{6}\,\mathit{b_1}
\,\sqrt{\mathit{N_2}\,\mathit{N_6}\,\mathit{N_5}\,\mathit{N_R}}
\nonumber \\ && +
162\,\mathit{N_2}^{2}\,\mathit{N_6}^{2}\,\mathit{N_R}\,\mathit{N_5}^{(7/
2)}\,\mathit{b_1}^{2} \mbox{} +
220\,\mathit{b_1}^{3}\,\sqrt{\mathit{N_R}}\,\sqrt{
\mathit{N_2}\,\mathit{N_6}\,\mathit{N_5}\,\mathit{N_R}} \nonumber
\\ && -\mathit{N_2}^{5}\,\mathit{N_6}^{5}\,\mathit{N_R}\,\mathit{N_5}^{(19/2)})
(\mathit{N_2}^{5}\,\mathit{N_6}^{5}\,\mathit{N_5}^{(19/2)}
\sqrt{\mathit{N_2}\,\mathit{N_6}\,\mathit{N_5}\,\mathit{N_R}})^{(-1)}. 
\end{eqnarray}

Il est aussi imm\'ediat que le d\'eterminant de la
m\'etrique tenseure $ g_{ij} $ est donn\'ee par:

\ba g&=& -16 \pi^4(N_2 N_6 N_5^2)^{-6} \tilde{g}(N_2, N_6, N_5), \ea

o\`u la fonction $\tilde{g}(N_2, N_6, N_5)$ est d\'efinie par

\ba \tilde{g}(N_2, N_6, N_5)&:=& 
(N_2 N_6 N_5^2)^6+ 6 b_1^2 (N_2 N_6 N_5^2)^3+ 100 b_1^4 \nn &&+ 
50 b_1^3 (N_2 N_6 N_5^2)^{3/2}+ 5 b_1 (N_2 N_6 N_5^2)^{9/2}.\ea

\begin{figure}
\hspace*{1.0cm}\vspace*{-6.0cm}
\includegraphics[width=12.0cm,angle=-0]{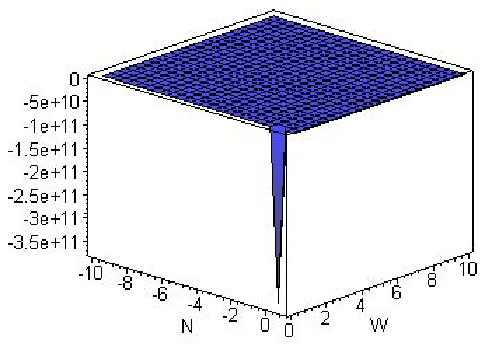}\vspace*{-3.0cm}
\caption{Le d\'eterminant de la m\'etrique tenseure trac\'ee 
comme la fonction de $\{N, W\}$, en d\'ecrivant les fluctuations dans 
la configuration des trous noirs charg\'es nonextremaux $D_2 D_6 NS_5$
\`a l'ordre de $\alpha^{\prime}$.} \label{d2d6ns5alphadetg}\vspace*{0.5cm}
\end{figure}

\begin{figure}
\hspace*{1.0cm}\vspace*{-6.0cm}
\includegraphics[width=12.0cm,angle=-0]{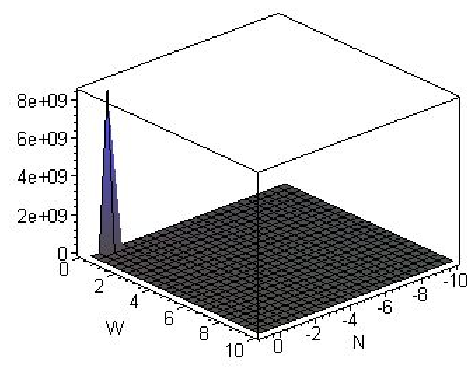}\vspace*{-3.0cm}
\caption{Le mineur d'hypersurface de la m\'etrique tenseure trac\'ee 
comme la fonction de $\{N, W\}$, en d\'ecrivant les fluctuations dans 
la configuration des trous noirs charg\'es nonextremaux $D_2 D_6 NS_5$
\`a l'ordre de $\alpha^{\prime}$.} \label{d2d6ns5alphaminor3}\vspace*{0.5cm}
\end{figure}

\begin{figure}
\hspace*{1.0cm}\vspace*{-6.0cm}
\includegraphics[width=12.0cm,angle=-0]{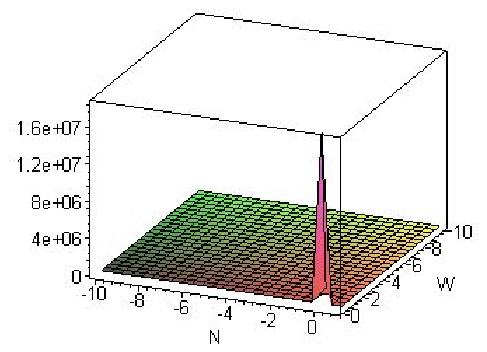}\vspace*{-3.0cm}
\caption{Le mineur de surface de la m\'etrique tenseure trac\'ee 
comme la fonction de $\{N, W\}$, en d\'ecrivant les fluctuations dans 
la configuration des trous noirs charg\'es nonextremaux $D_2 D_6 NS_5$
\`a l'ordre de $\alpha^{\prime}$.} \label{d2d6ns5alphaminor2}\vspace*{0.5cm}
\end{figure}

\begin{figure}
\hspace*{1.0cm}\vspace*{-6.0cm}
\includegraphics[width=12.0cm,angle=-0]{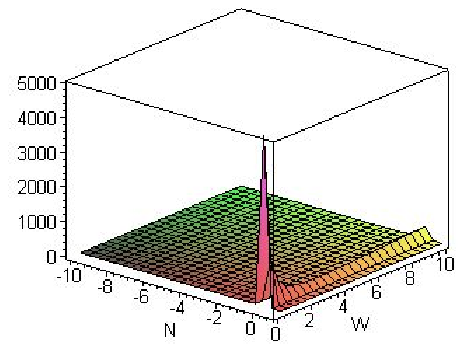}\vspace*{-3.0cm}
\caption{Le premier mineur de la m\'etrique tenseure trac\'ee 
comme la fonction de $\{N, W\}$, en d\'ecrivant les fluctuations dans 
la configuration des trous noirs charg\'es nonextremaux $D_2 D_6 NS_5$
\`a l'ordre de $\alpha^{\prime}$.} \label{d2d6ns5alphaminor1}\vspace*{0.5cm}
\end{figure}

Dans ce cas, sous les fluctuations de $\{N_2, N_6, N_5, N_R\}$,
la stabilit\'e d'un ensemble des trous noirs $D_2 D_6 NS_5 $ nonextremaux 
peut \^{e}tre d\'ecrite de la positivit\'e du d\'eterminant de la m\'etrique
tenseure $g$. Pour les trous noirs $D_2D_6 NS_5$ nonextremaux corrig\'ees par 
les termes de $\alpha^{\prime}$ avec $N_2= N$, $N_6= N$, $N_5= W$ et $N_R=W$,
nous constatons que le d\'eterminant de la m\'etrique tenseure $g$ prend une 
grande valeur n\'egative, quant \`a nous diminuons les param\`etres $\{|N|, W\}$.
Cela montre que les corrections d\'eriv\'ees sup\'erieures de la th\'eorie des cordes
\`a l'ordre dominant ne donnent pas de stabilit\'e statistique. En fait, nous observons
que $g \in (-4 \times 10^{+11}, 0)$. Pour une valeur typique de $N \in (-10, 1)$ et 
$W \in (0, 10)$, la Fig. (\ref{d2d6ns5alphadetg}) d\'ecrit la vue graphique du 
d\'eterminant de la m\'etrique tenseure $g$. Fait int\'eressant, la stabilit\'e 
de l'hypersurface d\'efinie par une valeur constante de $N_R$ est montr\'e dans 
la Fig. (\ref{d2d6ns5alphaminor3}). Notamment, nous voyons que le mineur de l'hypersurface 
$p_3$ r\'eside dans la gamme de $(0, 8 \times 10^{+09})$. Pour une petite valeur donn\'ee des 
param\`etres $|N|, W$, on constate que le pic de $p_3$ augmente \`a mesure que nous 
approchons \`a l'origine. D'autre part, la surface d\'efinies par les fluctuations de 
$\{N_2, N_6\}$ acquiert la signature positive sous les corrections de $\alpha^{\prime}$. 
Cette positivit\'e des mineurs principaux $\{p_2, p_3\}$ montre la stabilit\'e interne 
du syst\`eme de ces trous noirs. Pour $N \in (-10, 1)$ et $W \in (0, 10)$, la vue graphique 
du mineur $p_2$ est montr\'e dans la Fig. (\ref{d2d6ns5alphaminor2}). Dans ce cas, nous voyons
que $p_2$ r\'eside dans la gamme de $(0, 2 \times 10^{+07})$. En outre, nous remarquons par la 
Fig. (\ref{d2d6ns5alphaminor2}) que le mineur $p_2$ a un grand pic positif pr\`es de l'origine. 
Enfin, lorsque le seul param\`etre $N_2$ est autoris\'e \`a varier, la stabilit\'e du syst\`eme 
de ces trous noirs est donn\'ee par la positivit\'e de premier mineur principe $p_1:= g_{N_2 N_2}$.
Avec une orientation diff\'erente, la vue graphique de premier mineur principe $p_1$ est montr\'e 
dans la Fig. (\ref{d2d6ns5alphaminor1}). Le comportement qualitatif ci-dessus de la m\'etrique 
tenseure de l'espace d'\'etat et la positivit\'e des mineurs principaux d\'ecrirent les 
propri\'et\'es statistiques des fluctuations de trous noirs $D_2 D_6 NS_5$ nonextremaux 
corrig\'es de $\alpha^{\prime}$.

Donc, avec les valeurs des $\Gamma_{abc}$ comme nous les avons fournis dans l'annex $[A]$, 
on peut voir que la courbure scalaire avec ces corrections de $ \alpha^{\prime} $ est:

\ba R&=& \frac{3}{8 \pi} (N_2 N_6 N_5^{5/2} N_R^{-1/2}) \frac{r(N_2, N_6, N_5)}{\tilde{g}(N_2, N_6, N_5)^3},\ea

o\`u la fonction $r(N_2, N_6, N_5)$ est d\'efinie par

\ba r(N_2, N_6, N_5)&=& 
-114 b_1^3 (N_2 N_6 N_5^2)^{12}+ 26 b_1 (N_2 N_6 N_5^2)^{15}+ 
178980 b_1^7 (N_2 N_6 N_5^2)^6 \nn &&+ 3837 b_1^4 (N_2 N_6 N_5^2)^{21/2}+ 
604500 b_1^8 (N_2 N_6 N_5^2)^{9/2}+ 47472 b_1^6 (N_2 N_6 N_5^2)^{15/2} \nn &&+ 
17 b_1^2 (N_2 N_6 N_5^2)^{27/2}+ 565000 b_1^9 (N_2 N_6 N_5^2)^3+ 
2 (N_2 N_6 N_5^2)^{33/2} \nn &&+ 20280 b_1^5 (N_2 N_6 N_5^2)^9-
620000 b_1^{10} (N_2 N_6 N_5^2)^{3/2}- 800000b_1^{11}.\ea

Nous observons que cette curbure scalaire de Ruppenier est partout r\'eguli\`ere
pour chaque non-nulle $\tilde{g}(N_2, N_6, N_5)$. En ce cas, il est \'egalement simple 
\`a noter que cette configuration non-extr\'emale des branes $ D_2D_6NS_5 $ a toujours 
le m\^eme nombre des moments \`a gauche et \`a droite. De plus, la valeur correspondante 
de l'entropie est r\'eduite et cell de la curbure scalaire de Ruppenier est devenue 
certaine fraction des fonctions polyn\^omiales d'une variable $N$ d\'efinie comme 

\ba N&:=& N_2 N_6 N_5^2.\ea

En outre, dans le cas des branes noirs de trois charges avec $ N_R \rightarrow 0 $,
nous voyons sans aucune difficult\'e que, bien que la m\'etrique de Ruppenier est 
bien-d\'efinie, mais la courbure scalaire de Ruppenier agrandit jusqu'\`a l'infinie.
Dans ce cas, il s'agit donc un exemple du syst\`eme statistique de l'interaction infinie.

\begin{figure}
\hspace*{1.0cm}\vspace*{-6.0cm}
\includegraphics[width=12.0cm,angle=-0]{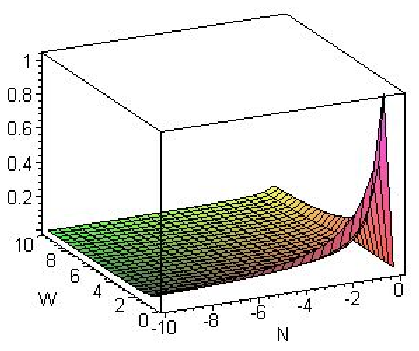}\vspace*{-3.0cm}
\caption{La courbure scalaire trac\'ee comme la fonction de $\{N, W\}$, 
en d\'ecrivant les fluctuations dans la configuration des trous noirs 
charg\'es nonextremaux $D_2 D_6 NS_5$ \`a l'ordre de $\alpha^{\prime}$ 
dans la gamme $n \in (-10,0)$ et $N \in (0,10)$.} \label{d2d6ns5alphaR}\vspace*{0.5cm}
\end{figure}

\begin{figure}
\hspace*{1.0cm}\vspace*{-6.0cm}
\includegraphics[width=12.0cm,angle=-0]{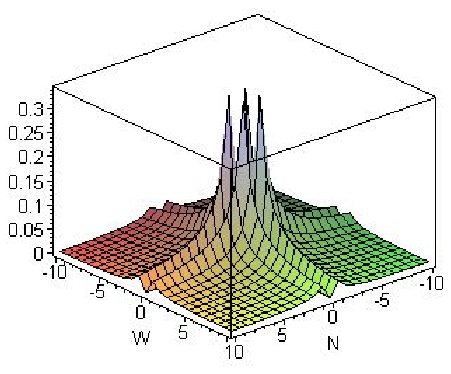}\vspace*{-3.0cm}
\caption{La courbure scalaire trac\'ee comme la fonction de $\{N, W\}$, 
en d\'ecrivant les fluctuations dans la configuration des trous noirs
charg\'es nonextremaux $D_2 D_6 NS_5$ \`a l'ordre de $\alpha^{\prime}$ 
dans la gamme $n, N \in (-10, 10)$.} \label{d2d6ns5alphaR10}\vspace*{0.5cm}
\end{figure}

Comme mentionn\'e dans la section pr\'ec\'edente, la description de l'espace
d'\'etat continue \`a la configuration des trous noirs $D_2 D_6 NS_5$ nonextremaux 
corrig\'es par les d\'eriv\'es sup\'erieures de $\alpha^{\prime}$. 
En particulier, la courbure scalaire sous-jacente de l'espace d'\'etat offre les 
propri\'et\'es globales des fluctuations statistiques. Dans la gamme de $N \in (-10, 0)$
et $W \in (0, 10)$, on observe de la Fig. (\ref{d2d6ns5alphaR}) que la courbure scalaire 
a un pic positif \`a l'ordre d'unit\'e. Dans cette gamme de param\`etres $\{N_2, N_6, N_5, N_R\}$,
les trous noirs $D_1 D_6 NS_5$ nonextremaux corrig\'ees par les termes de $\alpha^{\prime}$
correspondent \`a une configuration statistique avec certaines interactions non-nulles.
Comme mentionn\'e pr\'ec\'edemment, le signe positif de la courbure scalaire de l'espace 
d'\'etat signifie que les interactions statistiques sont r\'epulsives dans la leur nature. 
En outre, la Fig. (\ref{d2d6ns5alphaR10}) illustre le comportement de la courbure scalaire 
de l'espace d'\'etat dans la gamme des param\`etres $N, W \in (-10, 10)$. Justement, 
nous voyons qu'il y a d\'ecouvert de la bande des interactions. L'interaction augmente alors 
que nous approchons de l'origine. Nous constatons que la courbure scalaire correspondante
a un petit pic \`a l'ordre de $0.35$ pr\`es de l'origine $(N, W)= (0,0)$. Dans la gamme de 
$N, W \in (-10, 10) $, on observe de la Fig. (\ref{d2d6ns5alphaR10}) que les interactions 
statistiques globales sont sym\'etriques, dans chaqu'un de quadrant de plan $NW$. 
En contraire du sommet de la courbure scalaire apparaissant dans la gamme de 
$N, W \in (0, 10)$, le pic de la courbure scalaire s'av\'ere plus petit dans la gamme
$N, W \in (-10, 10) $. Par les Figs. (\ref{d2d6ns5alphaR}, \ref{d2d6ns5alphaR10}),
nous observons que la courbure scalaire de l'espace d'\'etat reste positif dans 
les deux intervalles ci-dessus. Qualitativement, les repr\'esentations de l'espace
d'\'etat indiquent que les trous noirs $D_2 D_6 NS_5 $ nonextremaux corrig\'es
par les termes de $\alpha^{\prime}$ sont globalement instables et correspondent 
\`a une configuration statistique faiblement interactive. En bref, les corrections 
des d\'eriv\'ees sup\'erieures de $\alpha^{\prime}$ \`a l'ordre dominant am\'eliorent 
la stabilit\'e des fluctuations dans la configuration des trous noirs $D_2D_6NS_5$
nonextremaux en offrant (i) des valeurs positives aux mineurs principaux et (ii) un petit 
pic \`a la courbure scalaire sous-jacente de l'espace d'\'etat.


\clearpage
\chapter{La g\'eom\'etrie de Ruppenier des trous noirs extr\'emaux en rotation
en quatre dimensions}

Dans ce chapitre, nous consid\'{e}rons la g\'eom\'etrie thermodynamique 
des trous noirs en rotation en ayant deux, trois et cinq param\`{e}tres. 
Il y a plusieurs syst\`emes thermodynamiques des trous noirs en rotation, 
en particulier le trou noir de Kerr-Newman et ceux de Kaluza-Klein charg\'e 
\'electriquement ou bien charg\'e \'electriquement et magn\'etiquement, 
qui sont les configurations consid\'{e}r\'{e}es r\'ec\'emnt par
Astefanesei, Goldstein, Jena, Sen et Trivedi \cite{Sen0606244}.
Bien s\^{u}r, dans le cas simple de la th\'eorie de gravit\'e des deux 
d\'eriv\'ees, on a des plusieurs exemples qui peuvent \^etre illustr\'es. 

Mais, ici, nous avons particuli\`erement int\'{e}ress\'{e} le cas des trous noirs 
en rotation par raport \`a la th\'eorie de la gravit\'e des d\'eriv\'ees sup\'erieures, 
c'est en tous cas, comme les trous noirs extr\'emaux de Kerr-Newman dans la th\'eorie 
d'Einstein-Maxwell, ou ceux de Kaluza-Klein dans la th\'eorie d'Einstein-Maxwell, 
ou bien aussi les trous noirs extr\'emaux d\'ecoulant naturellement dans la th\'eorie 
des cordes h\'et\'erotiques compactifi\'ee toroidalement.
Un premier niveau de la g\'eom\'etrie thermodynamique nous permet d'envisager 
la construction d'une m\'ethode pour qu'on puisse voir comme la nature des interactions 
thermodynamiques, la transition des phases et la branche d'ergonomie, $\ldots$ etc.
Pour voir clairement toutes ces \^id\'ees, il faut envisager la m\'etrique tenseure 
de Ruppenier dans la m\^eme forme ce que nous l'avons donn\'ee avant:

\ba g_{ij}:= -\partial_i \partial_j S(P_1, Q_2, P_3, Q_4, J).\ea

\section{Les trous noirs de Kerr-Newman dans la th\'eorie d'Einstein-Maxwell:}

Consid\'eons la théorie de la gravit\'e d'Einstein en quatre dimensions
associ\'ee \`a un seul champ de jauge de Maxwell $A_{\mu}$ avec la charge $q$. 
Puis l'entropie des trous noirs extr\'emaux sous-jacents de Kerr-Newman est: 

\ba S_{BH}= \sqrt{J^2+ (\frac{q^2}{8\pi})^2}\ea

et la g\'eom\'etrie de l'espace-temps pr\`es de l'horizon de ces trous noirs est une g\'eom\'etrie
de la gorge qu'elle \`a un analogie aven le vide $ AdS_2 \times S^2 $ de Bertotti-Robinson
\cite{BardeenHorowitz}. Ces trous noirs extr\'emaux de Kerr-Newman dans la th\'eorie 
d'Einstein-Maxwell ont la suivante densit\'e lagrangienne:

\ba \mathcal L= R- \frac{1}{4} F_{\mu\nu} F^{\mu\nu}, \ea

dont laquelle, on a cet entropie et cette g\'eom\'etrie de l'horizon \cite{Sen0606244}.
Avec cet entropie du trou noir, et ainsi en consid\`erant $\{q, J\}$ comme les variables
thermodynamiques, on peut voir  que les composantes de la m\'etrique tenseure de Ruppenier sont:

\ba g_{qq}&=& -\frac{J^2}{32 \pi} (J^2+ (\frac{q}{8 \pi})^2)^{-3/2},\nn
g_{qJ}&=& \frac{q J}{32 \pi} (J^2+ (\frac{q}{8 \pi})^2)^{-3/2},\nn
g_{JJ}&=& -\frac{q^2}{32 \pi} (J^2+ (\frac{q}{8 \pi})^2)^{-3/2}.\ea

\begin{figure}
\hspace*{1.0cm}\vspace*{-6.0cm}
\includegraphics[width=12.0cm,angle=-0]{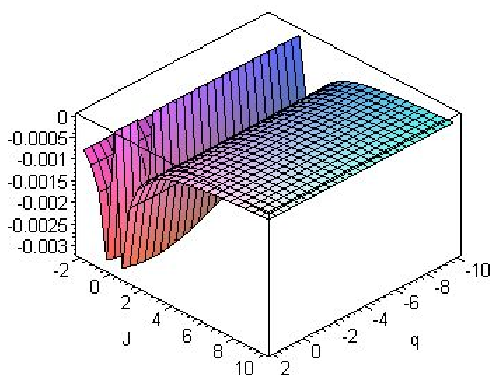}\vspace*{-3.0cm}
\caption{La composante $qq$ de la m\'etrique tenseure trac\'ee comme la
fonction de $\{q, J\}$, en d\'ecrivant les fluctuations dans la configuration
des trous noirs de Kerr-Neumann.} \label{KNqq}\vspace*{0.5cm}
\end{figure}

\begin{figure}
\hspace*{1.0cm}\vspace*{-6.0cm}
\includegraphics[width=12.0cm,angle=-0]{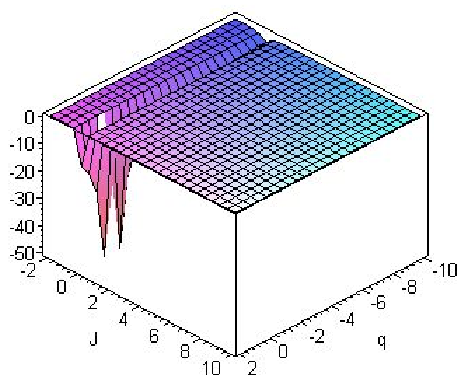}\vspace*{-3.0cm}
\caption{La composante $JJ$ de la m\'etrique tenseure trac\'ee comme la
fonction de $\{q, J\}$, en d\'ecrivant les fluctuations dans la configuration
des trous noirs de Kerr-Neumann.} \label{KNJJ}\vspace*{0.5cm}
\end{figure}

\begin{figure}
\hspace*{1.0cm}\vspace*{-6.0cm}
\includegraphics[width=12.0cm,angle=-0]{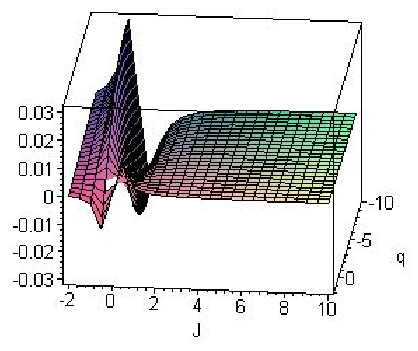}\vspace*{-3.0cm}
\caption{La composante $qJ$ de la m\'etrique tenseure trac\'ee comme la
fonction de $\{q, J\}$, en d\'ecrivant les fluctuations dans la configuration
des trous noirs de Kerr-Neumann.} \label{KNqJ}\vspace*{0.5cm}
\end{figure}

Sous la fluctuation des param\`etres $\{q, J\}$, les Figs. (\ref{KNqq}, \ref{KNJJ}) 
d\'epeignent la nature des composantes $\{g_{qq}, g_{JJ}\}$ de la m\'etrique tenseure 
thermodynamique. Dans le r\'egime de $q \in (-10, 2)$ et $J \in (-2, 10) $, 
nous constatons que l'amplitude des capacit\'es de chaleurs $\{g_{qq}\}$ prend 
une valeur dans l'intervalle $(-0.003, 0)$. Dans cette gamme des param\`etres $\{q, J\}$, 
on peur observer que la composante $\{g_{JJ}\}$ r\'eside dans la gamme de $(-50, 0)$. 
Dans ce cas, nous voyons que l'amplitudes des deux composantes $\{g_{qq}\}$ et $\{g_{JJ}\}$ 
reste la plupart du temps sur la ligne $J= 0$. Explicitement, les Figs. (\ref{KNqq}, 
\ref{KNJJ}) indiquent que la croissance de la composante $\{g_{qq}\}$ prend un place dans 
la limite d'une petite charge $q$. Le signe n\'egatif de $\{g_{qq}\}$ signifie 
que les trous noirs de Kerr-Neumann sont thermodynamiquement instables dans la limite d'une 
petite charge \'electrique. Du fait m\^{e}me, la Fig. (\ref{KNqJ}) montre que la nature de la 
composante $\{g_{qJ}\}$ de la m\'etrique tenseure thermodynamique. Nous constatons que la composante 
mixe $\{g_{qJ}\}$ reste dans l'intervalle $(-0.03, +0.03)$. Dans cette limite de $\{q, J\}$, 
les compressibilit\'es de chaleurs, qui sont pr\'es\'ent\'ees dans les Figs. (\ref{KNqq}, \ref{KNJJ}, \ref{KNqJ}) 
illustrent les propri\'et\'es param\'etriques au dessus des fluctuations de la configuration 
des trous noirs de Kerr-Neumann. Par la pr\'esente, nous pouvons voir que les fluctuations d'auto-paires, 
en impliquant $\{q, J\}$ tel qu'elles sont d\'efinies par les composantes de la m\'etrique tenseure 
$\{g_{ij} \ | \ i, j= q, J\}$, n'ont que des valeurs n\'egatives, tandis que la composante mixe 
$\{g_{qJ}\}$ \`a la fois a les deux signe. \`A ce stade, nous constatons que les trous noirs 
de Kerr-Neumann ont des capacit\'es de chaleurs n\'egatives, et donc on a une signature 
des instabilit\'es statistiques locales.

Maintenant, il est facile d'observer que le d\'eterminant de cette
m\'etrique tenseure de Ruppenier est z\'ero. C'est-\`a-dire que nous avons

\ba \Vert g_{ij} \Vert= S_{qq}S_{JJ}- S_{qJ}^2= 0.\ea

Avec cette propri\'et\'e de la  m\'etrique tenseure, pour 
toute g\'eom\'etrie thermodynamique bidimensionnelle, 
nous avons une observation suivante.

\subsection*{L'observation}

Pour n'importe quelle m\'etrique tenseure $ g_{ij} $ de Ruppenier 
d\'efinie par une fonction g\'en\'erale de profil comme 

\ba f(x_1,x_2)= k \sqrt{x_1^2- \alpha x_2^2},\ea 

on une observation simple que le d\'eterminant de la m\'etrique est z\'ero.
Donc, il est une corollaire simple que la vari\'et\'e de Ruppenier
pour laquelle la norme du d\'eterminant dispara\^ite, la courbure scalaire
thermodynamique est partout infinie dans l'espace d'\'etat \cite{Sar1},
parce que les vari\'et\'es thermodynamiques bidimensionnelles
ont une relation: 

\ba R= \frac{2}{\Vert g \Vert} R_{1212}.\ea

\begin{figure}
\hspace*{1.0cm}\vspace*{-6.0cm}
\includegraphics[width=12.0cm,angle=-0]{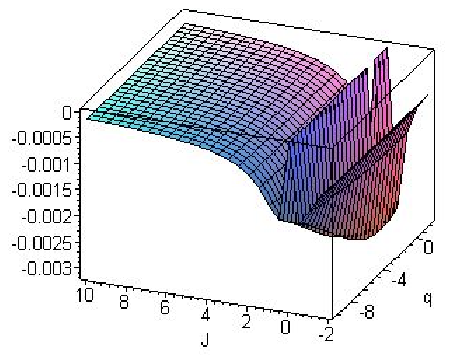}\vspace*{-3.0cm}
\caption{Le premier mineur de la m\'etrique tenseure trac\'ee comme la
fonction de $\{q, J\}$, en d\'ecrivant les fluctuations dans la configuration
des trous noirs de Kerr-Neumann.} \label{KNminor1}\vspace*{0.5cm}
\end{figure}

La stabilit\'e d'un ensemble des trous noirs de Kerr-Neumann peut \^{e}tre 
d\'etermin\'e en termes des param\`etres $\{q, J\}$. Cela d\'ecoule du comportement 
de la d\'eterminant de la m\'etrique tenseure thermodynamique. Dans ce cas, puisque
le d\'eterminant de la m\'etrique tenseure est identiquement nulle pour toute valeur 
des $\{q, J\}$. Ainsi, les fluctuations de $\{q, J\}$ introduissent une instabilit\'e
dans l'ensemble sous-jacent. Lorsque le seul param\`etre $q$ est autoris\'e \`a varier,
la stabilit\'e de la configuration des trous noirs de Kerr-Neumann est d\'etermin\'ee 
par la positivit\'e du premier mineur principe $p_1:= g_{qq}$. Une vue rotat\'ee de 
$p_1$ est montr\'ee dans la Fig. (\ref{KNminor1}). Les repr\'esentations graphiques 
ci-dessus de la g\'eom\'etrie thermodynamiques fournissent la propri\'et\'e qualitative
des fluctuations de la configuration statistique des trous noirs de Kerr-Neumann 
de deux param\`etres.

Notez bien que la stabilit\'e globale des trous noirs de Kerr-Neumann
d\'ecoule du comportement de la courbure scalaire thermodynamique.
Dans ce cas, pour toute valeur des param\`etres $\{q, J\}$, on trouve que 
la courbure scalaire est partout mal d\'efinie. Cela montre que la configuration 
des trous noirs de Kerr-Neumann correspond \`a mal comportement de base statistique
au niveau de la th\'eorie de gravit\'e des deux d\'eriv\'es. De plus, nous avons 
fourni les $\Gamma_{abc}$ de cette configuration des trous noirs sous-jacents 
dans l'annex $[A]$. En bref, la g\'eom\'etrie thermodynamique indiquent que les 
trous noirs de Kerr-Neumann correspondent \`a une configuration statistique 
instable en vertu de la fluctuation des param\`etres $\{q, J\}$.

\section{Les trous noirs extr\'emaux de Kaluza-Klein dans la th\'eorie d'Einstein-Maxwell:}

Dans le cas des trous noirs de Kaluza-Klein de $D=4$ en rotation, nous allons
maintenant examiner la th\'eorie de la gravit\'e qui est obtenue par la th\'eorie 
de la r\'eduction de la gravit\'e pure des cinq dimensions sur un cercle $S^1$.
Les champs importants en quatre dimensions sont la m\'etrique tenseure $g_{\mu\nu}$
de l'espace-temps, un champ scalaire $\phi$ associ\'es \`a la rayon de la cinqui\`eme 
dimension et un champ de jauge $A_{\mu} $ de $U(1)$. Il est bien connu \cite{Sen0606244} 
que l'entropie associ\'ee \`a ce syst\`eme des trous noirs est donn\'ee par: 

\ba S(P,Q,J)= 2 \pi \sqrt{P^2 Q^2- J^2}.\ea

En ce cas de l'entropie des trois param\`etres $\{P,Q,J\}$, la m\'etrique tenseure de Ruppenier,
le d\'eterminant de la m\'etrique tenseure et la courbure scalaire de Ruppenier peuvent \^etre \'ecrits 
facilement, comme nous les avons donn\'es en g\'en\'erale dans l'annexe [B]. En particulier,
nous voyons maintenant que les composantes de la m\'etrique de Ruppenier sont
simplement donn\'ees par:

\ba g_{PP}&=& \frac{2 \pi P^2 Q^4}{(P^2 Q^2- J^2)^{3/2}}- \frac{2 \pi Q^2}{(P^2 Q^2- J^2)^{1/2}},\nn
g_{PQ}&=& \frac{2 \pi P^3 Q^3}{(P^2 Q^2- J^2)^{3/2}}- \frac{4 \pi P Q}{(P^2 Q^2- J^2)^{1/2}},\nn
g_{PJ}&=& -\frac{2 \pi J P Q^2}{(P^2 Q^2- J^2)^{3/2}},\nn
g_{QQ}&=& \frac{2 \pi P^4 Q^2}{(P^2 Q^2- J^2)^{3/2}}- \frac{2 \pi P^2}{(P^2 Q^2- J^2)^{1/2}},\nn
g_{QJ}&=& -\frac{2 \pi J P^2 Q}{(P^2 Q^2- J^2)^{3/2}},\nn
g_{JJ}&=& \frac{2 \pi J^2}{(P^2 Q^2- J^2)^{3/2}}+ \frac{2 \pi}{(P^2 Q^2- J^2)^{1/2}}.\ea

\begin{figure}
\hspace*{1.0cm}\vspace*{-6.0cm}
\includegraphics[width=12.0cm,angle=-0]{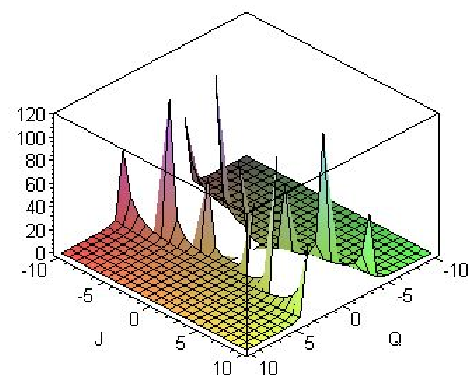}\vspace*{-3.0cm}
\caption{La composante $PP$ de la m\'etrique tenseure trac\'ee comme la
fonction de $\{Q, J\}$, en d\'ecrivant les fluctuations dans la configuration
des trous noirs de Kaluza-Klein.} \label{KKPP}\vspace*{0.5cm}
\end{figure}

\begin{figure}
\hspace*{1.0cm}\vspace*{-6.0cm}
\includegraphics[width=12.0cm,angle=-0]{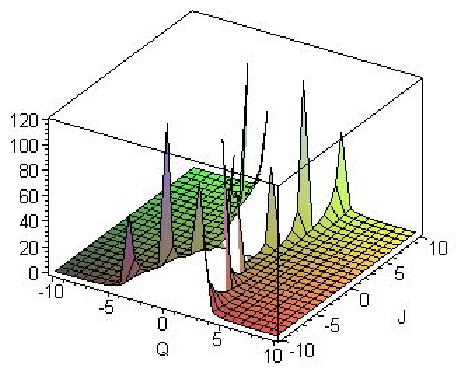}\vspace*{-3.0cm}
\caption{La composante $QQ$ de la m\'etrique tenseure trac\'ee comme la
fonction de $\{Q, J\}$, en d\'ecrivant les fluctuations dans la configuration
des trous noirs de Kaluza-Klein.} \label{KKQQ}\vspace*{0.5cm}
\end{figure}

\begin{figure}
\hspace*{1.0cm}\vspace*{-6.0cm}
\includegraphics[width=12.0cm,angle=-0]{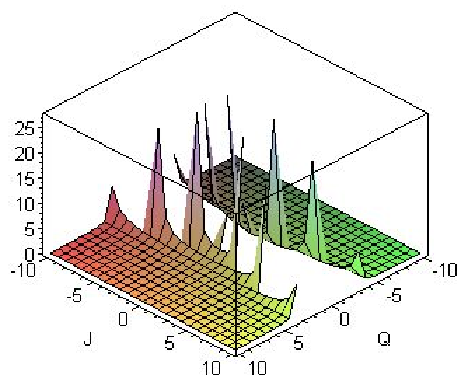}\vspace*{-3.0cm}
\caption{La composante $JJ$ de la m\'etrique tenseure trac\'ee comme la
fonction de $\{Q, J\}$, en d\'ecrivant les fluctuations dans la configuration
des trous noirs de Kaluza-Klein.} \label{KKJJ}\vspace*{0.5cm}
\end{figure}

\begin{figure}
\hspace*{1.0cm}\vspace*{-6.0cm}
\includegraphics[width=12.0cm,angle=-0]{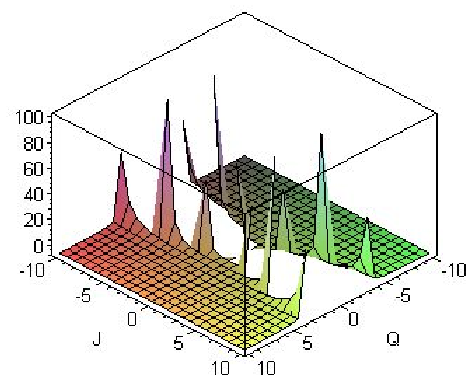}\vspace*{-3.0cm}
\caption{La composante $PQ$ de la m\'etrique tenseure trac\'ee comme la
fonction de $\{Q, J\}$, en d\'ecrivant les fluctuations dans la configuration
des trous noirs de Kaluza-Klein.} \label{KKPQ}\vspace*{0.5cm}
\end{figure}

\begin{figure}
\hspace*{1.0cm}\vspace*{-6.0cm}
\includegraphics[width=12.0cm,angle=-0]{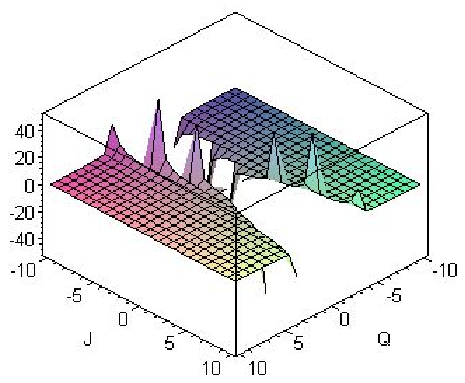}\vspace*{-3.0cm}
\caption{La composante $PJ$ de la m\'etrique tenseure trac\'ee comme la
fonction de $\{Q, J\}$, en d\'ecrivant les fluctuations dans la configuration
des trous noirs de Kaluza-Klein.} \label{KKPJ}\vspace*{0.5cm}
\end{figure}

\begin{figure}
\hspace*{1.0cm}\vspace*{-6.0cm}
\includegraphics[width=12.0cm,angle=-0]{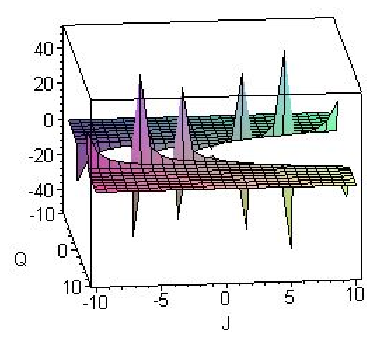}\vspace*{-3.0cm}
\caption{La composante $QJ$ de la m\'etrique tenseure trac\'ee comme la
fonction de $\{Q, J\}$, en d\'ecrivant les fluctuations dans la configuration
des trous noirs de Kaluza-Klein.} \label{KKQJ}\vspace*{0.5cm}
\end{figure}

Dans ce cas, puisqu'il y a trois variables ind\'ependantes $\{P, Q, J\}$,
donc nous sommes tenus \`a fixer l'une des variables afin d'offrir la vue graphique
de trois dimensions des fluctuations sous-jacentes. Nous le faisons en choisissant $P= Q$. 
Pour une configuration donn\'ee des trous noirs extr\'emalaux de Kaluza-Klein, 
les propri\'et\'es des fluctuations sont exprim\'ees comme les suivantes. 
Dans le r\'egime de $Q \in (-10, 10)$ et $J \in (-10,10)$, on observe que l'amplitude 
des capacit\'es de chaleurs $\{g_{QQ}, g_{PP}\}$ prend une valeur \`a l'ordre de $120$.
Dans cette gamme de $\{Q, J\}$, la capacit\'e de la chaleur $\{g_{JJ}\}$ r\'eside dans la
gamme de $(0, 30)$. En conformant \`a la pr\'ediction de la g\'eom\'etrie de l'espace d'\'etat,
nous constatons que les gammes de la premi\`ere ensemble et la deuxi\`eme ensemble des capacit\'es 
de chaleurs s'\'eloignent de l'origine, quand nous augmentons les valeurs de param\`etres $\{Q, J\}$. 
Plus pr\'ecis\'ement, pour $Q \simeq 5$, le pic des capacit\'es de chaleurs $\{g_{QQ}, g_{PP}\}$ 
augmente au hasard, \`a mesure que nous augmentons la valeur de $J$. De plus, pour $Q \simeq 5$, 
le pic de l'\'el\'ement $g_{JJ}$ diminue, quand nous augmentons la valeur de $J$. Dans ce cas, 
les Figs. (\ref{KKPP}, \ref{KKQQ}) montrent que la croissance des capacit\'es de chaleurs 
$\{g_{PP}, g_{QQ}\}$ a lieu dans un ensemble des amplitudes al\'eatoires sur la surface $QJ$. 
En outre, la Fig. (\ref{KKJJ}) montre que la croissance de la composante $\{g_{JJ}\}$ a lieu 
dans un ensemble des croissantes amplitudes sur la surface $QJ$. De m\^{e}me, les compressibilit\'es
de chaleurs en impliquant deux param\`etres distincts de la configuration des trous noirs de 
Kaluza-Klein sont repr\'esent\'es dans les Figs. (\ref{KKPQ}, \ref{KKPJ}, \ref{KKQJ}).
Par la pr\'esente, nous constatons que l'amplitude de (i) $g_{PQ}$ reste dans l'intervalle
$(0, 100)$, (ii) $g_{PJ}$ se situe dans la gamme de $(-50, 50)$ et (iii) $g_{QJ}$ r\'eside 
aussi dans l'intervalle $(-50, 50)$. De toute fa\c{c}on, les composantes de la m\'etrique 
tenseure de l'espace d'\'etat $\{g_{ij} \ | \ i, j = P, Q, J\}$, comme illustr\'e dans les
Figs. (\ref{KKPP}, \ref{KKQQ}, \ref{KKJJ}) montrent que la configuration des trous noirs de 
Kaluza-Klein correspond \`a un syst\`eme statistique localement stable avec un ensemble des 
capacit\'es thermiques positives par rapport \`a la g\'eom\'etrie de l'espace d'\'etat.

En ce cas, il s'ensuit que le mineur de surface a la forme suivante

\begin{eqnarray}
p_2:&=& - 4\,{\displaystyle \frac{\pi
^{2}\,Q^{2}\,P^{2}\,(P^{2}\, Q^{2} - 3\,J^{2})}{(P^{2}\,Q^{2} -
J^{2})^{2}}}.
\end{eqnarray}

Donc, il est pr\'ecise que le d\'eterminant de la m\'etrique tenseure est:

\ba g&=& -8 \pi^3 \frac{(PQ)^4}{(P^2 Q^2- J^2)^{13/2}} \tilde{g}(P,Q,J),\ea

o\`u la fonction $\tilde{g}(P,Q,J)$ est d\'efinie par

\ba \tilde{g}(P,Q,J):= (PQ)^8- 4 (PQ)^6 J^2+ 6 (PQ)^4 J^4- 4(PQ)^2 J^6+ J^8. \ea

\begin{figure}
\hspace*{1.0cm}\vspace*{-6.0cm}
\includegraphics[width=12.0cm,angle=-0]{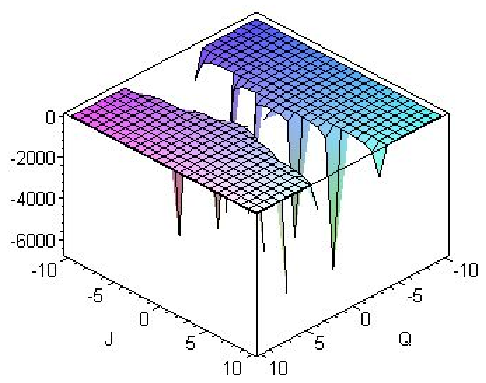}\vspace*{-3.0cm}
\caption{Le d\'eterminant de la m\'etrique tenseure trac\'ee comme la
fonction de $\{Q, J\}$, en d\'ecrivant les fluctuations dans la configuration
des trous noirs de Kaluza-Klein.} \label{KKdetg}\vspace*{0.5cm}
\end{figure}

\begin{figure}
\hspace*{1.0cm}\vspace*{-6.0cm}
\includegraphics[width=12.0cm,angle=-0]{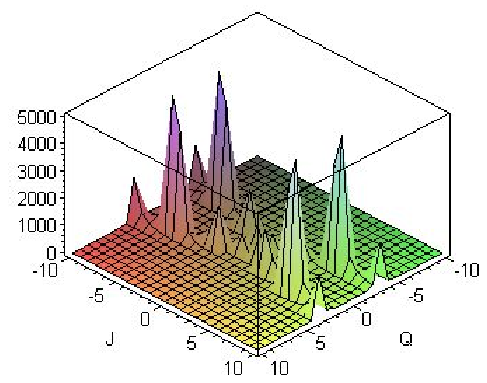}\vspace*{-3.0cm}
\caption{Le mineur de surface de la m\'etrique tenseure trac\'ee comme la
fonction de $\{Q, J\}$, en d\'ecrivant les fluctuations dans la configuration
des trous noirs de Kaluza-Klein.} \label{KKminor2}\vspace*{0.5cm}
\end{figure}

\begin{figure}
\hspace*{1.0cm}\vspace*{-6.0cm}
\includegraphics[width=12.0cm,angle=-0]{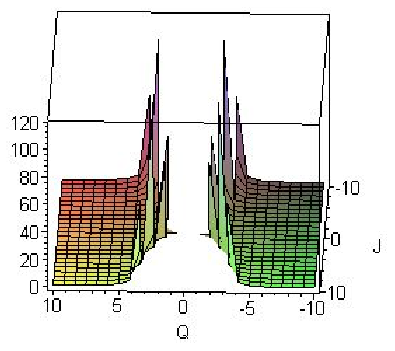}\vspace*{-3.0cm}
\caption{Le premier mineur de la m\'etrique tenseure trac\'ee comme la
fonction de $\{Q, J\}$, en d\'ecrivant les fluctuations dans la configuration
des trous noirs de Kaluza-Klein.} \label{KKminor1}\vspace*{0.5cm}
\end{figure}

Sous les fluctuations des param\`etres $\{P, Q, J\}$, la stabilit\'e 
d'un ensemble des trous noirs de Kaluza-Klein d\'ecoule de la positivit\'e
du d\'eterminant de la m\'etrique tenseure. Pour le choix du $P= Q$, 
nous constatons que le d\'eterminant de la m\'etrique tenseure $g$ prend une grande
valeur n\'egative, quand nous permetons \`a fluctuer les param\`etres $\{P, Q, J\}$.
Dans ce cas, nous voyons que $g \in (-7000, 0)$. Pour une valeur typique de
$Q \in (-10, 10)$ et $J \in (-10, 10)$, la Fig. (\ref{KKdetg}) offre la nature graphique 
de d\'eterminant de la m\'etrique tenseure $g$. Par la suite, les fluctuations de 
$\{P, Q\}$ correspond \`a une surface stable. De la Fig. (\ref{KKminor2}), 
ce qui suit par-ce que le mineur principe correspondant $p_2 \in (0, 5000)$. Enfin, 
lorsque le seul param\`etre $P$ est autoris\'e \`a varier, la stabilit\'e de la
configuration des trous noirs de Kaluza-Klein est d\'etermin\'ee par la positivit\'e 
du premier mineur principe $p_1:= g_{PP}$. Une vue rotat\'ee de $p_1$ est montr\'e 
dans la Fig. (\ref{KKminor1}). Les propri\'et\'es qualitative ci-dessus des mineurs
principaux d\'ecritent le comportement des fluctuations statistiques de l'espace 
d'\'etat des trous noirs de Kaluza-Klein.

Ensuite, comme nous avons fourni les $\Gamma_{abc}$ dans l'annex $[A]$, 
nous pouvons voir que la courbure scalaire de Ruppenier 
dans ce cas des trous noirs de Kaluza-Klein est:

\ba R&=& -(\frac{2 P^4 Q^4+ P^2 Q^2 J^2- 3J^4}{4 \pi P^2 Q^2 (P^2 Q^2- J^2)^{3/2}}).\ea

\`A la valeure $J=0$, nous voyons que le d\'eterminant de la m\'etrique tenseure est

\ba g\vert_{J=0}= -\frac{8 \pi^3}{PQ}\ea

et \'egalement la courbure scalaire de Ruppenier est donn\'ee par

\ba R\vert_{J=0}= -\frac{1}{ \pi \vert PQ \vert}.\ea

Donc, cette g\'eom\'etrie de Ruppenier est bien d\'efinie et
ainsi correspond \`a une syst\`eme statistique en interactions
pour chaque non-nulle charge \'electrique $Q$ et celle de la 
charge magn\'etique $P$, et encore ce qui reste la m\^eme aussi 
au point de $J=0$. En fait, il n'est pas aussi difficile de voir 
de la limite de $J=0$ que nous avons 

\ba R= \frac{2 P^2 Q^2 +3J^2}{4 \pi P^2 Q^2 (P^2 Q^2- J^2)^{1/2}}. \ea

Ensuite, on peut observer qu'il n'y a pas des divergences dans l'espace d'\'etat
et cette courbure de Ruppenier est partout r\'eguli\`ere, sauf d'une branche d'ergo.
Ceci est bien compatible avec le fait que l'espace de modules des trous noirs
extr\'emaux en rotation sont constitu\'es aux deux branches d'ergo,
comme discut\'e pour la premi\`ere fois dans le Ref. \cite{Rasheed}.
En d'autres termes, nous voyons que le d\'eterminant et la courbure scalaire 
de la g\'eom\'etrie de Ruppenier de la configurtion des trous noirs extr\'emaux 
de Kaluza-Klein dans la th\'eorie d'Einstein-Maxwell sont mal-d\'efinis sur 
la branche d'ergo. En outre, notez bien que la courbure scalaire de Ruppenier 
devient infinie, c'est-\`a-dire qu'il y des tr\`es fortes interactions statistiques, 
si on traverse la fronti\`ere d'une succursale de l'ergo.

\begin{figure}
\hspace*{1.0cm}\vspace*{-6.0cm}
\includegraphics[width=12.0cm,angle=-0]{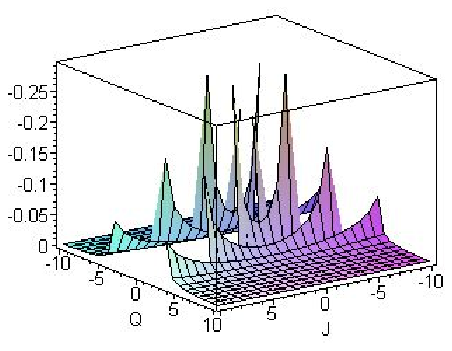}\vspace*{-3.0cm}
\caption{La courbure scalaire trac\'ee comme la fonction 
de $\{Q, J\}$, en d\'ecrivant les fluctuations dans la 
configuration des trous noirs de Kaluza-Klein dans la 
gamme de $Q, J \in (-10, 10)$.} \label{KKR}\vspace*{0.5cm}
\end{figure}

\begin{figure}
\hspace*{1.0cm}\vspace*{-6.0cm}
\includegraphics[width=12.0cm,angle=-0]{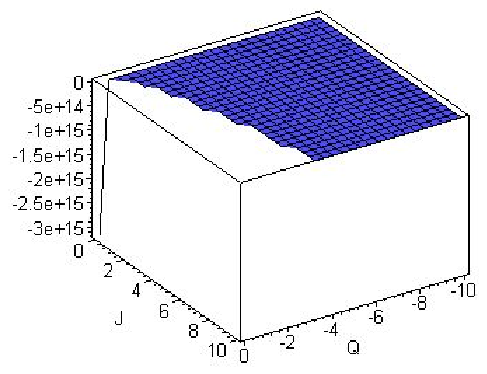}\vspace*{-3.0cm}
\caption{La courbure scalaire trac\'ee comme la fonction 
de $\{Q, J\}$, en d\'ecrivant les fluctuations dans la configuration 
des trous noirs de Kaluza-Klein dans la gamme de $Q \in (-10, 0)$
et $J \in (0, 10)$.} \label{KKRhalf}\vspace*{0.5cm}
\end{figure}

Comme mentionn\'e dans la section pr\'ec\'edente, la courbure scalaire de
l'espace d'\'etat d\'ecrit les propri\'et\'es globales d'un ensemble statistique
de ces trous noirs en fluctuant $\{P, Q, J\}$. Sous la fluctuation de $\{P, Q, J\}$,
la structure de la stabilit\'e globale des trous noirs de Kaluza-Klein
peut \^{e}tre d\'ecrit comme la suit. Dans la gamme de $Q, J \in (-10, 10)$,
la Fig. (\ref{KKR}) montre que la courbure scalaire prend une petite amplitude 
n\'egative \`a l'ordre de $-0.30$. Dans cette gamme de param\`etres, 
nous observons que la configuration de Kaluza-Klein est un syst\`eme statistique
faiblement interactif. Physiquement, le signe n\'egatif de la courbure scalaire
signifie que les interactions statistiques sont attrayantes dans la leur nature.
La Fig. (\ref{KKRhalf}) illustre le comportement de la courbure scalaire ci-dessus
pour la gamme des param\`etres $Q \in (-10, 0)$ et $J \in (0, 10)$. En fait, 
nous voyons que les interactions sont largement pr\'esentes pr\`es de l'origine.
La Fig. (\ref{KKRhalf}) montre que la courbure scalaire de l'espace d'\'etat 
acquiert un grand pic \`a l'ordre de $-3 \times 10^{+15}$ pr\`es de l'origine 
$(Q, J)= (0, 0)$. Par ailleurs, nous remarquons de la Fig. (\ref{KKRhalf}) que 
les interactions statistiques globales remplissent la gamme compl\`ete $J= 0$,
dans la limite de $Q= 0$. En comparaison des interactions apparaissent dans la 
gamme de $Q, J \in (-10,10)$, l'amplitude des interactions statistiques globales 
se r\'ev\`ele d'\^{e}tre beaucoup plus grande dans la gamme de $Q \in (-10, 0)$ et 
$J \in (0, 10)$. Pr\'ecis\'ement, nous constatons que le ratio des valeurs typiques 
de l'amplitude des courbures scalaire est \`a l'ordre de $10^{16}$. La vue graphique 
de cette comparaison ci-dessus d\'ecoule par les Figs. (\ref{KKR}, \ref{KKRhalf}). 
Qualitativement, de ce qui pr\'ec\`ede est mentionn\'es vers la repr\'esentation de
l'espace d'\'etat, nous observons que la configuration des trous noirs de Kaluza-Klein 
est globalement instable et correspond \`a un syst\`eme statistique int\'eractif, 
quand elle est pris en compte dans la th\'eorie de la gravit\'e des deux d\'eriv\'es.

\section{Les trous noirs extr\'emaux de la th\'eorie des cordes h\'et\'erotiques 
compactifi\'ee toroidalement:}

Dans cette section, nous examinons la g\'eom\'etrie thermodynamiques 
des trous noirs extr\'emaux de la th\'eorie de la gravit\'e en quatre 
dimensions coupl\'ee \`a un champ scalaire complexe 

\ba S = S_1 + iS_2, \ea

des quatrse champs de jauge de $ U(1) $ donn\'e par $ \lbrace A_{\mu}^{(i)} \rbrace_{i=1}^{4} $,
et un champ scalaire $ M $ faisant sa valeur de la matrice du type $ 4 \times 4 $ avec la contrainte: 

\ba M L M^{T}= L,\ea

o\`u

\ba L= \left( \begin{array}{rr} 
    0 & I_2 \\
     I_2 & 0 \\
\end{array} \right) \ea

et la $I_2$ signifie la matrice d'identit\'e de type $ 2 \times 2 $, 
voir \cite{Sen0606244, Larsen} pour des d\'etails de ces solutions de 
trous noirs en rotation avec un vecteur g\'en\'erale des charges \'electriques 
$ \overrightarrow{Q} $ et des charges magn\'etique $ \overrightarrow{P}$.
De plus, il existe une famille de ces solutions avec les m\^emes charges 
\'electriques et magn\'etiques, mais diff\'erentes valeurs asymptotiques 
des champs scalaires dans le cadre de la transformation des dualit\'es 
entr\'e les charges \'electriques et magn\'etiques. Ces trous noirs 
brisant la supersym\'etrie et peuvent \^etre construits en consid\'erant 
les vecteurs de charges comme:

\ba Q= \left( \begin{array}{rrrr}
0 \\
Q_2 \\
0 \\
Q_4\\
\end{array} \right),\ \
P= \left( \begin{array}{rrrr} 
P_1\\
0 \\
P_3 \\
0\\
\end{array} \right).\ea

Pour cette classe des solutions, la masse $M$ d'ADM, les charges \'electriques
et magn\'etiques $ \lbrace Q_i, P_i \rbrace $ et le moment cin\'etique $J$, et
l'entropie de trou noir associ\'e peut \^etre trouv\'ee en g\'en\'eral
dans les Refs. \cite{CveticYoum, Rasheed}. Il s'av\`ere que cette solution 
a deux type diff\'erent des limites extr\'emales qu'eles sont connues sous le 
nom de la succursale d'ergo et sans la succursale d'ergo. Dans la limite extr\'emale
correspondante \`a la succursale d'ergo, l'entropie des trous noirs extr\'emaux en rotation
dans la th\'eorie des cordes h\'et\'erotiques compactifi\'ee toroidalement \cite{Sen0606244} 
est donn\'ee par:

\ba S(P_1,Q_2,P_3,Q_4,J):= 2 \pi \sqrt{J^2+ P_1 Q_2 P_3 Q_4}.\ea

En outre, nous pouvons voir que la valeur absolue de cet entropie reste
la m\^eme dans la succursale d'ergo ainsi que dehors de la succursale d'ergo.

Selon le cas d'application, nous avons la m\^eme forme de la m\'etrique tenseure
de Ruppenier $ g_{ij}(P_1, Q_2, P_3, Q_4, J) $ ce que nous avons donn\'ee avant,
en particulier avec les variables thermodynamique $\{P_1, Q_2, P_3, Q_4, J\}$,
l'\'el\'ement de la ligne de Ruppenier peut \^etre don\'ee comme: 

\ba
ds^2&:=& 
-(\frac{\partial^2 S(P_1, Q_2, P_3, Q_4, J)}{\partial P_1^2})d P_1^2
-2(\frac{\partial^2 S(P_1, Q_2, P_3, Q_4, J)}{\partial Q_2 \partial P_1})d P_1 d Q_2\nn &&
-2(\frac{\partial^2 S(P_1, Q_2, P_3, Q_4, J)}{\partial P_1 \partial P_3})d P_1 d P_3
-2(\frac{\partial^2 S(P_1, Q_2, P_3, Q_4, J)}{\partial P_1 \partial Q_4})d P_1 d Q_4\nn &&
-2(\frac{\partial^2 S(P_1, Q_2, P_3, Q_4, J)}{\partial P_1 \partial J})d P_1 d J
-(\frac{\partial^2 S(P_1, Q_2, P_3, Q_4, J)}{\partial Q_2^2})d Q_2^2\nn &&
-2(\frac{\partial^2 S(P_1, Q_2, P_3, Q_4, J)}{\partial Q_2 \partial P_3})d Q_2 d P_3
-2(\frac{\partial^2 S(P_1, Q_2, P_3, Q_4, J)}{\partial Q_2 \partial Q_4})d Q_2 d Q_4\nn &&
-2(\frac{\partial^2 S(P_1, Q_2, P_3, Q_4, J)}{\partial Q_2 \partial J})d Q_2 d J
-(\frac{\partial^2 S(P_1, Q_2, P_3, Q_4, J)}{\partial P_3^2})d P_3^2\nn &&
-2(\frac{\partial^2 S(P_1, Q_2, P_3, Q_4, J)}{\partial Q_4 \partial P_3})d Q_4 d P_3
-2(\frac{\partial^2 S(P_1, Q_2, P_3, Q_4, J)}{\partial P_3 \partial J})d P_3 d J\nn &&
-(\frac{\partial^2 S(P_1, Q_2, P_3, Q_4, J)}{\partial Q_4^2})d Q_4^2
-2(\frac{\partial^2 S(P_1, Q_2, P_3, Q_4, J)}{\partial Q_4 \partial J})d Q_4 d J\nn &&
-(\frac{\partial^2 S(P_1, Q_2, P_3, Q_4, J)}{\partial J^2})d J^2.\ea

Dans ce cas, les formules du d\'eterminant et de la courbure scalaire de 
Ruppenier sont certaines grandes expressions. Mais, pour le cas particulier 
de cet entropie d'un trou noir extremal en rottaion en d\'ecoulant de la th\'eorie 
des cordes h\'et\'erotiques compactifi\'ee toroidalement, il est imm\'ediat d'obtenir 
que les composantes de la m\'etrique tenseure sont donn\'ees par:

\ba
g_{P_1 P_1}&=& \frac{\pi (Q_2 P_3 Q_4)^2}{2(J^2+ P_1 Q_2 P_3 Q_4)^{3/2}},\nn
g_{P_1 Q_2}&=& \frac{\pi P_1 Q_2 P_3^2 Q_4^2}{2 (J^2+ P_1 Q_2 P_3 Q_4)^{3/2}}
            - \frac{\pi P_3 Q_4}{ (J^2+ P_1 Q_2 P_3 Q_4)^{1/2}},\nn
g_{P_1 P_3}&=& \frac{\pi P_1 Q_2^2 P_3 Q_4^2} {2 (J^2+ P_1 Q_2 P_3 Q_4)^{3/2}}
            - \frac{\pi Q_2 Q_4}{ (J^2+ P_1 Q_2 P_3 Q_4)^{1/2}},\nn
g_{P_1 Q_4}&=& \frac{\pi P_1 Q_2^2 P_3^2 Q_4} {2 (J^2+ P_1 Q_2 P_3 Q_4)^{3/2}}
           - \frac{\pi Q_2 P_3}{(J^2+ P_1 Q_2 P_3 Q_4)^{1/2}},\nn
g_{P_1 J}&=& \frac{\pi J Q_2 P_3 Q_4}{(J^2+ P_1 Q_2 P_3 Q_4)^{3/2}},\ea
\ba g_{Q_2 Q_2}&=& \frac{\pi (P_1 P_3 Q_4)^2}{2 (J^2+ P_1 Q_2 P_3 Q_4)^{3/2}},\nn
g_{Q_2 P_3}&=& \frac{\pi P_1^2 Q_2 P_3 Q_4^2}{2 (J^2+ P_1 Q_2 P_3 Q_4)^{3/2}}
           - \frac{\pi P_1 Q_4}{(J^2+ P_1 Q_2 P_3 Q_4)^{1/2}},\nn
g_{Q_2 Q_4}&=& \frac{\pi P_1^2 Q_2 P_3^2 Q_4}{2 (J^2+ P_1 Q_2 P_3 Q_4)^{3/2}}
          - \frac{\pi P_1 P_3}{(J^2+ P_1 Q_2 P_3 Q_4)^{1/2}},\nn
g_{Q_2 J}&=& \frac{\pi J P_1 P_3 Q_4}{(J^2+ P_1 Q_2 P_3 Q_4)^{3/2}},\ea
\ba g_{P_3 P_3}&=& \frac{\pi (P_1 Q_2 Q_4)^2}{2 (J^2+ P_1 Q_2 P_3 Q_4)^{3/2}},\nn
g_{P_3 Q_4}&=& \frac{\pi P_1^2 Q_2^2 P_3 Q_4}{2 (J^2+ P_1 Q_2 P_3 Q_4)^{3/2}}
           -\frac{\pi P_1 Q_2}{(J^2+ P_1 Q_2 P_3 Q_4)^{1/2}},\nn
g_{P_3 J}&=& \frac{\pi J P_1 Q_2 Q_4}{(J^2+ P_1 Q_2 P_3 Q_4)^{3/2}},\nn
g_{Q_4 Q_4}&=& \frac{\pi (P_1 Q_2 P_3)^2}{2(J^2+ P_1 Q_2 P_3 Q_4)^{3/2}},\nn
g_{Q_4 J}&=& \frac{\pi J P_1 Q_2 P_3}{(J^2+ P_1 Q_2 P_3 Q_4)^{3/2}},\nn
g_{J J}&=& \frac{2 \pi J^2}{(J^2+ P_1 Q_2 P_3 Q_4)^{3/2}}
       - \frac{2 \pi}{ (J^2+ P_1 Q_2 P_3 Q_4)^{1/2}}.\ea

\begin{figure}
\hspace*{1.0cm}\vspace*{-6.0cm}
\includegraphics[width=12.0cm,angle=-0]{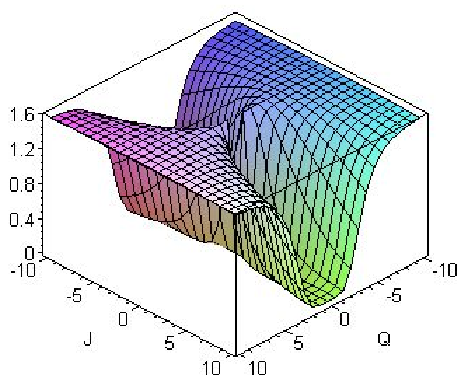}\vspace*{-3.0cm}
\caption{La composante $P_1P_1$ de la m\'etrique tenseure trac\'ee comme la
fonction de $\{Q, J\}$, en d\'ecrivant les fluctuations dans la configuration
des trous noirs de la th\'eorie des cordes h\'et\'erotiques.} \label{hetP1P1}\vspace*{0.5cm}
\end{figure}

\begin{figure}
\hspace*{1.0cm}\vspace*{-6.0cm}
\includegraphics[width=12.0cm,angle=-0]{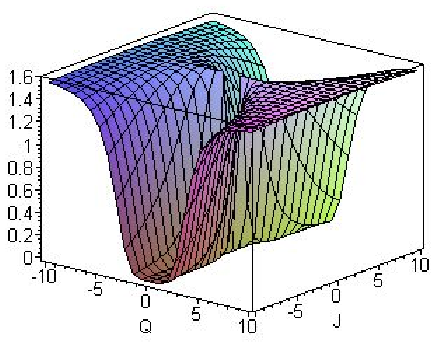}\vspace*{-3.0cm}
\caption{La composante $Q_2Q_2$ de la m\'etrique tenseure trac\'ee comme la
fonction de $\{Q, J\}$, en d\'ecrivant les fluctuations dans la configuration
des trous noirs de la th\'eorie des cordes h\'et\'erotiques.} \label{hetQ2Q2}\vspace*{0.5cm}
\end{figure}

\begin{figure}
\hspace*{1.0cm}\vspace*{-6.0cm}
\includegraphics[width=12.0cm,angle=-0]{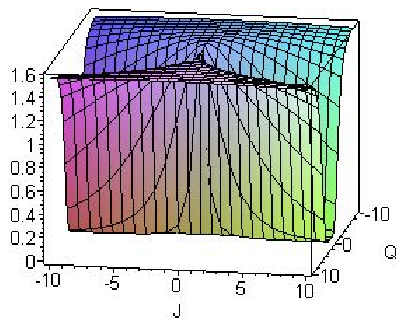}\vspace*{-3.0cm}
\caption{La composante $P_3P_3$ de la m\'etrique tenseure trac\'ee comme la
fonction de $\{Q, J\}$, en d\'ecrivant les fluctuations dans la configuration
des trous noirs de la th\'eorie des cordes h\'et\'erotiques.} \label{hetP3P3}\vspace*{0.5cm}
\end{figure}

\begin{figure}
\hspace*{1.0cm}\vspace*{-6.0cm}
\includegraphics[width=12.0cm,angle=-0]{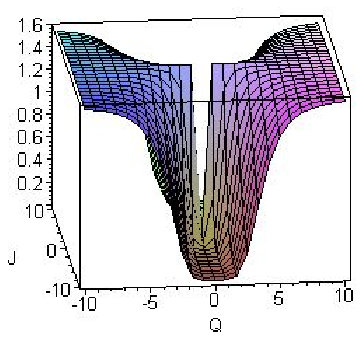}\vspace*{-3.0cm}
\caption{La composante $Q_4Q_4$ de la m\'etrique tenseure trac\'ee comme la
fonction de $\{Q, J\}$, en d\'ecrivant les fluctuations dans la configuration
des trous noirs de la th\'eorie des cordes h\'et\'erotiques.} \label{hetQ4Q4}\vspace*{0.5cm}
\end{figure}

\begin{figure}
\hspace*{1.0cm}\vspace*{-6.0cm}
\includegraphics[width=12.0cm,angle=-0]{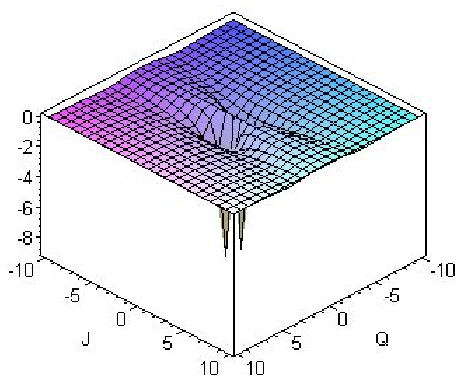}\vspace*{-3.0cm}
\caption{La composante $JJ$ de la m\'etrique tenseure trac\'ee comme la
fonction de $\{Q, J\}$, en d\'ecrivant les fluctuations dans la configuration
des trous noirs de la th\'eorie des cordes h\'et\'erotiques.} \label{hetJJ}\vspace*{0.5cm}
\end{figure}

\begin{figure}
\hspace*{1.0cm}\vspace*{-6.0cm}
\includegraphics[width=12.0cm,angle=-0]{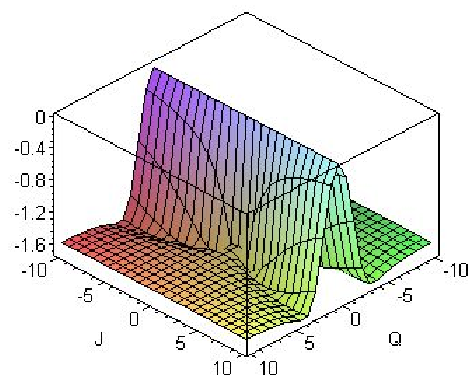}\vspace*{-3.0cm}
\caption{La composante $P_1Q_2$ de la m\'etrique tenseure trac\'ee comme la
fonction de $\{Q, J\}$, en d\'ecrivant les fluctuations dans la configuration
des trous noirs de la th\'eorie des cordes h\'et\'erotiques.} \label{hetP1Q2}\vspace*{0.5cm}
\end{figure}

\begin{figure}
\hspace*{1.0cm}\vspace*{-6.0cm}
\includegraphics[width=12.0cm,angle=-0]{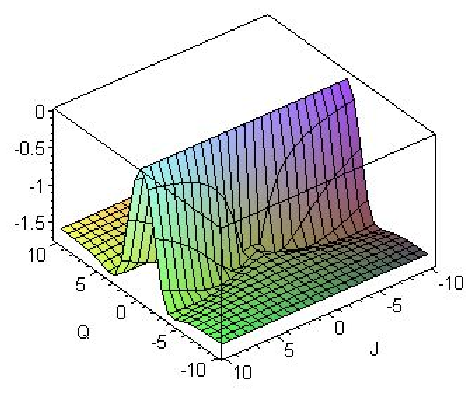}\vspace*{-3.0cm}
\caption{La composante $P_1P_3$ de la m\'etrique tenseure trac\'ee comme la
fonction de $\{Q, J\}$, en d\'ecrivant les fluctuations dans la configuration
des trous noirs de la th\'eorie des cordes h\'et\'erotiques.} \label{hetP1P3}\vspace*{0.5cm}
\end{figure}

\begin{figure}
\hspace*{1.0cm}\vspace*{-6.0cm}
\includegraphics[width=12.0cm,angle=-0]{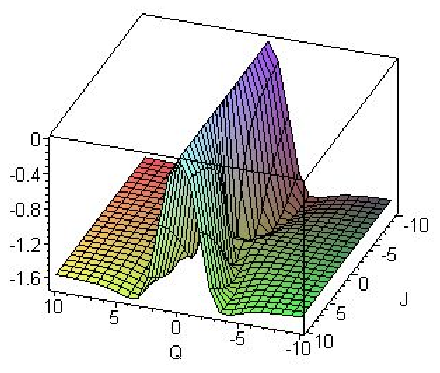}\vspace*{-3.0cm}
\caption{La composante $P_1Q_4$ de la m\'etrique tenseure trac\'ee comme la
fonction de $\{Q, J\}$, en d\'ecrivant les fluctuations dans la configuration
des trous noirs de la th\'eorie des cordes h\'et\'erotiques.} \label{hetP1Q4}\vspace*{0.5cm}
\end{figure}

\begin{figure}
\hspace*{1.0cm}\vspace*{-6.0cm}
\includegraphics[width=12.0cm,angle=-0]{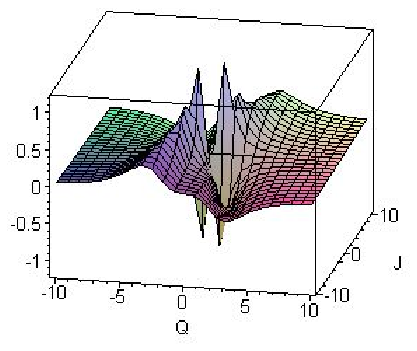}\vspace*{-3.0cm}
\caption{La composante $P_1J$ de la m\'etrique tenseure trac\'ee comme la
fonction de $\{Q, J\}$, en d\'ecrivant les fluctuations dans la configuration
des trous noirs de la th\'eorie des cordes h\'et\'erotiques.} \label{hetP1J}\vspace*{0.5cm}
\end{figure}

\begin{figure}
\hspace*{1.0cm}\vspace*{-6.0cm}
\includegraphics[width=12.0cm,angle=-0]{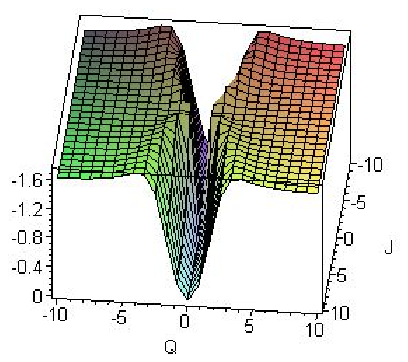}\vspace*{-3.0cm}
\caption{La composante $Q_2P_3$ de la m\'etrique tenseure trac\'ee comme la
fonction de $\{Q, J\}$, en d\'ecrivant les fluctuations dans la configuration
des trous noirs de la th\'eorie des cordes h\'et\'erotiques.} \label{hetQ2P3}\vspace*{0.5cm}
\end{figure}

\begin{figure}
\hspace*{1.0cm}\vspace*{-6.0cm}
\includegraphics[width=12.0cm,angle=-0]{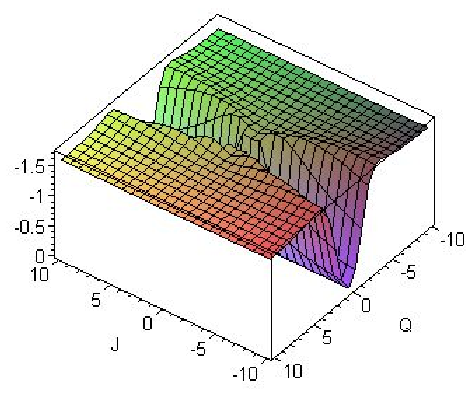}\vspace*{-3.0cm}
\caption{La composante $Q_2Q_4$ de la m\'etrique tenseure trac\'ee comme la
fonction de $\{Q, J\}$, en d\'ecrivant les fluctuations dans la configuration
des trous noirs de la th\'eorie des cordes h\'et\'erotiques.} \label{hetQ2Q4}\vspace*{0.5cm}
\end{figure}

\begin{figure}
\hspace*{1.0cm}\vspace*{-6.0cm}
\includegraphics[width=12.0cm,angle=-0]{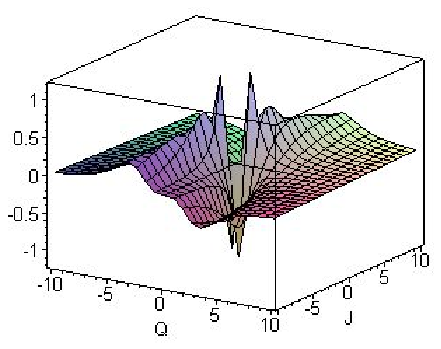}\vspace*{-3.0cm}
\caption{La composante $Q_2J$ de la m\'etrique tenseure trac\'ee comme la
fonction de $\{Q, J\}$, en d\'ecrivant les fluctuations dans la configuration
des trous noirs de la th\'eorie des cordes h\'et\'erotiques.} \label{hetQ2J}\vspace*{0.5cm}
\end{figure}

\begin{figure}
\hspace*{1.0cm}\vspace*{-6.0cm}
\includegraphics[width=12.0cm,angle=-0]{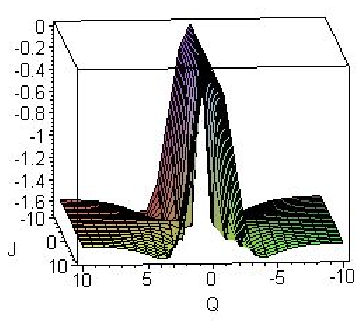}\vspace*{-3.0cm}
\caption{La composante $P_3Q_4$ de la m\'etrique tenseure trac\'ee comme la
fonction de $\{Q, J\}$, en d\'ecrivant les fluctuations dans la configuration
des trous noirs de la th\'eorie des cordes h\'et\'erotiques.} \label{hetP3Q4}\vspace*{0.5cm}
\end{figure}

\begin{figure}
\hspace*{1.0cm}\vspace*{-6.0cm}
\includegraphics[width=12.0cm,angle=-0]{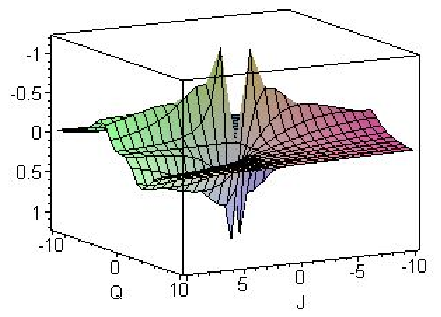}\vspace*{-3.0cm}
\caption{La composante $P_3J$ de la m\'etrique tenseure trac\'ee comme la
fonction de $\{Q, J\}$, en d\'ecrivant les fluctuations dans la configuration
des trous noirs de la th\'eorie des cordes h\'et\'erotiques.} \label{hetP3J}\vspace*{0.5cm}
\end{figure}

\begin{figure}
\hspace*{1.0cm}\vspace*{-6.0cm}
\includegraphics[width=12.0cm,angle=-0]{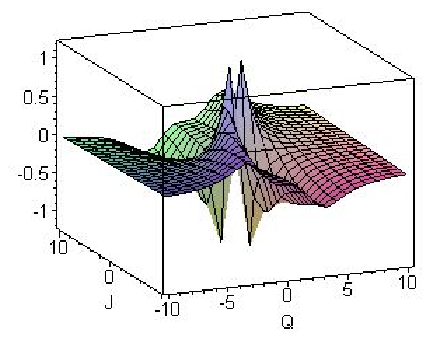}\vspace*{-3.0cm}
\caption{La composante $Q_4J$ de la m\'etrique tenseure trac\'ee comme la
fonction de $\{Q, J\}$, en d\'ecrivant les fluctuations dans la configuration
des trous noirs de la th\'eorie des cordes h\'et\'erotiques.} \label{hetQ4J}\vspace*{0.5cm}
\end{figure}

Dans ce cas, \`a l'avis qu'il y a cinq variables ind\'ependantes
$\{P_1, Q_2, P_3, Q_4, J\}$ des fluctuation, et ainsi, afin d'offrir
d'une vue tridimensionnelle des fluctuations statistiques, nous avons besoin de 
fixer les trois variables. Nous le faisons en consid\'erant $P_1= Q$, $Q_2 = Q$,
$P_3= Q$ et $Q_4 = Q $. Pour une configuration donn\'ee des trous noirs h\'et\'erotiques, 
les propri\'et\'es des fluctuations peuvent \^{e}tre d\'ecrites par un ensemble des 
graphiques suivantes de trois dimensions. Notamment, dans le r\'egime de $Q, J \in (-10, 10)$, 
nous voyons que l'amplitude des capacit\'es de chaleurs $\{g_{P_1 P_1}, g_{Q_2 Q_2},
g_{P_3 P_3}, g_{Q_4 Q_4}\}$ r\'eside dans l'intervalle $(0, 1.6)$. Dans cette gamme de 
$Q, J$, la capacit\'e de chaleur $\{g_{JJ}\}$ est comprise entre $(-10, 0)$. En conformant 
\`a la pr\'ediction de la g\'eom\'etrie de l'espace d'\'etat, nous observons que la gamme
de la croissance de la premi\`ere s\'erie et celle de la deuxi\`eme s\'erie des capacit\'es de 
chaleurs reste oppos\'ee \`a la sa nature-m\^eme. Plus pr\'ecis\'ement, pour une petite valeur 
donn\'ee de la charge $Q$, le premier ensemble des capacit\'es de chaleurs augment lorsque $Q$ augmente. 
Cependant, pour une petite valeur donn\'ee de la charge $Q$, la composante $\{g_{JJ}\}$ diminue,
lorsque la valeur de $Q$ est augment\'ee. Dans ce cas, les Figs. (\ref{hetP1P1}, 
\ref{hetQ2Q2}, \ref{hetP3P3}, \ref{hetQ4Q4} ) montrent que la croissance des capacit\'es de chaleurs 
$\{ g_{P_1 P_1}, g_{Q_2 Q_2}, g_{P_3 P_3}, g_{Q_4 Q_4}\}$ a lieu loin de l'origine et elles devient plat 
au sujet de la valeur de $|Q|= 5$. De plus, la Fig. (\ref{hetJJ}) montre que la croissance de la composante
$\{g_{JJ}\}$ a lieu dans la limite des petits $|Q|$. Fait int\'eressant, on trouve \`a rapporte de ces chiffres 
que l'ensemble des composantes diagonales de la m\'etrique tenseures de l'espace d'\'etat est sym\'etrique 
par rapport \`a la ligne $Q= 0$. D'autre part, les compressibilit\'es de chaleurs en comprenant des deux param\`etres 
distincts des trous noirs de la th\'eorie des cordes h\'et\'erotiques sont repr\'esent\'es dans les Figs. 
(\ref{hetP1Q2},\ref{hetP1P3}, \ref{hetP1Q4}, \ref{hetP1J},\ref{hetQ2P3}, \ref{hetQ2Q4}, \ref{hetQ2J}, 
\ref{hetP3Q4},\ref{hetP3J}, \ref{hetQ4J}). Dans ce cas, nous constatons que les capacit\'es de chaleurs,
en impliquant des charges \'electriques et magn\'etiques $\{ g_{P_1 Q_2}, g_{P_1 P_3}, g_{P_1 Q_4}, 
g_{Q_2 P_3}, g_{Q_2 Q_4}, g_{P_3 Q_4}\}$ appartiennent \`a l'intervalle $(-1.7, 0)$ et les capacit\'es 
de chaleurs, en impliquant le moment angulaire et l'une des charges \'electriques et magn\'etiques 
$\{ g_{P_1J}, g_{Q_2 J}, g_{P_3J}, g_{Q_4 J}\}$ appartiennent \`a l'intervalle $(-1.2, 1.2) $. 
De la repr\'esentation graphique de lam\'etrique tenseure de l'espace d'\'etat
$\{g_{ij}\ |\ i, j= P_1, Q_2, P_3, Q_4, J\}$, nous voyons que les trous noirs h\'et\'erotiques 
poss\`edent les capacit\'es de chaleurs positifs seulement \`a l'\'egard des charges \'electriques 
magn\'etiques. Notez bien que la composante en impliquant le moment angulaire $\{g_{JJ},\}$ a une 
valeur n\'egative. Ainsi, les fluctuations des trous noirs h\'et\'erotiques correspondent 
une configuration statistique localement instable.

Dans ce cas, nous trouvons que les mineurs principaux 
$\{ \mathit{p_i}\ |\ i= 1,2,3,4\}$ de la m\'etrique tenseure 
de l'espace d'\'etat peuvent \^{e}tre \'ecrite comme

\begin{eqnarray}
\mathit{p_1} := {\displaystyle \frac{1}{2}} \,{\displaystyle
\frac{\pi\,\mathit{Q_2}^{2}\,\mathit{P_3}^{2}\,
\mathit{Q_4}^{2}}{( J^{2}+ \mathit{P_1}\,\mathit{Q_2}\,\mathit{P_3}\,
\mathit{Q_4})^{(3/2 )}}},  \nonumber \\
\mathit{p_2} :=  - {\displaystyle \frac{\pi^{2}\,\mathit{P_3}^{2} \,\mathit{Q_4}^{2}\,J^{2}}{(J^{2} +
\mathit{P_1}\,\mathit{Q_2}\,
\mathit{P_3}\,\mathit{Q_4})^{2}}},  \nonumber \\
\mathit{p_3} :=  - {\displaystyle \frac{1}{2}} \,{\displaystyle
\frac{(4\,J^{2} + \mathit{P_1}\,\mathit{Q_2}\,\mathit{P_3}\,
\mathit{Q_4})\,\pi^{3}\,\mathit{Q_2}\,\mathit{P_3}\,\mathit{Q_4}^{3}\,
\mathit{P_1}}{(J^{2} +\mathit{P_1}\,\mathit{Q_2}\,\mathit{P_3}\,
\mathit{Q_4})^{(5/2)}}},  \nonumber \\
\mathit{p_4} :=  - {\displaystyle \frac{\pi^{4}\,\mathit{Q_2}^{2}\,
\mathit{P_3}^{2}\,\mathit{Q_4}^{2}\,\mathit{P_1}^{2}\,(3\,J^{2}
+ \mathit{P_1}\,\mathit{Q_2}\,\mathit{P_3}\,\mathit{Q_4})}{(J^{2}
+\mathit{P_1}\,\mathit{Q_2}\,\mathit{P_3}\,\mathit{Q_4})^{3}}}. 
\end{eqnarray}

Nous voyons donc que le d\'eterminant de cette m\'etrique tenseure,
comme le dernier mineur principe, $p_5:=g $ est donn\'e par:

\ba g= 2 \pi^5 \frac{(P_1 Q_2 P_3 Q_4)^3}{(J^2+ P_1 Q_2 P_3 Q_4)^{15/2}}
 \tilde{g}(P_1, Q_2, P_3, Q_4, J),\ea

o\`u la fonction $\tilde{g}(P_1, Q_2, P_3, Q_4, J)$ est d\'efinie par

\ba \tilde{g}(P_1, Q_2, P_3, Q_4, J)&:=&
J^8+ 4J^6 P_1 Q_2 P_3 Q_4+ 6J^4 (P_1 Q_2 P_3 Q_4)^2\nn &&+ 
4J^2(P_1 Q_2 P_3 Q_4)^3+ (P_1 Q_2 P_3 Q_4)^4).\ea

\begin{figure}
\hspace*{1.0cm}\vspace*{-6.0cm}
\includegraphics[width=12.0cm,angle=-0]{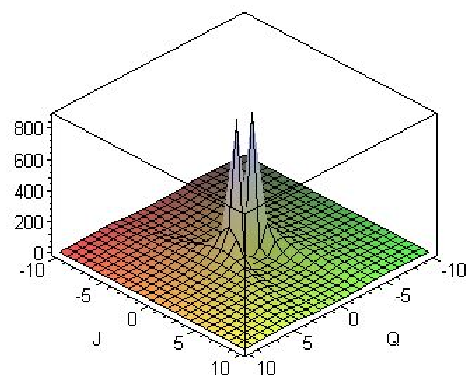}\vspace*{-3.0cm}
\caption{Le d\'eterminant de la m\'etrique tenseure trac\'ee comme la
fonction de $\{Q, J\}$, en d\'ecrivant les fluctuations dans la configuration
des trous noirs de la th\'eorie des cordes h\'et\'erotiques.} \label{hetdet}\vspace*{0.5cm}
\end{figure}

\begin{figure}
\hspace*{1.0cm}\vspace*{-6.0cm}
\includegraphics[width=12.0cm,angle=-0]{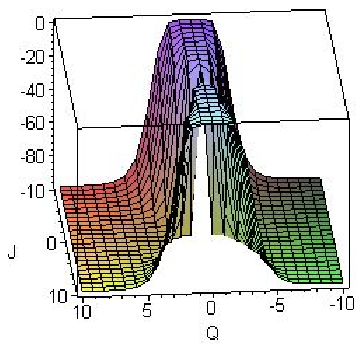}\vspace*{-3.0cm}
\caption{Le mineur d'hypersurface $p_4$ de la m\'etrique tenseure trac\'ee comme la
fonction de $\{Q, J\}$, en d\'ecrivant les fluctuations dans la configuration
des trous noirs de la th\'eorie des cordes h\'et\'erotiques.} \label{hetminor4}\vspace*{0.5cm}
\end{figure}

\begin{figure}
\hspace*{1.0cm}\vspace*{-6.0cm}
\includegraphics[width=12.0cm,angle=-0]{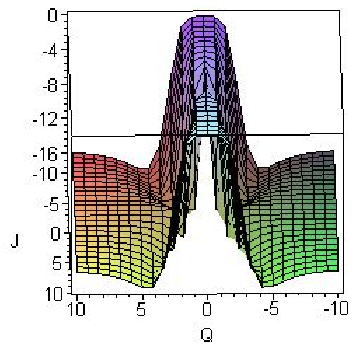}\vspace*{-3.0cm}
\caption{Le mineur d'hypersurface $p_3$ de la m\'etrique tenseure trac\'ee comme la
fonction de $\{Q, J\}$, en d\'ecrivant les fluctuations dans la configuration
des trous noirs de la th\'eorie des cordes h\'et\'erotiques.} \label{hetminor3}\vspace*{0.5cm}
\end{figure}

\begin{figure}
\hspace*{1.0cm}\vspace*{-6.0cm}
\includegraphics[width=12.0cm,angle=-0]{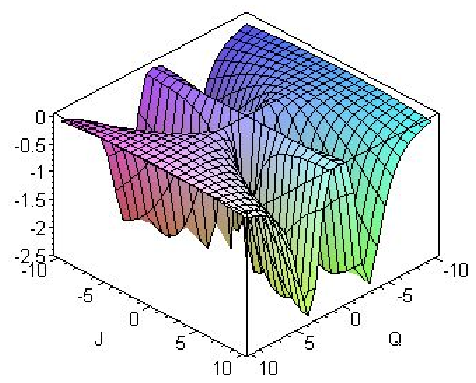}\vspace*{-3.0cm}
\caption{Le mineur de surface $p_2$ de la m\'etrique tenseure trac\'ee comme la
fonction de $\{Q, J\}$, en d\'ecrivant les fluctuations dans la configuration
des trous noirs de la th\'eorie des cordes h\'et\'erotiques.} \label{hetminor2}\vspace*{0.5cm}
\end{figure}

\begin{figure}
\hspace*{1.0cm}\vspace*{-6.0cm}
\includegraphics[width=12.0cm,angle=-0]{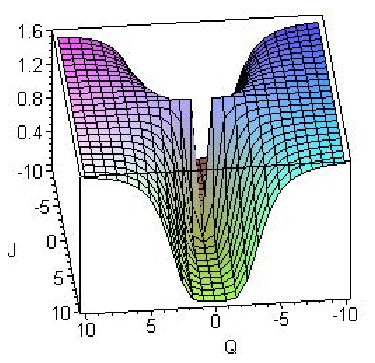}\vspace*{-3.0cm}
\caption{Le premier mineur de la m\'etrique tenseure trac\'ee comme la
fonction de $\{Q, J\}$, en d\'ecrivant les fluctuations dans la configuration
des trous noirs de la th\'eorie des cordes h\'et\'erotiques.} \label{hetminor1}\vspace*{0.5cm}
\end{figure}

Sous les fluctuations des param\`etres $\{P_1, Q_2, P_3, Q_4, J\}$,
la stabilit\'e d'un ensemble des trous noirs en rotation dans la th\'eorie des cordes 
h\'et\'erotiques d\'ecoule de la positivit\'e du d\'eterminant de la m\'etrique tenseure.
Pour le choix de $Q_i= Q$ et $P_i = Q$, nous constatons que le d\'eterminant de
la m\'etrique tenseure $g$ a le signe positif. En fait, la Fig. (\ref{hetdet}) montre
que la valeur num\'erique de $g$ est dans l'intervalle $(0, 900)$. Par ailleurs,
la stabilit\'e de l'hypersurface d\'efinie par une valeur constante de $J$ est indiqu\'ee
dans la Fig. (\ref{hetminor4}). Par la pr\'esente, nous voyons que le mineur $p_4$ r\'eside 
dans la gamme de $(-100, 0)$. Pour une valeur donn\'ee de $J$, nous remarquons que le mineur $p_4$ 
forme une sous porte ce qui est fa\c{c}onn\'e de la r\'egion sym\'etriques, comme le 
param\`etre $Q$ est pass\'ee de z\'ero \`a $5$. De plus, la stabilit\'e de l'hypersurface 
d\'efinie par une valeur constante de $J$ et $P_4$ est montr\'e dans la Fig. (\ref{hetminor3}). 
Dans ce cas, nous voyons que le mineur $p_3$ r\'eside dans la gamme de $(-20, 0)$. Le mineur $p_3$
aussi forme une sous porte en forme de la r\'egion sym\'etrique, mais l'hypersurface est moins lisse.
Sur la m\^{e}me ligne d'observation, la surface d\'efinie par les fluctuations de $\{P_1, Q_2\}$ 
est montr\'e dans la Fig. (\ref{hetminor2}). Dans ce cas, nous voyons que les mineurs $p_2$ r\'eside 
dans la gamme de $(-2.5, 0)$. Le mineur $p_2 $ est \'egalement forme une sous porte, mais la forme 
de la r\'egion sym\'etrique avec une hypersurface plus riugueuse des fluctuations. Enfin, lorsque 
le seul param\`etre $P_1$ est autoris\'e \`a varier, la stabilit\'e de la configuration de ces trous noirs 
est donn\'ee par la positivit\'e du premier mineur principe $p_1:= g_{P_1 P_1}$. Une vue rotat\'ee de 
$p_1$ est montr\'ee dans la Fig. (\ref{hetminor1}). Le comportement qualitatif ci-dessus de l'espace 
d'\'etat offrent des propri\'et\'es des fluctuations statistiques. Depuis les mineurs principaux
$\{p_i \ | \ i = 2, 3, 4\}$ sont moins que z\'ero pour certain $i$. Ainsi, l'exigence que tous les 
mineurs principaux doit \^{e}tre positif des fluctuations, en invoquant des param\`etres 
$\{P_1, Q_2, P_3, Q_4, J\}$, montre que la configuration des trous noirs en rotation 
de la th\'eorie des cordes h\'et\'erotiques correspondent \`a un ensemble statistique 
instables au niveau globale.

Avec les $\Gamma_{abc}$ ce que nous les avons fournis dans l'annex $[A]$, 
il est \'egalement facile de voir dans ce cas des trous noirs de la th\'eorie 
des cordes heterotiques compactifi\'ee toroidalement que la courbure scalaire 
correspondante de Ruppenier est r\'eduite \`a:

\ba R= \frac{1}{2\pi P_1 Q_2 P_3 Q_4} \frac{r_1r_2^{5/2}}{r_3},\ea

o\`u les fonctions $\{ r_i(P_1, Q_2, P_3, Q_4, J)\ | i=1,2,3 \}$ sont d\'efinies par

\ba r_1&:=& 9J^2+ 2P_1 Q_2 P_3 Q_4,\nn
r_2&:=& J^2+ P_1 Q_2 P_3 Q_4,\nn
r_3&:=& J^6+ 3J^4 (P_1 Q_2 P_3 Q_4)\nn &&
+ 3J^2 (P_1 Q_2 P_3 Q_4)^2+ (P_1 Q_2 P_3 Q_4)^3.\ea

On peut observer simplement que cette courbure scalaire de Ruppenier est 
non-nulle pour toutes valeurs physiquement autoris\'ees des charges \'electriques, 
magn\'etiques et le moment cin\'etique. Donc, la vari\'et\'e sous-jacente de l'espace d\'etat 
est partout r\'eguli\`ere, sauf la branche d'ergo. Ceci est bien compatible avec le fait 
que l'espace de modules des trous noirs extr\'emaux en rotation est constitu\'e des 
deux succursales de l'ergo, ce qu'on peuisse aussi voir facilement aux racines r\'eelles 
du d\'enominateur de cette courbure scalaire de Ruppenier. En ce cas quand il n'y pas de 
rotation, c'est-\`a-dire que le cas de $J=0$, nous voyons que le d\'eterminant de la 
m\'etrique tensuere est r\'eduite \`a la valeur: 

\ba  g\vert_{J=0}= \frac{2 \pi^5}{\sqrt{P_1 Q_2 P_3 Q_4}},\ea
 
et la courbure scalaire correspondante est donn\'ee par:

\ba R\vert_{J=0}= \frac{1}{ \pi \sqrt{P_1 Q_2 P_3 Q_4}}.\ea

Ainsi, cette g\'eom\'etrie thermodynamique de Ruppenier est bien d\'efinie et 
un syst\`eme statistique en interactions, ceci reste les m\^emes au point de $J = 0$.
En outre, pour le cas de $ Q_2= Q_4 $ et $ P_1= P_3 $ nous pouvons voir 
\'egalement que la valeur absolue du d\'eterminant de la m\'etrique tenseure
et celle de la courbure scalaire de Ruppenier deviennent les m\^emes que 
le d\'eterminant et la courbure scalaire de Ruppenier des trous noirs de 
Kaluza-Klein, comme nous les avons vu dans la section pr\'ec\'edente.
En bref, nous avons montr\'e que la g\'eom\'etrie thermodynamique
des trous noirs extr\'emaux en rotation en quatre dimensions de 
la th\'eorie des cordes et celle des trous noirs de Kluza-Klein 
et de Kerr-Newman dans la th\'eorie d'Einstein Maxwell a certaines 
instabilit\'es aux points de la brache d'ergo, sinon \`a un point
arbitaire de l'espace d'\'etat, elle est partout bien-d\'efinie.

\begin{figure}
\hspace*{1.0cm}\vspace*{-6.0cm}
\includegraphics[width=12.0cm,angle=-0]{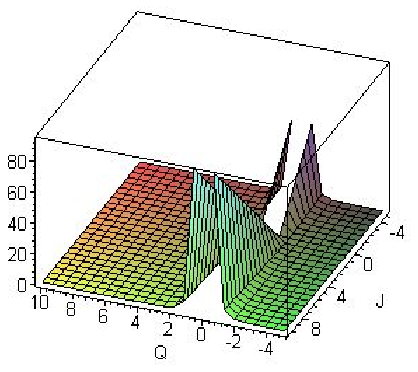}\vspace*{-3.0cm}
\caption{La courbure scalaire trac\'ee comme la fonction de $\{Q, J\}$, 
en d\'ecrivant les fluctuations dans la configuration des trous noirs 
de la th\'eorie des cordes h\'et\'erotiques dans la gamme de 
$Q, J \in (-5, 10) $.} \label{hetR}\vspace*{0.5cm}
\end{figure}

\begin{figure}
\hspace*{1.0cm}\vspace*{-6.0cm}
\includegraphics[width=12.0cm,angle=-0]{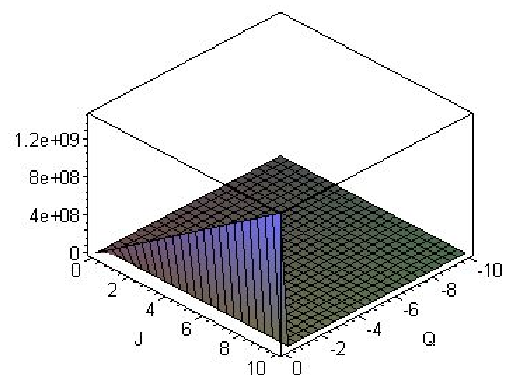}\vspace*{-3.0cm}
\caption{La courbure scalaire trac\'ee comme la fonction de $\{Q, J\}$, 
en d\'ecrivant les fluctuations dans la configuration des trous noirs 
de la th\'eorie des cordes h\'et\'erotiques dans la gamme de 
$Q \in (-10, 0.01)$ et $J \in (0, 10)$.} \label{hetRhalf}\vspace*{0.5cm}
\end{figure}

Comme mentionn\'e dans le cas des trous noirs de Kerr-Neumann et des trous noirs de Kaluza-Klein,
la courbure scalaire de l'espace d'\'etat signifie des interactions statistiques globales
de la fluctuation des charges et le moment angulaire. Ainsi, pour un ensemble donn\'e des 
param\`etres $\{P_1, Q_2, P_3, Q_4, J\}$, la structure de la stabilit\'e globale des trous noirs 
en rotation de la th\'eorie des cordes h\'et\'erotiques peut \^{e}tre d\'ecrite comme la suite.
Dans la gamme de $Q, J \in (-5, 10)$, la Fig. (\ref{hetR}) montre que la courbure scalaire a un 
pic positif \`a l'ordre de $100$. Dans cette gamme de param\`etres, on observe que la configuration 
sous-jacente de ces trous noirs correspondant \`a un syst\`eme statistique avec des interactions non-nulles. 
Physiquement, le signe positif de la courbure scalaire signifie que les interactions statistiques 
sont r\'epulsives dans la leur nature. La Fig. (\ref{hetRhalf}) illustre le comportement de la courbure 
scalaire ci-dessus pour la gamme des param\`etres $Q \in (-10, 0.01)$ et $J \in (0, 10) $. Dans la gamme 
de $Q, J \in (-5, 10)$, nous voyons que les interactions sont pr\'esentes pr\`esque sym\'etriquement 
sur les lignes $Q= 0$ et $J= 0$. Dans la gamme de $Q \in (-10, 0.01)$ et $J \in (0, 10)$, la courbure 
scalaire de l'espace d'\'etat acquiert un grand pic \`a l'ordre de $1.4 \times 10^{09}$ pr\`es 
de la valeur de $J= 10$. Dans une gamme petite de $Q$, nous constatons de la Fig. (\ref{hetRhalf}) 
que les interactions statistiques globales augmentent quand on augmente la valeur de $J$ de z\'ero \`a $10$. 
En comparaison des interactions apparaissant dans la gamme des param\`etres $Q, J \in (-5, 10)$, 
nous voyons que l'amplitude des interactions statistiques globales se r\'ev\`ele d'\^{e}tre beaucoup 
plus grande dans la gamme de $Q \in (-10, 0.01)$ et $J \in (0, 10)$. Pr\'ecis\'ement, il s'ensuit que 
le ratio des valeurs typiques de l'amplitude des courbures scalaires de l'espace d'\'etat
est \`a l'ordre de $10^{+07}$. La vue graphique de cette comparaison mentionn\'ee ci-dessus
d\'ecoule des Figs. (\ref{hetR}, \ref{hetRhalf}). Qualitativement, les repr\'esentations de 
l'espace d'\'etat indiquent que les trous noirs en rotation de la th\'eorie des cordes 
h\'et\'erotiques correspondent \`a une configuration statistique instable et fortement interactive.


\clearpage
\chapter{Remarques et conclusions}

Dans cette recherche, nous avons \'etudi\'e la g\'eom\'etrie thermodynamique de 
l'espace d'\'etat our une classe g\'en\'erale de trous noirs avec les corrections 
de $ \alpha^{\prime} $ ou bien avec le principe d'incertitude g\'en\'eralis\'ee.
Nous avons \'etabli que les inclusions des corrections de $ \alpha^{\prime} $ 
et du $l_p$ modifient les concepts g\'eom\'etriques thermodynamiques habituels
et chacune induise g\'en\'eralement une courbure scalaire diff\'erente
dans la g\'eom\'etrie de Ruppeiner o\`u celle de Wienhold.
Nous avons analys\'e les corrections de $ \alpha^{\prime} $ de la g\'eom\'etrie
de Ruppenier d'une classe de trous noirs, qui sont bien int\'egr\'ees dans
certains cas int\'eressants de l'entropie des trous noirs.
Cela nous motive \`a \'etudier la g\'eom\'etrie thermodynamique de Wienhold 
et celle de Ruppenier associ\'ee respectivement \`a la masse et l'entropie des trous noirs.
D'autre part, comme dans la perspective du m\'ecanisme d'attracteur,
la g\'eom\'etrie de Wienhold est proportionnelle \`a la g\'eom\'etrie de
l'espace des modules aux points fixes de l'attracteur,
laquelle nous avons analys\'e pour le cas de trous noirs dilatoniques topologiques.
\`A ce stade, nous avons constat\'e que la g\'eom\'etrie de Weinhold associ\'ee
\`a la masse du trou noir est bien d\'efinie et a non-nulle courbure de Weinhold.
Dans certains cas, les courbures de la thermodynamique sont partout r\'eguli\`{e}res dans
l'espace d'\'etat, afin qu'il n'y ait pas d'instabilit\'ees thermodynamiques dans
l'espace d'\'etat d'un ensemble des trous noirs consid\'er\'es.
Nous avons montr\'e que la g\'eom\'etrie thermodynamique du trou noir de 
Reissner-Nordstr\"om dans l'$AdS_4$ a des singularit\'es qui peuvent 
correspondre aux instabilit\'es tachyoniques et sont bien connues dans 
la th\'eorie de jauge supersym\'etrique.
Nos conclusions de l'\'etude de la g\'eom\'etrie de Weinhold sont donc clarifiantes,
compatibles avec l'existence de certaines instabilit\'es dans la vari\'et\'e
de sous-espace d'\'etat.

Nous avons constat\'e que les g\'eom\'etries de Ruppenier d'une classe
de trous noirs extr\'emaux sont en g\'en\'eral partout r\'eguliers,
except\'{e}s les cas o\`u il y a quelques instabilit\'es
dans la vari\'et\'e de l'espace d'\'etat ainsi que les cas o\`{u}
l'entropie des trous noirs consid\'er\'es ont besoin de plus de
contributions de $ \alpha^{\prime} $.
Ces r\'esultats sont v\'erifi\'es pour l'entropie des trous noirs
supersym\'etriques avec les corrections de Gauss-Bonnet et aussi pour
les divers ordres des corrections de $ \alpha^{\prime} $ \`a l' entropie
des trous noirs non-supersym\'etriques.
Nous avons montr\'e qu'il existe d'autres exemples comme les trous noirs 
de Kaluza-Klain en rotation, les trous noirs decoulant de la th\'eorie des 
cordes h\'et\'erotiques compactifi\'ee toroidalement, $\ldots$ etc, 
qui ont des divergences dans leur courbure de Ruppenier.
Ces divergences dans les courbures sclaires de Ruppenier sont en bon accord avec
la pr\'esence de branche de l'ergo dans les th\'eories des trous noirs
en rotation.
Les r\'esultats de l'\'etude g\'eom\'etrique thermodynamique des branes noirs
non-extremal des $ D_1D_5p $ et $ D_2D_6NS_5p $ sont tr\`es int\'eressants.
Ici, nous d\'ecouvrons que les g\'eom\'etries thermodynamiques
de Ruppenier associ\'ees aux vari\'et\'es d'espace d'\'etat 
des branes noirs non-extr\'emaux des $ D_1D_5p $ et $ D_2D_6NS_5p $
sont g\'en\'eralement bien d\'efinies et partout r\'eguli\`eres.

Cependant, nous avons \'egalement montr\'e que le trou noir de 
Reissner-Nordstr\"om n'est pas en interaction.
Ainsi, le plaisir du jeu g\'eom\'etrique est intact, et nous voyons 
que la conclusion reste la m\^eme bien qu'on ajoute les contributions 
du principe d'incertitude g\'en\'eralis\'ee \`a l'entropie du trou 
noir de Reissner-Nordstr\"om.
De plus, \`a partir de la th\'eorie des cordes, nous voyons dans 
certaines circonstances que les corrections de $ \alpha^{\prime} $ 
des d\'eriv\'ees sup\'erieures \`a l'entropie des trous noirs ont
modifi\'e la g\'eom\'etrie thermodynamique de mani\`{e}rs bien 
attendues.
Par exemple, la g\'eom\'etrie de Ruppenier \`a l'entropie des trous 
noirs non-supersym\'etriques extr\'emaux avec les corrections des 
d\'eriv\'es sup\'erieures est bien d\'efinie et courb\'ee.
En fait, dans ce cas, nous avons observ\'e une \'evolution de la 
courbure de Ruppenier avec l'ajout des corrections des d\'eriv\'es 
sup\'{e}rieures \`a l'entropie de ce trou noir.
Par cons\'equent, nous avons donn\'e une m\'ethode pour d\'eterminer 
le degr\'e du d\'eterminant de la m\'etrique tenseur, et ensuite, 
nous pouvons facilement d\'eterminer le degr\'e de l'autre facteur 
apparaissant dans la courbure de Ruppenier \`a l'ordre arbitraire 
$l$ des corrections de $ \alpha^{\prime} $.
Cette observation du degr\'e de la g\'eom\'etrie de Ruppenier est 
la mesure du formulaire pour toutes les sous-contributions \`a 
l'entropie du trou noir non-supersym\'etrique extr\'emal.
C'est-\`a-dire que cet observation de la g\'eom\'etrie thermodynamique 
reste vraie $ \forall l >1 $.

De plus, nous avons donn\'e une reformulation du probl\`eme en termes 
d'\'energie libre de la th\'eorie des cordes topologiques du trou noir
ainsi que de la fonction de partition d'un ensemble des trous noirs 
qui indique dans le cas des petits trous noirs que l'ensemble doit \^etre 
un ensemble m\'elang\'e. La g\'eom\'etrie de Ruppenier est peut \^etre l'un 
des concepts les plus importants pour comprendre les fonctions de corr\'elations 
de la th\'eorie des champs conformes \`a la fronti\`ere.
De toute mani\`ere, la courbure scalaire de Ruppenier est li\'ee aux 
fonctions de corr\'elations des deux points de la th\'eorie des champs 
conformes \`a la fronti\`ere.
Dans cette recherche, nous avons consid\'er\'e certains cas int\'eressants 
\`a l'\'egard de stabilit\'e des des trous noirs. 
Ensuite, nous avons vu que la g\'eom\'etrie thermodynamique de Ruppenier 
est bien d\'efinie et partout r\'eguli\`ere, avec ou sans les corrections 
de $\alpha^{\prime}$ des d\'eriv\'ees sup\'erieures.
Par exemple, c'est le cas des trous noirs extr\'emaux supersym\'etriques de BPS 
en $D= 4$ ou celui des non-supersym\'etriques ou les branes noirs 
$ D_1D_5 $ et $ D_2D_6NS_5 $ en $D = 10$ non-extr\'emaux.
En effet, ces g\'eom\'etries de l'espace d'\'etat sont partout r\'eguli\`eres.
Cela est parfaitement en accord avec un autre fait que ces trous noirs ou 
branes noirs de BPS sont stables, et restent les m\^emes objets avec les 
corrections de $\alpha^{\prime}$ des d\'eriv\'es sup\'erieures.
En revanche, la g\'eom\'etrie de Ruppenier des trous noirs en rotation en 
$D = 4$ diverge aux points de la branche d'ergo.
Donc, il devrait exister des fonctions de corr\'elations divergentes des 
deux points de la th\'eorie des champs conformes \`a la fronti\`ere 
correspondante \`a ces syst\`emes de trous noirs en rotation.

Du point de vue de la g\'eom\'etrie de Weinhold qui est directement associ\'ee
\`a la g\'eom\'etrie de l'espace des modules, on peut essayer de comprendre 
l'origine microscopique des singularit\'es tachyoniques des trous noirs de 
Reissner-Nordstr\"om dans l'$AdS_4$ en d\'ecoulant de la consid\'eration
d'un grand nombre de $M_2$-branes co\"incidantes.
Autrement dit, l'\'etude de la g\'eom\'etrie thermodynamique covariante peut 
\'eclairer sur la nature des fonctions de corr\'elations des deux 
points de la th\'eorie des champs conformes de la fronti\`ere correspondante.
De cette mani\`ere, on peut comprendre du point de vue de la th\'eorie des 
champs conformes que le syst\`eme thermique sous-jacent est stable ou instable.
Par cons\'equence, nous pouvons comprendre l'origine microscopique des 
singularit\'es thermodynamiques de certaines paires d'anti-branes et 
branes de la th\'eorie des champs conformes.
En particulier, le cas des trous noirs non-BPS d\'ecoulant de la th\'eorie 
des cordes peuvent \^etre \'etudi\'es des points de vu de la g\'eom\'etrie 
thermodynamique, et les interactions thermodynamiques apparaissent en pr\'esence 
des deux modes gauche et droite, de la duale th\'eorie des champs conformes.
En bref, notre \'etude g\'eom\'etrique peut pr\'evoir la nature des interactions 
pr\'esente dans la duale th\'eorie des champs conformes de certains trous noirs
ou branes noirs extr\'emaux ainsi que ceux de non-extr\'emaux,
d\'ecoulant de la th\'eorie des cordes ou bien de la M-th\'eorie fondamentale.

Finalement, il est \'egalement int\'eressant de g\'en\'eraliser ces \'etudes
aux th\'eories des d\'eriv\'ees sup\'erieures arbitraires pour les syst\`emes
de trous noirs.
Nous aimerions \'etudier les propri\'et\'es g\'eom\'etriques thermodynamiques
pour les solutions de certains branes, comme dans la th\'eorie des cordes ou
bien dans la M- th\'eorie, dans les diverses dimensions de l'espace-temps 
$ D \geq 4 $ et celles corrig\'ees par les d\'eriv\'ees sup\'erieures 
de $ \alpha^{\prime} $.
En outre, il est interessant d'enqu\^eter sur la g\'eom\'etrie thermodynamique
loins des points fixes de l'attracteur et aussi sur certaines extensions de
l'inclusion des situations de non-\'equilibre.
Dans les cas sp\'ecifiques, nous devrions aussi enqu\^eter pour les suites
de la g\'eom\'etrie thermodynamique loin des points fixes de l'attracteur.
Nous avons l'intention d'aborder ces questions dans une prochaine publication 
\cite{bntESC}.
Dans un proche avenir, nous sommes \'egalement int\'eress\'es \`a  \'etudier 
certaines propri\'et\'es g\'eom\'etriques et alg\'ebriques de ces syst\`emes 
de trous noirs dans la th\'eorie des supercordes ainsi que dans la M-th\'eorie.
%


\clearpage
%
\appendix 
\chapter{Les symboles de Christoffel du premier type}

Dans cet appendice, nous allons furnir les formules explicites des
symboles de Christoffel du premier type des trous noirs ce que nous
avons consid\'er\'e dans les chapitres $3-7$. 
Comme nous avons analys\'e la configuration de l'espace d'\'etat
d'un ensemble des trous noirs, dans cet appendice, nous pr\'esentons 
les symboles de Christoffel du premier type de tous ces configurations 
consid\'er\'ees pr\'ec\'edents de la g\'eom\'etrie thermodynamique 
de Ruppenier et de Wienhold.

\section{La gomtrie de Wienhold des trous noirs dilatoniques}

Avec la sym\'{e}trie dans les deux premiers indices, nous constatons que la
composante $SSS$ de symbole de Christoffel du premier type peut \^{e}tre
exprim\'{e}e comme

\begin{eqnarray}
{\Gamma_{S\,S\,S}} &=& - {\displaystyle \frac{1}{32}}
\frac{(a^{2} + 1)}{(S^{3}\,(n - 1) ^{3}\,\pi )}(32 \,\pi
^{2}\,b^{a^{2}}\,Q^{2}\,\mathrm{a_2}\,a^{4} + 2\,L\,b^{a
^{2}}\,\mathrm{a_1}\,a^{4} \nonumber \\ && - 3\,k\,b^{
-a^{2}}\,\mathrm{a_3}\,n\,a^{2}+2\,L\,b^{a^{2}}\,\mathrm{a_1}\,n\,a^{2}
+ k\,b^{-a^{2}}\,\mathrm{a_3}\,n^{2}\,a^{2} \nonumber \\ && -
6\,L\,b^{a^{2}}\,\mathrm{ a_1}\,a^{2} -
224\,\pi^{2}\,b^{a^{2}}\,Q^{2}\,\mathrm{a_2}\,a ^{2} \mbox{} +
160\,\pi ^{2}\,b^{a^{2}}\,Q^{2}\,\mathrm{a_2}\,n\,a^{ 2} \nonumber
\\ && + 2\,k\,b^{ -a^{2}}\,\mathrm{a_3}\,a^{2} + 3\,k\,b^{-a^{ 2}}\,\mathrm{a_3}\,n^{2}
- 2\,k\,b^{- a^{2}}\,\mathrm{a_3}\,n \mbox{} \nonumber \\ && -
k\,b^{-a^{2}}\,\mathrm{a_3}\,n^{3} + 4\,L\,b^{(a^{2
})}\,\mathrm{a_1} +192\,\pi ^{2}\,b^{a^{2}}\,Q^{2}\,\mathrm{ a_2}\,n^{2} \nonumber \\
&& + 384\,\pi ^{2}\,b^{a^{2}}\,Q^{2}\,\mathrm{a_2} \mbox{} -
544\,\pi ^{2}\,b^{a^{2}}\,Q^{2}\,\mathrm{a_2}\,n - 2
\,L\,b^{a^{2}}\,\mathrm{a_1}\,n),
\end{eqnarray}

o\`{u} les facteurs $ \{a_i, i = 1,2,3 \} $ sont donn\'{e}s par

\begin{eqnarray}
\mathrm{a_1}&=& (4\,S)^{(\frac{n - a^{2}}{n - 1})},  \nonumber \\
\mathrm{a_2}&=& (4\,S)^{( - \frac{a^{2} + n - 2}{n - 1})},  \nonumber \\
\mathrm{a_3}&=& (4\,S)^{(\frac{a^{2} + n - 2}{n - 1})}.
\end{eqnarray}

Dans ce cas, les autres symboles de Christoffel du premier type,
qui sont non-nulles, sont donn\'{e}s par

\begin{eqnarray}
{\Gamma_{S\,S\,Q}}&=& 2\,{\displaystyle \frac{\pi \,(a^{2} +
1)\,b ^{a^{2}}\,Q\,(4\,S)^{( - \frac{a^{2} + n - 2}{n -
1})}\,(a^{2} + 2\,n - 3)}{(n - 1)^{2}\,S^{2}}},  \nonumber \\
{\Gamma_{S\,Q\,S}}&=& 2\,{\displaystyle \frac{\pi \,(a^{2} +
1)\,b ^{a^{2}}\,Q\,(4\,S)^{( - \frac{a^{2} + n - 2}{n -
1})}\,(a^{2} + 2\,n - 3)}{(n - 1)^{2}\,S^{2}}},  \nonumber \\
{\Gamma_{S\,Q\,Q}}&=& - 2\,{\displaystyle \frac{\pi \,(a^{2} +
1) \,b^{a^{2}}\,(4\,S)^{( - \frac{a^{2} + n - 2}{n - 1})}}{(n -
1 )\,S}},  \nonumber \\
{\Gamma_{Q\,Q\,S}}&=& - 2\,{\displaystyle \frac{\pi \,(a^{2} +
1) \,b^{a^{2}}\,(4\,S)^{( - \frac{a^{2} + n - 2}{n - 1})}}{(n - 1
)\,S}}. \end{eqnarray}

\newpage
\section{La gomtrie de Wienhold des solutions de $ M_2$-branes}

Avec la sym\'{e}trie dans les deux premiers indices, nous constatons 
que les symboles de Christoffel du premier type sont donn'es par

\begin{eqnarray}
{\Gamma_{S\,S\,S}}&=& - {\displaystyle \frac{3}{32\pi^{4}\,L^{2}
\,S^{6}}} (S^{12} - S ^{6}\,\pi ^{6}\,L^{6}\,\mathit{Q_1}^{6} +
5\,\pi ^{12}\,L^{12}\, \mathit{Q_2}^{6}\,\mathit{Q_1}^{6} +
15\,S^{4}\,\pi ^{8}\,L^{8}\, \mathit{Q_1}^{6}\,\mathit{Q_2}^{2}
\nonumber \\ &&  + 13\,\pi
^{10}\,L^{10}\,\mathit{Q_2}^{6}\,S^{2}\,\mathit{ Q_1}^{4} +
13\,\pi ^{10}\,L^{10}\,\mathit{Q_2}^{4}\,\mathit{Q_1}^{6} \,S^{2}
+ 31\,S^{6}\,L^{6}\,\mathit{Q_1}^{4}\,\pi ^{6}\,\mathit{Q_2 }^{2}
\nonumber \\ &&  + 45\,S^{8}\,\pi^{4}\,L^{4}\,
\mathit{Q_1}^{2}\,\mathit{Q_2 }^{2} + 5\,S^{10}\,\pi^{2}\,L^{2}\,
\mathit{Q_1}^{2} + 5\,S^{10}\, \pi^{2}\,L^{2}\,\mathit{Q_2}^{2} -
5\,S^{8}\,\pi ^{4}\,L^{4}\, \mathit{Q_2}^{4} \nonumber \\ &&  -
5\,S^{8}\,\pi^{4}\,L^{4}\,\mathit{Q_1}^{4} - S^{6}\,\pi
^{6}\,L^{6}\,\mathit{Q_2}^{6} + 25\,S^{4}\,\pi^{8}\,L^{8}\,
\mathit{Q_1}^{4}\,\mathit{Q_2}^{4} + 15\,S^{4}\,\pi^{8}\,L^{8}\,
\mathit{Q_2}^{6}\,\mathit{Q_1}^{2} \nonumber \\ &&  +
31\,S^{6}\,\pi
^{6}\,L^{6}\,\mathit{Q_2}^{4}\,\mathit{Q_1 }^{2}) \times  
({\displaystyle \frac{S^{4} + \pi^{2}\,L^{2}\,S^{2}\,\mathit{Q_1
}^{2} +\pi^{2}\,L^{2}\, S^{2}\,\mathit{Q_2}^{2} + \pi^{4}\,L^{4}\,
\mathit{Q_1}^{2}\,\mathit{Q_2}^{2}}{\pi \,S}} )^{(-5/2)},
\nonumber \\
{\Gamma_{S\,S\,\mathit{Q_1}}} &=& {\displaystyle
\frac{1}{16\,\pi^{2}\,S^{5}}}
\mathit{Q_1}(3\,\pi^{10}\,L^{10}\,\mathit{Q_2}^{6}\,\mathit{Q_1}^{4
} + 9\,L^{8}\,\mathit{Q_1}^{4}\,\pi^{8}\,\mathit{Q_2}^{4}\,S^{2}
 - 6\,S^{6}\,\pi^{4}\,L^{4}\,\mathit{Q_1}^{2}\,\mathit{Q_2}^{2}
\nonumber \\ &&  + 5\,S^{4}\,\pi
^{6}\,L^{6}\,\mathit{Q_1}^{4}\,\mathit{Q_2}^{2} + 10\,S^{4}\,\pi
^{6}\,L^{6}\,\mathit{Q_2}^{4}\,\mathit{Q_1}^{2} - 10\,S^{8}\,\pi
^{2}\,L^{2}\,\mathit{Q_1}^{2} + 25\,S^{8}\,\pi^{2}\,
L^{2}\,\mathit{Q_2}^{2} \nonumber \\ &&  - S^{6}\,\pi
^{4}\,L^{4}\,\mathit{Q_1}^{4} + 37\,S^{6}\, \pi
^{4}\,L^{4}\,\mathit{Q_2}^{4} + 3\,S^{10} + 6\,\pi^{8}\,L^{8}
\,\mathit{Q_2}^{6}\,S^{2}\,\mathit{Q_1}^{2} + 15\,\pi^{6}\,L^{6}\,
\mathit{Q_2}^{6}\,S^{4}) \times  \nonumber \\ &&
({\displaystyle \frac{S^{4} + \pi^{2}\,L^{2}\,S
^{2}\,\mathit{Q_1}^{2} + \pi^{2}\,L^{2}\,S^{2}\,\mathit{Q_2}^{2} +
\pi^{4}\,L^{4}\,\mathit{Q_1}^{2}\,\mathit{Q_2}^{2}}{\pi
\,S}})^{(-5/2)}, \nonumber \\ 
{\Gamma_{S\,S\,\mathit{Q_2}}}&=&
{\displaystyle \frac{1}{16\pi^{2}\,S^{5}}}
\mathit{Q_2}(3\,\pi^{10}\,L^{10}\,\mathit{Q_2}^{4}\,\mathit{Q_1}^{6
} + 6\,\pi^{8}\,L^{8}\,\mathit{Q_1}^{6}\,S^{2}\,\mathit{Q_2}^{2}
 + 9\,L^{8}\,\mathit{Q_1}^{4}\,\pi^{8}\,\mathit{Q_2}^{4}\,S^{2}
\nonumber \\ &&  - 6\,S^{6}\,\pi
^{4}\,L^{4}\,\mathit{Q_1}^{2}\,\mathit{Q_2}^{2} + 10\,S^{4}\,\pi
^{6}\,L^{6}\,\mathit{Q_1}^{4}\,\mathit{Q_2}^{ 2} + 5\,S^{4}\,\pi
^{6}\,L^{6}\,\mathit{Q_2}^{4}\,\mathit{Q_1}^{2} + 3\,S^{10}
\nonumber \\ && + 25\,S^{8}\,\pi^{2}\,L^{2}\,\mathit{Q_1}^{2} -
10\,S^{8} \,\pi^{2}\,L^{2}\,\mathit{Q_2}^{2} + 37\,S^{6}\,
\pi^{4}\,L^{4}\,\mathit{Q_1}^{4} - S^{6}\,\pi^{4}\,L^{4}\,
\mathit{Q_2}^{4} \nonumber \\ && +
15\,\pi^{6}\,L^{6}\,\mathit{Q_1}^{6}\,S^{4})
\times ({\displaystyle \frac{S^{4} + \pi^{2}\,L^{2}\,S^{2}\,\mathit{Q_1
}^{2} + \pi^{2}\,L^{2}\,S^{2}\,\mathit{Q_2}^{2} + \pi^{4}\,L^{4}
\,\mathit{Q_1}^{2}\,\mathit{Q_2}^{2}}{\pi \,S}} )^{(-5/2)},
\nonumber\\ {\Gamma_{S\,\mathit{Q_1}\,S}}&=& {\displaystyle
\frac{1}{16\,\pi^{2}\,S^{5}}} \mathit{Q_1}(3\,\pi
^{10}\,L^{10}\,\mathit{Q_2}^{6}\,\mathit{Q_1}^{4 } +
9\,L^{8}\,\mathit{Q_1}^{4}\,\pi^{8}\,\mathit{Q_2}^{4}\,S^{2}
 - 6\,S^{6}\,\pi^{4}\,L^{4}\,\mathit{Q_1}^{2}\,\mathit{Q_2}^{2}
\nonumber \\ &&  + 5\,S^{4}\,\pi
^{6}\,L^{6}\,\mathit{Q_1}^{4}\,\mathit{Q_2}^{2} + 10\,S^{4}\,\pi
^{6}\,L^{6}\,\mathit{Q_2}^{4}\,\mathit{Q_1}^{ 2} - 10\,S^{8}\,\pi
^{2}\,L^{2}\,\mathit{Q_1}^{2} + 25\,S^{8}\,\pi
^{2}\,L^{2}\,\mathit{Q_2}^{2} \nonumber \\ &&
 - S^{6}\,\pi^{4}\,L^{4}\,\mathit{Q_1}^{4} + 37\,S^{6}\,
\pi^{4}\,L^{4}\,\mathit{Q_2}^{4} + 3\,S^{10} + 6\,\pi^{8}\,L^{8}
\,\mathit{Q_2}^{6}\,S^{2}\,\mathit{Q_1}^{2} + 15\,\pi^{6}\,L^{6}\,
\mathit{Q_2}^{6}\,S^{4}) \times  \nonumber \\ &&
({\displaystyle \frac{S^{4} + \pi^{2}\,L^{2}\,S^{2}\,
\mathit{Q_1}^{2} + \pi^{2}\,L^{2}\,S^{2}\,\mathit{Q_2}^{2} +
\pi^{4}\,L^{4}\,\mathit{Q_1}^{2}\,\mathit{Q_2}^{2}}{\pi \,S}})^{(-5/2)}, \nonumber \\
{\Gamma_{S\,\mathit{Q_1}\,\mathit{Q_1}}} &=& {\displaystyle \frac
{1}{8\, \pi^{2}\, S^{2}}}
(13\,S^{4}\,\pi^{4}\,L^{4}\,\mathit{Q_1}^{2}\,\mathit{Q_2}^{2} +
11\,\pi^{6}\,L^{6}\,S^{2}\,\mathit{Q_2}^{4}\,\mathit{Q_1}^{2} +
5\,S^{6}\,\pi^{2}\,L^{2}\,\mathit{Q_1}^{2}  \nonumber \\ && -
5\,S^{6}\,\pi^{2 }\,L^{2}\,\mathit{Q_2}^{2} - 7\,\pi^{4}\,
L^{4}\,S^{4}\,\mathit{Q_2}^{4} - S^{8} + 3 \,\pi^{8}\,L^{8}\,
\mathit{Q_2}^{6}\,\mathit{Q_1}^{2} - 3\,\pi^{6}\,L^{6}\,
\mathit{Q_2}^{6}\,S^{2}) \times \nonumber \\ &&
({\displaystyle \frac{S^{4} + \pi ^{2}\,L^{2}\,S^{2}\,\mathit{Q_1
}^{2} + \pi ^{2}\,L^{2}\,S^{2}\,\mathit{Q_2}^{2} + \pi^{4}\,L^{4}
\,\mathit{Q_1}^{2}\,\mathit{Q_2}^{2}}{\pi \,S}} )^{(-5/2)},
\nonumber \\
{\Gamma_{S\,\mathit{Q_1}\,\mathit{Q_2}}}&=& - {\displaystyle
\frac{ 1}{8\,S^{4}}}L^{2}\,\mathit{Q_1}\,\mathit{Q_2}
(12\,S^{4}\,\pi^{4}\,L^{4} \,\mathit{Q_1}^{2}\,\mathit{Q_2}^{2} +
5\,S^{8} + 4\,\pi^{6}\,L^{6}\,S^{2}\,\mathit{Q_1}^{4}\,
\mathit{Q_2}^{2}  \nonumber \\ && + 4\,\pi^{6}\,L^{6}\,S^{2}\,
\mathit{Q_2}^{4}\,\mathit{Q_1} ^{2} + 8\,S^{6}\,\pi^{2}\,L^{2}\,
\mathit{Q_1}^{2}  + 8\,S^{6}\,\pi ^{2}\,L^{2}\,\mathit{Q_2}^{2} +
3\,\pi^{4}\,L^{4}\,S^{4}\, \mathit{Q_1}^{4}  \nonumber \\ &&  +
3\,\pi ^{4}\,L^{4}\,S^{4}\,\mathit{Q_2}^{4} + \pi^{8}\,
L^{8}\,\mathit{Q_1}^{4}\,\mathit{Q_2}^{4}) \times   \nonumber \\
&& ({\displaystyle \frac{S^{4} + \pi^{2}\,L^{2}\,S^{2}\,\mathit{Q_1
}^{2} + \pi^{2}\,L^{2}\,S^{2}\,\mathit{Q_2}^{2} + \pi^{4}\,L^{4}
\,\mathit{Q_1}^{2}\,\mathit{Q_2}^{2}}{\pi \,S}} )^{(-5/2)},
\end{eqnarray}
\begin{eqnarray}
{\Gamma_{S\,\mathit{Q_2}\,S}}&=&{\displaystyle \frac{1}{16\,
\pi^{2} \,S^{5}}}\mathit{Q_2}(3\,\pi^{10}\,L^{10}\,
\mathit{Q_2}^{4}\,\mathit{Q_1}^{6 } + 6\,\pi^{8}\,L^{8}\,
\mathit{Q_1}^{6}\,S^{2}\,\mathit{Q_2}^{2} + 9\,L^{8}\,
\mathit{Q_1}^{4}\,\pi^{8}\,\mathit{Q_2}^{4}\,S^{2} \nonumber \\ &&
-6\,S^{6}\,\pi^{4}\,L^{4}\, \mathit{Q_1}^{2}\,\mathit{Q_2}^{2} +
10\,S^{4}\,\pi^{6}\, L^{6}\,\mathit{Q_1}^{4}\,\mathit{Q_2}^{ 2}
+5\,S^{4}\, \pi^{6}\,L^{6}\,\mathit{Q_2}^{4}\,\mathit{Q_1}^{2} +
3\,S^{10} \nonumber \\ && + 25\,S^{8}\,\pi^{2}\,L^{2}\,
\mathit{Q_1}^{2} - 10\,S^{8} \,\pi^{2}\,L^{2}\,\mathit{Q_2}^{2} +
37\,S^{6}\,\pi ^{4}\,L^{4}\, \mathit{Q_1}^{4} -
S^{6}\,\pi^{4}\,L^{4}\,\mathit{Q_2}^{4} \nonumber \\ &&  +
15\,\pi^{6}\,L^{6}\,\mathit{Q_1}^{6}\,S^{4}) \times
({\displaystyle \frac{S^{4} + \pi ^{2}\,L^{2}\,S^{2}\,\mathit{Q_1
}^{2} + \pi ^{2}\,L^{2}\,S^{2}\,\mathit{Q_2}^{2} + \pi^{4}\,L^{4}
\,\mathit{Q_1}^{2}\,\mathit{Q_2}^{2}}{\pi \,S}} )^{(-5/2)},
\nonumber \\
{\Gamma_{S\,\mathit{Q_2}\,\mathit{Q_1}}}&=& - {\displaystyle
\frac{ 1}{8\,S^{4}}} L^{2}\,\mathit{Q_1}\,\mathit{Q_2}
(12\,S^{4}\,\pi^{4}\,L^{4} \,\mathit{Q_1}^{2}\,\mathit{Q_2}^{2} +
5\,S^{8} + 4\,\pi^{6}\,L^{6}\,S^{2}\,\mathit{Q_1}^{4}\,
\mathit{Q_2}^{2} \nonumber \\ && + 4\,\pi^{6}\,L^{6}\,S^{2}\,
\mathit{Q_2}^{4}\,\mathit{Q_1}^{2} + 8\,S^{6}\,\pi^{2}\,L^{2}\,
\mathit{Q_1}^{2} + 8\,S^{6}\,\pi ^{2}\,L^{2}\,\mathit{Q_2}^{2} +
3\,\pi^{4}\,L^{4}\,S^{4}\, \mathit{Q_1}^{4} \nonumber \\ &&  +
3\,\pi^{4}\,L^{4}\,S^{4}\, \mathit{Q_2}^{4} + \pi^{8}\,
L^{8}\,\mathit{Q_1}^{4}\, \mathit{Q_2}^{4}) \times
\nonumber \\ && ({\displaystyle \frac{S^{4} +\pi^{2}\,L^{2}\,
S^{2}\,\mathit{Q_1 }^{2} + \pi^{2}\,L^{2}\,S^{2}\,\mathit{Q_2}^{2}
+ \pi^{4}\,L^{4} \,\mathit{Q_1}^{2}\,\mathit{Q_2}^{2}}{\pi
\,S}})^{(-5/2)},
\nonumber \\
{\Gamma_{S\,\mathit{Q_2}\,\mathit{Q_2}}}&=&{\displaystyle
\frac{1}{ 8\,\pi^{2}\, S^{2}}}
(3\,\pi^{8}\,L^{8}\,\mathit{Q_1}^{6}\, \mathit{Q_2}^{2} +
13\,S^{4}\,\pi^{4}\,L^{4}\,\mathit{Q_1}^{2}\, \mathit{Q_2}^{2} +
11\,\pi^{6}\,L^{6}\,S^{2}\,\mathit{Q_1}^{4}\, \mathit{Q_2}^{2}
\nonumber \\ &&  - 5\,S^{6}\,\pi^{2}\,L^{2}\, \mathit{Q_1}^{2} +
5\,S^{6}\, \pi^{2}\,L^{2}\,\mathit{Q_2}^{2} - 7\,\pi
^{4}\,L^{4}\,S^{4}\, \mathit{Q_1}^{4} - 3\,\pi^{6}\,L^{6}\,
\mathit{Q_1}^{6}\,S^{2} - S^{8}) \times
\nonumber \\ && ({\displaystyle \frac{S^{4} +
\pi^{2}\,L^{2}\,S^{2}\,\mathit{Q_1 }^{2} +
\pi^{2}\,L^{2}\,S^{2}\,\mathit{Q_2}^{2} + \pi^{4}\,L^{4}
\,\mathit{Q_1}^{2}\,\mathit{Q_2}^{2}}{\pi \,S}} )^{(-5/2)},
\nonumber\\
{\Gamma_{\mathit{Q_1}\,\mathit{Q_1}\,S}}&=&{\displaystyle \frac
{1}{ 8\,\pi^{2}\, S^{2}}} (13\,S^{4}\,\pi^{4}\,L^{4}\,
\mathit{Q_1}^{2}\,\mathit{Q_2}^{2} + 11\,\pi^{6}\,L^{6}\,S^{2}\,
\mathit{Q_2}^{4}\,\mathit{Q_1}^{2} + 5\,S^{6}\,\pi^{2}\,L^{2}\,
\mathit{Q_1}^{2} - S^{8} \nonumber \\ &&  - 5\,S^{6}\,\pi^{2
}\,L^{2}\,\mathit{Q_2}^{2} - 7\,\pi^{4}\,L^{4}\,S^{4}\,
\mathit{Q_2}^{4} +3 \,\pi^{8}\,L^{8}\,\mathit{Q_2}^{6}\,
\mathit{Q_1}^{2} - 3\,\pi^{6} \,L^{6}\,\mathit{Q_2}^{6}\,S^{2})
\nonumber \\ && ({\displaystyle \frac{S^{4} +
\pi^{2}\,L^{2}\,S^{2}\, \mathit{Q_1 }^{2} +
\pi^{2}\,L^{2}\,S^{2}\,\mathit{Q_2}^{2} + \pi ^{4}\,L^{4}
\,\mathit{Q_1}^{2}\,\mathit{Q_2}^{2}}{\pi \,S}} )^{(-5/2)},
\nonumber
\\
{\Gamma_{\mathit{Q_1}\,\mathit{Q_1}\,\mathit{Q_2}}}&=&{\displaystyle
\frac{1}{4\,S}} L^{2}\,\mathit{Q_2}(\pi^{6}\,L^{6}\,
\mathit{Q_2}^{4} \,\mathit{Q_1}^{2} + 2\,S^{2}\,\pi^{4}\,
L^{4}\,\mathit{Q_1}^{2}\, \mathit{Q_2}^{2} + \pi^{4}\,L^{4}\,
\mathit{Q_2}^{4}\,S^{2} + S^{4} \,\pi^{2}\,L^{2}\,\mathit{Q_1}^{2}
\nonumber \\ && + 2\,S^{4}\,\pi^{2}\,L^{2}\,\mathit{Q_2}^{2} +
S^{6}) \times ({\displaystyle \frac{S^{4} + \pi^{2}\,L^{2}\,S^{2}\,\mathit{Q_1
}^{2} + \pi^{2}\,L^{2}\,S^{2}\,\mathit{Q_2}^{2} + \pi^{4}\,L^{4}
\,\mathit{Q_1}^{2}\,\mathit{Q_2}^{2}}{\pi \,S}} )^{(-5/2)},
\nonumber \\
{\Gamma_{\mathit{Q_1}\,\mathit{Q_1}\,\mathit{Q_1}}}&=& -
{\displaystyle \frac{3}{4}} \,{\displaystyle \frac{L^{2}\,
\mathit{Q_1}\,(S^{2} + \pi^{2}\,L^{2}\,\mathit{Q_2}^{2})^{3}}{S\,(
{\displaystyle \frac{S^{4} +
\pi^{2}\,L^{2}\,S^{2}\,\mathit{Q_1}^{2} + \pi
^{2}\,L^{2}\,S^{2}\,\mathit{Q_2}^{2} + \pi^{4}\,L^{4}
\,\mathit{Q_1}^{2}\,\mathit{Q_2}^{2}}{\pi \,S}} )^{(5/2)}}},
\end{eqnarray}
\begin{eqnarray}
{\Gamma_{\mathit{Q_1}\,\mathit{Q_2}\,S}}&=& - {\displaystyle
\frac{ 1}{8\,S^{4}}} L^{2}\,\mathit{Q_1}\,\mathit{Q_2}
(12\,S^{4}\,\pi^{4}\,L^{4} \,\mathit{Q_1}^{2}\,\mathit{Q_2}^{2} +
5\,S^{8} + 4\,\pi^{6}\,L^{6}\,S^{2}\,\mathit{Q_1}^{4}\,
\mathit{Q_2}^{2} \nonumber \\ && + 4\,\pi^{6}\,L^{6}\,
S^{2}\,\mathit{Q_2}^{4}\,\mathit{Q_1}^{2} + 8\,S^{6}\,\pi^{2}\,
L^{2}\,\mathit{Q_1}^{2} +
8\,S^{6}\,\pi^{2}\,L^{2}\,\mathit{Q_2}^{2} +
3\,\pi^{4}\,L^{4}\,S^{4}\, \mathit{Q_1}^{4} \nonumber \\ &&  +
3\,\pi ^{4}\,L^{4}\,S^{4}\,\mathit{Q_2}^{4} + \pi^{8}\,
L^{8}\,\mathit{Q_1}^{4}\,\mathit{Q_2}^{4}) \times
\nonumber \\ && ({\displaystyle \frac{S^{4} + \pi
^{2}\,L^{2}\,S^{2}\,\mathit{Q_1 }^{2} + \pi
^{2}\,L^{2}\,S^{2}\,\mathit{Q_2}^{2} + \pi^{4}\,L^{4}
\,\mathit{Q_1}^{2}\,\mathit{Q_2}^{2}}{\pi \,S}} )^{(-5/2)},
\nonumber \\
{\Gamma_{\mathit{Q_1}\,\mathit{Q_2}\,\mathit{Q_1}}}&=&{\displaystyle
\frac{1}{4\, S}} L^{2}\,\mathit{Q_2}(\pi^{6}\,L^{6}\,
\mathit{Q_2}^{4} \,\mathit{Q_1}^{2} + 2\,S^{2}\,\pi^{4}\,L^{4}\,
\mathit{Q_1}^{2}\, \mathit{Q_2}^{2} + \pi^{4}\,L^{4}\,
\mathit{Q_2}^{4}\,S^{2} + S^{4} \,\pi^{2}\,L^{2}\,\mathit{Q_1}^{2}
\nonumber \\ && + 2\,S^{4}\,\pi ^{2}\,L^{2}\,\mathit{Q_2}^{2} +
S^{6}) ({\displaystyle \frac{S^{4} + \pi^{2}\,L^{2}\,S^{2}\,\mathit{Q_1
}^{2} + \pi^{2}\,L^{2}\,S^{2}\,\mathit{Q_2}^{2} + \pi^{4}\,L^{4}
\,\mathit{Q_1}^{2}\,\mathit{Q_2}^{2}}{\pi \,S}} )^{(-5/2)},
\nonumber\\
{\Gamma_{\mathit{Q_1}\,\mathit{Q_2}\,\mathit{Q_2}}}&=&{\displaystyle
\frac{1}{4\, S}} L^{2}\,\mathit{Q_1}(L^{6}\,
\mathit{Q_1}^{4}\,\pi^{6} \,\mathit{Q_2}^{2} +
2\,S^{2}\,\pi^{4}\,L^{4}\,\mathit{Q_1}^{2}\, \mathit{Q_2}^{2} +
S^{4}\,\pi^{2}\,L^{2}\,\mathit{Q_2}^{2} + \pi
^{4}\,L^{4}\,\mathit{Q_1}^{4}\,S^{2} \nonumber \\ &&  +
2\,S^{4}\,\pi^{2}\,L^{2}\,\mathit{Q_1}^{2} + S^{6}) \times
({\displaystyle \frac{S^{4} + \pi^{2}\,L^{2}\,S^{2}\,\mathit{Q_1
}^{2} + \pi^{2}\,L^{2}\,S^{2}\,\mathit{Q_2}^{2} + \pi^{4}\,L^{4}
\,\mathit{Q_1}^{2}\,\mathit{Q_2}^{2}}{\pi \,S}} )^{(-5/2)},
\nonumber \\
{\Gamma_{\mathit{Q_2}\,\mathit{Q_2}\,S}}&=&{\displaystyle
\frac{1}{8\,\pi^{2}\, S^{2}}} (3\,\pi^{8}\,L^{8}\,
\mathit{Q_1}^{6}\,\mathit{Q_2}^{2} + 13\,S^{4}\,\pi^{4}\,L^{4}\,
\mathit{Q_1}^{2}\,\mathit{Q_2}^{2} + 11\,\pi^{6}\,L^{6}\,S^{2}\,
\mathit{Q_1}^{4}\,\mathit{Q_2}^{2} \nonumber \\&& - 5\,S^{6}\,
\pi^{2}\,L^{2}\,\mathit{Q_1}^{2} + 5\,S^{6}\, \pi^{2}\,L^{2}\,
\mathit{Q_2}^{2} - 7\,\pi^{4}\,L^{4}\,S^{4}\, \mathit{Q_1}^{4} -
3\,\pi^{6}\,L^{6}\,\mathit{Q_1}^{6}\,S^{2} - S^{8}) \times
\nonumber \\ && ({\displaystyle \frac{S^{4} + \pi^{2}\,L^{2}\,S^{2}\,\mathit{Q_1
}^{2} + \pi^{2}\,L^{2}\,S^{2}\,\mathit{Q_2}^{2} + \pi^{4}\,L^{4}
\,\mathit{Q_1}^{2}\,\mathit{Q_2}^{2}}{\pi \,S}} )^{(-5/2)},
\nonumber \\
{\Gamma_{\mathit{Q_2}\,\mathit{Q_2}\,\mathit{Q_1}}}&=&{\displaystyle
\frac{1}{4\, S }} L^{2}\,\mathit{Q_1}(L^{6}\,
\mathit{Q_1}^{4}\,\pi^{6} \,\mathit{Q_2}^{2} +
2\,S^{2}\,\pi^{4}\,L^{4}\,\mathit{Q_1}^{2}\, \mathit{Q_2}^{2} +
S^{4}\,\pi^{2}\,L^{2}\,\mathit{Q_2}^{2} + \pi
^{4}\,L^{4}\,\mathit{Q_1}^{4}\,S^{2} \nonumber \\ &&  +
2\,S^{4}\,\pi^{2}\,L^{2}\,\mathit{Q_1}^{2} + S^{6})
({\displaystyle \frac{S^{4} + \pi^{2}\,L^{2}\,S^{2}\,\mathit{Q_1
}^{2} + \pi^{2}\,L^{2}\,S^{2}\,\mathit{Q_2}^{2} + \pi^{4}\,L^{4}
\,\mathit{Q_1}^{2}\,\mathit{Q_2}^{2}}{\pi \,S}} )^{(-5/2)},
\nonumber \\
{\Gamma_{\mathit{Q_2}\,\mathit{Q_2}\,\mathit{Q_2}}}&=& -
{\displaystyle \frac{3}{4}} \,{\displaystyle \frac{L^{2}\,
\mathit{Q_2}\,(S^{2} + \pi^{2}\,L^{2}\,\mathit{Q_1}^{2})^{3}}{S\,(
{\displaystyle \frac{S^{4} + \pi^{2}\,L^{2}\,S^{2}\,\mathit{Q_1}
^{2} + \pi^{2}\,L^{2}\,S^{2}\,\mathit{Q_2}^{2} + \pi^{4}\,L^{4}
\,\mathit{Q_1}^{2}\,\mathit{Q_2}^{2}}{\pi \,S}} )^{(5/2)}}}.
\end{eqnarray}

\newpage
\section{Les corrections de $ l_P $ dans la gomtrie thermodynamique:}

\subsection{La gomtrie de Ruppenier des trous noirs de Reissner-Nordstr\"om}

Avec la sym\'{e}trie dans les deux premiers indices, il s'av\`ere que les
symboles de Christoffel de premi\`ere esp\`{e}ce peuvent \^{e}tre exprim\'{e}s comme

\begin{eqnarray}
{\Gamma_{M\,M\,M}}&=& - 3\,{\displaystyle \frac{\pi
\,d\,Q^{4}}{L \,(M^{2} - Q^{2})^{(5/2)}}}, \nonumber \\ {\Gamma
_{M\,M\,Q}}&=& 3\,{\displaystyle \frac{\pi \,d\,M\,Q^{3}}{L
\,(M^{2} - Q^{2})^{(5/2)}}}, \nonumber \\ {\Gamma_{M\,Q\,M}}&=&
3\,{\displaystyle \frac{\pi \,d\,M\,Q^{3}}{L \,(M^{2} -
Q^{2})^{(5/2)}}}, \nonumber \\ {\Gamma_{M\,Q\,Q}}&=& -
3\,{\displaystyle \frac{\pi \,d\,M^{2}\,Q ^{2}}{L\,(M^{2} -
Q^{2})^{(5/2)}}}, \nonumber \\ {\Gamma_{Q\,Q\,M}}&=& -
3\,{\displaystyle \frac{\pi \,d\,M^{2}\,Q ^{2}}{L\,(M^{2} -
Q^{2})^{(5/2)}}}, \nonumber \\ {\Gamma_{Q\,Q\,Q}}&=&
3\,{\displaystyle \frac{\pi \,d\,M^{3}\,Q}{L \,(M^{2} -
Q^{2})^{(5/2)}}}.
\end{eqnarray}

\newpage
\subsection{La gomtrie de Ruppenier des trous noirs chargs magntiquement}

Avec la sym\'{e}trie dans les deux premiers indices, il s'ensuit que la
composantes non-nulles de symbole de Christoffel du premier type sont

\begin{eqnarray}
{\Gamma_{M\,M\,M}}&=&12\,{\displaystyle \frac{Q^{2}\,\zeta
(7)}{M ^{5}}}, \nonumber \\ {\Gamma_{M\,M\,Q}}&=& -
6\,{\displaystyle \frac{Q\,\zeta (7)}{M^{ 4}}}, \nonumber
\\ {\Gamma_{M\,Q\,M}}&=& - 6\,{\displaystyle \frac{Q\,\zeta (7)}{M^{
4}}}, \nonumber \\ {\Gamma_{M\,Q\,Q}}&=&2\,{\displaystyle \frac
{\zeta (7)}{M^{3}}}, \nonumber \\ {\Gamma
_{Q\,Q\,M}}&=&2\,{\displaystyle \frac{\zeta (7)}{M^{3}}}.
\end{eqnarray}

\newpage
\section{Les corrections de $ \alpha^{\prime} $ dans la gomtrie thermodynamique:}

\subsection{La gomtrie de Ruppenier des trous noirs dyoniques extr\'emaux supersymtriques en quatre dimensions}

Avec les propri\'{e}t\'{e}s de la sym\'{e}trie en deux premiers indices, 
nous constatons que les \'{e}l\'{e}ments de symbole de Christoffel du premier type sont
donn\'es par 

\begin{eqnarray}
{\Gamma_{n\,n\,n}}&=& - {\displaystyle \frac{3}{8}} \,
{\displaystyle \frac{\pi \,w^{3}\,(N\,\mathrm{W} + 4\, \widehat{\alpha})^{3}}{(n
\,w\,(N\,\mathrm{W} + 4\, \widehat{\alpha}))^{(5/2)}}},  \nonumber \\
{\Gamma_{n\,n\,w}}&=&{\displaystyle \frac{1}{8}} \,
{\displaystyle \frac{\pi \,w^{2}\,(N\,\mathrm{W} + 4\, \widehat{\alpha})^{3}\,n
}{(n\,w\,(N\,\mathrm{W} + 4\, \widehat{\alpha}))^{(5/2)}}},  \nonumber \\
{\Gamma_{n\,n\,N}}&=&{\displaystyle \frac{1}{8}} \,
{\displaystyle \frac{\pi \,w^{3}\,(N\,\mathrm{W} + 4\, \widehat{\alpha})^{2}\,n
\,\mathrm{W}}{(n\,w\,(N\,\mathrm{W} + 4\, \widehat{\alpha}))^{(5/2)}}},  \nonumber \\
{\Gamma_{n\,n\,\mathrm{W}}}&=&{\displaystyle \frac{1}{8}} \,
{\displaystyle \frac{\pi \,w^{3}\,(N\,\mathrm{W} + 4\, \widehat{\alpha})^{2}\,n
\,N}{(n\,w\,(N\,\mathrm{W} + 4\, \widehat{\alpha}))^{(5/2)}}},  \nonumber \\
{\Gamma_{n\,w\,n}}&=&{\displaystyle \frac{1}{8}} \,
{\displaystyle \frac{\pi \,w^{2}\,(N\,\mathrm{W} + 4\, \widehat{\alpha})^{3}\,n
}{(n\,w\,(N\,\mathrm{W} + 4\, \widehat{\alpha}))^{(5/2)}}},  
\end{eqnarray}
\begin{eqnarray}
{\Gamma_{n\,w\,w}}&=&{\displaystyle \frac{1}{8}} \,
{\displaystyle \frac{\pi \,w\,(N\,\mathrm{W} + 4\, \widehat{\alpha})^{3}\,n^{2}
}{(n\,w\,(N\,\mathrm{W} + 4\, \widehat{\alpha}))^{(5/2)}}},  \nonumber \\
{\Gamma_{n\,w\,N}}&=& - {\displaystyle \frac{1}{8}} \,
{\displaystyle \frac{\pi \,\mathrm{W}\,w^{2}\,n^{2}\,(N^{2}\,
\mathrm{W}^{2} + 8\,N\,\mathrm{W}\, \widehat{\alpha} + 16\, \widehat{\alpha}^{2})}{(n\,w\,(N\,
\mathrm{W} + 4\, \widehat{\alpha}))^{(5/2)}}},  \nonumber \\
{\Gamma_{n\,w\,\mathrm{W}}}&=& - {\displaystyle \frac{1}{8}} \,
{\displaystyle \frac{\pi \,N\,w^{2}\,n^{2}\,(N^{2}\,\mathrm{W}^{
2} + 8\,N\,\mathrm{W}\, \widehat{\alpha} + 16\, \widehat{\alpha}^{2})}{(n\,w\,(N\,\mathrm{W} + 4
\, \widehat{\alpha}))^{(5/2)}}},  \nonumber \\
{\Gamma_{n\,N\,n}}&=&{\displaystyle \frac{1}{8}} \,
{\displaystyle \frac{\pi \,w^{3}\,(N\,\mathrm{W} + 4\, \widehat{\alpha})^{2}\,n
\,\mathrm{W}}{(n\,w\,(N\,\mathrm{W} + 4\, \widehat{\alpha}))^{(5/2)}}},  \nonumber \\
{\Gamma_{n\,N\,w}}&=& - {\displaystyle \frac{1}{8}} \,
{\displaystyle \frac{\pi \,\mathrm{W}\,w^{2}\,n^{2}\,(N^{2}\,
\mathrm{W}^{2} + 8\,N\,\mathrm{W}\, \widehat{\alpha} + 16\, \widehat{\alpha}^{2})}{(n\,w\,(N\,
\mathrm{W} + 4\, \widehat{\alpha}))^{(5/2)}}},  
\end{eqnarray}
\begin{eqnarray}
{\Gamma_{n\,N\,N}}&=&{\displaystyle \frac{1}{8}} \,
{\displaystyle \frac{\pi \,n^{2}\,w^{3}\,\mathrm{W}^{2}\,(N\,
\mathrm{W} + 4\, \widehat{\alpha})}{(n\,w\,(N\,\mathrm{W} + 4\, \widehat{\alpha}))^{(5/2)}}},  \nonumber \\
{\Gamma_{n\,N\,\mathrm{W}}}&=& - {\displaystyle \frac{1}{8}} \,
{\displaystyle \frac{\pi \,w^{3}\,n^{2}\,(N^{2}\,\mathrm{W}^{2}
 + 12\,N\,\mathrm{W}\, \widehat{\alpha} + 32\, \widehat{\alpha}^{2})}{(n\,w\,(N\,\mathrm{W} + 4\,
\widehat{\alpha}))^{(5/2)}}},  \nonumber \\
{\Gamma_{n\,\mathrm{W}\,n}}&=&{\displaystyle \frac{1}{8}} \,
{\displaystyle \frac{\pi \,w^{3}\,(N\,\mathrm{W} + 4\, \widehat{\alpha})^{2}\,n
\,N}{(n\,w\,(N\,\mathrm{W} + 4\, \widehat{\alpha}))^{(5/2)}}},  \nonumber \\
{\Gamma_{n\,\mathrm{W}\,w}}&=& - {\displaystyle \frac{1}{8}} \,
{\displaystyle \frac{\pi \,N\,w^{2}\,n^{2}\,(N^{2}\,\mathrm{W}^{
2} + 8\,N\,\mathrm{W}\, \widehat{\alpha} + 16\, \widehat{\alpha}^{2})}{(n\,w\,(N\,\mathrm{W} + 4
\, \widehat{\alpha}))^{(5/2)}}},  \nonumber \\
{\Gamma_{n\,\mathrm{W}\,N}}&=& - {\displaystyle \frac{1}{8}} \,
{\displaystyle \frac{\pi \,w^{3}\,n^{2}\,(N^{2}\,\mathrm{W}^{2}
 + 12\,N\,\mathrm{W}\, \widehat{\alpha} + 32\, \widehat{\alpha}^{2})}{(n\,w\,(N\,\mathrm{W} + 4\,
\widehat{\alpha}))^{(5/2)}}},  
\end{eqnarray}
\begin{eqnarray}
{\Gamma_{n\,\mathrm{W}\,\mathrm{W}}}&=&{\displaystyle \frac{1}{8}
} \,{\displaystyle \frac{\pi \,n^{2}\,w^{3}\, N^{2}\,(N\,\mathrm{
W} + 4\, \widehat{\alpha})}{(n\,w\,(N\, \mathrm{W}
+ 4\, \widehat{\alpha}))^{(5/2)}}},  \nonumber \\
{\Gamma_{w\,w\,n}}&=&{\displaystyle \frac{1}{8}} \,
{\displaystyle \frac{\pi \,w\,(N\,\mathrm{W} + 4\, \widehat{\alpha})^{3}\,n^{2}
}{(n\,w\,(N\,\mathrm{W} + 4\, \widehat{\alpha}))^{(5/2)}}},  \nonumber \\
{\Gamma_{w\,w\,w}}&=& - {\displaystyle \frac{3}{8}} \,
{\displaystyle \frac{\pi \,n^{3}\,(N\,\mathrm{W} + 4\, \widehat{\alpha})^{3}}{(n
\,w\,(N\,\mathrm{W} + 4\, \widehat{\alpha}))^{(5/2)}}},  \nonumber \\
{\Gamma_{w\,w\,N}}&=&{\displaystyle \frac{1}{8}} \,
{\displaystyle \frac{\pi \,n^{3}\,(N\,\mathrm{W} + 4\, \widehat{\alpha})^{2}\,w
\,\mathrm{W}}{(n\,w\,(N\,\mathrm{W} + 4\, \widehat{\alpha}))^{(5/2)}}},  \nonumber \\
{\Gamma_{w\,w\,\mathrm{W}}}&=&{\displaystyle \frac{1}{8}} \,
{\displaystyle \frac{\pi \,n^{3}\,(N\,\mathrm{W} + 4\, \widehat{\alpha})^{2}\,w
\,N}{(n\,w\,(N\,\mathrm{W} + 4\, \widehat{\alpha}))^{(5/2)}}}, 
\end{eqnarray}
\begin{eqnarray}
{\Gamma_{w\,N\,n}}&=& - {\displaystyle \frac{1}{8}} \,
{\displaystyle \frac{\pi \,\mathrm{W}\,w^{2}\,n^{2}\,(N^{2}\,
\mathrm{W}^{2} + 8\,N\,\mathrm{W}\, \widehat{\alpha} + 16\, \widehat{\alpha}^{2})}{(n\,w\,(N\,
\mathrm{W} + 4\, \widehat{\alpha}))^{(5/2)}}},  \nonumber \\
{\Gamma_{w\,N\,w}}&=&{\displaystyle \frac{1}{8}} \,
{\displaystyle \frac{\pi \,n^{3}\,(N\,\mathrm{W} + 4\, \widehat{\alpha})^{2}\,w
\,\mathrm{W}}{(n\,w\,(N\,\mathrm{W} + 4\, \widehat{\alpha}))^{(5/2)}}},  \nonumber \\
{\Gamma_{w\,N\,N}}&=&{\displaystyle \frac{1}{8}} \,
{\displaystyle \frac{\pi \,n^{3}\,w^{2}\,\mathrm{W}^{2}\,(N\,
\mathrm{W} + 4\, \widehat{\alpha})}{(n\,w\,(N\,\mathrm{W} + 4\, \widehat{\alpha}))^{(5/2)}}},  \nonumber \\
{\Gamma_{w\,N\,\mathrm{W}}}&=& - {\displaystyle \frac{1}{8}} \,
{\displaystyle \frac{\pi \,n^{3}\,w^{2}\,(N^{2}\,\mathrm{W}^{2}
 + 12\,N\,\mathrm{W}\, \widehat{\alpha} + 32\, \widehat{\alpha}^{2})}{(n\,w\,(N\,\mathrm{W} + 4\,
\widehat{\alpha}))^{(5/2)}}},  \nonumber \\
{\Gamma_{w\,\mathrm{W}\,n}}&=& - {\displaystyle \frac{1}{8}} \,
{\displaystyle \frac{\pi \,N\,w^{2}\,n^{2}\,(N^{2}\,\mathrm{W}^{
2} + 8\,N\,\mathrm{W}\, \widehat{\alpha} + 16\, \widehat{\alpha}^{2})}{(n\,w\,(N\,\mathrm{W} + 4
\, \widehat{\alpha}))^{(5/2)}}}, 
\end{eqnarray}
\begin{eqnarray}
{\Gamma_{w\,\mathrm{W}\,w}}&=&{\displaystyle \frac{1}{8}} \,
{\displaystyle \frac{\pi \,n^{3}\,(N\,\mathrm{W} + 4\, \widehat{\alpha})^{2}\,w
\,N}{(n\,w\,(N\,\mathrm{W} + 4\, \widehat{\alpha}))^{(5/2)}}},  \nonumber \\
{\Gamma_{w\,\mathrm{W}\,N}}&=& - {\displaystyle \frac{1}{8}} \,
{\displaystyle \frac{\pi \,n^{3}\,w^{2}\,(N^{2}\,\mathrm{W}^{2}
 + 12\,N\,\mathrm{W}\, \widehat{\alpha} + 32\, \widehat{\alpha}^{2})}{(n\,w\,(N\,\mathrm{W} + 4\,
\widehat{\alpha}))^{(5/2)}}},  \nonumber \\
{\Gamma_{w\,\mathrm{W}\,\mathrm{W}}}&=&{\displaystyle \frac{1}{8}
} \,{\displaystyle \frac{\pi \,n^{3}\,w^{2}\,N^{2}\,(N\,\mathrm{
W} +
4\, \widehat{\alpha})}{(n\,w\,(N\,\mathrm{W} + 4\, \widehat{\alpha}))^{(5/2)}}},  \nonumber \\
{\Gamma_{N\,N\,n}}&=&{\displaystyle \frac{1}{8}} \,
{\displaystyle \frac{\pi \,n^{2}\,w^{3}\,\mathrm{W}^{2}\,(N\,
\mathrm{W} + 4\, \widehat{\alpha})}{(n\,w\,(N\,\mathrm{W} + 4\, \widehat{\alpha}))^{(5/2)}}},  \nonumber \\
{\Gamma_{N\,N\,w}}&=&{\displaystyle \frac{1}{8}} \,
{\displaystyle \frac{\pi \,n^{3}\,w^{2}\,\mathrm{W}^{2}\,(N\,
\mathrm{W} + 4\, \widehat{\alpha})}{(n\,w\,(N\,\mathrm{W} + 4\, \widehat{\alpha}))^{(5/2)}}},  
\end{eqnarray}
\begin{eqnarray}
{\Gamma_{N\,N\,N}}&=& - {\displaystyle \frac{3}{8}} \,
{\displaystyle \frac{\pi \,n^{3}\,w^{3}\,\mathrm{W}^{3}}{(n\,w\,
(N\,\mathrm{W} + 4\, \widehat{\alpha}))^{(5/2)}}},  \nonumber \\
{\Gamma_{N\,N\,\mathrm{W}}}&=&{\displaystyle \frac{1}{8}} \,
{\displaystyle \frac{\pi \,n^{3}\,w^{3}\,\mathrm{W}\,(N\,
\mathrm{W} + 16\, \widehat{\alpha})}{(n\,w\,(N\,\mathrm{W} + 4\, \widehat{\alpha}))^{(5/2)}}},  \nonumber \\
{\Gamma_{N\,\mathrm{W}\,n}}&=& - {\displaystyle \frac{1}{8}} \,
{\displaystyle \frac{\pi \,w^{3}\,n^{2}\,(N^{2}\,\mathrm{W}^{2}
 + 12\,N\,\mathrm{W}\, \widehat{\alpha} + 32\, \widehat{\alpha}^{2})}{(n\,w\,(N\,\mathrm{W} + 4\,
\widehat{\alpha}))^{(5/2)}}},  \nonumber \\
{\Gamma_{N\,\mathrm{W}\,w}}&=& - {\displaystyle \frac{1}{8}} \,
{\displaystyle \frac{\pi \,n^{3}\,w^{2}\,(N^{2}\,\mathrm{W}^{2}
 + 12\,N\,\mathrm{W}\, \widehat{\alpha} + 32\, \widehat{\alpha}^{2})}{(n\,w\,(N\,\mathrm{W} + 4\,
\widehat{\alpha}))^{(5/2)}}},  \nonumber \\
{\Gamma_{N\,\mathrm{W}\,N}}&=&{\displaystyle \frac{1}{8}} \,
{\displaystyle \frac{\pi \,n^{3}\,w^{3}\,\mathrm{W}\,(N\,
\mathrm{W} + 16\, \widehat{\alpha})}{(n\,w\,(N\,\mathrm{W} + 4\, \widehat{\alpha}))^{(5/2)}}}, 
\end{eqnarray}
\begin{eqnarray}
{\Gamma_{N\,\mathrm{W}\,\mathrm{W}}}&=&{\displaystyle \frac{1}{8}
} \,{\displaystyle \frac{\pi \,n^{3}\,w^{3}\,N\,(N\,\mathrm{W}
 + 16\, \widehat{\alpha})}{(n\,w\,(N\,\mathrm{W} + 4\, \widehat{\alpha}))^{(5/2)}}},  \nonumber \\
{\Gamma_{\mathrm{W}\,\mathrm{W}\,n}}&=&{\displaystyle \frac{1}{8}
} \,{\displaystyle \frac{\pi \,n^{2}\,w^{3}\,N^{2}\,(N\,\mathrm{
W} +
4\, \widehat{\alpha})}{(n\,w\,(N\,\mathrm{W} + 4\, \widehat{\alpha}))^{(5/2)}}},  \nonumber \\
{\Gamma_{\mathrm{W}\,\mathrm{W}\,w}}&=&{\displaystyle \frac{1}{8}
} \,{\displaystyle \frac{\pi \,n^{3}\,w^{2}\,N^{2}\,(N\,\mathrm{
W} +
4\, \widehat{\alpha})}{(n\,w\,(N\,\mathrm{W} + 4\, \widehat{\alpha}))^{(5/2)}}},  \nonumber \\
{\Gamma_{\mathrm{W}\,\mathrm{W}\,N}}&=&{\displaystyle \frac{1}{8}
} \,{\displaystyle \frac{\pi \,n^{3}\,w^{3}\,N\,(N\,\mathrm{W}
+ 16\, \widehat{\alpha})}{(n\,w\,(N\,\mathrm{W} + 4\, \widehat{\alpha}))^{(5/2)}}},  \nonumber \\
{\Gamma_{\mathrm{W}\,\mathrm{W}\,\mathrm{W}}}&=& - {\displaystyle
\frac{3}{8}} \,{\displaystyle \frac{\pi \,n^{3}\,w^{3}\,N^{3}}{
(n\,w\,(N\,\mathrm{W} + 4\, \widehat{\alpha}))^{(5/2)}}}.
\end{eqnarray}

\newpage
\subsection{La gomtrie de Ruppenier des trous noirs dyoniques extr\'emaux non-supersymtriques en quatre dimensions}

\subsubsection{\`A l'ordre de $(\alpha^{\prime})^0 $} 

Avec les propri\'{e}t\'{e}s de la sym\'{e}trie dans les deux premiers indices, 
nous constatons que les \'{e}l\'{e}ments de symbole de Christoffel du premier type sont

\begin{eqnarray}
{\Gamma_{n\,n\,n}}&=& - {\displaystyle \frac{3}{8}} \,
{\displaystyle \frac{\pi \,w^{3}\,N^{3}\,\mathrm{W}^{3}}{(n\,w\,
N\,\mathrm{W})^{(5/2)}}},  \nonumber \\
{\Gamma_{n\,n\,w}}&=&{\displaystyle \frac{1}{8}} \,
{\displaystyle \frac{\pi \,w^{2}\,N^{3}\,\mathrm{W}^{3}\,n}{(n\,
w\,N\,\mathrm{W})^{(5/2)}}},  \nonumber \\
{\Gamma_{n\,n\,N}}&=&{\displaystyle \frac{1}{8}} \,
{\displaystyle \frac{\pi \,w^{3}\,N^{2}\,\mathrm{W}^{3}\,n}{(n\,
w\,N\,\mathrm{W})^{(5/2)}}},  \nonumber \\
{\Gamma_{n\,n\,\mathrm{W}}}&=&{\displaystyle \frac{1}{8}} \,
{\displaystyle \frac{\pi \,w^{3}\,N^{3}\,\mathrm{W}^{2}\,n}{(n\,
w\,N\,\mathrm{W})^{(5/2)}}},  \nonumber \\
{\Gamma_{n\,w\,n}}&=&{\displaystyle \frac{1}{8}} \,
{\displaystyle \frac{\pi \,w^{2}\,N^{3}\,\mathrm{W}^{3}\,n}{(n\,
w\,N\,\mathrm{W})^{(5/2)}}}, 
\end{eqnarray}
\begin{eqnarray}
{\Gamma_{n\,w\,w}}&=&{\displaystyle \frac{1}{8}} \,
{\displaystyle \frac{\pi \,w\,N^{3}\,\mathrm{W}^{3}\,n^{2}}{(n\,
w\,N\,\mathrm{W})^{(5/2)}}},  \nonumber \\
{\Gamma_{n\,w\,N}}&=& - {\displaystyle \frac{1}{8}} \,
{\displaystyle \frac{\pi \,w^{2}\,N^{2}\,\mathrm{W}^{3}\,n^{2}}{
(n\,w\,N\,\mathrm{W})^{(5/2)}}},  \nonumber \\
{\Gamma_{n\,w\,\mathrm{W}}}&=& - {\displaystyle \frac{1}{8}} \,
{\displaystyle \frac{\pi \,w^{2}\,N^{3}\,\mathrm{W}^{2}\,n^{2}}{
(n\,w\,N\,\mathrm{W})^{(5/2)}}},  \nonumber \\
{\Gamma_{n\,N\,n}}&=&{\displaystyle \frac{1}{8}} \,
{\displaystyle \frac{\pi \,w^{3}\,N^{2}\,\mathrm{W}^{3}\,n}{(n\,
w\,N\,\mathrm{W})^{(5/2)}}},  \nonumber \\
{\Gamma_{n\,N\,w}}&=& - {\displaystyle \frac{1}{8}} \,
{\displaystyle \frac{\pi \,w^{2}\,N^{2}\,\mathrm{W}^{3}\,n^{2}}{
(n\,w\,N\,\mathrm{W})^{(5/2)}}}, 
\end{eqnarray}
\begin{eqnarray}
{\Gamma_{n\,N\,N}}&=&{\displaystyle \frac{1}{8}} \,
{\displaystyle \frac{\pi \,n^{2}\,w^{3}\,\mathrm{W}^{3}\,N}{(n\,
w\,N\,\mathrm{W})^{(5/2)}}},  \nonumber \\
{\Gamma_{n\,N\,\mathrm{W}}}&=& - {\displaystyle \frac{1}{8}} \,
{\displaystyle \frac{\pi \,n^{2}\,w^{3}\,\mathrm{W}^{2}\,N^{2}}{
(n\,w\,N\,\mathrm{W})^{(5/2)}}},  \nonumber \\
{\Gamma_{n\,\mathrm{W}\,n}}&=&{\displaystyle \frac{1}{8}} \,
{\displaystyle \frac{\pi \,w^{3}\,N^{3}\,\mathrm{W}^{2}\,n}{(n\,
w\,N\,\mathrm{W})^{(5/2)}}},  \nonumber \\
{\Gamma_{n\,\mathrm{W}\,w}}&=& - {\displaystyle \frac{1}{8}} \,
{\displaystyle \frac{\pi \,w^{2}\,N^{3}\,\mathrm{W}^{2}\,n^{2}}{
(n\,w\,N\,\mathrm{W})^{(5/2)}}},  \nonumber \\
{\Gamma_{n\,\mathrm{W}\,N}}&=& - {\displaystyle \frac{1}{8}} \,
{\displaystyle \frac{\pi \,n^{2}\,w^{3}\,\mathrm{W}^{2}\,N^{2}}{
(n\,w\,N\,\mathrm{W})^{(5/2)}}},  
\end{eqnarray}
\begin{eqnarray}
{\Gamma_{n\,\mathrm{W}\,\mathrm{W}}}&=&{\displaystyle \frac{1}{8}
} \,{\displaystyle \frac{\pi \,n^{2}\,w^{3}\,N^{3}\,\mathrm{W}}{
(n\,w\,N\,\mathrm{W})^{(5/2)}}},  \nonumber \\
{\Gamma_{w\,w\,n}}&=&{\displaystyle \frac{1}{8}} \,
{\displaystyle \frac{\pi \,w\,N^{3}\,\mathrm{W}^{3}\,n^{2}}{(n\,
w\,N\,\mathrm{W})^{(5/2)}}},  \nonumber \\
{\Gamma_{w\,w\,w}}&=& - {\displaystyle \frac{3}{8}} \,
{\displaystyle \frac{\pi \,n^{3}\,N^{3}\,\mathrm{W}^{3}}{(n\,w\,
N\,\mathrm{W})^{(5/2)}}},  \nonumber \\
{\Gamma_{w\,w\,N}}&=&{\displaystyle \frac{1}{8}} \,
{\displaystyle \frac{\pi \,n^{3}\,N^{2}\,\mathrm{W}^{3}\,w}{(n\,
w\,N\,\mathrm{W})^{(5/2)}}},  \nonumber \\
{\Gamma_{w\,w\,\mathrm{W}}}&=&{\displaystyle \frac{1}{8}} \,
{\displaystyle \frac{\pi \,n^{3}\,N^{3}\,\mathrm{W}^{2}\,w}{(n\,
w\,N\,\mathrm{W})^{(5/2)}}},  
\end{eqnarray}
\begin{eqnarray}
{\Gamma_{w\,N\,n}}&=& - {\displaystyle \frac{1}{8}} \,
{\displaystyle \frac{\pi \,w^{2}\,N^{2}\,\mathrm{W}^{3}\,n^{2}}{
(n\,w\,N\,\mathrm{W})^{(5/2)}}},  \nonumber \\
{\Gamma_{w\,N\,w}}&=&{\displaystyle \frac{1}{8}} \,
{\displaystyle \frac{\pi \,n^{3}\,N^{2}\,\mathrm{W}^{3}\,w}{(n\,
w\,N\,\mathrm{W})^{(5/2)}}},  \nonumber \\
{\Gamma_{w\,N\,N}}&=&{\displaystyle \frac{1}{8}} \,
{\displaystyle \frac{\pi \,n^{3}\,w^{2}\,\mathrm{W}^{3}\,N}{(n\,
w\,N\,\mathrm{W})^{(5/2)}}},  \nonumber \\
{\Gamma_{w\,N\,\mathrm{W}}}&=& - {\displaystyle \frac{1}{8}} \,
{\displaystyle \frac{\pi \,n^{3}\,w^{2}\,\mathrm{W}^{2}\,N^{2}}{
(n\,w\,N\,\mathrm{W})^{(5/2)}}},  \nonumber \\
{\Gamma_{w\,\mathrm{W}\,n}}&=& - {\displaystyle \frac{1}{8}} \,
{\displaystyle \frac{\pi \,w^{2}\,N^{3}\,\mathrm{W}^{2}\,n^{2}}{
(n\,w\,N\,\mathrm{W})^{(5/2)}}},  
\end{eqnarray}
\begin{eqnarray}
{\Gamma_{w\,\mathrm{W}\,w}}&=&{\displaystyle \frac{1}{8}} \,
{\displaystyle \frac{\pi \,n^{3}\,N^{3}\,\mathrm{W}^{2}\,w}{(n\,
w\,N\,\mathrm{W})^{(5/2)}}},  \nonumber \\
{\Gamma_{w\,\mathrm{W}\,N}}&=& - {\displaystyle \frac{1}{8}} \,
{\displaystyle \frac{\pi \,n^{3}\,w^{2}\,\mathrm{W}^{2}\,N^{2}}{
(n\,w\,N\,\mathrm{W})^{(5/2)}}},  \nonumber \\
{\Gamma_{w\,\mathrm{W}\,\mathrm{W}}}&=&{\displaystyle \frac{1}{8}
} \,{\displaystyle \frac{\pi \,n^{3}\,w^{2}\,N^{3}\,\mathrm{W}}{
(n\,w\,N\,\mathrm{W})^{(5/2)}}},  \nonumber \\
{\Gamma_{N\,N\,n}}&=&{\displaystyle \frac{1}{8}} \,
{\displaystyle \frac{\pi \,n^{2}\,w^{3}\,\mathrm{W}^{3}\,N}{(n\,
w\,N\,\mathrm{W})^{(5/2)}}},  \nonumber \\
{\Gamma_{N\,N\,w}}&=&{\displaystyle \frac{1}{8}} \,
{\displaystyle \frac{\pi \,n^{3}\,w^{2}\,\mathrm{W}^{3}\,N}{(n\,
w\,N\,\mathrm{W})^{(5/2)}}}, 
\end{eqnarray}
\begin{eqnarray}
{\Gamma_{N\,N\,N}}&=& - {\displaystyle \frac{3}{8}} \,
{\displaystyle \frac{\pi \,n^{3}\,w^{3}\,\mathrm{W}^{3}}{(n\,w\,
N\,\mathrm{W})^{(5/2)}}},  \nonumber \\
{\Gamma_{N\,N\,\mathrm{W}}}&=&{\displaystyle \frac{1}{8}} \,
{\displaystyle \frac{\pi \,n^{3}\,w^{3}\,\mathrm{W}^{2}\,N}{(n\,
w\,N\,\mathrm{W})^{(5/2)}}},  \nonumber \\
{\Gamma_{N\,\mathrm{W}\,n}}&=& - {\displaystyle \frac{1}{8}} \,
{\displaystyle \frac{\pi \,n^{2}\,w^{3}\,\mathrm{W}^{2}\,N^{2}}{
(n\,w\,N\,\mathrm{W})^{(5/2)}}},  \nonumber \\
{\Gamma_{N\,\mathrm{W}\,w}}&=& - {\displaystyle \frac{1}{8}} \,
{\displaystyle \frac{\pi \,n^{3}\,w^{2}\,\mathrm{W}^{2}\,N^{2}}{
(n\,w\,N\,\mathrm{W})^{(5/2)}}},  \nonumber \\
{\Gamma_{N\,\mathrm{W}\,N}}&=&{\displaystyle \frac{1}{8}} \,
{\displaystyle \frac{\pi \,n^{3}\,w^{3}\,\mathrm{W}^{2}\,N}{(n\,
w\,N\,\mathrm{W})^{(5/2)}}}, 
\end{eqnarray}
\begin{eqnarray}
{\Gamma_{N\,\mathrm{W}\,\mathrm{W}}}&=&{\displaystyle \frac{1}{8}
} \,{\displaystyle \frac{\pi \,n^{3}\,w^{3}\,\mathrm{W}\,N^{2}}{
(n\,w\,N\,\mathrm{W})^{(5/2)}}},  \nonumber \\
{\Gamma_{\mathrm{W}\,\mathrm{W}\,n}}&=&{\displaystyle \frac{1}{8}
} \,{\displaystyle \frac{\pi \,n^{2}\,w^{3}\,N^{3}\,\mathrm{W}}{
(n\,w\,N\,\mathrm{W})^{(5/2)}}},  \nonumber \\
{\Gamma_{\mathrm{W}\,\mathrm{W}\,w}}&=&{\displaystyle \frac{1}{8}
} \,{\displaystyle \frac{\pi \,n^{3}\,w^{2}\,N^{3}\,\mathrm{W}}{
(n\,w\,N\,\mathrm{W})^{(5/2)}}},  \nonumber \\
{\Gamma_{\mathrm{W}\,\mathrm{W}\,N}}&=&{\displaystyle \frac{1}{8}
} \,{\displaystyle \frac{\pi \,n^{3}\,w^{3}\,\mathrm{W}\,N^{2}}{
(n\,w\,N\,\mathrm{W})^{(5/2)}}},  \nonumber \\
{\Gamma_{\mathrm{W}\,\mathrm{W}\,\mathrm{W}}}&=& - {\displaystyle
\frac{3}{8}} \,{\displaystyle \frac{\pi \,n^{3}\,w^{3}\,N^{3}}{
(n\,w\,N\,\mathrm{W})^{(5/2)}}}.
\end{eqnarray}

\newpage
\subsubsection{\`A l'ordre de $(\alpha^{\prime})^1$}

Avec les propri\'{e}t\'{e}s de la sym\'{e}trie dans les deux premiers indices, 
nous constatons que les \'{e}l\'{e}ments de symbole de Christoffel du premier type sont

\begin{eqnarray}
{\Gamma_{\mathit{nnn}}}&=& - {\displaystyle \frac {3}{64}} \,
{\displaystyle \frac {\pi \,\sqrt{w}\,(8\,N\,\mathrm{W} + 5\, \widehat{\alpha})}{
n^{(5/2)}\,\sqrt{N}\,\sqrt{\mathrm{W}}}},  \nonumber \\ 
{\Gamma_{\mathit{nnw}}}&=&{\displaystyle \frac {1}{64}} \,
{\displaystyle \frac {\pi \,(8\,N\,\mathrm{W} + 5\, \widehat{\alpha})}{n^{(3/2)}
\,\sqrt{w}\,\sqrt{N}\,\sqrt{\mathrm{W}}}},  \nonumber \\ 
{\Gamma_{\mathit{nnN}}}&=& - {\displaystyle \frac {1}{64}} \,
{\displaystyle \frac {\pi \,\sqrt{w}\,( - 8\,N\,\mathrm{W} + 5\, \widehat{\alpha}
)}{n^{(3/2)}\,N^{(3/2)}\,\sqrt{\mathrm{W}}}},  \nonumber \\ 
{\Gamma_{\mathit{nnW}}}&=& - {\displaystyle \frac {1}{64}} \,
{\displaystyle \frac {\pi \,\sqrt{w}\,( - 8\,N\,\mathrm{W} + 5\, \widehat{\alpha}
)}{n^{(3/2)}\,\sqrt{N}\,\mathrm{W}^{(3/2)}}},  \nonumber \\ 
{\Gamma_{\mathit{nwn}}}&=&{\displaystyle \frac {1}{64}} \,
{\displaystyle \frac {\pi \,(8\,N\,\mathrm{W} + 5\, \widehat{\alpha})}{n^{(3/2)}
\,\sqrt{w}\,\sqrt{N}\,\sqrt{\mathrm{W}}}}, 
\end{eqnarray}
\begin{eqnarray}
{\Gamma_{\mathit{nww}}}&=&{\displaystyle \frac {1}{64}} \,
{\displaystyle \frac {\pi \,(8\,N\,\mathrm{W} + 5\, \widehat{\alpha})}{\sqrt{n}\,
w^{(3/2)}\,\sqrt{N}\,\sqrt{\mathrm{W}}}},  \nonumber \\ 
{\Gamma_{\mathit{nwN}}}&=&{\displaystyle \frac {1}{64}} \,
{\displaystyle \frac {\pi \,( - 8\,N\,\mathrm{W} + 5\, \widehat{\alpha})}{\sqrt{n
}\,\sqrt{w}\,N^{(3/2)}\,\sqrt{\mathrm{W}}}},  \nonumber \\ 
{\Gamma_{\mathit{nwW}}}&=&{\displaystyle \frac {1}{64}} \,
{\displaystyle \frac {\pi \,( - 8\,N\,\mathrm{W} + 5\, \widehat{\alpha})}{\sqrt{n
}\,\sqrt{w}\,\sqrt{N}\,\mathrm{W}^{(3/2)}}},  \nonumber \\ 
{\Gamma_{\mathit{nNn}}}&=& - {\displaystyle \frac {1}{64}} \,
{\displaystyle \frac {\pi \,\sqrt{w}\,( - 8\,N\,\mathrm{W} + 5\, \widehat{\alpha}
)}{n^{(3/2)}\,N^{(3/2)}\,\sqrt{\mathrm{W}}}},  \nonumber \\ 
{\Gamma_{\mathit{nNw}}}&=&{\displaystyle \frac {1}{64}} \,
{\displaystyle \frac {\pi \,( - 8\,N\,\mathrm{W} + 5\, \widehat{\alpha})}{\sqrt{n
}\,\sqrt{w}\,N^{(3/2)}\,\sqrt{\mathrm{W}}}}, 
\end{eqnarray}
\begin{eqnarray}
{\Gamma_{\mathit{nNN}}}&=& - {\displaystyle \frac {1}{64}} \,
{\displaystyle \frac {\pi \,\sqrt{w}\,( - 8\,N\,\mathrm{W} + 15\,
\widehat{\alpha})}{\sqrt{n}\,N^{(5/2)}\,\sqrt{\mathrm{W}}}},  \nonumber \\ 
{\Gamma_{\mathit{nNW}}}&=& - {\displaystyle \frac {1}{64}} \,
{\displaystyle \frac {\pi \,\sqrt{w}\,(8\,N\,\mathrm{W} + 5\, \widehat{\alpha})}{
\sqrt{n}\,N^{(3/2)}\,\mathrm{W}^{(3/2)}}},  \nonumber \\ 
{\Gamma_{\mathit{nWn}}}&=& - {\displaystyle \frac {1}{64}} \,
{\displaystyle \frac {\pi \,\sqrt{w}\,( - 8\,N\,\mathrm{W} + 5\, \widehat{\alpha}
)}{n^{(3/2)}\,\sqrt{N}\,\mathrm{W}^{(3/2)}}},  \nonumber \\ 
{\Gamma_{\mathit{nWw}}}&=&{\displaystyle \frac {1}{64}} \,
{\displaystyle \frac {\pi \,( - 8\,N\,\mathrm{W} + 5\, \widehat{\alpha})}{\sqrt{n
}\,\sqrt{w}\,\sqrt{N}\,\mathrm{W}^{(3/2)}}},  \nonumber \\ 
{\Gamma_{\mathit{nWN}}}&=& - {\displaystyle \frac {1}{64}} \,
{\displaystyle \frac {\pi \,\sqrt{w}\,(8\,N\,\mathrm{W} + 5\, \widehat{\alpha})}{
\sqrt{n}\,N^{(3/2)}\,\mathrm{W}^{(3/2)}}}, 
\end{eqnarray}
\begin{eqnarray}
{\Gamma_{\mathit{nWW}}}&=& - {\displaystyle \frac {1}{64}} \,
{\displaystyle \frac {\pi \,\sqrt{w}\,( - 8\,N\,\mathrm{W} + 15\,
\widehat{\alpha})}{\sqrt{n}\,\sqrt{N}\,\mathrm{W}^{(5/2)}}},  \nonumber \\ 
{\Gamma_{\mathit{wwn}}}&=&{\displaystyle \frac {1}{64}} \,
{\displaystyle \frac {\pi \,(8\,N\,\mathrm{W} + 5\, \widehat{\alpha})}{\sqrt{n}\,
w^{(3/2)}\,\sqrt{N}\,\sqrt{\mathrm{W}}}},  \nonumber \\ 
{\Gamma_{\mathit{www}}}&=& - {\displaystyle \frac {3}{64}} \,
{\displaystyle \frac {\pi \,\sqrt{n}\,(8\,N\,\mathrm{W} + 5\, \widehat{\alpha})}{
w^{(5/2)}\,\sqrt{N}\,\sqrt{\mathrm{W}}}},  \nonumber \\ 
{\Gamma_{\mathit{wwN}}}&=& - {\displaystyle \frac {1}{64}} \,
{\displaystyle \frac {\pi \,\sqrt{n}\,( - 8\,N\,\mathrm{W} + 5\, \widehat{\alpha}
)}{w^{(3/2)}\,N^{(3/2)}\,\sqrt{\mathrm{W}}}},  \nonumber \\ 
{\Gamma_{\mathit{wwW}}}&=& - {\displaystyle \frac {1}{64}} \,
{\displaystyle \frac {\pi \,\sqrt{n}\,( - 8\,N\,\mathrm{W} + 5\, \widehat{\alpha}
)}{w^{(3/2)}\,\sqrt{N}\,\mathrm{W}^{(3/2)}}}, 
\end{eqnarray}
\begin{eqnarray}
{\Gamma_{\mathit{wNn}}}&=&{\displaystyle \frac {1}{64}} \,
{\displaystyle \frac {\pi \,( - 8\,N\,\mathrm{W} + 5\, \widehat{\alpha})}{\sqrt{n
}\,\sqrt{w}\,N^{(3/2)}\,\sqrt{\mathrm{W}}}},  \nonumber \\ 
{\Gamma_{\mathit{wNw}}}&=& - {\displaystyle \frac {1}{64}} \,
{\displaystyle \frac {\pi \,\sqrt{n}\,( - 8\,N\,\mathrm{W} + 5\, \widehat{\alpha}
)}{w^{(3/2)}\,N^{(3/2)}\,\sqrt{\mathrm{W}}}},  \nonumber \\ 
{\Gamma_{\mathit{wNN}}}&=& - {\displaystyle \frac {1}{64}} \,
{\displaystyle \frac {\pi \,\sqrt{n}\,( - 8\,N\,\mathrm{W} + 15\,
\widehat{\alpha})}{\sqrt{w}\,N^{(5/2)}\,\sqrt{\mathrm{W}}}},  \nonumber \\ 
{\Gamma_{\mathit{wNW}}}&=& - {\displaystyle \frac {1}{64}} \,
{\displaystyle \frac {\pi \,\sqrt{n}\,(8\,N\,\mathrm{W} + 5\, \widehat{\alpha})}{
\sqrt{w}\,N^{(3/2)}\,\mathrm{W}^{(3/2)}}},  \nonumber \\ 
{\Gamma_{\mathit{wWn}}}&=&{\displaystyle \frac {1}{64}} \,
{\displaystyle \frac {\pi \,( - 8\,N\,\mathrm{W} + 5\, \widehat{\alpha})}{\sqrt{n
}\,\sqrt{w}\,\sqrt{N}\,\mathrm{W}^{(3/2)}}}, 
\end{eqnarray}
\begin{eqnarray}
{\Gamma_{\mathit{wWw}}}&=& - {\displaystyle \frac {1}{64}} \,
{\displaystyle \frac {\pi \,\sqrt{n}\,( - 8\,N\,\mathrm{W} + 5\, \widehat{\alpha}
)}{w^{(3/2)}\,\sqrt{N}\,\mathrm{W}^{(3/2)}}},  \nonumber \\ 
{\Gamma_{\mathit{wWN}}}&=& - {\displaystyle \frac {1}{64}} \,
{\displaystyle \frac {\pi \,\sqrt{n}\,(8\,N\,\mathrm{W} + 5\, \widehat{\alpha})}{
\sqrt{w}\,N^{(3/2)}\,\mathrm{W}^{(3/2)}}},  \nonumber \\ 
{\Gamma_{\mathit{wWW}}}&=& - {\displaystyle \frac {1}{64}} \,
{\displaystyle \frac {\pi \,\sqrt{n}\,( - 8\,N\,\mathrm{W} + 15\,
\widehat{\alpha})}{\sqrt{w}\,\sqrt{N}\,\mathrm{W}^{(5/2)}}},  \nonumber \\ 
{\Gamma_{\mathit{NNn}}}&=& - {\displaystyle \frac {1}{64}} \,
{\displaystyle \frac {\pi \,\sqrt{w}\,( - 8\,N\,\mathrm{W} + 15\,
\widehat{\alpha})}{\sqrt{n}\,N^{(5/2)}\,\sqrt{\mathrm{W}}}},  \nonumber \\ 
{\Gamma_{\mathit{NNw}}}&=& - {\displaystyle \frac {1}{64}} \,
{\displaystyle \frac {\pi \,\sqrt{n}\,( - 8\,N\,\mathrm{W} + 15\,
\widehat{\alpha})}{\sqrt{w}\,N^{(5/2)}\,\sqrt{\mathrm{W}}}}, 
\end{eqnarray}
\begin{eqnarray}
{\Gamma_{\mathit{NNN}}}&=&{\displaystyle \frac {3}{64}} \,
{\displaystyle \frac {\pi \,\sqrt{n}\,\sqrt{w}\,( - 8\,N\,
\mathrm{W} + 25\, \widehat{\alpha})}{N^{(7/2)}\,\sqrt{\mathrm{W}}}},  \nonumber \\ 
{\Gamma_{\mathit{NNW}}}&=&{\displaystyle \frac {1}{64}} \,
{\displaystyle \frac {\pi \,\sqrt{n}\,\sqrt{w}\,(8\,N\,\mathrm{W}
 + 15\, \widehat{\alpha})}{N^{(5/2)}\,\mathrm{W}^{(3/2)}}},  \nonumber \\ 
{\Gamma_{\mathit{NWn}}}&=& - {\displaystyle \frac {1}{64}} \,
{\displaystyle \frac {\pi \,\sqrt{w}\,(8\,N\,\mathrm{W} + 5\, \widehat{\alpha})}{
\sqrt{n}\,N^{(3/2)}\,\mathrm{W}^{(3/2)}}},  \nonumber \\ 
{\Gamma_{\mathit{NWw}}}&=& - {\displaystyle \frac {1}{64}} \,
{\displaystyle \frac {\pi \,\sqrt{n}\,(8\,N\,\mathrm{W} + 5\, \widehat{\alpha})}{
\sqrt{w}\,N^{(3/2)}\,\mathrm{W}^{(3/2)}}},  \nonumber \\ 
{\Gamma_{\mathit{NWN}}}&=&{\displaystyle \frac {1}{64}} \,
{\displaystyle \frac {\pi \,\sqrt{n}\,\sqrt{w}\,(8\,N\,\mathrm{W}
 + 15\, \widehat{\alpha})}{N^{(5/2)}\,\mathrm{W}^{(3/2)}}}, 
\end{eqnarray}
\begin{eqnarray}
{\Gamma_{\mathit{NWW}}}&=&{\displaystyle \frac {1}{64}} \,
{\displaystyle \frac {\pi \,\sqrt{n}\,\sqrt{w}\,(8\,N\,\mathrm{W}
 + 15\, \widehat{\alpha})}{N^{(3/2)}\,\mathrm{W}^{(5/2)}}},  \nonumber \\ 
{\Gamma_{\mathit{WWn}}}&=& - {\displaystyle \frac {1}{64}} \,
{\displaystyle \frac {\pi \,\sqrt{w}\,( - 8\,N\,\mathrm{W} + 15\,
\widehat{\alpha})}{\sqrt{n}\,\sqrt{N}\,\mathrm{W}^{(5/2)}}},  \nonumber \\ 
{\Gamma_{\mathit{WWw}}}&=& - {\displaystyle \frac {1}{64}} \,
{\displaystyle \frac {\pi \,\sqrt{n}\,( - 8\,N\,\mathrm{W} + 15\,
\widehat{\alpha})}{\sqrt{w}\,\sqrt{N}\,\mathrm{W}^{(5/2)}}},  \nonumber \\ 
{\Gamma_{\mathit{WWN}}}&=&{\displaystyle \frac {1}{64}} \,
{\displaystyle \frac {\pi \,\sqrt{n}\,\sqrt{w}\,(8\,N\,\mathrm{W}
 + 15\, \widehat{\alpha})}{N^{(3/2)}\,\mathrm{W}^{(5/2)}}},  \nonumber \\ 
{\Gamma_{\mathit{WWW}}}&=&{\displaystyle \frac {3}{64}} \,
{\displaystyle \frac {\pi \,\sqrt{n}\,\sqrt{w}\,( - 8\,N\,
\mathrm{W} + 25\, \widehat{\alpha})}{\sqrt{N}\,\mathrm{W}^{(7/2)}}}.
\end{eqnarray}

\newpage
\subsubsection{\`A l'ordre de $(\alpha^{\prime})^2 $}

Avec les propri\'{e}t\'{e}s de le sym\'{e}trie dans les deux premiers indices, 
nous constatons que les \'{e}l\'{e}ments de symbole de Christoffel du premier type sont

\begin{eqnarray}
{\Gamma_{\mathit{nnn}}}&=&{\displaystyle \frac {3}{1024}} \,
{\displaystyle \frac {\pi \,\sqrt{w}\,( - 128\,N^{2}\,\mathrm{W}
^{2} - 80\, \widehat{\alpha}\,N\,\mathrm{W} + 29\, \widehat{\alpha}^{2})}{n^{(5/2)}\,N^{(3/2)}\,
\mathrm{W}^{(3/2)}}},  \nonumber \\ 
{\Gamma_{\mathit{nnw}}}&=& - {\displaystyle \frac {1}{1024}} \,
{\displaystyle \frac {\pi \,( - 128\,N^{2}\,\mathrm{W}^{2} - 80\,
\widehat{\alpha}\,N\,\mathrm{W} + 29\, \widehat{\alpha}^{2})}{n^{(3/2)}\,\sqrt{w}\,N^{(3/2)}\,
\mathrm{W}^{(3/2)}}},  \nonumber \\ 
{\Gamma_{\mathit{nnN}}}&=&{\displaystyle \frac {1}{1024}} \,
{\displaystyle \frac {\pi \,\sqrt{w}\,(128\,N^{2}\,\mathrm{W}^{2}
 - 80\, \widehat{\alpha}\,N\,\mathrm{W} + 87\, \widehat{\alpha}^{2})}{n^{(3/2)}\,N^{(5/2)}\,
\mathrm{W}^{(3/2)}}},  \nonumber \\ 
{\Gamma_{\mathit{nnW}}}&=&{\displaystyle \frac {1}{1024}} \,
{\displaystyle \frac {\pi \,\sqrt{w}\,(128\,N^{2}\,\mathrm{W}^{2}
 - 80\, \widehat{\alpha}\,N\,\mathrm{W} + 87\, \widehat{\alpha}^{2})}{n^{(3/2)}\,N^{(3/2)}\,
\mathrm{W}^{(5/2)}}},  \nonumber \\ 
{\Gamma_{\mathit{nwn}}}&=& - {\displaystyle \frac {1}{1024}} \,
{\displaystyle \frac {\pi \,( - 128\,N^{2}\,\mathrm{W}^{2} - 80\,
\widehat{\alpha}\,N\,\mathrm{W} + 29\, \widehat{\alpha}^{2})}{n^{(3/2)}\,\sqrt{w}\,N^{(3/2)}\,
\mathrm{W}^{(3/2)}}}, 
\end{eqnarray}
\begin{eqnarray}
{\Gamma_{\mathit{nww}}}&=& - {\displaystyle \frac {1}{1024}} \,
{\displaystyle \frac {\pi \,( - 128\,N^{2}\,\mathrm{W}^{2} - 80\,
\widehat{\alpha}\,N\,\mathrm{W} + 29\, \widehat{\alpha}^{2})}{\sqrt{n}\,w^{(3/2)}\,N^{(3/2)}\,
\mathrm{W}^{(3/2)}}},  \nonumber \\ 
{\Gamma_{\mathit{nwN}}}&=& - {\displaystyle \frac {1}{1024}} \,
{\displaystyle \frac {\pi \,(128\,N^{2}\,\mathrm{W}^{2} - 80\, \widehat{\alpha}\,
N\,\mathrm{W} + 87\, \widehat{\alpha}^{2})}{\sqrt{n}\,\sqrt{w}\,N^{(5/2)}\,
\mathrm{W}^{(3/2)}}},  \nonumber \\ 
{\Gamma_{\mathit{nwW}}}&=& - {\displaystyle \frac {1}{1024}} \,
{\displaystyle \frac {\pi \,(128\,N^{2}\,\mathrm{W}^{2} - 80\, \widehat{\alpha}\,
N\,\mathrm{W} + 87\, \widehat{\alpha}^{2})}{\sqrt{n}\,\sqrt{w}\,N^{(3/2)}\,
\mathrm{W}^{(5/2)}}},  \nonumber \\ 
{\Gamma_{\mathit{nNn}}}&=&{\displaystyle \frac {1}{1024}} \,
{\displaystyle \frac {\pi \,\sqrt{w}\,(128\,N^{2}\,\mathrm{W}^{2}
 - 80\, \widehat{\alpha}\,N\,\mathrm{W} + 87\, \widehat{\alpha}^{2})}{n^{(3/2)}\,N^{(5/2)}\,
\mathrm{W}^{(3/2)}}},  \nonumber \\ 
{\Gamma_{\mathit{nNw}}}&=& - {\displaystyle \frac {1}{1024}} \,
{\displaystyle \frac {\pi \,(128\,N^{2}\,\mathrm{W}^{2} - 80\, \widehat{\alpha}\,
N\,\mathrm{W} + 87\, \widehat{\alpha}^{2})}{\sqrt{n}\,\sqrt{w}\,N^{(5/2)}\,
\mathrm{W}^{(3/2)}}}, 
\end{eqnarray}
\begin{eqnarray}
{\Gamma_{\mathit{nNN}}}&=&{\displaystyle \frac {1}{1024}} \,
{\displaystyle \frac {\pi \,\sqrt{w}\,(128\,N^{2}\,\mathrm{W}^{2}
 - 240\, \widehat{\alpha}\,N\,\mathrm{W} + 435\, \widehat{\alpha}^{2})}{\sqrt{n}\,\mathrm{W}^{(3/
2)}\,N^{(7/2)}}},  \nonumber \\ 
{\Gamma_{\mathit{nNW}}}&=&{\displaystyle \frac {1}{1024}} \,
{\displaystyle \frac {\pi \,\sqrt{w}\,( - 128\,N^{2}\,\mathrm{W}
^{2} - 80\, \widehat{\alpha}\,N\,\mathrm{W} + 261\, \widehat{\alpha}^{2})}{\sqrt{n}\,\mathrm{W}^{
(5/2)}\,N^{(5/2)}}},  \nonumber \\ 
{\Gamma_{\mathit{nWn}}}&=&{\displaystyle \frac {1}{1024}} \,
{\displaystyle \frac {\pi \,\sqrt{w}\,(128\,N^{2}\,\mathrm{W}^{2}
 - 80\, \widehat{\alpha}\,N\,\mathrm{W} + 87\, \widehat{\alpha}^{2})}{n^{(3/2)}\,N^{(3/2)}\,
\mathrm{W}^{(5/2)}}},  \nonumber \\ 
{\Gamma_{\mathit{nWw}}}&=& - {\displaystyle \frac {1}{1024}} \,
{\displaystyle \frac {\pi \,(128\,N^{2}\,\mathrm{W}^{2} - 80\, \widehat{\alpha}\,
N\,\mathrm{W} + 87\, \widehat{\alpha}^{2})}{\sqrt{n}\,\sqrt{w}\,N^{(3/2)}\,
\mathrm{W}^{(5/2)}}},  \nonumber \\ 
{\Gamma_{\mathit{nWN}}}&=&{\displaystyle \frac {1}{1024}} \,
{\displaystyle \frac {\pi \,\sqrt{w}\,( - 128\,N^{2}\,\mathrm{W}
^{2} - 80\, \widehat{\alpha}\,N\,\mathrm{W} + 261\, \widehat{\alpha}^{2})}{\sqrt{n}\,\mathrm{W}^{
(5/2)}\,N^{(5/2)}}}, 
\end{eqnarray}
\begin{eqnarray}
{\Gamma_{\mathit{nWW}}}&=&{\displaystyle \frac {1}{1024}} \,
{\displaystyle \frac {\pi \,\sqrt{w}\,(128\,N^{2}\,\mathrm{W}^{2}
 - 240\, \widehat{\alpha}\,N\,\mathrm{W} + 435\, \widehat{\alpha}^{2})}{\sqrt{n}\,\mathrm{W}^{(7/
2)}\,N^{(3/2)}}},  \nonumber \\ 
{\Gamma_{\mathit{wwn}}}&=& - {\displaystyle \frac {1}{1024}} \,
{\displaystyle \frac {\pi \,( - 128\,N^{2}\,\mathrm{W}^{2} - 80\,
\widehat{\alpha}\,N\,\mathrm{W} + 29\, \widehat{\alpha}^{2})}{\sqrt{n}\,w^{(3/2)}\,N^{(3/2)}\,
\mathrm{W}^{(3/2)}}},  \nonumber \\ 
{\Gamma_{w^{3}}}&=&{\displaystyle \frac {3}{1024}} \,
{\displaystyle \frac {\pi \,\sqrt{n}\,( - 128\,N^{2}\,\mathrm{W}
^{2} - 80\, \widehat{\alpha}\,N\,\mathrm{W} + 29\, \widehat{\alpha}^{2})}{w^{(5/2)}\,N^{(3/2)}\,
\mathrm{W}^{(3/2)}}},  \nonumber \\ 
{\Gamma_{\mathit{wwN}}}&=&{\displaystyle \frac {1}{1024}} \,
{\displaystyle \frac {\pi \,\sqrt{n}\,(128\,N^{2}\,\mathrm{W}^{2}
 - 80\, \widehat{\alpha}\,N\,\mathrm{W} + 87\, \widehat{\alpha}^{2})}{w^{(3/2)}\,N^{(5/2)}\,
\mathrm{W}^{(3/2)}}},  \nonumber \\ 
{\Gamma_{\mathit{wwW}}}&=&{\displaystyle \frac {1}{1024}} \,
{\displaystyle \frac {\pi \,\sqrt{n}\,(128\,N^{2}\,\mathrm{W}^{2}
 - 80\, \widehat{\alpha}\,N\,\mathrm{W} + 87\, \widehat{\alpha}^{2})}{w^{(3/2)}\,N^{(3/2)}\,
\mathrm{W}^{(5/2)}}}, 
\end{eqnarray}
\begin{eqnarray}
{\Gamma_{\mathit{wNn}}}&=& - {\displaystyle \frac {1}{1024}} \,
{\displaystyle \frac {\pi \,(128\,N^{2}\,\mathrm{W}^{2} - 80\, \widehat{\alpha}\,
N\,\mathrm{W} + 87\, \widehat{\alpha}^{2})}{\sqrt{n}\,\sqrt{w}\,N^{(5/2)}\,
\mathrm{W}^{(3/2)}}},  \nonumber \\ 
{\Gamma_{\mathit{wNw}}}&=&{\displaystyle \frac {1}{1024}} \,
{\displaystyle \frac {\pi \,\sqrt{n}\,(128\,N^{2}\,\mathrm{W}^{2}
 - 80\, \widehat{\alpha}\,N\,\mathrm{W} + 87\, \widehat{\alpha}^{2})}{w^{(3/2)}\,N^{(5/2)}\,
\mathrm{W}^{(3/2)}}},  \nonumber \\ 
{\Gamma_{\mathit{wNN}}}&=&{\displaystyle \frac {1}{1024}} \,
{\displaystyle \frac {\pi \,\sqrt{n}\,(128\,N^{2}\,\mathrm{W}^{2}
 - 240\, \widehat{\alpha}\,N\,\mathrm{W} + 435\, \widehat{\alpha}^{2})}{\sqrt{w}\,\mathrm{W}^{(3/
2)}\,N^{(7/2)}}},  \nonumber \\ 
{\Gamma_{\mathit{wNW}}}&=&{\displaystyle \frac {1}{1024}} \,
{\displaystyle \frac {\pi \,\sqrt{n}\,( - 128\,N^{2}\,\mathrm{W}
^{2} - 80\, \widehat{\alpha}\,N\,\mathrm{W} + 261\, \widehat{\alpha}^{2})}{\sqrt{w}\,\mathrm{W}^{
(5/2)}\,N^{(5/2)}}},  \nonumber \\ 
{\Gamma_{\mathit{wWn}}}&=& - {\displaystyle \frac {1}{1024}} \,
{\displaystyle \frac {\pi \,(128\,N^{2}\,\mathrm{W}^{2} - 80\, \widehat{\alpha}\,
N\,\mathrm{W} + 87\, \widehat{\alpha}^{2})}{\sqrt{n}\,\sqrt{w}\,N^{(3/2)}\,
\mathrm{W}^{(5/2)}}}, 
\end{eqnarray}
\begin{eqnarray}
{\Gamma_{\mathit{wWw}}}&=&{\displaystyle \frac {1}{1024}} \,
{\displaystyle \frac {\pi \,\sqrt{n}\,(128\,N^{2}\,\mathrm{W}^{2}
 - 80\, \widehat{\alpha}\,N\,\mathrm{W} + 87\, \widehat{\alpha}^{2})}{w^{(3/2)}\,N^{(3/2)}\,
\mathrm{W}^{(5/2)}}},  \nonumber \\ 
{\Gamma_{\mathit{wWN}}}&=&{\displaystyle \frac {1}{1024}} \,
{\displaystyle \frac {\pi \,\sqrt{n}\,( - 128\,N^{2}\,\mathrm{W}
^{2} - 80\, \widehat{\alpha}\,N\,\mathrm{W} + 261\, \widehat{\alpha}^{2})}{\sqrt{w}\,\mathrm{W}^{
(5/2)}\,N^{(5/2)}}},  \nonumber \\ 
{\Gamma_{\mathit{wWW}}}&=&{\displaystyle \frac {1}{1024}} \,
{\displaystyle \frac {\pi \,\sqrt{n}\,(128\,N^{2}\,\mathrm{W}^{2}
 - 240\, \widehat{\alpha}\,N\,\mathrm{W} + 435\, \widehat{\alpha}^{2})}{\sqrt{w}\,\mathrm{W}^{(7/
2)}\,N^{(3/2)}}},  \nonumber \\ 
{\Gamma_{\mathit{NNn}}}&=&{\displaystyle \frac {1}{1024}} \,
{\displaystyle \frac {\pi \,\sqrt{w}\,(128\,N^{2}\,\mathrm{W}^{2}
 - 240\, \widehat{\alpha}\,N\,\mathrm{W} + 435\, \widehat{\alpha}^{2})}{\sqrt{n}\,\mathrm{W}^{(3/
2)}\,N^{(7/2)}}},  \nonumber \\ 
{\Gamma_{\mathit{NNw}}}&=&{\displaystyle \frac {1}{1024}} \,
{\displaystyle \frac {\pi \,\sqrt{n}\,(128\,N^{2}\,\mathrm{W}^{2}
 - 240\, \widehat{\alpha}\,N\,\mathrm{W} + 435\, \widehat{\alpha}^{2})}{\sqrt{w}\,\mathrm{W}^{(3/
2)}\,N^{(7/2)}}}, 
\end{eqnarray}
\begin{eqnarray}
{\Gamma_{\mathit{NNN}}}&=& - {\displaystyle \frac {3}{1024}} \,
{\displaystyle \frac {\pi \,\sqrt{n}\,\sqrt{w}\,(128\,N^{2}\,
\mathrm{W}^{2} - 400\, \widehat{\alpha}\,N\,\mathrm{W} + 1015\, \widehat{\alpha}^{2})}{\mathrm{W}
^{(3/2)}\,N^{(9/2)}}},  \nonumber \\ 
{\Gamma_{\mathit{NNW}}}&=& - {\displaystyle \frac {1}{1024}} \,
{\displaystyle \frac {\pi \,\sqrt{n}\,\sqrt{w}\,( - 128\,N^{2}\,
\mathrm{W}^{2} - 240\, \widehat{\alpha}\,N\,\mathrm{W} + 1305\, \widehat{\alpha}^{2})}{\mathrm{W}
^{(5/2)}\,N^{(7/2)}}},  \nonumber \\ 
{\Gamma_{\mathit{NWn}}}&=&{\displaystyle \frac {1}{1024}} \,
{\displaystyle \frac {\pi \,\sqrt{w}\,( - 128\,N^{2}\,\mathrm{W}
^{2} - 80\, \widehat{\alpha}\,N\,\mathrm{W} + 261\, \widehat{\alpha}^{2})}{\sqrt{n}\,\mathrm{W}^{
(5/2)}\,N^{(5/2)}}},  \nonumber \\ 
{\Gamma_{\mathit{NWw}}}&=&{\displaystyle \frac {1}{1024}} \,
{\displaystyle \frac {\pi \,\sqrt{n}\,( - 128\,N^{2}\,\mathrm{W}
^{2} - 80\, \widehat{\alpha}\,N\,\mathrm{W} + 261\, \widehat{\alpha}^{2})}{\sqrt{w}\,\mathrm{W}^{
(5/2)}\,N^{(5/2)}}},  \nonumber \\ 
{\Gamma_{\mathit{NWN}}}&=& - {\displaystyle \frac {1}{1024}} \,
{\displaystyle \frac {\pi \,\sqrt{n}\,\sqrt{w}\,( - 128\,N^{2}\,
\mathrm{W}^{2} - 240\, \widehat{\alpha}\,N\,\mathrm{W} + 1305\, \widehat{\alpha}^{2})}{\mathrm{W}
^{(5/2)}\,N^{(7/2)}}}, 
\end{eqnarray}
\begin{eqnarray}
{\Gamma_{\mathit{NWW}}}&=& - {\displaystyle \frac {1}{1024}} \,
{\displaystyle \frac {\pi \,\sqrt{n}\,\sqrt{w}\,( - 128\,N^{2}\,
\mathrm{W}^{2} - 240\, \widehat{\alpha}\,N\,\mathrm{W} + 1305\, \widehat{\alpha}^{2})}{\mathrm{W}
^{(7/2)}\,N^{(5/2)}}},  \nonumber \\ 
{\Gamma_{\mathit{WWn}}}&=&{\displaystyle \frac {1}{1024}} \,
{\displaystyle \frac {\pi \,\sqrt{w}\,(128\,N^{2}\,\mathrm{W}^{2}
 - 240\, \widehat{\alpha}\,N\,\mathrm{W} + 435\, \widehat{\alpha}^{2})}{\sqrt{n}\,\mathrm{W}^{(7/
2)}\,N^{(3/2)}}},  \nonumber \\ 
{\Gamma_{\mathit{WWw}}}&=&{\displaystyle \frac {1}{1024}} \,
{\displaystyle \frac {\pi \,\sqrt{n}\,(128\,N^{2}\,\mathrm{W}^{2}
 - 240\, \widehat{\alpha}\,N\,\mathrm{W} + 435\, \widehat{\alpha}^{2})}{\sqrt{w}\,\mathrm{W}^{(7/
2)}\,N^{(3/2)}}},  \nonumber \\ 
{\Gamma_{\mathit{WWN}}}&=& - {\displaystyle \frac {1}{1024}} \,
{\displaystyle \frac {\pi \,\sqrt{n}\,\sqrt{w}\,( - 128\,N^{2}\,
\mathrm{W}^{2} - 240\, \widehat{\alpha}\,N\,\mathrm{W} + 1305\, \widehat{\alpha}^{2})}{\mathrm{W}
^{(7/2)}\,N^{(5/2)}}},  \nonumber \\ 
{\Gamma_{\mathit{WWW}}}&=& - {\displaystyle \frac {3}{1024}} \,
{\displaystyle \frac {\pi \,\sqrt{n}\,\sqrt{w}\,(128\,N^{2}\,
\mathrm{W}^{2} - 400\, \widehat{\alpha}\,N\,\mathrm{W} + 1015\, \widehat{\alpha}^{2})}{\mathrm{W}
^{(9/2)}\,N^{(3/2)}}}.
\end{eqnarray}

\newpage
\subsubsection{\`A l'ordre de $(\alpha^{\prime})^3 $}

Avec les propri\'et\'es de la sym\'etrie dans les deux premiers indices, 
\`a troisi\`eme ordre des corrections d\'eriv\'ees sup\'erieures de la th\'eorie des cordes, 
nous constatons que les \'el\'ements de symbole de Christoffel du premier type sont

\begin{eqnarray}
{\Gamma_{\mathit{nnn}}}&=&{\displaystyle \frac{3}{8192}} \,
{\displaystyle \frac{\pi \,\sqrt{w}\,( - 1024\,N^{3}\,\mathrm{W}
^{3} - 640\, \widehat{\alpha}\,N^{2}\,\mathrm{W}^{2} + 232\, \widehat{\alpha}^{2}\,N\,\mathrm{W}
 + 119\, \widehat{\alpha}^{3})}{n^{(5/2)}\,N^{(5/2)}\,\mathrm{W}^{(5/2)}}},  \nonumber \\  
{\Gamma_{\mathit{nnw}}}&=& - {\displaystyle \frac{1}{8192}} \,
{\displaystyle \frac{\pi \,( - 1024\,N^{3}\,\mathrm{W}^{3} - 640
\, \widehat{\alpha}\,N^{2}\,\mathrm{W}^{2} + 232\, \widehat{\alpha}^{2}\,N\,\mathrm{W} + 119\, 
\widehat{\alpha}^{3})}{n^{(3/2)}\,\sqrt{w}\,N^{(5/2)}\,\mathrm{W}^{(5/2)}}},  \nonumber \\  
{\Gamma_{\mathit{nnN}}}&=&{\displaystyle \frac{1}{8192}} \,
{\displaystyle \frac{\pi \,\sqrt{w}\,(1024\,N^{3}\,\mathrm{W}^{3
} - 640\, \widehat{\alpha}\,N^{2}\,\mathrm{W}^{2} + 696\, \widehat{\alpha}^{2}\,N\,\mathrm{W} + 
595\, \widehat{\alpha}^{3})}{n^{(3/2)}\,N^{(7/2)}\,\mathrm{W}^{(5/2)}}},  \nonumber \\  
{\Gamma_{\mathit{nnW}}}&=&{\displaystyle \frac{1}{8192}} \,
{\displaystyle \frac{\pi \,\sqrt{w}\,(1024\,N^{3}\,\mathrm{W}^{3
} - 640\, \widehat{\alpha}\,N^{2}\,\mathrm{W}^{2} + 696\, \widehat{\alpha}^{2}\,N\,\mathrm{W} + 
595\, \widehat{\alpha}^{3})}{n^{(3/2)}\,N^{(5/2)}\,\mathrm{W}^{(7/2)}}},  \nonumber \\  
{\Gamma_{\mathit{nwn}}}&=& - {\displaystyle \frac{1}{8192}} \,
{\displaystyle \frac{\pi \,( - 1024\,N^{3}\,\mathrm{W}^{3} - 640
\, \widehat{\alpha}\,N^{2}\,\mathrm{W}^{2} + 232\, \widehat{\alpha}^{2}\,N\,\mathrm{W} + 119\, \widehat{\alpha}^{
3})}{n^{(3/2)}\,\sqrt{w}\,N^{(5/2)}\,\mathrm{W}^{(5/2)}}}, 
\end{eqnarray}
\begin{eqnarray}
{\Gamma_{\mathit{nww}}}&=& - {\displaystyle \frac{1}{8192}} \,
{\displaystyle \frac{\pi \,( - 1024\,N^{3}\,\mathrm{W}^{3} - 640
\, \widehat{\alpha}\,N^{2}\,\mathrm{W}^{2} + 232\, \widehat{\alpha}^{2}\,N\,\mathrm{W} + 119\, \widehat{\alpha}^{
3})}{\sqrt{n}\,w^{(3/2)}\,N^{(5/2)}\,\mathrm{W}^{(5/2)}}},  \nonumber \\  
{\Gamma_{\mathit{nwN}}}&=& - {\displaystyle \frac{1}{8192}} \,
{\displaystyle \frac{\pi \,(1024\,N^{3}\,\mathrm{W}^{3} - 640\, \widehat{\alpha}
\,N^{2}\,\mathrm{W}^{2} + 696\, \widehat{\alpha}^{2}\,N\,\mathrm{W} + 595\, \widehat{\alpha}^{3})
}{\sqrt{n}\,\sqrt{w}\,N^{(7/2)}\,\mathrm{W}^{(5/2)}}},  \nonumber \\  
{\Gamma_{\mathit{nwW}}}&=& - {\displaystyle \frac{1}{8192}} \,
{\displaystyle \frac{\pi \,(1024\,N^{3}\,\mathrm{W}^{3} - 640\, \widehat{\alpha}
\,N^{2}\,\mathrm{W}^{2} + 696\, \widehat{\alpha}^{2}\,N\,\mathrm{W} + 595\, \widehat{\alpha}^{3})
}{\sqrt{n}\,\sqrt{w}\,N^{(5/2)}\,\mathrm{W}^{(7/2)}}},  \nonumber \\  
{\Gamma_{\mathit{nNn}}}&=&{\displaystyle \frac{1}{8192}} \,
{\displaystyle \frac{\pi \,\sqrt{w}\,(1024\,N^{3}\,\mathrm{W}^{3
} - 640\, \widehat{\alpha}\,N^{2}\,\mathrm{W}^{2} + 696\, \widehat{\alpha}^{2}\,N\,\mathrm{W} + 
595\, \widehat{\alpha}^{3})}{n^{(3/2)}\,N^{(7/2)}\,\mathrm{W}^{(5/2)}}},  \nonumber \\   
{\Gamma_{\mathit{nNw}}}&=& - {\displaystyle \frac{1}{8192}} \,
{\displaystyle \frac{\pi \,(1024\,N^{3}\,\mathrm{W}^{3} - 640\, \widehat{\alpha}
\,N^{2}\,\mathrm{W}^{2} + 696\, \widehat{\alpha}^{2}\,N\,\mathrm{W} + 595\, \widehat{\alpha}^{3})}
{\sqrt{n}\,\sqrt{w}\,N^{(7/2)}\,\mathrm{W}^{(5/2)}}}, 
\end{eqnarray}
\begin{eqnarray}
{\Gamma_{\mathit{nNN}}}&=&{\displaystyle \frac{1}{8192}} \,
{\displaystyle \frac{\pi \,\sqrt{w}\,(1024\,N^{3}\,\mathrm{W}^{3
} - 1920\, \widehat{\alpha}\,N^{2}\,\mathrm{W}^{2} + 3480\, \widehat{\alpha}^{2}\,N\,\mathrm{W}
 + 4165\, \widehat{\alpha}^{3})}{\sqrt{n}\,\mathrm{W}^{(5/2)}\,N^{(9/2)}}},  \nonumber \\  
{\Gamma_{\mathit{nNW}}}&=&{\displaystyle \frac{1}{8192}} \,
{\displaystyle \frac{\pi \,\sqrt{w}\,( - 1024\,N^{3}\,\mathrm{W}
^{3} - 640\, \widehat{\alpha}\,N^{2}\,\mathrm{W}^{2} + 2088\, \widehat{\alpha}^{2}\,N\,\mathrm{W}
 + 2975\, \widehat{\alpha}^{3})}{\sqrt{n}\,\mathrm{W}^{(7/2)}\,N^{(7/2)}}},  \nonumber \\   
{\Gamma_{\mathit{nWn}}}&=&{\displaystyle \frac{1}{8192}} \,
{\displaystyle \frac{\pi \,\sqrt{w}\,(1024\,N^{3}\,\mathrm{W}^{3
} - 640\, \widehat{\alpha}\,N^{2}\,\mathrm{W}^{2} + 696\, \widehat{\alpha}^{2}\,N\,\mathrm{W} + 
595\, \widehat{\alpha}^{3})}{n^{(3/2)}\,N^{(5/2)}\,\mathrm{W}^{(7/2)}}},  \nonumber \\  
{\Gamma_{\mathit{nWw}}}&=& - {\displaystyle \frac{1}{8192}} \,
{\displaystyle \frac{\pi \,(1024\,N^{3}\,\mathrm{W}^{3} - 640\, \widehat{\alpha}
\,N^{2}\,\mathrm{W}^{2} + 696\, \widehat{\alpha}^{2}\,N\,\mathrm{W} + 595\, \widehat{\alpha}^{3})
}{\sqrt{n}\,\sqrt{w}\,N^{(5/2)}\,\mathrm{W}^{(7/2)}}},  \nonumber \\  
{\Gamma_{\mathit{nWN}}}&=&{\displaystyle \frac{1}{8192}} \,
{\displaystyle \frac{\pi \,\sqrt{w}\,( - 1024\,N^{3}\,\mathrm{W}
^{3} - 640\, \widehat{\alpha}\,N^{2}\,\mathrm{W}^{2} + 2088\, \widehat{\alpha}^{2}\,N\,\mathrm{W}
 + 2975\, \widehat{\alpha}^{3})}{\sqrt{n}\,\mathrm{W}^{(7/2)}\,N^{(7/2)}}}, 
\end{eqnarray}
\begin{eqnarray}
{\Gamma_{\mathit{nWW}}}&=&{\displaystyle \frac{1}{8192}} \,
{\displaystyle \frac{\pi \,\sqrt{w}\,(1024\,N^{3}\,\mathrm{W}^{3
} - 1920\, \widehat{\alpha}\,N^{2}\,\mathrm{W}^{2} + 3480\, \widehat{\alpha}^{2}\,N\,\mathrm{W}
 + 4165\, \widehat{\alpha}^{3})}{\sqrt{n}\,\mathrm{W}^{(9/2)}\,N^{(5/2)}}},  \nonumber \\  
{\Gamma_{\mathit{wwn}}}&=& - {\displaystyle \frac{1}{8192}} \,
{\displaystyle \frac{\pi \,( - 1024\,N^{3}\,\mathrm{W}^{3} - 640
\, \widehat{\alpha}\,N^{2}\,\mathrm{W}^{2} + 232\, \widehat{\alpha}^{2}\,N\,\mathrm{W} + 119\, \widehat{\alpha}^{
3})}{\sqrt{n}\,w^{(3/2)}\,N^{(5/2)}\,\mathrm{W}^{(5/2)}}},  \nonumber \\  
{\Gamma_{\mathit{www}}}&=&{\displaystyle \frac{3}{8192}} \,
{\displaystyle \frac{\pi \,\sqrt{n}\,( - 1024\,N^{3}\,\mathrm{W}
^{3} - 640\, \widehat{\alpha}\,N^{2}\,\mathrm{W}^{2} + 232\, \widehat{\alpha}^{2}\,N\,\mathrm{W}
 + 119\, \widehat{\alpha}^{3})}{w^{(5/2)}\,N^{(5/2)}\,\mathrm{W}^{(5/2)}}},  \nonumber \\  
{\Gamma_{\mathit{wwN}}}&=&{\displaystyle \frac{1}{8192}} \,
{\displaystyle \frac{\pi \,\sqrt{n}\,(1024\,N^{3}\,\mathrm{W}^{3
} - 640\, \widehat{\alpha}\,N^{2}\,\mathrm{W}^{2} + 696\, \widehat{\alpha}^{2}\,N\,\mathrm{W} + 
595\, \widehat{\alpha}^{3})}{w^{(3/2)}\,N^{(7/2)}\,\mathrm{W}^{(5/2)}}},  \nonumber \\  
{\Gamma_{\mathit{wwW}}}&=&{\displaystyle \frac{1}{8192}} \,
{\displaystyle \frac{\pi \,\sqrt{n}\,(1024\,N^{3}\,\mathrm{W}^{3
} - 640\, \widehat{\alpha}\,N^{2}\,\mathrm{W}^{2} + 696\, \widehat{\alpha}^{2}\,N\,\mathrm{W} + 
595\, \widehat{\alpha}^{3})}{w^{(3/2)}\,N^{(5/2)}\,\mathrm{W}^{(7/2)}}}, 
\end{eqnarray}
\begin{eqnarray}
{\Gamma_{\mathit{wNn}}}&=& - {\displaystyle \frac{1}{8192}} \,
{\displaystyle \frac{\pi \,(1024\,N^{3}\,\mathrm{W}^{3} - 640\, \widehat{\alpha}
\,N^{2}\,\mathrm{W}^{2} + 696\, \widehat{\alpha}^{2}\,N\,\mathrm{W} + 595\, \widehat{\alpha}^{3})
}{\sqrt{n}\,\sqrt{w}\,N^{(7/2)}\,\mathrm{W}^{(5/2)}}},  \nonumber \\  
{\Gamma_{\mathit{wNw}}}&=&{\displaystyle \frac{1}{8192}} \,
{\displaystyle \frac{\pi \,\sqrt{n}\,(1024\,N^{3}\,\mathrm{W}^{3
} - 640\, \widehat{\alpha}\,N^{2}\,\mathrm{W}^{2} + 696\, \widehat{\alpha}^{2}\,N\,\mathrm{W} + 
595\, \widehat{\alpha}^{3})}{w^{(3/2)}\,N^{(7/2)}\,\mathrm{W}^{(5/2)}}},  \nonumber \\  
{\Gamma_{\mathit{wNN}}}&=&{\displaystyle \frac{1}{8192}} \,
{\displaystyle \frac{\pi \,\sqrt{n}\,(1024\,N^{3}\,\mathrm{W}^{3
} - 1920\, \widehat{\alpha}\,N^{2}\,\mathrm{W}^{2} + 3480\, \widehat{\alpha}^{2}\,N\,\mathrm{W}
 + 4165\, \widehat{\alpha}^{3})}{\sqrt{w}\,\mathrm{W}^{(5/2)}\,N^{(9/2)}}},  \nonumber \\  
{\Gamma_{\mathit{wNW}}}&=&{\displaystyle \frac{1}{8192}} \,
{\displaystyle \frac{\pi \,\sqrt{n}\,( - 1024\,N^{3}\,\mathrm{W}
^{3} - 640\, \widehat{\alpha}\,N^{2}\,\mathrm{W}^{2} + 2088\, \widehat{\alpha}^{2}\,N\,\mathrm{W}
 + 2975\, \widehat{\alpha}^{3})}{\sqrt{w}\,\mathrm{W}^{(7/2)}\,N^{(7/2)}}},  \nonumber \\  
{\Gamma_{\mathit{wWn}}}&=& - {\displaystyle \frac{1}{8192}} \,
{\displaystyle \frac{\pi \,(1024\,N^{3}\,\mathrm{W}^{3} - 640\, \widehat{\alpha}
\,N^{2}\,\mathrm{W}^{2} + 696\, \widehat{\alpha}^{2}\,N\,\mathrm{W} + 595\, \widehat{\alpha}^{3})
}{\sqrt{n}\,\sqrt{w}\,N^{(5/2)}\,\mathrm{W}^{(7/2)}}}, 
\end{eqnarray}
\begin{eqnarray}
{\Gamma_{\mathit{wWw}}}&=&{\displaystyle \frac{1}{8192}} \,
{\displaystyle \frac{\pi \,\sqrt{n}\,(1024\,N^{3}\,\mathrm{W}^{3
} - 640\, \widehat{\alpha}\,N^{2}\,\mathrm{W}^{2} + 696\, \widehat{\alpha}^{2}\,N\,\mathrm{W} + 
595\, \widehat{\alpha}^{3})}{w^{(3/2)}\,N^{(5/2)}\,\mathrm{W}^{(7/2)}}},  \nonumber \\  
{\Gamma_{\mathit{wWN}}}&=&{\displaystyle \frac{1}{8192}} \,
{\displaystyle \frac{\pi \,\sqrt{n}\,( - 1024\,N^{3}\,\mathrm{W}
^{3} - 640\, \widehat{\alpha}\,N^{2}\,\mathrm{W}^{2} + 2088\, \widehat{\alpha}^{2}\,N\,\mathrm{W}
 + 2975\, \widehat{\alpha}^{3})}{\sqrt{w}\,\mathrm{W}^{(7/2)}\,N^{(7/2)}}},  \nonumber \\  
{\Gamma_{\mathit{wWW}}}&=&{\displaystyle \frac{1}{8192}} \,
{\displaystyle \frac{\pi \,\sqrt{n}\,(1024\,N^{3}\,\mathrm{W}^{3
} - 1920\, \widehat{\alpha}\,N^{2}\,\mathrm{W}^{2} + 3480\, \widehat{\alpha}^{2}\,N\,\mathrm{W}
 + 4165\, \widehat{\alpha}^{3})}{\sqrt{w}\,\mathrm{W}^{(9/2)}\,N^{(5/2)}}},  \nonumber \\  
{\Gamma_{\mathit{NNn}}}&=&{\displaystyle \frac{1}{8192}} \,
{\displaystyle \frac{\pi \,\sqrt{w}\,(1024\,N^{3}\,\mathrm{W}^{3
} - 1920\, \widehat{\alpha}\,N^{2}\,\mathrm{W}^{2} + 3480\, \widehat{\alpha}^{2}\,N\,\mathrm{W}
 + 4165\, \widehat{\alpha}^{3})}{\sqrt{n}\,\mathrm{W}^{(5/2)}\,N^{(9/2)}}},  \nonumber \\  
{\Gamma_{\mathit{NNw}}}&=&{\displaystyle \frac{1}{8192}} \,
{\displaystyle \frac{\pi \,\sqrt{n}\,(1024\,N^{3}\,\mathrm{W}^{3
} - 1920\, \widehat{\alpha}\,N^{2}\,\mathrm{W}^{2} + 3480\, \widehat{\alpha}^{2}\,N\,\mathrm{W}
 + 4165\, \widehat{\alpha}^{3})}{\sqrt{w}\,\mathrm{W}^{(5/2)}\,N^{(9/2)}}}, 
\end{eqnarray}
\begin{eqnarray}
{\Gamma_{\mathit{NNN}}}&=& - {\displaystyle \frac{3}{8192}} \,
{\displaystyle \frac{\pi \,\sqrt{n}\,\sqrt{w}\,(1024\,N^{3}\,
\mathrm{W}^{3} - 3200\, \widehat{\alpha}\,N^{2}\,\mathrm{W}^{2} + 8120\, \widehat{\alpha}^{2}\,N
\,\mathrm{W} + 12495\, \widehat{\alpha}^{3})}{\mathrm{W}^{(5/2)}\,N^{(11/2)}}},  \nonumber \\  
{\Gamma_{\mathit{NNW}}}&=& - {\displaystyle \frac{1}{8192}} \,
{\displaystyle \frac{\pi \,\sqrt{n}\,\sqrt{w}\,( - 1024\,N^{3}\,
\mathrm{W}^{3} - 1920\, \widehat{\alpha}\,N^{2}\,\mathrm{W}^{2} + 10440\, \widehat{\alpha}^{2}\,N
\,\mathrm{W} + 20825\, \widehat{\alpha}^{3})}{\mathrm{W}^{(7/2)}\,N^{(9/2)}}},  \nonumber \\  
{\Gamma_{\mathit{NWn}}}&=&{\displaystyle \frac{1}{8192}} \,
{\displaystyle \frac{\pi \,\sqrt{w}\,( - 1024\,N^{3}\,\mathrm{W}
^{3} - 640\, \widehat{\alpha}\,N^{2}\,\mathrm{W}^{2} + 2088\, \widehat{\alpha}^{2}\,N\,\mathrm{W}
 + 2975\, \widehat{\alpha}^{3})}{\sqrt{n}\,\mathrm{W}^{(7/2)}\,N^{(7/2)}}},  \nonumber \\  
{\Gamma_{\mathit{NWw}}}&=&{\displaystyle \frac{1}{8192}} \,
{\displaystyle \frac{\pi \,\sqrt{n}\,( - 1024\,N^{3}\,\mathrm{W}
^{3} - 640\, \widehat{\alpha}\,N^{2}\,\mathrm{W}^{2} + 2088\, \widehat{\alpha}^{2}\,N\,\mathrm{W}
 + 2975\, \widehat{\alpha}^{3})}{\sqrt{w}\,\mathrm{W}^{(7/2)}\,N^{(7/2)}}},  \nonumber \\  
{\Gamma_{\mathit{NWN}}}&=& - {\displaystyle \frac{1}{8192}} \,
{\displaystyle \frac{\pi \,\sqrt{n}\,\sqrt{w}\,( - 1024\,N^{3}\,
\mathrm{W}^{3} - 1920\, \widehat{\alpha}\,N^{2}\,\mathrm{W}^{2} + 10440\, \widehat{\alpha}^{2}\,N
\,\mathrm{W} + 20825\, \widehat{\alpha}^{3})}{\mathrm{W}^{(7/2)}\,N^{(9/2)}}}, 
\end{eqnarray}
\begin{eqnarray}
{\Gamma_{\mathit{NWW}}}&=& - {\displaystyle \frac{1}{8192}} \,
{\displaystyle \frac{\pi \,\sqrt{n}\,\sqrt{w}\,( - 1024\,N^{3}\,
\mathrm{W}^{3} - 1920\, \widehat{\alpha}\,N^{2}\,\mathrm{W}^{2} + 10440\, \widehat{\alpha}^{2}\,N
\,\mathrm{W} + 20825\, \widehat{\alpha}^{3})}{\mathrm{W}^{(9/2)}\,N^{(7/2)}}},  \nonumber \\  
{\Gamma_{\mathit{WWn}}}&=&{\displaystyle \frac{1}{8192}} \,
{\displaystyle \frac{\pi \,\sqrt{w}\,(1024\,N^{3}\,\mathrm{W}^{3
} - 1920\, \widehat{\alpha}\,N^{2}\,\mathrm{W}^{2} + 3480\, \widehat{\alpha}^{2}\,N\,\mathrm{W}
 + 4165\, \widehat{\alpha}^{3})}{\sqrt{n}\,\mathrm{W}^{(9/2)}\,N^{(5/2)}}},  \nonumber \\  
{\Gamma_{\mathit{WWw}}}&=&{\displaystyle \frac{1}{8192}} \,
{\displaystyle \frac{\pi \,\sqrt{n}\,(1024\,N^{3}\,\mathrm{W}^{3
} - 1920\, \widehat{\alpha}\,N^{2}\,\mathrm{W}^{2} + 3480\, \widehat{\alpha}^{2}\,N\,\mathrm{W}
 + 4165\, \widehat{\alpha}^{3})}{\sqrt{w}\,\mathrm{W}^{(9/2)}\,N^{(5/2)}}},  \nonumber \\  
{\Gamma_{\mathit{WWN}}}&=& - {\displaystyle \frac{1}{8192}} \,
{\displaystyle \frac{\pi \,\sqrt{n}\,\sqrt{w}\,( - 1024\,N^{3}\,
\mathrm{W}^{3} - 1920\, \widehat{\alpha}\,N^{2}\,\mathrm{W}^{2} + 10440\, \widehat{\alpha}^{2}\,N
\,\mathrm{W} + 20825\, \widehat{\alpha}^{3})}{\mathrm{W}^{(9/2)}\,N^{(7/2)}}},  \nonumber \\  
{\Gamma_{\mathit{WWW}}}&=& - {\displaystyle \frac{3}{8192}} \,
{\displaystyle \frac{\pi \,\sqrt{n}\,\sqrt{w}\,(1024\,N^{3}\,
\mathrm{W}^{3} - 3200\, \widehat{\alpha}\,N^{2}\,\mathrm{W}^{2} + 8120\, \widehat{\alpha}^{2}\,N
\,\mathrm{W} + 12495\, \widehat{\alpha}^{3})}{\mathrm{W}^{(11/2)}\,N^{(5/2)}}}. 
\end{eqnarray}

\subsubsection{\`A l'ordre de $(\alpha^{\prime})^4 $}

Avec les propri\'{e}t\'{e}s de la sym\'{e}trie dans les deux premiers indices, 
nous constatons que les \'{e}l\'{e}ments de symbole de Christoffel du premier type sont

\begin{eqnarray}
{\Gamma_{\mathit{nnn}}}&=& {\displaystyle \frac{3}{262144}} \,
{\displaystyle \frac{\pi \,\sqrt{w}\,}{n^{(5
/2)}\,N^{(7/2)}\,\mathrm{W}^{(7/2)}}}( - 32768\,N^{4}\,\mathrm{W
}^{4} - 20480\, \widehat{\alpha}\,N^{3}\,\mathrm{W}^{3}  \nonumber \\ && + 7424\, \widehat{\alpha}^{2}\,N^{2}\,
\mathrm{W}^{2} + 3808\, \widehat{\alpha}^{3}\,N\,\mathrm{W} + 2237\, \widehat{\alpha}^{4}), \nonumber \\  
{\Gamma_{\mathit{nnw}}}&=&  - {\displaystyle \frac{1}{262144}} 
{\displaystyle \frac{\pi \,}{n^{(3/2)}\,
\sqrt{w}\,N^{(7/2)}\,\mathrm{W}^{(7/2)}}}( - 32768\,N^{4}\,\mathrm{W}^{4} - 
20480\, \widehat{\alpha}\,N^{3}\,\mathrm{W}^{3}  \nonumber \\ && + 7424\, \widehat{\alpha}^{2}\,N^{2}\,\mathrm{W}
^{2} + 3808\, \widehat{\alpha}^{3}\,N\,\mathrm{W} + 2237\, \widehat{\alpha}^{4}), \nonumber \\   
{\Gamma_{\mathit{nnN}}}&=& {\displaystyle \frac{1}{262144}} \,
{\displaystyle \frac{\pi \,\sqrt{w}}{n^{
(3/2)}\,N^{(9/2)}\,\mathrm{W}^{(7/2)}}}\,(32768\,N^{4}\,\mathrm{W}^{
4} - 20480\, \widehat{\alpha}\,N^{3}\,\mathrm{W}^{3}  \nonumber \\ && + 22272\, \widehat{\alpha}^{2}\,N^{2}\,
\mathrm{W}^{2} + 19040\, \widehat{\alpha}^{3}\,N\,\mathrm{W} + 15659\, \widehat{\alpha}^{4}), \nonumber \\  
{\Gamma_{\mathit{nnW}}}&=& {\displaystyle \frac{1}{262144}} \,
{\displaystyle \frac{\pi \,\sqrt{w}}{n^{
(3/2)}\,N^{(7/2)}\,\mathrm{W}^{(9/2)}}}\,(32768\,N^{4}\,\mathrm{W}^{
4} - 20480\, \widehat{\alpha}\,N^{3}\,\mathrm{W}^{3}  \nonumber \\ && + 22272\, \widehat{\alpha}^{2}\,N^{2}\,
\mathrm{W}^{2} + 19040\, \widehat{\alpha}^{3}\,N\,\mathrm{W} + 15659\, \widehat{\alpha}^{4}), \nonumber \\  
{\Gamma_{\mathit{nwn}}}&=&  - {\displaystyle \frac{1}{262144}} 
{\displaystyle \frac{\pi }{n^{(3/2)}\,
\sqrt{w}\,N^{(7/2)}\,\mathrm{W}^{(7/2)}}}\,( - 32768\,N^{4}\,\mathrm{W}^{4} - 
20480\, \widehat{\alpha}\,N^{3}\,\mathrm{W}^{3}  \nonumber \\ && + 7424\, \widehat{\alpha}^{2}\,N^{2}\,\mathrm{W}
^{2} + 3808\, \widehat{\alpha}^{3}\,N\,\mathrm{W} + 2237\, \widehat{\alpha}^{4}), 
\end{eqnarray}
\begin{eqnarray}
{\Gamma_{\mathit{nww}}}&=&  - {\displaystyle \frac{1}{262144}} 
{\displaystyle \frac{\pi }{\sqrt{n}\,w^{(3
/2)}\,N^{(7/2)}\,\mathrm{W}^{(7/2)}}}\,( - 32768\,N^{4}\,\mathrm{W}^{4} - 
20480\, \widehat{\alpha}\,N^{3}\,\mathrm{W}^{3}  \nonumber \\ && + 7424\, \widehat{\alpha}^{2}\,N^{2}\,\mathrm{W}
^{2} + 3808\, \widehat{\alpha}^{3}\,N\,\mathrm{W} + 2237\, \widehat{\alpha}^{4}), \nonumber \\   
{\Gamma_{\mathit{nwN}}}&=&  - {\displaystyle \frac{1}{262144}} 
{\displaystyle \frac{\pi }{\sqrt{n}\,\sqrt{w
}\,N^{(9/2)}\,\mathrm{W}^{(7/2)}}}\,(32768\,N^{4}\,\mathrm{W}^{4} - 20480
\, \widehat{\alpha}\,N^{3}\,\mathrm{W}^{3}  \nonumber \\ && + 22272\, \widehat{\alpha}^{2}\,N^{2}\,\mathrm{W}^{2}
 + 19040\, \widehat{\alpha}^{3}\,N\,\mathrm{W} + 15659\, \widehat{\alpha}^{4}), \nonumber \\  
{\Gamma_{\mathit{nwW}}}&=&  - {\displaystyle \frac{1}{262144}} 
{\displaystyle \frac{\pi }{\sqrt{n}\,\sqrt{w
}\,N^{(7/2)}\,\mathrm{W}^{(9/2)}}}\,(32768\,N^{4}\,\mathrm{W}^{4} - 20480
\, \widehat{\alpha}\,N^{3}\,\mathrm{W}^{3}  \nonumber \\ && + 22272\, \widehat{\alpha}^{2}\,N^{2}\,\mathrm{W}^{2}
 + 19040\, \widehat{\alpha}^{3}\,N\,\mathrm{W} + 15659\, \widehat{\alpha}^{4}), \nonumber \\   
{\Gamma_{\mathit{nNn}}}&=& {\displaystyle \frac{1}{262144}} \,
{\displaystyle \frac{\pi \,\sqrt{w}}{n^{
(3/2)}\,N^{(9/2)}\,\mathrm{W}^{(7/2)}}}\,(32768\,N^{4}\,\mathrm{W}^{
4} - 20480\, \widehat{\alpha}\,N^{3}\,\mathrm{W}^{3}  \nonumber \\ && + 22272\, \widehat{\alpha}^{2}\,N^{2}\,
\mathrm{W}^{2} + 19040\, \widehat{\alpha}^{3}\,N\,\mathrm{W} + 15659\, \widehat{\alpha}^{4}), \nonumber \\  
{\Gamma_{\mathit{nNw}}}&=&  - {\displaystyle \frac{1}{262144}} 
{\displaystyle \frac{\pi }{\sqrt{n}\,\sqrt{w
}\,N^{(9/2)}\,\mathrm{W}^{(7/2)}}}\,(32768\,N^{4}\,\mathrm{W}^{4} - 20480
\, \widehat{\alpha}\,N^{3}\,\mathrm{W}^{3}  \nonumber \\ && + 22272\, \widehat{\alpha}^{2}\,N^{2}\,\mathrm{W}^{2}
 + 19040\, \widehat{\alpha}^{3}\,N\,\mathrm{W} + 15659\, \widehat{\alpha}^{4}), 
\end{eqnarray}
\begin{eqnarray}
{\Gamma_{\mathit{nNN}}}&=& {\displaystyle \frac{1}{262144}} \,
{\displaystyle \frac{\pi \,\sqrt{w}}{
\sqrt{n}\,\mathrm{W}^{(7/2)}\,N^{(11/2)}}}\,(32768\,N^{4}\,\mathrm{W}^{
4} - 61440\, \widehat{\alpha}\,N^{3}\,\mathrm{W}^{3}  \nonumber \\ && + 111360\, \widehat{\alpha}^{2}\,N^{2}\,
\mathrm{W}^{2} + 133280\, \widehat{\alpha}^{3}\,N\,\mathrm{W} + 140931\, \widehat{\alpha}^{4}), \nonumber \\  
{\Gamma_{\mathit{nNW}}}&=& {\displaystyle \frac{1}{262144}} \,
{\displaystyle \frac{\pi \,\sqrt{w}}{
\sqrt{n}\,\mathrm{W}^{(9/2)}\,N^{(9/2)}}}\,( - 32768\,N^{4}\,\mathrm{W
}^{4} - 20480\, \widehat{\alpha}\,N^{3}\,\mathrm{W}^{3}  \nonumber \\ && + 66816\, \widehat{\alpha}^{2}\,N^{2}\,
\mathrm{W}^{2} + 95200\, \widehat{\alpha}^{3}\,N\,\mathrm{W} + 109613\, \widehat{\alpha}^{4}), \nonumber \\  
{\Gamma_{\mathit{nWn}}}&=& {\displaystyle \frac{1}{262144}} \,
{\displaystyle \frac{\pi \,\sqrt{w}}{n^{
(3/2)}\,N^{(7/2)}\,\mathrm{W}^{(9/2)}}}\,(32768\,N^{4}\,\mathrm{W}^{
4} - 20480\, \widehat{\alpha}\,N^{3}\,\mathrm{W}^{3}  \nonumber \\ && + 22272\, \widehat{\alpha}^{2}\,N^{2}\,
\mathrm{W}^{2} + 19040\, \widehat{\alpha}^{3}\,N\,\mathrm{W} + 15659\, \widehat{\alpha}^{4}), \nonumber \\  
{\Gamma_{\mathit{nWw}}}&=&  - {\displaystyle \frac{1}{262144}} 
{\displaystyle \frac{\pi }{\sqrt{n}\,\sqrt{w
}\,N^{(7/2)}\,\mathrm{W}^{(9/2)}}}\,(32768\,N^{4}\,\mathrm{W}^{4} - 20480
\, \widehat{\alpha}\,N^{3}\,\mathrm{W}^{3}  \nonumber \\ && + 22272\, \widehat{\alpha}^{2}\,N^{2}\,\mathrm{W}^{2}
 + 19040\, \widehat{\alpha}^{3}\,N\,\mathrm{W} + 15659\, \widehat{\alpha}^{4}), \nonumber \\   
{\Gamma_{\mathit{nWN}}}&=& {\displaystyle \frac{1}{262144}} \,
{\displaystyle \frac{\pi \,\sqrt{w}}{
\sqrt{n}\,\mathrm{W}^{(9/2)}\,N^{(9/2)}}}\,( - 32768\,N^{4}\,\mathrm{W
}^{4} - 20480\, \widehat{\alpha}\,N^{3}\,\mathrm{W}^{3}  \nonumber \\ && + 66816\, \widehat{\alpha}^{2}\,N^{2}\,
\mathrm{W}^{2} + 95200\, \widehat{\alpha}^{3}\,N\,\mathrm{W} + 109613\, \widehat{\alpha}^{4}), 
\end{eqnarray}
\begin{eqnarray}
{\Gamma_{\mathit{nWW}}}&=& {\displaystyle \frac{1}{262144}} \,
{\displaystyle \frac{\pi \,\sqrt{w}}{
\sqrt{n}\,\mathrm{W}^{(11/2)}\,N^{(7/2)}}}\,(32768\,N^{4}\,\mathrm{W}^{
4} - 61440\, \widehat{\alpha}\,N^{3}\,\mathrm{W}^{3}  \nonumber \\ && + 111360\, \widehat{\alpha}^{2}\,N^{2}\,
\mathrm{W}^{2} + 133280\, \widehat{\alpha}^{3}\,N\,\mathrm{W} + 140931\, \widehat{\alpha}^{4}), \nonumber \\  
{\Gamma_{\mathit{wwn}}}&=&  - {\displaystyle \frac{1}{262144}} 
{\displaystyle \frac{\pi }{\sqrt{n}\,w^{(3
/2)}\,N^{(7/2)}\,\mathrm{W}^{(7/2)}}}\,( - 32768\,N^{4}\,\mathrm{W}^{4} - 
20480\, \widehat{\alpha}\,N^{3}\,\mathrm{W}^{3}  \nonumber \\ && + 7424\, \widehat{\alpha}^{2}\,N^{2}\,\mathrm{W}
^{2} + 3808\, \widehat{\alpha}^{3}\,N\,\mathrm{W} + 2237\, \widehat{\alpha}^{4}), \nonumber \\   
{\Gamma_{\mathit{www}}}&=& {\displaystyle \frac{3}{262144}} \,
{\displaystyle \frac{\pi \,\sqrt{n}}{w^{(5
/2)}\,N^{(7/2)}\,\mathrm{W}^{(7/2)}}}\,( - 32768\,N^{4}\,\mathrm{W
}^{4} - 20480\, \widehat{\alpha}\,N^{3}\,\mathrm{W}^{3}  \nonumber \\ && + 7424\, \widehat{\alpha}^{2}\,N^{2}\,
\mathrm{W}^{2} + 3808\, \widehat{\alpha}^{3}\,N\,\mathrm{W} + 2237\, \widehat{\alpha}^{4}), \nonumber \\  
{\Gamma_{\mathit{wwN}}}&=& {\displaystyle \frac{1}{262144}} \,
{\displaystyle \frac{\pi \,\sqrt{n}}{w^{
(3/2)}\,N^{(9/2)}\,\mathrm{W}^{(7/2)}}}\,(32768\,N^{4}\,\mathrm{W}^{
4} - 20480\, \widehat{\alpha}\,N^{3}\,\mathrm{W}^{3}  \nonumber \\ && + 22272\, \widehat{\alpha}^{2}\,N^{2}\,
\mathrm{W}^{2} + 19040\, \widehat{\alpha}^{3}\,N\,\mathrm{W} + 15659\, \widehat{\alpha}^{4}), \nonumber \\  
{\Gamma_{\mathit{wwW}}}&=& {\displaystyle \frac{1}{262144}} \,
{\displaystyle \frac{\pi \,\sqrt{n}}{w^{
(3/2)}\,N^{(7/2)}\,\mathrm{W}^{(9/2)}}}\,(32768\,N^{4}\,\mathrm{W}^{
4} - 20480\, \widehat{\alpha}\,N^{3}\,\mathrm{W}^{3}  \nonumber \\ && + 22272\, \widehat{\alpha}^{2}\,N^{2}\,
\mathrm{W}^{2} + 19040\, \widehat{\alpha}^{3}\,N\,\mathrm{W} + 15659\, \widehat{\alpha}^{4}), 
\end{eqnarray}
\begin{eqnarray}
{\Gamma_{\mathit{wNn}}}&=&  - {\displaystyle \frac{1}{262144}} 
{\displaystyle \frac{\pi }{\sqrt{n}\,\sqrt{w
}\,N^{(9/2)}\,\mathrm{W}^{(7/2)}}}\,(32768\,N^{4}\,\mathrm{W}^{4} - 20480
\, \widehat{\alpha}\,N^{3}\,\mathrm{W}^{3}  \nonumber \\ && + 22272\, \widehat{\alpha}^{2}\,N^{2}\,\mathrm{W}^{2}
 + 19040\, \widehat{\alpha}^{3}\,N\,\mathrm{W} + 15659\, \widehat{\alpha}^{4}), \nonumber \\   
{\Gamma_{\mathit{wNw}}}&=& {\displaystyle \frac{1}{262144}} \,
{\displaystyle \frac{\pi \,\sqrt{n}}{w^{
(3/2)}\,N^{(9/2)}\,\mathrm{W}^{(7/2)}}}\,(32768\,N^{4}\,\mathrm{W}^{
4} - 20480\, \widehat{\alpha}\,N^{3}\,\mathrm{W}^{3}  \nonumber \\ && + 22272\, \widehat{\alpha}^{2}\,N^{2}\,
\mathrm{W}^{2} + 19040\, \widehat{\alpha}^{3}\,N\,\mathrm{W} + 15659\, \widehat{\alpha}^{4}), \nonumber \\  
{\Gamma_{\mathit{wNN}}}&=& {\displaystyle \frac{1}{262144}} \,
{\displaystyle \frac{\pi \,\sqrt{n}}{
\sqrt{w}\,\mathrm{W}^{(7/2)}\,N^{(11/2)}}}\,(32768\,N^{4}\,\mathrm{W}^{
4} - 61440\, \widehat{\alpha}\,N^{3}\,\mathrm{W}^{3}  \nonumber \\ && + 111360\, \widehat{\alpha}^{2}\,N^{2}\,
\mathrm{W}^{2} + 133280\, \widehat{\alpha}^{3}\,N\,\mathrm{W} + 140931\, \widehat{\alpha}^{4}), \nonumber \\  
{\Gamma_{\mathit{wNW}}}&=& {\displaystyle \frac{1}{262144}} \,
{\displaystyle \frac{\pi \,\sqrt{n}}{
\sqrt{w}\,\mathrm{W}^{(9/2)}\,N^{(9/2)}}}\,( - 32768\,N^{4}\,\mathrm{W
}^{4} - 20480\, \widehat{\alpha}\,N^{3}\,\mathrm{W}^{3}  \nonumber \\ && + 66816\, \widehat{\alpha}^{2}\,N^{2}\,
\mathrm{W}^{2} + 95200\, \widehat{\alpha}^{3}\,N\,\mathrm{W} + 109613\, \widehat{\alpha}^{4}), \nonumber \\  
{\Gamma_{\mathit{wWn}}}&=&  - {\displaystyle \frac{1}{262144}} 
{\displaystyle \frac{\pi }{\sqrt{n}\,\sqrt{w
}\,N^{(7/2)}\,\mathrm{W}^{(9/2)}}}\,(32768\,N^{4}\,\mathrm{W}^{4} - 20480
\, \widehat{\alpha}\,N^{3}\,\mathrm{W}^{3}  \nonumber \\ && + 22272\, \widehat{\alpha}^{2}\,N^{2}\,\mathrm{W}^{2}
 + 19040\, \widehat{\alpha}^{3}\,N\,\mathrm{W} + 15659\, \widehat{\alpha}^{4}), 
\end{eqnarray}
\begin{eqnarray}
{\Gamma_{\mathit{wWw}}}&=& {\displaystyle \frac{1}{262144}} \,
{\displaystyle \frac{\pi \,\sqrt{n}}{w^{
(3/2)}\,N^{(7/2)}\,\mathrm{W}^{(9/2)}}}\,(32768\,N^{4}\,\mathrm{W}^{
4} - 20480\, \widehat{\alpha}\,N^{3}\,\mathrm{W}^{3}  \nonumber \\ && + 22272\, \widehat{\alpha}^{2}\,N^{2}\,
\mathrm{W}^{2} + 19040\, \widehat{\alpha}^{3}\,N\,\mathrm{W} + 15659\, \widehat{\alpha}^{4}), \nonumber \\  
{\Gamma_{\mathit{wWN}}}&=& {\displaystyle \frac{1}{262144}} \,
{\displaystyle \frac{\pi \,\sqrt{n}}{
\sqrt{w}\,\mathrm{W}^{(9/2)}\,N^{(9/2)}}}\,( - 32768\,N^{4}\,\mathrm{W
}^{4} - 20480\, \widehat{\alpha}\,N^{3}\,\mathrm{W}^{3}  \nonumber \\ && + 66816\, \widehat{\alpha}^{2}\,N^{2}\,
\mathrm{W}^{2} + 95200\, \widehat{\alpha}^{3}\,N\,\mathrm{W} + 109613\, \widehat{\alpha}^{4}), \nonumber \\  
{\Gamma_{\mathit{wWW}}}&=& {\displaystyle \frac{1}{262144}} \,
{\displaystyle \frac{\pi \,\sqrt{n}}{
\sqrt{w}\,\mathrm{W}^{(11/2)}\,N^{(7/2)}}}\,(32768\,N^{4}\,\mathrm{W}^{
4} - 61440\, \widehat{\alpha}\,N^{3}\,\mathrm{W}^{3}  \nonumber \\ && + 111360\, \widehat{\alpha}^{2}\,N^{2}\,
\mathrm{W}^{2} + 133280\, \widehat{\alpha}^{3}\,N\,\mathrm{W} + 140931\, \widehat{\alpha}^{4}), \nonumber \\  
{\Gamma_{\mathit{NNn}}}&=& {\displaystyle \frac{1}{262144}} \,
{\displaystyle \frac{\pi \,\sqrt{w}}{
\sqrt{n}\,\mathrm{W}^{(7/2)}\,N^{(11/2)}}}\,(32768\,N^{4}\,\mathrm{W}^{
4} - 61440\, \widehat{\alpha}\,N^{3}\,\mathrm{W}^{3}  \nonumber \\ && + 111360\, \widehat{\alpha}^{2}\,N^{2}\,
\mathrm{W}^{2} + 133280\, \widehat{\alpha}^{3}\,N\,\mathrm{W} + 140931\, \widehat{\alpha}^{4}), \nonumber \\  
{\Gamma_{\mathit{NNw}}}&=& {\displaystyle \frac{1}{262144}} \,
{\displaystyle \frac{\pi \,\sqrt{n}}{
\sqrt{w}\,\mathrm{W}^{(7/2)}\,N^{(11/2)}}}\,(32768\,N^{4}\,\mathrm{W}^{
4} - 61440\, \widehat{\alpha}\,N^{3}\,\mathrm{W}^{3}  \nonumber \\ && + 111360\, \widehat{\alpha}^{2}\,N^{2}\,
\mathrm{W}^{2} + 133280\, \widehat{\alpha}^{3}\,N\,\mathrm{W} + 140931\, \widehat{\alpha}^{4}), 
\end{eqnarray}
\begin{eqnarray}
{\Gamma_{\mathit{NNN}}}&=&  - {\displaystyle \frac{3}{262144}} \frac{\pi
 \,\sqrt{n}\,\sqrt{w}}{\mathrm{W}^{(7/2)}\,N^{(13/2)}}\, (32768\,N^{4}\,\mathrm{W}^{4} 
 - 102400\, \widehat{\alpha}\,N^{3}\,\mathrm{W}^{3}  \nonumber \\ && 
+ 259840\, \widehat{\alpha}^{2}\,N^{2}\,\mathrm{W}^{2} 
\mbox{} + 399840\, \widehat{\alpha}^{3}\,N\,\mathrm{W} + 516747\, \widehat{\alpha}^{4}), \nonumber \\ 
{\Gamma_{\mathit{NNW}}}&=&  - {\displaystyle \frac{1}{262144}} 
\frac{\pi\,\sqrt{n}\,\sqrt{w}}{\mathrm{W}^{(9/2)}\,N^{(11/2)}}\,
( - 32768\,N^{4}\,\mathrm{W}^{4} - 61440\, \widehat{\alpha}
\,N^{3}\,\mathrm{W}^{3}  \nonumber \\ && + 334080\, \widehat{\alpha}^{2}\,N^{2}\,\mathrm{W}^{2}
\mbox{} + 666400\, \widehat{\alpha}^{3}\,N\,\mathrm{W} + 986517\, \widehat{\alpha}^{4}), \nonumber \\ 
{\Gamma_{\mathit{NWn}}}&=& {\displaystyle \frac{1}{262144}} \,
{\displaystyle \frac{\pi \,\sqrt{w}}{\sqrt{n}\,\mathrm{W}^{(9/2)}\,N^{(9/2)}}}
\,( - 32768\,N^{4}\,\mathrm{W}^{4} - 20480\, \widehat{\alpha}\,N^{3}\,\mathrm{W}^{3}  \nonumber \\ && 
+ 66816\, \widehat{\alpha}^{2}\,N^{2}\,
\mathrm{W}^{2} + 95200\, \widehat{\alpha}^{3}\,N\,\mathrm{W} + 109613\, \widehat{\alpha}^{4}), \nonumber \\ 
{\Gamma_{\mathit{NWw}}}&=& {\displaystyle \frac{1}{262144}} \,
{\displaystyle \frac{\pi \,\sqrt{n}}{\sqrt{w}\,\mathrm{W}^{(9/2)}\,N^{(9/2)}}}
\,( - 32768\,N^{4}\,\mathrm{W
}^{4} - 20480\, \widehat{\alpha}\,N^{3}\,\mathrm{W}^{3}  \nonumber \\ && + 66816\, \widehat{\alpha}^{2}\,N^{2}\,
\mathrm{W}^{2} + 95200\, \widehat{\alpha}^{3}\,N\,\mathrm{W} + 109613\, \widehat{\alpha}^{4}), \nonumber \\  
{\Gamma_{\mathit{NWN}}}&=&  - {\displaystyle \frac{1}{262144}} 
\frac{\pi \,\sqrt{n}\,\sqrt{w}}{\mathrm{W}^{(9/2)}\,N^{(11/2)}}
( - 32768\,N^{4}\,\mathrm{W}^{4} - 61440\, \widehat{\alpha}
\,N^{3}\,\mathrm{W}^{3}  \nonumber \\ && + 334080\, \widehat{\alpha}^{2}\,N^{2}\,\mathrm{W}^{2}
\mbox{} + 666400\, \widehat{\alpha}^{3}\,N\,\mathrm{W} + 986517\, \widehat{\alpha}^{4}), 
\end{eqnarray}
\begin{eqnarray}
{\Gamma_{\mathit{NWW}}}&=&  - {\displaystyle \frac{1}{262144}} \frac{\pi
\,\sqrt{n}\,\sqrt{w}}{\mathrm{W}^{(11/2)}\,N^{(9/2)}}
( - 32768\,N^{4}\,\mathrm{W}^{4} - 61440\, \widehat{\alpha}
\,N^{3}\,\mathrm{W}^{3}  \nonumber \\ && + 334080\, \widehat{\alpha}^{2}\,N^{2}\,\mathrm{W}^{2}
\mbox{} + 666400\, \widehat{\alpha}^{3}\,N\,\mathrm{W} + 986517\, \widehat{\alpha}^{4}), \nonumber \\ 
{\Gamma_{\mathit{WWn}}}&=& {\displaystyle \frac{1}{262144}} \,
{\displaystyle \frac{\pi \,\sqrt{w}}{\sqrt{n}\,\mathrm{W}^{(11/2)}\,N^{(7/2)}}}
\,(32768\,N^{4}\,\mathrm{W}^{
4} - 61440\, \widehat{\alpha}\,N^{3}\,\mathrm{W}^{3}  \nonumber \\ && + 111360\, \widehat{\alpha}^{2}\,N^{2}\,
\mathrm{W}^{2} + 133280\, \widehat{\alpha}^{3}\,N\,\mathrm{W} + 140931\, \widehat{\alpha}^{4}), \nonumber \\  
{\Gamma_{\mathit{WWw}}}&=& {\displaystyle \frac{1}{262144}} \,
{\displaystyle \frac{\pi \,\sqrt{n}}{\sqrt{w}\,\mathrm{W}^{(11/2)}\,N^{(7/2)}}}
\,(32768\,N^{4}\,\mathrm{W}^{
4} - 61440\, \widehat{\alpha}\,N^{3}\,\mathrm{W}^{3}  \nonumber \\ && + 111360\, \widehat{\alpha}^{2}\,N^{2}\,
\mathrm{W}^{2} + 133280\, \widehat{\alpha}^{3}\,N\,\mathrm{W} + 140931\, \widehat{\alpha}^{4}), \nonumber \\  
{\Gamma_{\mathit{WWN}}}&=&  - {\displaystyle \frac{1}{262144}} \frac{\pi
\,\sqrt{n}\,\sqrt{w}}{\mathrm{W}^{(11/2)}\,N^{(9/2)}}
( - 32768\,N^{4}\,\mathrm{W}^{4} - 61440\, \widehat{\alpha}
\,N^{3}\,\mathrm{W}^{3}  \nonumber \\ && + 334080\, \widehat{\alpha}^{2}\,N^{2}\,\mathrm{W}^{2}
\mbox{} + 666400\, \widehat{\alpha}^{3}\,N\,\mathrm{W} + 986517\, \widehat{\alpha}^{4}), \nonumber \\ 
{\Gamma_{\mathit{WWW}}}&=&  - {\displaystyle \frac{3}{262144}} \frac{\pi
\,\sqrt{n}\,\sqrt{w}}{\mathrm{W}^{(13/2)}\,N^{(7/2)}}
(32768\,N^{4}\,\mathrm{W}^{4} - 102400\, \widehat{\alpha}\,N^{3}\,
\mathrm{W}^{3}  \nonumber \\ && + 259840\, \widehat{\alpha}^{2}\,N^{2}\,\mathrm{W}^{2}
\mbox{} + 399840\, \widehat{\alpha}^{3}\,N\,\mathrm{W} + 516747\, \widehat{\alpha}^{4}).
\end{eqnarray}

\subsubsection{\`A l'ordre arbitraire de $\alpha^{\prime} $}

Avec la sym\'etrie dans les deux premiers indices,
le symbole de Christoffel du premier type sont donn\'es par

\begin{eqnarray}
{\Gamma _{\mathit{nnn}}}&=&  - {\displaystyle \frac {3}{16}} \,
{\displaystyle \frac {\pi \,w\,N\,\mathrm{W}\, \left(  \! 
{\displaystyle \sum _{k=0}^{r}} \,{c_{k}}\,({\displaystyle 
\frac {\widehat{\alpha}}{N\,\mathrm{W}}} )^{k} \!  \right) }{n^{2}\,\sqrt{n\,w\,
N\,\mathrm{W}}}}, \nonumber  \\ 
{\Gamma _{\mathit{nnw}}}&=& {\displaystyle \frac {1}{16}} \,
{\displaystyle \frac {\pi \,N\,\mathrm{W}\, \left(  \! 
{\displaystyle \sum _{k=0}^{r}} \,{c_{k}}\,({\displaystyle 
\frac {\widehat{\alpha}}{N\,\mathrm{W}}} )^{k} \!  \right) }{n\,\sqrt{n\,w\,N\,
\mathrm{W}}}}, \nonumber  \\ 
{\Gamma _{\mathit{nnN}}}&=&  - {\displaystyle \frac {1}{16}} \,
{\displaystyle \frac {\pi \,w\,\mathrm{W}\, \left(  \!  - 
 \left(  \! {\displaystyle \sum _{k=0}^{r}} \,{c_{k}}\,(
{\displaystyle \frac {\widehat{\alpha}}{N\,\mathrm{W}}} )^{k} \!  \right)  + 2\,
 \left(  \! {\displaystyle \sum _{k=0}^{r}} \,{c_{k}}\,(
{\displaystyle \frac {\widehat{\alpha}}{N\,\mathrm{W}}} )^{k}\,k \!  \right) 
 \!  \right) }{n\,\sqrt{n\,w\,N\,\mathrm{W}}}}, \nonumber  \\ 
{\Gamma _{\mathit{nnW}}}&=&  - {\displaystyle \frac {1}{16}} \,
{\displaystyle \frac {\pi \,w\,N\, \left(  \!  -  \left(  \! 
{\displaystyle \sum _{k=0}^{r}} \,{c_{k}}\,({\displaystyle 
\frac {\widehat{\alpha}}{N\,\mathrm{W}}} )^{k} \!  \right)  + 2\, \left(  \! 
{\displaystyle \sum _{k=0}^{r}} \,{c_{k}}\,({\displaystyle 
\frac {\widehat{\alpha}}{N\,\mathrm{W}}} )^{k}\,k \!  \right)  \!  \right) }{n\,
\sqrt{n\,w\,N\,\mathrm{W}}}}, \nonumber  \\ 
{\Gamma _{\mathit{nwn}}}&=& {\displaystyle \frac {1}{16}} \,
{\displaystyle \frac {\pi \,N\,\mathrm{W}\, \left(  \! 
{\displaystyle \sum _{k=0}^{r}} \,{c_{k}}\,({\displaystyle 
\frac {\widehat{\alpha}}{N\,\mathrm{W}}} )^{k} \!  \right) }{n\,\sqrt{n\,w\,N\,
\mathrm{W}}}}, 
\end{eqnarray}
\begin{eqnarray}
{\Gamma _{\mathit{nww}}}&=& {\displaystyle \frac {1}{16}} \,
{\displaystyle \frac {\pi \,N\,\mathrm{W}\, \left(  \! 
{\displaystyle \sum _{k=0}^{r}} \,{c_{k}}\,({\displaystyle 
\frac {\widehat{\alpha}}{N\,\mathrm{W}}} )^{k} \!  \right) }{w\,\sqrt{n\,w\,N\,
\mathrm{W}}}}, \nonumber  \\ 
{\Gamma _{\mathit{nwN}}}&=& {\displaystyle \frac {1}{16}} \,
{\displaystyle \frac {\pi \,\mathrm{W}\, \left(  \!  -  \left( 
 \! {\displaystyle \sum _{k=0}^{r}} \,{c_{k}}\,({\displaystyle 
\frac {\widehat{\alpha}}{N\,\mathrm{W}}} )^{k} \!  \right)  + 2\, \left(  \! 
{\displaystyle \sum _{k=0}^{r}} \,{c_{k}}\,({\displaystyle 
\frac {\widehat{\alpha}}{N\,\mathrm{W}}} )^{k}\,k \!  \right)  \!  \right) }{
\sqrt{n\,w\,N\,\mathrm{W}}}}, \nonumber  \\ 
{\Gamma _{\mathit{nwW}}}&=& {\displaystyle \frac {1}{16}} \,
{\displaystyle \frac {\pi \,N\, \left(  \!  -  \left(  \! 
{\displaystyle \sum _{k=0}^{r}} \,{c_{k}}\,({\displaystyle 
\frac {\widehat{\alpha}}{N\,\mathrm{W}}} )^{k} \!  \right)  + 2\, \left(  \! 
{\displaystyle \sum _{k=0}^{r}} \,{c_{k}}\,({\displaystyle 
\frac {\widehat{\alpha}}{N\,\mathrm{W}}} )^{k}\,k \!  \right)  \!  \right) }{
\sqrt{n\,w\,N\,\mathrm{W}}}}, \nonumber  \\ 
{\Gamma _{\mathit{nNn}}}&=&  - {\displaystyle \frac {1}{16}} \,
{\displaystyle \frac {\pi \,w\,\mathrm{W}\, \left(  \!  - 
\left(  \! {\displaystyle \sum _{k=0}^{r}} \,{c_{k}}\,(
{\displaystyle \frac {\widehat{\alpha}}{N\,\mathrm{W}}} )^{k} \!  \right)  + 2\,
\left(  \! {\displaystyle \sum _{k=0}^{r}} \,{c_{k}}\,(
{\displaystyle \frac {\widehat{\alpha}}{N\,\mathrm{W}}} )^{k}\,k \!  \right) 
\!  \right) }{n\,\sqrt{n\,w\,N\,\mathrm{W}}}}, \nonumber  \\ 
{\Gamma _{\mathit{nNw}}}&=& {\displaystyle \frac {1}{16}} \,
{\displaystyle \frac {\pi \,\mathrm{W}\, \left(  \!  -  \left( 
 \! {\displaystyle \sum _{k=0}^{r}} \,{c_{k}}\,({\displaystyle 
\frac {\widehat{\alpha}}{N\,\mathrm{W}}} )^{k} \!  \right)  + 2\, \left(  \! 
{\displaystyle \sum _{k=0}^{r}} \,{c_{k}}\,({\displaystyle 
\frac {\widehat{\alpha}}{N\,\mathrm{W}}} )^{k}\,k \!  \right)  \!  \right) }{
\sqrt{n\,w\,N\,\mathrm{W}}}}, 
\end{eqnarray}
\begin{eqnarray}
{\Gamma _{\mathit{nNN}}}&=& - {\displaystyle \frac {1}{16}} \,
{\displaystyle \frac {\pi \,w\,\mathrm{W}\, \left(  \!  - 
\left(  \! {\displaystyle \sum _{k=0}^{r}} \,{c_{k}}\,(
{\displaystyle \frac {\widehat{\alpha}}{N\,\mathrm{W}}} )^{k} \!  \right)  - 4\,
\left(  \! {\displaystyle \sum _{k=0}^{r}} \,{c_{k}}\,(
{\displaystyle \frac {\widehat{\alpha}}{N\,\mathrm{W}}} )^{k}\,k \!  \right)  + 
4\, \left(  \! {\displaystyle \sum _{k=0}^{r}} \,{c_{k}}\,(
{\displaystyle \frac {\widehat{\alpha}}{N\,\mathrm{W}}} )^{k}\,k\,(k + 1) \! 
\right)  \!  \right) }{N\,\sqrt{n\,w\,N\,\mathrm{W}}}}, \nonumber  \\ 
{\Gamma _{\mathit{nNW}}}&=& - {\displaystyle \frac {1}{16}} \,
{\displaystyle \frac {\pi \,w\, \left(  \!  \left(  \! 
{\displaystyle \sum _{k=0}^{r}} \,{c_{k}}\,({\displaystyle 
\frac {\widehat{\alpha}}{N\,\mathrm{W}}} )^{k} \!  \right)  - 4\, \left(  \! 
{\displaystyle \sum _{k=0}^{r}} \,{c_{k}}\,({\displaystyle 
\frac {\widehat{\alpha}}{N\,\mathrm{W}}} )^{k}\,k \!  \right)  + 4\, \left(  \! 
{\displaystyle \sum _{k=0}^{r}} \,{c_{k}}\,({\displaystyle 
\frac {\widehat{\alpha}}{N\,\mathrm{W}}} )^{k}\,k^{2} \!  \right)  \!  \right) 
}{\sqrt{n\,w\,N\,\mathrm{W}}}}, \nonumber  \\ 
{\Gamma _{\mathit{nWn}}}&=& - {\displaystyle \frac {1}{16}} \,
{\displaystyle \frac {\pi \,w\,N\, \left(  \!  -  \left(  \! 
{\displaystyle \sum _{k=0}^{r}} \,{c_{k}}\,({\displaystyle 
\frac {\widehat{\alpha}}{N\,\mathrm{W}}} )^{k} \!  \right)  + 2\, \left(  \! 
{\displaystyle \sum _{k=0}^{r}} \,{c_{k}}\,({\displaystyle 
\frac {\widehat{\alpha}}{N\,\mathrm{W}}} )^{k}\,k \!  \right)  \!  \right) }{n\,
\sqrt{n\,w\,N\,\mathrm{W}}}}, \nonumber  \\ 
{\Gamma _{\mathit{nWw}}}&=& {\displaystyle \frac {1}{16}} \,
{\displaystyle \frac {\pi \,N\, \left(  \!  -  \left(  \! 
{\displaystyle \sum _{k=0}^{r}} \,{c_{k}}\,({\displaystyle 
\frac {\widehat{\alpha}}{N\,\mathrm{W}}} )^{k} \!  \right)  + 2\, \left(  \! 
{\displaystyle \sum _{k=0}^{r}} \,{c_{k}}\,({\displaystyle 
\frac {\widehat{\alpha}}{N\,\mathrm{W}}} )^{k}\,k \!  \right)  \!  \right) }{
\sqrt{n\,w\,N\,\mathrm{W}}}}, \nonumber  \\ 
{\Gamma _{\mathit{nWN}}}&=& - {\displaystyle \frac {1}{16}} \,
{\displaystyle \frac {\pi \,w\, \left(  \!  \left(  \! 
{\displaystyle \sum _{k=0}^{r}} \,{c_{k}}\,({\displaystyle 
\frac {\widehat{\alpha}}{N\,\mathrm{W}}} )^{k} \!  \right)  - 4\, \left(  \! 
{\displaystyle \sum _{k=0}^{r}} \,{c_{k}}\,({\displaystyle 
\frac {\widehat{\alpha}}{N\,\mathrm{W}}} )^{k}\,k \!  \right)  + 4\, \left(  \! 
{\displaystyle \sum _{k=0}^{r}} \,{c_{k}}\,({\displaystyle 
\frac {\widehat{\alpha}}{N\,\mathrm{W}}} )^{k}\,k^{2} \!  \right)  \!  \right) 
}{\sqrt{n\,w\,N\,\mathrm{W}}}}, 
\end{eqnarray}
\begin{eqnarray}
{\Gamma _{\mathit{nWW}}}&=& - {\displaystyle \frac {1}{16}} \,
{\displaystyle \frac {\pi \,w\,N\, \left(  \!  -  \left(  \! 
{\displaystyle \sum _{k=0}^{r}} \,{c_{k}}\,({\displaystyle 
\frac {\widehat{\alpha}}{N\,\mathrm{W}}} )^{k} \!  \right)  - 4\, \left(  \! 
{\displaystyle \sum _{k=0}^{r}} \,{c_{k}}\,({\displaystyle 
\frac {\widehat{\alpha}}{N\,\mathrm{W}}} )^{k}\,k \!  \right)  + 4\, \left(  \! 
{\displaystyle \sum _{k=0}^{r}} \,{c_{k}}\,({\displaystyle 
\frac {\widehat{\alpha}}{N\,\mathrm{W}}} )^{k}\,k\,(k + 1) \!  \right)  \! 
\right) }{\mathrm{W}\,\sqrt{n\,w\,N\,\mathrm{W}}}}, \nonumber  \\ 
{\Gamma _{\mathit{wwn}}}&=& {\displaystyle \frac {1}{16}} \,
{\displaystyle \frac {\pi \,N\,\mathrm{W}\, \left(  \! 
{\displaystyle \sum _{k=0}^{r}} \,{c_{k}}\,({\displaystyle 
\frac {\widehat{\alpha}}{N\,\mathrm{W}}} )^{k} \!  \right) }{w\,\sqrt{n\,w\,N\,
\mathrm{W}}}}, \nonumber  \\ 
{\Gamma _{\mathit{www}}}&=& - {\displaystyle \frac {3}{16}} \,
{\displaystyle \frac {\pi \,n\,N\,\mathrm{W}\, \left(  \! 
{\displaystyle \sum _{k=0}^{r}} \,{c_{k}}\,({\displaystyle 
\frac {\widehat{\alpha}}{N\,\mathrm{W}}} )^{k} \!  \right) }{w^{2}\,\sqrt{n\,w\,
N\,\mathrm{W}}}}, \nonumber  \\ 
{\Gamma _{\mathit{wwN}}}&=& - {\displaystyle \frac {1}{16}} \,
{\displaystyle \frac {\pi \,n\,\mathrm{W}\, \left(  \!  - 
 \left(  \! {\displaystyle \sum _{k=0}^{r}} \,{c_{k}}\,(
{\displaystyle \frac {\widehat{\alpha}}{N\,\mathrm{W}}} )^{k} \!  \right)  + 2\,
 \left(  \! {\displaystyle \sum _{k=0}^{r}} \,{c_{k}}\,(
{\displaystyle \frac {\widehat{\alpha}}{N\,\mathrm{W}}} )^{k}\,k \!  \right) 
 \!  \right) }{w\,\sqrt{n\,w\,N\,\mathrm{W}}}}, \nonumber  \\ 
{\Gamma _{\mathit{wwW}}}&=&  - {\displaystyle \frac {1}{16}} \,
{\displaystyle \frac {\pi \,n\,N\, \left(  \!  -  \left(  \! 
{\displaystyle \sum _{k=0}^{r}} \,{c_{k}}\,({\displaystyle 
\frac {\widehat{\alpha}}{N\,\mathrm{W}}} )^{k} \!  \right)  + 2\, \left(  \! 
{\displaystyle \sum _{k=0}^{r}} \,{c_{k}}\,({\displaystyle 
\frac {\widehat{\alpha}}{N\,\mathrm{W}}} )^{k}\,k \!  \right)  \!  \right) }{w\,
\sqrt{n\,w\,N\,\mathrm{W}}}}, 
\end{eqnarray}
\begin{eqnarray}
{\Gamma _{\mathit{wNn}}}&=& {\displaystyle \frac {1}{16}} \,
{\displaystyle \frac {\pi \,\mathrm{W}\, \left(  \!  -  \left( 
 \! {\displaystyle \sum _{k=0}^{r}} \,{c_{k}}\,({\displaystyle 
\frac {\widehat{\alpha}}{N\,\mathrm{W}}} )^{k} \!  \right)  + 2\, \left(  \! 
{\displaystyle \sum _{k=0}^{r}} \,{c_{k}}\,({\displaystyle 
\frac {\widehat{\alpha}}{N\,\mathrm{W}}} )^{k}\,k \!  \right)  \!  \right) }{
\sqrt{n\,w\,N\,\mathrm{W}}}}, \nonumber  \\ 
{\Gamma _{\mathit{wNw}}}&=& - {\displaystyle \frac {1}{16}} \,
{\displaystyle \frac {\pi \,n\,\mathrm{W}\, \left(  \!  - 
 \left(  \! {\displaystyle \sum _{k=0}^{r}} \,{c_{k}}\,(
{\displaystyle \frac {\widehat{\alpha}}{N\,\mathrm{W}}} )^{k} \!  \right)  + 2\,
 \left(  \! {\displaystyle \sum _{k=0}^{r}} \,{c_{k}}\,(
{\displaystyle \frac {\widehat{\alpha}}{N\,\mathrm{W}}} )^{k}\,k \!  \right) 
 \!  \right) }{w\,\sqrt{n\,w\,N\,\mathrm{W}}}}, \nonumber  \\ 
{\Gamma _{\mathit{wNN}}}&=& - {\displaystyle \frac {1}{16}} \,
{\displaystyle \frac {\pi \,n\,\mathrm{W}\, \left(  \!  - 
 \left(  \! {\displaystyle \sum _{k=0}^{r}} \,{c_{k}}\,(
{\displaystyle \frac {\widehat{\alpha}}{N\,\mathrm{W}}} )^{k} \!  \right)  - 4\,
 \left(  \! {\displaystyle \sum _{k=0}^{r}} \,{c_{k}}\,(
{\displaystyle \frac {\widehat{\alpha}}{N\,\mathrm{W}}} )^{k}\,k \!  \right)  + 
4\, \left(  \! {\displaystyle \sum _{k=0}^{r}} \,{c_{k}}\,(
{\displaystyle \frac {\widehat{\alpha}}{N\,\mathrm{W}}} )^{k}\,k\,(k + 1) \! 
 \right)  \!  \right) }{N\,\sqrt{n\,w\,N\,\mathrm{W}}}}, \nonumber  \\ 
{\Gamma _{\mathit{wNW}}}&=& - {\displaystyle \frac {1}{16}} \,
{\displaystyle \frac {\pi \,n\, \left(  \!  \left(  \! 
{\displaystyle \sum _{k=0}^{r}} \,{c_{k}}\,({\displaystyle 
\frac {\widehat{\alpha}}{N\,\mathrm{W}}} )^{k} \!  \right)  - 4\, \left(  \! 
{\displaystyle \sum _{k=0}^{r}} \,{c_{k}}\,({\displaystyle 
\frac {\widehat{\alpha}}{N\,\mathrm{W}}} )^{k}\,k \!  \right)  + 4\, \left(  \! 
{\displaystyle \sum _{k=0}^{r}} \,{c_{k}}\,({\displaystyle 
\frac {\widehat{\alpha}}{N\,\mathrm{W}}} )^{k}\,k^{2} \!  \right)  \!  \right) 
}{\sqrt{n\,w\,N\,\mathrm{W}}}}, \nonumber  \\ 
{\Gamma _{\mathit{wWn}}}&=& {\displaystyle \frac {1}{16}} \,
{\displaystyle \frac {\pi \,N\, \left(  \!  -  \left(  \! 
{\displaystyle \sum _{k=0}^{r}} \,{c_{k}}\,({\displaystyle 
\frac {\widehat{\alpha}}{N\,\mathrm{W}}} )^{k} \!  \right)  + 2\, \left(  \! 
{\displaystyle \sum _{k=0}^{r}} \,{c_{k}}\,({\displaystyle 
\frac {\widehat{\alpha}}{N\,\mathrm{W}}} )^{k}\,k \!  \right)  \!  \right) }{
\sqrt{n\,w\,N\,\mathrm{W}}}}, 
\end{eqnarray}
\begin{eqnarray}
{\Gamma _{\mathit{wWw}}}&=& - {\displaystyle \frac {1}{16}} \,
{\displaystyle \frac {\pi \,n\,N\, \left(  \!  -  \left(  \! 
{\displaystyle \sum _{k=0}^{r}} \,{c_{k}}\,({\displaystyle 
\frac {\widehat{\alpha}}{N\,\mathrm{W}}} )^{k} \!  \right)  + 2\, \left(  \! 
{\displaystyle \sum _{k=0}^{r}} \,{c_{k}}\,({\displaystyle 
\frac {\widehat{\alpha}}{N\,\mathrm{W}}} )^{k}\,k \!  \right)  \!  \right) }{w\,
\sqrt{n\,w\,N\,\mathrm{W}}}}, \nonumber  \\ 
{\Gamma _{\mathit{wWN}}}&=& - {\displaystyle \frac {1}{16}} \,
{\displaystyle \frac {\pi \,n\, \left(  \!  \left(  \! 
{\displaystyle \sum _{k=0}^{r}} \,{c_{k}}\,({\displaystyle 
\frac {\widehat{\alpha}}{N\,\mathrm{W}}} )^{k} \!  \right)  - 4\, \left(  \! 
{\displaystyle \sum _{k=0}^{r}} \,{c_{k}}\,({\displaystyle 
\frac {\widehat{\alpha}}{N\,\mathrm{W}}} )^{k}\,k \!  \right)  + 4\, \left(  \! 
{\displaystyle \sum _{k=0}^{r}} \,{c_{k}}\,({\displaystyle 
\frac {\widehat{\alpha}}{N\,\mathrm{W}}} )^{k}\,k^{2} \!  \right)  \!  \right) 
}{\sqrt{n\,w\,N\,\mathrm{W}}}}, \nonumber  \\ 
{\Gamma _{\mathit{wWW}}}&=& - {\displaystyle \frac {1}{16}} \,
{\displaystyle \frac {\pi \,n\,N\, \left(  \!  -  \left(  \! 
{\displaystyle \sum _{k=0}^{r}} \,{c_{k}}\,({\displaystyle 
\frac {\widehat{\alpha}}{N\,\mathrm{W}}} )^{k} \!  \right)  - 4\, \left(  \! 
{\displaystyle \sum _{k=0}^{r}} \,{c_{k}}\,({\displaystyle 
\frac {\widehat{\alpha}}{N\,\mathrm{W}}} )^{k}\,k \!  \right)  + 4\, \left(  \! 
{\displaystyle \sum _{k=0}^{r}} \,{c_{k}}\,({\displaystyle 
\frac {\widehat{\alpha}}{N\,\mathrm{W}}} )^{k}\,k\,(k + 1) \!  \right)  \! 
\right) }{\mathrm{W}\,\sqrt{n\,w\,N\,\mathrm{W}}}}, \nonumber  \\ 
{\Gamma _{\mathit{NNn}}}&=& - {\displaystyle \frac {1}{16}} \,
{\displaystyle \frac {\pi \,w\,\mathrm{W}\, \left(  \!  - 
\left(  \! {\displaystyle \sum _{k=0}^{r}} \,{c_{k}}\,(
{\displaystyle \frac {\widehat{\alpha}}{N\,\mathrm{W}}} )^{k} \!  \right)  - 4\,
\left(  \! {\displaystyle \sum _{k=0}^{r}} \,{c_{k}}\,(
{\displaystyle \frac {\widehat{\alpha}}{N\,\mathrm{W}}} )^{k}\,k \!  \right)  + 
4\, \left(  \! {\displaystyle \sum _{k=0}^{r}} \,{c_{k}}\,(
{\displaystyle \frac {\widehat{\alpha}}{N\,\mathrm{W}}} )^{k}\,k\,(k + 1) \! 
\right)  \!  \right) }{N\,\sqrt{n\,w\,N\,\mathrm{W}}}}, \nonumber  \\ 
{\Gamma _{\mathit{NNw}}}&=& - {\displaystyle \frac {1}{16}} \,
{\displaystyle \frac {\pi \,n\,\mathrm{W}\, \left(  \!  - 
\left(  \! {\displaystyle \sum _{k=0}^{r}} \,{c_{k}}\,(
{\displaystyle \frac {\widehat{\alpha}}{N\,\mathrm{W}}} )^{k} \!  \right)  - 4\,
\left(  \! {\displaystyle \sum _{k=0}^{r}} \,{c_{k}}\,(
{\displaystyle \frac {\widehat{\alpha}}{N\,\mathrm{W}}} )^{k}\,k \!  \right)  + 
4\, \left(  \! {\displaystyle \sum _{k=0}^{r}} \,{c_{k}}\,(
{\displaystyle \frac {\widehat{\alpha}}{N\,\mathrm{W}}} )^{k}\,k\,(k + 1) \! 
\right)  \!  \right) }{N\,\sqrt{n\,w\,N\,\mathrm{W}}}}, 
\end{eqnarray}
\begin{eqnarray}
{\Gamma _{\mathit{NNN}}}&=& - {\displaystyle \frac {1}{16}} \pi \,n
\,w\,\mathrm{W} \left( {\vrule height1.36em width0em depth1.36em}
 \right. \!  \! 3\, \left(  \! {\displaystyle \sum _{k=0}^{r}} \,
{c_{k}}\,({\displaystyle \frac {\widehat{\alpha}}{N\,\mathrm{W}}} )^{k} \! 
 \right)  + 6\, \left(  \! {\displaystyle \sum _{k=0}^{r}} \,{c_{
k}}\,({\displaystyle \frac {\widehat{\alpha}}{N\,\mathrm{W}}} )^{k}\,k \! 
 \right)   \nonumber \\ &&
\mbox{} + 12\, \left(  \! {\displaystyle \sum _{k=0}^{r}} \,{c_{k
}}\,({\displaystyle \frac {\widehat{\alpha}}{N\,\mathrm{W}}} )^{k}\,k\,(k + 1)
 \!  \right)  - 8\, \left(  \! {\displaystyle \sum _{k=0}^{r}} \,
{c_{k}}\,({\displaystyle \frac {\widehat{\alpha}}{N\,\mathrm{W}}} )^{k}\,k\,(k^{
2} + 3\,k + 2) \!  \right)  \! \! \left. {\vrule 
height1.36em width0em depth1.36em} \right)  \left/ {\vrule 
height0.41em width0em depth0.41em} \right. \!  \! (N^{2}  \nonumber \\ &&
\sqrt{n\,w\,N\,\mathrm{W}}), \nonumber  \\
{\Gamma _{\mathit{NNW}}}&=& - {\displaystyle \frac {1}{16}} \pi \,n
\,w \left( {\vrule height1.36em width0em depth1.36em} \right. \! 
 \!  -  \left(  \! {\displaystyle \sum _{k=0}^{r}} \,{c_{k}}\,(
{\displaystyle \frac {\widehat{\alpha}}{N\,\mathrm{W}}} )^{k} \!  \right)  - 2\,
 \left(  \! {\displaystyle \sum _{k=0}^{r}} \,{c_{k}}\,(
{\displaystyle \frac {\widehat{\alpha}}{N\,\mathrm{W}}} )^{k}\,k \!  \right)  + 
8\, \left(  \! {\displaystyle \sum _{k=0}^{r}} \,{c_{k}}\,(
{\displaystyle \frac {\widehat{\alpha}}{N\,\mathrm{W}}} )^{k}\,k^{2} \! 
 \right)   \nonumber \\ &&
\mbox{} + 4\, \left(  \! {\displaystyle \sum _{k=0}^{r}} \,{c_{k}
}\,({\displaystyle \frac {\widehat{\alpha}}{N\,\mathrm{W}}} )^{k}\,k\,(k + 1)
 \!  \right)  - 8\, \left(  \! {\displaystyle \sum _{k=0}^{r}} \,
{c_{k}}\,({\displaystyle \frac {\widehat{\alpha}}{N\,\mathrm{W}}} )^{k}\,k^{2}\,
(k + 1) \!  \right)  \! \! \left. {\vrule 
height1.36em width0em depth1.36em} \right)  \left/ {\vrule 
height0.41em width0em depth0.41em} \right. \!  \! (N\,\sqrt{n\,w
\,N\,\mathrm{W}}) ,\nonumber \\
{\Gamma _{\mathit{NWn}}}&=& - {\displaystyle \frac {1}{16}} \,
{\displaystyle \frac {\pi \,w\, \left(  \!  \left(  \! 
{\displaystyle \sum _{k=0}^{r}} \,{c_{k}}\,({\displaystyle 
\frac {\widehat{\alpha}}{N\,\mathrm{W}}} )^{k} \!  \right)  - 4\, \left(  \! 
{\displaystyle \sum _{k=0}^{r}} \,{c_{k}}\,({\displaystyle 
\frac {\widehat{\alpha}}{N\,\mathrm{W}}} )^{k}\,k \!  \right)  + 4\, \left(  \! 
{\displaystyle \sum _{k=0}^{r}} \,{c_{k}}\,({\displaystyle 
\frac {\widehat{\alpha}}{N\,\mathrm{W}}} )^{k}\,k^{2} \!  \right)  \!  \right) 
}{\sqrt{n\,w\,N\,\mathrm{W}}}}, \nonumber  \\ 
{\Gamma _{\mathit{NWw}}}&=& - {\displaystyle \frac {1}{16}} \,
{\displaystyle \frac {\pi \,n\, \left(  \!  \left(  \! 
{\displaystyle \sum _{k=0}^{r}} \,{c_{k}}\,({\displaystyle 
\frac {\widehat{\alpha}}{N\,\mathrm{W}}} )^{k} \!  \right)  - 4\, \left(  \! 
{\displaystyle \sum _{k=0}^{r}} \,{c_{k}}\,({\displaystyle 
\frac {\widehat{\alpha}}{N\,\mathrm{W}}} )^{k}\,k \!  \right)  + 4\, \left(  \! 
{\displaystyle \sum _{k=0}^{r}} \,{c_{k}}\,({\displaystyle 
\frac {\widehat{\alpha}}{N\,\mathrm{W}}} )^{k}\,k^{2} \!  \right)  \!  \right) 
}{\sqrt{n\,w\,N\,\mathrm{W}}}}, \nonumber  \\ 
{\Gamma _{\mathit{NWN}}}&=& - {\displaystyle \frac {1}{16}} \pi \,n
\,w \left( {\vrule height1.36em width0em depth1.36em} \right. \! 
 \!  -  \left(  \! {\displaystyle \sum _{k=0}^{r}} \,{c_{k}}\,(
{\displaystyle \frac {\widehat{\alpha}}{N\,\mathrm{W}}} )^{k} \!  \right)  - 2\,
 \left(  \! {\displaystyle \sum _{k=0}^{r}} \,{c_{k}}\,(
{\displaystyle \frac {\widehat{\alpha}}{N\,\mathrm{W}}} )^{k}\,k \!  \right)  + 
8\, \left(  \! {\displaystyle \sum _{k=0}^{r}} \,{c_{k}}\,(
{\displaystyle \frac {\widehat{\alpha}}{N\,\mathrm{W}}} )^{k}\,k^{2} \! 
 \right)   \nonumber \\ &&
\mbox{} + 4\, \left(  \! {\displaystyle \sum _{k=0}^{r}} \,{c_{k}
}\,({\displaystyle \frac {\widehat{\alpha}}{N\,\mathrm{W}}} )^{k}\,k\,(k + 1)
 \!  \right)  - 8\, \left(  \! {\displaystyle \sum _{k=0}^{r}} \,
{c_{k}}\,({\displaystyle \frac {\widehat{\alpha}}{N\,\mathrm{W}}} )^{k}\,k^{2}\,
(k + 1) \!  \right)  \! \! \left. {\vrule 
height1.36em width0em depth1.36em} \right)  \left/ {\vrule 
height0.41em width0em depth0.41em} \right. \!  \! (N\,\sqrt{n\,w
\,N\,\mathrm{W}}), 
\end{eqnarray}
\begin{eqnarray}
{\Gamma _{\mathit{NWW}}}&=& - {\displaystyle \frac {1}{16}} \pi \,n
\,w \left( {\vrule height1.36em width0em depth1.36em} \right. \! 
 \!  -  \left(  \! {\displaystyle \sum _{k=0}^{r}} \,{c_{k}}\,(
{\displaystyle \frac {\widehat{\alpha}}{N\,\mathrm{W}}} )^{k} \!  \right)  - 2\,
 \left(  \! {\displaystyle \sum _{k=0}^{r}} \,{c_{k}}\,(
{\displaystyle \frac {\widehat{\alpha}}{N\,\mathrm{W}}} )^{k}\,k \!  \right)  + 
8\, \left(  \! {\displaystyle \sum _{k=0}^{r}} \,{c_{k}}\,(
{\displaystyle \frac {\widehat{\alpha}}{N\,\mathrm{W}}} )^{k}\,k^{2} \! 
 \right)   \nonumber \\ &&
\mbox{} + 4\, \left(  \! {\displaystyle \sum _{k=0}^{r}} \,{c_{k}
}\,({\displaystyle \frac {\widehat{\alpha}}{N\,\mathrm{W}}} )^{k}\,k\,(k + 1)
 \!  \right)  - 8\, \left(  \! {\displaystyle \sum _{k=0}^{r}} \,
{c_{k}}\,({\displaystyle \frac {\widehat{\alpha}}{N\,\mathrm{W}}} )^{k}\,k^{2}\,
(k + 1) \!  \right)  \! \! \left. {\vrule 
height1.36em width0em depth1.36em} \right)  \left/ {\vrule 
height0.41em width0em depth0.41em} \right. \!  \! (\mathrm{W}  \nonumber \\ &&
\sqrt{n\,w\,N\,\mathrm{W}}), \nonumber  \\
{\Gamma _{\mathit{WWn}}}&=& - {\displaystyle \frac {1}{16}} \,
{\displaystyle \frac {\pi \,w\,N\, \left(  \!  -  \left(  \! 
{\displaystyle \sum _{k=0}^{r}} \,{c_{k}}\,({\displaystyle 
\frac {\widehat{\alpha}}{N\,\mathrm{W}}} )^{k} \!  \right)  - 4\, \left(  \! 
{\displaystyle \sum _{k=0}^{r}} \,{c_{k}}\,({\displaystyle 
\frac {\widehat{\alpha}}{N\,\mathrm{W}}} )^{k}\,k \!  \right)  + 4\, \left(  \! 
{\displaystyle \sum _{k=0}^{r}} \,{c_{k}}\,({\displaystyle 
\frac {\widehat{\alpha}}{N\,\mathrm{W}}} )^{k}\,k\,(k + 1) \!  \right)  \! 
\right) }{\mathrm{W}\,\sqrt{n\,w\,N\,\mathrm{W}}}}, \nonumber  \\ 
{\Gamma _{\mathit{WWw}}}&=& - {\displaystyle \frac {1}{16}} \,
{\displaystyle \frac {\pi \,n\,N\, \left(  \!  -  \left(  \! 
{\displaystyle \sum _{k=0}^{r}} \,{c_{k}}\,({\displaystyle 
\frac {\widehat{\alpha}}{N\,\mathrm{W}}} )^{k} \!  \right)  - 4\, \left(  \! 
{\displaystyle \sum _{k=0}^{r}} \,{c_{k}}\,({\displaystyle 
\frac {\widehat{\alpha}}{N\,\mathrm{W}}} )^{k}\,k \!  \right)  + 4\, \left(  \! 
{\displaystyle \sum _{k=0}^{r}} \,{c_{k}}\,({\displaystyle 
\frac {\widehat{\alpha}}{N\,\mathrm{W}}} )^{k}\,k\,(k + 1) \!  \right)  \! 
\right) }{\mathrm{W}\,\sqrt{n\,w\,N\,\mathrm{W}}}}, \nonumber  \\ 
{\Gamma _{\mathit{WWN}}}&=& - {\displaystyle \frac {1}{16}} \pi \,n
\,w \left( {\vrule height1.36em width0em depth1.36em} \right. \! 
 \!  -  \left(  \! {\displaystyle \sum _{k=0}^{r}} \,{c_{k}}\,(
{\displaystyle \frac {\widehat{\alpha}}{N\,\mathrm{W}}} )^{k} \!  \right)  - 2\,
 \left(  \! {\displaystyle \sum _{k=0}^{r}} \,{c_{k}}\,(
{\displaystyle \frac {\widehat{\alpha}}{N\,\mathrm{W}}} )^{k}\,k \!  \right)  + 
8\, \left(  \! {\displaystyle \sum _{k=0}^{r}} \,{c_{k}}\,(
{\displaystyle \frac {\widehat{\alpha}}{N\,\mathrm{W}}} )^{k}\,k^{2} \! 
 \right)   \nonumber \\ &&
\mbox{} + 4\, \left(  \! {\displaystyle \sum _{k=0}^{r}} \,{c_{k}
}\,({\displaystyle \frac {\widehat{\alpha}}{N\,\mathrm{W}}} )^{k}\,k\,(k + 1)
 \!  \right)  - 8\, \left(  \! {\displaystyle \sum _{k=0}^{r}} \,
{c_{k}}\,({\displaystyle \frac {\widehat{\alpha}}{N\,\mathrm{W}}} )^{k}\,k^{2}\,
(k + 1) \!  \right)  \! \! \left. {\vrule 
height1.36em width0em depth1.36em} \right)  \left/ {\vrule 
height0.41em width0em depth0.41em} \right. \!  \! (\mathrm{W}  \nonumber \\ &&
\sqrt{n\,w\,N\,\mathrm{W}}), \nonumber  \\
{\Gamma _{\mathit{WWW}}}&=& - {\displaystyle \frac {1}{16}} \pi \,n
\,w\,N \left( {\vrule height1.36em width0em depth1.36em}
 \right. \!  \! 3\, \left(  \! {\displaystyle \sum _{k=0}^{r}} \,
{c_{k}}\,({\displaystyle \frac {\widehat{\alpha}}{N\,\mathrm{W}}} )^{k} \! 
 \right)  + 6\, \left(  \! {\displaystyle \sum _{k=0}^{r}} \,{c_{
k}}\,({\displaystyle \frac {\widehat{\alpha}}{N\,\mathrm{W}}} )^{k}\,k \! 
 \right)   \nonumber \\ &&
\mbox{} + 12\, \left(  \! {\displaystyle \sum _{k=0}^{r}} \,{c_{k
}}\,({\displaystyle \frac {\widehat{\alpha}}{N\,\mathrm{W}}} )^{k}\,k\,(k + 1)
 \!  \right)  - 8\, \left(  \! {\displaystyle \sum _{k=0}^{r}} \,
{c_{k}}\,({\displaystyle \frac {\widehat{\alpha}}{N\,\mathrm{W}}} )^{k}\,k\,(k^{
2} + 3\,k + 2) \!  \right)  \! \! \left. {\vrule 
height1.36em width0em depth1.36em} \right)  \left/ {\vrule 
height0.41em width0em depth0.41em} \right. \!  \! (\mathrm{W}^{2}
 \nonumber \\ && \sqrt{n\,w\,N\,\mathrm{W}}).
\end{eqnarray}

\newpage
\section{La gomtrie de Ruppenier des solutions non-extrmales de branes $ D_1D_5 $ et $ D_2D_6NS_5 $ en dimensions $ D = 10 $:}

\subsection{La gomtrie de Ruppenier des solutions non-extrmales de branes $ D_1D_5 $}

\subsubsection{\`A l'ordre de $(\alpha^{\prime})^0$}

Avec la sym\'etrie dans les deux premiers indices, les \'el\'ements de symbole
de Christoffel du premier type peuvent \^{e}tre r\'eduite comme les suivants

\begin{eqnarray}
{\Gamma_{\mathit{N1}\,\mathit{N_1}\,\mathit{N_1}}}&=& -
{\displaystyle \frac{3}{4}} \,{\displaystyle \frac{\pi \,
\mathit{N_5}^{3}\,\mathit{N_R}^{3}}{(\mathit{N_1}\,\mathit{N_5}\,
\mathit{N_R})^{(5/2)}}},  \nonumber \\
{\Gamma_{\mathit{N_1}\,\mathit{N_1}\,\mathit{N_5}}}&=&{\displaystyle
\frac{1}{4}} \,{\displaystyle \frac{\pi \,\mathit{N_5}^{2}\,
\mathit{N_R}^{3}\,\mathit{N_1}}{(\mathit{N_1}\,\mathit{N_5}\,\mathit{
N_R})^{(5/2)}}},  \nonumber \\
{\Gamma_{\mathit{N_1}\,\mathit{N_1}\,\mathit{N_R}}}&=&{\displaystyle
\frac{1}{4}} \,{\displaystyle \frac{\pi \,\mathit{N_5}^{3}\,
\mathit{N_R}^{2}\,\mathit{N_1}}{(\mathit{N_1}\,\mathit{N_5}\,\mathit{
N_R})^{(5/2)}}},  \nonumber \\
{\Gamma_{\mathit{N_1}\,\mathit{N_5}\,\mathit{N_1}}}&=&{\displaystyle
\frac{1}{4}} \,{\displaystyle \frac{\pi \,\mathit{N_5}^{2}\,
\mathit{N_R}^{3}\,\mathit{N_1}}{(\mathit{N_1}\,\mathit{N_5}\,\mathit{
N_R})^{(5/2)}}},  \nonumber \\
{\Gamma_{\mathit{N_1}\,\mathit{N_5}\,\mathit{N_5}}}&=&{\displaystyle
\frac{1}{4}} \,{\displaystyle \frac{\pi \,\mathit{N_5}\,\mathit{
N_R}^{3}\,\mathit{N_1}^{2}}{(\mathit{N_1}\,\mathit{N_5}\,\mathit{N_R})
^{(5/2)}}},  \nonumber \\
{\Gamma_{\mathit{N_1}\,\mathit{N_5}\,\mathit{N_R}}}&=& -
{\displaystyle \frac{1}{4}} \,{\displaystyle \frac{\pi \,
\mathit{N_5}^{2}\,\mathit{N_R}^{2}\,\mathit{N_1}^{2}}{(\mathit{N_1}\,
\mathit{N_5}\,\mathit{N_R})^{(5/2)}}},  
\end{eqnarray}
\begin{eqnarray}
{\Gamma_{\mathit{N_1}\,\mathit{N_R}\,\mathit{N_1}}}&=&{\displaystyle
\frac{1}{4}} \,{\displaystyle \frac{\pi \,\mathit{N_5}^{3}\,
\mathit{N_R}^{2}\,\mathit{N_1}}{(\mathit{N_1}\,\mathit{N_5}\,\mathit{
N_R})^{(5/2)}}},  \nonumber \\
{\Gamma_{\mathit{N_1}\,\mathit{N_R}\,\mathit{N_5}}}&=& -
{\displaystyle \frac{1}{4}} \,{\displaystyle \frac{\pi \,
\mathit{N_5}^{2}\,\mathit{N_R}^{2}\,\mathit{N_1}^{2}}{(\mathit{N_1}\,
\mathit{N_5}\,\mathit{N_R})^{(5/2)}}},  \nonumber \\
{\Gamma_{\mathit{N_1}\,\mathit{N_R}\,\mathit{N_R}}}&=&{\displaystyle
\frac{1}{4}} \,{\displaystyle \frac{\pi \,\mathit{N_5}^{3}\,
\mathit{N_R}\,\mathit{N_1}^{2}}{(\mathit{N_1}\,\mathit{N_5}\,\mathit{
N_R})^{(5/2)}}},  \nonumber \\
{\Gamma_{\mathit{N_5}\,\mathit{N_5}\,\mathit{N_1}}}&=&{\displaystyle
\frac{1}{4}} \,{\displaystyle \frac{\pi \,\mathit{N_5}\,\mathit{
N_R}^{3}\,\mathit{N_1}^{2}}{(\mathit{N_1}\,\mathit{N_5}\,\mathit{N_R})
^{(5/2)}}},  \nonumber \\
{\Gamma_{\mathit{N_5}\,\mathit{N_5}\,\mathit{N_5}}}&=& -
{\displaystyle \frac{3}{4}} \,{\displaystyle \frac{\pi \,
\mathit{N_1}^{3}\,\mathit{N_R}^{3}}{(\mathit{N_1}\,\mathit{N_5}\,
\mathit{N_R})^{(5/2)}}},  \nonumber \\
{\Gamma_{\mathit{N_5}\,\mathit{N_5}\,\mathit{N_R}}}&=&{\displaystyle
\frac{1}{4}} \,{\displaystyle \frac{\pi \,\mathit{N_1}^{3}\,
\mathit{N_R}^{2}\,\mathit{N_5}}{(\mathit{N_1}\,\mathit{N_5}\,\mathit{
N_R})^{(5/2)}}}, 
\end{eqnarray}
\begin{eqnarray}
{\Gamma_{\mathit{N_5}\,\mathit{N_R}\,\mathit{N_1}}}&=& -
{\displaystyle \frac{1}{4}} \,{\displaystyle \frac{\pi \,
\mathit{N_5}^{2}\,\mathit{N_R}^{2}\,\mathit{N_1}^{2}}{(\mathit{N_1}\,
\mathit{N_5}\,\mathit{N_R})^{(5/2)}}},  \nonumber \\
{\Gamma_{\mathit{N_5}\,\mathit{N_R}\,\mathit{N_5}}}&=&{\displaystyle
\frac{1}{4}} \,{\displaystyle \frac{\pi \,\mathit{N_1}^{3}\,
\mathit{N_R}^{2}\,\mathit{N_5}}{(\mathit{N_1}\,\mathit{N_5}\,\mathit{
N_R})^{(5/2)}}},  \nonumber \\
{\Gamma_{\mathit{N_5}\,\mathit{N_R}\,\mathit{N_R}}}&=&{\displaystyle
\frac{1}{4}} \,{\displaystyle \frac{\pi \,\mathit{N_1}^{3}\,
\mathit{N_R}\,\mathit{N_5}^{2}}{(\mathit{N_1}\,\mathit{N_5}\,\mathit{
N_R})^{(5/2)}}},  \nonumber \\
{\Gamma_{\mathit{N_R}\,\mathit{N_R}\,\mathit{N_1}}}&=&{\displaystyle
\frac{1}{4}} \,{\displaystyle \frac{\pi \,\mathit{N_5}^{3}\,
\mathit{N_R}\,\mathit{N_1}^{2}}{(\mathit{N_1}\,\mathit{N_5}\,\mathit{
N_R})^{(5/2)}}},  \nonumber \\
{\Gamma_{\mathit{N_R}\,\mathit{N_R}\,\mathit{N_5}}}&=&{\displaystyle
\frac{1}{4}} \,{\displaystyle \frac{\pi \,\mathit{N_1}^{3}\,
\mathit{N_R}\,\mathit{N_5}^{2}}{(\mathit{N_1}\,\mathit{N_5}\,\mathit{
N_R})^{(5/2)}}},  \nonumber \\
{\Gamma_{\mathit{N_R}\,\mathit{N_R}\,\mathit{N_R}}}&=& -
{\displaystyle \frac{3}{4}} \,{\displaystyle \frac{\pi \,
\mathit{N_1}^{3}\,\mathit{N_5}^{3}}{(\mathit{N_1}\,\mathit{N_5}\,
\mathit{N_R})^{(5/2)}}}.
\end{eqnarray}

\newpage
\subsubsection{\`A l'ordre de $(\alpha^{\prime})^1$}

Avec la sym\'etrie dans les deux premiers indices, il s'ensuit que 
les \'el\'ements de symbole de Christoffel du premier type sont

\begin{eqnarray}
{\Gamma_{\mathit{N_1}\,\mathit{N_1}\,\mathit{N_1}}}&=& -
{\displaystyle \frac{3}{4}} \,{\displaystyle \frac{\pi \,(
\mathit{N_5}^{4}\,\mathit{N_R}^{3}\,\mathit{N_1}^{4} +
16\,b\,\sqrt{
\mathit{N_R}}\,(\mathit{N_1}\,\mathit{N_5}\,\mathit{N_R})^{(5/2)})}{(
\mathit{N_1}\,\mathit{N_5}\,\mathit{N_R})^{(5/2)}\,\mathit{N_1}^{4}\,
\mathit{N_5}}},  \nonumber \\
{\Gamma_{\mathit{N_1}\,\mathit{N_1}\,\mathit{N_5}}}&=&{\displaystyle
\frac{1}{4}} \,{\displaystyle \frac{\pi \,(\mathit{N_5}^{4}\,
\mathit{N_R}^{3}\,\mathit{N_1}^{4} - 16\,b\,\sqrt{\mathit{N_R}}\,(
\mathit{N_1}\,\mathit{N_5}\,\mathit{N_R})^{(5/2)})}{(\mathit{N_1}\,
\mathit{N_5}\,\mathit{N_R})^{(5/2)}\,\mathit{N_1}^{3}\,\mathit{N_5}^{
2}}},  \nonumber \\
{\Gamma_{\mathit{N_1}\,\mathit{N_1}\,\mathit{N_R}}}&=&{\displaystyle
\frac{1}{4}} \,{\displaystyle \frac{\pi \,(\mathit{N_5}^{4}\,
\mathit{N_R}^{(5/2)}\,\mathit{N_1}^{4} + 8\,b\,(\mathit{N_1}\,
\mathit{N_5}\,\mathit{N_R})^{(5/2)})}{(\mathit{N_1}\,\mathit{N_5}\,
\mathit{N_R})^{(5/2)}\,\sqrt{\mathit{N_R}}\,\mathit{N_1}^{3}\,
\mathit{N_5}}},  \nonumber \\
{\Gamma_{\mathit{N_1}\,\mathit{N_5}\,\mathit{N_1}}}&=&{\displaystyle
\frac{1}{4}} \,{\displaystyle \frac{\pi \,(\mathit{N_5}^{4}\,
\mathit{N_R}^{3}\,\mathit{N_1}^{4} - 16\,b\,\sqrt{\mathit{N_R}}\,(
\mathit{N_1}\,\mathit{N_5}\,\mathit{N_R})^{(5/2)})}{(\mathit{N_1}\,
\mathit{N_5}\,\mathit{N_R})^{(5/2)}\,\mathit{N_1}^{3}\,\mathit{N_5}^{
2}}},  \nonumber \\
{\Gamma_{\mathit{N_1}\,\mathit{N_5}\,\mathit{N_5}}}&=&{\displaystyle
\frac{1}{4}} \,{\displaystyle \frac{\pi \,(\mathit{N_5}^{4}\,
\mathit{N_R}^{3}\,\mathit{N_1}^{4} - 16\,b\,\sqrt{\mathit{N_R}}\,(
\mathit{N_1}\,\mathit{N_5}\,\mathit{N_R})^{(5/2)})}{(\mathit{N_1}\,
\mathit{N_5}\,\mathit{N_R})^{(5/2)}\,\mathit{N_1}^{2}\,\mathit{N_5}^{
3}}},  \nonumber \\
{\Gamma_{\mathit{N_1}\,\mathit{N_5}\,\mathit{N_R}}}&=&{\displaystyle
\frac{1}{4}} \,{\displaystyle \frac{\pi \,( - \mathit{N_5}^{4}\,
\mathit{N_R}^{(5/2)}\,\mathit{N_1}^{4} + 4\,b\,(\mathit{N_1}\,
\mathit{N_5}\,\mathit{N_R})^{(5/2)})}{(\mathit{N_1}\,\mathit{N_5}\,
\mathit{N_R})^{(5/2)}\,\mathit{N_5}^{2}\,\mathit{N_1}^{2}\,\sqrt{
\mathit{N_R}}}}, 
\end{eqnarray}
\begin{eqnarray}
{\Gamma_{\mathit{N_1}\,\mathit{N_R}\,\mathit{N_1}}}&=&{\displaystyle
\frac{1}{4}} \,{\displaystyle \frac{\pi \,(\mathit{N_5}^{4}\,
\mathit{N_R}^{(5/2)}\,\mathit{N_1}^{4} + 8\,b\,(\mathit{N_1}\,
\mathit{N_5}\,\mathit{N_R})^{(5/2)})}{(\mathit{N_1}\,\mathit{N_5}\,
\mathit{N_R})^{(5/2)}\,\sqrt{\mathit{N_R}}\,\mathit{N_1}^{3}\,
\mathit{N_5}}},  \nonumber \\
{\Gamma_{\mathit{N_1}\,\mathit{N_R}\,\mathit{N_5}}}&=&{\displaystyle
\frac{1}{4}} \,{\displaystyle \frac{\pi \,( - \mathit{N_5}^{4}\,
\mathit{N_R}^{(5/2)}\,\mathit{N_1}^{4} + 4\,b\,(\mathit{N_1}\,
\mathit{N_5}\,\mathit{N_R})^{(5/2)})}{(\mathit{N_1}\,\mathit{N_5}\,
\mathit{N_R})^{(5/2)}\,\mathit{N_5}^{2}\,\mathit{N_1}^{2}\,\sqrt{
\mathit{N_R}}}},  \nonumber \\
{\Gamma_{\mathit{N_1}\,\mathit{N_R}\,\mathit{N_R}}}&=&{\displaystyle
\frac{1}{4}} \,{\displaystyle \frac{\pi \,(\mathit{N_5}^{4}\,
\mathit{N_R}^{(5/2)}\,\mathit{N_1}^{4} + 2\,b\,(\mathit{N_1}\,
\mathit{N_5}\,\mathit{N_R})^{(5/2)})}{(\mathit{N_1}\,\mathit{N_5}\,
\mathit{N_R})^{(5/2)}\,\mathit{N_R}^{(3/2)}\,\mathit{N_1}^{2}\,
\mathit{N_5}}},  \nonumber \\
{\Gamma_{\mathit{N_5}\,\mathit{N_5}\,\mathit{N_1}}}&=&{\displaystyle
\frac{1}{4}} \,{\displaystyle \frac{\pi \,(\mathit{N_5}^{4}\,
\mathit{N_R}^{3}\,\mathit{N_1}^{4} - 16\,b\,\sqrt{\mathit{N_R}}\,(
\mathit{N_1}\,\mathit{N_5}\,\mathit{N_R})^{(5/2)})}{(\mathit{N_1}\,
\mathit{N_5}\,\mathit{N_R})^{(5/2)}\,\mathit{N_1}^{2}\,\mathit{N_5}^{
3}}},  \nonumber \\
{\Gamma_{\mathit{N_5}\,\mathit{N_5}\,\mathit{N_5}}}&=& -
{\displaystyle \frac{3}{4}} \,{\displaystyle \frac{\pi \,(
\mathit{N_5}^{4}\,\mathit{N_R}^{3}\,\mathit{N_1}^{4} +
16\,b\,\sqrt{
\mathit{N_R}}\,(\mathit{N_1}\,\mathit{N_5}\,\mathit{N_R})^{(5/2)})}{(
\mathit{N_1}\,\mathit{N_5}\,\mathit{N_R})^{(5/2)}\,\mathit{N_1}\,
\mathit{N_5}^{4}}},  \nonumber \\
{\Gamma_{\mathit{N_5}\,\mathit{N_5}\,\mathit{N_R}}}&=&{\displaystyle
\frac{1}{4}} \,{\displaystyle \frac{\pi \,(\mathit{N_5}^{4}\,
\mathit{N_R}^{(5/2)}\,\mathit{N_1}^{4} + 8\,b\,(\mathit{N_1}\,
\mathit{N_5}\,\mathit{N_R})^{(5/2)})}{(\mathit{N_1}\,\mathit{N_5}\,
\mathit{N_R})^{(5/2)}\,\sqrt{\mathit{N_R}}\,\mathit{N_1}\,\mathit{N_5
}^{3}}}, 
\end{eqnarray}
\begin{eqnarray}
{\Gamma_{\mathit{N_5}\,\mathit{N_R}\,\mathit{N_1}}}&=&{\displaystyle
\frac{1}{4}} \,{\displaystyle \frac{\pi \,( - \mathit{N_5}^{4}\,
\mathit{N_R}^{(5/2)}\,\mathit{N_1}^{4} + 4\,b\,(\mathit{N_1}\,
\mathit{N_5}\,\mathit{N_R})^{(5/2)})}{(\mathit{N_1}\,\mathit{N_5}\,
\mathit{N_R})^{(5/2)}\,\mathit{N_5}^{2}\,\mathit{N_1}^{2}\,\sqrt{
\mathit{N_R}}}},  \nonumber \\
{\Gamma_{\mathit{N_5}\,\mathit{N_R}\,\mathit{N_5}}}&=&{\displaystyle
\frac{1}{4}} \,{\displaystyle \frac{\pi \,(\mathit{N_5}^{4}\,
\mathit{N_R}^{(5/2)}\,\mathit{N_1}^{4} + 8\,b\,(\mathit{N_1}\,
\mathit{N_5}\,\mathit{N_R})^{(5/2)})}{(\mathit{N_1}\,\mathit{N_5}\,
\mathit{N_R})^{(5/2)}\,\sqrt{\mathit{N_R}}\,\mathit{N_1}\,\mathit{N_5
}^{3}}},  \nonumber \\
{\Gamma_{\mathit{N_5}\,\mathit{N_R}\,\mathit{N_R}}}&=&{\displaystyle
\frac{1}{4}} \,{\displaystyle \frac{\pi \,(\mathit{N_5}^{4}\,
\mathit{N_R}^{(5/2)}\,\mathit{N_1}^{4} + 2\,b\,(\mathit{N_1}\,
\mathit{N_5}\,\mathit{N_R})^{(5/2)})}{(\mathit{N_1}\,\mathit{N_5}\,
\mathit{N_R})^{(5/2)}\,\mathit{N_R}^{(3/2)}\,\mathit{N_1}\,\mathit{
N_5}^{2}}},  \nonumber \\
{\Gamma_{\mathit{N_R}\,\mathit{N_R}\,\mathit{N_1}}}&=&{\displaystyle
\frac{1}{4}} \,{\displaystyle \frac{\pi \,(\mathit{N_5}^{4}\,
\mathit{N_R}^{(5/2)}\,\mathit{N_1}^{4} + 2\,b\,(\mathit{N_1}\,
\mathit{N_5}\,\mathit{N_R})^{(5/2)})}{(\mathit{N_1}\,\mathit{N_5}\,
\mathit{N_R})^{(5/2)}\,\mathit{N_R}^{(3/2)}\,\mathit{N_1}^{2}\,
\mathit{N_5}}},  \nonumber \\
{\Gamma_{\mathit{N_R}\,\mathit{N_R}\,\mathit{N_5}}}&=&{\displaystyle
\frac{1}{4}} \,{\displaystyle \frac{\pi \,(\mathit{N_5}^{4}\,
\mathit{N_R}^{(5/2)}\,\mathit{N_1}^{4} + 2\,b\,(\mathit{N_1}\,
\mathit{N_5}\,\mathit{N_R})^{(5/2)})}{(\mathit{N_1}\,\mathit{N_5}\,
\mathit{N_R})^{(5/2)}\,\mathit{N_R}^{(3/2)}\,\mathit{N_1}\,\mathit{
N_5}^{2}}},  \nonumber \\
{\Gamma_{\mathit{N_R}\,\mathit{N_R}\,\mathit{N_R}}}&=&{\displaystyle
\frac{3}{4}} \,{\displaystyle \frac{\pi \,( - \mathit{N_5}^{4}\,
\mathit{N_R}^{(5/2)}\,\mathit{N_1}^{4} +
b\,(\mathit{N_1}\,\mathit{
N_5}\,\mathit{N_R})^{(5/2)})}{(\mathit{N_1}\,\mathit{N_5}\,\mathit{N_R
})^{(5/2)}\,\mathit{N_R}^{(5/2)}\,\mathit{N_1}\,\mathit{N_5}}}.
\end{eqnarray}

\newpage
\subsection{La gomtrie de Ruppenier des solutions non-extrmales de branes $ D_2D_6NS_5 $}

\subsubsection{\`A l'ordre de $(\alpha^{\prime})^0$}

Avec la sym\'etrie dans les deux premiers indices, 
les \'el\'ements de symbole de Christoffel du premier type sont

\begin{eqnarray}
{\Gamma_{\mathit{N2}\,\mathit{N_2}\,\mathit{N_2}}}&=&  -
{\displaystyle \frac{3}{4}} \,{\displaystyle \frac{\pi \,
\mathit{N_6}^{3}\,\mathit{N_5}^{3}\,\mathit{N_R}^{3}}{(\mathit{N_2}\,
\mathit{N_6}\,\mathit{N_5}\,\mathit{N_R})^{(5/2)}}},  \nonumber \\
{\Gamma_{\mathit{N_2}\,\mathit{N_2}\,\mathit{N_6}}}&=&
{\displaystyle \frac{1}{4}} \,{\displaystyle \frac{\pi
\,\mathit{N_6}^{2}\,
\mathit{N_5}^{3}\,\mathit{N_R}^{3}\,\mathit{N_2}}{(\mathit{N_2}\,
\mathit{N_6}\,\mathit{N_5}\,\mathit{N_R})^{(5/2)}}},  \nonumber \\
{\Gamma_{\mathit{N_2}\,\mathit{N_2}\,\mathit{N_5}}}&=&
{\displaystyle \frac{1}{4}} \,{\displaystyle \frac{\pi
\,\mathit{N_6}^{3}\,
\mathit{N_5}^{2}\,\mathit{N_R}^{3}\,\mathit{N_2}}{(\mathit{N_2}\,
\mathit{N_6}\,\mathit{N_5}\,\mathit{N_R})^{(5/2)}}},  \nonumber \\
{\Gamma_{\mathit{N_2}\,\mathit{N_2}\,\mathit{N_R}}}&=&
{\displaystyle \frac{1}{4}} \,{\displaystyle \frac{\pi
\,\mathit{N_6}^{3}\,
\mathit{N_5}^{3}\,\mathit{N_R}^{2}\,\mathit{N_2}}{(\mathit{N_2}\,
\mathit{N_6}\,\mathit{N_5}\,\mathit{N_R})^{(5/2)}}},  \nonumber \\
{\Gamma_{\mathit{N_2}\,\mathit{N_6}\,\mathit{N_2}}}&=&
{\displaystyle \frac{1}{4}} \,{\displaystyle \frac{\pi
\,\mathit{N_6}^{2}\,
\mathit{N_5}^{3}\,\mathit{N_R}^{3}\,\mathit{N_2}}{(\mathit{N_2}\,
\mathit{N_6}\,\mathit{N_5}\,\mathit{N_R})^{(5/2)}}}, 
\end{eqnarray}
\begin{eqnarray}
{\Gamma_{\mathit{N_2}\,\mathit{N_6}\,\mathit{N_6}}}&=&
{\displaystyle \frac{1}{4}} \,{\displaystyle \frac{\pi
\,\mathit{N_6}\,\mathit{
N_5}^{3}\,\mathit{N_R}^{3}\,\mathit{N_2}^{2}}{(\mathit{N_2}\,\mathit{
N_6}\,\mathit{N_5}\,\mathit{N_R})^{(5/2)}}},  \nonumber \\
{\Gamma_{\mathit{N_2}\,\mathit{N_6}\,\mathit{N_5}}}&=&  -
{\displaystyle \frac{1}{4}} \,{\displaystyle \frac{\pi \,
\mathit{N_6}^{2}\,\mathit{N_5}^{2}\,\mathit{N_R}^{3}\,\mathit{N_2}^{2
}}{(\mathit{N_2}\,\mathit{N_6}\,\mathit{N_5}\,\mathit{N_R})^{(5/2)}} },  \nonumber \\
{\Gamma_{\mathit{N_2}\,\mathit{N_6}\,\mathit{N_R}}}&=&  -
{\displaystyle \frac{1}{4}} \,{\displaystyle \frac{\pi \,
\mathit{N_6}^{2}\,\mathit{N_5}^{3}\,\mathit{N_R}^{2}\,\mathit{N_2}^{2
}}{(\mathit{N_2}\,\mathit{N_6}\,\mathit{N_5}\,\mathit{N_R})^{(5/2)}} },  \nonumber \\
{\Gamma_{\mathit{N_2}\,\mathit{N_5}\,\mathit{N_2}}}&=&
{\displaystyle \frac{1}{4}} \,{\displaystyle \frac{\pi
\,\mathit{N_6}^{3}\,
\mathit{N_5}^{2}\,\mathit{N_R}^{3}\,\mathit{N_2}}{(\mathit{N_2}\,
\mathit{N_6}\,\mathit{N_5}\,\mathit{N_R})^{(5/2)}}},  \nonumber \\
{\Gamma_{\mathit{N_2}\,\mathit{N_5}\,\mathit{N_6}}}&=&  -
{\displaystyle \frac{1}{4}} \,{\displaystyle \frac{\pi \,
\mathit{N_6}^{2}\,\mathit{N_5}^{2}\,\mathit{N_R}^{3}\,\mathit{N_2}^{2
}}{(\mathit{N_2}\,\mathit{N_6}\,\mathit{N_5}\,\mathit{N_R})^{(5/2)}} }, 
\end{eqnarray}
\begin{eqnarray}
{\Gamma_{\mathit{N_2}\,\mathit{N_5}\,\mathit{N_5}}}&=&
{\displaystyle \frac{1}{4}} \,{\displaystyle \frac{\pi
\,\mathit{N_6}^{3}\,
\mathit{N_5}\,\mathit{N_R}^{3}\,\mathit{N_2}^{2}}{(\mathit{N_2}\,
\mathit{N_6}\,\mathit{N_5}\,\mathit{N_R})^{(5/2)}}},  \nonumber \\
{\Gamma_{\mathit{N_2}\,\mathit{N_5}\,\mathit{N_R}}}&=&  -
{\displaystyle \frac{1}{4}} \,{\displaystyle \frac{\pi \,
\mathit{N_6}^{3}\,\mathit{N_5}^{2}\,\mathit{N_R}^{2}\,\mathit{N_2}^{2
}}{(\mathit{N_2}\,\mathit{N_6}\,\mathit{N_5}\,\mathit{N_R})^{(5/2)}} },  \nonumber \\
{\Gamma_{\mathit{N_2}\,\mathit{N_R}\,\mathit{N_2}}}&=&
{\displaystyle \frac{1}{4}} \,{\displaystyle \frac{\pi
\,\mathit{N_6}^{3}\,
\mathit{N_5}^{3}\,\mathit{N_R}^{2}\,\mathit{N_2}}{(\mathit{N_2}\,
\mathit{N_6}\,\mathit{N_5}\,\mathit{N_R})^{(5/2)}}},  \nonumber \\
{\Gamma_{\mathit{N_2}\,\mathit{N_R}\,\mathit{N_6}}}&=&  -
{\displaystyle \frac{1}{4}} \,{\displaystyle \frac{\pi \,
\mathit{N_6}^{2}\,\mathit{N_5}^{3}\,\mathit{N_R}^{2}\,\mathit{N_2}^{2
}}{(\mathit{N_2}\,\mathit{N_6}\,\mathit{N_5}\,\mathit{N_R})^{(5/2)}} },  \nonumber \\
{\Gamma_{\mathit{N_2}\,\mathit{N_R}\,\mathit{N_5}}}&=&  -
{\displaystyle \frac{1}{4}} \,{\displaystyle \frac{\pi \,
\mathit{N_6}^{3}\,\mathit{N_5}^{2}\,\mathit{N_R}^{2}\,\mathit{N_2}^{2
}}{(\mathit{N_2}\,\mathit{N_6}\,\mathit{N_5}\,\mathit{N_R})^{(5/2)}} }, 
\end{eqnarray}
\begin{eqnarray}
{\Gamma_{\mathit{N_2}\,\mathit{N_R}\,\mathit{N_R}}}&=&
{\displaystyle \frac{1}{4}} \,{\displaystyle \frac{\pi
\,\mathit{N_6}^{3}\,
\mathit{N_5}^{3}\,\mathit{N_R}\,\mathit{N_2}^{2}}{(\mathit{N_2}\,
\mathit{N_6}\,\mathit{N_5}\,\mathit{N_R})^{(5/2)}}},  \nonumber \\
{\Gamma_{\mathit{N_6}\,\mathit{N_6}\,\mathit{N_2}}}&=&
{\displaystyle \frac{1}{4}} \,{\displaystyle \frac{\pi
\,\mathit{N_6}\,\mathit{
N_5}^{3}\,\mathit{N_R}^{3}\,\mathit{N_2}^{2}}{(\mathit{N_2}\,\mathit{
N_6}\,\mathit{N_5}\,\mathit{N_R})^{(5/2)}}},  \nonumber \\
{\Gamma_{\mathit{N_6}\,\mathit{N_6}\,\mathit{N_6}}}&=&  -
{\displaystyle \frac{3}{4}} \,{\displaystyle \frac{\pi \,
\mathit{N_2}^{3}\,\mathit{N_5}^{3}\,\mathit{N_R}^{3}}{(\mathit{N_2}\,
\mathit{N_6}\,\mathit{N_5}\,\mathit{N_R})^{(5/2)}}},  \nonumber \\
{\Gamma_{\mathit{N_6}\,\mathit{N_6}\,\mathit{N_5}}}&=&
{\displaystyle \frac{1}{4}} \,{\displaystyle \frac{\pi
\,\mathit{N_2}^{3}\,
\mathit{N_5}^{2}\,\mathit{N_R}^{3}\,\mathit{N_6}}{(\mathit{N_2}\,
\mathit{N_6}\,\mathit{N_5}\,\mathit{N_R})^{(5/2)}}},  \nonumber \\
{\Gamma_{\mathit{N_6}\,\mathit{N_6}\,\mathit{N_R}}}&=&
{\displaystyle \frac{1}{4}} \,{\displaystyle \frac{\pi
\,\mathit{N_2}^{3}\,
\mathit{N_5}^{3}\,\mathit{N_R}^{2}\,\mathit{N_6}}{(\mathit{N_2}\,
\mathit{N_6}\,\mathit{N_5}\,\mathit{N_R})^{(5/2)}}}, 
\end{eqnarray}
\begin{eqnarray}
{\Gamma_{\mathit{N_6}\,\mathit{N_5}\,\mathit{N_2}}}&=&  -
{\displaystyle \frac{1}{4}} \,{\displaystyle \frac{\pi \,
\mathit{N_6}^{2}\,\mathit{N_5}^{2}\,\mathit{N_R}^{3}\,\mathit{N_2}^{2
}}{(\mathit{N_2}\,\mathit{N_6}\,\mathit{N_5}\,\mathit{N_R})^{(5/2)}} },  \nonumber \\
{\Gamma_{\mathit{N_6}\,\mathit{N_5}\,\mathit{N_6}}}&=&
{\displaystyle \frac{1}{4}} \,{\displaystyle \frac{\pi
\,\mathit{N_2}^{3}\,
\mathit{N_5}^{2}\,\mathit{N_R}^{3}\,\mathit{N_6}}{(\mathit{N_2}\,
\mathit{N_6}\,\mathit{N_5}\,\mathit{N_R})^{(5/2)}}},  \nonumber \\
{\Gamma_{\mathit{N_6}\,\mathit{N_5}\,\mathit{N_5}}}&=&
{\displaystyle \frac{1}{4}} \,{\displaystyle \frac{\pi
\,\mathit{N_2}^{3}\,
\mathit{N_6}^{2}\,\mathit{N_R}^{3}\,\mathit{N_5}}{(\mathit{N_2}\,
\mathit{N_6}\,\mathit{N_5}\,\mathit{N_R})^{(5/2)}}},  \nonumber \\
{\Gamma_{\mathit{N_6}\,\mathit{N_5}\,\mathit{N_R}}}&=&  -
{\displaystyle \frac{1}{4}} \,{\displaystyle \frac{\pi \,
\mathit{N_2}^{3}\,\mathit{N_6}^{2}\,\mathit{N_R}^{2}\,\mathit{N_5}^{2
}}{(\mathit{N_2}\,\mathit{N_6}\,\mathit{N_5}\,\mathit{N_R})^{(5/2)}} },  \nonumber \\
{\Gamma_{\mathit{N_6}\,\mathit{N_R}\,\mathit{N_2}}}&=&  -
{\displaystyle \frac{1}{4}} \,{\displaystyle \frac{\pi \,
\mathit{N_6}^{2}\,\mathit{N_5}^{3}\,\mathit{N_R}^{2}\,\mathit{N_2}^{2
}}{(\mathit{N_2}\,\mathit{N_6}\,\mathit{N_5}\,\mathit{N_R})^{(5/2)}} }, 
\end{eqnarray}
\begin{eqnarray}
{\Gamma_{\mathit{N_6}\,\mathit{N_R}\,\mathit{N_6}}}&=&
{\displaystyle \frac{1}{4}} \,{\displaystyle \frac{\pi
\,\mathit{N_2}^{3}\,
\mathit{N_5}^{3}\,\mathit{N_R}^{2}\,\mathit{N_6}}{(\mathit{N_2}\,
\mathit{N_6}\,\mathit{N_5}\,\mathit{N_R})^{(5/2)}}},  \nonumber \\
{\Gamma_{\mathit{N_6}\,\mathit{N_R}\,\mathit{N_5}}}&=&  -
{\displaystyle \frac{1}{4}} \,{\displaystyle \frac{\pi \,
\mathit{N_2}^{3}\,\mathit{N_6}^{2}\,\mathit{N_R}^{2}\,\mathit{N_5}^{2
}}{(\mathit{N_2}\,\mathit{N_6}\,\mathit{N_5}\,\mathit{N_R})^{(5/2)}} },  \nonumber \\
{\Gamma_{\mathit{N_6}\,\mathit{N_R}\,\mathit{N_R}}}&=&
{\displaystyle \frac{1}{4}} \,{\displaystyle \frac{\pi
\,\mathit{N_2}^{3}\,
\mathit{N_5}^{3}\,\mathit{N_R}\,\mathit{N_6}^{2}}{(\mathit{N_2}\,
\mathit{N_6}\,\mathit{N_5}\,\mathit{N_R})^{(5/2)}}},  \nonumber \\
{\Gamma_{\mathit{N_5}\,\mathit{N_5}\,\mathit{N_2}}}&=&
{\displaystyle \frac{1}{4}} \,{\displaystyle \frac{\pi
\,\mathit{N_6}^{3}\,
\mathit{N_5}\,\mathit{N_R}^{3}\,\mathit{N_2}^{2}}{(\mathit{N_2}\,
\mathit{N_6}\,\mathit{N_5}\,\mathit{N_R})^{(5/2)}}},  \nonumber \\
{\Gamma_{\mathit{N_5}\,\mathit{N_5}\,\mathit{N_6}}}&=&
{\displaystyle \frac{1}{4}} \,{\displaystyle \frac{\pi
\,\mathit{N_2}^{3}\,
\mathit{N_6}^{2}\,\mathit{N_R}^{3}\,\mathit{N_5}}{(\mathit{N_2}\,
\mathit{N_6}\,\mathit{N_5}\,\mathit{N_R})^{(5/2)}}}, 
\end{eqnarray}
\begin{eqnarray}
{\Gamma_{\mathit{N_5}\,\mathit{N_5}\,\mathit{N_5}}}&=&  -
{\displaystyle \frac{3}{4}} \,{\displaystyle \frac{\pi \,
\mathit{N_2}^{3}\,\mathit{N_6}^{3}\,\mathit{N_R}^{3}}{(\mathit{N_2}\,
\mathit{N_6}\,\mathit{N_5}\,\mathit{N_R})^{(5/2)}}},  \nonumber \\
{\Gamma_{\mathit{N_5}\,\mathit{N_5}\,\mathit{N_R}}}&=&
{\displaystyle \frac{1}{4}} \,{\displaystyle \frac{\pi
\,\mathit{N_2}^{3}\,
\mathit{N_6}^{3}\,\mathit{N_R}^{2}\,\mathit{N_5}}{(\mathit{N_2}\,
\mathit{N_6}\,\mathit{N_5}\,\mathit{N_R})^{(5/2)}}},  \nonumber \\
{\Gamma_{\mathit{N_5}\,\mathit{N_R}\,\mathit{N_2}}}&=&  -
{\displaystyle \frac{1}{4}} \,{\displaystyle \frac{\pi \,
\mathit{N_6}^{3}\,\mathit{N_5}^{2}\,\mathit{N_R}^{2}\,\mathit{N_2}^{2
}}{(\mathit{N_2}\,\mathit{N_6}\,\mathit{N_5}\,\mathit{N_R})^{(5/2)}} },  \nonumber \\
{\Gamma_{\mathit{N_5}\,\mathit{N_R}\,\mathit{N_6}}}&=&  -
{\displaystyle \frac{1}{4}} \,{\displaystyle \frac{\pi \,
\mathit{N_2}^{3}\,\mathit{N_6}^{2}\,\mathit{N_R}^{2}\,\mathit{N_5}^{2
}}{(\mathit{N_2}\,\mathit{N_6}\,\mathit{N_5}\,\mathit{N_R})^{(5/2)}} },  \nonumber \\
{\Gamma_{\mathit{N_5}\,\mathit{N_R}\,\mathit{N_5}}}&=&
{\displaystyle \frac{1}{4}} \,{\displaystyle \frac{\pi
\,\mathit{N_2}^{3}\,
\mathit{N_6}^{3}\,\mathit{N_R}^{2}\,\mathit{N_5}}{(\mathit{N_2}\,
\mathit{N_6}\,\mathit{N_5}\,\mathit{N_R})^{(5/2)}}}, 
\end{eqnarray}
\begin{eqnarray}
{\Gamma_{\mathit{N_5}\,\mathit{N_R}\,\mathit{N_R}}}&=&
{\displaystyle \frac{1}{4}} \,{\displaystyle \frac{\pi
\,\mathit{N_2}^{3}\,
\mathit{N_6}^{3}\,\mathit{N_R}\,\mathit{N_5}^{2}}{(\mathit{N_2}\,
\mathit{N_6}\,\mathit{N_5}\,\mathit{N_R})^{(5/2)}}},  \nonumber \\
{\Gamma_{\mathit{N_R}\,\mathit{N_R}\,\mathit{N_2}}}&=&
{\displaystyle \frac{1}{4}} \,{\displaystyle \frac{\pi
\,\mathit{N_6}^{3}\,
\mathit{N_5}^{3}\,\mathit{N_R}\,\mathit{N_2}^{2}}{(\mathit{N_2}\,
\mathit{N_6}\,\mathit{N_5}\,\mathit{N_R})^{(5/2)}}},  \nonumber \\
{\Gamma_{\mathit{N_R}\,\mathit{N_R}\,\mathit{N_6}}}&=&
{\displaystyle \frac{1}{4}} \,{\displaystyle \frac{\pi
\,\mathit{N_2}^{3}\,
\mathit{N_5}^{3}\,\mathit{N_R}\,\mathit{N_6}^{2}}{(\mathit{N_2}\,
\mathit{N_6}\,\mathit{N_5}\,\mathit{N_R})^{(5/2)}}},  \nonumber \\
{\Gamma_{\mathit{N_R}\,\mathit{N_R}\,\mathit{N_5}}}&=&
{\displaystyle \frac{1}{4}} \,{\displaystyle \frac{\pi
\,\mathit{N_2}^{3}\,
\mathit{N_6}^{3}\,\mathit{N_R}\,\mathit{N_5}^{2}}{(\mathit{N_2}\,
\mathit{N_6}\,\mathit{N_5}\,\mathit{N_R})^{(5/2)}}},  \nonumber \\
{\Gamma_{\mathit{N_R}\,\mathit{N_R}\,\mathit{N_R}}}&=&  -
{\displaystyle \frac{3}{4}} \,{\displaystyle \frac{\pi \,
\mathit{N_2}^{3}\,\mathit{N_6}^{3}\,\mathit{N_5}^{3}}{(\mathit{N_2}\,
\mathit{N_6}\,\mathit{N_5}\,\mathit{N_R})^{(5/2)}}}.
\end{eqnarray}

\newpage
\subsubsection{\`A l'ordre de $(\alpha^{\prime})^1$}

Dans ce cas, en suivant les propri\'et\'es de sym\'etrie dans les deux premiers indices, 
nous observons que les composantes de symbole de Christoffel du premier type sont

\begin{eqnarray}
{\Gamma_{\mathit{N_2}\,\mathit{N_2}\,\mathit{N_2}}}&=&  -
{\displaystyle \frac{3}{4}} \,{\displaystyle \frac{\pi \,(
\mathit{N_6}^{4}\,\mathit{N_5}^{(11/2)}\,\mathit{N_R}^{3}\,\mathit{
N_2}^{4} + 16\,\mathit{b_1}\,\sqrt{\mathit{N_R}}\,(\mathit{N_2}\,
\mathit{N_6}\,\mathit{N_5}\,\mathit{N_R})^{(5/2)})}{(\mathit{N_2}\,
\mathit{N_6}\,\mathit{N_5}\,\mathit{N_R})^{(5/2)}\,\mathit{N_2}^{4}\,
\mathit{N_6}\,\mathit{N_5}^{(5/2)}}},  \nonumber \\
{\Gamma_{\mathit{N_2}\,\mathit{N_2}\,\mathit{N_6}}}&=&  -
{\displaystyle \frac{1}{4}} \,{\displaystyle \frac{\pi \,( -
\mathit{N_6}^{4}\,\mathit{N_5}^{(11/2)}\,\mathit{N_R}^{3}\,\mathit{
N_2}^{4} + 16\,\mathit{b_1}\,\sqrt{\mathit{N_R}}\,(\mathit{N_2}\,
\mathit{N_6}\,\mathit{N_5}\,\mathit{N_R})^{(5/2)})}{(\mathit{N_2}\,
\mathit{N_6}\,\mathit{N_5}\,\mathit{N_R})^{(5/2)}\,\mathit{N_2}^{3}\,
\mathit{N_6}^{2}\,\mathit{N_5}^{(5/2)}}},  \nonumber \\
{\Gamma_{\mathit{N_2}\,\mathit{N_2}\,\mathit{N_5}}}&=&  -
{\displaystyle \frac{1}{4}} \,{\displaystyle \frac{\pi \,( -
\mathit{N_6}^{4}\,\mathit{N_5}^{(11/2)}\,\mathit{N_R}^{3}\,\mathit{
N_2}^{4} + 40\,\mathit{b_1}\,\sqrt{\mathit{N_R}}\,(\mathit{N_2}\,
\mathit{N_6}\,\mathit{N_5}\,\mathit{N_R})^{(5/2)})}{(\mathit{N_2}\,
\mathit{N_6}\,\mathit{N_5}\,\mathit{N_R})^{(5/2)}\,\mathit{N_2}^{3}\,
\mathit{N_6}\,\mathit{N_5}^{(7/2)}}},  \nonumber \\
{\Gamma_{\mathit{N_2}\,\mathit{N_2}\,\mathit{N_R}}}&=&
{\displaystyle \frac{1}{4}} \,{\displaystyle \frac{\pi
\,(\mathit{N_6}^{4}\,\mathit{N_5}^{(11/2)}\,\mathit{N_R}^{(5/2)}\,
\mathit{N_2}^{4} + 8\, \mathit{b_1}\,(\mathit{N_2}\,\mathit{N_6}\,
\mathit{N_5}\,\mathit{N_R})^{(5/2)})}{(\mathit{N_2}\,\mathit{N_6}\,
\mathit{N_5}\,\mathit{N_R})^{(5/2)}\,\sqrt{\mathit{N_R}}\,
\mathit{N_2}^{3}\,\mathit{N_6}\,\mathit{N_5}^{(5/2)}}},  \nonumber \\
{\Gamma_{\mathit{N_2}\,\mathit{N_6}\,\mathit{N_2}}}&=&  -
{\displaystyle \frac{1}{4}} \,{\displaystyle \frac{\pi \,( -
\mathit{N_6}^{4}\,\mathit{N_5}^{(11/2)}\,\mathit{N_R}^{3}\,\mathit{
N_2}^{4} + 16\,\mathit{b_1}\,\sqrt{\mathit{N_R}}\,(\mathit{N_2}\,
\mathit{N_6}\,\mathit{N_5}\,\mathit{N_R})^{(5/2)})}{(\mathit{N_2}\,
\mathit{N_6}\,\mathit{N_5}\,\mathit{N_R})^{(5/2)}\,\mathit{N_2}^{3}\,
\mathit{N_6}^{2}\,\mathit{N_5}^{(5/2)}}}, 
\end{eqnarray}
\begin{eqnarray}
{\Gamma_{\mathit{N_2}\,\mathit{N_6}\,\mathit{N_6}}}&=&  -
{\displaystyle \frac{1}{4}} \,{\displaystyle \frac{\pi \,( -
\mathit{N_6}^{4}\,\mathit{N_5}^{(11/2)}\,\mathit{N_R}^{3}\,\mathit{
N_2}^{4} + 16\,\mathit{b_1}\,\sqrt{\mathit{N_R}}\,(\mathit{N_2}\,
\mathit{N_6}\,\mathit{N_5}\,\mathit{N_R})^{(5/2)})}{(\mathit{N_2}\,
\mathit{N_6}\,\mathit{N_5}\,\mathit{N_R})^{(5/2)}\,\mathit{N_2}^{2}\,
\mathit{N_6}^{3}\,\mathit{N_5}^{(5/2)}}},  \nonumber \\
{\Gamma_{\mathit{N_2}\,\mathit{N_6}\,\mathit{N_5}}}&=&  -
{\displaystyle \frac{1}{4}} \,{\displaystyle \frac{\pi \,(
\mathit{N_6}^{4}\,\mathit{N_5}^{(11/2)}\,\mathit{N_R}^{3}\,\mathit{
N_2}^{4} + 20\,\mathit{b_1}\,\sqrt{\mathit{N_R}}\,(\mathit{N_2}\,
\mathit{N_6}\,\mathit{N_5}\,\mathit{N_R})^{(5/2)})}{(\mathit{N_2}\,
\mathit{N_6}\,\mathit{N_5}\,\mathit{N_R})^{(5/2)}\,\mathit{N_2}^{2}\,
\mathit{N_6}^{2}\,\mathit{N_5}^{(7/2)}}},  \nonumber \\
{\Gamma_{\mathit{N_2}\,\mathit{N_6}\,\mathit{N_R}}}&=&
{\displaystyle \frac{1}{4}} \,{\displaystyle \frac{\pi \,( -
\mathit{N_6}^{4}\, \mathit{N_5}^{(11/2)}\,\mathit{N_R}^{(5/2)}\,
\mathit{N_2}^{4} + 4\,\mathit{b_1}\,(\mathit{N_2}\,\mathit{N_6}\,
\mathit{N_5}\,\mathit{N_R})^{(5/2)})}{(\mathit{N_2}\,\mathit{N_6}\,
\mathit{N_5}\,\mathit{N_R})^{(5/2)}\,\sqrt{\mathit{N_R}}\,
\mathit{N_2}^{2}\,\mathit{N_6}^{2}\,\mathit{N_5}^{(5/2)}}},  \nonumber \\
{\Gamma_{\mathit{N_2}\,\mathit{N_5}\,\mathit{N_2}}}&=&  -
{\displaystyle \frac{1}{4}} \,{\displaystyle \frac{\pi \,( -
\mathit{N_6}^{4}\,\mathit{N_5}^{(11/2)}\,\mathit{N_R}^{3}\,\mathit{
N_2}^{4} + 40\,\mathit{b_1}\,\sqrt{\mathit{N_R}}\,(\mathit{N_2}\,
\mathit{N_6}\,\mathit{N_5}\,\mathit{N_R})^{(5/2)})}{(\mathit{N_2}\,
\mathit{N_6}\,\mathit{N_5}\,\mathit{N_R})^{(5/2)}\,\mathit{N_2}^{3}\,
\mathit{N_6}\,\mathit{N_5}^{(7/2)}}},  \nonumber \\
{\Gamma_{\mathit{N_2}\,\mathit{N_5}\,\mathit{N_6}}}&=&  -
{\displaystyle \frac{1}{4}} \,{\displaystyle \frac{\pi \,(
\mathit{N_6}^{4}\,\mathit{N_5}^{(11/2)}\,\mathit{N_R}^{3}\,\mathit{
N_2}^{4} + 20\,\mathit{b_1}\,\sqrt{\mathit{N_R}}\,(\mathit{N_2}\,
\mathit{N_6}\,\mathit{N_5}\,\mathit{N_R})^{(5/2)})}{(\mathit{N_2}\,
\mathit{N_6}\,\mathit{N_5}\,\mathit{N_R})^{(5/2)}\,\mathit{N_2}^{2}\,
\mathit{N_6}^{2}\,\mathit{N_5}^{(7/2)}}}, 
\end{eqnarray}
\begin{eqnarray}
{\Gamma_{\mathit{N_2}\,\mathit{N_5}\,\mathit{N_5}}}&=&  -
{\displaystyle \frac{1}{4}} \,{\displaystyle \frac{\pi \,( -
\mathit{N_6}^{4}\,\mathit{N_5}^{(11/2)}\,\mathit{N_R}^{3}\,\mathit{
N_2}^{4} + 70\,\mathit{b_1}\,\sqrt{\mathit{N_R}}\,(\mathit{N_2}\,
\mathit{N_6}\,\mathit{N_5}\,\mathit{N_R})^{(5/2)})}{(\mathit{N_2}\,
\mathit{N_6}\,\mathit{N_5}\,\mathit{N_R})^{(5/2)}\,\mathit{N_2}^{2}\,
\mathit{N_6}\,\mathit{N_5}^{(9/2)}}},  \nonumber \\
{\Gamma_{\mathit{N_2}\,\mathit{N_5}\,\mathit{N_R}}}&=&
{\displaystyle \frac{1}{4}} \,{\displaystyle \frac{\pi \,( -
\mathit{N_6}^{4}\,\mathit{N_5}^{(11/2)}\,\mathit{N_R}^{(5/2)}\,
\mathit{N_2}^{4}+ 10\,\mathit{b_1}\,(\mathit{N_2}\,\mathit{N_6}\,
\mathit{N_5}\,\mathit{N_R})^{(5/2)})}{(\mathit{N_2}\,\mathit{N_6}\,
\mathit{N_5}\,\mathit{N_R})^{(5/2)}\,\sqrt{\mathit{N_R}}\,
\mathit{N_2}^{2}\,\mathit{N_6}\,\mathit{N_5}^{(7/2)}}},  \nonumber \\
{\Gamma_{\mathit{N_2}\,\mathit{N_R}\,\mathit{N_2}}}&=&
{\displaystyle \frac{1}{4}} \,{\displaystyle \frac{\pi
\,(\mathit{N_6}^{4}\,\mathit{N_5}^{(11/2)}\,\mathit{N_R}^{(5/2)}\,
\mathit{N_2}^{4}+ 8\,\mathit{b_1}\,(\mathit{N_2}\,\mathit{N_6}\,
\mathit{N_5}\,\mathit{N_R})^{(5/2)})}{(\mathit{N_2}\,\mathit{N_6}\,
\mathit{N_5}\,\mathit{N_R})^{(5/2)}\,\sqrt{\mathit{N_R}}\,
\mathit{N_2}^{3}\,\mathit{N_6}\, \mathit{N_5}^{(5/2)}}},  \nonumber \\
{\Gamma_{\mathit{N_2}\,\mathit{N_R}\,\mathit{N_6}}}&=&
{\displaystyle \frac{1}{4}} \,{\displaystyle \frac{\pi \,( -
\mathit{N_6}^{4}\, \mathit{N_5}^{(11/2)}\,\mathit{N_R}^{(5/2)}\,
\mathit{N_2}^{4} + 4\,\mathit{b_1}\,(\mathit{N_2}\,\mathit{N_6}\,
\mathit{N_5}\,\mathit{N_R})^{(5/2)})}{(\mathit{N_2}\,\mathit{N_6}\,
\mathit{N_5}\,\mathit{N_R})^{(5/2)}\,\sqrt{\mathit{N_R}}\,
\mathit{N_2}^{2}\,\mathit{N_6}^{2}\, \mathit{N_5}^{(5/2)}}},  \nonumber \\
{\Gamma_{\mathit{N_2}\,\mathit{N_R}\,\mathit{N_5}}}&=&
{\displaystyle \frac{1}{4}} \,{\displaystyle \frac{\pi \,( -
\mathit{N_6}^{4}\, \mathit{N_5}^{(11/2)}\,\mathit{N_R}^{(5/2)}\,
\mathit{N_2}^{4} + 10\, \mathit{b_1}\,(\mathit{N_2}\,\mathit{N_6}\,
\mathit{N_5}\,\mathit{N_R})^{(5/2)})}{(\mathit{N_2}\,\mathit{N_6}\,
\mathit{N_5}\,\mathit{N_R})^{(5/2)}\,\sqrt{\mathit{N_R}}\,
\mathit{N_2}^{2}\,\mathit{N_6}\,\mathit{N_5}^{(7/2)}}}, 
\end{eqnarray}
\begin{eqnarray}
{\Gamma_{\mathit{N_2}\,\mathit{N_R}\,\mathit{N_R}}}&=&
{\displaystyle \frac{1}{4}} \,{\displaystyle \frac{\pi
\,(\mathit{N_6}^{4}\,\mathit{N_5}^{(11/2)}\,\mathit{N_R}^{(5/2)}\,
\mathit{N_2}^{4}+ 2\,\mathit{b_1}\,(\mathit{N_2}\,\mathit{N_6}\,
\mathit{N_5}\,\mathit{N_R})^{(5/2)})}{(\mathit{N_2}\,
\mathit{N_6}\,\mathit{N_5}\,\mathit{N_R})^{(5/2)}\,
\mathit{N_R}^{(3/2)}\,\mathit{N_2}^{2}\,\mathit{N_6}\,
\mathit{N_5}^{(5/2)}}},  \nonumber \\
{\Gamma_{\mathit{N_6}\,\mathit{N_6}\,\mathit{N_2}}}&=&  -
{\displaystyle \frac{1}{4}} \,{\displaystyle \frac{\pi \,( -
\mathit{N_6}^{4}\,\mathit{N_5}^{(11/2)}\,\mathit{N_R}^{3}\,\mathit{
N_2}^{4} + 16\,\mathit{b_1}\,\sqrt{\mathit{N_R}}\,(\mathit{N_2}\,
\mathit{N_6}\,\mathit{N_5}\,\mathit{N_R})^{(5/2)})}{(\mathit{N_2}\,
\mathit{N_6}\,\mathit{N_5}\,\mathit{N_R})^{(5/2)}\,\mathit{N_2}^{2}\,
\mathit{N_6}^{3}\,\mathit{N_5}^{(5/2)}}},  \nonumber \\
{\Gamma_{\mathit{N_6}\,\mathit{N_6}\,\mathit{N_6}}}&=&  -
{\displaystyle \frac{3}{4}} \,{\displaystyle \frac{\pi \,(
\mathit{N_6}^{4}\,\mathit{N_5}^{(11/2)}\,\mathit{N_R}^{3}\,\mathit{
N_2}^{4} + 16\,\mathit{b_1}\,\sqrt{\mathit{N_R}}\,(\mathit{N_2}\,
\mathit{N_6}\,\mathit{N_5}\,\mathit{N_R})^{(5/2)})}{(\mathit{N_2}\,
\mathit{N_6}\,\mathit{N_5}\,\mathit{N_R})^{(5/2)}\,\mathit{N_2}\,
\mathit{N_6}^{4}\,\mathit{N_5}^{(5/2)}}},  \nonumber \\
{\Gamma_{\mathit{N_6}\,\mathit{N_6}\,\mathit{N_5}}}&=&  -
{\displaystyle \frac{1}{4}} \,{\displaystyle \frac{\pi \,( -
\mathit{N_6}^{4}\,\mathit{N_5}^{(11/2)}\,\mathit{N_R}^{3}\,\mathit{
N_2}^{4} + 40\,\mathit{b_1}\,\sqrt{\mathit{N_R}}\,(\mathit{N_2}\,
\mathit{N_6}\,\mathit{N_5}\,\mathit{N_R})^{(5/2)})}{(\mathit{N_2}\,
\mathit{N_6}\,\mathit{N_5}\,\mathit{N_R})^{(5/2)}\,\mathit{N_2}\,
\mathit{N_6}^{3}\,\mathit{N_5}^{(7/2)}}},  \nonumber \\
{\Gamma_{\mathit{N_6}\,\mathit{N_6}\,\mathit{N_R}}}&=&
{\displaystyle \frac{1}{4}} \,{\displaystyle \frac{\pi
\,(\mathit{N_6}^{4}\, \mathit{N_5}^{(11/2)}\,
\mathit{N_R}^{(5/2)}\,\mathit{N_2}^{4} + 8\,
\mathit{b_1}\,(\mathit{N_2}\,\mathit{N_6}\,\mathit{N_5}\,\mathit{N_R})
^{(5/2)})}{(\mathit{N_2}\,\mathit{N_6}\,\mathit{N_5}\,\mathit{N_R})^{
(5/2)}\,\sqrt{\mathit{N_R}}\,\mathit{N_2}\,\mathit{N_6}^{3}\,
\mathit{N_5}^{(5/2)}}}, 
\end{eqnarray}
\begin{eqnarray}
{\Gamma_{\mathit{N_6}\,\mathit{N_5}\,\mathit{N_2}}}&=&  -
{\displaystyle \frac{1}{4}} \,{\displaystyle \frac{\pi \,(
\mathit{N_6}^{4}\,\mathit{N_5}^{(11/2)}\,\mathit{N_R}^{3}\,\mathit{
N_2}^{4} + 20\,\mathit{b_1}\,\sqrt{\mathit{N_R}}\,(\mathit{N_2}\,
\mathit{N_6}\,\mathit{N_5}\,\mathit{N_R})^{(5/2)})}{(\mathit{N_2}\,
\mathit{N_6}\,\mathit{N_5}\,\mathit{N_R})^{(5/2)}\,\mathit{N_2}^{2}\,
\mathit{N_6}^{2}\,\mathit{N_5}^{(7/2)}}},  \nonumber \\
{\Gamma_{\mathit{N_6}\,\mathit{N_5}\,\mathit{N_6}}}&=&  -
{\displaystyle \frac{1}{4}} \,{\displaystyle \frac{\pi \,( -
\mathit{N_6}^{4}\,\mathit{N_5}^{(11/2)}\,\mathit{N_R}^{3}\,\mathit{
N_2}^{4} + 40\,\mathit{b_1}\,\sqrt{\mathit{N_R}}\,(\mathit{N_2}\,
\mathit{N_6}\,\mathit{N_5}\,\mathit{N_R})^{(5/2)})}{(\mathit{N_2}\,
\mathit{N_6}\,\mathit{N_5}\,\mathit{N_R})^{(5/2)}\,\mathit{N_2}\,
\mathit{N_6}^{3}\,\mathit{N_5}^{(7/2)}}},  \nonumber \\
{\Gamma_{\mathit{N_6}\,\mathit{N_5}\,\mathit{N_5}}}&=&  -
{\displaystyle \frac{1}{4}} \,{\displaystyle \frac{\pi \,( -
\mathit{N_6}^{4}\,\mathit{N_5}^{(11/2)}\,\mathit{N_R}^{3}\,\mathit{
N_2}^{4} + 70\,\mathit{b_1}\,\sqrt{\mathit{N_R}}\,(\mathit{N_2}\,
\mathit{N_6}\,\mathit{N_5}\,\mathit{N_R})^{(5/2)})}{(\mathit{N_2}\,
\mathit{N_6}\,\mathit{N_5}\,\mathit{N_R})^{(5/2)}\,\mathit{N_2}\,
\mathit{N_6}^{2}\,\mathit{N_5}^{(9/2)}}},  \nonumber \\
{\Gamma_{\mathit{N_6}\,\mathit{N_5}\,\mathit{N_R}}}&=&
{\displaystyle \frac{1}{4}} \,{\displaystyle \frac{\pi \,( -
\mathit{N_6}^{4}\, \mathit{N_5}^{(11/2)}\,\mathit{N_R}^{(5/2)}\,
\mathit{N_2}^{4} + 10\, \mathit{b_1}\,(\mathit{N_2}\,
\mathit{N_6}\,\mathit{N_5}\,\mathit{N_R})^{(5/2)})}{(\mathit{N_2}\,
\mathit{N_6}\,\mathit{N_5}\,\mathit{N_R})^{(5/2)}\,\sqrt{\mathit{N_R}}\,
\mathit{N_2}\,\mathit{N_6}^{2}\,\mathit{N_5}^{(7/2)}}},  \nonumber \\
{\Gamma_{\mathit{N_6}\,\mathit{N_R}\,\mathit{N_2}}}&=&
{\displaystyle \frac{1}{4}} \,{\displaystyle \frac{\pi \,( -
\mathit{N_6}^{4}\, \mathit{N_5}^{(11/2)}\,\mathit{N_R}^{(5/2)}\,
\mathit{N_2}^{4} + 4\, \mathit{b_1}\,(\mathit{N_2}\,
\mathit{N_6}\,\mathit{N_5}\,\mathit{N_R})^{(5/2)})}{(\mathit{N_2}\,
\mathit{N_6}\,\mathit{N_5}\,\mathit{N_R})^{(5/2)}\,\sqrt{\mathit{N_R}}\,
\mathit{N_2}^{2}\,\mathit{N_6}^{2}\,\mathit{N_5}^{(5/2)}}}, 
\end{eqnarray}
\begin{eqnarray}
{\Gamma_{\mathit{N_6}\,\mathit{N_R}\,\mathit{N_6}}}&=&
{\displaystyle \frac{1}{4}} \,{\displaystyle \frac{\pi
\,(\mathit{N_6}^{4}\, \mathit{N_5}^{(11/2)}\,
\mathit{N_R}^{(5/2)}\,\mathit{N_2}^{4} + 8\,
\mathit{b_1}\,(\mathit{N_2}\,\mathit{N_6}\,\mathit{N_5}\,\mathit{N_R})
^{(5/2)})}{(\mathit{N_2}\,\mathit{N_6}\,\mathit{N_5}\,\mathit{N_R})^{
(5/2)}\,\sqrt{\mathit{N_R}}\,\mathit{N_2}\,\mathit{N_6}^{3}\,
\mathit{N_5}^{(5/2)}}},  \nonumber \\
{\Gamma_{\mathit{N_6}\,\mathit{N_R}\,\mathit{N_5}}}&=&
{\displaystyle \frac{1}{4}} \,{\displaystyle \frac{\pi \,( -
\mathit{N_6}^{4}\, \mathit{N_5}^{(11/2)}\,
\mathit{N_R}^{(5/2)}\,\mathit{N_2}^{4} + 10\,
\mathit{b_1}\,(\mathit{N_2}\,\mathit{N_6}\,\mathit{N_5}\,\mathit{N_R})
^{(5/2)})}{(\mathit{N_2}\,\mathit{N_6}\,\mathit{N_5}\,\mathit{N_R})^{
(5/2)}\,\sqrt{\mathit{N_R}}\,\mathit{N_2}\,\mathit{N_6}^{2}\,
\mathit{N_5}^{(7/2)}}},  \nonumber \\
{\Gamma_{\mathit{N_6}\,\mathit{N_R}\,\mathit{N_R}}}&=&
{\displaystyle \frac{1}{4}} \,{\displaystyle \frac{\pi
\,(\mathit{N_6}^{4}\, \mathit{N_5}^{(11/2)}\,
\mathit{N_R}^{(5/2)}\,\mathit{N_2}^{4} + 2\,
\mathit{b_1}\,(\mathit{N_2}\,\mathit{N_6}\,\mathit{N_5}\,\mathit{N_R})
^{(5/2)})}{(\mathit{N_2}\,\mathit{N_6}\,\mathit{N_5}\,\mathit{N_R})^{
(5/2)}\,\mathit{N_R}^{(3/2)}\,\mathit{N_2}\,\mathit{N_6}^{2}\,
\mathit{N_5}^{(5/2)}}},  \nonumber \\
{\Gamma_{\mathit{N_5}\,\mathit{N_5}\,\mathit{N_2}}}&=&  -
{\displaystyle \frac{1}{4}} \,{\displaystyle \frac{\pi \,( -
\mathit{N_6}^{4}\,\mathit{N_5}^{(11/2)}\,\mathit{N_R}^{3}\,\mathit{
N_2}^{4} + 70\,\mathit{b_1}\,\sqrt{\mathit{N_R}}\,(\mathit{N_2}\,
\mathit{N_6}\,\mathit{N_5}\,\mathit{N_R})^{(5/2)})}{(\mathit{N_2}\,
\mathit{N_6}\,\mathit{N_5}\,\mathit{N_R})^{(5/2)}\,\mathit{N_2}^{2}\,
\mathit{N_6}\,\mathit{N_5}^{(9/2)}}},  \nonumber \\
{\Gamma_{\mathit{N_5}\,\mathit{N_5}\,\mathit{N_6}}}&=&  -
{\displaystyle \frac{1}{4}} \,{\displaystyle \frac{\pi \,( -
\mathit{N_6}^{4}\,\mathit{N_5}^{(11/2)}\,\mathit{N_R}^{3}\,\mathit{
N_2}^{4} + 70\,\mathit{b_1}\,\sqrt{\mathit{N_R}}\,(\mathit{N_2}\,
\mathit{N_6}\,\mathit{N_5}\,\mathit{N_R})^{(5/2)})}{(\mathit{N_2}\,
\mathit{N_6}\,\mathit{N_5}\,\mathit{N_R})^{(5/2)}\,\mathit{N_2}\,
\mathit{N_6}^{2}\,\mathit{N_5}^{(9/2)}}}, 
\end{eqnarray}
\begin{eqnarray}
{\Gamma_{\mathit{N_5}\,\mathit{N_5}\,\mathit{N_5}}}&=&  -
{\displaystyle \frac{3}{4}} \,{\displaystyle \frac{\pi \,(
\mathit{N_6}^{4}\,\mathit{N_5}^{(11/2)}\,\mathit{N_R}^{3}\,\mathit{
N_2}^{4} + 105\,\mathit{b_1}\,\sqrt{\mathit{N_R}}\,(\mathit{N_2}\,
\mathit{N_6}\,\mathit{N_5}\,\mathit{N_R})^{(5/2)})}{(\mathit{N_2}\,
\mathit{N_6}\,\mathit{N_5}\,\mathit{N_R})^{(5/2)}\,\mathit{N_2}\,
\mathit{N_6}\,\mathit{N_5}^{(11/2)}}},  \nonumber \\
{\Gamma_{\mathit{N_5}\,\mathit{N_5}\,\mathit{N_R}}}&=&
{\displaystyle \frac{1}{4}} \,{\displaystyle \frac{\pi
\,(\mathit{N_6}^{4}\, \mathit{N_5}^{(11/2)}\,
\mathit{N_R}^{(5/2)}\,\mathit{N_2}^{4} + 35\,
\mathit{b_1}\,(\mathit{N_2}\,\mathit{N_6}\,\mathit{N_5}\,\mathit{N_R})
^{(5/2)})}{(\mathit{N_2}\,\mathit{N_6}\,\mathit{N_5}\,\mathit{N_R})^{
(5/2)}\,\sqrt{\mathit{N_R}}\,\mathit{N_2}\,\mathit{N_6}\,
\mathit{N_5}^{(9/2)}}},  \nonumber \\
{\Gamma_{\mathit{N_5}\,\mathit{N_R}\,\mathit{N_2}}}&=&
{\displaystyle \frac{1}{4}} \,{\displaystyle \frac{\pi \,( -
\mathit{N_6}^{4}\, \mathit{N_5}^{(11/2)}\,
\mathit{N_R}^{(5/2)}\,\mathit{N_2}^{4} + 10\,
\mathit{b_1}\,(\mathit{N_2}\,\mathit{N_6}\,\mathit{N_5}\,\mathit{N_R})
^{(5/2)})}{(\mathit{N_2}\,\mathit{N_6}\,\mathit{N_5}\,\mathit{N_R})^{
(5/2)}\,\sqrt{\mathit{N_R}}\,\mathit{N_2}^{2}\,\mathit{N_6}\,
\mathit{N_5}^{(7/2)}}},  \nonumber \\
{\Gamma_{\mathit{N_5}\,\mathit{N_R}\,\mathit{N_6}}}&=&
{\displaystyle \frac{1}{4}} \,{\displaystyle \frac{\pi \,( -
\mathit{N_6}^{4}\, \mathit{N_5}^{(11/2)}\,
\mathit{N_R}^{(5/2)}\,\mathit{N_2}^{4} + 10\,
\mathit{b_1}\,(\mathit{N_2}\,\mathit{N_6}\,\mathit{N_5}\,\mathit{N_R})
^{(5/2)})}{(\mathit{N_2}\,\mathit{N_6}\,\mathit{N_5}\,\mathit{N_R})^{
(5/2)}\,\sqrt{\mathit{N_R}}\,\mathit{N_2}\,\mathit{N_6}^{2}\,
\mathit{N_5}^{(7/2)}}},  \nonumber \\
{\Gamma_{\mathit{N_5}\,\mathit{N_R}\,\mathit{N_5}}}&=&
{\displaystyle \frac{1}{4}} \,{\displaystyle \frac{\pi
\,(\mathit{N_6}^{4}\, \mathit{N_5}^{(11/2)}\,
\mathit{N_R}^{(5/2)}\,\mathit{N_2}^{4} + 35\,
\mathit{b_1}\,(\mathit{N_2}\,\mathit{N_6}\,\mathit{N_5}\,\mathit{N_R})
^{(5/2)})}{(\mathit{N_2}\,\mathit{N_6}\,\mathit{N_5}\,\mathit{N_R})^{
(5/2)}\,\sqrt{\mathit{N_R}}\,\mathit{N_2}\,\mathit{N_6}\,
\mathit{N_5}^{(9/2)}}}, 
\end{eqnarray}
\begin{eqnarray}
{\Gamma_{\mathit{N_5}\,\mathit{N_R}\,\mathit{N_R}}}&=&
{\displaystyle \frac{1}{4}} \,{\displaystyle \frac{\pi
\,(\mathit{N_6}^{4}\, \mathit{N_5}^{(11/2)}\,
\mathit{N_R}^{(5/2)}\,\mathit{N_2}^{4} + 5\,
\mathit{b_1}\,(\mathit{N_2}\,\mathit{N_6}\,\mathit{N_5}\,
\mathit{N_R})^{(5/2)})}{(\mathit{N_2}\,\mathit{N_6}\,
\mathit{N_5}\,\mathit{N_R})^{(5/2)}\,\mathit{N_R}^{(3/2)}\,
\mathit{N_2}\,\mathit{N_6}\,\mathit{N_5 }^{(7/2)}}},  \nonumber \\
{\Gamma_{\mathit{N_R}\,\mathit{N_R}\,\mathit{N_2}}}&=&
{\displaystyle \frac{1}{4}} \,{\displaystyle \frac{\pi
\,(\mathit{N_6}^{4}\, \mathit{N_5}^{(11/2)}\,
\mathit{N_R}^{(5/2)}\,\mathit{N_2}^{4} + 2\,
\mathit{b_1}\,(\mathit{N_2}\,\mathit{N_6}\,\mathit{N_5}\,\mathit{N_R})
^{(5/2)})}{(\mathit{N_2}\,\mathit{N_6}\,\mathit{N_5}\,\mathit{N_R})^{
(5/2)}\,\mathit{N_R}^{(3/2)}\,\mathit{N_2}^{2}\,\mathit{N_6}\,
\mathit{N_5}^{(5/2)}}},  \nonumber \\
{\Gamma_{\mathit{N_R}\,\mathit{N_R}\,\mathit{N_6}}}&=&
{\displaystyle \frac{1}{4}} \,{\displaystyle \frac{\pi
\,(\mathit{N_6}^{4}\, \mathit{N_5}^{(11/2)}\,
\mathit{N_R}^{(5/2)}\,\mathit{N_2}^{4} + 2\,
\mathit{b_1}\,(\mathit{N_2}\,\mathit{N_6}\,\mathit{N_5}\,\mathit{N_R})
^{(5/2)})}{(\mathit{N_2}\,\mathit{N_6}\,\mathit{N_5}\,\mathit{N_R})^{
(5/2)}\,\mathit{N_R}^{(3/2)}\,\mathit{N_2}\,\mathit{N_6}^{2}\,
\mathit{N_5}^{(5/2)}}},  \nonumber \\
{\Gamma_{\mathit{N_R}\,\mathit{N_R}\,\mathit{N_5}}}&=&
{\displaystyle \frac{1}{4}} \,{\displaystyle \frac{\pi
\,(\mathit{N_6}^{4}\, \mathit{N_5}^{(11/2)}\,
\mathit{N_R}^{(5/2)}\,\mathit{N_2}^{4} + 5\,
\mathit{b_1}\,(\mathit{N_2}\,\mathit{N_6}\,\mathit{N_5}\,
\mathit{N_R})^{(5/2)})}{(\mathit{N_2}\,\mathit{N_6}\,
\mathit{N_5}\,\mathit{N_R})^{(5/2)}\,\mathit{N_R}^{(3/2)}\,
\mathit{N_2}\,\mathit{N_6}\,\mathit{N_5 }^{(7/2)}}},  \nonumber \\
{\Gamma_{\mathit{N_R}\,\mathit{N_R}\,\mathit{N_R}}}&=&
{\displaystyle \frac{3}{4}} \,{\displaystyle \frac{\pi \,( -
\mathit{N_6}^{4}\, \mathit{N_5}^{(11/2)}\,
\mathit{N_R}^{(5/2)}\,\mathit{N_2}^{4} +
\mathit{b_1}\,(\mathit{N_2}\,\mathit{N_6}\,\mathit{N_5}\,
\mathit{N_R})^{(5/2)})}{(\mathit{N_2}\,\mathit{N_6}\,
\mathit{N_5}\,\mathit{N_R})^{(5/2)}\,\mathit{N_R}^{(5/2)}\,
\mathit{N_2}\,\mathit{N_6}\,\mathit{N_5 }^{(5/2)}}}.
\end{eqnarray}

\newpage
\section{La gomtrie de Ruppenier des trous noirs extr\'emaux en rotation en quatre dimensions}

\subsection{Les trous noirs de Kerr-Newman dans la thorie d'Einstein-Maxwell:}

Avec la sym\'etrie dans les deux premiers indices, les \'el\'ements de symbole 
de Christoffel du premier type sont donn\'es par 

\begin{eqnarray}
{\Gamma_{\mathit{qqq}}}&=&12\,{\displaystyle \frac{q\,J^{2}}{(64
\,J^{2}\,\pi^{2} + q^{2})^{2}\,\sqrt{{\displaystyle \frac{64\,J
^{2}\,\pi^{2} + q^{2}}{\pi^{2}}} }}},  \nonumber \\
{\Gamma_{\mathit{qqJ}}}&=&8\,{\displaystyle \frac{J\,( - q^{2} +
32\,J^{2}\,\pi^{2})}{(64\,J^{2}\,\pi^{2} + q^{2})^{2}\,\sqrt{
{\displaystyle \frac{64\,J^{2}\,\pi^{2} + q^{2}}{\pi^{2}}} }} },  \nonumber \\
{\Gamma_{\mathit{qJq}}}&=&8\,{\displaystyle \frac{J\,( - q^{2} +
32\,J^{2}\,\pi^{2})}{(64\,J^{2}\,\pi^{2} + q^{2})^{2}\,\sqrt{
{\displaystyle \frac{64\,J^{2}\,\pi^{2} + q^{2}}{\pi^{2}}} }} },  \nonumber \\
{\Gamma_{\mathit{qJJ}}}&=& - 4\,{\displaystyle \frac
{q\,(128\,J^{ 2}\,\pi^{2} - q^{2})}{(64\,J^{2}\,\pi^{2} +
q^{2})^{2}\,\sqrt{ {\displaystyle \frac{64\,J^{2}\,\pi^{2} +
q^{2}}{\pi^{2}}} }} },  \nonumber \\
{\Gamma_{\mathit{JJq}}}&=& - 4\,{\displaystyle \frac
{q\,(128\,J^{ 2}\,\pi^{2} - q^{2})}{(64\,J^{2}\,\pi^{2} +
q^{2})^{2}\,\sqrt{ {\displaystyle \frac{64\,J^{2}\,\pi^{2} +
q^{2}}{\pi^{2}}} }} },  \nonumber \\
{\Gamma_{\mathit{JJJ}}}&=&768\,{\displaystyle \frac
{J\,q^{2}\,\pi^{2}}{(64\,J^{2}\,\pi^{2} +
q^{2})^{2}\,\sqrt{{\displaystyle \frac{64\,J^{2}\,\pi^{2} +
q^{2}}{\pi^{2}}} }}}.
\end{eqnarray}

\newpage
\subsection{Les trous noirs de Kaluza-Klein dans la thorie d'Einstein-Maxwell:}

Avec la sym\'etrie dans les deux premiers indices, nous voyons que les 
\'el\'ements de symbole de Christoffel du premier type sont donn'es par

\begin{eqnarray}
{\Gamma_{P\,P\,P}}&=& - 3\,{\displaystyle \frac{\pi
\,P\,Q^{4}\,J ^{2}}{(P^{2}\,Q^{2} - J^{2})^{(5/2)}}},
\nonumber \\
{\Gamma_{P\,P\,Q}}&=& - {\displaystyle \frac{\pi
\,Q\,J^{2}\,(P^{ 2}\,Q^{2} + 2\,J^{2})}{(P^{2}\,Q^{2} -
J^{2})^{(5/2)}}},  \nonumber \\
{\Gamma_{P\,P\,J}}&=&{\displaystyle \frac{\pi
\,Q^{2}\,J\,(2\,P^{ 2}\,Q^{2} + J^{2})}{(P^{2}\,Q^{2} -
J^{2})^{(5/2)}}},  \nonumber \\
{\Gamma_{P\,Q\,P}}&=& - {\displaystyle \frac{\pi
\,Q\,J^{2}\,(P^{ 2}\,Q^{2} + 2\,J^{2})}{(P^{2}\,Q^{2} -
J^{2})^{(5/2)}}},  \nonumber \\
{\Gamma_{P\,Q\,Q}}&=& - {\displaystyle \frac{\pi
\,P\,J^{2}\,(P^{ 2}\,Q^{2} + 2\,J^{2})}{(P^{2}\,Q^{2} -
J^{2})^{(5/2)}}},  \nonumber \\
{\Gamma_{P\,Q\,J}}&=&{\displaystyle \frac{\pi
\,P\,Q\,J\,(P^{2}\, Q^{2} + 2\,J^{2})}{(P^{2}\,Q^{2} -
J^{2})^{(5/2)}}}, 
\end{eqnarray}
\begin{eqnarray}
{\Gamma_{P\,J\,P}}&=&{\displaystyle \frac{\pi
\,Q^{2}\,J\,(2\,P^{ 2}\,Q^{2} + J^{2})}{(P^{2}\,Q^{2} -
J^{2})^{(5/2)}}},  \nonumber \\
{\Gamma_{P\,J\,Q}}&=&{\displaystyle \frac{\pi
\,P\,Q\,J\,(P^{2}\, Q^{2} + 2\,J^{2})}{(P^{2}\,Q^{2} -
J^{2})^{(5/2)}}},  \nonumber \\
{\Gamma_{P\,J\,J}}&=& - {\displaystyle \frac{\pi
\,P\,Q^{2}\,(P^{ 2}\,Q^{2} + 2\,J^{2})}{(P^{2}\,Q^{2} -
J^{2})^{(5/2)}}},  \nonumber \\
{\Gamma_{Q\,Q\,P}}&=& - {\displaystyle \frac{\pi
\,P\,J^{2}\,(P^{ 2}\,Q^{2} + 2\,J^{2})}{(P^{2}\,Q^{2} -
J^{2})^{(5/2)}}},  \nonumber \\
{\Gamma_{Q\,Q\,Q}}&=& - 3\,{\displaystyle \frac{\pi
\,P^{4}\,Q\,J ^{2}}{(P^{2}\,Q^{2} - J^{2})^{(5/2)}}},  \nonumber \\
{\Gamma_{Q\,Q\,J}}&=&{\displaystyle \frac{\pi
\,P^{2}\,J\,(2\,P^{ 2}\,Q^{2} + J^{2})}{(P^{2}\,Q^{2} -
J^{2})^{(5/2)}}}, 
\end{eqnarray}
\begin{eqnarray}
{\Gamma_{Q\,J\,P}}&=&{\displaystyle \frac{\pi
\,P\,Q\,J\,(P^{2}\, Q^{2} + 2\,J^{2})}{(P^{2}\,Q^{2} -
J^{2})^{(5/2)}}},  \nonumber \\
{\Gamma_{Q\,J\,Q}}&=&{\displaystyle \frac{\pi
\,P^{2}\,J\,(2\,P^{ 2}\,Q^{2} + J^{2})}{(P^{2}\,Q^{2} -
J^{2})^{(5/2)}}},  \nonumber \\
{\Gamma_{Q\,J\,J}}&=& - {\displaystyle \frac{\pi
\,P^{2}\,Q\,(P^{ 2}\,Q^{2} + 2\,J^{2})}{(P^{2}\,Q^{2} -
J^{2})^{(5/2)}}},  \nonumber \\
{\Gamma_{J\,J\,P}}&=& - {\displaystyle \frac{\pi
\,P\,Q^{2}\,(P^{ 2}\,Q^{2} + 2\,J^{2})}{(P^{2}\,Q^{2} -
J^{2})^{(5/2)}}},  \nonumber \\
{\Gamma_{J\,J\,Q}}&=& - {\displaystyle \frac{\pi
\,P^{2}\,Q\,(P^{ 2}\,Q^{2} + 2\,J^{2})}{(P^{2}\,Q^{2} -
J^{2})^{(5/2)}}},  \nonumber \\
{\Gamma_{J\,J\,J}}&=&3\,{\displaystyle \frac{\pi
\,J\,P^{2}\,Q^{2 }}{(P^{2}\,Q^{2} - J^{2})^{(5/2)}}}.
\end{eqnarray}

\newpage
\subsection{Les trous noirs de la thorie des cordes h\'et\'erotiques compactifie toroidalement:}

Avec la sym\'etrie dans les deux premiers indices, 
les \'el\'ements de symbole de Christoffel du premier type 
sont donn\'es comme les suivants

\begin{eqnarray}
{\Gamma_{\mathit{P1}\,\mathit{P_1}\,\mathit{P_1}}}&=& -
{\displaystyle \frac{3}{8}} \,{\displaystyle \frac{\pi \,
\mathit{Q_2}^{3}\,\mathit{P_3}^{3}\,\mathit{Q_4}^{3}}{(J^{2} +
\mathit{P_1}\,\mathit{Q_2}\,\mathit{P_3}\,\mathit{Q_4})^{(5/2)}}},  \nonumber \\
{\Gamma_{\mathit{P_1}\,\mathit{P_1}\,\mathit{Q_2}}}&=&{\displaystyle
\frac{1}{8}} \,{\displaystyle \frac{\pi \,\mathit{Q_2}\,\mathit{
P_3}^{2}\,\mathit{Q_4}^{2}\,(\mathit{P_1}\,\mathit{Q_2}\,\mathit{P_3}
\,\mathit{Q_4} + 4\,J^{2})}{(J^{2} + \mathit{P_1}\,\mathit{Q_2}\,
\mathit{P_3}\,\mathit{Q_4})^{(5/2)}}},  \nonumber \\
{\Gamma_{\mathit{P_1}\,\mathit{P_1}\,\mathit{P_3}}}&=&{\displaystyle
\frac{1}{8}} \,{\displaystyle \frac{\pi \,\mathit{Q_2}^{2}\,
\mathit{P_3}\,\mathit{Q_4}^{2}\,(\mathit{P_1}\,\mathit{Q_2}\,\mathit{
P_3}\,\mathit{Q_4} + 4\,J^{2})}{(J^{2} +
\mathit{P_1}\,\mathit{Q_2}\,
\mathit{P_3}\,\mathit{Q_4})^{(5/2)}}},  \nonumber \\
{\Gamma_{\mathit{P_1}\,\mathit{P_1}\,\mathit{Q_4}}}&=&{\displaystyle
\frac{1}{8}} \,{\displaystyle \frac{\pi \,\mathit{Q_2}^{2}\,
\mathit{P_3}^{2}\,\mathit{Q_4}\,(\mathit{P_1}\,\mathit{Q_2}\,\mathit{
P_3}\,\mathit{Q_4} + 4\,J^{2})}{(J^{2} +
\mathit{P_1}\,\mathit{Q_2}\,
\mathit{P_3}\,\mathit{Q_4})^{(5/2)}}},  \nonumber \\
{\Gamma_{\mathit{P_1}\,\mathit{P_1}\,J}}&=& - {\displaystyle
\frac{ 3}{4}} \,{\displaystyle \frac{\pi
\,\mathit{Q_2}^{2}\,\mathit{P_3} ^{2}\,\mathit{Q_4}^{2}\,J}{(J^{2}
+ \mathit{P_1}\,\mathit{Q_2}\,
\mathit{P_3}\,\mathit{Q_4})^{(5/2)}}}, 
\end{eqnarray}
\begin{eqnarray}
{\Gamma_{\mathit{P_1}\,\mathit{Q_2}\,\mathit{P_1}}}&=&{\displaystyle
\frac{1}{8}} \,{\displaystyle \frac{\pi \,\mathit{Q_2}\,\mathit{
P_3}^{2}\,\mathit{Q_4}^{2}\,(\mathit{P_1}\,\mathit{Q_2}\,\mathit{P_3}
\,\mathit{Q_4} + 4\,J^{2})}{(J^{2} + \mathit{P_1}\,\mathit{Q_2}\,
\mathit{P_3}\,\mathit{Q_4})^{(5/2)}}},  \nonumber \\
{\Gamma_{\mathit{P_1}\,\mathit{Q_2}\,\mathit{Q_2}}}&=&{\displaystyle
\frac{1}{8}} \,{\displaystyle \frac{\pi \,\mathit{P_3}^{2}\,
\mathit{Q_4}^{2}\,\mathit{P_1}\,(\mathit{P_1}\,\mathit{Q_2}\,\mathit{
P_3}\,\mathit{Q_4} + 4\,J^{2})}{(J^{2} +
\mathit{P_1}\,\mathit{Q_2}\,
\mathit{P_3}\,\mathit{Q_4})^{(5/2)}}},  \nonumber \\
{\Gamma_{\mathit{P_1}\,\mathit{Q_2}\,\mathit{P_3}}}&=& -
{\displaystyle \frac{1}{8}} \,{\displaystyle \frac{\pi \,
\mathit{Q_4}\,(\mathit{P_1}^{2}\,\mathit{Q_2}^{2}\,\mathit{P_3}^{2}\,
\mathit{Q_4}^{2} +
2\,J^{2}\,\mathit{P_1}\,\mathit{Q_2}\,\mathit{P_3} \,\mathit{Q_4}
+ 4\,J^{4})}{(J^{2} + \mathit{P_1}\,\mathit{Q_2}\,
\mathit{P_3}\,\mathit{Q_4})^{(5/2)}}},  \nonumber \\
{\Gamma_{\mathit{P_1}\,\mathit{Q_2}\,\mathit{Q_4}}}&=& -
{\displaystyle \frac{1}{8}} \,{\displaystyle \frac{\pi \,
\mathit{P_3}\,(\mathit{P_1}^{2}\,\mathit{Q_2}^{2}\,\mathit{P_3}^{2}\,
\mathit{Q_4}^{2} +
2\,J^{2}\,\mathit{P_1}\,\mathit{Q_2}\,\mathit{P_3} \,\mathit{Q_4}
+ 4\,J^{4})}{(J^{2} + \mathit{P_1}\,\mathit{Q_2}\,
\mathit{P_3}\,\mathit{Q_4})^{(5/2)}}},  \nonumber \\
{\Gamma_{\mathit{P_1}\,\mathit{Q_2}\,J}}&=& - {\displaystyle
\frac{ 1}{4}} \,{\displaystyle \frac{\pi
\,\mathit{P_3}\,\mathit{Q_4}\,J
\,(\mathit{P_1}\,\mathit{Q_2}\,\mathit{P_3}\,\mathit{Q_4} -
2\,J^{2}) }{(J^{2} +
\mathit{P_1}\,\mathit{Q_2}\,\mathit{P_3}\,\mathit{Q_4})^{(
5/2)}}}, 
\end{eqnarray}
\begin{eqnarray}
{\Gamma_{\mathit{P_1}\,\mathit{P_3}\,\mathit{P_1}}}&=&{\displaystyle
\frac{1}{8}} \,{\displaystyle \frac{\pi \,\mathit{Q_2}^{2}\,
\mathit{P_3}\,\mathit{Q_4}^{2}\,(\mathit{P_1}\,\mathit{Q_2}\,\mathit{
P_3}\,\mathit{Q_4} + 4\,J^{2})}{(J^{2} +
\mathit{P_1}\,\mathit{Q_2}\,
\mathit{P_3}\,\mathit{Q_4})^{(5/2)}}},  \nonumber \\
{\Gamma_{\mathit{P_1}\,\mathit{P_3}\,\mathit{Q_2}}}&=& -
{\displaystyle \frac{1}{8}} \,{\displaystyle \frac{\pi \,
\mathit{Q_4}\,(\mathit{P_1}^{2}\,\mathit{Q_2}^{2}\,\mathit{P_3}^{2}\,
\mathit{Q_4}^{2} +
2\,J^{2}\,\mathit{P_1}\,\mathit{Q_2}\,\mathit{P_3} \,\mathit{Q_4}
+ 4\,J^{4})}{(J^{2} + \mathit{P_1}\,\mathit{Q_2}\,
\mathit{P_3}\,\mathit{Q_4})^{(5/2)}}},  \nonumber \\
{\Gamma
_{\mathit{P_1}\,\mathit{P_3}\,\mathit{P_3}}}&=&{\displaystyle
\frac{1}{8}} \,{\displaystyle \frac{\pi \,\mathit{Q_2}^{2}\,
\mathit{Q_4}^{2}\,\mathit{P_1}\,(\mathit{P_1}\,\mathit{Q_2}\,\mathit{
P_3}\,\mathit{Q_4} + 4\,J^{2})}{(J^{2} +
\mathit{P_1}\,\mathit{Q_2}\,
\mathit{P_3}\,\mathit{Q_4})^{(5/2)}}},  \nonumber \\
{\Gamma_{\mathit{P_1}\,\mathit{P_3}\,\mathit{Q_4}}}&=& -
{\displaystyle \frac{1}{8}} \,{\displaystyle \frac{\mathit{Q_2}
\,\pi \,(\mathit{P_1}^{2}\,\mathit{Q_2}^{2}\,\mathit{P_3}^{2}\,
\mathit{Q_4}^{2} +
2\,J^{2}\,\mathit{P_1}\,\mathit{Q_2}\,\mathit{P_3} \,\mathit{Q_4}
+ 4\,J^{4})}{(J^{2} + \mathit{P_1}\,\mathit{Q_2}\,
\mathit{P_3}\,\mathit{Q_4})^{(5/2)}}},  \nonumber \\
{\Gamma_{\mathit{P_1}\,\mathit{P_3}\,J}}&=& - {\displaystyle
\frac{ 1}{4}} \,{\displaystyle \frac{\pi
\,\mathit{Q_2}\,\mathit{Q_4}\,J
\,(\mathit{P_1}\,\mathit{Q_2}\,\mathit{P_3}\,\mathit{Q_4} -
2\,J^{2}) }{(J^{2} +
\mathit{P_1}\,\mathit{Q_2}\,\mathit{P_3}\,\mathit{Q_4})^{(
5/2)}}}, 
\end{eqnarray}
\begin{eqnarray}
{\Gamma_{\mathit{P_1}\,\mathit{Q_4}\,\mathit{P_1}}}&=&{\displaystyle
\frac{1}{8}} \,{\displaystyle \frac{\pi \,\mathit{Q_2}^{2}\,
\mathit{P_3}^{2}\,\mathit{Q_4}\,(\mathit{P_1}\,\mathit{Q_2}\,\mathit{
P_3}\,\mathit{Q_4} + 4\,J^{2})}{(J^{2} +
\mathit{P_1}\,\mathit{Q_2}\,
\mathit{P_3}\,\mathit{Q_4})^{(5/2)}}},  \nonumber \\
{\Gamma_{\mathit{P_1}\,\mathit{Q_4}\,\mathit{Q_2}}}&=& -
{\displaystyle \frac{1}{8}} \,{\displaystyle \frac{\pi \,
\mathit{P_3}\,(\mathit{P_1}^{2}\,\mathit{Q_2}^{2}\,\mathit{P_3}^{2}\,
\mathit{Q_4}^{2} +
2\,J^{2}\,\mathit{P_1}\,\mathit{Q_2}\,\mathit{P_3} \,\mathit{Q_4}
+ 4\,J^{4})}{(J^{2} + \mathit{P_1}\,\mathit{Q_2}\,
\mathit{P_3}\,\mathit{Q_4})^{(5/2)}}},  \nonumber \\
{\Gamma_{\mathit{P_1}\,\mathit{Q_4}\,\mathit{P_3}}}&=& -
{\displaystyle \frac{1}{8}} \,{\displaystyle \frac{\mathit{Q_2}
\,\pi \,(\mathit{P_1}^{2}\,\mathit{Q_2}^{2}\,\mathit{P_3}^{2}\,
\mathit{Q_4}^{2} +
2\,J^{2}\,\mathit{P_1}\,\mathit{Q_2}\,\mathit{P_3} \,\mathit{Q_4}
+ 4\,J^{4})}{(J^{2} + \mathit{P_1}\,\mathit{Q_2}\,
\mathit{P_3}\,\mathit{Q_4})^{(5/2)}}},  \nonumber \\
{\Gamma
_{\mathit{P_1}\,\mathit{Q_4}\,\mathit{Q_4}}}&=&{\displaystyle
\frac{1}{8}} \,{\displaystyle \frac{\pi \,\mathit{Q_2}^{2}\,
\mathit{P_3}^{2}\,\mathit{P_1}\,(\mathit{P_1}\,\mathit{Q_2}\,\mathit{
P_3}\,\mathit{Q_4} + 4\,J^{2})}{(J^{2} +
\mathit{P_1}\,\mathit{Q_2}\,
\mathit{P_3}\,\mathit{Q_4})^{(5/2)}}},  \nonumber \\
{\Gamma_{\mathit{P_1}\,\mathit{Q_4}\,J}}&=& - {\displaystyle
\frac{ 1}{4}} \,{\displaystyle \frac{\pi
\,\mathit{Q_2}\,\mathit{P_3}\,J
\,(\mathit{P_1}\,\mathit{Q_2}\,\mathit{P_3}\,\mathit{Q_4} -
2\,J^{2}) }{(J^{2} +
\mathit{P_1}\,\mathit{Q_2}\,\mathit{P_3}\,\mathit{Q_4})^{(
5/2)}}}, 
\end{eqnarray}
\begin{eqnarray}
{\Gamma_{\mathit{P_1}\,J\,\mathit{P_1}}}&=& - {\displaystyle
\frac{ 3}{4}} \,{\displaystyle \frac{\pi
\,\mathit{Q_2}^{2}\,\mathit{P_3} ^{2}\,\mathit{Q_4}^{2}\,J}{(J^{2}
+ \mathit{P_1}\,\mathit{Q_2}\,
\mathit{P_3}\,\mathit{Q_4})^{(5/2)}}},  \nonumber \\
{\Gamma_{\mathit{P_1}\,J\,\mathit{Q_2}}}&=& - {\displaystyle
\frac{ 1}{4}} \,{\displaystyle \frac{\pi
\,\mathit{P_3}\,\mathit{Q_4}\,J
\,(\mathit{P_1}\,\mathit{Q_2}\,\mathit{P_3}\,\mathit{Q_4} -
2\,J^{2}) }{(J^{2} +
\mathit{P_1}\,\mathit{Q_2}\,\mathit{P_3}\,\mathit{Q_4})^{(
5/2)}}},  \nonumber \\
{\Gamma_{\mathit{P_1}\,J\,\mathit{P_3}}}&=& - {\displaystyle
\frac{ 1}{4}} \,{\displaystyle \frac{\pi
\,\mathit{Q_2}\,\mathit{Q_4}\,J
\,(\mathit{P_1}\,\mathit{Q_2}\,\mathit{P_3}\,\mathit{Q_4} -
2\,J^{2}) }{(J^{2} +
\mathit{P_1}\,\mathit{Q_2}\,\mathit{P_3}\,\mathit{Q_4})^{(
5/2)}}},  \nonumber \\
{\Gamma_{\mathit{P_1}\,J\,\mathit{Q_4}}}&=& - {\displaystyle
\frac{ 1}{4}} \,{\displaystyle \frac{\pi
\,\mathit{Q_2}\,\mathit{P_3}\,J
\,(\mathit{P_1}\,\mathit{Q_2}\,\mathit{P_3}\,\mathit{Q_4} -
2\,J^{2}) }{(J^{2} +
\mathit{P_1}\,\mathit{Q_2}\,\mathit{P_3}\,\mathit{Q_4})^{(
5/2)}}},  \nonumber \\
{\Gamma_{\mathit{P_1}\,J\,J}}&=&{\displaystyle \frac{1}{2}} \,
{\displaystyle \frac{\mathit{Q_4}\,\pi
\,\mathit{Q_2}\,\mathit{P_3}
\,(\mathit{P_1}\,\mathit{Q_2}\,\mathit{P_3}\,\mathit{Q_4} -
2\,J^{2}) }{(J^{2} +
\mathit{P_1}\,\mathit{Q_2}\,\mathit{P_3}\,\mathit{Q_4})^{(
5/2)}}}, 
\end{eqnarray}
\begin{eqnarray}
{\Gamma_{\mathit{Q_2}\,\mathit{Q_2}\,\mathit{P_1}}}&=&{\displaystyle
\frac{1}{8}} \,{\displaystyle \frac{\pi \,\mathit{P_3}^{2}\,
\mathit{Q_4}^{2}\,\mathit{P_1}\,(\mathit{P_1}\,\mathit{Q_2}\,\mathit{
P_3}\,\mathit{Q_4} + 4\,J^{2})}{(J^{2} +
\mathit{P_1}\,\mathit{Q_2}\,
\mathit{P_3}\,\mathit{Q_4})^{(5/2)}}},  \nonumber \\
{\Gamma_{\mathit{Q_2}\,\mathit{Q_2}\,\mathit{Q_2}}}&=& -
{\displaystyle \frac{3}{8}} \,{\displaystyle \frac{\pi \,
\mathit{P_1}^{3}\,\mathit{P_3}^{3}\,\mathit{Q_4}^{3}}{(J^{2} +
\mathit{P_1}\,\mathit{Q_2}\,\mathit{P_3}\,\mathit{Q_4})^{(5/2)}}},  \nonumber \\
{\Gamma
_{\mathit{Q_2}\,\mathit{Q_2}\,\mathit{P_3}}}&=&{\displaystyle
\frac{1}{8}} \,{\displaystyle \frac{\pi \,\mathit{P_1}^{2}\,
\mathit{P_3}\,\mathit{Q_4}^{2}\,(\mathit{P_1}\,\mathit{Q_2}\,\mathit{
P_3}\,\mathit{Q_4} + 4\,J^{2})}{(J^{2} +
\mathit{P_1}\,\mathit{Q_2}\,
\mathit{P_3}\,\mathit{Q_4})^{(5/2)}}},  \nonumber \\
{\Gamma
_{\mathit{Q_2}\,\mathit{Q_2}\,\mathit{Q_4}}}&=&{\displaystyle
\frac{1}{8}} \,{\displaystyle \frac{\pi \,\mathit{P_1}^{2}\,
\mathit{P_3}^{2}\,\mathit{Q_4}\,(\mathit{P_1}\,\mathit{Q_2}\,\mathit{
P_3}\,\mathit{Q_4} + 4\,J^{2})}{(J^{2} +
\mathit{P_1}\,\mathit{Q_2}\,
\mathit{P_3}\,\mathit{Q_4})^{(5/2)}}},  \nonumber \\
{\Gamma_{\mathit{Q_2}\,\mathit{Q_2}\,J}}&=& - {\displaystyle
\frac{ 3}{4}} \,{\displaystyle \frac{\pi
\,\mathit{P_1}^{2}\,\mathit{P_3} ^{2}\,\mathit{Q_4}^{2}\,J}{(J^{2}
+ \mathit{P_1}\,\mathit{Q_2}\,
\mathit{P_3}\,\mathit{Q_4})^{(5/2)}}}, 
\end{eqnarray}
\begin{eqnarray}
{\Gamma_{\mathit{Q_2}\,\mathit{P_3}\,\mathit{P_1}}}&=& -
{\displaystyle \frac{1}{8}} \,{\displaystyle \frac{\pi \,
\mathit{Q_4}\,(\mathit{P_1}^{2}\,\mathit{Q_2}^{2}\,\mathit{P_3}^{2}\,
\mathit{Q_4}^{2} +
2\,J^{2}\,\mathit{P_1}\,\mathit{Q_2}\,\mathit{P_3} \,\mathit{Q_4}
+ 4\,J^{4})}{(J^{2} + \mathit{P_1}\,\mathit{Q_2}\,
\mathit{P_3}\,\mathit{Q_4})^{(5/2)}}},  \nonumber \\
{\Gamma
_{\mathit{Q_2}\,\mathit{P_3}\,\mathit{Q_2}}}&=&{\displaystyle
\frac{1}{8}} \,{\displaystyle \frac{\pi \,\mathit{P_1}^{2}\,
\mathit{P_3}\,\mathit{Q_4}^{2}\,(\mathit{P_1}\,\mathit{Q_2}\,\mathit{
P_3}\,\mathit{Q_4} + 4\,J^{2})}{(J^{2} +
\mathit{P_1}\,\mathit{Q_2}\,
\mathit{P_3}\,\mathit{Q_4})^{(5/2)}}},  \nonumber \\
{\Gamma
_{\mathit{Q_2}\,\mathit{P_3}\,\mathit{P_3}}}&=&{\displaystyle
\frac{1}{8}} \,{\displaystyle \frac{\pi \,\mathit{P_1}^{2}\,
\mathit{Q_2}\,\mathit{Q_4}^{2}\,(\mathit{P_1}\,\mathit{Q_2}\,\mathit{
P_3}\,\mathit{Q_4} + 4\,J^{2})}{(J^{2} +
\mathit{P_1}\,\mathit{Q_2}\,
\mathit{P_3}\,\mathit{Q_4})^{(5/2)}}},  \nonumber \\
{\Gamma_{\mathit{Q_2}\,\mathit{P_3}\,\mathit{Q_4}}}&=& -
{\displaystyle \frac{1}{8}} \,{\displaystyle \frac{\pi \,
\mathit{P_1}\,(\mathit{P_1}^{2}\,\mathit{Q_2}^{2}\,\mathit{P_3}^{2}\,
\mathit{Q_4}^{2} +
2\,J^{2}\,\mathit{P_1}\,\mathit{Q_2}\,\mathit{P_3} \,\mathit{Q_4}
+ 4\,J^{4})}{(J^{2} + \mathit{P_1}\,\mathit{Q_2}\,
\mathit{P_3}\,\mathit{Q_4})^{(5/2)}}},  \nonumber \\
{\Gamma_{\mathit{Q_2}\,\mathit{P_3}\,J}}&=& - {\displaystyle
\frac{ 1}{4}} \,{\displaystyle \frac{\pi
\,\mathit{P_1}\,\mathit{Q_4}\,J
\,(\mathit{P_1}\,\mathit{Q_2}\,\mathit{P_3}\,\mathit{Q_4} -
2\,J^{2}) }{(J^{2} +
\mathit{P_1}\,\mathit{Q_2}\,\mathit{P_3}\,\mathit{Q_4})^{(
5/2)}}}, 
\end{eqnarray}
\begin{eqnarray}
{\Gamma_{\mathit{Q_2}\,\mathit{Q_4}\,\mathit{P_1}}}&=& -
{\displaystyle \frac{1}{8}} \,{\displaystyle \frac{\pi \,
\mathit{P_3}\,(\mathit{P_1}^{2}\,\mathit{Q_2}^{2}\,\mathit{P_3}^{2}\,
\mathit{Q_4}^{2} +
2\,J^{2}\,\mathit{P_1}\,\mathit{Q_2}\,\mathit{P_3} \,\mathit{Q_4}
+ 4\,J^{4})}{(J^{2} + \mathit{P_1}\,\mathit{Q_2}\,
\mathit{P_3}\,\mathit{Q_4})^{(5/2)}}},  \nonumber \\
{\Gamma
_{\mathit{Q_2}\,\mathit{Q_4}\,\mathit{Q_2}}}&=&{\displaystyle
\frac{1}{8}} \,{\displaystyle \frac{\pi \,\mathit{P_1}^{2}\,
\mathit{P_3}^{2}\,\mathit{Q_4}\,(\mathit{P_1}\,\mathit{Q_2}\,\mathit{
P_3}\,\mathit{Q_4} + 4\,J^{2})}{(J^{2} +
\mathit{P_1}\,\mathit{Q_2}\,
\mathit{P_3}\,\mathit{Q_4})^{(5/2)}}},  \nonumber \\
{\Gamma_{\mathit{Q_2}\,\mathit{Q_4}\,\mathit{P_3}}}&=& -
{\displaystyle \frac{1}{8}} \,{\displaystyle \frac{\pi \,
\mathit{P_1}\,(\mathit{P_1}^{2}\,\mathit{Q_2}^{2}\,\mathit{P_3}^{2}\,
\mathit{Q_4}^{2} +
2\,J^{2}\,\mathit{P_1}\,\mathit{Q_2}\,\mathit{P_3} \,\mathit{Q_4}
+ 4\,J^{4})}{(J^{2} + \mathit{P_1}\,\mathit{Q_2}\,
\mathit{P_3}\,\mathit{Q_4})^{(5/2)}}},  \nonumber \\
{\Gamma
_{\mathit{Q_2}\,\mathit{Q_4}\,\mathit{Q_4}}}&=&{\displaystyle
\frac{1}{8}} \,{\displaystyle \frac{\pi \,\mathit{P_1}^{2}\,
\mathit{P_3}^{2}\,\mathit{Q_2}\,(\mathit{P_1}\,\mathit{Q_2}\,\mathit{
P_3}\,\mathit{Q_4} + 4\,J^{2})}{(J^{2} +
\mathit{P_1}\,\mathit{Q_2}\,
\mathit{P_3}\,\mathit{Q_4})^{(5/2)}}},  \nonumber \\
{\Gamma_{\mathit{Q_2}\,\mathit{Q_4}\,J}}&=& - {\displaystyle
\frac{ 1}{4}} \,{\displaystyle \frac{\pi
\,\mathit{P_1}\,\mathit{P_3}\,J
\,(\mathit{P_1}\,\mathit{Q_2}\,\mathit{P_3}\,\mathit{Q_4} -
2\,J^{2}) }{(J^{2} +
\mathit{P_1}\,\mathit{Q_2}\,\mathit{P_3}\,\mathit{Q_4})^{(
5/2)}}}, 
\end{eqnarray}
\begin{eqnarray}
{\Gamma_{\mathit{Q_2}\,J\,\mathit{P_1}}}&=& - {\displaystyle
\frac{ 1}{4}} \,{\displaystyle \frac{\pi
\,\mathit{P_3}\,\mathit{Q_4}\,J
\,(\mathit{P_1}\,\mathit{Q_2}\,\mathit{P_3}\,\mathit{Q_4} -
2\,J^{2}) }{(J^{2} +
\mathit{P_1}\,\mathit{Q_2}\,\mathit{P_3}\,\mathit{Q_4})^{(
5/2)}}},  \nonumber \\
{\Gamma_{\mathit{Q_2}\,J\,\mathit{Q_2}}}&=& - {\displaystyle
\frac{ 3}{4}} \,{\displaystyle \frac{\pi
\,\mathit{P_1}^{2}\,\mathit{P_3} ^{2}\,\mathit{Q_4}^{2}\,J}{(J^{2}
+ \mathit{P_1}\,\mathit{Q_2}\,
\mathit{P_3}\,\mathit{Q_4})^{(5/2)}}},  \nonumber \\
{\Gamma_{\mathit{Q_2}\,J\,\mathit{P_3}}}&=& - {\displaystyle
\frac{ 1}{4}} \,{\displaystyle \frac{\pi
\,\mathit{P_1}\,\mathit{Q_4}\,J
\,(\mathit{P_1}\,\mathit{Q_2}\,\mathit{P_3}\,\mathit{Q_4} -
2\,J^{2}) }{(J^{2} +
\mathit{P_1}\,\mathit{Q_2}\,\mathit{P_3}\,\mathit{Q_4})^{(
5/2)}}},  \nonumber \\
{\Gamma_{\mathit{Q_2}\,J\,\mathit{Q_4}}}&=& - {\displaystyle
\frac{ 1}{4}} \,{\displaystyle \frac{\pi
\,\mathit{P_1}\,\mathit{P_3}\,J
\,(\mathit{P_1}\,\mathit{Q_2}\,\mathit{P_3}\,\mathit{Q_4} -
2\,J^{2}) }{(J^{2} +
\mathit{P_1}\,\mathit{Q_2}\,\mathit{P_3}\,\mathit{Q_4})^{(
5/2)}}},  \nonumber \\
{\Gamma_{\mathit{Q_2}\,J\,J}}&=&{\displaystyle \frac{1}{2}} \,
{\displaystyle \frac{\pi
\,\mathit{P_1}\,\mathit{P_3}\,\mathit{Q_4}
\,(\mathit{P_1}\,\mathit{Q_2}\,\mathit{P_3}\,\mathit{Q_4} -
2\,J^{2}) }{(J^{2} +
\mathit{P_1}\,\mathit{Q_2}\,\mathit{P_3}\,\mathit{Q_4})^{(
5/2)}}}, 
\end{eqnarray}
\begin{eqnarray}
{\Gamma_{\mathit{P_3}\,\mathit{P_3}\,\mathit{P_1}}}&=&{\displaystyle
\frac{1}{8}} \,{\displaystyle \frac{\pi \,\mathit{Q_2}^{2}\,
\mathit{Q_4}^{2}\,\mathit{P_1}\,(\mathit{P_1}\,\mathit{Q_2}\,\mathit{
P_3}\,\mathit{Q_4} + 4\,J^{2})}{(J^{2} +
\mathit{P_1}\,\mathit{Q_2}\,
\mathit{P_3}\,\mathit{Q_4})^{(5/2)}}},  \nonumber \\
{\Gamma
_{\mathit{P_3}\,\mathit{P_3}\,\mathit{Q_2}}}&=&{\displaystyle
\frac{1}{8}} \,{\displaystyle \frac{\pi \,\mathit{P_1}^{2}\,
\mathit{Q_2}\,\mathit{Q_4}^{2}\,(\mathit{P_1}\,\mathit{Q_2}\,\mathit{
P_3}\,\mathit{Q_4} + 4\,J^{2})}{(J^{2} +
\mathit{P_1}\,\mathit{Q_2}\,
\mathit{P_3}\,\mathit{Q_4})^{(5/2)}}},  \nonumber \\
{\Gamma_{\mathit{P_3}\,\mathit{P_3}\,\mathit{P_3}}}&=& -
{\displaystyle \frac{3}{8}} \,{\displaystyle \frac{\pi \,
\mathit{P_1}^{3}\,\mathit{Q_2}^{3}\,\mathit{Q_4}^{3}}{(J^{2} +
\mathit{P_1}\,\mathit{Q_2}\,\mathit{P_3}\,\mathit{Q_4})^{(5/2)}}},  \nonumber \\
{\Gamma
_{\mathit{P_3}\,\mathit{P_3}\,\mathit{Q_4}}}&=&{\displaystyle
\frac{1}{8}} \,{\displaystyle \frac{\pi \,\mathit{P_1}^{2}\,
\mathit{Q_2}^{2}\,\mathit{Q_4}\,(\mathit{P_1}\,\mathit{Q_2}\,\mathit{
P_3}\,\mathit{Q_4} + 4\,J^{2})}{(J^{2} +
\mathit{P_1}\,\mathit{Q_2}\,
\mathit{P_3}\,\mathit{Q_4})^{(5/2)}}},  \nonumber \\
{\Gamma_{\mathit{P_3}\,\mathit{P_3}\,J}}&=& - {\displaystyle
\frac{ 3}{4}} \,{\displaystyle \frac{\pi
\,\mathit{P_1}^{2}\,\mathit{Q_2} ^{2}\,\mathit{Q_4}^{2}\,J}{(J^{2}
+ \mathit{P_1}\,\mathit{Q_2}\,
\mathit{P_3}\,\mathit{Q_4})^{(5/2)}}}, 
\end{eqnarray}
\begin{eqnarray}
{\Gamma_{\mathit{P_3}\,\mathit{Q_4}\,\mathit{P_1}}}&=& -
{\displaystyle \frac{1}{8}} \,{\displaystyle \frac{\mathit{Q_2}
\,\pi \,(\mathit{P_1}^{2}\,\mathit{Q_2}^{2}\,\mathit{P_3}^{2}\,
\mathit{Q_4}^{2} +
2\,J^{2}\,\mathit{P_1}\,\mathit{Q_2}\,\mathit{P_3} \,\mathit{Q_4}
+ 4\,J^{4})}{(J^{2} + \mathit{P_1}\,\mathit{Q_2}\,
\mathit{P_3}\,\mathit{Q_4})^{(5/2)}}},  \nonumber \\
{\Gamma_{\mathit{P_3}\,\mathit{Q_4}\,\mathit{Q_2}}}&=& -
{\displaystyle \frac{1}{8}} \,{\displaystyle \frac{\pi \,
\mathit{P_1}\,(\mathit{P_1}^{2}\,\mathit{Q_2}^{2}\,\mathit{P_3}^{2}\,
\mathit{Q_4}^{2} +
2\,J^{2}\,\mathit{P_1}\,\mathit{Q_2}\,\mathit{P_3} \,\mathit{Q_4}
+ 4\,J^{4})}{(J^{2} + \mathit{P_1}\,\mathit{Q_2}\,
\mathit{P_3}\,\mathit{Q_4})^{(5/2)}}},  \nonumber \\
{\Gamma
_{\mathit{P_3}\,\mathit{Q_4}\,\mathit{P_3}}}&=&{\displaystyle
\frac{1}{8}} \,{\displaystyle \frac{\pi \,\mathit{P_1}^{2}\,
\mathit{Q_2}^{2}\,\mathit{Q_4}\,(\mathit{P_1}\,\mathit{Q_2}\,\mathit{
P_3}\,\mathit{Q_4} + 4\,J^{2})}{(J^{2} +
\mathit{P_1}\,\mathit{Q_2}\,
\mathit{P_3}\,\mathit{Q_4})^{(5/2)}}},  \nonumber \\
{\Gamma
_{\mathit{P_3}\,\mathit{Q_4}\,\mathit{Q_4}}}&=&{\displaystyle
\frac{1}{8}} \,{\displaystyle \frac{\pi \,\mathit{P_1}^{2}\,
\mathit{Q_2}^{2}\,\mathit{P_3}\,(\mathit{P_1}\,\mathit{Q_2}\,\mathit{
P_3}\,\mathit{Q_4} + 4\,J^{2})}{(J^{2} +
\mathit{P_1}\,\mathit{Q_2}\,
\mathit{P_3}\,\mathit{Q_4})^{(5/2)}}},  \nonumber \\
{\Gamma_{\mathit{P_3}\,\mathit{Q_4}\,J}}&=& - {\displaystyle
\frac{ 1}{4}} \,{\displaystyle \frac{\pi
\,\mathit{P_1}\,\mathit{Q_2}\,J
\,(\mathit{P_1}\,\mathit{Q_2}\,\mathit{P_3}\,\mathit{Q_4} -
2\,J^{2}) }{(J^{2} +
\mathit{P_1}\,\mathit{Q_2}\,\mathit{P_3}\,\mathit{Q_4})^{(
5/2)}}}, 
\end{eqnarray}
\begin{eqnarray}
{\Gamma_{\mathit{P_3}\,J\,\mathit{P_1}}}&=& - {\displaystyle
\frac{ 1}{4}} \,{\displaystyle \frac{\pi
\,\mathit{Q_2}\,\mathit{Q_4}\,J
\,(\mathit{P_1}\,\mathit{Q_2}\,\mathit{P_3}\,\mathit{Q_4} -
2\,J^{2}) }{(J^{2} +
\mathit{P_1}\,\mathit{Q_2}\,\mathit{P_3}\,\mathit{Q_4})^{(
5/2)}}},  \nonumber \\
{\Gamma_{\mathit{P_3}\,J\,\mathit{Q_2}}}&=& - {\displaystyle
\frac{ 1}{4}} \,{\displaystyle \frac{\pi
\,\mathit{P_1}\,\mathit{Q_4}\,J
\,(\mathit{P_1}\,\mathit{Q_2}\,\mathit{P_3}\,\mathit{Q_4} -
2\,J^{2}) }{(J^{2} +
\mathit{P_1}\,\mathit{Q_2}\,\mathit{P_3}\,\mathit{Q_4})^{(
5/2)}}},  \nonumber \\
{\Gamma_{\mathit{P_3}\,J\,\mathit{P_3}}}&=& - {\displaystyle
\frac{ 3}{4}} \,{\displaystyle \frac{\pi
\,\mathit{P_1}^{2}\,\mathit{Q_2} ^{2}\,\mathit{Q_4}^{2}\,J}{(J^{2}
+ \mathit{P_1}\,\mathit{Q_2}\,
\mathit{P_3}\,\mathit{Q_4})^{(5/2)}}},  \nonumber \\
{\Gamma_{\mathit{P_3}\,J\,\mathit{Q_4}}}&=& - {\displaystyle
\frac{ 1}{4}} \,{\displaystyle \frac{\pi
\,\mathit{P_1}\,\mathit{Q_2}\,J
\,(\mathit{P_1}\,\mathit{Q_2}\,\mathit{P_3}\,\mathit{Q_4} -
2\,J^{2}) }{(J^{2} +
\mathit{P_1}\,\mathit{Q_2}\,\mathit{P_3}\,\mathit{Q_4})^{(
5/2)}}},  \nonumber \\
{\Gamma_{\mathit{P_3}\,J\,J}}&=&{\displaystyle \frac{1}{2}} \,
{\displaystyle \frac{\pi
\,\mathit{P_1}\,\mathit{Q_2}\,\mathit{Q_4}
\,(\mathit{P_1}\,\mathit{Q_2}\,\mathit{P_3}\,\mathit{Q_4} -
2\,J^{2}) }{(J^{2} +
\mathit{P_1}\,\mathit{Q_2}\,\mathit{P_3}\,\mathit{Q_4})^{(
5/2)}}}, 
\end{eqnarray}
\begin{eqnarray}
{\Gamma_{\mathit{Q_4}\,\mathit{Q_4}\,\mathit{P_1}}}&=&{\displaystyle
\frac{1}{8}} \,{\displaystyle \frac{\pi \,\mathit{Q_2}^{2}\,
\mathit{P_3}^{2}\,\mathit{P_1}\,(\mathit{P_1}\,\mathit{Q_2}\,\mathit{
P_3}\,\mathit{Q_4} + 4\,J^{2})}{(J^{2} +
\mathit{P_1}\,\mathit{Q_2}\,
\mathit{P_3}\,\mathit{Q_4})^{(5/2)}}},  \nonumber \\
{\Gamma
_{\mathit{Q_4}\,\mathit{Q_4}\,\mathit{Q_2}}}&=&{\displaystyle
\frac{1}{8}} \,{\displaystyle \frac{\pi \,\mathit{P_1}^{2}\,
\mathit{P_3}^{2}\,\mathit{Q_2}\,(\mathit{P_1}\,\mathit{Q_2}\,\mathit{
P_3}\,\mathit{Q_4} + 4\,J^{2})}{(J^{2} +
\mathit{P_1}\,\mathit{Q_2}\,
\mathit{P_3}\,\mathit{Q_4})^{(5/2)}}},  \nonumber \\
{\Gamma
_{\mathit{Q_4}\,\mathit{Q_4}\,\mathit{P_3}}}&=&{\displaystyle
\frac{1}{8}} \,{\displaystyle \frac{\pi \,\mathit{P_1}^{2}\,
\mathit{Q_2}^{2}\,\mathit{P_3}\,(\mathit{P_1}\,\mathit{Q_2}\,\mathit{
P_3}\,\mathit{Q_4} + 4\,J^{2})}{(J^{2} +
\mathit{P_1}\,\mathit{Q_2}\,
\mathit{P_3}\,\mathit{Q_4})^{(5/2)}}},  \nonumber \\
{\Gamma_{\mathit{Q_4}\,\mathit{Q_4}\,\mathit{Q_4}}}&=& -
{\displaystyle \frac{3}{8}} \,{\displaystyle \frac{\pi \,
\mathit{P_1}^{3}\,\mathit{Q_2}^{3}\,\mathit{P_3}^{3}}{(J^{2} +
\mathit{P_1}\,\mathit{Q_2}\,\mathit{P_3}\,\mathit{Q_4})^{(5/2)}}},  \nonumber \\
{\Gamma_{\mathit{Q_4}\,\mathit{Q_4}\,J}}&=& - {\displaystyle
\frac{ 3}{4}} \,{\displaystyle \frac{\pi
\,\mathit{P_1}^{2}\,\mathit{Q_2} ^{2}\,\mathit{P_3}^{2}\,J}{(J^{2}
+ \mathit{P_1}\,\mathit{Q_2}\,
\mathit{P_3}\,\mathit{Q_4})^{(5/2)}}}, 
\end{eqnarray}
\begin{eqnarray}
{\Gamma_{\mathit{Q_4}\,J\,\mathit{P_1}}}&=& - {\displaystyle
\frac{ 1}{4}} \,{\displaystyle \frac{\pi
\,\mathit{Q_2}\,\mathit{P_3}\,J
\,(\mathit{P_1}\,\mathit{Q_2}\,\mathit{P_3}\,\mathit{Q_4} -
2\,J^{2}) }{(J^{2} +
\mathit{P_1}\,\mathit{Q_2}\,\mathit{P_3}\,\mathit{Q_4})^{(
5/2)}}},  \nonumber \\
{\Gamma_{\mathit{Q_4}\,J\,\mathit{Q_2}}}&=& - {\displaystyle
\frac{ 1}{4}} \,{\displaystyle \frac{\pi
\,\mathit{P_1}\,\mathit{P_3}\,J
\,(\mathit{P_1}\,\mathit{Q_2}\,\mathit{P_3}\,\mathit{Q_4} -
2\,J^{2}) }{(J^{2} +
\mathit{P_1}\,\mathit{Q_2}\,\mathit{P_3}\,\mathit{Q_4})^{(
5/2)}}},  \nonumber \\
{\Gamma_{\mathit{Q_4}\,J\,\mathit{P_3}}}&=& - {\displaystyle
\frac{ 1}{4}} \,{\displaystyle \frac{\pi
\,\mathit{P_1}\,\mathit{Q_2}\,J
\,(\mathit{P_1}\,\mathit{Q_2}\,\mathit{P_3}\,\mathit{Q_4} -
2\,J^{2}) }{(J^{2} +
\mathit{P_1}\,\mathit{Q_2}\,\mathit{P_3}\,\mathit{Q_4})^{(
5/2)}}},  \nonumber \\
{\Gamma_{\mathit{Q_4}\,J\,\mathit{Q_4}}}&=& - {\displaystyle
\frac{ 3}{4}} \,{\displaystyle \frac{\pi
\,\mathit{P_1}^{2}\,\mathit{Q_2} ^{2}\,\mathit{P_3}^{2}\,J}{(J^{2}
+ \mathit{P_1}\,\mathit{Q_2}\,
\mathit{P_3}\,\mathit{Q_4})^{(5/2)}}},  \nonumber \\
{\Gamma_{\mathit{Q_4}\,J\,J}}&=&{\displaystyle \frac{1}{2}} \,
{\displaystyle \frac{\pi
\,\mathit{P_1}\,\mathit{Q_2}\,\mathit{P_3}
\,(\mathit{P_1}\,\mathit{Q_2}\,\mathit{P_3}\,\mathit{Q_4} -
2\,J^{2}) }{(J^{2} +
\mathit{P_1}\,\mathit{Q_2}\,\mathit{P_3}\,\mathit{Q_4})^{(
5/2)}}}, 
\end{eqnarray}
\begin{eqnarray}
{\Gamma_{J\,J\,\mathit{P_1}}}&=&{\displaystyle \frac{1}{2}} \,
{\displaystyle \frac{\mathit{Q_4}\,\pi\,\mathit{Q_2}\,\mathit{P_3}\,
(\mathit{P_1}\,\mathit{Q_2}\,\mathit{P_3}\,\mathit{Q_4} -2\,J^{2}) }{(J^{2} +
\mathit{P_1}\,\mathit{Q_2}\,\mathit{P_3}\,\mathit{Q_4})^{(5/2)}}},  \nonumber \\
{\Gamma_{J\,J\,\mathit{Q_2}}}&=&{\displaystyle \frac{1}{2}} \,
{\displaystyle \frac{\pi \,\mathit{P_1}\,\mathit{P_3}\,\mathit{Q_4}
\,(\mathit{P_1}\,\mathit{Q_2}\,\mathit{P_3}\,\mathit{Q_4} -2\,J^{2})}{(J^{2} 
+ \mathit{P_1}\,\mathit{Q_2}\,\mathit{P_3}\,\mathit{Q_4})^{(5/2)}}},  \nonumber \\
{\Gamma_{J\,J\,\mathit{P_3}}}&=&{\displaystyle \frac{1}{2}} \,
{\displaystyle \frac{\pi\,\mathit{P_1}\,\mathit{Q_2}\,\mathit{Q_4}\,
(\mathit{P_1}\,\mathit{Q_2}\,\mathit{P_3}\,\mathit{Q_4} -2\,J^{2}) }{(J^{2} +
\mathit{P_1}\,\mathit{Q_2}\,\mathit{P_3}\,\mathit{Q_4})^{(5/2)}}}, \nonumber \\ 
{\Gamma_{J\,J\,\mathit{Q_4}}}&=&{\displaystyle \frac{1}{2}} \,
{\displaystyle \frac{\pi\,\mathit{P_1}\,\mathit{Q_2}\,
\mathit{P_3}\,(\mathit{P_1}\,\mathit{Q_2}\,\mathit{P_3}\,\mathit{Q_4} -
2\,J^{2}) }{(J^{2} +\mathit{P_1}\,\mathit{Q_2}\,\mathit{P_3}\,
\mathit{Q_4})^{(5/2)}}},  \nonumber \\
{\Gamma_{J\,J\,J}}&=&3\,{\displaystyle \frac{\pi \,J\,\mathit{P_1}\,
\mathit{Q_2}\,\mathit{P_3}\,\mathit{Q_4}}{(J^{2} +
\mathit{P_1}\, \mathit{Q_2}\,\mathit{P_3}\,\mathit{Q_4})^{(5/2)}}}.
\end{eqnarray}



\clearpage
\chapter{La g\'eom\'etrie thermodynamique de Ruppenier 
des trous noirs de trois param\`etres}
Dans cet appendice, nous allons furnir les coeficients de la courbure scalaire 
de la g\'eom\'etrie thermodynamique de Ruppenier d'un ensemble des trous noirs 
de trois param\`etres $\{P, Q, J \} \in \mathcal{M}_3$. Il s'av\`ere que les fonctions 
$\{ r_i, \tilde{r}_j\ |\ i=1,2,3,4,6, j=1,2,3 \}$ appara\^issant dans l'Eqn. (\ref{rup3dg})
de num\'erateur de la courbure scalaire de Ruppenier sont donn\'ees par:

\ba
r_1&=&
\frac{\partial^2 S}{\partial P^2}
\frac{\partial^2 S}{\partial Q^2}
\frac{\partial^3 S}{\partial Q^2 \partial P}
\frac{\partial^3 S}{\partial P \partial J^2}
\frac{\partial^2 S}{\partial J^2}
+
\frac{\partial^2 S}{\partial P^2}
\frac{\partial^3 S}{\partial Q^3}
\frac{\partial^2 S}{\partial Q \partial P}
\frac{\partial^2 S}{\partial P \partial J}
\frac{\partial^3 S}{\partial J^3} \nn &&
+
\frac{\partial^2 S}{\partial P^2}
\frac{\partial^3 S}{\partial P^2 \partial J}
\frac{\partial^3 S}{\partial Q^2 \partial J}
\frac{\partial^2 S}{\partial Q^2}
\frac{\partial^2 S}{\partial J^2}
+ 
\frac{\partial^2 S}{\partial P^2}
\frac{\partial^3 S}{\partial Q \partial P^2}
\frac{\partial^3 S}{\partial Q \partial J^2}
\frac{\partial^2 S}{\partial Q^2}
\frac{\partial^2 S}{\partial J^2} \nn &&
+
\frac{\partial^2 S}{\partial P^2}
(\frac{\partial^3 S}{\partial Q^2 \partial J})^2
(\frac{\partial^2 S}{\partial P \partial J})^2
+
\frac{\partial^2 S}{\partial P^2}
(\frac{\partial^3 S}{\partial Q \partial J^2})^2
(\frac{\partial^2 S}{\partial Q \partial P})^2 \nn &&
+
\frac{\partial^3 S}{\partial P^3}
\frac{\partial^2 S}{\partial Q \partial P}
\frac{\partial^2 S}{\partial Q \partial J}
\frac{\partial^3 S}{\partial J^3}
\frac{\partial^2 S}{\partial Q^2}
+
(\frac{\partial^3 S}{\partial Q^2 \partial P})^2
\frac{\partial^2 S}{\partial J^2}
(\frac{\partial^2 S}{\partial P \partial J})^2 \nn &&
+
(\frac{\partial^3 S}{\partial P^2 \partial J})^2
\frac{\partial^2 S}{\partial Q^2}
(\frac{\partial^2 S}{\partial Q \partial J})^2
+
(\frac{\partial^3 S}{\partial P \partial J^2})^2
\frac{\partial^2 S}{\partial Q^2}
(\frac{\partial^2 S}{\partial Q \partial P})^2 \nn &&
+
(\frac{\partial^3 S}{\partial Q \partial P^2})^2
\frac{\partial^2 S}{\partial J^2}
(\frac{\partial^2 S}{\partial Q \partial J})^2
+
\frac{\partial^3 S}{\partial P^3}
\frac{\partial^2 S}{\partial P \partial J}
\frac{\partial^2 S}{\partial Q \partial J}
\frac{\partial^3 S}{\partial Q^3}
\frac{\partial^2 S}{\partial J^2} \nn &&
+
\frac{\partial^3 S}{\partial P^2 \partial J}
(\frac{\partial^2 S}{\partial Q^2})^2
\frac{\partial^3 S}{\partial P \partial J^2}
\frac{\partial^2 S}{\partial P \partial J}
+
\frac{\partial^3 S}{\partial Q \partial P^2}
(\frac{\partial^2 S}{\partial J^2})^2
\frac{\partial^3 S}{\partial Q^2 \partial P}
\frac{\partial^2 S}{\partial Q \partial P} \nn &&
+
\frac{\partial^3 S}{\partial P^3}
\frac{\partial^3 S}{\partial P \partial J^2}
(\frac{\partial^2 S}{\partial Q^2})^2
\frac{\partial^2 S}{\partial J^2}
+
\frac{\partial^3 S}{\partial P^3}
\frac{\partial^3 S}{\partial Q^2 \partial P}
(\frac{\partial^2 S}{\partial J^2})^2
\frac{\partial^2 S}{\partial Q^2} \nn &&
+
(\frac{\partial^2 S}{\partial P^2})^2
\frac{\partial^3 S}{\partial Q^2 \partial J}
\frac{\partial^3 S}{\partial Q \partial J^2}
\frac{\partial^2 S}{\partial Q \partial J}
+
(\frac{\partial^2 S}{\partial P^2})^2
\frac{\partial^3 S}{\partial Q^3}
\frac{\partial^3 S}{\partial Q \partial J^2}
\frac{\partial^2 S}{\partial J^2} \nn &&
+
(\frac{\partial^2 S}{\partial P^2})^2
\frac{\partial^2 S}{\partial Q^2}
\frac{\partial^3 S}{\partial Q^2 \partial J} 
\frac{\partial^3 S}{\partial J^3}
+
\frac{\partial^2 S}{\partial P^2}
\frac{\partial^3 S}{\partial P^2 \partial J}
(\frac{\partial^2 S}{\partial Q^2})^2
\frac{\partial^3 S}{\partial J^3} \nn &&
+
\frac{\partial^2 S}{\partial P^2}
\frac{\partial^3 S}{\partial Q \partial P^2}
(\frac{\partial^2 S}{\partial J^2})^2
\frac{\partial^3 S}{\partial Q^3},\ea
%

\ba
\tilde{r}_1&=&
(\frac{\partial^3 S}{\partial P^2 \partial J})^2
(\frac{\partial^2 S}{\partial Q^2})^2
\frac{\partial^2 S}{\partial J^2}
+
(\frac{\partial^3 S}{\partial Q \partial P^2})^2
(\frac{\partial^2 S}{\partial J^2})^2
\frac{\partial^2 S}{\partial Q^2} \nn &&
+
\frac{\partial^2 S}{\partial P^2}
\frac{\partial^2 S}{\partial Q^2}
\frac{\partial^3 S}{\partial Q^2 \partial P}
\frac{\partial^2 S}{\partial P \partial J}
\frac{\partial^3 S}{\partial J^3}
+
\frac{\partial^2 S}{\partial P^2}
\frac{\partial^3 S}{\partial Q^2 \partial J}
\frac{\partial^3 S}{\partial Q \partial J^2}
\frac{\partial^2 S}{\partial Q \partial P}
\frac{\partial^2 S}{\partial P \partial J} \nn &&
+
\frac{\partial^2 S}{\partial P^2}
\frac{\partial^3 S}{\partial Q^2 \partial P}
\frac{\partial^2 S}{\partial P \partial J}
\frac{\partial^2 S}{\partial Q \partial J}
\frac{\partial^3 S}{\partial Q \partial J^2}
+ 
\frac{\partial^2 S}{\partial P^2}
\frac{\partial^3 S}{\partial Q^2 \partial J}
\frac{\partial^2 S}{\partial Q \partial P}
\frac{\partial^2 S}{\partial Q \partial J}
\frac{\partial^3 S}{\partial P \partial J^2} \nn &&
+ 
\frac{\partial^2 S}{\partial P^2}
\frac{\partial^3 S}{\partial Q^3}
\frac{\partial^2 S}{\partial Q \partial P}
\frac{\partial^3 S}{\partial P \partial J^2}
\frac{\partial^2 S}{\partial J^2}
+
\frac{\partial^2 S}{\partial P^2}
\frac{\partial^3 S}{\partial Q \partial P^2}
\frac{\partial^2 S}{\partial Q \partial J}
\frac{\partial^3 S}{\partial J^3}
\frac{\partial^2 S}{\partial Q^2} \nn &&
+ 
\frac{\partial^2 S}{\partial P^2}
\frac{\partial^3 S}{\partial P^2 \partial J}
\frac{\partial^2 S}{\partial Q \partial J}
\frac{\partial^3 S}{\partial Q^3}
\frac{\partial^2 S}{\partial J^2}
+
\frac{\partial^3 S}{\partial Q \partial P^2}
\frac{\partial^2 S}{\partial J^2}
\frac{\partial^3 S}{\partial Q^2 \partial P}
\frac{\partial^2 S}{\partial P \partial J}
\frac{\partial^2 S}{\partial Q \partial J} \nn &&
+
\frac{\partial^3 S}{\partial P^2 \partial J}
\frac{\partial^2 S}{\partial Q^2}
\frac{\partial^3 S}{\partial P \partial J^2}
\frac{\partial^2 S}{\partial Q \partial P}
\frac{\partial^2 S}{\partial Q \partial J}
+
\frac{\partial^3 S}{\partial Q \partial P^2}
\frac{\partial^2 S}{\partial P \partial J}
\frac{\partial^2 S}{\partial Q \partial J}
\frac{\partial^3 S}{\partial P \partial J^2}
\frac{\partial^2 S}{\partial Q^2} \nn &&
+
 \frac{\partial^3 S}{\partial P^3}
\frac{\partial^2 S}{\partial P \partial J}
\frac{\partial^3 S}{\partial Q^2 \partial J}
\frac{\partial^2 S}{\partial Q^2}
\frac{\partial^2 S}{\partial J^2}
+
\frac{\partial^3 S}{\partial P^3}
\frac{\partial^2 S}{\partial Q \partial P}
\frac{\partial^3 S}{\partial Q \partial J^2}
\frac{\partial^2 S}{\partial Q^2}
\frac{\partial^2 S}{\partial J^2} \nn &&
+
\frac{\partial^3 S}{\partial P^2 \partial J}
\frac{\partial^2 S}{\partial Q \partial J}
\frac{\partial^3 S}{\partial Q^2 \partial P}
\frac{\partial^2 S}{\partial Q \partial P}
\frac{\partial^2 S}{\partial J^2}
+
\frac{\partial^3 S}{\partial P^2 \partial J}
\frac{\partial^2 S}{\partial Q \partial P}
\frac{\partial^2 S}{\partial P \partial J}
\frac{\partial^3 S}{\partial Q \partial J^2}
\frac{\partial^2 S}{\partial Q^2} \nn &&
+
\frac{\partial^3 S}{\partial Q \partial P^2}
\frac{\partial^2 S}{\partial J^2}
\frac{\partial^3 S}{\partial Q^2 \partial J}
\frac{\partial^2 S}{\partial Q \partial P}
\frac{\partial^2 S}{\partial P \partial J}
+
\frac{\partial^2 S}{\partial P^2}
(\frac{\partial^3 S}{\partial Q \partial P \partial J})^2
(\frac{\partial^2 S}{\partial Q \partial J})^2 \nn &&
+
\frac{\partial^2 S}{\partial P^2}
(\frac{\partial^3 S}{\partial Q^2 \partial P})^2
(\frac{\partial^2 S}{\partial J^2})^2
+
(\frac{\partial^2 S}{\partial P^2})^2
(\frac{\partial^3 S}{\partial Q^2 \partial J})^2
\frac{\partial^2 S}{\partial J^2} \nn &&
+
(\frac{\partial^2 S}{\partial P^2})^2
\frac{\partial^2 S}{\partial Q^2}
(\frac{\partial^3 S}{\partial Q \partial J^2})^2
+
(\frac{\partial^3 S}{\partial Q \partial P \partial J})^2
(\frac{\partial^2 S}{\partial Q \partial P})^2
\frac{\partial^2 S}{\partial J^2} \nn &&
+
(\frac{\partial^3 S}{\partial Q \partial P \partial J})^2
(\frac{\partial^2 S}{\partial P \partial J})^2
\frac{\partial^2 S}{\partial Q^2}
+
\frac{\partial^2 S}{\partial P^2}
(\frac{\partial^3 S}{\partial P \partial J^2})^2
(\frac{\partial^2 S}{\partial Q^2})^2 \nn &&
+
\frac{\partial^3 S}{\partial P^3}
\frac{\partial^3 S}{\partial P \partial J^2}
\frac{\partial^2 S}{\partial Q^2}
(\frac{\partial^2 S}{\partial Q \partial J})^2
+
\frac{\partial^3 S}{\partial Q \partial P^2}
\frac{\partial^2 S}{\partial J^2}
\frac{\partial^3 S}{\partial Q^3}
(\frac{\partial^2 S}{\partial P \partial J})^2 \nn &&
+
\frac{\partial^3 S}{\partial P^3}
\frac{\partial^2 S}{\partial P \partial J}
(\frac{\partial^2 S}{\partial Q^2})^2
\frac{\partial^3 S}{\partial J^3}
+
\frac{\partial^3 S}{\partial P^3}
\frac{\partial^3 S}{\partial Q^2 \partial P}
\frac{\partial^2 S}{\partial J^2}
(\frac{\partial^2 S}{\partial Q \partial J})^2 \nn &&
+
\frac{\partial^3 S}{\partial P^3}
\frac{\partial^2 S}{\partial Q \partial P}
(\frac{\partial^2 S}{\partial J^2})^2
\frac{\partial^3 S}{\partial Q^3}
+
\frac{\partial^3 S}{\partial P^2 \partial J}
\frac{\partial^2 S}{\partial Q^2}
\frac{\partial^3 S}{\partial J^3}
(\frac{\partial^2 S}{\partial Q \partial P})^2 \nn &&
+
(\frac{\partial^2 S}{\partial P^2})^2
\frac{\partial^3 S}{\partial Q^3}
\frac{\partial^2 S}{\partial Q \partial J}
\frac{\partial^3 S}{\partial J^3}
+
\frac{\partial^2 S}{\partial P^2}
\frac{\partial^3 S}{\partial Q^2 \partial J}
\frac{\partial^3 S}{\partial J^3}
(\frac{\partial^2 S}{\partial Q \partial P})^2 \nn &&
+
\frac{\partial^2 S}{\partial P^2}
\frac{\partial^3 S}{\partial Q^3}
\frac{\partial^3 S}{\partial Q \partial J^2}
(\frac{\partial^2 S}{\partial P \partial J})^2,\ea

%
\ba
r_2&=&
\frac{\partial^2 S}{\partial P^2}
\frac{\partial^2 S}{\partial Q^2}
\frac{\partial^3 S}{\partial Q \partial J^2}
\frac{\partial^2 S}{\partial Q \partial P}
\frac{\partial^3 S}{\partial P \partial J^2}
+
\frac{\partial^3 S}{\partial P^2 \partial J}
\frac{\partial^3 S}{\partial Q \partial P^2}
\frac{\partial^2 S}{\partial Q \partial J}
\frac{\partial^2 S}{\partial Q^2}
\frac{\partial^2 S}{\partial J^2} \nn &&
+
\frac{\partial^2 S}{\partial P^2}
\frac{\partial^3 S}{\partial Q^2 \partial P}
\frac{\partial^2 S}{\partial P \partial J}
\frac{\partial^3 S}{\partial Q^2 \partial J}
\frac{\partial^2 S}{\partial J^2}
+
\frac{\partial^3 S}{\partial Q \partial P \partial J}
(\frac{\partial^2 S}{\partial Q \partial P})^3
\frac{\partial^3 S}{\partial J^3} \nn &&
+
\frac{\partial^3 S}{\partial Q \partial P \partial J}
(\frac{\partial^2 S}{\partial P \partial J})^3
\frac{\partial^3 S}{\partial Q^3}
+
\frac{\partial^3 S}{\partial P^3}
\frac{\partial^3 S}{\partial Q \partial P \partial J}
(\frac{\partial^2 S}{\partial Q \partial J})^3 \nn &&
+
\frac{\partial^3 S}{\partial Q \partial P^2}
\frac{\partial^3 S}{\partial Q \partial J^2}
(\frac{\partial^2 S}{\partial Q \partial P})^2
\frac{\partial^2 S}{\partial J^2}
+
\frac{\partial^3 S}{\partial P^2 \partial J}
\frac{\partial^3 S}{\partial Q^2 \partial J}
(\frac{\partial^2 S}{\partial P \partial J})^2
\frac{\partial^2 S}{\partial Q^2} \nn &&
+
\frac{\partial^2 S}{\partial Q^2}
\frac{\partial^3 S}{\partial Q^2 \partial P}
(\frac{\partial^2 S}{\partial P \partial J})^2
\frac{\partial^3 S}{\partial P \partial J^2}
+
\frac{\partial^3 S}{\partial Q^2 \partial P}
\frac{\partial^3 S}{\partial P \partial J^2}
(\frac{\partial^2 S}{\partial Q \partial P})^2
\frac{\partial^2 S}{\partial J^2} \nn &&
+
\frac{\partial^2 S}{\partial P^2}
\frac{\partial^3 S}{\partial P^2 \partial J}
\frac{\partial^3 S}{\partial Q^2 \partial J}
(\frac{\partial^2 S}{\partial Q \partial J})^2
+
\frac{\partial^2 S}{\partial P^2}
\frac{\partial^3 S}{\partial Q \partial P^2}
\frac{\partial^3 S}{\partial Q \partial J^2}
(\frac{\partial^2 S}{\partial Q \partial J})^2, \ea

\ba
\tilde{r}_2&=&
\frac{\partial^2 S}{\partial P^2}
\frac{\partial^2 S}{\partial Q^2}
\frac{\partial^3 S}{\partial Q \partial P \partial J}
\frac{\partial^2 S}{\partial Q \partial P}
\frac{\partial^3 S}{\partial J^3}
+
\frac{\partial^2 S}{\partial P^2}
\frac{\partial^3 S}{\partial Q \partial P \partial J}
\frac{\partial^2 S}{\partial Q \partial P}
\frac{\partial^2 S}{\partial Q \partial J}
\frac{\partial^3 S}{\partial Q \partial J^2} \nn &&
+
\frac{\partial^2 S}{\partial P^2}
\frac{\partial^3 S}{\partial Q^2 \partial J}
\frac{\partial^2 S}{\partial P \partial J}
\frac{\partial^3 S}{\partial Q \partial P \partial J}
\frac{\partial^2 S}{\partial Q \partial J}
+
\frac{\partial^3 S}{\partial Q^2 \partial P}
\frac{\partial^2 S}{\partial J^2}
\frac{\partial^3 S}{\partial Q \partial P \partial J}
\frac{\partial^2 S}{\partial Q \partial P}
\frac{\partial^2 S}{\partial P \partial J} \nn &&
+
\frac{\partial^3 S}{\partial Q \partial P \partial J}
\frac{\partial^2 S}{\partial Q \partial P}
\frac{\partial^2 S}{\partial P \partial J}
\frac{\partial^3 S}{\partial P \partial J^2}
\frac{\partial^2 S}{\partial Q^2}
+
\frac{\partial^3 S}{\partial Q^2 \partial P}
\frac{\partial^3 S}{\partial P \partial J^2}
\frac{\partial^2 S}{\partial Q \partial P}
\frac{\partial^2 S}{\partial P \partial J}
\frac{\partial^2 S}{\partial Q \partial J} \nn &&
+
\frac{\partial^3 S}{\partial P^3}
\frac{\partial^3 S}{\partial Q \partial P \partial J}
\frac{\partial^2 S}{\partial Q \partial J}
\frac{\partial^2 S}{\partial Q^2}
\frac{\partial^2 S}{\partial J^2}
+
\frac{\partial^3 S}{\partial P^2 \partial J}
\frac{\partial^2 S}{\partial Q \partial J}
\frac{\partial^3 S}{\partial Q \partial P \partial J}
\frac{\partial^2 S}{\partial P \partial J}
\frac{\partial^2 S}{\partial Q^2} \nn &&
+
\frac{\partial^3 S}{\partial Q \partial P^2}
\frac{\partial^2 S}{\partial J^2}
\frac{\partial^3 S}{\partial Q \partial P \partial J}
\frac{\partial^2 S}{\partial Q \partial P}
\frac{\partial^2 S}{\partial Q \partial J}
+
\frac{\partial^3 S}{\partial P^2 \partial J}
\frac{\partial^3 S}{\partial Q^2 \partial J}
\frac{\partial^2 S}{\partial Q \partial P}
\frac{\partial^2 S}{\partial P \partial J}
\frac{\partial^2 S}{\partial Q \partial J} \nn &&
+
\frac{\partial^3 S}{\partial Q \partial P^2}
\frac{\partial^3 S}{\partial Q \partial J^2}
\frac{\partial^2 S}{\partial Q \partial P}
\frac{\partial^2 S}{\partial P \partial J}
\frac{\partial^2 S}{\partial Q \partial J}
+
\frac{\partial^3 S}{\partial Q \partial J^2}
(\frac{\partial^2 S}{\partial Q \partial P})^3
\frac{\partial^3 S}{\partial P \partial J^2} \nn &&
+
\frac{\partial^3 S}{\partial Q^2 \partial P}
(\frac{\partial^2 S}{\partial P \partial J})^3
\frac{\partial^3 S}{\partial Q^2 \partial J}
+
\frac{\partial^3 S}{\partial P^2 \partial J}
\frac{\partial^3 S}{\partial Q \partial P^2}
(\frac{\partial^2 S}{\partial Q \partial J})^3 \nn &&
+
\frac{\partial^3 S}{\partial Q \partial P \partial J}
\frac{\partial^2 S}{\partial Q \partial J}
\frac{\partial^3 S}{\partial P \partial J^2}
(\frac{\partial^2 S}{\partial Q \partial P})^2
+
\frac{\partial^3 S}{\partial Q^2 \partial P}
(\frac{\partial^2 S}{\partial P \partial J})^2
\frac{\partial^3 S}{\partial Q \partial P \partial J}
\frac{\partial^2 S}{\partial Q \partial J} \nn &&
+
\frac{\partial^3 S}{\partial P^3}
\frac{\partial^2 S}{\partial P \partial J}
\frac{\partial^3 S}{\partial Q^2 \partial J}
(\frac{\partial^2 S}{\partial Q \partial J})^2
+
\frac{\partial^3 S}{\partial P^3}
\frac{\partial^2 S}{\partial Q \partial P}
\frac{\partial^3 S}{\partial Q \partial J^2}
(\frac{\partial^2 S}{\partial Q \partial J})^2  \nn &&
+
\frac{\partial^3 S}{\partial P^2 \partial J}
\frac{\partial^2 S}{\partial Q \partial J}
\frac{\partial^3 S}{\partial Q^3}
(\frac{\partial^2 S}{\partial P \partial J})^2
+
\frac{\partial^3 S}{\partial P^2 \partial J}
(\frac{\partial^2 S}{\partial Q \partial J})^2
\frac{\partial^3 S}{\partial Q \partial P \partial J}
\frac{\partial^2 S}{\partial Q \partial P} \nn &&
+
\frac{\partial^3 S}{\partial Q \partial P \partial J}
\frac{\partial^3 S}{\partial Q^2 \partial J}
(\frac{\partial^2 S}{\partial P \partial J})^2
\frac{\partial^2 S}{\partial Q \partial P}
+
\frac{\partial^3 S}{\partial Q^2 \partial P}
\frac{\partial^2 S}{\partial P \partial J}
\frac{\partial^3 S}{\partial J^3}
(\frac{\partial^2 S}{\partial Q \partial P})^2 \nn &&
+
\frac{\partial^3 S}{\partial Q \partial P \partial J}
\frac{\partial^2 S}{\partial P \partial J}
\frac{\partial^3 S}{\partial Q \partial J^2}
(\frac{\partial^2 S}{\partial Q \partial P})^2
+
\frac{\partial^3 S}{\partial Q \partial P^2}
(\frac{\partial^2 S}{\partial Q \partial J})^2
\frac{\partial^3 S}{\partial Q \partial P \partial J}
\frac{\partial^2 S}{\partial P \partial J} \nn &&
+
\frac{\partial^3 S}{\partial Q^3}
\frac{\partial^2 S}{\partial Q \partial P}
\frac{\partial^3 S}{\partial P \partial J^2}
(\frac{\partial^2 S}{\partial P \partial J})^2
+
\frac{\partial^3 S}{\partial Q \partial P^2}
\frac{\partial^2 S}{\partial Q \partial J}
\frac{\partial^3 S}{\partial J^3}
(\frac{\partial^2 S}{\partial Q \partial P})^2 \nn &&
+ 
\frac{\partial^2 S}{\partial P^2}
\frac{\partial^3 S}{\partial Q \partial P \partial J}
\frac{\partial^2 S}{\partial P \partial J}
\frac{\partial^3 S}{\partial Q^3}
\frac{\partial^2 S}{\partial J^2},\ea

\ba
r_3&=&
\frac{\partial^2 S}{\partial P^2}
\frac{\partial^3 S}{\partial Q^3}
\frac{\partial^2 S}{\partial P \partial J}
\frac{\partial^2 S}{\partial Q \partial J}
\frac{\partial^3 S}{\partial P \partial J^2}
+
\frac{\partial^2 S}{\partial P^2}
\frac{\partial^3 S}{\partial Q^2 \partial P}
\frac{\partial^2 S}{\partial Q \partial P}
\frac{\partial^2 S}{\partial Q \partial J}
\frac{\partial^3 S}{\partial J^3} \nn &&
+
\frac{\partial^3 S}{\partial P^3}
\frac{\partial^2 S}{\partial P \partial J}
\frac{\partial^2 S}{\partial Q \partial J}
\frac{\partial^3 S}{\partial Q \partial J^2}
\frac{\partial^2 S}{\partial Q^2}
+
\frac{\partial^3 S}{\partial P^3}
\frac{\partial^2 S}{\partial Q \partial P}
\frac{\partial^2 S}{\partial J^2}
\frac{\partial^3 S}{\partial Q^2 \partial J}
\frac{\partial^2 S}{\partial Q \partial J} \nn &&
+
\frac{\partial^3 S}{\partial P^2 \partial J}
\frac{\partial^2 S}{\partial Q \partial P}
\frac{\partial^2 S}{\partial P \partial J}
\frac{\partial^3 S}{\partial Q^3}
\frac{\partial^2 S}{\partial J^2}
+
\frac{\partial^3 S}{\partial Q \partial P^2}
\frac{\partial^23 S}{\partial Q \partial P}
\frac{\partial^2 S}{\partial P \partial J}
\frac{\partial^3 S}{\partial J^3}
\frac{\partial^2 S}{\partial Q^2}, \ea

\ba
\tilde{r}_3&=&
\frac{\partial^2 S}{\partial P^2}
\frac{\partial^2 S}{\partial Q^2}
\frac{\partial^3 S}{\partial P \partial J^2}
\frac{\partial^3 S}{\partial Q^2 \partial J}
\frac{\partial^2 S}{\partial P \partial J}
+
\frac{\partial^2 S}{\partial P^2}
\frac{\partial^3 S}{\partial P^2 \partial J}
\frac{\partial^2 S}{\partial Q \partial J}
\frac{\partial^3 S}{\partial Q \partial J^2}
\frac{\partial^2 S}{\partial Q^2} \nn &&
+
\frac{\partial^2 S}{\partial P^2}
\frac{\partial^3 S}{\partial Q \partial P^2}
\frac{\partial^2 S}{\partial J^2}
\frac{\partial^3 S}{\partial Q^2 \partial J}
\frac{\partial^2 S}{\partial Q \partial J}
+
\frac{\partial^3 S}{\partial Q \partial P^2}
\frac{\partial^2 S}{\partial Q \partial P}
\frac{\partial^3 S}{\partial P \partial J^2}
\frac{\partial^2 S}{\partial Q^2}
\frac{\partial^2 S}{\partial J^2} \nn &&
+
\frac{\partial^3 S}{\partial P^2 \partial J}
\frac{\partial^3 S}{\partial Q^2 \partial P}
\frac{\partial^2 S}{\partial P \partial J}
\frac{\partial^2 S}{\partial Q^2}
\frac{\partial^2 S}{\partial J^2}
+
\frac{\partial^3 S}{\partial P^2 \partial J}
\frac{\partial^3 S}{\partial Q^2 \partial J}
(\frac{\partial^2 S}{\partial Q \partial P})^2
\frac{\partial^2 S}{\partial J^2} \nn &&
+
\frac{\partial^3 S}{\partial Q \partial P^2}
\frac{\partial^3 S}{\partial Q \partial J^2}
(\frac{\partial^2 S}{\partial P \partial J})^2
\frac{\partial^2 S}{\partial Q^2}
+
\frac{\partial^2 S}{\partial P^2}
\frac{\partial^3 S}{\partial Q^2 \partial P}
\frac{\partial^3 S}{\partial P \partial J^2}
(\frac{\partial^2 S}{\partial Q \partial J})^2 \nn &&
+
\frac{\partial^2 S}{\partial P^2}
(\frac{\partial^3 S}{\partial Q \partial P \partial J})^2
\frac{\partial^2 S}{\partial Q^2}
\frac{\partial^2 S}{\partial J^2}
+ 
\frac{\partial^2 S}{\partial P^2}
\frac{\partial^3 S}{\partial Q^2 \partial P}
\frac{\partial^2 S}{\partial Q \partial P}
\frac{\partial^3 S}{\partial Q \partial J^2}
\frac{\partial^2 S}{\partial J^2},\ea

\ba
r_4&=&
\frac{\partial^2 S}{\partial P^2}
\frac{\partial^2 S}{\partial Q^2} 
\frac{\partial^3 S}{\partial Q \partial P \partial J}
\frac{\partial^3 S}{\partial Q \partial J^2}
\frac{\partial^2 S}{\partial P \partial J}
+
\frac{\partial^2 S}{\partial P^2}
\frac{\partial^3 S}{\partial Q^2 \partial J}
\frac{\partial^2 S}{\partial Q \partial P}
\frac{\partial^3 S}{\partial Q \partial P \partial J}
\frac{\partial^2 S}{\partial J^2} \nn &&
+
\frac{\partial^2 S(}{\partial P^2}
\frac{\partial^3 S}{\partial Q \partial P \partial J}
\frac{\partial^2 S}{\partial Q \partial J}
\frac{\partial^3 S}{\partial P \partial J^2}
\frac{\partial^2 S}{\partial Q^2}
+
\frac{\partial^2 S}{\partial P^2}
\frac{\partial^3 S}{\partial Q^2 \partial P}
\frac{\partial^2 S}{\partial J^2}
\frac{\partial^3 S}{\partial Q \partial P \partial J}
\frac{\partial^2 S}{\partial Q \partial J} \nn &&
+
\frac{\partial^3 S}{\partial P^2 \partial J}
\frac{\partial^2 S}{\partial Q \partial P}
\frac{\partial^3 S}{\partial Q \partial P \partial J}
\frac{\partial^2 S}{\partial Q^2}
\frac{\partial^2 S}{\partial J^2}
+
\frac{\partial^3 S}{\partial Q \partial P^2}
\frac{\partial^2 S}{\partial Q \partial P}
\frac{\partial^3 S}{\partial P \partial J^2}
(\frac{\partial^2 S}{\partial Q \partial J})^2 \nn &&
+
\frac{\partial^3 S}{\partial Q \partial P^2}
\frac{\partial^2 S}{\partial P \partial J}
\frac{\partial^3 S}{\partial Q \partial P \partial J}
\frac{\partial^2 S}{\partial Q^2}
\frac{\partial^2 S}{\partial J^2}
+
\frac{\partial^3 S}{\partial P^2 \partial J}
\frac{\partial^3 S}{\partial Q^2 \partial P}
\frac{\partial^2 S}{\partial P \partial J}
(\frac{\partial^2 S}{\partial Q \partial J})^2 \nn &&
+
\frac{\partial^3 S}{\partial Q^2 \partial P}
\frac{\partial^2 S}{\partial Q \partial P}
\frac{\partial^3 S}{\partial Q \partial J^2}
(\frac{\partial^2 S}{\partial P \partial J})^2
+
\frac{\partial^3 S}{\partial Q^2 \partial J}
\frac{\partial^2 S}{\partial P \partial J}
\frac{\partial^3 S}{\partial P \partial J^2}
(\frac{\partial^2 S}{\partial Q \partial P})^2 \nn &&
+
\frac{\partial^3 S}{\partial P^2 \partial J}
\frac{\partial^2 S}{\partial Q \partial J}
\frac{\partial^3 S}{\partial Q \partial J^2}
(\frac{\partial^2 S}{\partial Q \partial P})^2
+
\frac{\partial^3 S}{\partial Q \partial P^2}
(\frac{\partial^2 S}{\partial P \partial J})^2
\frac{\partial^3 S}{\partial Q^2 \partial J}
\frac{\partial^2 S}{\partial Q \partial J}, \ea

\ba
r_6&=&
(\frac{\partial^3 S}{\partial Q \partial P \partial J})^2
\frac{\partial^2 S}{\partial Q \partial P}
\frac{\partial^2 S}{\partial P \partial J}
\frac{\partial^2 S}{\partial Q \partial J}.
\ea
\clearpage
\chapter{La g\'eom\'etrie thermodynamique de Ruppenier 
des trous noirs dyoniques extr\'emaux non-supersym\'etriques}
\`{A} la fin, dans cet appendice, nous terminons en donnant seulement 
les coeficients du d\'eterminant de la m\'etrique tenseure de Ruppenier et 
celles de la num\'erateur et d\'enominateur de la courbure scalaire de Ruppenier 
pour les trous noirs dyoniques extr\'emaux non-supersym\'etriques en quatre 
dimensions avec quatre param\`etres $\{n, w, N, W\}$. Ces r\'esultats sont 
donn\'es comme ci-dessous:

\section{\`A l'ordre de $(\alpha^{\prime})^2$}

L'entropie de trou noir dyonique non-supersym\'etrique extr\'emal
\cite{AshokeSen} est modifi\'ee par les corrections de $\alpha^{\prime}$ \`{a}:

\ba S_{BH}^{ns}&=& 2 \pi \sqrt{nw N W}+ \frac{5 \pi \widehat{\alpha}}{4} \sqrt{\frac{nw}{N W}}
- \frac{29 \pi \widehat{\alpha}^2}{64} \frac{\sqrt{nw}}{(NW)^{3/2}}.\ea

Les coeficients du d\'eterminant de la m\'etrique tenseure de Ruppenier sont:

\ba 
a^{(2)}_0 &=& 268435465,\nn
a^{(2)}_1 &=& 335544320,\nn
a^{(2)}_2 &=& 486539264,\nn
a^{(2)}_3 &=& 96993280,\nn
a^{(2)}_4 &=& 78020608, \nn
a^{(2)}_5 &=& 303493120,\nn
a^{(2)}_6 &=& 193766400,\nn
a^{(2)}_7 &=& 105360480,\nn
a^{(2)}_8 &=& 19096587.
\ea

Les coeficients de la num\'erateur de la courbure scalaire sont donn\'ees par:

\ba
b^{(2)}_{0}&=& 1511 1572 7451 8286 4683 8272,\nn
b^{(2)}_{1}&=& 5666 8397 7944 3574 2564 3520,\nn
b^{(2)}_{2}&=& 15548 3916 4484 8306 8660 9408,\nn
b^{(2)}_{3}&=& 26917 4889 5235 6977 7180 6720,\nn
b^{(2)}_{4}&=& 36446 8002 3967 4226 4396 1856,\nn
b^{(2)}_{5}&=& 32385 2191 8795 4949 5017 4720,\nn
b^{(2)}_{6}&=& 16365 7157 1386 2610 1693 7348,\nn
b^{(2)}_{7}&=& 10350 7275 0709 6965 9523 0720,\nn
b^{(2)}_{8}&=& 31144 9521 1372 0107 3618 9440,\nn
b^{(2)}_{9}&=& 40453 1215 4131 2465 6005 1200,\nn
b^{(2)}_{10}&=& 32842 7122 0150 2909 3937 9712,\nn
b^{(2)}_{11}&=& 17449 4837 6062 5376 5598 9760,
\ea
\ba 
b^{(2)}_{12}&=& 2076 4236 9296 4568 2307 8912,\nn
b^{(2)}_{13}&=& 6928 1710 4454 6225 6877 5680,\nn
b^{(2)}_{14}&=& 9556 1281 9719 8292 5456 9984,\nn
b^{(2)}_{15}&=& 7675 7590 4726 1572 3532 2880,\nn
b^{(2)}_{16}&=& 4715 4467 6657 5055 9765 2992,\nn
b^{(2)}_{17}&=& 2267 5345 3528 8395 7399 5520,\nn
b^{(2)}_{18}&=& 884 7458 4103 0067 7070 8480,\nn
b^{(2)}_{19}&=& 270 8626 3837 5487 5371 5200,\nn
b^{(2)}_{20}&=& 63 4204 1422 8318 4987 9680,\nn
b^{(2)}_{21}&=& 10 3049 7125 5348 4433 7440,\nn
b^{(2)}_{22}&=& 8004 7544 5727 9594 4069.\ea

\section{\`A l'ordre de $(\alpha^{\prime})^3$}

Pour le cas d'un trou noir dyonique non-supersym\'etrique extr\'emal \cite{AshokeSen},
l'entropie est modifi\'e par les corrections de $\alpha^{\prime}$ \`{a}:

\ba S_{BH}^{ns}&=& 2 \pi \sqrt{nw N W}+ \frac{5 \pi \widehat{\alpha}}{4} \sqrt{\frac{nw}{N W}}
               - \frac{29 \pi \widehat{\alpha}^2}{64} \frac{\sqrt{nw}}{(NW)^{3/2}}
               - \frac{119 \pi \widehat{\alpha}^3}{512} \frac{\sqrt{nw}}{(NW)^{5/2}}.\ea

Avec cet entropie d'un trou noir dyonique non-supersym\'etrique extr\'emal, 
les coeficients du d\'eterminant de la m\'etrique tenseure de Ruppenier sont donn\'ees par:

\ba
a^{(3)}_0&=& 1 0995 1162 7776,\nn 
a^{(3)}_1&=& 1 3743 8953 4720,\nn
a^{(3)}_2&=&1 9928 6482 5344, \nn
a^{(3)}_3&=& 1 3915 6940 3904,\nn
a^{(3)}_4&=& 1 2788 8698 4288,\nn 
a^{(3)}_5&=& 2 2553 2759 2548,\nn
a^{(3)}_6&=& 1 4648 6067 2000,\nn 
a^{(3)}_7&=& 4611 0788 8128,\nn
a^{(3)}_8&=& 6346 0873 4208,\nn
a^{(3)}_9&=& 6655 7637 5296,\nn
a^{(3)}_{10}&=& 586 9224 7040, \nn
a^{(3)}_{11}&=& 1368 3491 0800,\nn
a^{(3)}_{12}&=& 250 6674 0125. \ea

Bien aussi, il n'est pas difficile de voir que les coeficients 
de la num\'erateur de la courbure scalaire de Ruppenier sont donn\'ees par:

\ba
b^{(3)}_{0}&=& 6 4903 7107 3168 5345 3566 3120 4115 2512,\nn
b^{(3)}_{1}&=& 24 3388 9152 4382 0045 0873 6701 5432 1920,\nn
b^{(3)}_{2}&=& 66 7798 3362 0023 1248 7084 7224 8592 0768,\nn
b^{(3)}_{3}&=& 85 4396 5045 5382 6616 6087 7796 0423 4240,\nn
b^{(3)}_{4}&=& 63 1991 1781 5190 9481 3330 6106 0247 7216,\nn
b^{(3)}_{5}&=& 87 0878 9334 1288 7883 7408 9485 9991 4496,\nn
b^{(3)}_{6}&=& 182 3010 0923 5277 2683 9673 0061 0903 2348,\nn
b^{(3)}_{7}&=& 228 0981 9400 5851 9965 9705 7788 6701 1584,\nn
b^{(3)}_{8}&=& 27 4635 1837 7630 4920 7854 0793 8333 9008,\nn
b^{(3)}_{9}&=& 169 3751 2378 0501 3605 8388 6957 5919 2064,\nn
b^{(3)}_{10}&=& 305 1768 8854 2439 9126 3286 9363 6464 6400,\nn
b^{(3)}_{11}&=& 25 0873 7031 6500 8481 6936 4152 6267 0848,\ea

\ba
b^{(3)}_{12}&=& 96 2929 7583 9805 3950 5359 1694 7774 8496,\nn
b^{(3)}_{13}&=& 252 1165 6500 3280 8547 1145 7768 5381 1200,\nn
b^{(3)}_{14}&=& 54 0781 9854 5303 1999 8576 7887 2897 1264,\nn
b^{(3)}_{15}&=& 37 4988 2437 8843 4247 7964 0406 5927 1680,\nn
b^{(3)}_{16}&=& 138 5175 0383 6273 5282 3985 2205 9312 3328,\nn
b^{(3)}_{17}&=& 57 6632 8052 2380 5963 0405 4615 3404 8256,\nn
b^{(3)}_{18}&=& 10 3834 4759 8708 6011 2999 1237 4078 6688,\nn
b^{(3)}_{19}&=& 46 6104 8705 5018 9443 4971 8726 3538 7904,\nn
b^{(3)}_{20}&=& 29 6184 6280 7601 9769 8290 1721 4831 8208,\nn
b^{(3)}_{21}&=& 2 6205 2090 9141 2466 3980 7734 3550 6688,\nn
b^{(3)}_{22}&=& 5 3214 1532 1569 2767 6742 0732 1902 2848,\ea

\ba
b^{(3)}_{23}&=& 6 5302 6803 8150 8055 6684 8244 6083 1680,\nn
b^{(3)}_{24}&=& 9046 8645 2914 0016 4777 5025 4681 5360,\nn
b^{(3)}_{25}&=& 1 3400 2466 4792 2433 5657 8013 2600 2176,\nn
b^{(3)}_{26}&=& 902 1455 7271 5446 7926 5653 9090 9440,\nn
b^{(3)}_{27}&=& 402 0769 2677 8265 0664 8790 6614 7200,\nn
b^{(3)}_{28}&=& 3143 0414 3791 2341 9878 6842 5420 8000,\nn
b^{(3)}_{29}&=& 820 0267 0683 0856 9622 3178 7520 0000,\nn
b^{(3)}_{30}&=& 284 7141 1796 7361 4813 1813 2080 0000,\nn
b^{(3)}_{31}&=& 191 0264 9283 2515 4051 3125 7800 0000,\nn
b^{(3)}_{32}&=& 37 6738 9244 9000 2189 2992 8656 2500,\nn
b^{(3)}_{33}&=& 2 6471 3816 8062 7221 0661 4023 4375.
\ea

\section{\`A l'ordre de $(\alpha^{\prime})^4$}

Pour le cas de ces trous noirs dyoniques non-supersym\'etriques \cite{AshokeSen},
la prochaine correction de $\alpha^{\prime}$ donne que l'entropie est modifi\'e \`{a}: 

\ba S_{BH}^{ns}&=&
2 \pi \sqrt{nw N W}+ \frac{5 \pi \widehat{\alpha}}{4} \sqrt{\frac{nw}{N W}} 
- \frac{29 \pi \widehat{\alpha}^2}{64} \frac{\sqrt{nw}}{(NW)^{3/2}},\nn &&
- \frac{119 \pi \widehat{\alpha}^3}{512} \frac{\sqrt{nw}}{(NW)^{5/2}} 
- \frac{2237 \pi \widehat{\alpha}^4}{16384} \frac{\sqrt{nw}}{(NW)^{7/2}}.\ea

Cette fois, on voit que les coeficients du d\'eterminant de la m\'etrique 
tenseure de Ruppenier sont donn\'ees par:

\ba 
a^{(4)}_0&=& 11 52921 5046 0684 6976, \nn
a^{(4)}_1&=& 144 1151 8807 5855 8720,\nn
a^{(4)}_2&=& 208 9670 2270 9991 0144, \nn
a^{(4)}_3&=& 97 9532 9189 5308 2880,\nn
a^{(4)}_4&=& 23 4187 1806 2326 5792, \nn
a^{(4)}_5&=& 147 9423 6814 9818 5728,\nn
a^{(4)}_6&=& 251 0022 3184 9169 7152, \nn 
a^{(4)}_7&=& 219 9463 9535 5630 7968,\nn
a^{(4)}_8&=& 81 2853 4093 7060 3220, 
\ea
\ba 
a^{(4)}_9&=& 124 6334 4349 0752 1024,\nn
a^{(4)}_{10}&=& 121 5464 5516 1346 4576, \nn 
a^{(4)}_{11}&=& 72 6005 3312 3169 8944,\nn
a^{(4)}_{12}&=& 03 4314 9504 4405 6576, \nn 
a^{(4)}_{13}&=& 221 5248 5891 4376 0896,\nn
a^{(4)}_{14}&=& 15 6587 2498 3663 8208, \nn
a^{(4)}_{15}&=&  4 5952 9771 1139 0272,\nn
a^{(4)}_{16}&=&  8589 3056 3153 2423.\ea

Et \`a la fin, les coeficients de num\'erateur de la courbure
scalaire de Ruppenier sont donn\'ees par les chiffres suivants:

\ba 
b^{(4)}_0&=& 46 7680 5239 4588 8933 8251 7914 6469 2105 6628 9898 4137 5232,\nn
b^{(4)}_1&=& 175 3801 9647 9708 3501 8444 2179 9259 5396 2358 7119 0515 7120,\nn
b^{(4)}_2&=& 481 1994 1409 1199 7858 1856 3231 1718 3618 4221 7157 8977 4848,\nn
b^{(4)}_3&=& 615 6575 6462 5642 8542 9330 2235 7817 3422 2030 0615 8372 8640,\nn
b^{(4)}_4&=& 637 3846 5635 1705 6815 6265 2523 0946 6388 6491 2161 6558 5792,\nn
b^{(4)}_5&=& 060 8207 3317 6971 7744 7141 7808 3021 2078 5742 8143 6563 5136,\nn
b^{(4)}_6&=& 0077 0489 5220 8353 0023 9425 5318 3084 6083 8088 2890 9315 6864,\nn
b^{(4)}_7&=& 044 3744 18634 2635 0784 5588 5344 7791 5049 2330 2656 2909 1840,\nn
b^{(4)}_8&=& 0864 0403 8520 1805 0678 8213 0888 1019 4503 7607 7622 0010 0864,\nn
b^{(4)}_9&=& 0855 7439 6207 2769 3991 8376 2667 3505 7262 6065 2615 8209 0240,\nn
b^{(4)}_{10}&=& 0802 5292 3676 2125 0019 8815 9352 9158 4658 0454 8298 5481 8304,\nn
b^{(4)}_{11}&=& 2026 7751 9877 3356 9902 6355 6464 2803 4839 6045 5207 8144 3072,\ea

\ba
b^{(4)}_{12}&=& 2096 7319 0686 2568 6996 5171 7922 2594 4442 9118 8371 8243 1232,\nn
b^{(4)}_{13}&=& 2324 1098 8727 3868 8387 1289 4878 6934 7660 8217 5460 9139 7120,\nn
b^{(4)}_{14}&=& 2797 7777 6981 1171 0644 0613 0693 9116 2563 9461 5185 4630 5024,\nn
b^{(4)}_{15}&=& 3622 3076 0347 0438 7241 8859 0741 4014 2801 8893 7059 4240 1024,\nn
b^{(4)}_{16}&=& 3665 6983 5641 8889 1185 7921 9304 5238 1961 1771 2815 7734 5024,\nn
b^{(4)}_{17}&=& 3624 3893 1153 2945 8637 1615 3745 0370 7179 4279 1692 0809 0624,\nn
b^{(4)}_{18}&=& 3831 6216 4226 9134 7828 7278 0673 1108 2229 4119 9176 2082 2016,\nn
b^{(4)}_{19}&=& 3843 0120 3393 1158 9551 9296 3205 0783 9121 2165 8609 6889 0368,\nn
b^{(4)}_{20}&=& 3450 8885 7553 7611 0677 1727 3382 6300 7177 7087 8737 0609 8688,\nn
b^{(4)}_{21}&=& 2920 0946 8895 0407 2025 7668 1898 5687 8736 0082 0460 1262 0800,\nn
b^{(4)}_{22}&=& 2530 3141 8124 4638 6884 4318 5152 8398 5536 8523 6186 1083 9552,\ea

\ba b^{(4)}_{23}&=& 2060 2625 2186 2868 2091 7408 7375 5085 3496 4309 2865 9165 9032,\nn
b^{(4)}_{24}&=& 1506 5023 6220 6011 4521 8212 3478 3663 0405 7101 9567 9087 8208,\nn
b^{(4)}_{25}&=& 984 0725 3755 0755 4640 1533 4416 8683 0870 6081 4123 4261 1968,\nn
b^{(4)}_{26}&=& 621 7654 7831 2283 2299 1153 3088 8284 9549 2879 0787 8397 5424,\nn
b^{(4)}_{27}&=& 352 0839 6338 0280 0940 6249 7709 8873 3631 7749 3576 6819 7376,\nn
b^{(4)}_{28}&=& 146 3307 5237 3459 7796 9834 9421 0156 6600 3863 5632 9001 7792,\nn
b^{(4)}_{29}&=& 09 1281 7225 9270 6872 2150 7807 6554 4839 9941 0118 5327 9232,\nn\
b^{(4)}_{30}&=& 48 0233 1224 4500 1567 0818 0631 3823 6600 7707 7505 4548 9920,\nn
b^{(4)}_{31}&=& 56 3703 8254 8564 1305 8198 1727 3284 2451 6260 0040 0357 7856,\nn
b^{(4)}_{32}&=& 46 3558 2316 8213 2717 4053 0182 2495 8787 5436 1482 1813 4520,\nn
b^{(4)}_{33}&=& 33 0832 0664 0482 0653 2474 8063 8831 9683 2504 4436 6459 6992,\ea

\ba b^{(4)}_{34}&=& 19 3217 8001 8124 8191 6067 8130 8012 4645 1249 3275 4210 8160,\nn
b^{(4)}_{35}&=& 7 9201 3812 8575 9412 6564 9284 6735 2314 8226 0805 9105 2800,\nn
b^{(4)}_{36}&=& 10823 0269 7430 4996 8728 4237 3130 2591 3398 8387 2754 0736,\nn
b^{(4)}_{37}&=& 1 3187 7098 4813 4846 4275 1201 2104 8810 6765 9191 3546 5472,\nn
b^{(4)}_{38}&=& 1 3148 0704 0648 9078 4909 0453 1953 2734 2265 9650 4797 1840,\nn
b^{(4)}_{39}&=& 6999 3602 0368 2980 9359 8774 3379 3530 3126 5715 7383 7824,\nn
b^{(4)}_{40}&=& 2594 4277 4629 6833 9700 1379 1616 1498 0699 0083 3498 7264,\nn
b^{(4)}_{41}&=& 698 4259 2165 7597 2484 0743 2711 7849 1287 1037 0650 5216,\nn
b^{(4)}_{42}&=& 135 2878 1821 2951 6853 7413 7648 1347 7641 4106 5624 5504,\nn
b^{(4)}_{43}&=& 17 3203 1270 4434 9843 7016 1437 7821 5638 4563 4977 9456,\nn
b^{(4)}_{44}&=& 1214 0355 3809 5997 45266 7043 4591 8335 8714 9652 5839.\ea

\clearpage


\clearpage
\end{document}